\documentclass[12pt]{article}
\usepackage[section] {placeins}
\usepackage{amssymb}
\usepackage{amsmath}
\usepackage{graphicx}
\usepackage{subfigure}
\usepackage{color}
\usepackage{ifpdf}
\numberwithin{equation}{section}

\begin{document}

\title{Anomalous Localization in Low-Dimensional Systems
with Correlated Disorder}

\author{F.~M.~Izrailev${}^{1}$, A.~A.~Krokhin${}^{2}$ and N.~M.~Makarov${}^{3}$ \\
{\it ${}^{1}$ Instituto de F\'{\i}sica, Universidad Aut\'{o}noma de Puebla} \\
{\it Puebla, Pue., 72570, Mexico}\\
{\it ${}^{2}$ Department of Physics, University of North Texas}\\
{\it P.O. Box 311427, Denton, TX 76203}\\
{\it ${}^{3}$ Instituto de Ciencias, Universidad Aut\'{o}noma de Puebla} \\
{\it Priv. 17 Norte No. 3417, Col. San Miguel Hueyotlipan, Puebla 72050, M\'{e}xico}}

\maketitle
\begin{abstract}
This review presents a unified view on the problem of Anderson localization in one-dimensional weakly disordered systems with short-range and long-range statistical correlations in random potentials. The following models are analyzed: the models with continuous potentials, the tight-binding models of the Anderson type, and various Kronig-Penney models with different types of perturbations. Main attention is payed to the methods of obtaining the localization length in dependence on the controlling parameters of the models. Specific interest is in an emergence of effective mobility edges due to certain long-range correlations in a disorder. The predictions of the theoretical and numerical analysis are compared to recent experiments on microwave transmission through randomly filled waveguides.
\end{abstract}



\tableofcontents



\newpage

\section{Introduction}
\label{Intro}

To date, the theory of localization in one-dimensional disordered systems is developed in great detail. Comparatively, one can speak about three types of models thoroughly studied in the literature. The most developed theories refer to the models with weak {\it continuous} potentials for which the unperturbed solution is a plane wave in infinite coordinate space. Here the standard assumptions are the validity of the Born approximation and statistical homogeneity of random potentials. A particular case is the white-noise disorder that can be treated as a good approximation for the realistic situations where the correlation length $R_c$ defined by the binary correlator, is the smallest length scale.

The key quantity for the scattering processes at the continuous potentials is the {\it backscattering length} that emerges from the solution for the Green function of the corresponding wave equation, see, for example, Ref.~\cite{M99}. For one-dimensional systems this length determines the scale on which the propagating wave essentially decreases. The remarkable result is that the same length emerges in the description of the structure of the eigenstates that turn out to be exponentially localized in infinite geometry. This property of eigenstates is the core of famous Anderson localization occurring for {\it any} weak disorder in one-dimensional potentials.

One of the main predictions of the theory of Anderson localization is the {\it single parameter scaling} (SPS) according to which all statistical properties of transport through finite disordered samples are defined by the ratio between the {\it localization length} $L_{loc}$ of eigenstates and the sample size $L$. This highly non-trivial result makes the problem of finding the value of $L_{loc}$ the key problem in the theory of disordered systems. If one knows the localization length $L_{loc}$ (defined in the limit $L\rightarrow\infty$), all transport properties of finite disordered samples of size $L$ can be obtained via the corresponding analytical expressions (see, for example, \cite{LGP88,M99}). For the continuous one-dimensional potentials the SPS is somewhat trivial since according to rigorous analytical results all moments of the transmission coefficient depend solely on the parameter, $L_{loc}/L$.

The second type of disordered systems refers to the {\it tight-binding models} that are essentially discrete models since they are described by finite-difference equations for $\psi_n$-function defined at a set of discrete points $n$. Correspondingly, all properties of such models depend on a {\it site potential} $V_n$. For the uncorrelated $V_n$ all states are known to be exponentially localized in an infinite sample (see, for example, Refs.\cite{MI70a,I73,T74} and references therein). However, a rigorous analysis of the tight-biding models meets serious mathematical problems due to the existence of an infinite set of energy values (so-called {\it resonant energies}) for which the distribution of the phase of the wave function is singular in the absence of disorder.

So far, a rigorous analysis of scattering properties in tight-binding models is performed for the non-resonant values of energy and for few lowest resonances. Thus, still there is no general analytical solution of the tight-binding Anderson model with weak disorder even for basic quantities such as the mean transmission coefficient and its variance. For this reason, in literature the SPS is often treated as a {\it hypothesis} to which a plenty of studies is devoted, both analytical and numerical, see, for example, \cite{P86,DLA00,DLA01,DEL03,DEL03a,DEL03b,DELA03,ST03,TS03,TS05,A06}. Still, the problem of the SPS for tight-binding models remains open. There are analytical and numerical data demonstrating that at least at the center of energy band, $E=0$, (and in its vicinity) the SPS is violated due to failure of the Born approximation. This fact can be treated as an indication that for other resonances the SPS is also not valid. An additional problem arises when studying the vicinity of the band edges,  where analysis of the localization length requires an application of quite sophisticated methods, see Refs.~\cite{HL89,IRT98,DG84,GNS93,GNS94,TI00,DEL03,DEL03a}.

The third type of one-dimensional models is related to the famous Kronig-Penney model. The main difference from the above two types of models is that in the models of Kronig-Penney type the potential is periodic in the absence of disorder. Therefore, all eigenstates are extended Bloch states characterized by the Bloch number. This fact results in an infinite set of allowed energy bands, in contrast with one {\it infinite} band for $0<E<\infty$ for the models with continuous potentials, and with one {\it finite} band for $-2 < E < 2$ (in rescaled units) for the tight-binding Anderson model. This difference results in many specific scattering properties of the Kronig-Penney models.

One of the problems is the influence of weak disorder that can be imposed by random variations of the parameters in some of the versions of the Kronig-Penny model. For example, it could be the fluctuations of widths of barriers and their amplitudes, or the fluctuations of refractive index inside the barriers, etc. One can see that with weak disorder such models can be treated as {\it periodic on average}, the term typically used in the literature. The important consequence of the underlying periodic structure of Kronig-Penney models is the presence of resonances of the Fabry-Perrot type. These resonances survive in the presence of weak disorder, thus making the problem of finding the localization length quite complicated. Note that such {\it regular} resonances are principally different from those emerging {\it randomly} in disordered continuous or discrete potentials (see, for example, \cite{A83,AS83,LGP88,G05,So07a,Bo08}).

Since the early studies of the disordered tight-binding systems it is believed that all eigenstates are localized in one dimension, whatever the magnitude and type of the disorder is (see, for example, review \cite{I73} and references therein). This belief is based on an enormous number of numerical results, although only for particular models with certain types of disorder it was analytically proved that all eigenstates are, indeed, exponentially localized. For a long time the analysis of Anderson localization was mainly restricted by the systems with white-noise disorder. Until the late 1990s random potentials with statistical correlations (colored noise) were studied scarcely, probably, due to the seeming lack of physical applications. Besides the white-noise disorder, some quasi-periodic potentials have been studied numerically and analytically, demonstrating a possibility of metal-insulator transition (see Ref.~\cite{BFLT82} and references therein). One should also mention the studies \cite{GF88,SHX88,T88} of the tight-binding models with pseudo-random potentials defined by deterministic functions with slowly varying period. It was shown that depending on how slowly the period changes one can observe the transition to the situation when the eigenstates are extended. Another case was suggested in Ref.~\cite{C89} where some square-well-like potentials with various types of correlations were studied both analytically and numerically in the tight-binding model. It was found that under some condition imposed on the deterministic correlations, the transition from localized to delocalized eigenstates emerges. Note, however, that all these results refer to the deterministic potentials with a kind of pseudo-randomness, however, not to truly random potentials. The conditions under which the delocalization emerges in deterministic and random potentials has been discussed in Ref.~\cite{GC09} (see also references therein.)

To the best of our knowledge, a detailed study of a statistically correlated disorder was started in Ref.~\cite{JK86}. The authors asked a question concerning the dependence of the localization length on the type of correlations in disordered potentials. The analytical and numerical analysis for few types of correlations has led to the conclusion that the localization length generally increases when the correlation function is positive. However, in the presence of negative correlation the localization length may be shorter than in the uncorrelated case. These results have initiated a further analysis of random potentials with underlying correlations, see, for example, Refs.~\cite{F89,C89}. The principal question of these and the following studies was regarding a possibility of coexistence of the localized states in the energy spectrum together with delocalized (or extended) states, resulting from statistical correlations in random potential. In particular, in Ref.~\cite{F89} the tight-binding model with both diagonal and off-diagonal disorder was considered, and it was shown that for specific correlations between these two types of disorder, there are eigenstates that are, indeed, delocalized at some energy.

The burst of interest to random potentials with specific correlations was triggered  by the observation that certain {\it short-range} correlations result in an emergence of extended states at {\it discrete} values of energy \cite{DWP90}. The correlations were introduced in the standard tight-binding model with the binary potential where the potential at $n$-th site randomly takes one of two values, $\epsilon_1$ or $\epsilon_2$ (the so-called {\it dimer model}). To enforce the short-range correlations, the next $(n+1)$-th site also takes the same value. Thus, the non-zero correlations emerge between adjacent $n$-th and $n+1$-th sites only. This type of correlations was suggested in order to explain anomalously high conductivity observed in certain organic polymers \cite{PW91}. Although truly extended states emerge for two discrete values of energy only, in a finite size samples the bands of effectively delocalized states appear in the vicinity of these resonant energies. Here the localization length can be larger than any sample size. A practical implementation of the dimer model was done in semiconductor superlattices \cite{Bo99}, thus conforming a feasibility of the effect of short-range correlations. Similar models with short-range correlations have been analyzed in Refs.\cite{DSD94,DSD94a,DSD95} in connection with their practical realizations in semiconductor superlattices (see also discussion in Ref.~\cite{G05,G05a}).

Since the early studies of the random dimer models in Refs.~\cite{ESC88,CFPV89,DWP90,WP91,WP91a,B92,EK92,GS92,DGK93a,DGK93,EW93,P93,EE93,FH93}, the short-range correlations were also considered in Kronig-Penney models \cite{SMD94,SDBI95},  as well as for other models, see, for example, Refs.~\cite{WGP92,IKT95,IKT96,HWG97,S04,H97,KTI97,Do00,LO01,LO01a,Ho01,Fo03}.
In all above models the effect of short-range correlations results in a discrete set of isolated extended states of null measure.

Nowadays, the growing interest in the theory of disordered systems is related to {\it long-range} correlations. In Refs.~\cite{ML98,ML99,ML00,CBIS02,ZX02,SNN04,CBI04,DRDM05,NYS09} the disordered potentials constructed from different {\it self-affine} sequences have been studied numerically, with a clear indication of an emergence of delocalized states in a finite range of energies. Note, however, that the fluctuations in such potentials increase with the systems size thus making self-affine potentials very difficult for a rigorous analysis (see discussion in Refs.~\cite{Ko00,RKBH01,L05}).

Although a general expression for the localization length in one-dimensional continuous potentials is known to be expressed in terms of the binary correlator \cite{LGP88} (for any kind, however, weak disorder), the problem of the effects produced by long-range correlations in tight-binding models has remained open for a long time. As mentioned above, a rigorous analysis of both the standard Anderson model and the models of the Kronig-Penney type meets serious problems because of an infinite set of energies for which standard perturbation theory is not valid.  For this reason the problem of long-range correlations in such models requires special approaches. One of these approaches was suggested in Refs.~\cite{IK99,KI99} where it was shown how to derive the expression for the localization length in the standard Anderson model for potentials with any kind of correlations. As a result, the conclusion was made that specific long-range correlations in random potentials can give rise to an appearance of the {\it effective mobility edges} for 1D finite structures. In particular, an interval of electron energy (or wave frequency) arises on one side of the mobility edge where the eigenstates turn out to be extended, in contrast to the other side where the eigenstates remain to be exponentially localized. The position and the width of the windows of transparency can be controlled by the form of the binary correlator of a scattering potential. A quite simple {\it convolution method} was suggested for constructing random potentials that result in any predefined energy window of a perfect transparency.

We have to stress that here and below by the mobility edge we mean the situation when on one side of this "edge", the inverse localization length $L_{loc}^{-1}$ (that is associated with the Lyapunov exponent) is determined by the first order of perturbation theory and, therefore, is proportional to the variance $V_0^2$ of a weak disordered potential. However, on the other side $L_{loc}^{-1}$ abruptly decreases to the value determined by the next-order terms which are proportional to the squared variance, $V_0^4$, or higher. Thus, there is no conflict with the common statement that in {\it infinite} samples the localization length is always finite in a low-dimensional geometry with random potentials. As one can see, a separate problem of the theoretical and experimental interest (see Refs.~\cite{T02,GK09,Lo09}) refers to the next orders of perturbation theory, if in the first order the Lyapunov exponent vanishes within some energy region.

The feasibility of (effective) mobility edges has been verified experimentally \cite{KIKS00,KIKSU02,KIK08} by studying transport properties of a single-mode electromagnetic waveguide with point-like scatterers. These scatterers were intentionally inserted into the waveguide in order to provide an arrangement of random potential with a given binary correlator. In the experiment the disorder was created by an array of 100-500 screws with random heights obtained numerically in accordance with the analytical expressions \cite{IK99,KI99}. The agreement between experimental and numerical data was found to be very good. In particular, the predicted windows of a complete reflection alternated by those of a high transparency were clearly observed, in spite of many experimental imperfections.

The analytical method used in Refs.~\cite{IK99,KI99} is based on the representation of the original problem (either of the propagation of electormagnetic waves in waveguides and disordered photonic crystals, or of the electron transport in superlattices) in the form of classical Hamiltonians describing linear oscillators with a noisy frequency. In such a description the disordered potential of an initial problem is treated as a parametric noise perturbing a harmonic oscillator. Thus, the problem of finding the localization length reduces to the calculation of the increment of parametric instability expressed via classical Lyapunov exponent. In fact, this method can be considered as a kind of the transfer matrix approach widely used in the theory of disordered systems \cite{LGP88}. However, in many cases the Hamiltonian approach is more transparent and simple in comparison with standard transfer matrix methods. In particular, a possibility to treat the product of local transfer matrices as the trajectory in the classical phase space allows one to use the canonical transformation to new variables in which the analytical derivation of the Lyapunov exponents turns out to be strongly simplified for the standard Anderson model  \cite{IRT98,TI00,TI01,IDKT04,DIK04,DIK04a,K07a,K09}, for the dimer model  \cite{IKT95}, for delta-like Kronig-Penney models \cite{KTI97,IKU01,HIT10,HIT10a} and for the Kronig-Penney models with finite barriers \cite{IM09,IMT10}.

Apart from an obvious theoretical interest to the correlated disorder, a remarkable progress in manufacturing various devices with controllable low-dimensional potentials has triggered the study of anomalous transport that arises due to long-range correlations in random potentials. In the first line one has to mention a possibility to strongly suppress or enhance the localization length in desired windows of energy spectra, which is important for the creation of selective transmission and reflection of waves of different nature. It should be stressed that the analytical results obtained with the use of perturbation theories turn out to be very robust with respect to many experimental imperfections, such as a relatively small number (100-200)  of scatterers (or finite width barriers) and the presence of absorption which can not be avoided in practice. This fact makes the perturbation theory a quite good tool in the analysis of anomalous transport characteristics for the wave (electron) scattering through finite samples with statistically correlated random potentials.

In view of experimental applications one of the theoretical problems is the influence of disorder in the systems which are periodic on average rather than homogeneous \cite{MCMM93}. The origin of a disorder could be due to unavoidable fluctuations of the width of layers, or due to the variation of medium parameters such as the dielectric constant and magnetic permeability, or the barrier hight (for electrons) (see, for example, Refs.~\cite{MCMM93,So00,SSS01,Po03,Eo06,No07,Po07}). A particular attention in recent years was paid to the periodic structures with elementary cells consisting of two components (see, for example, Ref.~\cite{MS08} and references therein). In particular, each of the cells can be composed by two optical or electromagnetic materials (or metamaterials) of finite width, or by a pair of quantum wells and barriers in electronics \cite{Ao07,Ao10,Ao10a,Mo10,Mo10a}. The interest to such {\it bi-layer structures} is related to the creation of semiconductor superlattices (or the lattices with materials and metamaterials) with given transmission properties.

Experimentally, localization of light has been observed in diffusive media \cite{WBLR97,SGAM06} and photonic crystals \cite{CSG00,SBFS07,Lo08}, and localization of sound - in disordered elastic media \cite{HM86,Ho08,Fo09}. As for the experiments with correlated disorder, one has to mention Refs. \cite{KIKS00,KIKSU02,KIK08,Do11} where the systematic study of localization effects has been performed in microwave structures. In particular, a strong localization of microwaves at distances shorter than the localization length for white-noise disorder has been observed due to specific long-range correlations in a disordered array of relatively small number of scatterers [KIK08]. Recently, the Anderson localization was found in the experiments with ultracold atoms (see, for example, Refs. \cite{So07,Bo08,Co08,Ro08,Lo09,M10} and references therein). A distinctive peculiarity of these experiments is that long-range correlations in experimentally induced disorder appear naturally. The reason is that speckle potentials are produced with the use of finite-size diffusive plates \cite{G75}. As a result, the effective mobility edges have been observed in these experiments.

The structure of this review is as follows. First, the general theory of the transport in one-dimensional {\it continuous} systems with random potentials is shortly presented in Section 2. The main attention is paid to the rigorous derivation of basic relations for transport characteristics that are known in literature, however, distributed in different publications. One should stress that this derivation includes possible correlations that may be present in a disorder. Here, we formulate the inverse problem of finding the random potential with prescribed correlation properties and introduce the continuous version of the convolution method allowing to solve this problem. Then we compare the features of widely used Gaussian correlations with those of long-range correlations that give rise to effective mobility edges. In the next Section 3 we shortly consider the quasi-one dimensional models with a disorder depending on the longitudinal coordinate only. In this specific case the solution of the scattering problem reduces to the analysis of the transport in each transverse channel along which the propagation is independent. In this way all transport characteristics can be understood via the combination of partial transmission coefficients obtained for individual channels. The main interest in this model is that with specific long-range correlations some channels can be intentionally closed and other open. The very recent experiments \cite{Do11} in quasi-one-dimensional electromagnetic waveguides confirm the theoretical predictions discussed in this Section.

The standard tight-binding Anderson model is analyzed in Sections 4-7. First, we start with a white-noise disorder (Section 4) and present main results for the localization length valid inside in the energy band. With the use of the Hamiltonian map approach we show how relatively easy one can treat the anomalies at the center of energy band as well as close to the band edges. In the second part of Section 4 we focus on such transport characteristics as the transmittance and resistance in the ballistic and localized regimes. All results are compared with those obtained in Section 2 for continuous potentials. Specific attention is paid to the point that the localization problem in disordered potentials can be exactly mapped onto the classical problem of instability of linear oscillators with a noisy parametric force. This analogy is directly used in Section 6 where the relevance of the tight-binding model to the model with continuous potentials is discussed in detail.

The correlated disorder in tight-binding models is considered in Section 6. Here we derive the basic expression for the Lyapunov exponent valid for statistical correlations of any kind. Some examples of both short-range and long-range correlations are given in order to demonstrate a principal difference between these two types of correlations. Specifically, it is shown that long-range correlations can result in an emergence of effective mobility edges. Shorty we discuss what is known about higher order terms in the weak disorder expansion of the Lyapunov exponent. At the end of this Section we show how the inverse problem of constructing the potentials with given properties of the Lyapunov exponent can be numerically solved. The discrete version of the convolution method we present here turns out to be quite effective in experimental realizations of a disorder with specific long-range correlations. In Section 7 we briefly discuss the problem of off-diagonal disorder, paying the main attention to the correlations between diagonal and off-diagonal random entries.

Next part of our review (Sections 8-12) is devoted to the models of the Kronig-Penny type. First, in Section 8 we consider the models with delta-like barriers that can be formally reduced to tight-binding models, however, with infinite number of energy bands. In this Section we analyze the general properties of the Lyapunov exponent by deriving the expression which is valid everywhere in the energy band, including the center of energy bands and the vicinity of band-edges. Two types of weak disorder are mainly discussed here, the amplitude and positional disorder, that typically emerge in experimental situations. The predictions of theoretical and numerical data are compared in Section 9 with experimental results obtained in single-mode waveguides with intentionally inserted scatterers. The remarkable result is that in spite of many experimental imperfections, the correspondence between the theory and experiment turned out to be unexpectedly good. The use of specific long-range correlations has allowed to observe an emergence of strongly localized states in the case of weak disorder and small number of scatterers.

Another practically important Kronig-Penney model with finite barriers was analyzed in Section 9, both theoretically and experimentally. It was shown that the theory gives a correct description of transport characteristics in the device with alternating air-teflon bars in the circular geometry with only 26 teflon barriers. The experimental data manifest an emergence of quite unexpected resonances that were explained theoretically.

In Section 11 we discuss recent results obtained for the bi-layer arrays consisting of two types of barriers emerging in the potential in pairs. The analysis was performed for the case when both types of barriers are slightly perturbed in their widths. The theoretical treatment was done for the general case of any kind of correlations inside both types of disorder, as well as for the correlations between two disorders. The analytical expression for the Lyapunov exponent is derived with the use of the Hamiltonian map approach which was found to be very helpful. The specific attention in this Section is paid to possible applications, namely, to the photonic crystals, metamaterials and semiconductor superlattices.

A quite interesting model that recently attracted much attention, was considered in Section 12. This model represents a stack with different widths of basic slabs consisting of either material-material or material-metamaterial layers of finite width. It was recently found \cite{Ao07,Ao10,Ao10a} that in the presence of weak fluctuations of impedances the localization length diverges with a decrease of frequency in an anomalous way for material-metamaterial slabs. The Hamiltonian map approach has allowed to explain what is the basic mechanism of such an anomaly. We show that in the case of equal widths of both types of layers (and only in this case) the Lyapunov exponent vanishes in the Born approximation. Due to this fact the dependence of the localization length on the model parameters (described beyond the Born approximation) turns out to be a very specific.

The next two sections 13 - 14 are related to the anomalous electron transport in DNA demonstrating the windows of high transparency in the frequency domain. In order to see whether possible long-range correlations in DNA may lead to this effect, the theoretical analysis of the tight-binding model with two chains is discussed in Section 13. It is shown that, indeed, for the parameters of the models taken from the experiment, the frequency dependence of the Lyapunov exponent has a quite specific form, which indicate that in DNA the long-range correlations may play an important role in the electron transmission. In Section 14 a very brief summary of the study of statistical correlations in DNA is done.

In Section 15, we shorty discuss the localization in self-affine potentials with long-range correlations. As shown in many numerical studies, for this type of potentials the mobility edges can arise irrespectively of the strength of disorder. This fact is quite interesting from both the theoretical viewpoint and due to possible experimental realizations. In last Section 16 we shortly discuss main results obtained for the disordered potentials with Bloch-like oscillations. Main attention is paid to the prediction that such oscillations can emerge in random potentials with long-range correlations resulting in effective mobility edges.

\section{One-dimensional random potentials: Continuous systems}
\label{2}

\subsection{Description of the model}
\label{2.1}

The model we discuss in this Section is a one-dimensional (1D) isotropic Fermi-liquid model for an electron in the static potential $V(x)$ described by stationary Hamiltonian,
\begin{equation}\label{1DCP-Ham}
{\cal H}=-\frac{\hbar^2}{2m}\,\frac{d^2}{dx^2}+V(x).
\end{equation}
Here $\hbar$ is the Planck constant and $m$ is the electron effective mass. We assume the potential $V(x)$ to be a \emph{continuous} and homogeneous random process, determined by the statistical properties,
\begin{equation}\label{1DCP-VCor}
\langle V(x)\rangle=0,\qquad\langle V^2(x)\rangle=V_0^2,\qquad\langle V(x)V(x')\rangle=V_0^2\,K(x-x').
\end{equation}
Hereinafter, the angular brackets $\langle\ldots \rangle $ stand for the ensemble averaging over the disorder, i. e. over different realizations of the random function $V(x)$, or for a spatial average over the coordinate $x$ of any prescribed realization. These two types of averaging are assumed to be equivalent due to ergodicity. The function $K(x-x')$ is the \emph{binary correlator} normalized at $x=x'$ to one, $K(0)=1$. It is characterized by the \emph{correlation length} $R_c$ that is the scale on which the correlator effectively decreases. In the limit of extremely small value of $R_c$ the potential $V(x)$ can be regarded as the Gaussian white noise. As a result, the two-point correlator can be replaced by the Dirac delta-function,
\begin{equation}\label{1DCP-Wdelta}
K(x-x')\to R_c\delta(x-x')\qquad\mathrm{when}\quad R_c\to0,\quad V_0^2R_c=\mathrm{const}.
\end{equation}
For example, this is the case of electron-impurity scattering. In Eq.~\eqref{1DCP-Wdelta} we have emphasized that within a continuous model the formal transition to the white noise, or delta-correlated disorder, has to be performed for a fixed value of the product $V_0^2R_c$ in order to conserve the finite value of the integral $\langle V(x)V(x')\rangle$. In reality this means that the white noise limit asymptotically occurs for finite values of $V_0$, provided $R_c$ is much smaller than any other length. The variance $V_0^2$ of the potential $V(x)$ does not depend on the coordinate $x$, while the correlator $K(x-x')$ depends only on the distance $|x-x'|$ between $x$ and $x'$. These properties are the direct consequences of the statistical homogeneity of $V(x)$.

In what follows, we consider weak over-barrier scattering, for which random potential $V(x)$ can be considered as a perturbation to the electron energy $E$,
\begin{equation}\label{1DCP-WD}
V_{0}\ll E\,.
\end{equation}
It will be shown that in this case all transport properties are entirely determined by the \emph{randomness power spectrum} ${\cal K}(k_x)$,
\begin{subequations}\label{1DCP-FTW}
\begin{eqnarray}
{\cal K}(k_x)&=&\int_{-\infty}^{\infty}dx\,K(x)\exp(-ik_xx),\\[6pt]
K(x)&=&\int_{-\infty}^{\infty}\frac{dk_x}{2\pi}\, {\cal K}(k_x)\exp\left(ik_xx\right).
\end{eqnarray}
\end{subequations}
Since the correlator $K(x)$ is a real and even function of the coordinate $x$, its Fourier transform ${\cal K}(k_x)$ is an even and real function of the wave number $k_x$. It should be stressed that according to rigorous mathematical theorem, the power spectrum ${\cal K}(k_x)$ is a non-negative function of $k_x$ for \emph{any} real function $V(x)$. This fact ensures a non-negativeness of the attenuation length as well as the Lyapunov exponent, see below. The condition $K(0)=1$ results in the following normalization for ${\cal K}(k_x)$:
\begin{equation}\label{1DCP-Wnorm}
\int_{-\infty}^{\infty}\frac{dk_x}{2\pi}\,{\cal K}(k_x)=1.
\end{equation}

Our main interest is two-fold. First, we are interested in the global spectral properties of electron eigenstates $\psi_E(x)$ satisfied to the Schr\"{o}dinger equation,
\begin{equation}\label{1DCP-psi(E)}
{\cal H}\psi_E(x)=E\psi_E(x),
\end{equation}
on an infinite axis, $-\infty<x<\infty$. Second, we plan to explore the transport properties in 1D conductors of finite length $L$, i.e. in open systems for which the random potential $V(x)$ occupies a domain $-L/2\leq x\leq L/2$, and $V(x)=0$ otherwise. Note that our analysis can be equally applied to the problem of propagation of classical waves in 1D random media. Indeed, with the use of the units with $\hbar^2/2m=1$, the Schr\"{o}dinger equation \eqref{1DCP-psi(E)} can be rewritten in the form,
\begin{equation}\label{1DCP-Schreq}
\left[\frac{d^{2}}{dx^{2}}+k^{2}-V(x)\right]\psi_k(x)=0,
\end{equation}
where $k=\sqrt{E}$ is the wave number for electrons, or wave number $k=\omega\sqrt{\varepsilon_0}/c$ for an electromagnetic wave of frequency $\omega$ propagating in a medium with fluctuating dielectric constant $\varepsilon(x)=\varepsilon_0-\Delta\varepsilon(x)$. In the latter case the scattering potential is $V(x)=k^2\Delta\varepsilon(x)/\varepsilon_0$, and its variance is related to the variance of the dielectric constant, $V_0^2=k^4\langle\Delta^2\varepsilon(x)\rangle/\varepsilon_0^2$.

\subsection{Green function}
\label{2.2}

As known, the single-particle spectral problem for infinitely long 1D structures is convenient to analyze in terms of the retarded Green function ${\cal G}(x,x';k)$ that is governed by the equation
\begin{equation}\label{1DCP-SchrG}
\left[\frac{d^{2}}{dx^{2}}+k^{2}-V(x)\right]{\cal G}(x,x';k)=\delta(x-x')
\end{equation}
with the standard choice of the radiative boundary condition, $|{\cal G}(x=\pm \infty,x')|<\infty$ (see details in Ref.~\cite{M99}). One should keep in mind that for the retarded Green function the wave number $k$ acquires infinitesimally small positive imaginary part, $k \rightarrow k+i0$.

Without disorder, $V(x)=0$, the unperturbed Green function ${\cal G}_0(x,x';k)$ can be shown to have the form,
\begin{equation}\label{1DCP-G0}
{\cal G}_0(x-x';k)=\int_{-\infty}^\infty\frac{dk_x}{2\pi}\,\frac{\exp[ik_x(x-x')]}{k^2-k_x^2}=\frac{\exp(ik|x-x'|)}{2ik}.
\end{equation}
This expression corresponds to the electron eigenfunctions in the form of two plane waves $\exp(\pm ikx)$ traveling in opposite directions.

In order to define the perturbation procedure over the scattering potential $V(x)$, we relate the perturbed Green function ${\cal G}(x,x';k)$ to the unperturbed one \eqref{1DCP-G0} using the Green formula,
\begin{equation}\label{1DCP-GF}
{\cal G}(x,x';k)={\cal G}_0(x-x';k)+\int_{-\infty}^\infty dx_1{\cal G}_0(x-x_1;k)V(x_1){\cal G}(x_1,x';k).
\end{equation}
The integral relation \eqref{1DCP-GF} is a straightforward counterpart of the differential equation \eqref{1DCP-SchrG}. Due to the statistical homogeneity of the random potential, it is sufficient to find the Green function $\langle{\cal G}(x-x';k)\rangle$ averaged over disorder. To this end, we should perform the averaging of Eq.~\eqref{1DCP-GF}. This can be done with different perturbative methods, one of which is based on the diagrammatic technique similar to that introduced by Feynman in quantum electrodynamics (see, e.g., Ref.~\cite{M92}). Another technique which is simpler and more elegant, was developed in Refs.~\cite{MM84,BCHMM85} and discussed in detail in Ref.~\cite{M99}. Both methods allow us to develop a consistent perturbative approach with respect to the random potential $V(x)$ taking into account the effects of multiple scattering. The result of the averaging procedure is the Dyson equation for the average Green function,
\begin{equation}\label{1DCP-Deq}
\langle{\cal G}(x-x';k)\rangle={\cal G}_0(x-x';k)+
\int_{-\infty}^{\infty}dx_1dx_2\,{\cal G}_0(x-x_1;k)\mathcal{M}(x_1,x_2)\langle{\cal G}(x_2-x';k)\rangle.
\end{equation}
The integral operator $\hat{\mathcal{M}}$ that describes the electron interaction with the random potential, is called the \emph{self-energy} or \emph{mass operator}. The exact Dyson equation \eqref{1DCP-Deq} can be solved only for a weak scattering, when the self-energy is written in the first non-vanishing (i.e., quadratic) order in the scattering potential $V(x)$. Specifically, we write,
\begin{equation}\label{1DCP-MBA}
\mathcal{M}_B(x-x')=\langle V(x){\cal G}_0(x-x')V(x')\rangle=V_0^2K(x-x'){\cal G}_0(x-x').
\end{equation}
This approximation is equivalent to the so-called Bourret approximation~\cite{B62} in the diagrammatic technique which includes only the first simplest term of the diagrammatic series for the self-energy. In the scattering theory such an approximation is known as the Born approximation.

Since the integral equation \eqref{1DCP-Deq} contains the difference kernel \eqref{1DCP-MBA}, it can be readily solved by the Fourier transformation. As a result, the average Green function reads,
\begin{equation}\label{1DCP-GAF}
\langle{\cal G}(x-x';k)\rangle= \int_{-\infty}^\infty\frac{dk_x}{2\pi}\,\frac{\exp[ik_x(x-x')]}{k^2-k_x^2-M_B(k)}.
\end{equation}
Here $M_B(k)$ is the Fourier transform of the Born self-energy. Note that actually Eq.~\eqref{1DCP-GAF} contains $M_B(k_x)$ rather than $M_B(k)$. Nevertheless, within the same Born approximation one should regard $M_B(k_x)$ as a small perturbation to the energy $k^2$ in the denominator, $|M_B(k_x)|\ll k^2$. Consequently, the integration variable $k_x$ can be replaced by the wave number $k$ in the argument of the self-energy $M_B$. Taking into account the explicit expression \eqref{1DCP-MBA} for the self-energy, together with the Fourier transforms for the binary correlator \eqref{1DCP-FTW} and unperturbed Green function \eqref{1DCP-G0}, one can get
\begin{equation}\label{1DCP-MBk}
M_B(k)=V_0^2\int_{-\infty}^\infty\frac{dk_x}{2\pi}\,\frac{{\cal K}(k-k_x)}{k^2 - k_x^2}.
\end{equation}

Thus, we have obtained the average Green function $\langle{\cal G}(x-x';k)\rangle$ in the Fourier representation \eqref{1DCP-GAF}. Note that this expression differs from the Fourier representation \eqref{1DCP-G0} for the unperturbed Green function ${\cal G}_0(x-x';k)$ only by the change $k^2\to k^2-M_B(k)$. This remark gives rise to the important conclusion: within the Born approximation the perturbed average Green function is always equal to the unperturbed Green function with the wave number $k$ reduced by the value $M_B(k)/2k$,
\begin{equation}\label{1DCP-GAB-G0}
\langle{\cal G}(x-x';k)\rangle={\cal G}_0(x-x';k-M_B(k)/2k).
\end{equation}
With the use of Eqs.~\eqref{1DCP-G0} and \eqref{1DCP-GAB-G0} one can immediately write down the explicit expression for the perturbed Green function,
\begin{equation}\label{1DCP-GAB}
\langle{\cal G}(x-x';k)\rangle=\frac{\exp[i(k+\delta k)|x-x'|]}{2ik} \exp\left(-\frac{|x-x'|}{L_{ts}}\right),
\end{equation}
\begin{equation}\label{1DCP-dkLts}
\delta k=-\mathrm{Re}M_B(k)/2k,\qquad L_{ts}^{-1}=-\mathrm{Im}M_B(k)/2k.
\end{equation}

As one can see, the electron over-barrier scattering by a random potential gives rise to the complex correction $-M_B(k)/2k$ of the electron wave number in comparison with the unperturbed value $k$. This fact causes the phase renormalization and attenuation of the average Green function \eqref{1DCP-GAB} along the conductor. The real part $\delta k$ of the complex shift results in the disorder-induced modification of the phase of the Green function, and does not play any role in transport properties. To the contrary, the quantity $L_{ts}$ that is specified by the imaginary part of the self-energy $M_B(k)$, is responsible for a strong change of transport properties. The quantity $L_{ts}$ known as the \emph{attenuation length}, determines the degree of localization of eigenstates in random potential, and is of the main interest in the theory of localization.

The detailed analysis shows that the Born approximation used above is valid as long as the attenuation length $L_{ts}$ is much larger than both the electron wave length and the correlation length,
\begin{equation}\label{1DCP-WS}
kL_{ts}\gg1,\qquad\qquad L_{ts}\gg R_c.
\end{equation}
These two inequalities formulate explicitly the conditions of weak scattering. It is clear that the first requirement allows us to think of electrons as ``quasifree" (but not quasiclassical!) particles. The second inequality is, in fact, the necessary and sufficient condition for the statistical approach to the problem of electron interaction with random potential. Note that Eqs.~\eqref{1DCP-WS} do not imply any restriction for the relation between the electron wave length $k^{-1}$ and the correlation length $R_c$.

\subsection{Mean free path}
\label{2.3}

In order to derive the expressions for the phase shift $\delta k$ and attenuation length $L_{ts}$, i.e. to extract the real and imaginary parts of Eq.~\eqref{1DCP-MBk}, it is suitable to employ the Dirac identity,
\begin{equation}\label{1DCP-Dirac}
\frac{1}{k-k_x\pm i0}={\cal P.V.}\frac{1}{k-k_x}\mp\pi i\delta(k-k_x).
\end{equation}
As a result, we get,
\begin{equation}\label{1DCP-deltak}
\delta k=-\frac{V_0^2}{2k}\,{\cal P.V.}\int_{-\infty}^\infty \frac{dk_x}{2\pi}\,\frac{{\cal K}(k-k_x)}{k^2-k_x^2},
\end{equation}
\begin{equation}\label{1DCP-Lts}
\frac{1}{L_{ts}}=\frac{V_0^2}{8k^2}\left[{\cal K}(0)+{\cal K}(2k)\right],
\end{equation}
where ${\cal P.V.}$ stands for the principal value of the integral. Eqs.~\eqref{1DCP-deltak} and \eqref{1DCP-Lts} associate $\delta k$ and $L_{ts}$ directly with random scattering potential via its variance $V_0^2$ and power spectrum ${\cal K}(k_x)$. Since the real shift $\delta k$ does not contribute to the electron transport, here we do not discuss its properties. We note only that $\delta k=0$ for a white-noise disorder because it has constant power spectrum, ${\cal K}(k_x)=R_c$.

Now, let us discuss the meaning of the relation \eqref{1DCP-Lts} for the attenuation length $L_{ts}$. From the quantum scattering theory it is known that the attenuation length $L_{ts}$ of the average single-particle Green function is the ``outgoing'' mean free path of electrons, formed by scattering from a given state into all possible states (including a given one). Since the outgoing mean free path is inversely proportional to the total scattering cross section, the attenuation length $L_{ts}$ is often referred to as the \emph{total mean free path}.

In one-dimensional systems an electron is scattered either forward or backward. For elastic scattering for which the energy of electrons conserves, there are only two possibilities: the scattering without the change of $\vec{k}$ (\emph{forward scattering}) and with the change $\vec{k}$ to $-\vec{k}$ (\emph{backward scattering}). Correspondingly, the change of the wave vector is either zero, $|\Delta\vec k|=0$, or $|\Delta\vec k|=2k$. These two scattering processes give additive contributions to $L_{ts}^{-1}$ in Eq.~(\ref{1DCP-Lts}). Thus,
the total electron mean free path can be rewritten as follows:
\begin{equation}\label{1DCP-LtsLfLb}
L_{ts}^{-1}=L_{fs}^{-1}+L_{bs}^{-1},
\end{equation}
where
\begin{equation}\label{1DCP-LfLb}
L_{fs}^{-1}=\frac{V_0^2}{8k^2}{\cal K}(0),\qquad\qquad L_{bs}^{-1}=\frac{V_0^2}{8k^2}{\cal K}(2k).
\end{equation}
The lengths $L_{fs}$ and $L_{bs}$ account for the forward and backward electron scattering, respectively. Often, the length $L_{fs}$ is called ``forward scattering mean free path", and the length $L_{bs}$ is referred to as the ``backscattering mean free path".

Note that for the white-noise disorder we have $L_{fs}=L_{bs}$. Nevertheless, typically, the backscattering length $L_{bs}$ is much larger than the length $L_{fs}$ because of the relation, ${\cal K}(2k)<{\cal K}(0)$. Therefore, $L_{ts}^{-1}\approx L_{fs}^{-1}$ and the average Green function attenuates along the conductor mainly on the scale of the mean free path $L_{fs}$ of {\it forward} electron scattering. However, as we shall see below, the conductance of 1D disordered structures is fully determined by the {\it backscattering} length $L_{bs}$ alone and \emph{not} by $L_{fs}$.

\subsection{Transport problem: Two-scale approach}
\label{2.4}

Now we consider the transport properties of 1D disordered systems of finite length $L$. As before, the model is described by the same Hamiltonian \eqref{1DCP-Ham}, however, the coordinate $x$ is now confined within the interval
\begin{equation}\label{1DCP-x}
-L/2\leq x\leq L/2.
\end{equation}

Our main interest is in the \emph{dimensionless conductance} or \emph{transmittance} $T_L$ of the system. In the case of the electron transport, $T_L$ is defined as the conductance $G_L$ measured in the units of the conductance quantum $e^2/\pi\hbar$,
\begin{equation}\label{1DCP-Tdef}
T_L=\frac{G_L}{e^2/\pi\hbar},
\end{equation}
where $e$ is the elementary charge.

The transmittance $T_L$ can be obtained with the use of well developed methods, such as the perturbative diagrammatic technique of Berezinski \cite{B73,AR78}, the invariant imbedding method \cite{BW75,K86}, or the two-scale approach \cite{MT98,MT01,M99}. In the latter approach the starting point of calculations is the expression,
\begin{equation}\label{1DCP-TKubo}
T_L=-\frac{4}{L^2}\int_{-L/2}^{L/2}dx\;\int_{-L/2}^{L/2}dx'\,
\frac{\partial{\cal G}(x,x';k)}{\partial x}\,\frac{\partial{\cal G}^*(x,x';k)}{\partial x'}\,
\end{equation}
that follows from the standard linear response theory developed by R.~Kubo~\cite{K57,AR78,E83}. Here the asterisk stands for the complex conjugation. As before, the retarded Green function ${\cal G}(x,x';k)$ obeys the equation \eqref{1DCP-SchrG} in which $k$ is the Fermi wave number, assuming that the electron gas obeys Fermi-Dirac statistics. In order to obtain the average transmittance one has to find the two-point correlator between Green functions. Clearly, this problem is much more difficult in comparison with the derivation of the average Green function.

All our analytical results are obtained in the weak-scattering limit,
\begin{equation}\label{1DCP-TwoScale}
\max\{k^{-1},R_c\}\ll\min\{L_{ts},L\}\,.
\end{equation}
In comparison with Eq.~\eqref{1DCP-WS} the above conditions involve the structure length $L$ in order to perform ensemble averaging to be reasonable for finite systems. It should be stressed that we do not assume any restriction for the relation between the wavelength $k^{-1}$ and correlation length $R_c$, as well as between the sample size $L$ and mean free path $L_{ts}$. It is also important to note that according to the assumption \eqref{1DCP-TwoScale}, there are two sets of substantially different spatial scales. One set consists of the {\it macroscopic} lengths $L_{ts}$ and $L$, in contrast with the other pair of {\it microscopic} lengths, $k^{-1}$ and $R_c$. The existence of two different scales gives rise to the \emph{two-scale approach}.

We start with the well-known representation for one-dimensional Green function,
\begin{equation}\label{1DCP-Gpsi}
{\cal G}(x,x')=\mathbb{W}^{-1}\big[\psi_+(x)\psi_-(x')\Theta(x-x')+\psi_+(x')\psi_-(x)\Theta(x'-x)\big],
\end{equation}
where the functions $\psi_{\pm}(x)$ are two linearly independent solutions of the homogeneous equation \eqref{1DCP-Schreq} that satisfy the boundary conditions at the right/left open ends $x=\pm L/2$, respectively. The Wronskian of these functions is $\mathbb{W}=\psi_{-}(x)\psi'_{+}(x)-\psi_{+}(x)\psi'_{-}(x)$, and $\Theta(x)$ is the Heaviside unit-step function, $\Theta(x<0)=0$ and $\Theta(x>0)=1$. Note that for the wave equation \eqref{1DCP-Schreq} the Wronskian does not depend on the coordinate $x$. Therefore, it can be calculated at any point of the structure. It is clear that the most suitable points are the ends $x=\pm L/2$.

The wave functions $\psi_{\pm}(x)$ can be presented as a superposition of modulated plane waves propagating in opposite directions, i.e. as a sum of transmitted and reflected waves,
\begin{equation}\label{1DCP-psipm}
\psi_{\pm}(x)=\pi_{\pm}(x)\exp(\pm ikx)-i\gamma_{\pm}(x)\exp(\mp ikx).
\end{equation}
The reflectionless boundary conditions for the wave functions $\psi_{\pm}(x)$ at the open ends $x=\pm L/2$ imply the following \emph{initial conditions} for the amplitudes $\pi_{\pm}(x)$ and $\gamma_{\pm}(x)$:
\begin{equation}\label{1DCP-IC}
\pi_{\pm}(\pm L/2)=1,\qquad\qquad\gamma_{\pm}(\pm L/2)=0.
\end{equation}
We emphasize that in line with the two-scale concept \eqref{1DCP-TwoScale}, the amplitudes $\pi_{\pm}(x)$ and $\gamma_{\pm}(x)$ are assumed to vary at the macroscopic scales $L_{ts}$, or $L$. Therefore they are smooth functions of $x$ in comparison with rapidly oscillating exponents $\exp(\pm ikx)$ and the disorder correlator $K(x)$. The appropriate equations for them are obtained by the standard method of averaging over rapid phases (see, e.g., Ref.~\cite{BM74}). In this method Eq.~\eqref{1DCP-psipm} is substituted into Eq.~\eqref{1DCP-Schreq} and both sides are multiplied from the left by $\exp(\mp ikx)$. The obtained equation is averaged (integrated) over the spatial interval $(x-l,x+l)$ of length $2l$, intermediate between the macroscopic and microscopic scales, $\max\{k^{-1},R_c\}\ll l\ll\min\{L_{ts},L\}$. The same procedure is repeated using the multiplier $\exp(\pm ikx)$. Certainly, the final results should be independent of the choice of the averaging interval. After averaging, we get a set of four first order differential equations,
\begin{subequations}\label{1DCP-PiGam}
\begin{eqnarray}
&&\pi'_{\pm}(x)\pm i\eta(x)\pi_{\pm}(x)\pm\zeta_{\pm}^*(x)\gamma_{\pm}(x)=0,\\[6pt]
&&\gamma'_{\pm}(x)\mp i\eta(x)\gamma_{\pm}(x)\pm\zeta_{\pm}(x)\pi_{\pm}(x)=0,
\end{eqnarray}
\end{subequations}
that have to be complemented by four initial conditions \eqref{1DCP-IC}. In such a way we have reduced the boundary-value problem to the ``dynamical" one. The variable coefficients $\eta(x)$ and $\zeta_{\pm}(x)$ are the space-averaged random fields associated with random scattering potential $V(x)$. In what follows, only the correlation properties of these fields are important. The function $\eta(x)$ is real while $\zeta_{\pm}(x)$ are mutually conjugated, $\zeta_{\pm}^*(x)=\zeta_{\mp}(x)$. Within the two-scale approach \eqref{1DCP-TwoScale} all these functions can be regarded as delta-correlated Gaussian random processes,
\begin{subequations}\label{1DCP-etazeta-corr}
\begin{eqnarray}
\langle\eta(x)\rangle=0,\qquad\langle\eta(x)\zeta_{\pm}(x')\rangle=0,\qquad \langle\eta(x)\eta(x')\rangle=2L_{fs}^{-1}\delta(x-x'),\\[6pt]
\langle\zeta_{\pm}(x)\rangle=0,\qquad\langle\zeta_{\pm}(x)\zeta_{\pm}(x')\rangle=0,\qquad \langle\zeta_{\pm}(x)\zeta_{\pm}^*(x')\rangle=2L_{bs}^{-1}\delta(x-x').
\end{eqnarray}
\end{subequations}
Remarkably, the coefficients $L_{fs}^{-1}$ and $L_{bs}^{-1}$ are exactly the forward and backward scattering lengths \eqref{1DCP-LfLb} emerging in the average Green function. Hence, the conclusion is: the fields $\eta(x)$ and $\zeta_{\pm}(x)$ are responsible for the forward and backward electron/wave scattering, respectively. The fact that the random fields $\eta(x)$ and $\zeta_{\pm}(x)$ turn out to be the independent white noises, plays a dominating role in subsequent averaging procedures. It is also useful to note that the ``dynamical" problem \eqref{1DCP-PiGam}, \eqref{1DCP-IC} results in the ``unimodularity condition" for the smooth amplitude,
\begin{equation}\label{1DCP-unim}
|\pi_\pm(x)|^2 -|\gamma_\pm(x)|^2=1.
\end{equation}

The next step is to express the transmittance $T_L$ via the smooth amplitudes $\pi_{\pm}$ and $\gamma_{\pm}$, with further averaging over random fields $\eta(x)$ and $\zeta_{\pm}(x)$. In correspondence with the two-scale approximation \eqref{1DCP-TwoScale}, when substituting the Green function \eqref{1DCP-Gpsi} into the Kubo's formula \eqref{1DCP-TKubo}, it is sufficient to differentiate only the rapid factors $\exp(\pm ikx)$ entering Eqs.~\eqref{1DCP-psipm}, and keep the terms in which these rapid exponentials cancel out. Taking into account the condition \eqref{1DCP-unim}, after some algebra, we get the formula,
\begin{equation}\label{1DCP-Tpi}
T_L=|\pi_{\pm}^{-1}(\mp L/2)|^2.
\end{equation}
This equality means that the quantity $\pi_{\pm}^{-1}(\mp L/2)$ is the {\it transmission amplitude} for a 1D disordered structure of length $L$. On the other hand, the form of the wave functions \eqref{1DCP-psipm} provides the ratio
\begin{equation}\label{1DCP-GammaDef}
\Gamma_{\pm}(x)=\gamma_{\pm}(x)/\pi_{\pm}(x)
\end{equation}
to be the {\it reflection amplitude} for the plane waves entering with the unit amplitude into the interval $(x,L/2)$/$(-L/2,x)$ from the left-hand/right-hand side, respectively. In agreement with this definition, the unimodularity condition \eqref{1DCP-unim} is immediately rewritten in the form of the flow conservation law for a non-dissipative medium,
\begin{equation}\label{1DCP-FlowCons}
|\Gamma_{\pm}(x)|^2+|\pi_{\pm}^{-1}(x)|^2=1.
\end{equation}
This fact confirms once more that $\pi_{\pm}^{-1}(x)$ is the local transmission amplitude, in accordance with the expression \eqref{1DCP-Tpi} for the transmittance. Thus, the transition from Eq.~\eqref{1DCP-TKubo} to Eq.~\eqref{1DCP-Tpi} establishes the correspondence between Kubo's and Landauer's concepts.

The equation for $\Gamma_{\pm}(x)$ is obtained by a straightforward differentiation of Eq.~\eqref{1DCP-GammaDef} using Eqs.~\eqref{1DCP-PiGam} for the smooth amplitudes. The result is the Riccati-type equation with the zero initial condition, see Eq.~\eqref{1DCP-IC},
\begin{equation}\label{1DCP-GammaEqIC}
\pm\frac{d\Gamma_{\pm}(x)}{dx}=2i\eta(x)\Gamma_{\pm}(x)+\zeta_{\pm}^*(x)\Gamma_{\pm}^2(x)-\zeta_{\pm}(x),\qquad\qquad \Gamma_{\pm}(\pm L/2)=0.
\end{equation}
In comparison with Eqs.~\eqref{1DCP-PiGam} the above equation is much more convenient for the analysis. By expressing the transmittance \eqref{1DCP-Tpi} through the \emph{reflectance} $|\Gamma_{\pm}(\mp L/2)|^2$ in agreement with the conservation law \eqref{1DCP-FlowCons}, we perform further calculations in terms of the reflection amplitude $\Gamma_{\pm}(x)$ rather than the transmission one, $\pi_{\pm}^{-1}(x)$.

We emphasize that the forward-scattering random field the function $\eta(x)$ can be eliminated from Eq.~\eqref{1DCP-GammaEqIC} by the concurrent phase transformation of the reflection amplitude $\Gamma_{\pm}(x)$ and the backscattering fields $\zeta_{\pm}(x)$. This transformation conserves the absolute values of both transmission and reflection amplitudes, and consequently, does not affect the definitions for the transmittance and reflectance. Clearly, the conservation law also holds its form \eqref{1DCP-FlowCons}. Moreover, since the random fields $\eta(x)$ and $\zeta_{\pm}(x)$ are statistically independent, $\langle\eta(x)\zeta_{\pm}(x')\rangle=0$, the phase transformation keeps the correlation relations \eqref{1DCP-etazeta-corr} for new renormalized fields $\zeta_{\pm}(x)$ unaffected. Because we are interested solely in the absolute values of transmission and reflection amplitudes, all these facts indicate that without loss of generality one can assume that $\eta(x)$ in Eq.~\eqref{1DCP-GammaEqIC} equals zero \cite{M99},
\begin{equation}\label{1DCP-eta0}
\eta(x)\to0.
\end{equation}
Thus, we conclude that the average reflectance or transmittance (and their moments) are specified by the backscattering processes only, while the forward scattering has no effect. In other words, unlike the spectral properties that are characterized by the total mean free path $L_{ts}$, the transport through a 1D disordered structure is determined exclusively by the backscattering length $L_{bs}$ defined by Eq.~\eqref{1DCP-LfLb}.

The Cauchy problem \eqref{1DCP-GammaEqIC} for $\Gamma_{\pm}(x)$ is equivalent to the following integral equation:
\begin{equation}\label{1DCP-GammaIntEq}
\Gamma_\pm(x)=\pm\int_{\pm L/2}^{x}dx'\left[\zeta_{\pm}^{*}(x')\Gamma_\pm^2(x')-\zeta_{\pm}(x')\right].
\end{equation}
This equation displays that the local reflection amplitude $\Gamma_\pm(x)$ belongs to a class of causal functionals of the renormalized complex field $\zeta_{\pm}(x)$. Indeed, in agreement with Eq.~\eqref{1DCP-GammaIntEq}, the amplitude $\Gamma_{+}(x)$ at a given point $x$ is specified by the values of the random field $\zeta_{+}(x)$ only within the interval $(x,L/2)$, whereas $\Gamma_{-}(x)$ at the same point $x$ is determined by the random field $\zeta_{-}(x)$ only within the complementary interval $(-L/2,x)$.

\subsection{Average logarithm of transmittance}
\label{2.5}

From the equations obtained in previous Section it is relatively easy to derive the average logarithm of the transmittance, $\langle\ln|\pi_{\pm}^{-1}(\mp L/2)|^2\rangle$. By differentiating the quantity $\ln|\pi_{\pm}^{-1}(x)|^2=-\ln[\pi_{\pm}(x)\pi_{\pm}^*(x)]$ and using the dynamical equations \eqref{1DCP-PiGam}, the following result is obtained after averaging:
\begin{equation} \label{1DCP-lnTEq}
\pm\frac{d}{dx}\langle\ln|\pi_{\pm}^{-1}(x)|^2\rangle=\langle\zeta_{\pm}^*(x)\Gamma_{\pm}(x)\rangle+\langle\zeta_{\pm}(x)\Gamma_{\pm}^*(x)\rangle.
\end{equation}
Thus, the problem is reduced to the calculation of the correlators $\langle\zeta_{\pm}^*(x)\Gamma_{\pm}(x)\rangle$ and $\langle\zeta_{\pm}(x)\Gamma_{\pm}^*(x)\rangle$. Because the reflection amplitude $\Gamma_\pm(x)$ is the causal functional of the Gaussian random field $\zeta_{\pm}(x)$ with zero average, we can apply Furutsu-Novikov formula \cite{K86},
\begin{equation}\label{1DCP-AvGammaFN}
\langle\zeta_{\pm}^*(x)\Gamma_{\pm}(x)\rangle=
\pm\int_x^{\pm L/2}dx'\langle\zeta_{\pm}^*(x)\zeta_{\pm}(x')\rangle\Big\langle\frac{\delta\Gamma_{\pm}(x)}{\delta\zeta_{\pm}(x')}\Big\rangle=
\frac{1}{L_{bs}}\left.\Big\langle\frac{\delta\Gamma_{\pm}(x)}{\delta\zeta_{\pm}(x')}\Big\rangle\right|_{x'\to x\pm0},
\end{equation}
where we have used the delta-correlation of $\zeta_{\pm}(x)$ and $\zeta_{\pm}^*(x)$, see Eq.~\eqref{1DCP-etazeta-corr}. The variational derivative in Eq.~\eqref{1DCP-AvGammaFN} can be readily calculated from the integral equation \eqref{1DCP-GammaIntEq}, resulting in unity. Since the second term in the right-hand side of Eq.~(\ref{1DCP-lnTEq}) is the complex conjugation of the first one, we have,
\begin{equation} \label{1DCP-lnTEq2}
\pm\frac{d}{dx}\langle\ln|\pi_{\pm}^{-1}(x)|^2\rangle=2/L_{bs}.
\end{equation}
By integrating this expression over $x$ within the interval $(-L/2,L/2)$ and applying the initial conditions \eqref{1DCP-IC}, we immediately obtain the well-known result for the average logarithm of the transmittance for 1D disordered systems,
\begin{equation}\label{1DCP-lnT}
\langle\ln|\pi_{\pm}^{-1}(\mp L/2)|^2\rangle=-2L/L_{bs}.
\end{equation}

\subsection{Distribution functions}
\label{2.6}

Now let us introduce the $n$th moment of the local reflectance $|\Gamma_\pm(x)|^2$,
\begin{equation}\label{1DCP-RnDef}
R_n^\pm(x)\equiv\langle|\Gamma_\pm(x)|^{2n}\rangle,\qquad \qquad n=0,1,2,3\ldots\,.
\end{equation}
One can see that the moment \eqref{1DCP-RnDef} being a single-point average, depends on the coordinate $x$. At the first sight, this assumption contradicts to the statistical homogeneity of the random potential $V(x)$. Indeed, if the condition of statistical homogeneity holds true, then single-point moments should be independent of the coordinate. However, the homogeneity is assumed to occur on the scale of \emph{microscopic} correlation length $R_c$ only, and can be broken on \emph{macroscopic} scales, either $L_{bs}$ or $L$. Therefore, the moments of the reflectance $R_n^\pm(x)$ do depend, while the variance $\langle V^2(x)\rangle=V_0^2$ of the potential $V(x)$ does not depend on the coordinate $x$.

In order to obtain the equation for $R_n^\pm(x)$ one should differentiate Eq.~\eqref{1DCP-RnDef} over $x$ taking into account
Eq.~\eqref{1DCP-GammaEqIC} for $\Gamma_\pm(x)$. The averaging is performed via the Furutsu-Novikov formalism with the use of the integral presentation \eqref{1DCP-GammaIntEq} and correlation relations \eqref{1DCP-etazeta-corr}. As a result, we arrive at the recurrence relation with the corresponding initial condition on $x$,
\begin{equation}\label{1DCP-RnEq}
\pm\frac{dR_n^{\pm}(x)}{dx}=-\frac{2n^2}{L_{bs}}\left[R_{n+1}^{\pm}(x)-2R_n^{\pm}(x)+R_{n-1}^{\pm}(x)\right],\qquad R_n^{\pm}(\pm L/2)=\delta_{n0}.
\end{equation}
Two necessary boundary conditions with respect to index $n$, namely, $R_0^{\pm}(x)=1$ and $R_n^\pm(x)\to0$ for $n\to\infty$, directly follow from definition \eqref{1DCP-RnDef} at $n=0$ and $n\to\infty$.

The solution of Eq.~\eqref{1DCP-RnEq} can be obtained by the distribution function $P_L^\pm(u|x)$ of the random variable $u$, through to which the moments of reflectance are expressed as follows:
\begin{equation}\label{1DCP-RnPL}
R_n^{\pm}(x)=\int_1^\infty duP_L^\pm(u|x)\left(\frac{u-1}{u+1}\right)^n,\qquad n=0,1,2,3\ldots\,.
\end{equation}
Correspondingly, the same probability density provides the statistical moments of the transmittance \eqref{1DCP-Tpi},
\begin{eqnarray}\label{1DCP-TnPL}
\langle T_L^n\rangle=\langle\left[1-|\Gamma_{\pm}(\mp L/2)|^2\right]^n\rangle=
\int_1^\infty duP_L^\pm(u|\mp L/2)\left(\frac{2}{u+1}\right)^n,\\[6pt]
n=0,\pm1,\pm2,\pm3,\ldots\,.\nonumber
\end{eqnarray}
Here the relation between the transmittance $T_L$ and random variable $u$ reads
\begin{equation}\label{1DCP-TLu}
T_L=2/(u+1).
\end{equation}
Note that the transmittance moment $\langle T_L^n\rangle$ with $n$ of either sign can be calculated with the same distribution function $P_L^\pm (u|x)$ since the integral in Eq.~(\ref{1DCP-TnPL}) converges for both positive and negative values of $n$. The normalization of the distribution function $P_L^\pm(u|x)$ to unity ensures the boundary condition $R_0^{\pm}(x)=1$. On the other hand, such a normalization implies $P_L^\pm(u|x)$ to be integrable with respect to the variable $u$, in particular, at $u\to1$ and $u\to\infty$. This integrability should be used when establishing the equation for the probability density.

The equation for $P_L^\pm (u|x)$ can be obtained by substituting the integral \eqref{1DCP-RnPL} into Eq.~\eqref{1DCP-RnEq} with further integration by parts two times. This leads to the Fokker-Plank equation (see, for example, \cite{LGP88}),
\begin{equation}\label{1DCP-FokkerPlank}
\pm\frac{L_{bs}}{2}\frac{\partial P_L^\pm(u|x)}{\partial x}=-\frac{\partial}{\partial u}(u^2-1)\frac{\partial P_L^\pm(u|x)}{\partial u},\qquad
P_L^\pm(u|\pm L/2)=\delta(u-1-\epsilon)\,,
\end{equation}
where $\epsilon \rightarrow 0$. The initial condition with respect to $x$ is derived from the condition at $x=\pm L/2$ in Eq.~\eqref{1DCP-RnEq}. Note that the normalization condition for $P_L^\pm(u|x)$ serves as a boundary condition with respect to $u$.

The solution of Eq.~\eqref{1DCP-FokkerPlank} can be found by using the standard method of the Mehler-Fock transformation \cite{M881,F43}. The resulting expression has the following form:
\begin{eqnarray}\label{1DCP-PL}
&&P_L^\pm(\cosh2\alpha|x)=\frac{1}{8\sqrt{\pi}}\left(\frac{L\mp2x}{4L_{bs}}\right)^{-3/2}\exp\left(-\frac{L\mp2x}{4L_{bs}}\right)\nonumber\\[6pt]
&&\times\int_{\alpha}^{\infty}\frac{vdv}{(\cosh^2v-\cosh^2\alpha)^{1/2}}\exp\left[-v^2\left(\frac{L\mp2x}{L_{bs}}\right)^{-1}\right],\qquad u=\cosh2\alpha,\,\alpha\geqslant0.\qquad
\end{eqnarray}

The probability density \eqref{1DCP-PL} of the random variable $u$ gives rise to the distribution functions of all measurable transport characteristics of the problem. Specifically, in accordance with Eqs.~\eqref{1DCP-TnPL} and \eqref{1DCP-TLu} the transition from $P_L^\pm(\cosh2\alpha|x)$ to the \emph{transmittance distribution function} $P_L(T_L)$ is realized with the use of the evident rule,
\begin{equation}\label{1DCP-PL-PLTL}
P_L(T_L)=\left|du/dT_L\right|P_L^\pm(u|\mp L/2)= 2\cosh^4\alpha\,P_L^\pm(\cosh2\alpha|\mp L/2).
\end{equation}
In such a manner, one can readily obtain the explicit expression for the transmittance distribution function \cite{LGP88},
\begin{eqnarray}\label{1DCP-PLTL}
&&P_L(T_L)=\frac{1}{4\sqrt{\pi}}\left(\frac{L}{2L_{bs}}\right)^{-3/2}\exp\left(-\frac{L}{2L_{bs}}\right) \cosh^4\alpha\nonumber\\[6pt]
&&\times\int_{\alpha}^{\infty}\frac{vdv}{(\cosh^2v-\cosh^2\alpha)^{1/2}}
\exp\left(-v^2\frac{L_{bs}}{2L}\right),\qquad
T_L=\cosh^{-2}\alpha,\, 0\leqslant\alpha<\infty.\qquad
\end{eqnarray}
This expression determines the transport characteristics of a 1D disordered structure. In particular, it gives an exact formula for the $n$th moment of the transmittance. First results for the distribution of the resistance can be found, for example, in Refs.~\cite{AR78,A81a,M86}.

\subsection{Average transmittance and its moments}
\label{2.7}

The final expression for the average transmittance $\langle T_L\rangle$ and all its moments $\langle T_L^n\rangle$ can be written in the form,
\begin{eqnarray}\label{1DCP-1DTn}
\langle T_L^n\rangle&=&\frac{1}{2\sqrt{\pi}}\left(\frac{L}{2L_{loc}}\right)^{-3/2}\exp{\left(-\frac{L}{2L_{loc}}\right)}\nonumber\\[6pt]
&\times&\int_0^\infty\frac{vdv}{\cosh^{2n-1}v}\exp\left(-v^2\frac{L_{loc}}{2L}\right)\int_0^vd\beta\cosh^{2(n-1)}\beta\,,\\[6pt]
&&\ n=0,\pm1,\pm2,\pm3,\ldots\,.\nonumber
\end{eqnarray}
Here we have introduced the so-called \emph{localization length} $L_{loc}$ that in the \emph{transfer matrix approach} \cite{LGP88} is associated with the \emph{Lyapunov exponent} $\lambda=L_{loc}^{-1}$. The latter describes how fast the average wave function $\langle\psi_{\pm}(x)\rangle$ decreases away from the center of localization in the coordinate space. As follows from our analysis, the localization length is nothing but the
\emph{backscattering length} emerging in the scattering problem for infinite samples with a given disorder,
\begin{equation}\label{1DCP-LlocLbs}
L_{loc}=L_{bs}.
\end{equation}

The r.h.s. of Eq.~\eqref{1DCP-1DTn} depends solely on the ratio $L/L_{loc}$ between the sample size $L$ and localization length $L_{loc}$. This is a manifestation of the so-called \emph{single parameter scaling} that constitutes the phenomenon of the 1D Anderson localization. According to this scaling, if we know the localization length $L_{loc}$ determined in \emph{infinite} system, we can describe all transport properties
for \emph{finite} samples. Therefore, the knowledge of $L_{loc}$ is of great importance in many applications. We remind that Eq.~\eqref{1DCP-1DTn} is valid for arbitrary relation between the localization length $L_{loc}$ and sample length $L$.

From Eq.~\eqref{1DCP-1DTn} one can find relatively easily the low moments of the distribution of the transmittance $T_L$. Specifically, for $n=1$ and $n=2$ one gets the average transmittance $\langle T_L\rangle$ and its second moment $\langle T_L^2\rangle$ that defines the transmittance variance,
\begin{subequations}\label{1DCP-AvVarT}
\begin{eqnarray}
&&\langle T_L\rangle=\frac{1}{2\sqrt{\pi}} \left(\frac{L}{2L_{loc}}\right)^{-3/2}\exp{\left(-\frac{L}{2L_{loc}}\right)}
\int_0^\infty\frac{v^2dv}{\cosh v}\exp\left(-v^2\frac{L_{loc}}{2L}\right),\label{1DCP-AvT}\\[6pt]
&&\langle T_L^2\rangle=\frac{1}{2\sqrt{\pi}} \left(\frac{L}{2L_{loc}}\right)^{-3/2}\exp{\left(-\frac{L}{2L_{loc}}\right)}
\int_0^\infty\frac{vdv}{\cosh^3v}\frac{2v+\sinh2v}{4}\exp\left(-v^2\frac{L_{loc}}{2L}\right),\qquad\quad\\[6pt]
&&\mathrm{Var}\{T_L\}\equiv\langle T_L^{2}\rangle-\langle T_L\rangle^2.
\end{eqnarray}
\end{subequations}
Eqs.~\eqref{1DCP-AvVarT} manifest an exponential decrease of the transmittance and its moments with an increase of the sample length $L$.

The average dimensionless resistance $\langle T_L^{-1}\rangle$ and its variance are described by Eq.~\eqref{1DCP-1DTn} with $n=-1$ and $n=-2$. They are given by the following simple formulae:
\begin{subequations}\label{1DCP-AvVarRes}
\begin{eqnarray}
\langle T_L^{-1}\rangle&=&\frac{1}{2}\left[1+\exp\left(\frac{4L}{L_{loc}}\right)\right],\\[6pt]
\langle T_L^{-2}\rangle&=&\frac{1}{6}\left[2+3\exp\left(\frac{4L}{L_{loc}}\right)+ \exp\left(\frac{12L}{L_{loc}}\right)\right],\\[6pt]
\mathrm{Var}\{T_L^{-1}\}&\equiv&\langle T_L^{-2}\rangle-\langle T_L^{-1}\rangle^2=
\frac{1}{12}\left[1+2\exp\left(\frac{12L}{L_{loc}}\right)-3\exp\left(\frac{8L}{L_{loc}}\right)\right].
\end{eqnarray}
\end{subequations}
The above expressions demonstrate an exponential increase of the resistance and its second moment, as well as the variance, with the size $L$. Note that in literature one can find another definition of the resistance, $T_L^{-1}-1$, which is the so-called Landauer resistance.

In agreement with the concept of single parameter scaling, there are only two characteristic transport regimes in the 1D disordered structures, namely, the regimes of the ballistic and localized transport.

\textbf{(i)} The \emph{ballistic transport} occurs if the localization length $L_{loc}$ is much larger than the sample length $L$. In this case the samples are almost fully transparent,
\begin{subequations}\label{1DCP-TResBal}
\begin{eqnarray}
&&\langle T_L\rangle\approx1-2L/L_{loc},\qquad\mathrm{Var}\{T_L\}\approx(2L/L_{loc})^2;\label{1DCP-TBal}\\[6pt]
&&\langle T_L^{-1}\rangle\approx1+2L/L_{loc},\qquad\mathrm{Var}\{T_L^{-1}\}\approx12(2L/L_{loc})^2\qquad\mbox{for}\quad L\ll L_{loc}. \label{1DCP-ResBal}
\end{eqnarray}
\end{subequations}
In this situation the average transmittance and resistance are consistent with each other up to linear terms $L/L_{loc}$, i.e. $\langle T_L\rangle^{-1}=\langle T_L^{-1}\rangle$. Moreover, in the same approximation the transmittance and resistance are the self-averaging quantities, since both variances are sufficiently small.

\textbf{(ii)} In the opposite case, when the localization length $L_{loc}$ is small enough in comparison with the sample length $L$, the 1D disordered structures exhibit \emph{localized transport}. In this case the average transmittance is exponentially small whereas the average resistance is exponentially large,
\begin{subequations}\label{1DCP-TResLoc}
\begin{eqnarray}
&&\langle T_L\rangle\approx\frac{\pi^3}{16\sqrt{\pi}}\left(\frac{L}{2L_{loc}}\right)^{-3/2}\exp\left(-\frac{L}{2L_{loc}}\right),\qquad
\mathrm{Var}\{T_L\}\approx\langle T_L^{2}\rangle\approx\frac{1}{4}\langle T_L\rangle;\label{1DCP-TLoc}\\[6pt]
&&\langle T_L^{-1}\rangle\approx \frac{1}{2}\exp\left(\frac{4L}{L_{loc}}\right),\quad
\mathrm{Var}\{T_L^{-1}\}\approx\langle T_L^{-2}\rangle\approx\frac{4}{3}\langle T_L^{-1}\rangle^3\qquad\mbox{for}\quad L_{loc}\ll L.\qquad
\label{1DCP-ResLoc}
\end{eqnarray}
\end{subequations}
As one can see, in the localization regime a 1D disordered system perfectly (with an exponential accuracy) reflects the electrons (or waves) due to strong localization of all eigenstates.

Here we would like to stress the following. First, in the localization regime the average transmittance $\langle T_L\rangle$ and resistance $\langle T_L^{-1}\rangle$ as functions of the system length $L$ have sufficiently different scales of decrease or increase, namely, $2L_{loc}$ for the former and $L_{loc}/4$ for the latter. Second, the average resistance does not coincide with the inverse value of the average transmittance, $\langle
T_L^{-1}\rangle\neq\langle T_L\rangle^{-1}$. And, third, the variance of the transmittance normalized to the square of its average, is proportional to $\exp(L/2L_{loc})\gg1$. In comparison, the variance of the resistance normalized to the squared average, is proportional to $\exp(4L/L_{loc})\gg1$. These facts mean that in the case of strong localization both the transmittance and resistance are not self-averaging quantities. Hence, by changing the length $L$ of the sample, or the disorder itself for the same $L$, one should expect very large fluctuations of $T_L$ and $T_L^{-1}$. These fluctuations are known as the \emph{mesoscopic fluctuations} that are characteristic of strong quantum (interference) effects on a macroscopic scale.

\subsection{Average logarithm of transmittance and its moments}
\label{2.8}

As we have revealed in previous Subsection, the transmittance $T_L$ of 1D disordered systems is not self-averaging in the regime of strong localization, $L_{loc}\ll L$. In this situation, for a proper description of the transport one should refer to the logarithm of the transmittance, $\ln T_L$, that is the self-averaging quantity for strong localization. The use of the probability density \eqref{1DCP-PLTL} yields the following expression for the moments of the transmittance logarithm:
\begin{eqnarray}\label{1DCP-nLnT}
\langle\ln^nT_L\rangle&=&\frac{1}{2\sqrt{\pi}}\left(\frac{L}{2L_{loc}}\right)^{-3/2}\exp{\left(-\frac{L}{2L_{loc}}\right)}\nonumber\\[6pt]
&\times&\int_0^\infty vdv\cosh v\exp\left(-v^2\frac{L_{loc}}{2L}\right)
\int_0^v\frac{d\beta}{\cosh^2\beta}\ln^n\left(\frac{\cosh\beta}{\cosh v}\right)^2\,.
\end{eqnarray}
With $n=1$ this expression provides the exact relation \eqref{1DCP-lnT} for the average logarithm that has been already obtained in Subsection~\ref{2.5},
\begin{equation}\label{1DCP-AvLn}
\langle\ln T_L\rangle=-2L/L_{loc}.
\end{equation}
This famous result is consistent with an exponential decrease of the transmittance averaged over the so-called representative (non-resonant) realizations of disorder \cite{LGP88},
\begin{equation}\label{1DCP-repavT}
\langle T_L\rangle_{rep}=\exp\langle\ln T_L\rangle=\exp(-2L/L_{loc}).
\end{equation}
According to the asymptotic expression \eqref{1DCP-TLoc} the transmission exponentially decreases on the scale $L\approx2L_{loc}$, with an additional power prefactor. In contrast, the transmittance \eqref{1DCP-repavT} has an exponential dependence with a much faster decrease on the scale $L\approx L_{loc}/2$. This fact can be explained as follows. The main contribution to Eq.~\eqref{1DCP-TLoc} for the average transmittance \eqref{1DCP-AvVarT} is given by the resonant realizations of the random potential $V(x)$. For these realizations the transmittance is almost equal to one, however, they have an exponentially small probability. On the other hand, for the representative realizations (most probable, but non-resonant) the transmittance is described by Eq.~\eqref{1DCP-repavT}. This effect is peculiar to the mesoscopic nature of Anderson localization. Note also that both expressions for the average transmittance, \eqref{1DCP-AvVarT} and \eqref{1DCP-repavT}, give the same result \eqref{1DCP-TBal} in the case of ballistic transport, $L\ll L_{loc}$.

It is important to stress that Eq.~\eqref{1DCP-AvLn} is valid for \emph{any} ratio between the localization length $L_{loc}$ and
sample length $L$. For this reason and due to the self-averaging nature of $\langle\ln T_L\rangle$, the relation \eqref{1DCP-AvLn} is often used as the definition of the localization length,
\begin{equation}\label{1DCP-LlocLnT}
\lambda=L_{loc}^{-1}=-\frac{1}{2}\lim_{L\to\infty}\frac{\ln T_L}{L}\,
\end{equation}
Remarkably, the same definition emerges within the Lyapunov exponent concept in the transfer matrix method, where $T_L$ is defined through the eigenvalues of the product of individual transfer matrices relating the $\psi-$function in two successive points, see for example, \cite{P86,LGP88}.

The expression \eqref{1DCP-nLnT} with $n=2$ results in the second moment of the transmittance logarithm, $\langle\ln^2T_L\rangle$, that defines its variance,
\begin{equation}\label{1DCP-VarLn}
\mathrm{Var}\{\ln T_L\}\equiv\langle\ln^{2}T_L\rangle-\langle\ln T_L\rangle^2.
\end{equation}
For two characteristic transport regimes, the ballistic and localized transport, one can derive, respectively,
\begin{subequations}\label{1DCP-VarLnBalLoc}
\begin{eqnarray}
&&\mathrm{Var}\{\ln T_L\}\approx\langle\ln T_L\rangle^2\qquad\mbox{for}\quad L\ll L_{loc}\,;\label{1DCP-VarLnBal}\\[6pt]
&&\mathrm{Var}\{\ln T_L\}\approx-2\langle\ln T_L\rangle\qquad\mbox{for}\quad L_{loc}\ll L.\label{1DCP-VarLnLoc}
\end{eqnarray}
\end{subequations}
One can see that in the regime of strong localization, the variance \eqref{1DCP-VarLnLoc} is much smaller than the squared
average logarithm of transmittance, $\mathrm{Var}\{\ln T_L\}/\langle\ln T_L\rangle^2=-2/\langle\ln T_L\rangle=L_{loc}/L\ll1$. This fact confirms the self-averaging nature of the transmittance logarithm.

The expression \eqref{1DCP-VarLnLoc} is widely discussed in connection with the single parameter scaling conjecture. As we have mentioned in Subsection~\ref{2.7}, in the context of our study this scaling means that \emph{all} statistical properties of scattering through a finite sample of size $L$ depend on a single parameter, which is the ratio $L/L_{loc}$. Note that this result has been rigorously proved only for the model \eqref{1DCP-Ham} with a \emph{continuous} random potential $V(x)$. For other models, such as the tight-binding Anderson model or Kronig-Penney
models (see Sections~\ref{4} and \ref{8}), the single parameter scaling is not supported by rigorous analysis. Thus, the question about this scaling arises every time when considering a specific random model. Moreover, our results deal with weak disorder. As for strong disorder, it is still not clear whether one-parameter scaling remains valid.

In application to solid state physics the meaning of a single parameter scaling is entirely related to the hypothesis according to which all observable characteristics of transport depend on the dimensionless conductance or, equivalently, transmission coefficient $\langle T_{L}\rangle$ given by Eq.~\eqref{1DCP-AvT} only. In view of our rigorous results, this statement automatically stems from the fact that the distribution
function \eqref{1DCP-PLTL} for $T_{L}$ depends on the ratio $L/L_{loc}$. Indeed, according to Eq.~\eqref{1DCP-AvT} instead of this scaling ratio one can introduce a new scaling parameter which is the average value $\langle T_{L}\rangle$ itself. Then, all moments of $T_{L}$ can be expressed via the \emph{only} parameter $\langle T_{L}\rangle$. This is the core of the scaling theory of localization.

\subsection{Probability density of transmittance logarithm}
\label{2.8.2}

It is a common belief that in the limit of strong localization, $L_{loc}\ll L$, the transmittance logarithm, $\ln T_L$, obeys the Gaussian distribution. In general, the Gaussian probability density $P(z)$ of random variable $z$ is described by two independent parameters, the average value $\langle z\rangle$ and variance $\mathrm{Var}\{z\}$. In our case only {\it one} parameter comes into play, due to relation \eqref{1DCP-VarLnLoc}. That is why when studying scaling properties of various random models, the relation between the variance of transmittance logarithm and average logarithm of transmittance itself is often under detailed studies \cite{DLA00,DLA01,DEL03,DEL03a,DELA03}.

Taking into account the expressions \eqref{1DCP-AvLn} for the average logarithm and \eqref{1DCP-VarLnLoc} for its variance, the Gaussian probability density $P_G(\ln T_L)$ of the transmittance logarithm $\ln T_L$ in the regime of strong localization has the form,
\begin{eqnarray}\label{1DCP-GDF-LnT}
P_G(\ln T_L)&=&\frac{1}{2\sqrt{2\pi L/L_{loc}}}\, \exp\left[-\frac{(\ln T_L+2L/L_{loc})^2}{8L/L_{loc}}\right],\\[6pt]
&&-\infty<\ln T_L\leqslant0,\qquad 2L/L_{loc}\gg1.\nonumber
\end{eqnarray}

The relation between the distribution functions of the transmittance and its logarithm is,
\begin{equation}\label{1DCP-DF-Ln-T}
P_L(\ln T_L)=\left|\frac{dT_L}{d\ln T_L}\right|P_L(T_L)=T_LP_L(T_L).
\end{equation}
The exact expression for $P_L(T_L)$ is given by Eq.~\eqref{1DCP-PLTL}, therefore, the probability density $P_L(\ln T_L)$  can be written as follows,
\begin{eqnarray}\label{1DCP-PLlnTL}
&&P_L(\ln T_L)=\frac{2}{\sqrt{\pi}}
\left(\frac{2L}{L_{loc}}\right)^{-3/2}\exp\left(-\frac{L}{2L_{loc}}\right) \cosh^2\alpha\nonumber\\[6pt]
&&\times\int_{\alpha}^{\infty}\frac{vdv}{(\cosh^2v-\cosh^2\alpha)^{1/2}}
\exp\left(-v^2\frac{L_{loc}}{2L}\right),\quad
\ln T_L=-2\ln\cosh\alpha,\,0\leqslant\alpha<\infty.\qquad
\end{eqnarray}

To proceed further, we take into account the conditions that are valid in the regime of strong localization,
\begin{eqnarray}\label{1DCP-LocCond}
-\ln T_L=2\alpha,\qquad\alpha\gg1;\qquad -\langle\ln T_L\rangle=2L/L_{loc}=2\langle\alpha\rangle,\qquad\langle\alpha\rangle\gg1.
\end{eqnarray}
Due to these inequalities, Eq.~\eqref{1DCP-PLlnTL} can be expanded in large parameters $\alpha$ and $v$. As a result, the distribution function $P_L(\ln T_L)$ is appropriate to present in the form of the Gaussian probability density \eqref{1DCP-GDF-LnT} multiplied by the form-factor $F(\alpha/\langle\alpha\rangle)$ that is assumed to be a smooth function of its argument,
\begin{equation}\label{1DCP-PLn-F}
P_L(\ln T_L)\approx\frac{\langle\alpha\rangle^{-1/2}}{2\sqrt{2\pi}}
\exp\left[-\frac{(\alpha-\langle\alpha\rangle)^2}{2\langle\alpha\rangle}\right]
F\left(\frac{\alpha}{\langle\alpha\rangle}\right).
\end{equation}

The probability density \eqref{1DCP-PLn-F} should asymptotically meet the normalization condition,
\begin{equation}\label{1DCP-PLn-NormDef}
\int_{-\infty}^{0}d(\ln T_L)P_L(\ln T_L)=1.
\end{equation}
By substitution of Eq.~\eqref{1DCP-PLn-F}, the normalization takes the explicit form in the variable $\alpha$,
\begin{equation}\label{1DCP-PLn-Norm01}
\frac{\langle\alpha\rangle^{-1/2}}{\sqrt{2\pi}}\int_{0}^{\infty}d\alpha F\left(\frac{\alpha}{\langle\alpha\rangle}\right)
\exp\left[-\frac{(\alpha-\langle\alpha\rangle)^2}{2\langle\alpha\rangle}\right]=1.
\end{equation}
By changing the integration variable $\alpha=x\langle\alpha\rangle$, we come to the relation,
\begin{equation}\label{1DCP-PLn-Norm02}
\frac{\langle\alpha\rangle^{1/2}}{\sqrt{2\pi}}\int_{0}^{\infty}dx F\left(x\right)\exp\left[-\frac{\langle\alpha\rangle}{2}(x-1)^2\right]=1.
\end{equation}
Since $\langle\alpha\rangle/2\gg1$, the exponent in the integrand is a sharp function. Therefore, one can simply relate the integral to the value of smooth function $F(x)$ at $x=1$. As a result, the normalization condition \eqref{1DCP-PLn-NormDef} is reduced to the requirement
\begin{equation}\label{1DCP-F-Norm}
F(1)=1.
\end{equation}
Thus, the form-factor $F(\alpha/\langle\alpha\rangle)$ should be equal to unity when the transmittance logarithm is equal to its average, $\alpha=\langle\alpha\rangle$.

In accordance with Eqs.~\eqref{1DCP-PLlnTL} and \eqref{1DCP-PLn-F} the form-factor $F(\alpha/\langle\alpha\rangle)$ is defined by
\begin{equation}\label{1DCP-F-def02}
F\left(\frac{\alpha}{\langle\alpha\rangle}\right)=\frac{\alpha^2}{\langle\alpha\rangle}
\int_{1}^{\infty}\frac{xdx}{\sqrt{\exp[2\alpha(x-1)]-1}}\,
\exp\left[-\frac{\alpha^2}{2\langle\alpha\rangle}(x^2-1)\right].
\end{equation}
Note that $\alpha\gg1$ and $\alpha^2/\langle\alpha\rangle\sim\alpha\gg1$. On the other hand, one can see that the integrand in Eq.~\eqref{1DCP-F-def02} diverges at $x=1$. Moreover, the sharp exponent in the numerator achieves its maximal value. Due to these facts, it seems to be reasonable to conclude that the main contribution to the integral over $x$ comes from a narrow region near the lower limit of integration. Therefore, we can replace $x^2-1$ with $2(x-1)$. After the change of the integration variable $x-1=z/2\alpha$, we get,
\begin{equation}\label{1DCP-F-As02}
F\left(\frac{\alpha}{\langle\alpha\rangle}\right)= \frac{\alpha}{2\langle\alpha\rangle}
\int_{0}^{\infty}\frac{dz}{\sqrt{\exp(z)-1}}\,\exp\left(-\frac{\alpha}{2\langle\alpha\rangle}z\right).
\end{equation}
This expression displays that the form-factor is, indeed, a function of the ratio $\alpha/\langle\alpha\rangle$. The exact calculation of the integral in Eq.~\eqref{1DCP-F-As02} yields,
\begin{equation}\label{1DCP-F-AsFin}
F\left(\frac{\alpha}{\langle\alpha\rangle}\right)=\frac{\alpha}{2\langle\alpha\rangle}\,
\textrm{B}\left(\frac{1}{2},\frac{\alpha+\langle\alpha\rangle}{2\langle\alpha\rangle}\right).
\end{equation}
Here $\textrm{B}(x,y)$ is the beta-function,
\begin{equation}\label{Beta-def}
\textrm{B}(x,y)=\frac{\Gamma(x)\Gamma(y)}{\Gamma(x+y)}=\textrm{B}(y,x),
\end{equation}
and $\Gamma(x)$ is the gamma-function. Note that Eq.~\eqref{1DCP-F-AsFin} automatically obeys the normalization condition \eqref{1DCP-F-Norm}.

Let us briefly analyze the behavior of the form-factor \eqref{1DCP-F-AsFin} with the variation of its argument $\alpha/\langle\alpha\rangle$. This behavior is characterized by corresponding asymptotics of $F(\alpha/\langle\alpha\rangle)$. It can be simply recognized that at small values of the argument one gets,
\begin{equation}\label{1DCP-F-smallAlpha}
F\left(\frac{\alpha}{\langle\alpha\rangle}\right)\approx \frac{\pi\alpha}{2\langle\alpha\rangle}\,,\qquad\alpha\ll\langle\alpha\rangle.
\end{equation}
As for large values of the argument, one can readily get,
\begin{equation}\label{1DCP-F-largeAlpha}
F\left(\frac{\alpha}{\langle\alpha\rangle}\right)\approx\sqrt{\frac{\pi\alpha}{2\langle\alpha\rangle}}
\,,\qquad\langle\alpha\rangle\ll\alpha.
\end{equation}
One can see that the form-factor increases with an increase of its argument.

In accordance with Eqs.~\eqref{1DCP-PLn-F} and \eqref{1DCP-F-AsFin}, the distribution function of the transmittance logarithm can be presented in the following form,
\begin{eqnarray}\label{1DCP-PLn-As02}
P_L(\ln T_L)=\frac{\alpha\langle\alpha\rangle^{-3/2}}{4\sqrt{2\pi}}\,
\textrm{B}\left[\frac{1}{2},\frac{1}{2}\left(\frac{\alpha}{\langle\alpha\rangle}+1\right)\right]\,
\exp\left[-\frac{\langle\alpha\rangle}{2}\left(\frac{\alpha}{\langle\alpha\rangle}-1\right)^2\right],\\[6pt]
0\leqslant2\alpha=-\ln T_L<\infty,\qquad 2\langle\alpha\rangle=-\langle\ln T_L\rangle=2L/L_{loc}\gg1.\nonumber
\end{eqnarray}
This asymptotics is valid for any relation between the random variable $\alpha$ and its average $\langle\alpha\rangle$, provided $\alpha\gg1$ and $\langle\alpha\rangle\gg1$. Strictly speaking, this probability density differs from the Gaussian one. However, the main factor in Eq.~\eqref{1DCP-PLn-As02} is the Gaussian exponent. Indeed, although the Gaussian exponent is a quite smooth function of $\alpha$ with the scale of decrease $\langle\alpha\rangle^{1/2}\gg1$, it is a sharp function of the ratio $\alpha/\langle\alpha\rangle$ that rapidly decreases from its symmetry center $\alpha/\langle\alpha\rangle=1$ on a parametrically small scale $|1-\alpha/\langle\alpha\rangle|\sim\langle\alpha\rangle^{-1/2}\ll1$. Such a dualism is a direct consequence of the relation \eqref{1DCP-VarLnLoc} which gives $\mathrm{Var}\{\alpha\}=\langle\alpha\rangle$ for random variable $\alpha$. In comparison with the Gaussian exponent, the form-factor is a smooth function of $\alpha/\langle\alpha\rangle$ since it varies on the non-parametrical scale of the order of unity. Nevertheless, as results from asymptotics \eqref{1DCP-F-smallAlpha} and \eqref{1DCP-F-largeAlpha}, the form-factor in Eq.~\eqref{1DCP-PLn-As02} gives rise to the asymmetrical line-shape of the transmittance-logarithm distribution function.

Now we give the explicit form for the probability density of the transmittance logarithm in the regime of strong localization,
\begin{eqnarray}\label{1DCP-PLn-AsFin}
P_L(\ln T_L)=-\frac{\ln T_L}{4L/L_{loc}}\,\textrm{B}\left(\frac{1}{2},\frac{1}{2}-\frac{\ln T_L}{4L/L_{loc}}\right)\nonumber\\[6pt]
\times\frac{1}{2\sqrt{2\pi L/L_{loc}}}\,\exp\left[-\frac{(\ln T_L+2L/L_{loc})^2}{8L/L_{loc}}\right],\\[6pt]
-\infty<\ln T_L\leqslant0,\qquad 2L/L_{loc}\gg1.\nonumber
\end{eqnarray}
From the analysis performed above one can conclude that the averaging with this distribution function should be performed asymptotically, i.e. exactly as with the corresponding Gaussian probability density \eqref{1DCP-GDF-LnT}. Such a specific property is inherent in the concept of \emph{log-normal distribution} of the transmittance.

It is interesting that the logarithm distribution \eqref{1DCP-PLn-AsFin} gives rise to the following ``magic" asymptotics for the moments of transmittance logarithm and for even moments of its random deviation in the regime of strong localization,
\begin{subequations}\label{{1DCP-nLnTLoc}}
\begin{eqnarray}
&&\langle(\ln T_L-\langle\ln T_L\rangle)^{2n}\rangle=1\cdot3\cdot5\ldots(2n-3)(2n-1)\mathrm{Var}^{n}\{\ln T_L\},\\[6pt]
&&\langle\ln^nT_L\rangle\approx\langle\ln T_L\rangle^n,\qquad\mbox{for}\quad L_{loc}\ll L.
\end{eqnarray}
\end{subequations}
Again, these relations are valid due to Eqs.~\eqref{1DCP-AvLn} and \eqref{1DCP-VarLnLoc}.

\subsection{Localization length and convolution method}
\label{2.9}

The expressions \eqref{1DCP-1DTn} -- \eqref{1DCP-VarLnBalLoc} are universal and applicable for any continuous 1D system with weak static disorder. As we stressed, the knowledge of the localization length allows us to fully describe the transport properties of finite disordered samples. Our analysis is in accordance with different approaches \cite{B73,AR78,LGP88,BW75,K86,M99} in which the inverse localization length for any kind of weak disorder is determined by the $2k$-harmonic in the power spectrum ${\cal K}(k_x)$ of the random scattering potential $V(x)$, see Eqs.~\eqref{1DCP-LlocLbs}, \eqref{1DCP-LfLb},
\begin{equation}\label{1DCP-Lloc}
L_{loc}^{-1}(k)=\frac{V_0^2}{8k^2}\,{\cal K}(2k).
\end{equation}
It is instructive to discuss the dependence of the localization length $L_{loc}$ on the electron energy $E=k^2$ or, in the case of classical electromagnetic waves, the dependence of $L_{loc}$ on the wave frequency $\omega$. For electron structures, the perturbation potential $V(x)$ typically does not depend on $E$. Therefore, the energy dependence in $L_{loc}$ results from the power spectrum ${\cal K}(2k)$ that is due to the correlations, apart from the term $k^2$ in the denominator. For a sufficiently small energy and finite ${\cal K}(0)$, the localization length turns out to be proportional to $E$. Therefore, $L_{loc}$ vanishes as the electron energy tends to zero,
\begin{equation}\label{1DCP-LlocE}
L_{loc}\propto E=k^2,\qquad\mathrm{when}\qquad E\to0.
\end{equation}
Another situation emerges for photonic random media. Here the wave number is $k=\omega\sqrt{\varepsilon_0}/c$, and the variance of the scattering potential is estimated as $V_0^2\propto k^4\propto\omega^4$, see Eq.~\eqref{1DCP-Schreq} and the corresponding comments. Therefore, the frequency dependence of the Lyapunov exponent $\lambda$ is quadratic for long wavelengths,
\begin{equation}\label{1DCP-LlocOmega}
\lambda=L_{loc}^{-1}\propto k^2\propto\omega^2,\qquad\mathrm{when}\qquad\omega\to0.
\end{equation}
Thus, the localization length diverges as the wave frequency vanishes. As one can see, the localization is enhanced for electrons and suppressed for photons (as well as for phonons) with an energy/frequency decrease.

Before we start with the discussion of implementation of the expression \eqref{1DCP-Lloc} for the inverse localization length, it is worthwhile make some comments. First, one should stress that the general result \eqref{1DCP-Lloc} also arising in other approaches, is an asymptotic one. This means that the higher terms are non-controlled, however, they can be neglected in the limit $V_0^2\to0$. Second, the main assumptions used in the derivation of this expression are based on the validity of ensemble averaging over different realizations of disorder. In particular, a possibility of introducing fast and slow variables is assumed, allowing one to perform such an averaging. As noted above, the condition \eqref{1DCP-TwoScale} justifying these approaches, states that two macroscopic (slow) lengths, the localization length $L_{loc}$ and sample size $L$, are much larger than two microscopic (fast) lengths, the wavelength $k^{-1}$ and the correlation length $R_c$ determining the maximal value of the power spectrum ${\cal K}(2k)$. Note that for any fixed value of $k$ and $R_c$, this condition can be always fulfilled due to the asymptotic character of the expression \eqref{1DCP-Lloc}, i.e. in the limit $V_0^2 \rightarrow 0$.

The canonical expression \eqref{1DCP-Lloc} indicates that all features of the electron/wave transmission through the 1D disordered media depend on the two-point correlations in the random scattering potential. In particular, if the power spectrum ${\cal K}(2k)$ is very small or vanishes within some interval of the wave number $k$, the localization length $L_{loc}$ is very large ($L_{loc}\gg L$) or diverges. Evidently, the localization effects can be neglected in this case and the disordered structure even of a very large length will be fully transparent. This means that, in principle, by a proper choice of disorder one can design the structures with a selective (\emph{anomalous}) ballistic transport within a prescribed range of $k$.

Thus, the important practical problem arises of how to construct random potentials $V(x)$ from the prescribed power spectrum ${\cal K}(k_x)$. This problem can be solved by employing a widely used \emph{convolution method} that was originally proposed by Rice \cite{R54}. The modern applications of this method for the generation of random structures with specific short- or long-range correlations can be found in Refs.~\cite{S88,F88,WO95,CMHS95,MHSS96,RS99,GS99,CGK06,IKMU07,Ao08} (for other references see along this paper).

The convolution method consists of the following steps. First, starting from a desirable form of the power spectrum ${\cal K}(k_x)$, one has to derive the \emph{modulation function} $G(x)$ whose Fourier transform is ${\cal K}^{1/2}(k_x)$,
\begin{equation}\label{1DCP-MF}
G(x)=\int_{-\infty}^{\infty}\frac{dk_x}{2\pi}\,{\cal K}^{1/2}(k_x)\exp\left(ik_xx\right).
\end{equation}
After, the random profile of the scattering potential $V(x)$ is generated as a convolution of the delta-correlated random process $\alpha(x)$ with the modulation function $G(x)$,
\begin{equation}\label{1DCP-VG}
V(x)=V_0\,\int_{-\infty}^\infty\,dx'\,\alpha(x-x')\,G(x').
\end{equation}
Here the white noise $\alpha(x)$ is determined by the standard properties,
\begin{equation}\label{1DCP-AlphaCorr}
\langle\alpha(x)\rangle=0, \qquad\qquad \langle\alpha(x)\alpha(x')\rangle=\delta(x-x'),
\end{equation}
and in practice can be easily created with the use of random number generators.

The above expressions allow us to solve the inverse scattering problem of constructing random potentials from their power spectrum (or, the same, from the binary correlator of disorder). Note that this construction is possible in the case of weak disorder only. That is why only the binary correlator is involved in the reconstruction of $V(x)$ while the higher correlators do not contribute. Note also that the potential obtained by the proposed method is not unique. Indeed, there is an infinite set of delta-correlated random processes $\alpha(x)$. Therefore, substituting various realizations of $\alpha(x)$ into Eq.~\eqref{1DCP-VG}, we obtain different realizations of the potential with the same power spectrum ${\cal K}(k_x)$.

\subsection{Gaussian correlations}
\label{2.10}

In order to demonstrate how to realize the properties of correlated disorder, let us first consider a widely used Gaussian correlations,
\begin{subequations}\label{1DCP-GausCorPS}
\begin{eqnarray}
K(x)&=&\exp\left(-k_c^2x^2\right),\label{1DCP-GausCor}\\[6pt]
{\cal K}(k_x)&=&\sqrt{\pi}\,k_c^{-1}\exp\left(-k_x^2/4k_c^2\right).\label{1DCP-GausPS}
\end{eqnarray}
\end{subequations}
The Gaussian correlator $K(x)$ and power spectrum ${\cal K}(k_x)$ decrease exponentially on the scale of the correlation length $R_c\sim k_c^{-1}$ and correlation wave number $k_c$, respectively.

Using the convolution method \eqref{1DCP-MF} -- \eqref{1DCP-AlphaCorr} one can obtain that the potential profile $V(x)$ with the Gaussian correlation properties \eqref{1DCP-GausCorPS} is described by the function,
\begin{equation}\label{1DCP-VGaus}
V(x)=\frac{V_0\sqrt{2k_c}}{\pi^{1/4}}\,\int_{-\infty}^\infty\,dx'\,\alpha(x-x')\,\exp\left(-2k_c^2x'^2\right).
\end{equation}
Correspondingly, the inverse localization length \eqref{1DCP-Lloc} takes the following explicit form,
\begin{equation}\label{1DCP-LlocGaus}
L_{loc}^{-1}(k)=\frac{V_0^2\sqrt{\pi}}{8k_c}\, \frac{\exp\left(-k^2/k_c^2\right)}{k^2}.
\end{equation}
For electrons, when the variance $V_0^2$ does not depend on the wave number $k$, the localization length increases exponentially with $k$ from the zero value at $k=0$, see Eq.~\eqref{1DCP-LlocE}. Clearly, in the vicinity of $k=0$ the localization length is always much smaller than the length $L$, thus, the structure is non-transparent. Thus, with an increase of $k$ one can observe the crossover from the localized transport \eqref{1DCP-TResLoc} to the ballistic one \eqref{1DCP-TResBal}. For the Gaussian correlations \eqref{1DCP-GausCorPS}, both the crossing point where $L_{loc}(k)=L$, and the crossover width depend on the values of $k_c$ and $L$. Because of this dependence and a smooth character of the crossover, the crossing point cannot be regarded as a true mobility edge. However, one can see that the longer the correlation length $k_c^{-1}$, the larger the localization length $L_{loc}(k)$. Hence, the smaller is the localization region and the narrower is the crossover. One can conclude that, in general, \emph{Gaussian correlations suppress localization}.

The potential profile $V(x)$ with Gaussian correlations admits the uncorrelated disorder of the white-noise type. Indeed, from Eqs.~\eqref{1DCP-GausCorPS} one can readily obtain the delta-like correlator and constant power spectrum, compare with Eq.~\eqref{1DCP-Wdelta},
\begin{subequations}\label{1DCP-WNCorPS}
\begin{eqnarray}
K(x)&=&\sqrt{\pi}\,k_c^{-1}\delta(x),\\[6pt]
{\cal K}(k_x)&=&\sqrt{\pi}\,k_c^{-1},\qquad\mathrm{when}\quad k_c^{-1}\to0,\quad V_0^2k_c^{-1}=\mathrm{const}.
\end{eqnarray}
\end{subequations}
The convolution method \eqref{1DCP-MF} -- \eqref{1DCP-AlphaCorr} results in the following expression for the scattering potential $V(x)$:
\begin{equation}\label{1DCP-V-WN}
V(x)=\frac{V_0\pi^{1/4}}{\sqrt{k_c}}\,\alpha(x).
\end{equation}
According to Eq.~\eqref{1DCP-Lloc}, the localization length reads
\begin{equation}\label{1DCP-LlocWN}
L_{loc}^{-1}(k)=\frac{V_0^2\sqrt{\pi}}{8k_ck^2}.
\end{equation}
The expressions \eqref{1DCP-WNCorPS} -- \eqref{1DCP-LlocWN} can be considered as the asymptotic limits of the corresponding equations \eqref{1DCP-GausCorPS} -- \eqref{1DCP-LlocGaus} in the case of small-scale potential, $(k/k_c)^2\ll1$.

The comparison of Eqs.~\eqref{1DCP-LlocWN} and \eqref{1DCP-LlocGaus} leads to the conclusion: the best way to observe localized transport in continuous random potentials is to employ an uncorrelated disordered potential. Indeed, the condition $L_{loc}(k)\ll L$ is stronger for the Gaussian correlations than for the white-noise case. On the other hand, for large-scale Gaussian correlations with $(k/k_c)^2\gg1$, the ballistic regime \eqref{1DCP-TResBal} can be realized even for such structure lengths $L$ and wave numbers $k$ for which the strong localization \eqref{1DCP-TResLoc} takes place for delta-like correlations. Again, this fact confirms that the Gaussian correlations suppress the localization.

\subsection{Correlated disorder and mobility edges}
\label{2.11}

The important point in the theory of correlated disorder is that one can arrange a very \emph{sharp transition} between ballistic and localized regimes of electron/wave transport at prescribed points on $k$-axis. It is clear that in this case the power spectrum ${\cal K}(k_x)$ has to vanish abruptly at these points. This means that the binary correlator $K(x-x')$ has to be a slowly decaying function of the distance $x-x'$. In other words, the random scattering potentials $V(x)$ that give rise to the combination of ballistic and localized transport windows, should have specific long-range correlations along the structure. Because of the abrupt character, the transition points can be regarded as \emph{effective mobility edges}.

As pointed out above, a statistical treatment is meaningful if the scale of decrease $R_c\sim k_c^{-1}$ of the correlator $K(x)$ is much smaller than both the sample length $L$ and the localization length $L_{loc}$. In this connection one should stress that the long-range correlations we speak about, actually do not assume large values of $R_c$. Indeed, the sharpness of the transition is defined by the form of the binary correlator rather than by the value of its correlation length.

Let us now demonstrate how the suggested approach can be applied to the problem of emergence of the (effective) mobility edges in one-dimensional disordered models. For this, we consider the power spectrum of the following rectangular form:
\begin{equation}\label{1DCP-WOD-FTW}
{\cal K}(k_x)=\frac{\pi}{2(k_+-k_-)}\left[\Theta(2k_+-|k_x|)-\Theta(2k_--|k_x|)\right],\qquad k_+>k_->0.
\end{equation}
Here $\Theta(x)$ is the Heaviside unit-step function, $\Theta(x<0)=0$ and $\Theta(x>0)=1$. The characteristic wave numbers $k_\pm$ are the correlation parameters to be specified. Note that such a power spectrum has been recently employed to create specific rough surfaces in the experimental study of enhanced backscattering \cite{WO95}. This spectrum was also used in the theoretical analysis of light scattering from the amplifying media \cite{SLM01}, as well as, in the study of localization of plasmon polaritons on random surfaces \cite{MSLM01}.

According to definition \eqref{1DCP-FTW} the binary correlator $K(x)$ corresponding to the power spectrum \eqref{1DCP-WOD-FTW}, has the following form:
\begin{equation}\label{1DCP-WOD-W}
K(x)=\frac{\sin(2k_+x)-\sin(2k_-x)}{2(k_+-k_-)x}.
\end{equation}
In Eqs.~\eqref{1DCP-WOD-FTW} and \eqref{1DCP-WOD-W} the factor $1/2(k_+-k_-)$ provides the normalization requirement \eqref{1DCP-Wnorm}, or equivalently, $K(0)=1$.

Following the recipe \eqref{1DCP-MF} -- \eqref{1DCP-AlphaCorr}, one can construct the scattering potential $V(x)$ with the correlation properties \eqref{1DCP-WOD-W},
\begin{equation}\label{1DCP-WOD-V}
V(x)=\frac{V_0}{\sqrt{2\pi}}\,\int_{-\infty}^\infty\,dx'\,\alpha(x-x')\,
\frac{\sin(2k_+x')-\sin(2k_-x')}{(k_+-k_-)^{1/2}x'}.
\end{equation}
One should note that the potential \eqref{1DCP-WOD-V} with the binary correlator \eqref{1DCP-WOD-W} and power spectrum \eqref{1DCP-WOD-FTW} is substantially different from the commonly used delta-correlated noise \eqref{1DCP-V-WN} or random process \eqref{1DCP-VGaus} with fast-decaying Gaussian correlations \eqref{1DCP-GausCorPS}. Here the potential \eqref{1DCP-WOD-V} is specified by two parameters, $(2k_+)^{-1}$ and $(2k_-)^{-1}$, and has long tails in the expression for the two-point correlator \eqref{1DCP-WOD-W}. The existence of such tails manifests the long-range correlations in the potential originated from the step-wise discontinuities at $k_x=\pm 2k_\pm$ in the power spectrum \eqref{1DCP-WOD-FTW}. In fact, these two parameters, $(2k_+)^{-1}$ and $(2k_-)^{-1}$, are nothing but two correlation lengths of the two-point correlator. Thus, the form of the correlator determines the existence of the mobility edges, while the correlation lengths specify their positions.

From Eqs.~\eqref{1DCP-Lloc} and \eqref{1DCP-WOD-FTW} one can find the inverse localization length,
\begin{equation}\label{1DCP-WOD-Lloc}
L_{loc}^{-1}(k)=\frac{\pi V_0^2}{16(k_+-k_-)}\frac{\Theta(k_+-k)\Theta(k-k_-)}{k^2}.
\end{equation}
As one can see, there are two mobility edges at $k=k_-$ and $k=k_+$. The localization length $L_{loc}(k)$ diverges below the first point, $k=k_-$, and above the second one, $k=k_+$. Between these points, for $k_-<k<k_+$, the localization length \eqref{1DCP-WOD-Lloc} has a finite value and smoothly increases with an increase of the wave number $k$.

Let us now choose the parameters for which the regime of strong localization occurs at the upper transition point $k=k_+$. This automatically provides strong localization within the whole interval $k_-<k<k_+$. The condition of such situation reads
\begin{equation}\label{1DCP-WOD-LocTr}
\frac{L}{L_{loc}(k_+)}=\frac{\pi V_0^2L}{16k_+^2(k_+-k_-)}\gg1.
\end{equation}
As a result, there are two regions of transparency for the samples of finite length $L$ with random potentials of chosen type. Between these regions the average transmittance $\langle T_L\rangle$ is exponentially small according to the expression \eqref{1DCP-TResLoc}. Due to this fact, the system exhibits localized transport within the interval $k_-<k<k_+$ and the ballistic regime with $\langle T_L\rangle=1$ outside this interval. Experimentally, one can observe that with an increase of the wave number $k$ the perfect transparency below $k=k_-$ abruptly alternates with a complete reflection, and recovers at $k=k_+$. From Eqs.~\eqref{1DCP-WOD-Lloc} and \eqref{1DCP-WOD-LocTr}, one can see that the smaller the value $k_+-k_-$ of the reflecting region, the smaller the localization length $L_{loc}(k)$ and, consequently, the stronger is the localization within this region. Thus, the correlations not only suppress the localization, as mentioned in previous Section. In contrast, they can also enhance it.

\section{Quasi-one-dimensional stratified media}
\label{3}

In order to better understand the impact of long-range correlations, in this Section we extend our study to quasi-one-dimensional (quasi-1D) geometry. Specifically, we shall discuss the anomalous transport and mobility edges in multi-mode waveguides (or many-channel conducting electron wires) with the so-called \emph{stratified} or \emph{layered disorder} \cite{IM04,MI04,IM05}. In spite of apparent simplicity, this model is quite typical in many applications and allows one to reveal the general peculiarities of the anomalous ballistic transport in quasi-1D systems. The model describes the scattering of classical waves or electrons through the structures with disorder depending on the longitudinal coordinate only. Therefore, it absorbs both the properties of 1D correlated media and those specific for quasi-1D systems. On the one hand, the transport through any of open channels is independent from the others. On the other hand, since the localization length in each channel strongly depends on its number, the total transmittance is a complicated combination of partial 1D transmittances. Thus, the concept of the single parameter scaling turns out to be broken for the total waveguide transport in contrast with quasi-1D bulk-disordered models. As a result, the effect of long-range correlations turns is much more sophisticated, leading to quite unexpected phenomena. In particular, it is shown that with a proper choice of the binary correlator one can arrange the situation when some of the channels are completely transparent and the others are closed. This interplay between localized and transparent channels gives rise to the effect of a \emph{non-monotonic} step-wise dependence of the transmittance (dimensionless conductance) on the wave number. The results may find practical applications for fabrication of electromagnetic/acoustic waveguides, optic fibers and electron nanoconductors with non-conventional selective transport.

\subsection{Basic relations}
\label{3.1}

\begin{figure}[ht]
\includegraphics[width=\textwidth]{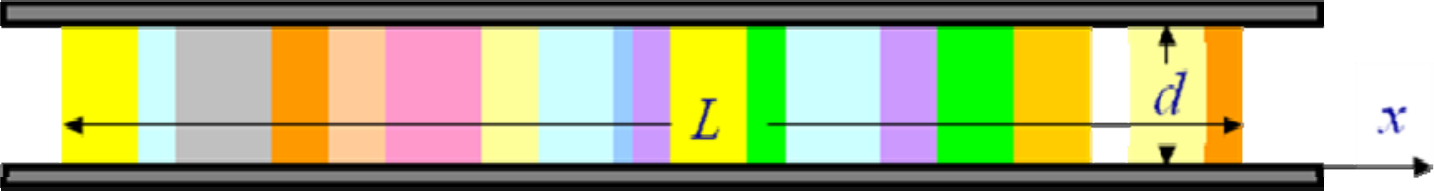}
\caption{(color online). Waveguiding structure with longitudinal stratification.}\label{Q1D-Fig01}
\end{figure}

We consider a planar waveguide (or conducting electron wire) of width $d$, stretched along the $x$-axis. The $z$-axis is directed in transverse to the waveguide direction so that one (lower) edge of the waveguide is $z=0$ and the other (upper) edge is $z=d$. The waveguide is assumed to have a \emph{stratified disorder} inside the finite region $|x|<L/2$ of length $L$, which depends on $x$ only (see Fig.~\ref{Q1D-Fig01}). Thus, the scattering is confined within a region of the $(x,z)$-plane defined by
\begin{equation}\label{Q1D-Domain}
-L/2<x<L/2,\qquad\qquad 0\leq z\leq d.
\end{equation}
As a physically plausible model for the stratified disorder we employ the continuous and statistically homogeneous random potential $V(x)$ with the correlation properties described by Eqs.~\eqref{1DCP-VCor}, \eqref{1DCP-FTW} and \eqref{1DCP-Wnorm}. One should note that in this case the scale $R_c$ of a decrease of the binary correlator $K(x)$ is of the order of the characteristic scale of stratification.

The transmittance (dimensionless conductance) \eqref{1DCP-Tdef} is defined by the expression,
\begin{equation}\label{Q1D-TKubo}
T_L=-\frac{4}{L^2}\int_{-L/2}^{L/2}dxdx'\;\int_{0}^{d}dzdz'\,\frac{\partial{\cal G}(x,x';z,z';k)}{\partial x}\,
\frac{\partial{\cal G}^*(x,x';z,z';k)}{\partial x'}.
\end{equation}
The retarded Green function ${\cal G}(x,x';z,z';k)$ obeys the equation,
\begin{equation}\label{Q1D-SchrG}
\left[\frac{\partial^{2}}{\partial x^{2}}+\frac{\partial^{2}}{\partial z^{2}}+ k^{2}-V(x)\right]{\cal G}(x,x';z,z';k)=\delta(x-x')\delta(z-z').
\end{equation}
This equation is complemented by the zero Dirichlet boundary conditions at both walls of the waveguide, ${\cal G}(x,x';z=0,z';k)={\cal G}(x,x';z=d,z';k)=0$, whereas in the $x$-direction the structure is open. It is noteworthy that the equations \eqref{Q1D-TKubo} and \eqref{Q1D-SchrG} for the quasi-1D problem are considered as a natural generalization of the one-dimensional model described by Eqs.~\eqref{1DCP-TKubo} and \eqref{1DCP-SchrG}.

Since the scattering potential $V(x)$ depends only on the longitudinal coordinate $x$, the quasi-1D Green function ${\cal G}(x,x';z,z';k)$ can be presented in terms of the normal waveguide modes,
\begin{equation}\label{Q1D-GNM}
{\cal G}(x,x';z,z';k)=\frac{2}{d}\sum_{n=1}^{N_d} \sin\left(\frac{\pi nz}{d}\right)\sin\left(\frac{\pi nz'}{d}\right)\,{\cal G}(x,x';k_n).
\end{equation}
The longitudinal mode-component ${\cal G}(x,x';k_n)$ is governed by the quasi-one-dimensional wave equation that directly follows from
Eqs.~\eqref{Q1D-SchrG} and \eqref{Q1D-GNM}. It differs from the one-dimensional equation \eqref{1DCP-SchrG}, due to the change of total wave number $k$ by the quantum value $k_n$ of longitudinal wave number $k_x$ for the $n$th waveguide mode,
\begin{equation}\label{Q1D-kn}
k_n=\sqrt{k^2-(\pi n/d)^2}.
\end{equation}
Evidently, the transport properties are contributed only by those waveguide modes that can propagate along the structure, i.e. have a real value of $k_n$. These \emph{propagating modes} are associated with \emph{conducting channels}. As follows from Eq.~\eqref{Q1D-kn}, the total number $N_d$ of propagating modes is equal to the integer part $[...]$ of the ratio $kd/\pi$,
\begin{equation}\label{Q1D-Nd}
N_d=[kd/\pi].
\end{equation}
The waveguide modes with indices $n>N_d$ corresponding to imaginary values of $k_n$ and not contributing to the electron/wave transmission are called the \emph{evanescent modes}.

From Eqs.~\eqref{Q1D-SchrG} and \eqref{Q1D-GNM} it can be seen that since the stratified disorder does not depend on the transverse coordinate $z$, there is no coupling between the propagating modes. Therefore, the stratified waveguide is equivalent to a set of $N_d$ one-dimensional non-interacting conducting channels. That is why in full agreement with the Landauer concept \cite{L92}, the \emph{total transmittance} \eqref{Q1D-TKubo} of the stratified structure is expressed as a sum of independent \emph{partial transmittances} $T_n$ corresponding to the $n$th propagating mode,
\begin{equation}\label{Q1D-Ttot}
T_L=\sum_{n=1}^{N_d}T_n(L).
\end{equation}

In such a way we have reduced the transport problem for the quasi-1D disordered media to the consideration of the transport through a number of 1D channels with common random potential $V(x)$. The scattering inside every of these channels is described by Eqs.~\eqref{1DCP-TKubo} and \eqref{1DCP-SchrG} with $k_n$ instead of $k$, and hence, it is entirely consistent with the phenomenon of the 1D Anderson localization. Specifically, the average mode-transmittance $\langle T_n\rangle$ is described by the universal expressions \eqref{1DCP-1DTn} -- \eqref{1DCP-VarLnBalLoc}. The mode localization length $L_{loc}(k_n)$ associated with specific $n$th channel, is determined by the canonical expression \eqref{1DCP-Lloc} in which the total wave number $k$ should be replaced by the longitudinal wave number $k_n$,
\begin{equation}\label{Q1D-nLloc}
L_{loc}^{-1}(k_n)=\frac{V_0^2}{8k_n^2}\,{\cal K}(2k_n).
\end{equation}
Thus, the partial transport in each $n$th conducting channel obeys the concept of the single parameter scaling. In particular, for
$L_{loc}(k_n)/L\gg1$, the average transmittance $\langle T_n\rangle$ exhibits the ballistic behavior, therefore, the corresponding $n$th normal modes are fully transparent, see Eq.~\eqref{1DCP-TResBal}. In contrast, the transmittance $\langle T_n\rangle$ is exponentially small in line with
Eqs.~\eqref{1DCP-TResLoc} and \eqref{1DCP-repavT}, when the mode localization length is much smaller than the length of the waveguide, $L_{loc}(k_n)/L\ll1$. This implies a strong electron/wave localization in the $n$th channel.

\subsection{Coexistence of ballistic and localized transport}
\label{3.2}

The main feature of the mode localization length $L_{loc}(k_n)$ is its a rather strong dependence on the channel index $n$. One can see
from the definition \eqref{Q1D-nLloc} that the larger $n$ is, the smaller the mode localization length $L_{loc}(k_n)$ and, consequently, the stronger is the coherent scattering within this mode. This strong dependence is due to the squared wave number $k_n$ in the denominator of Eq.~\eqref{Q1D-nLloc}. Evidently, with an increase of the mode index $n$ the value of $k_n$ decreases. An additional dependence appears because of the stratification power spectrum ${\cal K}(2k_n)$. Since the binary correlator $K(x)$ of the random stratification is a decreasing function of $|x|$, the numerator ${\cal K}(2k_n)$ increases with $n$ (note that it is a constant for the delta-correlated stratification only). Therefore, both the numerator and denominator contribute in the same direction for the dependence of $L_{loc}(k_n)$ on $n$. As a result, we arrive at the hierarchy of mode localization lengths,
\begin{equation}\label{Q1D-nLlocHierarchy}
L_{loc}(k_{N_d})<L_{loc}(k_{N_d-1})<...<L_{loc}(k_2)<L_{loc}(k_1).
\end{equation}
The smallest mode localization length $L_{loc}(k_{N_d})$ corresponds to the highest (last) channel with the mode index $n=N_d$, while
the largest mode localization length $L_{loc}(k_1)$ is associated with the lowest (first) channel with $n=1$. A similar hierarchy was found in Refs.~\cite{GTSN98,SFYM98,SFMY99} for the attenuation lengths in the model of quasi-1D waveguides with rough surfaces.

Thus, a remarkable phenomenon arises. On the one hand, the concept of the single parameter scaling holds for any of $N_d$ conducting channels whose partial transport is characterized solely by the ratio $L/L_{loc}(k_n)$. On the other hand, this concept turns out to be broken for the total waveguide transport. Indeed, due to the hierarchy \eqref{Q1D-nLlocHierarchy} of the mode localization lengths $L_{loc}(k_n)$, the total average transmittance \eqref{Q1D-Ttot} depends on the whole set of scaling parameters $L/L_{loc}(k_n)$. This fact is in contrast with the quasi-1D
bulk-disordered models, for which all transport properties are shown to be described by one parameter only.

The interplay between the hierarchy of the mode localization lengths $L_{loc}(k_n)$ on the one hand, and the single parameter scaling for every partial mode-transmittance $\langle T_n\rangle$ on the other hand, gives rise to a new phenomenon of the intermediate coexistence regime, in addition to the regimes of ballistic or localized transport known in the 1D geometry.

{\bf(i)} \emph{Ballistic transport}. If the \emph{smallest} mode localization length $L_{loc}(k_{N_d})$ is much \emph{larger} than the scattering region of size $L$, all conducting channels are open. They have almost unit partial transmittance, $T_n(L)\approx1$, hence, the stratified waveguide is fully transparent. In this case, the total average transmittance \eqref{Q1D-Ttot} is equal to the total number of the propagating modes,
\begin{equation}\label{Q1D-Tbal}
\langle T_L\rangle\approx N_d\qquad\mbox{for}\qquad L\ll L_{loc}(k_{N_d}).
\end{equation}

{\bf(ii)} \emph{Localized transport}. In contrast, if the \emph{largest} of the lengths $L_{loc}(k_1)$ is much \emph{smaller} than the waveguide length $L$, all the propagating modes are strongly localized and the waveguide is non-transparent. Its total average transmittance is exponentially small, in accordance with Eqs.~\eqref{Q1D-Ttot} and \eqref{1DCP-TResLoc},
\begin{equation}\label{Q1D-Tloc}
\langle T_L\rangle\approx\frac{\pi^3}{16\sqrt{\pi}}\left[L/2L_{loc}(k_1)\right]^{-3/2}\exp\left[-L/2L_{loc}(k_1)\right]\qquad
\mbox{for}\qquad L_{loc}(k_1)\ll L.
\end{equation}

{\bf(iii)} \emph{Coexistence transport}. The intermediate situation arises when the \emph{smallest} localization length $L_{loc}(k_{N_d})$ of the last (highest) $N_d$th mode is \emph{smaller}, while the \emph{largest} localization length $L_{loc}(k_1)$ of the first (lowest) mode is \emph{larger} than the waveguide length $L$,
\begin{equation}\label{Q1D-CoTr}
L_{loc}(k_{N_d})\ll L\ll L_{loc}(k_1).
\end{equation}
In this case a remarkable phenomenon of the coexistence of ballistic and localized transport occurs. Namely, while the \emph{lower} modes are in the ballistic regime, the \emph{higher} modes display strong localization.

These transport regimes can be observed experimentally when, for example, the stratification is either the random delta-correlated process, Eqs.~\eqref{1DCP-WNCorPS} -- \eqref{1DCP-LlocWN} of the white-noise type with a constant power spectrum $V_0^2{\cal K}(k_x)={\cal K}_0=\mathrm{const}$, or a colored noise, Eqs.~\eqref{1DCP-GausCorPS} -- \eqref{1DCP-LlocGaus} with the Gaussian correlations. As a demonstration, let us consider the stratified waveguide with a large number of conducting channels,
\begin{equation}\label{Q1D-NdLarge}
N_d=[kd/\pi]\approx kd/\pi\gg1,
\end{equation}
and with a random potential $V(x)$ having the Gaussian correlator with the power spectrum \eqref{1DCP-GausCorPS}. In this case it is convenient to introduce two parameters,
\begin{equation}\label{Q1D-AlphaDelta}
\alpha=\frac{L}{L_{loc}^{w}(k_1)}=\frac{{\cal K}_0L}{8k^2},\qquad\qquad
\delta=\frac{L_{loc}^{w}(k_{N_d})}{L_{loc}^{w}(k_1)}=\frac{2\{kd/\pi\}}{(kd/\pi)}\ll1,
\end{equation}
where $L_{loc}^{w}(k_1)$ and $L_{loc}^{w}(k_{N_d})$ refer, respectively, to the largest and smallest mode localization lengths in the limit case of the white-noise potential, i.e., when $k_c\to\infty$, $\sqrt{\pi}\,V_0^2k_c^{-1}={\cal K}_0=\mathrm{const}$. Here $\{kd/\pi\}$ is the fractional part of the mode parameter $kd/\pi$.

One can find that for the Gaussian correlations \eqref{1DCP-GausCorPS}, all propagating modes are strongly localized when
\begin{equation}\label{Q1d-GCTloc}
\exp(k^2/k_c^2)\ll\alpha.
\end{equation}
Note that this inequality is stronger than that valid for the white-noise case, for which $\alpha\gg1$.

The intermediate situation with the coexistence transport occurs when
\begin{equation}\label{Q1D-GCTcoex}
\delta\exp(\delta k^2/k_c^2)\ll\alpha\ll\exp(k^2/k_c^2).
\end{equation}
Therefore, the longer range $k_c^{-1}$ of the correlated disorder, the simpler the conditions \eqref{Q1D-GCTcoex} of the coexistence of ballistic and localized transport.

Finally, the waveguide is almost perfectly transparent in the case when
\begin{equation}\label{Q1D-GC-BalTr}
\alpha\ll\delta\exp(\delta k^2/k_c^2).
\end{equation}
From this brief analysis one can conclude that the total localization is easily achieved for the white-noise stratification when the condition $\alpha\gg 1$ (which is weaker than \eqref{Q1d-GCTloc}) is fulfilled. However, at any given value of $\alpha\gg 1$ (fixed values of the waveguide length $L$, the wave number $k$ and the disorder strength ${\cal K}_0$), the intermediate or ballistic transport can be realized by a proper choice of the stratified correlated disorder, i.e. by a proper choice of the correlation scale $k_c^{-1}$. From Eq.~\eqref{Q1D-GCTcoex}, one can also understand that for the standard white-noise case the coexistence of ballistic and localized modes can be observed under the condition $\delta\ll\alpha\ll1$. Contrary, if $\alpha\ll\delta$, the quasi-1D structure with the white-noise stratification displays the ballistic transport.

\subsection{Correlated stratification and step-wise non-monotonic transmittance}
\label{3.3}

From the above analysis it becomes clear that in the quasi-1D stratified guiding structures with the delta-correlated or Gaussian correlations, the crossover from the ballistic to localized transport is realized through the successive localization of highest propagating modes. Otherwise, if we start from the totally localized transport regime, the crossover to the ballistic transport is realized via the successive opening (delocalization) of the lowest conducting channels.

A fundamentally different situation arises when random stratified media have specific long-range correlations. To show this, we have to note that the mode localization length $L_{loc}(k_n)$ of any $n$th conducting channel is entirely determined by the stratification power spectrum ${\cal K}(k_x)$, see Eq.~\eqref{Q1D-nLloc}. Therefore, if ${\cal K}(2k_n)$ abruptly vanishes for some wave numbers $k_n$, then $L_{loc}(k_n)$ diverges and the corresponding propagating mode appears to be fully transparent even for a large length of the waveguide.

Let us take the following binary correlator $K(x)$ with the corresponding step-wise power spectrum ${\cal K}(k_x)$:
\begin{subequations}\label{Q1D-WPScor}
\begin{eqnarray}
K(x)&=&\pi\delta(2k_cx)-\frac{\sin(2k_cx)}{2k_cx},\label{Q1D-Wcor}\\[6pt]
{\cal K}(k_x)&=&\frac{\pi}{2k_c}\,\Theta(|k_x|-2k_c).\label{Q1D-PScor}
\end{eqnarray}
\end{subequations}
Here $\delta(x)$ is the Dirac delta-function and the characteristic wave number $k_c>0$ is the correlation parameter to be specified, $R_c\sim k_c^{-1}$. Applying the convolution method defined by Eqs.~\eqref{1DCP-MF} -- \eqref{1DCP-AlphaCorr}, one can find that the profile of the random stratification having such correlations can be fabricated with the use of the potential,
\begin{equation}\label{Q1D-Vcor}
V(x)=\frac{V_0}{\sqrt{2\pi k_c}}\left[\pi\alpha(x)-\int_{-\infty}^{\infty}dx'\alpha(x-x')\frac{\sin(2k_cx')}{x'}\right].
\end{equation}
This potential has a quite sophisticated form. It is a superposition of the white noise and long-range correlated potential.

For the case under consideration, the inverse value of the mode localization length \eqref{Q1D-nLloc} takes the following explicit form:
\begin{equation}\label{Q1D-nLloc-cor}
L_{loc}^{-1}(k_n)=\frac{\pi V_0^2}{16k_c}\,\frac{\Theta(k_n-k_c)}{k_n^2}.
\end{equation}
This expression results in very interesting conclusions.

{\bf(i)} All \emph{low} propagating modes with the longitudinal wave numbers $k_n$ that exceed the correlation parameter $k_c$ ($k_n>k_c$), have finite mode localization length,
\begin{equation}\label{Q1D-Lnlow}
L_{loc}^{-1}(k_n>k_c)=\pi V_0^2/16k_ck_n^2.
\end{equation}
For a large enough waveguide length,
\begin{equation}\label{Q1D-Lnlowloc}
\frac{L}{L_{loc}(k_1)}=\frac{\pi V_0^2L}{16k_ck_1^2}\gg1,
\end{equation}
all of these modes are strongly localized. The requirement $k_n>k_c$ implies that the mode indices $n$ of localized channels are restricted from above by the condition,
\begin{equation}\label{Q1D-Nloc}
n\leq N_{loc}=[(kd/\pi)(1-k_c^2/k^2)^{1/2}]\Theta(k-k_c).
\end{equation}
Therefore, the integer $N_{loc}$ should be regarded as the total number of localized and non-transparent modes.

{\bf(ii)} For \emph{high} conducting channels with $k_n<k_c$ the mode localization length diverges,
\begin{equation}\label{Q1D-Lnhigh}
L_{loc}^{-1}(k_n<k_c)=0.
\end{equation}
Therefore, each of the modes with the index $n>N_{loc}$ has the unit partial transmittance, $\langle T_n\rangle=1$, and displays the ballistic transport. These modes form a subset of ballistic, therefore, completely transparent, channels.

{\bf(iii)} The value of $N_{loc}$ determines the total number of localized modes. The total number $N_{bal}$ of ballistic modes is evidently equal to $N_{bal}=N_d-N_{loc}$. Since the localized modes do not contribute to the total average waveguide transmittance \eqref{Q1D-Ttot}, the latter is equal to the number of completely transparent modes $N_{bal}$ and do not depend on the waveguide length $L$,
\begin{equation}\label{Q1D-TtotCor}
\langle T_L\rangle=[kd/\pi]-[\sqrt{(kd/\pi)^2-(k_cd/\pi)^2}\,]\Theta(k-k_c).
\end{equation}
We remind that the square brackets stand for the integer part of the inner expression. It should be emphasized once more that in contrast with the usual situation characterized by the hierarchy of mode localization lengths \eqref{Q1D-nLlocHierarchy}, now \emph{low} propagating modes are \emph{localized} and non-transparent, while \emph{high} modes are \emph{ballistic} with a perfect transparency.

For the correlation parameter $k_c\ll k$, the number of localized modes is of the order of $N_d$,
\begin{equation}\label{Q1D-Nloc-kc>k}
N_{loc}\approx[(kd/\pi)(1-k_c^2/2k^2)]\quad\mbox{for}\quad k_c/k\ll1.
\end{equation}
Consequently, the number of ballistic modes $N_{bal}$ is small, or there are no such modes at all. Otherwise, if $k_c\to k$, the integer $N_{loc}$ is much smaller than the total number of propagating modes $N_d$,
\begin{equation}\label{Q1D-Nloc-kctok}
N_{loc}\approx[\sqrt{2}(kd/\pi)(1-k_c/k)^{1/2}]\ll N_d\quad\mbox{for}\quad 1-k_c/k\ll1.
\end{equation}
Here the number of transparent modes $N_{loc}$ is large. When $k_c>k_1$, the number $N_{loc}$ vanishes and all modes become fully transparent, in spite of their scattering by random stratification. In this case the correlated disorder results in a perfect transmission of quantum/classical waves. One can see that the point $k_1=k_c$ at which the wave number of the first mode $k_1$ is equal to the correlation wave number $k_c$, is, in fact, the total mobility edge that separates the region of a complete transparency from that where the lower modes are localized.

\begin{figure}[ht]
\begin{center}
\includegraphics[width=4.2in,height=2.4in,angle=0]{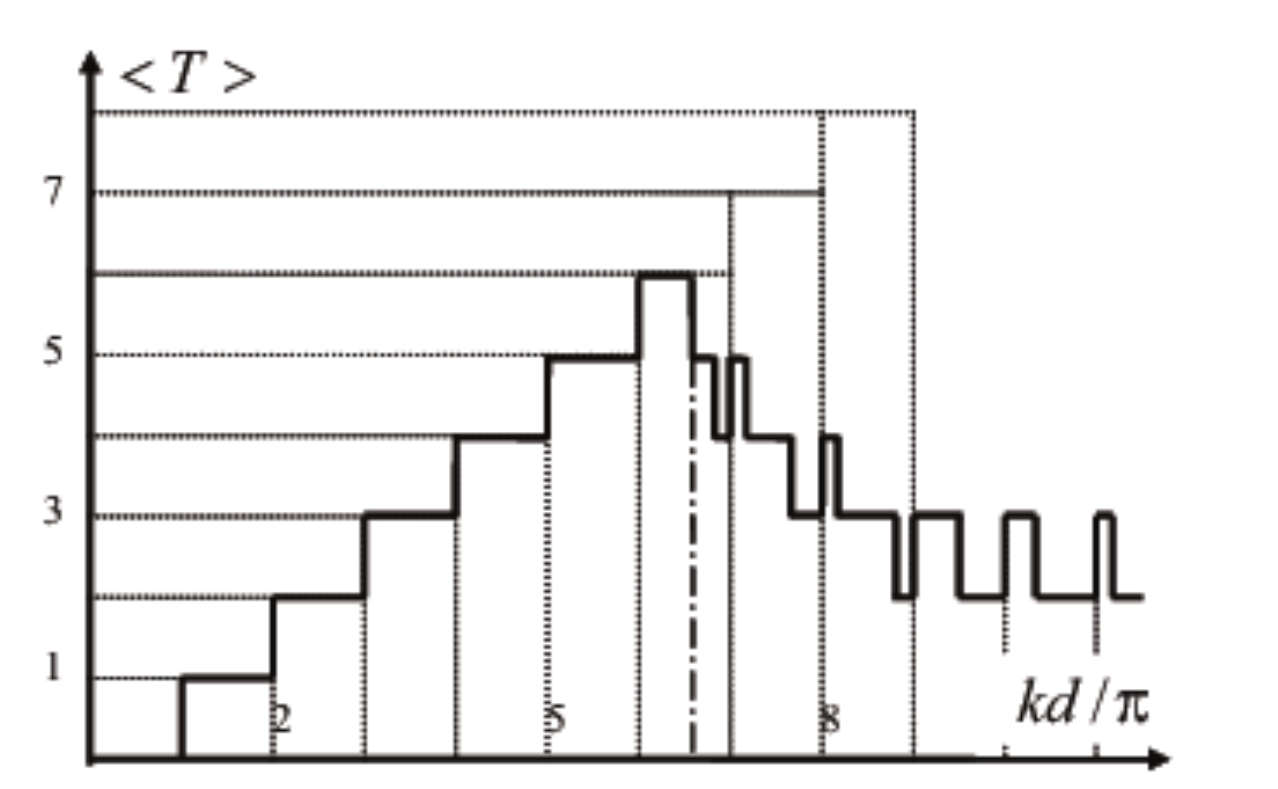}
\caption{Non-monotonic step-wise dependence of the transmittance \eqref{Q1D-TtotCor} of stratified structure versus the normalized wave number $kd/\pi$ for the dimensionless correlation parameter $k_cd/\pi=6.5$ (after \cite{IM05}).}\label{Q1D-Fig02}
\end{center}
\end{figure}

From the above analysis one can conclude that the transmittance \eqref{Q1D-TtotCor} of a quasi-1D multimode structure with the long-range correlated stratification reveals a quite unexpected {\it non-monotonic} step-wise dependence on the total wave number $k$ that is controlled by the correlation parameter $k_c$. An example of such a dependence is shown in Fig.~\ref{Q1D-Fig02}.

Within the region where the wave number of the lowest mode $k_1$ is smaller than the correlation wave number $k_c$ ($k_1<k_c$ or, the same, $kd/\pi<\sqrt{(k_cd/\pi)^2+1}$), the second term in Eq.~\eqref{Q1D-TtotCor} is absent, and all propagating modes are completely transparent. Here, the average transmittance exhibits a ballistic step-wise increase with an increase of the total wave number $k$. Each step ``up" arises for the integer value of the parameter $kd/\pi$, manifesting an emergence of a new conducting channel. Such a step-wise behavior of the total transmittance is similar to that known to occur in the quasi-1D \emph{non-disordered} structures (see, e.g., \cite{Wo88}).

With a subsequent increase of $k$, the first-mode wave number $k_1$ becomes larger than $k_c$ ($k_1\geq k_c$ or $kd/\pi\geq\sqrt{(k_cd/\pi)^2+1}$). Here, the transmittance shows not only standard steps ``up" associated with the first term in Eq.~\eqref{Q1D-TtotCor}) but also steps ``down" that are originated from the second term. In contrast with the steps ``up", the steps ``down" are formed by the correlated scattering and occur due to the successive abrupt localization of low modes at the corresponding mobility edges. The first step ``down" occurs at the total mobility edge $k_1=k_c$, where the first mode is localized. The second step ``down" is due to the mobility edge $k_2=k_c$ of the second mode, etc. The positions of the steps ``down" are at the integer values of the square root $\sqrt{(kd/\pi)^2-(k_cd/\pi)^2}$, see Eq.~\eqref{Q1D-TtotCor}, and, in general, do not coincide with the integer values of the mode parameter $kd/\pi$. The interplay between steps ``up" and ``down" results in a new kind of step-wise non-monotonic dependence of the total quasi-1D transmittance.

These results may find practical applications for the fabrication of electromagnetic/acoustic waveguides, optic fibers and electron nanodevices with a selective transport. For an abrupt dependence of the power spectrum at some energy, the transition from the ballistic to localized transport is expected to be sharp enough in order to observe them experimentally. For example, for the GaAs quasi-1D quantum-well structures with the effective electron mass $m_e=6.7\times10^{-2}m_0$, the Fermi-energy $E_F=7$ meV and $d=500$ nm, the number of channels is about $17$. Therefore, for the potential with $\pi/k_c\approx 80$ nm, with an increase of $k$ one can observe about $16$ local mobility edges characterized by the steps ``down", after initial $6$ steps ``up" in the conductance dependence on the wave number $k$.

\section{Tight-binding Anderson model: White noise}
\label{4}

\subsection{Definitions and general remarks}
\label{4.1}

In statistical physics many phenomena are described with the use of discrete lattices since in comparison with continuous systems they are more simpler for the rigorous mathematical analysis. On the other hand, the discreteness is the origin of various physical effects and has to be taken into account in physical models explicitly. Starting from the seminal paper of P.W. Anderson \cite{A58}, many of important results on quantum diffusion and localization have been obtained with the use of various discrete models.

In one-dimensional discrete models the potential energy of a quantum particle occupying the site with coordinate $x_n$ is associated with on-site energy $\epsilon_n$. The kinetic energy appears as a result of quantum transitions (hoppings) between the sites. By introducing the matrix elements $\vartheta_{mn}$ of the operator of kinetic energy, the stationary discrete Schrodinger equation can be written as follows,
\begin{equation}
\label{disShr}
\epsilon_n \psi_n + \sum_{m\neq n} {\vartheta_{nm}\psi_m} = E \psi_{n}\,,
\end{equation}
were $\psi_n$ is an eigenstate with energy $E$. This equation can be justified via different procedures, including the standard tight-binding approximation (see, for example, \cite{K05,LGP88}). Usually the matrix elements $\vartheta_{mn}$ decay fast away from the diagonal, $n=m$, therefore, the main contribution to the sum in Eq. (\ref{disShr}) comes from the nearest neighbors, i.e. from the cites $m=n \pm 1$. If only these two interactions are hold, Eq. (\ref{disShr}) simplifies to the well-known form,
\begin{equation}
\label{tb}
\vartheta_{n+1,n}\psi_{n+1} + \vartheta_{n-1,n}\psi_{n-1} + \epsilon_n \psi_n = E \psi_{n} .
\end{equation}
Being much easier for the analysis than the Schrodinger equation with the continuous potential, it keeps all the important properties of the medium. As one can see, they are entirely defined by two sets, off-diagonal matrix elements $\vartheta_{n\pm 1,n}$ and on-site energies $\epsilon_n$. It is worth mentioning that any continuous 1D Schrodinger equation can be written in the above form, giving rise to a linear relation between the values of the wave function at three points on the axis $x$ \cite{K86a,SMD94}.

If the lattice is periodic, $x_n=nd$, and both the off-diagonal elements and site energies are coordinate-independent, $\vartheta_{n\pm1,n}=\vartheta$ and $\epsilon_n = \epsilon$, the solution of Eq.~(\ref{tb}) takes the form of plane waves, $\psi_n \sim \exp(iqdn)$. Therefore, the dispersion relation determining the energy spectrum has the form,
\begin{equation}
\label{disp1}
E= \epsilon + 2\vartheta \cos \mu\,.
\end{equation}
Here the total phase shift $\mu=qd$ of the wave function in the period $d$ of the lattice determines the Bloch vector $q$ ($0 \leq q \leq \pi/d$), giving rise to
the symmetric band of allowed energies, $-2\vartheta < E-\epsilon < 2\vartheta$. In what follows, we assume $d=1$.

In many physical realizations of the model, such as doped semiconductors or alloys, the parameters $\vartheta_{n\pm1,n}$ and $\epsilon_n$ can fluctuate randomly due to various reasons. In such a case the properties of eigenstates and energy spectrum are strongly affected by disorder, and the main problem is to find these properties in dependence on the model parameters. One can speak about two limiting cases of discrete tight-binding models. They are classified according to a different structure of the Hamiltonian represented by tri-diagonal matrix,
\begin{equation}
{\cal H}_{mn} = \vartheta_{mn} (\delta_{m+1,n} + \delta_{m-1,n}) + \epsilon_n \delta_{mn}
\label{tri-diagonal}\,.
\end{equation}
If the hopping amplitudes are coordinate-independent and isotropic, $\vartheta_{n\pm 1,n}= \vartheta$, this is the so-called {\it diagonal disorder}. In this case the random elements of the matrix ${\cal H}_{mn}$ appear only on the diagonal, ${\cal H}_{nn}=\epsilon_n$. This Hamiltonian describes quantum diffusion of a particle on a lattice with fluctuating site energies, however, with equal hopping amplitudes. Another case is the {\it off-diagonal disorder}, when the on-site energies are the same, $\epsilon_n = \epsilon$, however, the hopping amplitudes are random numbers.

Our main interest in this Section is the diagonal disorder for which the model can be written in the following form,
\begin{equation}
\label{tb diagonal}
\psi_{n+1} + \psi_{n-1}=(E - \epsilon_n) \psi_n.
\end{equation}
Here and below the energies $E$ and $\epsilon_n$ are measured in the units of the constant hopping parameter $\vartheta$. If random potential $\epsilon_n$ has nonzero mean value $\langle\epsilon_n\rangle$, it can be included in $E$, therefore, the eigenenergy is counted from the mean value of the random potential. For a white-noise potential, $\langle\epsilon_m \epsilon_n \rangle = \sigma^2 \delta_{mn}$, all eigenstates are known to be localized \cite{I73,T74,M78,P80,KS80}, independently of the value of the variance $\sigma^2$ of disorder. A lot of efforts, both analytical and numerical, have been spent in order to obtain the localization length that is the most important characteristic of eigenstates. A summary of some results for 1D uncorrelated disorder can be found in a number of reviews (see, for example, \cite{I73,P94}).

\subsection{Hamiltonian map approach}
\label{4.2}

In this Section we describe an approach \cite{IKT95} based on the representation of the Schr\"{o}dinger equation (\ref{tb diagonal}) in the form of a classical two-dimensional map,
\begin{eqnarray}x_{n+1} &=& x_n \cos \mu - (p_n + A_n x_n) \sin \mu \nonumber \\
p_{n+1} &=& x_n \sin \mu + (p_n + A_n x_n) \cos \mu~.
\label{Ham}
\end{eqnarray}
It can be obtained with the use of canonical variables,
\begin{equation}
\label{p-x}
x_n=\psi_n, \qquad p_{n}=(\psi_{n}\cos\mu-\psi_{n-1})/\sin\mu
\end{equation}
where
\begin{equation}
A_n=\frac{\epsilon_n}{\sin \mu}\,.
\label{A-n}
\end{equation}
Here the ``coordinate" $x_n=\psi_{n}$ stands for the $\psi$-function at site $n$, and $p_n$ can be treated as the ``momentum" in the Hamiltonian description. The equivalence of the map (\ref{Ham}) to the Anderson model (\ref{tb diagonal}) can be easily seen by eliminating $p_{n+1}$ and $p_{n}$ from Eqs.~(\ref{Ham}).

The map (\ref{Ham}) describes the behavior of a harmonic oscillator subjected to periodic delta kicks of amplitude $A_{n}$. Without the disorder, $\epsilon_n=0$, all trajectories of this map in the phase space $(p, x)$ are circles specified by initial conditions $(p_0, x_0)$. In such an approach, the localized quantum states correspond to trajectories that are unbounded in the phase space $p,x$, when ``time" $n$ increases, $n \rightarrow \infty$. Contrary, the extended states are represented by bounded trajectories.

In fact, the analysis of the time evolution of the map (\ref{Ham}) is similar to the transfer matrix approach well developed in the theory of disordered solids \cite{LGP88}. In the Hamiltonian map approach the problem of localization of eigenstates of the stationary Schr\"{o}dinger equation (\ref{tb diagonal}) with fixed {\it boundary conditions} is treated through the evolution of $\psi_n$ with fixed {\it initial conditions} $\psi_0, \psi_1$. Specifically, an instability of trajectories $\psi_n$ in ``time" $n$ can be measured in terms of the classical Lyapunov exponent $\lambda$, and its inverse value determines the localization length, $\lambda^{-1}=L_{loc}$. Thus, the problem of localization of quantum eigenstates can be considered via local instability of classical trajectories of the map (\ref{Ham}).

The advantage of the map (\ref{Ham}) is due to a possibility to use the canonical variables $\left(r,\theta\right)$ introduced by the
transformation, $x=r\sin\theta$ and $p=r\cos\theta$. Then the map (\ref{Ham}) takes the following form,
\begin{equation}
\begin{array}{l}
\sin\theta_{n+1} = D_{n}^{-1} \left[ \sin\left(\theta_{n}-\mu\right)
- A_{n}\sin\theta_{n}\sin\mu \right], \\ \\
\cos\theta_{n+1} = D_{n}^{-1} \left[ \cos\left(\theta_{n}-\mu\right)
+ A_{n}\sin\theta_{n}\cos\mu \right]
\end{array}
\label{map-theta}
\end{equation}
where
\begin{equation}
D_{n} = \frac{r_{n+1}}{r_n} =
\sqrt{1 + A_{n}\sin\left(2\theta_{n}\right) + A_{n}^{2}\sin^{2}\theta_{n}}
\label{Dn}
\end{equation}
and
\begin{equation}
E=2 \cos \mu\,.
\label{E-mu}
\end{equation}
According to the standard definition of the localization length (see, for example, \cite{LGP88}), one can write,
\begin{equation}
\label{loclength}
L_{loc}^{-1}(E) = \lim_{N\rightarrow \infty}{\frac{1}{N} \sum_{n=1}^{N}{ \ln{\left| \frac{x_{n+1}}{x_n}\right|}}} = \lim_{N\rightarrow \infty}{\frac{1}{2N} \sum_{n=1}^{N}{ \ln{\frac{r_{n+1}}{r_n}}}} .
\end{equation}
The last relation in Eq.~(\ref{loclength}) emerges since the average of $\ln{\left|{\sin{\theta_{n+1}}/\sin{\theta_n}}\right|}$ vanishes when $N\rightarrow \infty$.
One can write the above definition in the following form,
\begin{equation}
\label{locAnd}\lambda = L_{loc}^{-1}= \frac{1}{2} \left \langle \ln \left( \frac{r_{n+1}}{r_n}\right)
\right \rangle .
\end{equation}
Here the angular brackets stay for the average over ``time" $n$, therefore, along the trajectory computed in accordance with the map (\ref{Ham}). Quite often an additional average over different realizations $\epsilon_n$ of disorder is used in numerical simulations. However, one has to perform such an ensemble average  {\it after} the average along ``time" $n$, due to possible correlations between the phases $\theta_{n+1}$ and $\theta_n$, see below.

It is interesting to note that due to Eq.(\ref{Dn}) the expression (\ref{locAnd}) involves the phases $\theta_n$ only. We have to stress that the phase $\theta_n$ has clear physical meaning: it is the phase of $\psi$-function at the site $n$. The exact map for the phase $\theta_n$ can be obtained from Eq.(\ref{map-theta}),
\begin{equation}
\label{map1D}
\cot ({\theta_{n+1}}+\mu) = \cot \theta_n + A_n .
\end{equation}
It is interesting to note that from this expression one can get,
\begin{equation}
\label{dtheta}
\frac{d \theta_n}{d \theta_{n+1}} = 1 + A_{n}\sin\left(2\theta_{n}\right) + A_{n}^{2}\sin^{2}\theta_{n} = D_n^2 .
\end{equation}
Therefore, in order to find the localization length one can find all phases $\theta_n$ by iterating the 1D-map (\ref{map1D}) rather than the 2D-map (\ref{Ham}).
Then, the localization length can be expressed through the local instability of phases $\theta_n$,
\begin{equation}
\label{loc-phases}
\lambda=L_{loc}^{-1} = \frac{1}{2} \langle \ln D_n^2 \rangle = - \frac{1}{2} \left \langle \ln \left | \frac{d \theta_{n+1}}{d \theta_{n}} \right | \right \rangle .
\end{equation}
As one can see, the problem of quantum localization in one-dimensional random potentials can be reduced to the analysis of local (exponential) instability of the classical nonlinear one-dimensional map (\ref{map1D}) with the {\it effective disorder} $A_n=\epsilon_n / \sin\mu$ playing the role of time dependent external force. It is important to stress that such a representation of the problem of Anderson localization is valid for {\it any} kind and strength of disorder in Eq.~(\ref{map1D}). In this way the solution for the Lyapunov exponent is defined by the stationary distribution of phase $\theta$ needed to know in order to perform the average in Eq.~(\ref{loc-phases}).

\subsubsection{Thouless formula}
\label{4.2.1}

Let us see how to obtain the localization length of eigenstates with the use of the Hamiltonian map approach. We have to stress that here the analytical derivation is restricted by weak disorder, $\sigma \ll 1$, and inside the energy band $-2<E<2$ defined by the zero disorder. The case for which the energy $E$ is close to the band edges, $E=\pm 2$, will be considered separately in Section~\ref{4.2.3}.

According to the expression (\ref{loc-phases}), one can write,
 \begin{equation}
\label{Measure}L_{loc}^{-1}=\int P(\epsilon )\int_0^{2\pi }\ln (D(\epsilon ,\theta
))\,\rho (\theta )\,{d\theta }{d\epsilon }~.
\end{equation}
Here $P(\epsilon )$ is the density of distribution of random entries $\epsilon_n$, and $\rho (\theta )$ stands
for the invariant measure of the distribution of phase $\theta$. As one can see, the average of the logarithm $\ln (D_n)$ along the ``time" $n$ is substituted by the ensemble average (over phases $\theta_n$ and over disorder $\epsilon_n$). Therefore, we suggest an ergodic character of the dynamical process, which is a typical assumption in the theory of weakly disordered systems.

An additional assumption made in Eq.~(\ref{Measure}) is that
the distribution $\rho (\theta )$ does not depend on the specific sequence $\epsilon _n$. Therefore, the integration in Eq.~(\ref{Measure}) can be done separately over phases and disorder, which is correct in the first order of perturbation theory. Thus, the main problem is to find the distribution $\rho (\theta )$ that emerges provided the phases $\theta_n$ are randomized. The problem of phase randomization in the Anderson model has attracted much attention since early papers on this subject, see for example, \cite{LT82,SAJ83}. As was pointed out in Ref.~\cite{ATAF80} and later discussed in the literature (see, for example, Refs.\cite{TS05,DZL98} and references therein), the phase randomization is the key ingredient of the single parameter scaling, and is of special interest in the theory of localization.

By weak disorder we mean that $A_n$ is small. Therefore, even close to the band edges the standard perturbation theory can work if the disorder is sufficiently weak.
Specifically, the disorder $\epsilon _n$ must vanish faster than $\mu $ when approaching the energy bands (see Section~\ref{4.2.3}). Retaining only terms up to $O(A_n^2)$ in the map (\ref{map1D}) for $\theta _n$ one gets,
\begin{equation}
\label{smallmap}\theta _{n+1}=\theta _n-\mu +A_n\sin {}^2\theta _n+A_n^2\sin
{}^3\theta _n\cos \,\theta _n\,\,\,\,\,\,\,\,\,\,\,\,\,\,\,\{\mbox{mod}
\,\,2\pi \}\,,
\end{equation}
where the phase $\theta_n$ is taken modulo $2\pi$ (reduced to the interval $(0 \leq \theta _n < 2\pi))$.

In the weak disorder limit the expression (\ref{Measure}) for $L_{loc}^{-1}$ can be written as follows,
\begin{equation}
\label{smalllyap}L_{loc}^{-1}=\frac{1}{2\sin {}^2\mu }\int \epsilon ^2 P(\epsilon
)\,d\epsilon \int\limits_0^{2\pi }\rho (\theta )\left( \frac 14-\frac 12\cos
(2\theta )+\frac 14\cos (4\theta )\right) d{\theta }~,
\end{equation}
which is valid over all the spectrum except at the band edges where the specific evaluation has to be made. For the standard case of flat distribution of $\epsilon_n$ between $-W/2$ and $W/2$ the variance $\sigma^2$ gets,
\begin{equation}
\sigma^2= \int \epsilon ^2 P(\epsilon)\,d\epsilon =\frac{W^2}{12}\,.
\label{sigma-And}
\end{equation}
Concerning the phase distribution, normalized in the standard way,
\begin{equation}
\int\limits_0^{2\pi }\rho (\theta ) d{\theta }=1
\label{theta-norm}\,,
\end{equation}
it is reasonable to assume that it is also constant for {\it irrational} values of the unperturbed phase shift, $\mu \neq \pi S/2Q <\pi/2$ where $S,Q$ are integers. More details on how the distribution of phases $\theta_n$ changes with an increase of disorder can be found in Ref.\cite{K09} for both the white noise disorder and correlated disorder defined via the dichotomic Poisson process. Note that for the {\it rational} values of energy (for the so-called resonances, $\mu = \pi S/2Q$ with $\mu < \pi/2$) the unperturbed density $\rho_0 (\theta )$ is a set of $\delta$-peaks, that does not allow to use standard methods of the perturbation theory \cite{KW81,DG84}.

For the non-resonant values of $E$ the perturbed distribution $\rho(\theta)$ in the first order of the perturbation theory turns out to be flat, therefore, one can easily obtain the relation,
\begin{equation}
\label{standard}L_{loc}^{-1}=\frac{\sigma ^2}{8\sin {}^2\mu }=\frac{W^2}{96\left(
1-\frac{E^2}4\right) },
\end{equation}
that is known as the Thouless expression \cite{T79}. Here the localization length is given in units $d$ of the distance between neighboring sites.
This formula was found to work quite well over all energies, however, the numerical data of Ref.~\cite{CKM81} manifested a small but
clear deviation at the band center, $E=0$.

In order to explain the band center anomaly, the non-standard
perturbation theory methods were developed in Refs.~\cite{KW81,DG84,L84,E84,GNS93,GNS94}, where
the correct value for $L_{loc}^{-1}$ at the band center was obtained. Also, a quite specific expression for the localization length was analytically obtained at the band edges. Below, within the frame of the Hamiltonian map approach, we show how to resolve the band center and band edges anomalies in a relatively simple way.

\subsubsection{Band center}
\label{4.2.2}

In order to derive the correct expression for $L_{loc}^{-1}$ at the
band center $E=0$ , one has to find the form of the invariant
probability measure $\rho (\theta )$. The latter arises from the map (\ref
{smallmap}) with $\mu =\pi /2$. For vanishing disorder
the trajectory is a period four, therefore, the phase distribution is highly degenerate since it consists of four delta-peaks equally spaced in the $\theta$-space with the period $\pi/2$, specified by the initial phase $\theta_0$.
For a weak disorder any orbit diffuses around the period four, with an
additional drift in $\theta $. Asymptotically, any initial condition gives
rise to the same invariant distribution, which can now be expected to be
different from the flat one. To find this distribution, in Ref.\cite{IRT98} it was suggested to study the fourth iterate of the map (\ref{smallmap})
\begin{equation}
\label{iterate}\theta _{n+4}=\theta _n+\xi _n^{(1)}\sin {}^2\theta _n+\xi
_n^{(2)}\cos {}^2\theta _n-\frac{\sigma ^2}2\sin \,(4\theta _n)~,
\end{equation}
where $\xi _n^{(1)}=\epsilon _n+\epsilon _{n+2}$ and $\xi _n^{(2)}=\epsilon
_{n+1}+\epsilon _{n+3}$ are uncorrelated random variables with the zero mean and
variance $2\sigma ^2$. Here, we have neglected the mixed
terms of the kind $\epsilon _n\epsilon _m$ ($m\neq n$) and approximated $%
(\xi _n^{(1)})^2$ and $ (\xi _n^{(2)})^2 $ by
their common variance $2\sigma ^2$, which is meaningful in a perturbative
calculation at first order in $\epsilon _n$.

Thus, the invariant distribution can be determined analytically in the
continuum limit for which $\theta _{n+4}-\theta _n$ is replaced with $d\theta $. Correspondingly, the random variables $\xi _n^{(1)},\xi _n^{(2)}$ are treated as independent random processes $W_1$ and $W_2$ with the properties,
\begin{eqnarray*}
\langle dW_i \rangle &=& 0 \\
\langle dW_i dW_j \rangle &=& 2 \delta_{ij} \sigma^2 dt \qquad i,j=1,2 .
\end{eqnarray*}
Therefore, one can write the It\^{o} equation \cite{I51},
\begin{equation}
d\theta =dW_1\sin {}^2\theta +dW_2\cos {}^2\theta -\frac{\sigma ^2}2\sin
\,(4\theta )\,dt~,
\end{equation}
that can be associated with the Fokker-Planck equation~\cite{G04},
\begin{equation}
\label{FP}\frac{\partial P(\theta ,t)}{\partial t}=\frac{\sigma ^2}2\frac
\partial {\partial \theta }\left( \sin (4\theta )P(\theta ,t)\right) +\frac{%
\sigma ^2}4\frac{\partial ^2}{\partial \theta ^2}\left[ (3+\cos (4\theta
))P(\theta ,t)\right] ~.
\end{equation}
According to Eq.~(\ref{smalllyap}), in order to find the Lyapunov exponent one needs to know the distribution of phases emerging in the limit $t=N \rightarrow \infty$. Therefore, rather than the time-dependent solution of the Fokker-Planck equation, we have to find the stationary solution $\rho (\theta )$ of Eq.~(\ref{FP}), which is not a difficult task. This solution, satisfying the
condition of periodicity $\rho (0)=\rho (2\pi )$ and the normalization condition (\ref{theta-norm}), has the form \cite{IRT98},
\begin{equation}
\label{stationary}\rho (\theta )=\left[ 2{\bf K}\left( \frac 1{\sqrt{2}
}\right) \,\sqrt{3+\cos (4\theta )}\right] ^{-1}~,
\end{equation}
where ${\bf K}(1/\sqrt{2})$ is the complete elliptic integral of the first kind. This expression allows one to derive an explicit formula for the localization length at the band center. The same form of the distribution was found to emerge for the phase of a reflected wave function in the Anderson model with zero boundary condition at one edge (see details in Ref.~\cite{OKG00}). Inserting Eq.~(\ref{stationary}) into (\ref{smalllyap}) with $\mu =\pi /2$ we finally obtain,
\begin{equation}
\label{elle}L_{loc}^{-1}=\frac{\sigma ^2}8\left( 1+\int\limits_0^{2\pi }\rho
(\theta )\cos (4\theta )\,d\theta \right) =\sigma ^2\left( \frac{\Gamma (3/4)
}{\Gamma (1/4)}\right) ^2 \approx \frac{W^2}{105.2 }.
\end{equation}
This result agrees with that obtained in Refs.~\cite{E84,KW81,DG84}, however, here it is derived with a different approach that seems to be more elegant and involves much simpler calculations.

The standard perturbation theory used by Thouless neglects the contribution of the non-zero average of the $\cos (4\theta )$ term in (\ref{elle}), meaning that the
stationary solution (\ref{stationary}) is approximated with a flat
distribution. This approximation works very well for all energies $E=2\cos \,\mu $, with $\mu =\alpha \pi/2 $ and $\alpha $ irrational, but does not work for the band center. It is important to note that the expression (\ref{smalllyap}) determining the localization length involves only two harmonics, $\cos (2\theta)$ and $\cos (4 \theta)$. This means that for higher resonances with
other rational values of $\alpha =S/Q<1$ (with $S$ and $Q$ integers)
one should expect an appearance of higher than fourth-harmonics in the distribution of $\theta$. Since they are not present in Eq.(\ref{smalllyap}), the localization length for higher resonance values of $E$ is given by the standard Thouless expression (\ref{standard}). As one can see, the localization length anomaly occurs only at the band center. However, the moments of finite length Lyapunov exponents (defined for finite samples) are sensitive to higher harmonics in the distribution of $\theta$, see, for example, Ref.\cite{ST03}. We would like to remind that all the results we discuss here are restricted by weak disorder conditions.

\subsubsection{Band edges}
\label{4.2.3}

For weak disorder, the band edges correspond to $\mu = 0, \pi$, or the same, to $E=\pm 2$. Due to the symmetry with respect to $\mu=\pi/2$ (therefore, with respect to $E=0$ in the energy spectrum), it is enough to consider the vicinity of the band edge corresponding to $\mu=0$. In the second order of perturbation theory the term $A_n^2$ in the map (\ref{smallmap}) can be replaced by its average, which is the same approximation we did in previous Section. Then the map (\ref{smallmap}) reduces to
\begin{equation}
\label{bandmap}\theta _{n+1}=\theta _n-\mu +\frac{\epsilon _n}\mu \sin
{}^2\theta _n+\frac{\sigma ^2}{\mu ^2}\sin {}^3\theta _n\cos \,\theta
_n\,\,\,\,\,\,\,\,\,\{ \mbox{mod}\,\,2\pi \}~,
\end{equation}
where $\sigma ^2$ is the variance of the noise $\epsilon _n$, and we assume that $\sigma^2 \ll \mu^2$ in order to evaluate Eq.~(\ref{bandmap}) perturbatively. From the exact map, as well from the above one, one can see that for vanishing
disorder and $\mu \to 0$ the orbits (in $\theta$-space) are fixed points. Moving away from the band edge produces a quasi-periodic motion and switching on the disorder
gives rise to diffusion. Following the procedure discussed in previous Section
(however, here we do not need to go to the four-step map), we obtain the
corresponding Fokker-Planck equation \cite{IRT98},
\begin{equation}
\label{FP2}\frac{\partial P(\theta ,t)}{\partial t}=\frac \partial {\partial
\theta }\left[ \left(\mu -\frac{\sigma ^2}{\mu ^2}\sin {}^3 \theta
\cos \theta \right)
P(\theta ,t)\right] +\frac{\sigma ^2}{2\mu ^2}\frac{\partial ^2}{\partial
\theta ^2}\left( P(\theta ,t)\,\sin {}^4\theta \right) ~.
\end{equation}
In the vicinity of the band edge there are two small quantities, the noise $\epsilon _n$ and the distance from the band edge $\Delta =2-2\cos \,\mu \approx \mu ^2$. Below we consider the double limit $\Delta \to 0$, $\sigma ^2\to 0$. From the expression (\ref{FP2}) one can see that there is a competition between two parameters, $\mu$ and $\sigma^2/\mu^2$, therefore, the control parameter is the ratio
\begin{equation}
\varkappa=\frac{\mu ^3}{\sigma ^2}.
\label{varkappa-def}
\end{equation}
By keeping the value of $\varkappa$ fixed, the time scale of the drift term in Eq.~(\ref{FP2}) is unique. We can thus rescale the time, $\tau =t\mu $,
and obtain the following stationary Fokker-Planck equation,
\begin{equation}
\label{FP3}\frac \partial {\partial \theta }\left[ \left(\varkappa-\sin {}^3\theta \cos
\theta \right)\rho (\theta )\right] +\frac 12\frac{\partial ^2}{\partial \theta
^2}\left( \sin {}^4\theta \,\rho (\theta )\right) =0~,
\end{equation}
which depends only on the control parameter $\varkappa\,$. Its general solution satisfying to the normalization condition (\ref{theta-norm}), is
\begin{equation}
\rho (\theta )=\frac{f(\theta )}{\sin {}^2\theta }\left[
C+\int\limits_0^\theta dx\frac{2J}{f(x)\sin {}^2 x }\right] ~,
\end{equation}
where
\begin{equation}
f(\theta )=\exp \left[ 2\varkappa\left( \frac 13\cot {}^3\theta +\cot \,\theta
\right) \right] ~,
\end{equation}
and $C,J$ are the integration constants. In order to have $\rho $ normalizable, the constant $C$ must vanish and $J$ is then fixed by the normalization condition,
\begin{equation}
J^{-1}=\frac{\sqrt{8\pi }}{\varkappa^{2/3}}\int\limits_0^\infty dx\frac 1{\sqrt{x}%
}\exp \left( -\frac{x^3}6-2\varkappa^{2/3}x\right) ~.
\end{equation}
In the considered case the distribution of phases turns out to be highly non-homogeneous. In this case it is convenient to evaluate the Lyapunov exponent in the following form of Eq.~(\ref{locAnd}) \cite{IRT98},
\begin{equation}
\label{full-loc}
L_{loc}^{-1}=\left\langle \ln \left| D_n\frac{\sin \theta _{n+1}}{\sin
\theta _n}\right| \right\rangle .
\end{equation}
In the limit of weak disorder, by expanding the logarithm in Eq.~(\ref{full-loc}) for $\mu \to 0$, one gets,
\begin{equation}
L_{loc}^{-1}=-\mu \,\langle \cot \,\theta _n\rangle =-2\mu
\int\limits_0^\pi \cot \,\theta \,\rho (\theta )\,d\theta ~.
\end{equation}
After straightforward calculations one
obtains,
\begin{equation}
L_{loc}^{-1}=\frac{\sigma^{2/3}}2\frac{\int_0^\infty dx\,x^{1/2}\exp \left( -
\frac{x^3}6-2\varkappa^{2/3}x\right) }{\int_0^\infty dx\,x^{-1/2}\exp \left( -\frac{
x^3}6-2\varkappa^{2/3}x\right) }~,
\label{Kgeneral}
\end{equation}
which coincides with Eq.~(36) in Ref.~\cite{DG84}. This expression indicates that in contrast with the cases with the energy $E$ inside the band spectrum, in the vicinity of the band edges the localization length essentially depends on two independent parameters. This means that the single parameter scaling conjecture may not valid at the band edges (for details see, for example, \cite{DEL03,DEL03a}).

It is instructive to analyze two limiting cases, $\varkappa\to \infty $ and $\varkappa\to 0$. One can show that in the first case the standard perturbation theory works well, thus giving the conventional Thouless expression (\ref{standard}). Indeed, for this case $\mu \gg \sigma^2/\mu^2$ in Eq.~(\ref{FP2}), therefore, the distance $\Delta=2-|E|$ is much larger than the disorder and one can expect no influence of the band edges. On the contrary, when the distance from band edges is very small in comparison with the effective parameter of disorder, $\Delta \ll \sigma^{4/3}$, the situation corresponds to the limit $\varkappa\to 0$. In this case the evaluation of Eq.~(\ref{Kgeneral}) leads to the expression, \cite{DG84,IRT98},
\begin{equation}
L_{loc}^{-1}=\frac{6^{1/3}\sqrt{\pi }}{2\Gamma \left( \frac 16\right) }\sigma
^{2/3}\approx 0.289 \,\sigma^{2/3}
\label{an-scale}
\end{equation}
with the dependence $\lambda =L_{loc}^{-1} \sim \sigma^{2/3}$ which is known as the {\it anomalous scaling} at band edges.
It is interesting to note that similar scaling law was also found for
chaotic classical billiards, when studying the dependence of the
Lyapunov exponent $\lambda$ on the parameter of closeness to the integrable limit~\cite{B84}.

\subsubsection{Strong disorder}
\label{4.2.4}

For comparison, let us see how the localization length can be evaluated for strong disorder. It is interesting that the relation (\ref{Measure}) allows one to derive an approximate expression which is also good for a quite large disorder. Indeed, if the energy is far enough from the band edge and the disorder is not extremely strong, one can expect a fast rotation of the phase $\theta_n$. Therefore, the invariant measure $\rho (\theta )$ can be approximately taken as constant, $\rho (\theta )=\left( 2\pi \right) ^{-1}$ (see Eq.~(\ref{theta-norm})), and one can explicitly integrate Eq.(\ref{Measure}), first over the phase $\theta $,
\begin{equation}
\label{theta}\frac 1{4\pi }\int\limits_0^{2\pi }\ln \left( 1+A\sin (2\theta
)+A^2\sin {}^2\theta \right) d\theta =\frac 12\ln \left( 1+\frac{A^2}%
4\right) ;\,\,\,\,\,\,\,\,A^2=\epsilon ^2/\sin {}^2\mu \,
\end{equation}
and after, over the disorder $\epsilon $ \cite{IRT98},
\begin{equation}
\label{l1}L_{loc}^{-1}=\frac 12 \int P(\epsilon )\ln
\left( 1+\frac{\epsilon ^2}{4\sin
{}^2\mu }\right) d\epsilon =\frac 12\ln \left( 1+\frac{W^2}{16\sin {}^2\mu }
\right) +\frac{\arctan \left(\frac W{4\sin \mu }\right) }{\frac W{4\sin \mu }}
-1\,.
\end{equation}
This expression has been derived in Ref.\cite{P86} with the use of standard transfer matrix approach. The direct numerical simulations show that this expression gives quite good agreement with the data for the disorder $W\leq 1\div 3$ inside the
energy range $\left| E\right| \leq 1.85$ .
Therefore, the above expression can serve as a generalization of the
weak disorder formula (\ref{standard}), since it is also valid for
relatively strong disorder $W \approx 1$. However, for a very strong
disorder $W \gg 1$, Eq.~(\ref{l1}) gives incorrect results. The reason
is that in this case the invariant measure $\rho (\theta)$ is strongly
non-uniform, hence the expression (\ref{theta}) is no more valid.

Instead, for a much stronger disorder one can use another approach.
Note, that for the unstable region,
\begin{equation}
\label{unstable}|E-\epsilon _n|>\,2
\end{equation}
both eigenvalues
$\lambda _n^{(1,2)}$ of the one-step Hamiltonian map (\ref{Ham}) are real,
\begin{equation}
\label{eigen0}\lambda _n^{(1,2)}=\frac 12\left( (E-\epsilon _n)\,\pm \,\sqrt{
(E-\epsilon _n)^2-4}\right)
\end{equation}
with $\lambda _n^{(1)}\lambda _n^{(2)}=1$. Therefore, for $
W\gg 1\,$ one can compute the inverse localization length directly via the
largest value $\lambda _{+}$ of these two eigenvalues, by neglecting the
region $|E-\epsilon| < 2$,
\begin{equation}
\label{l2}
L_{loc}^{-1}=\left\langle \ln \left| \lambda _{+}\right| \right\rangle =
\int \ln \left( \frac12 \left( x+ \sqrt{x^2-4}\right)\right) dx=
F(z_1)+F(z_2)
\end{equation}
Here, $x=|E-\epsilon |$ and $z_1=W/2+E, \, z_2=W/2-E$, and the function $F$ is
defined by
\begin{equation}
\label{F}
F(z)= \frac 2W \left( z \ln \left( z+\sqrt{z^2-4} \right)-\sqrt{z^2-4}-z\ln 2
\right).
\end{equation}
This expression fits the data more accurately than the well known expression,
\begin{equation}
\label{limit}L_{loc}^{-1}=\ln \frac W2-1,
\end{equation}
derived for a very strong disorder (see, for example, \cite{P86}).

\subsection{Anderson localization as a parametric instability}
\label{4.3}

As we already mentioned, the Anderson localization can be treated as a parametric instability of a linear oscillator whose frequency randomly depends on time in accordance with the disorder in the quantum model. This analogy allows one to obtain the general expression for the localization length
which is valid both inside the energy band and at the band edges \cite{TI00}. We write here the Hamiltonian that corresponds to the Anderson model,
\begin{equation}
{\cal H} = \omega \left( \frac{p^2}{2} + \frac{x^2}{2} \right) +
\frac{x^2}{2} \xi (t),
\label{kickosc}
\end{equation}
where
\begin{equation}
\xi \left( t \right) = \sum_{n = -\infty}^{+\infty} A_{n} \; \delta
\left( t - n T \right).
\label{xi}
\end{equation}
By integrating the
Hamilton equations of motion of the oscillator~(\ref{kickosc}) over one
period between two successive kicks, one gets the map,
\begin{equation}
\begin{array}{ccc}
x_{n+1} & = & x_{n} \cos \mu + (p_{n} - A_{n} x_{n})
\sin \mu \\
p_{n+1} & = & -x_{n} \sin \mu + (p_{n} - A_{n} x_{n})
\cos \mu
\end{array}
\label{mapp}
\end{equation}
where $x_{n}$ and $p_{n}$ stand for the position and momentum of the
oscillator immediately {\it before} the $n$th kick, and $\mu$ is the phase shift between two successive kicks,
\begin{equation}
\mu=\omega T\,.
\label{def-mu}
\end{equation}
One can see that this map becomes equivalent to the Hamiltonian \eqref{Ham} after the replacement $A_n\rightarrow-A_n$ and $\omega\rightarrow-\omega$. This replacement does not affect our further consideration and we will not discriminate between Eqs.~\eqref{mapp} and \eqref{Ham}. Indeed, by eliminating the momentum from Eqs.~\eqref{mapp}, one gets the following relation,
\begin{displaymath}
x_{n+1} + x_{n-1} + A_{n}  x_{n}\sin \mu = 2x_{n}
\cos \mu
\label{map-T}
\end{displaymath}
which coincides with the tight-binding Anderson model ~\eqref{tb diagonal}, provided that the position $x_{n}$ of the oscillator at time $t=nT$ is identified with the electron amplitude $\psi_{n}$ at the $n$th site, and $T=d=1$. In such a way the parameters of the kicked oscillator are related to those of the Anderson model by the identities,
\begin{equation}
\begin{array}{ccc}
\varepsilon_{n} = A_{n} \sin \mu & \mbox{and} & E = 2 \cos \mu.
\end{array}
\label{corres}
\end{equation}

Let us now turn to the kicked oscillator~(\ref{kickosc}). Its dynamics is determined by the
Hamiltonian equations of motion,
\begin{equation}
\left\{
\begin{array}{ccl}
\dot{p} & = & - \left( \omega + \xi \left( t \right) \right) x \\
\dot{x} & = & \omega p
\end{array} .
\right.
\label{dyneq-1}
\end{equation}
In order to study the behavior of the kicked oscillator, it is convenient
to replace the couple of differential equations~(\ref{dyneq-1}) by
the system of stochastic It\^{o} equations \cite{I51},
\begin{equation}
\left\{
\begin{array}{ccl}
dp & = & - \omega x \; dt - x \sqrt{\langle A_{n}^{2} \rangle/T}
\; d{\mathbb W}(t) \\
dx & = & \omega p \; dt
\end{array}
\right.
\label{itoeq}
\end{equation}
where ${\mathbb W}(t)$ is a white-noise Wiener process with $\langle d{\mathbb W}(t) \rangle = 0$ and $\langle [d{\mathbb W}(t)]^{2} \rangle = dt$. This process can be regarded as the limiting case of random walk for very small and fast-spaced steps \cite{G04}. Therefore, the ``jump process" $\xi(t)$ can be described by a ``diffusion process" ${\mathbb W}(t)$ when the condition
\begin{equation}
\label{smallness}
\langle A_n^{2} \rangle \ll 1
\end{equation}
is satisfied. By introducing the variance
\begin{equation}
\sigma^2=\langle \epsilon_n^{2} \rangle
\label{def-variance}
\end{equation}
of disorder, the above condition can be written as follows,
\begin{equation}
\frac{\sigma^2}{\sin^2 \mu} \ll 1.
\label{def-weakness}
\end{equation}
As one can see, it is violated close to the band edges where $\mu \rightarrow 0$.

Our interest is in deriving the
Lyapunov exponent which gives the inverse localization length for the
eigenstates of Eq.~(\ref{tb diagonal}). Since an exponential
decrease of tails of these eigenstates can be studied in terms of
trajectories of the stochastic oscillator~(\ref{itoeq}),
the Lyapunov exponent is naturally redefined as the exponential
divergence rate of neighboring trajectories, i.e. through the limit
\begin{equation}
\lambda = \lim_{t \to \infty} \; \lim_{\delta \to 0}
\frac{1}{t} \int_{0}^{t} d\tauñ
 \;\; \frac{1}{\delta} \ln \left|
\frac{x(\tau+\delta)}{x(\tau)}\right| ,
\label{divrate}
\end{equation}
which corresponds to the standard expression (\ref{loclength})
for the Anderson model~(\ref{tb diagonal}).
By taking the limit $\delta \to 0$ first, the expression~(\ref{divrate})
can be written in the simpler form,
\begin{equation}
\lambda = \langle z \rangle = \lim_{t \to \infty} \frac{1}{t}
\int_{0}^{t} d\tau \;\; z(\tau) ,
\label{lyapdef}
\end{equation}
where the Riccati variable $z = \dot{x}/x$ is introduced, and
the symbol $\langle z \rangle$ stands for the (time) average of $z$.

To compute the Lyapunov exponent, as defined by Eq.~(\ref{lyapdef}), it
is necessary to analyze the dynamics of the variable $z = \omega p/x$.
The time evolution of this quantity is determined by the It\^{o}
stochastic equation \cite{TI00},
\begin{equation}
dz = -\left( \omega^{2} + z^{2} \right) dt - \omega \sqrt{ \langle
A_{n}^{2} \rangle/T} \; d{\mathbb W}(t) \; ,
\label{dz}
\end{equation}
which can be easily derived using Eqs.~(\ref{itoeq}).
Notice that while the position and momentum of the
oscillator~(\ref{itoeq}) do not evolve independently from each other,
the dynamics of their ratio $z =\omega p/x$ is totally autonomous from
that of any other variable. As a consequence, one deals with the single
differential equation (\ref{dz}) instead of the
coupled equations (\ref{itoeq}).
As known, the It\^{o} stochastic differential equation~(\ref{dz}) is
equivalent to the Fokker-Planck equation~\cite{G04,K92},
\begin{equation}
\frac{\partial p}{\partial t} (z,t) = \frac{\partial}{\partial z}
\left[ \left( \omega^{2} + z^{2} \right) p(z,t) \right] +
\frac{\mathfrak{D} \omega^{3}}{2} \frac{\partial^{2} p}
{\partial z^{2}}(z,t) \; ,
\label{fokpl}
\end{equation}
where we introduced the control parameter $\mathfrak{D}$,
\begin{equation}
\mathfrak{D} = \frac{\langle A_{n}^{2} \rangle}{\omega T} =\frac{\sigma^2}{\mu \sin^2 \mu}.
\label{kappa}
\end{equation}
As one can see, the evolution of $z(t)$ dictated by Eq.~(\ref{dz}) is a
diffusion process with the deterministic drift coefficient $\left( \omega^{2}
+ z^{2} \right)$ and noise-induced diffusion coefficient $\mathfrak{D} \omega^3$.
The equation (\ref{fokpl}) gives the time evolution of the probability density $p(z,t)$ of the stochastic process $z$ at any time $t$. However, in order to find the Lyapunov exponent (\ref{lyapdef}), one needs to know the stationary solution of Eq.~(\ref{fokpl}) that gives the distribution of the variable $z$. In such a way, instead of averaging over a single trajectory we pass to the ensemble average,
\begin{equation}
\lambda = \int_{-\infty}^{\infty} dz \; z \, p(z) \, ,
\label{averag}
\end{equation}
that is in accordance with our assumption of an ergodic character of $z(t)$.

The stationary solution of Eq.~(\ref{fokpl}) can be easily found by direct integration,
\begin{displaymath}
p(z) = \left[ C_{1} + C_{2} \int_{-\infty}^{z} dx \; \exp \left\{
\Phi(x/\omega) \right\} \right] \, \exp \left\{ -\Phi(z/\omega) \right\} \; ,
\end{displaymath}
where $C_{1}$ and $C_{2}$ are the integration constants and the function
$\Phi(x)$ is given by the relation
\begin{equation}
\Phi (x) = \frac{2}{\mathfrak{D}} \left( x + \frac{x^{3}}{3} \right) \; .
\label{phi}
\end{equation}
Since $p(z)$ must satisfy to the normalization condition $\int_{-\infty}^{\infty} p(z) \; dz = 1$, the constant $C_{1}$ vanishes. The resulting distribution has the form,
\begin{equation}
p(z) = \frac{1}{N \omega^{2}}
\int_{-\infty}^{z} dx \; \exp \left\{ \Phi \left( x/\omega \right)
- \Phi \left( z/\omega \right) \right\},
\label{misinv}
\end{equation}
with
\begin{equation}
N = \sqrt{\frac{\pi \mathfrak{D}}{2}}
\int_{0}^{\infty} dx \; \frac{1}{\sqrt{x}} \exp \left[ -\frac{2}{\mathfrak{D}}
\left(x + \frac{x^{3}}{12} \right) \right] \;.
\label{norm}
\end{equation}
Once the steady-state probability distribution~(\ref{misinv}) is known, one can compute the Lyapunov
exponent~(\ref{lyapdef}),
\begin{equation}
\lambda = \frac{\omega}{2N}\sqrt{\frac{\pi \mathfrak{D}}{2}} \int_{0}^{\infty} dx \; \sqrt{x} \exp
\left[ -\frac{2}{\mathfrak{D}} \left( x + \frac{x^{3}}{12} \right) \right].
\label{lyap}
\end{equation}
With the change of variable $x=\mathfrak{D}^{2/3} y$, the Lyapunov exponent takes the form,
\begin{equation}
\label{lyap-total}
\lambda = \frac{1}{T} \frac{\sigma^{2/3}}{2}
\left( \frac{\mu}{\sin \mu} \right)^{2/3}
\frac{\int_0^\infty dy\,y^{1/2}\exp \left( -
\frac{y^3}6-2\mathfrak{D}^{-2/3}y\right) }{\int_0^\infty dy\,y^{-1/2}\exp \left( -\frac{
y^3}6-2\mathfrak{D}^{-2/3}y\right) }\,.
\end{equation}
Comparing the above expression with Eq.(\ref{Kgeneral}) obtained for the Anderson model for energies close to the band edges, $|E| \lesssim 2$, we can see that they coincide when $\omega T =\mu \rightarrow 0$. Indeed, in this case the relation
\begin{equation}
\mathfrak{D}=\sigma^2/\mu^3=\varkappa^{-1}
\label{rel-D-kappa}
\end{equation}
holds between $\mathfrak{D}$ (see Eq.(\ref{kappa})) and the parameter $\varkappa=\mu^3/\sigma^2$ used in Section~\ref{4.2.3}.

Let us analyze the expression (\ref{lyap-total}) inside the energy band. By
making use of expressions~(\ref{lyap}) and~(\ref{norm}) it is easy to
verify that for $\mathfrak{D} \rightarrow 0$ the Lyapunov exponent can be written
in the form,
\begin{equation}
\lambda = \frac{\sigma^2}{4 T \sin^2 \mu} \;
\frac{\displaystyle \sum_{n=0}^{\infty} (-1)^{n}
\frac{\Gamma \left( 3n + 3/2 \right)}{n!}
\left( \frac{\mathfrak{D}^{2}}{48} \right)^{n}}
{\displaystyle \sum_{n=0}^{\infty} (-1)^{n}
\frac{\Gamma \left( 3n + 1/2 \right)}{n!}
\left( \frac{\mathfrak{D}^{2}}{48} \right)^{n}} .
\label{expan}
\end{equation}
To the lowest order in $\mathfrak{D}$ this expression reduces to the following one,
\begin{equation}
\lambda = \frac{\langle A_{n}^{2} \rangle}{8 T}=\frac{\sigma^2}
{8T(1 - E^{2}/4)} \, ,
\label{wklyap}
\end{equation}
which represents the basic approximation for the inverse localization length in the weak disorder case. The expression~(\ref{wklyap}) corresponds to the result derived by Thouless using standard perturbation methods~\cite{T79}. Therefore, the expression (\ref{lyap-total}) for the inverse localization length turns out to be valid {\it both} inside the energy band and at the band edges (although it fails to reproduce the anomaly of the Lyapunov exponent at the band center). The extended validity range of Eq.~(\ref{lyap-total}) is a remarkable feature, since the behavior of the localization length at the band edge is known to be anomalous~\cite{DG84,IRT98} and
has been previously derived with the methods well distinct (and more complicated) than the ones used to study the localization length inside the band.

It is important to stress that, in general, the Lyapunov exponent (\ref{lyap-total}) depends on the control parameter $\mathfrak{D}$ which can have {\it any} value when approaching the band edge. This value depends on how the {\it two limits},
\begin{equation}
\begin{array}{ccc}
\sigma^2 = \langle \epsilon_{n}^{2} \rangle \rightarrow 0 & \mbox{and} & \mu=\omega T \rightarrow 0\,,
\end{array}
\label{2-limits}
\end{equation}
are related to each other. For example, besides the $\mathfrak{D} \rightarrow 0$ limit discussed above for weak disorder and inside the energy band, one can also study the behavior of the Lyapunov exponent~(\ref{lyap})
in the complementary case $\mathfrak{D} \rightarrow \infty$.
Physically speaking, the limit $\mathfrak{D} \rightarrow \infty$ can be interpreted
in different ways, depending on the reference model. If one bears in mind
the kicked oscillator~(\ref{kickosc}), then taking the limit $\mathfrak{D}
\rightarrow \infty$ is tantamount to studying the case of a very strong
noise. More precisely, the condition $\mathfrak{D} \gg 1$ implies that the kicks
play a predominant role in the oscillator dynamics with respect to the
elastic force. Notice that this is not in contrast with the requirement
that each single kick be weak, as established by the
condition~(\ref{smallness}). In fact, regardless of how weak the individual
kicks are, their collective effect can be arbitrarily enhanced by making
the interval $T$ between two successive kicks sufficiently shorter than
the fraction $\langle A_{n}^{2} \rangle / \omega$ of the period of the
unperturbed oscillator. From the point of view of the Anderson model~(\ref{tb diagonal}), the analysis of the case $\mathfrak{D} \gg 1$ is equivalent to the study of the localization length in a neighborhood of the band edge, where $\mu \rightarrow 0$ with the disorder {\it fixed}.

To conclude our discussion of the $\mathfrak{D} \rightarrow \infty$ limit, we observe
that in this case it is appropriate to expand the integrals that
appear in expressions~(\ref{lyap}) and~(\ref{norm}) in series of the
inverse powers of $\mathfrak{D}$. One thus obtains,
\begin{equation}
\lambda = \frac{1}{T} \left( \frac{3}{4} \right)^{1/3}
\left(\frac{\mu}{\sin \mu}\right)^2
\frac{\displaystyle \sum_{n=0}^{\infty}
\frac{(-2\sqrt[3]{6})^{n}}{n!}
\Gamma \left( \frac{2n+3}{6} \right)
\mathfrak{D}^{-{2n/3}}}
{\displaystyle \sum_{n=0}^{\infty}
\frac{(-2\sqrt[3]{6})^{n}}{n!}
\Gamma \left( \frac{2n+1}{6} \right)
\mathfrak{D}^{-{2n/3}}} \;,
\label{lambda-final}
\end{equation}
which is the counterpart of the expansion~(\ref{expan}).
To the lowest order in $\mathfrak{D}^{-1}$ this expression reduces to the expression,
\begin{equation}
\lambda = \frac{1}{T} \frac{\sqrt[3]{6}}{2} \frac{\sqrt{\pi}}{ \Gamma
\left( 1/6 \right)} \left( \frac{\mu}{\sin \mu}
\right)^{2/3} \sigma^{2/3} \; ,
\label{lambda-final1}
\end{equation}
so that for $E \rightarrow 2$ (i.e., for $\mu \rightarrow 0$), the Lyapunov exponent turns out to be the same as given by Eq.(\ref{an-scale}).

\subsection{Transmittance and resistance}
\label{4.4}

Now we address a problem of the transport properties for {\it finite} samples. As shown in Ref.\cite{KTI97} (see also Ref.~\cite{LGP88}), the transmission coefficient $T_L$ (transmittance, or dimensionless conductance, see Eq.~(\ref{1DCP-Tdef})) for samples of finite size $L$ can be expressed in terms of the trajectories of the Hamiltonian map (\ref{Ham}) through the following expression,
\begin{equation}
T_L = \frac{4}{2 + (r_{L}^{(1)})^2 + (r_{L}^{(2)})^2}.
\label{eq:6}
\end{equation}
Here $r_{L}^{(1,2)}$ are the radii of two trajectories at ``time" $n=L$, that start from $\left(
r_{0}^{(1)},\theta_{0}^{(1)} \right) = \left( 1,0 \right)$ and
\mbox{$\left( r_{0}^{(2)},\theta_{0}^{(2)}
\right)= \left( 1,\pi/2 \right)$} respectively. The values $(r_{L}^{(1)})^2 $ and $(r_{L}^{(2)})^2$
are entirely determined by
Eq.~(\ref{Dn}),
\begin{equation}
\label{radii}
(r_{L}^{(i)})^2 =\prod_{n=1}^{L}\left( D_{n}^{(i)} \right)^2 =
\prod_{n=1}^{L}\left( 1 + A_{n}\sin 2\theta_{n}^{(i)} +
A_{n}^{2}\sin^{2}\theta_{n}^{(i)} \right)
\end{equation}
with $i=1,2$.
Therefore, we have an explicit expression for the transmission
coefficient $T_L$ in terms of the site potential ${\epsilon_n}$
and the phases ${\theta_n}$. The relation (\ref{eq:6}) reflects the fact
that the transmission coefficient $T_L$ depends on two fundamental solutions
of Eq.(\ref{tb diagonal}), that in the Hamiltonian approach are associated with
two classical trajectories $\theta_n^{(i)}$. In spite of being linearly independent, these trajectories
are statistically
correlated since they both are determined by the same potential
$\epsilon_n$. These statistical correlations have to be taken into account in the analytical treatment of the transmission coefficient, that is a hard task. This problem partially was solved in Refs.\cite{DIK04,DIK04a} where analytical expressions for the transmission coefficient $T_L$ as well as for the resistance $T_{L}^{-1}$, were obtained in two limiting cases of ballistic and localized transport.

For our further consideration it is convenient to introduce the following quantity,
\begin{equation}
Z_{L} = \frac{(r_{L}^{(1)})^2 + (r_{L}^{(2)})^2}{2},
\label{eq:8}
\end{equation}
and express it in the form,
\begin{equation}
Z_{L} = \frac{1}{2} \left [
\exp \left( 2\sum_{n=1}^{L} \ln D_{n}^{(1)} \right) +
\exp \left( 2\sum_{n=1}^{L} \ln D_{n}^{(2)} \right)
\right ],
\label{Zdetail}
\end{equation}
therefore,
\begin{equation}
T_L=\frac{2}{1+Z_L} \,.
\label{T-L}
\end{equation}

\subsubsection{Ballistic regime: Transmittance}
\label{4.4.1}

We start with the ballistic regime for which the size $L$ of a
sample is much less than the localization length $L_{loc}$,
\begin{equation}\label{ballistic}
\frac{L}{L_{loc}} \ll 1.
\end{equation}
In this case the value of $Z_L$
is close to unity since both $r_{L}^{(1)}$ and $r_{L}^{(2)}$ increase
in $L$ very slowly, $r_L \sim \exp(L/L_{loc})$. Therefore,
it is more convenient to write the transmission coefficient in the form,
\begin{equation}
T_{L} = \frac{2}{1+\exp(\ln Z_L)} \,,
\end{equation}
which can be evaluated perturbatively in terms of $\ln Z_L$.

In the first line we are interested in the mean value
$\left<T_L\right>$ of the transmission coefficient, and in its
second moment $\left<T_L^2\right>$. Keeping the terms up to
$\ln^2 Z_L$ in the expansion of $T_L$, we obtain
that the quadratic term does not
contribute to $T_L$,
\begin{equation}\label{eq:27x}
T_{L} \approx 1- \frac{1}{2} \ln Z_L + O \left( \ln^{3} Z_L
\right) \; \; .
\end{equation}
Similarly, we get,
\begin{equation}
T_{L}^2 \approx 1 - \ln Z_L + \frac{1}{4}\ln^{2}Z_L + O
\left(\ln^3 Z_L \right) \; \; . \label{eq:28x}
\end{equation}
After some straightforward calculations that involve
the map (\ref{Ham}), the following expression for the mean value
of $\ln Z_L$ is obtained \cite{DIK04},
\begin{eqnarray}
\left< \ln Z_L \right> & \approx & 3 \left( \frac{L}{L_{loc}}\right) - \frac{1}{2} \left(
\frac{1}{2}-\frac{L}{L_{loc}} \right) S_2 - \left( \frac{L}{L_{loc}}\right)^2 - \frac{1}{8} S_4,
\label{eq:29x}
\end{eqnarray}
where
\begin{equation}\label{S2}
S_2 = \left< \sum_{n=1}^{L} A_{n}^{2} \sin \left(
2\theta_{n}^{(1)} \right)\sin \left( 2\theta_{n}^{(2)} \right)
\right>,
\end{equation}
and
\begin{equation}
S_4 = \Biggl<
\sum_{n>k}^{L} \sum_{k=1}^{L} A_{n}^{2} \, A_{k}^{2} \, \sin
\left( 2\theta_{n}^{(1)} \right) \sin \left( 2\theta_{n}^{(2)}
\right) \sin \left( 2\theta_{k}^{(1)} \right) \sin \left(
2\theta_{k}^{(2)} \right)
\Biggr>\,\,\,. \label{S4}
\end{equation}
Here the average is taken over the disorder for a {\it fixed}
number of kicks $n = 1,...,L$.
The terms $S_2$ and $S_4$ describe the correlations between
the phases $\theta_{n}^{(1)}$ and $\theta_{n}^{(2)}$ of two
classical trajectories that start from two complementary initial conditions, see above.

The correlations between $\theta_{n}^{(1)}$ and $\theta_{n}^{(2)}$ turn out to be quite specific. First, it is instructive to analyze the term $S_2$. Taking into account that in the first order of perturbation theory the values of $A_n$ and $\sin
\theta_n$ are uncorrelated (for large $L$), one can average the relation (\ref{S2}) over disorder and angle $\theta_n$ independently. As a result, we get,
\begin{equation}\label{S2R2}
S_2 = 8 L_{loc}^{-1} \sum_{n=1}^{L} R_2(\zeta_n),
\end{equation}
where we introduced the two-point correlator $R_2$,
\begin{equation}\label{R2}
R_2(\zeta_n) = \left<\sin \left( 2\theta_{n}^{(1)} \right) \sin
\left( 2\theta_{n}^{(2)} \right) \right>\,,
\end{equation}
which strongly depends on the ratio $\zeta_n=n/L_{loc}$.
Here the average is taken over the disorder for a {\it fixed}
number of kicks $n$.

\begin{figure}[ht]
\begin{center}
\includegraphics[width=8.6cm,height=6.6cm]{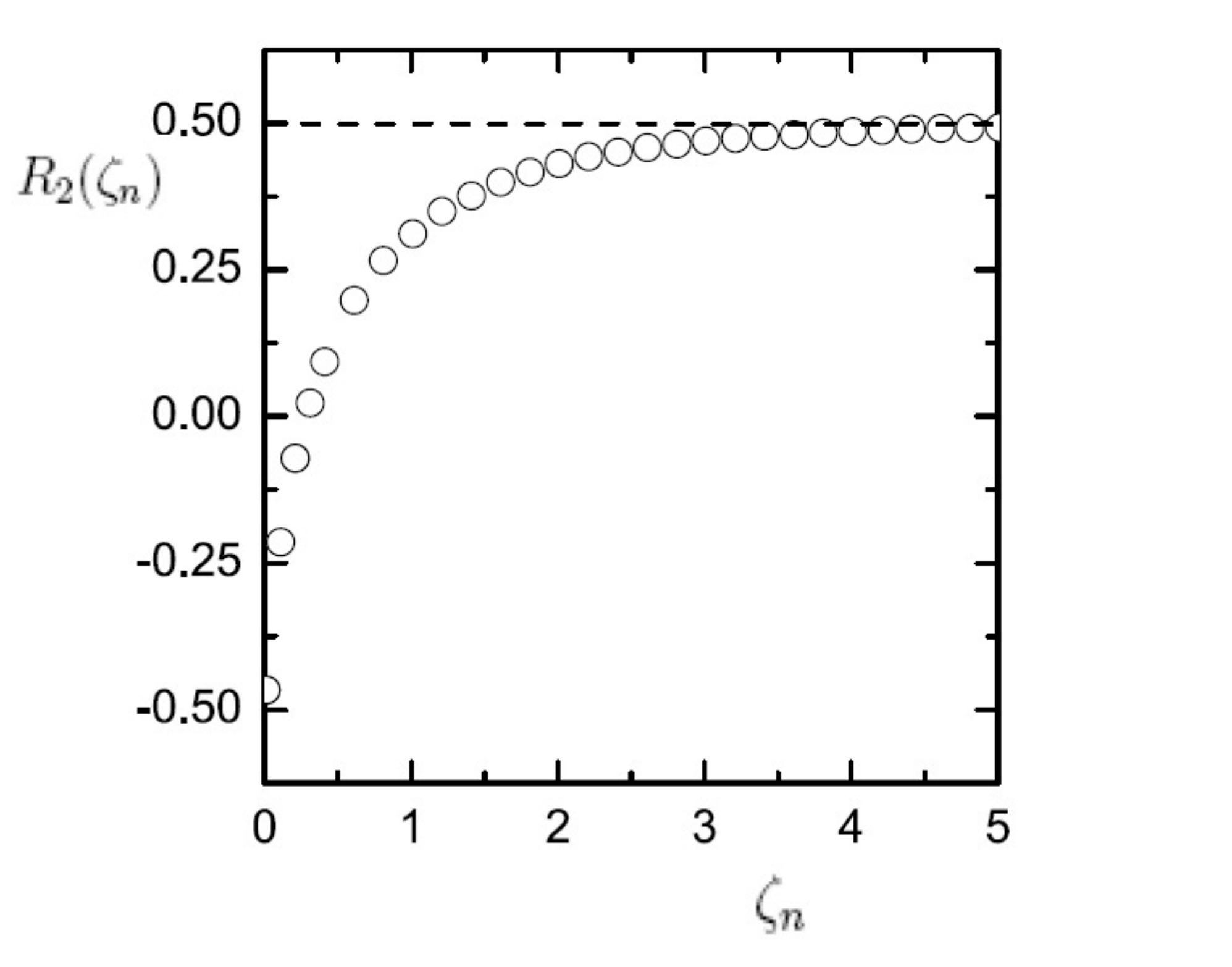}
\end{center}
\caption{Numerical data for $R_2$ versus $\zeta_n = n/L_{loc} $ for $E=1.5$ and $W=0.1$. The average was done over $10^4$ realizations of the
disorder. An additional ``window moving" average was performed in
order to reduce fluctuations (after \cite{DIK04}).}
\label{fig1}
\end{figure}

In Fig.\ref{fig1} we show the numerical data for the correlator $R_2$ in a
wide range of the parameter $\zeta_n$. One can see that the correlator $R_2$ changes from $-1/2$ for the
ballistic regime, $\zeta_n \ll 1$, to $1/2$ for the localized regime, $\zeta_n \gg 1$,. This graph
demonstrates that the correlations give very different contributions in the ballistic and localized regimes.

In order to evaluate analytically the correlator $R_2$, one can express the phases $\theta_{n}^{(1)}$ and $\theta_{n}^{(2)}$ through the initial values $\theta_{0}^{(1)}\,,\theta_{0}^{(2)}$ and the sequence $\{\epsilon_{1}, ..., \epsilon_{n}\}$ of disorder. This can be done with the use of approximate relation (\ref{smallmap}) between $\theta_{n+1}$ with $\theta_n$, see Ref~\cite{DIK04}. As a result, one gets,
\begin{equation}
R_2(\zeta_n) = - \frac{1}{2} + 4 \zeta_n + O(\zeta_n^3).
\label{eq:12}
\end{equation}
This result leads to the expression for the term $S_2$,
\begin{equation}
S_2 = - 4\lambda + 16\lambda^2 + O\left(\lambda^3\right).
\label{S2an}
\end{equation}
The four-point correlator $S_4$, see Eq.\ (\ref{S4}), can be
calculated analytically in a similar way \cite{DIK04},
\begin{equation}
\label{S4an} S_4 = 8\left( \frac{L}{L_{loc}} \right)^2 + O\left( \frac{L}{L_{loc}} \right)^3.
\end{equation}

Thus, in the lowest order of perturbation theory the following estimates can be obtained,
\begin{equation}\label{mean}
\left< T_{L} \right> = 1 - 2 \left( \frac{L}{L_{loc}} \right) + 4
\left(\frac{L}{L_{loc}} \right)^2 + O \left(\frac{L}{L_{loc}} \right)^3,
\end{equation}
and
\begin{equation}\label{square}
\left< T_{L}^2 \right> = 1 - 4 \left( \frac{L}{L_{loc}} \right) + 16
\left(\frac{L}{L_{loc}} \right)^2 + O \left(\frac{L}{L_{loc}} \right)^3 .
\end{equation}
Correspondingly, the variance of $T_L$ can be written as follows,
\begin{equation}\label{var}
\text{Var} \{T_L\} = \left< T_{L}^2 \right> - \left< T_{L} \right>^2
= 4 \left( \frac{L}{L_{loc}} \right)^2 + O \left(\frac{L}{L_{loc}} \right)^3 \,.
\end{equation}
This relation for $ \text{Var} \{T_L\} $ corresponds to Eq.(\ref{1DCP-TBal}) obtained in Section~\ref{2.7} for the continuous model. In Ref.\cite{DIK04} a
more accurate expressions that depend on higher powers of
$\zeta=L/L_{loc}$ were suggested,
\begin{equation}
\label{eq:35xb} \left< T_{L} \right> = \frac{1}{1 +2
\zeta}, \qquad \left< T_{L}^2 \right> = \frac{1}{1+ 4 \zeta} .
\end{equation}
The numerical data \cite{DIK04}, indeed, confirm that these expressions are much better than Eqs.~(\ref{mean}) and (\ref{square}). With the use of Eq.~(\ref{eq:35xb}) one can obtain the following expression for the variance of the transmission coefficient,
\begin{equation}
\text{Var} \{T_L\} = \frac{4\zeta^2}{(1+4\zeta)(1+2\zeta)^2}.
\label{eq:37xb}
\end{equation}
\begin{figure}[t!]
\begin{center}
\includegraphics[width=8.6cm,height=6.6cm]{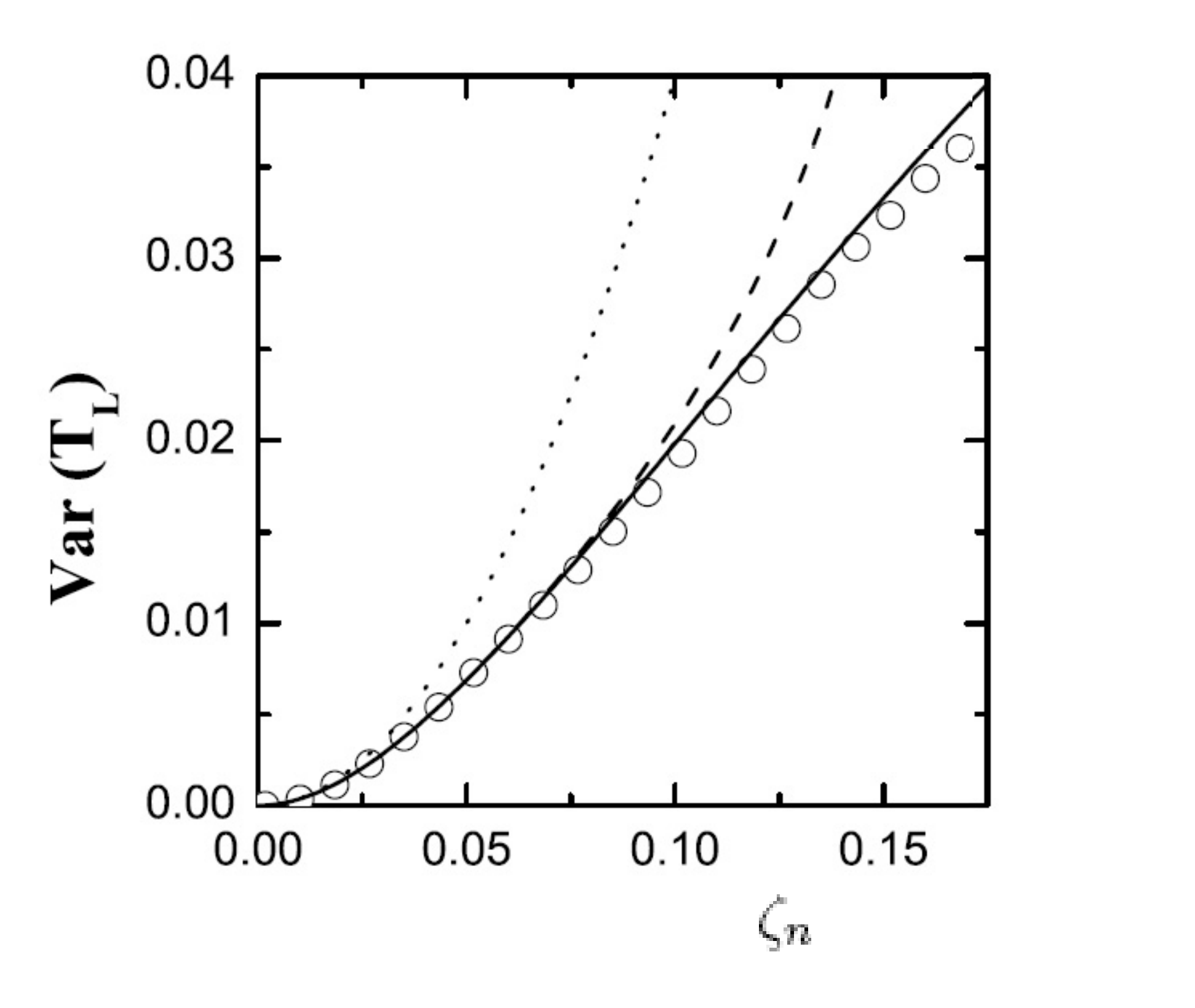}
\end{center}
\caption{Analytical estimates of the mean variance $\text{Var}
\{T_L\}$ are plotted against numerical data (open circles) for
$E=1.5$ and $W=0.1$, with the average over $10^4$ realizations of
disorder. Dots refer to the standard expression
(\ref{var}), dashed lines represent the estimate that takes into
account terms up to the sixth power of $\zeta$ in the expansions
for $\left< \mbox{ln} Z_{L} \right>$ and $\left< \mbox{ln}^2 Z_{L}
\right>$, namely, $\text{Var} \{T_L\} = 4\zeta^2 - 32\zeta^3 +
176 \zeta^4 - 832 \zeta^5 + 3648\zeta^6$. Full curve
corresponds to Eq.~(\ref{eq:37xb}) (after \cite{DIK04}).}
\label{fig2}
\end{figure}
In Fig.\ref{fig2} different approximations for $\text{Var}\{T_L\}$ with the numerical data are compared. Clearly, the expression \ref{eq:37xb} gives a very good agreement with the data. Note that the region of validity of the standard formula (\ref{var}) obtained in the quadratic approximation is very narrow because the coefficients at higher terms ($\zeta^3,\,\zeta^4$, etc) in the expansion of $ {\text {Var}}\{T_L\}$ grow rapidly.

\subsubsection{Ballistic regime: Resistance}
\label{4.4.2}

According to Eq.(\ref{eq:6}) the resistance $T_L^{-1}$ of a
sample of length $L$ is given by the following formula,
\begin{equation}
T_L^{-1} = \frac{1}{2} + \frac{1}{4}(r_{L}^{(1)})^2 + \frac{1}{4}(r_{L}^{(2)})^2 \,,
\label{eq:Re1}
\end{equation}
where the radii $(r_{L}^{(i)})$ are defined by Eq.(\ref{radii}).
It is convenient to represent the mean value of $(r_{L}^{(i)})^2$ in the following form,
\begin{equation}
\left<(r_{L}^{(i)})^2\right> = \left<\exp\left\{\sum_{n=1}^{L}\mbox{ln}\left(
1 + A_{n}^{2}\sin^{2}\theta_{n}^{(i)} + A_{n}\sin2\theta_{n}^{(i)} \right)
\right\}\right> .
\label{eq:mrad1}
\end{equation}
Then, in the limit of weak disorder, $\mid A_n\mid \ll 1$, one can
expand the logarithm in Eq. (\ref{eq:mrad1}). By separating the quadratic and linear terms, we represent $\left <r_{i, L}^{2}\right >$ as a product of two factors,
\begin{equation}
\left<(r_{L}^{(i)})^2\right> =
\left<\exp\left\{\sum_{n=1}^{L}A_{n}^{2}\sin^{2}\theta_{n}^{(i)}
- \frac{1}{2} \sum_{n=1}^{L}A_{n}^{2}\sin^{2}2\theta_{n}^{(i)}\right\}\right> \left<\exp\left\{\sum_{n=1}^{L}A_{n}\sin2\theta_{n}^{(i)}\right\}\right> .
\label{eq:mrad2}
\end{equation}
The first factor $\left< ... \right>$ contains the sums
$\sum_{n=1}^{L}A_{n}^{2}\sin^{2}\phi_{n}^{(i)}$ where $\phi_{n}^{(i)}$ stand for
either $\theta_{n}^{(i)}$ or $2\theta_{n}^{(i)}$. Since $L >> 1$, these sums are the self-averaging quantities with a small variance. Thus, they can be substituted by their mean values,
\begin{equation}
\sum_{n=1}^{L}A_{n}^{2}\sin^{2}\phi_{i}^{(n)}
\approx L\left<A_{n}^{2}\right>\left<\sin^{2}\phi_{n}^{(i)}\right>
= 4\left( \frac{L}{L_{loc}} \right) \; .
\label{eq:msum1}
\end{equation}
Therefore, the first factor of Eq.(\ref{eq:mrad2}) takes the form,
\begin{equation}
\left<\exp\left\{\sum_{n=1}^{L}A_{n}^{2}\sin^{2}\theta_{n}^{(i)}
-\frac{1}{2} \sum_{n=1}^{L}A_{n}^{2}\sin^{2}2\theta_{n}^{(i)}\right\}\right>
\approx \exp\left(2L/L_{loc}\right) \; .
\label{eq:1stfact}
\end{equation}
The second factor in Eq.(\ref{eq:mrad2}) exhibits strong fluctuations and can be calculated with the use of the
distribution function ${\cal P(S_A)}$,
\begin{equation}
\left<f \right> = \int_{-\infty}^{\infty}f\left(S_A\right)
\, {\cal P}\left(S_A\right) \, dS_A \; ,
\label{eq:mvdef}
\end{equation}
where $f(S_A)=\exp \left(S_A \right)$ with
$S_A=\sum_{n=1}^{L}A_{n}\sin2\theta_{n}^{(i)}$. The variable
$S_A$ is a sum of random independent numbers, therefore, according
to the central limit theorem ${\cal P}(S_A)$ is a Gaussian function with the
variance $4 L/L_{loc}$. As a result, the second
factor of Eq.(\ref{eq:mrad2}) can be calculated explicitly,
\begin{equation}
\left<\exp\left(S_A\right)\right> =
\frac{1}{\sqrt{8\pi}}\left( \frac{L}{L_{loc}}\right)^{-1/2}
\int_{-\infty}^{\infty} \exp\left(S_A - 8(L/L_{loc})^{-1}S_A^{2}\right) \, dS_A
= \exp\left(2L/L_{loc}\right).
\label{eq:2ndfact}
\end{equation}
Substituting Eqs.(\ref{eq:1stfact}) and (\ref{eq:2ndfact}) into Eq.(\ref{eq:mrad2}),
we obtain the following expression for the mean value of $(r_{L}^{(i)})^2$,
\begin{equation}
\left<(r_{L}^{(i)})^2\right> = \exp\left(4L/L_{loc}\right) \; .
\label{eq:mrad3}
\end{equation}
This leads to the final formula for the mean value of the resistance
(\ref{eq:Re1}),
\begin{equation}
\left<T_L^{-1}\right> = \frac{1}{2}\left[1 + \exp\left(\frac{4L}{L_{loc}} \right)\right] \; .
\label{eq:Re2}
\end{equation}

Remarkably, this expression is the same as Eq.~(\ref{1DCP-AvVarRes}) derived for continuous potentials for {\it any} value of the control parameter $L/L_{loc}$. This means that Eq.~(\ref{eq:Re2}) gives the resistance both in the ballistic $(L/L_{loc} \ll 1)$ and localized regimes $(L/L_{loc} \gg 1 )$, as well as in the crossover $(L/L_{loc} \simeq 1 )$. Note that the only condition used here is the condition of weak disorder.

The second moment of the resistance can be expressed from Eq.(\ref{eq:Re1}) as follows,
\begin{equation}
T_{L}^{-2} = \frac{1}{4} + \frac{1}{4}\left( (r_{L}^{(1)})^2 + (r_{L}^{(2)})^2 \right)
+ \frac{1}{16}\left( (r_{L}^{(1)})^4+(r_{L}^{(2)})^4\right)
+ \frac{1}{8}(r_{L}^{(1)})^2 (r_{L}^{(2)})^2 \; .
\label{eq:2ndmomre}
\end{equation}
One can see that apart from the second and fourth moments of
$r_{L}^{(i)}$, this expression contains the product $(r_{L}^{(1)})^2 (r_{L}^{(2)})^2 $. The mean value of this term depends on the correlations between two complementary trajectories of the classical map (\ref{Ham}). This fact strongly complicates the analytical treatment.

In the ballistic regime the mean value of the fourth
moments $r_{i, L}^{4}$ can be explicitly evaluated in the same way as described above,
\begin{equation}
\left<r_{i, L}^{4}\right> =
\left<\exp\left(2\sum_{n=1}^{L}A_{n}^{2}\sin^{2}\theta_{n}^{(i)}
- \sum_{n=1}^{L}A_{n}^{2}\sin^{2}2\theta_{n}^{(i)}
\right)\right>
\left<\exp\left(2\sum_{n=1}^{L}A_{n}\sin2\theta_{n}^{(i)}
\right)\right>,
\label{eq:mrads1}
\end{equation}
where,
\begin{equation}
\left<\exp\left(2\sum_{n=1}^{L}A_{n}^{2}\sin^{2}\theta_{n}^{(i)}
- \sum_{n=1}^{L}A_{n}^{2}\sin^{2}2\theta_{n}^{(i)}\right)\right>
\approx \exp\left(4L/L_{loc}\right),
\label{eq:1stfact2}
\end{equation}
and
\begin{equation}
\left<\exp\left(2\sum_{n=1}^{L}A_{n}\sin2\theta_{n}^{(i)}
\right)\right>
\approx \exp\left(8L/L_{loc}\right) \; .
\label{eq:2ndfact2}
\end{equation}
Therefore, for the mean value of the second moment $r_{i, L}^{4}$ we have,
\begin{equation}
\left<r_{i, L}^{4}\right> = \exp\left(12L/L_{loc}\right) \; .
\label{eq:mrads2}
\end{equation}

Let us now consider the term $(r_{L}^{(1)})^2 (r_{L}^{(2)})^2$. For weak disorder one can write the
mean value of this
quantity in the following form,
\begin{equation}
\left<(r_{L}^{(1)})^2 (r_{L}^{(2)})^2\right> \approx
\left<\exp\left\{\sum_{n=1}^{L}A_{n}^{2} Z_n \right\} \right>
\left<\exp\left(S_{P}\right)\right> \approx \exp\left(4L/L_{loc}\right)
\left<\exp\left(S_{P}\right)\right>,
\label{eq:r1r2a}
\end{equation}
where
\begin{equation}
Z_n = \left(\sin^{2}\theta_{n}^{(1)}
+ \sin^{2}\theta_{n}^{(2)}\right)
- \frac{1}{2}\left(\sin^{2}2\theta_{n}^{(1)}
+ \sin^{2}2\theta_{n}^{(2)}\right),
\end{equation}
and $S_{P}$ is a new random variable,
\begin{equation}
S_{P}=\sum_{n=1}^{L}A_{n}\left(\sin2\theta_{n}^{(1)} + \sin2\theta_{n}^{(2)}\right).
\label{eq:S_P}
\end{equation}
As discussed above, the correlations between phases
$\theta_{n}^{(1)}$ and $\theta_{n}^{(2)}$
are very different in the ballistic ($L/L_{loc} \ll 1$) and localized ($L/L_{loc} \gg 1$)
regimes. For the ballistic regime we can safely to write,
\begin{equation}
\left<\exp\left(S_{P}\right)\right> \approx 1
+ \frac{1}{2}\left[\left<\sum_{n=1}^{L}A_{n}^{2}\sin^{2}2\theta_{n}^{(1)}\right> +
\left<\sum_{n=1}^{L}A_{n}^{2}\sin^{2}2\theta_{n}^{(2)}\right>
\right]+ S_2 \approx 1 + O\left( \frac{L}{L_{loc}}\right ) ^2,
\label{eq:2ndfact3}
\end{equation}
where $S_{2}$ is determined by Eq.(\ref{S2}).
Thus, we obtain,
\begin{equation}
\left<(r_{L}^{(1)})^2 (r_{L}^{(2)})^2\right> \approx \exp\left(4L/L_{loc}\right)
 \; .
\label{eq:r1r2d}
\end{equation}
As a result, the second moment of the resistance in the ballistic regime has the form,
\begin{equation}
\left<T_{L}^{-2}\right> \approx 1 + 4 \left( \frac{L}{L_{loc}} \right) + 16 \left(\frac{L}{L_{loc}}\right)^2 +O \left(\frac{L}{L_{loc}}\right)^3\; .
\label{eq:Re2MR}
\end{equation}
This expression is in a good agreement with the numerical data \cite{DIK04a} up to $L/L_{loc} \approx 0.3$, and with the expression (\ref{1DCP-AvVarRes}) for continuous potentials.

\subsubsection{Localized regime: Transmittance}
\label{4.4.3}

Now let us consider another limiting case of strong localization, when the localization length $L_{loc}$ is much less than the sample size, $L/L_{loc}\ll 1$. In this case all trajectories of the classical map (\ref{Ham})
move off the origin $p=x=0$ very fast with an increase of ``time" $n$ (for $n \rightarrow \infty$). This means that in Eq.~(\ref{eq:6}) the radii $r_{L}^{(1)}$ and $r_{L}^{(2)}$ are rapidly growing functions of the final ``time" $L$. Therefore, in this case $Z_L \gg 1$ and in the lowest
approximation we have,
\begin{equation}
T_{L} \approx \frac{2}{Z_{L}}.
\label{eq:39x}
\end{equation}
Since in a strongly localized regime the transmission coefficient reveals very large
fluctuations, it is not a self-averaged quantity, and the appropriate quantity for averaging is the logarithm of $T_L$. Its average value is given by
\begin{equation}
\left< \mbox{ln}\,T_{L} \right> = \mbox{ln} \, 2
- \left< \mbox{ln}Z_{L} \right>,
\label{eq:40x}
\end{equation}
and the variance is,
\begin{equation}
\text{Var} \, \{ \ln T_L\} = \left< \ln^2 T_L \right> - \left< \ln T_L \right> ^2
= \text{Var} \,\{\ln Z_L\}.
\label{varTdef}
\end{equation}
Using the expression (\ref{Zdetail}) for $Z_L$, it can be shown that
the mean value of $\ln Z_L$ can be expressed as follows,
\begin{equation}
\left< \mbox{ln}Z_{L} \right> = 2 \left( \frac{L}{L_{loc}}\right) - \ln 2 +
R_\infty.
\label{eq:41x}
\end{equation}
Here the effect of correlations between the phases
$\theta_{n}^{(1)}$ and $\theta_{n}^{(2)}$ enters in the last term,
\begin{equation}
\label{term}
R_\infty =
\left< \mbox{ln}
{\left\{ 1 + \exp{ \left [\sum_{n=0}^{L-1} A_{n}
\left( \sin{ 2\theta_{n}^{(2)}}
- \sin {2\theta_{n}^{(1)}} \right) \right ]}
\right\} } \right>.
\end{equation}

The detailed analysis \cite{DIK04} of this expression shows that the term $R_\infty$ is independent of the sample length
$L$. Indeed, in the
limit $\zeta_n\rightarrow \infty$ the correlator (\ref{R2})
shown in Fig.\ref{fig1} approaches $1/2$, that is the mean value of
$\sin^2\theta$. Therefore, the phases $\theta_{0}^{(1)}$ and
$\theta_{0}^{(2)}$ fluctuate coherently in such a way that
\begin{equation}
\sin \left( 2 \theta_{n}^{(2)} \right) -
\sin \left( 2 \theta_{n}^{(1)} \right) \rightarrow 0
\label{eq:25}
\end{equation}
for $n\rightarrow \infty$. For this reason the upper limit in the sum in Eq.~(\ref{term}) can be replaced by $L =\infty$ and the correlation term $R_\infty$
becomes $L-$independent. This means that $R_\infty$ is independent of $L_{loc}$ as well, due to the scaling dependence of $T_L$ on the ratio $L/L_{loc}$ . As a result, the correlation term $R_\infty$ is a constant. Analytically, it is not possible to evaluate $R_\infty$ due to a quite complicated character of the correlations between phases $\theta_{n}^{(2)}$ and $\theta_{n}^{(1)}$ on the initial stage of evolution, when the two trajectories are neither statistically independent nor fully coherent.
This initial stage gives a finite contribution and does not
depend on $L$ provided $L$ is larger than $L_{loc}$ (from Fig.1 it is seen that practically one needs to have $\zeta_n \geq 5$). The numerical evaluation of $R_\infty$ performed in Ref.\cite{DIK04}, has shown that the relation, $R_\infty = 2\ln2$, holds with a high accuracy. This fact significantly simplifies further analytical estimates.

First, one can obtain the famous expression (\ref{1DCP-AvLn}) derived for continuous disordered potentials,
\begin{equation}
\left< \mbox{ln}\,T_{L} \right> = -2\frac{L}{L_{loc}} .
\label{eq:43}
\end{equation}
This formula is accurate up to the linear term with
respect to $L/L_{loc}$, and it takes into account the correlations along the whole trajectory.

It is instructive that even by neglecting specific phase correlations on the initial time scale of the evolution of classical trajectories, one can obtain important results. Indeed, by taking into account the effect of attraction of phases, see Eq.(\ref{eq:25}), one can write, $r_{L}^{(1)}=r_{L}^{(2)}=r_{L}$, therefore,
\begin{equation}
\label{lnZ}
\ln Z_L \approx \ln r_L^2 =
\sum\limits_{n=1}^{L} \ln D_{n}^2 \approx \sum\limits_{n=1}^{L}
A_n \sin 2\theta_n + Y_L
\,\,\,,
\end{equation}
where
\begin{equation}\label{Sigma2}
Y_L = \sum\limits_{n=1}^{L}
\left(- \frac{1}{2} A_n^2 \sin^2 2\theta_n
+ A_n^2 \sin^2 \theta_n \right)\,\,\,.
\end{equation}

The first term on the r.h.s. of Eq.\ (\ref{lnZ}) is a sum of
$L>>1$ random independent numbers. Therefore, the statistical
distribution of $\ln Z_L$ is a Gaussian provided the quadratic term
$Y_L$ is neglected. Thus, we can see that the log-norm
distribution for the transmission coefficient in the localized
regime occurs in the lowest approximation in weak disorder, together with an ignorance of the difference (\ref{eq:25}) between the phases $\theta_n^{(1)}$ and $\theta_n^{(2)}$ along the two trajectories.

From Eq.\ (\ref{lnZ}) one can easily obtain the expressions for the first
and second moments of $\ln Z_L$. The first non-zero term for the
mean value of $\ln Z_L$ is due to the second order term $Y_L$,
therefore,
\begin{equation}
\left<\ln Z_L \right> = \left< Y_L \right> \approx
\frac{1}{4} L \left<A_n^2\right> = 2L/L_{loc}
\label{ln-ZL}
\end{equation}
which is consistent with Eq.\ (\ref{eq:43}).
Calculating the second moment of $\ln Z_L$, we can neglect the
quadratic term in Eq.\ (\ref{lnZ}) and get,
\begin{equation}\label{lnZ2}
\left< \ln^2 Z_L \right> = 4\frac{L}{L_{loc}} + 4 \left( \frac{L}{L_{loc}}\right )^2.
\end{equation}
Therefore, for the variance of $\ln Z_L$ and $\ln T_L$ we have,
\begin{equation}\label{sigma2}
\mbox{Var} \, \{ \ln T_L\} = 4\frac{L}{L_{loc}}.
\end{equation}
By comparing this expression with Eq.~(\ref{ln-ZL}) one gets the famous relation (\ref{1DCP-VarLnLoc}),
\begin{equation}
\mbox{Var} (\ln T_L)= -2 \left< \ln T_{L} \right> .
\label{ratio}
\end{equation}
As one can see, the effective width of the distribution of $P (\ln T_L)$ is of the square root of the mean value of $\ln T_L$ itself. Since the latter is very large for $L \gg 1$, the approximation according to which the term $Y_L$ in Eq.(\ref{lnZ}) can be neglected, is justified.

The obtained expressions (\ref{eq:43}) and (\ref{ratio}) are widely discussed in connection with the single parameter scaling (see, for example, Refs.~\cite{AALR79,DLA00,DLA01,DEL03,ST03} and references therein). In solid state application by this scaling one means the conjecture that all the moments of the dimensional conductance $T_L$ depend on its average value $\left< T_{L} \right>$ only. For the continuous potentials considered in Section~\ref{2} this fact trivially stems from the exact result according to which the distribution function of $T_L$ depends on one parameter which is the ratio $L/L_{loc}$. The expression (\ref{ratio}), together with the log-norm distribution for $\ln T_L$ (see Section~\ref{2.8}) is the cornerstone of the single parameter scaling conjecture.

Thus, our analysis of the classical map (\ref{Ham})
shows that the single parameter scaling is equivalent to the regime when two phases
$\theta_n^{(1)}$ and $\theta_n^{(2)}$ behave in a coherent way.
This allows to relate the transmission coefficient to one trajectory
only, in contrast to the fact that formally the solution of the Schr\"odinger
equation (\ref{tb diagonal}) depends on two independent particular solutions.

\subsubsection{Localized regime: Resistance}
\label{4.4.4}

As shown above, in a strong localized regime ($L/L_{loc} \gg 1 $) the phases fluctuate coherently, i.e. $\theta_{n}^{(1)} \approx \theta_{n}^{(2)}$.
This property helps to evaluate the mean value $\left<S_P\right>$, see
Eq. (\ref{eq:S_P}). Using the same procedure as in the case of
ballistic regime, one can find
that in the localized regime the distribution function of random
variable $S_{P}$ is the Gaussian with the zero mean and variance given by the relation,
\begin{equation}
\left<S_{P}^{2}\right> - \left<S_{P}\right>^{2} = 16L/L_{loc} \; .
\label{eq:varsum}
\end{equation}
The numerical analysis performed in Ref.\cite{DIK04} confirms this conclusion.

The mean value of the normally distributed variable $S_P$ can be easily evaluated,
\begin{equation}
\left<\exp\left(S_{P}\right)\right> = \exp\left(8L/L_{loc}\right),
\label{eq:2ndfact5}
\end{equation}
that leads to the expression,
\begin{equation}
\left<\left(r_{L}^{(1)}\right)^{2} \left(r_{L}^{(2)}\right)^{2}\right> = \exp\left(12L/L_{loc}\right) \; .
\label{eq:r1r2e}
\end{equation}
As one can see, in the localized regime this correlator is exponentially large, reflecting the fact that $r_{L}^{(1)}$ and $r_{L}^{(2)}$ are strongly correlated.
Thus, the average square of resistance $R_L^2$ also grows exponentially,
\begin{equation}
\left<T_L^{-2}\right> = \frac{1}{4} \exp \left(12L/L_{loc} \right)
+\frac{1}{2} \exp\left(4L/L_{loc}\right)
+ \frac{1}{4} \; .
\label{eq:Re2SLR}
\end{equation}
The above result is slightly different from the exact expression (\ref{1DCP-AvVarRes}) obtained for continuous potentials, and the origin of this discrepancy remains unclear.

The direct numerical computation of both $ \left<T_L^{-1}\right> $ and $ \left<T_L^{-2}\right> $
is not possible due to enormously large fluctuations. This confirms the fact that the resistance
$R_L=T_L^{-1}$ (as well as the conductance $T_L$) is not a self-averaged quantity. For this reason, one has to
consider the average of logarithm of the moments of resistance.

Due to a simple relation between the resistance and conductance, it is clear that in the strongly localized regime some of the results obtained for $T_L$
can be also used for $T_L^{-1}$. In particular, since $\ln(T_L^{-1}) = - \ln(T_L)$, the following relations hold,
\begin{equation}
\label{loglog}
\left <\ln T_L^{-1} \right > = 2\left( \frac{L}{L_{loc}} \right); \,\,\,\,\,\,\,\,\,\,
\left < \ln ^2 T_L^{-1} \right> = 4\left( \frac{L}{L_{loc}} \right) + 4\left( \frac{L}{L_{loc}} \right)^2 \; ,
\end{equation}
\begin{equation}
\mbox{Var} \{ \ln T_L^{-1} \} = \left < \ln ^2 T_L^{-1} \right>
- \left <\ln T_L^{-1} \right > ^2 =
2 \left <\ln T_L^{-1} \right > \; .
\label{spsR}
\end{equation}
The last expression can be also treated as the manifestation of the single parameter scaling.

\section{Correlated disorder in the Anderson model}
\label{5}

\subsection{Random dimers}
\label{5.1}

For a long time it was believed that in one-dimensional random potentials {\it all} eigenstates are localized in infinitely large samples, regardless how weak is disorder \cite{A58,MT61,I73,E83}. Strictly speaking it is not true; the existence of fully extended states (the so-called Azbel resonances) for specific values of energy was known to occur, according to the rigorous proof in Refs.\cite{A81,A83}. Since the measure of these resonances is zero, and they appear randomly in the energy spectrum, generally they can be neglected. About two decades ago, it was, however, discovered \cite{DWP90,WP91,WP91a} the presence of fully transparent states that appear in tight-binding disordered models (called {\it dimers}), when specific short range correlations are imposed. The crucial difference from the Azbel resonances is that the found resonances, although occurring for specific energies, exist for any specific realization of a random potential. This fact triggered an extensive discussion, see, for example, \cite{EK92,GS92,DGK93a,DGK93,EW93,EE93}. A practical importance of the dimer resonances is that they may emerge in realistic finite one-dimensional systems, such as conducting polymers. The resonances of a similar nature have been extensively studied in Ref.~\cite{H94}.

Let us discuss the origin of delocalized states in random dimers. The random dimer model is specified by two site energies $\bar\epsilon_1$ and $\bar\epsilon_2$ in the Shr\"odinger equation (\ref{tb diagonal}), that appear in the sequence $\{\epsilon_n \}$ in pairs, specifically, $\epsilon_n=\epsilon_{n+1} = \bar\epsilon_1$ or $\bar \epsilon_2$. These pairs (dimers), $(\bar\epsilon_1,\bar\epsilon_1)$ and $(\bar\epsilon_2,\bar\epsilon_2)$ appear in the potential $\epsilon_n$ {\it randomly}, with the probability $1-Q$ and $Q$, respectively. Without loss of generality, we consider below the most random case, $Q=0.5$.

In order to understand the mechanism related to the resonance transmission in the dimer potential, we start with the situation when only one dimer, say, $\bar\epsilon_2$-dimer is placed on two adjacent sites $m$ and $m+1$. All other sites are occupied by the other $\bar\epsilon_1$-dimer. Asking about the eigenvalues $\varsigma_n$ of the classical map (\ref{Ham}), they are given by the expression,
\begin{equation}
\varsigma_n^{\pm}=\frac{v_n \pm i \sqrt{4-v_n^2}}{2}=e^{\pm i \mu_n},
\label{eigen1}
\end{equation}
where $v_n=E-\epsilon_n$ and $n=m,m+1$. Since $\mu_m=\mu_{m+1}$, this single dimer with energy $\bar\epsilon_2$ is ``invisible" for a wave function if the total phase advance, $2\mu_m$ is equal to $\pi$ or $2 \pi$. However, the value $\mu_m=\pi$ is forbidden due to the stability condition,
\begin{equation}
|v_m|=|2 \cos \mu_m| < 2
\label{stab}
\end{equation}
therefore, the resonant energy $E_c$ is defined by $\mu_m=\pi/2$ only, thus giving $E_c=\bar \epsilon_2$. This happens for {\it all} $\bar \epsilon_2$-dimers embedded in the lattice with $\epsilon_n=\bar\epsilon_1$. The same resonance effect occurs for other resonant energy, $E_c=\bar\epsilon_1$. In this case, the $\bar\epsilon_1$-dimers are ``invisible", and the wave propagation is the same as in the lattice with constant values of $\epsilon_n=\bar\epsilon_1$. As a result, for a general case of randomly distributed dimers $\bar \epsilon_1$ and $\bar \epsilon_2$ there are two resonant values,
\begin{equation}
E_c = \bar \epsilon_1 , \, \bar \epsilon_2 ,
\label{2-res}
\end{equation}
for which either the $\bar \epsilon_1$-dimers or $\bar \epsilon_2$-dimers have no any influence on the scattering states. One should note that in the case of only one type of dimer, say, $\bar \epsilon_2$, the transparent states appear only for one resonant energy $E_c=\bar\epsilon_2$.

Based on the above analysis, it is easy to understand that in the general case of $N$-mer (two values $\bar \epsilon_1$ and $\bar \epsilon_2$ appear in the site potential in blocks of length $N$) the resonant energies are defined by the following condition,
\begin{equation}
\mu = \frac{(s+1)\pi}{N},
\label{Nmer}
\end{equation}
with $s=0,1,2,...,N-2$.
For example, for trimers the resonant energies are given by the relations,
\begin{equation}
E_c-\bar \epsilon_{1,2}=2 \cos \mu_m\,, \qquad \mu_m= \frac{\pi}{3}\, ; \frac{2\pi}{3},
\label{3mer}
\end{equation}
which results in four values for $E_c$,
\begin{equation}
E_c= \bar\epsilon_1-1\,, \bar\epsilon_1+1\,, \bar\epsilon_2-1\,, \bar\epsilon_2+1,
\label{trim-E}
\end{equation}
see Refs.\cite{EE93,IKT95}.

An important question is how fast the localization length $L_{loc}$ diverges when approaching the resonance with $E_c$. Since the phases $\theta_n$ and $\theta_{n+1}$ can not be treated fully uncorrelated, in Ref.\cite{IKT95} it was suggested that instead of considering the one-step map (\ref{loclength}) is it more convenient to deal with the two-step map that relates $x_{n+2}, p_{n+2}$ to $x_n, p_n$. Then the expression for the Lyapunov exponent takes the form (compare with Eq.(\ref{locAnd})),
\begin{equation}
\lambda = L_{loc}^{-1}= \lim_{N\rightarrow \infty} \frac{1}{2N} \sum _{n=1}^{N} \ln \left( \frac{r_{n+2}}{r_n} \right) .
\label{lyap-2R}
\end{equation}
It is clear that the phases $\theta_{n+2}$ and $\theta_n$ are uncorrelated, and the averaging of the logarithm in Eq.(\ref{lyap-2R}) can be easily performed. The relation between $r_{n+2}$ and $r_n$ can be found by constructing the two-step map, based on Eqs.(\ref{map-theta}) and (\ref{Dn}). For a weak disorder, $|\epsilon_n|\ll 1$, in the vicinity of the resonance, $E=E_c-\delta$ with $\delta \ll 1$, the following relation has been obtained \cite{IKT95},
\begin{equation}
\label{r-n2}
r_{n+2}=r_n \{ 1- \epsilon_n \left[ F_1(\theta_n) - 2\delta F_2(\theta_n) \right] \} ,
\end{equation}
where the functions $F_1$ and $F_2$ are given by,
\begin{equation}
F_1(\theta_n) = \sin ^2 \theta_n - \sin ^2 (\theta_n + \mu)\,, \qquad
F_2(\theta_n) = \sin \theta_n \sin (\theta_n+\mu),
\label{F1}
\end{equation}
with $2\cos \mu =E$ and $\epsilon_n= \bar \epsilon_1\,,\bar \epsilon_2$.

Near the resonance, $\delta \ll 1$, the Lyapunov exponent can be estimated by making use of expansion of $\ln \left(r_{n+2}/r_n \right)$ with the successive average over $\theta_n$. As a result, one can obtain,
\begin{equation}
\label{lam-gen}
\lambda \approx \frac{Q}{4} \frac {\delta^2 \cos^2 \mu}{\sin^2 \mu} ,
\end{equation}
where the factor $Q$ stands for the probability for the $\bar \epsilon_2$-dimer to appear. From the above expression the dependence of the localization length on the distance $\delta$ from the resonance can be easily found for two extreme cases. The first one is when the value of $\bar \epsilon_2$ is far enough from the stability border $E_b=2$, however, being inside the energy spectrum. If the distance $\Delta = E_b -\bar \epsilon_2 = 2-\bar \epsilon_2$ is large in comparison with $\delta =\bar \epsilon_2 - E$, then one gets,
\begin{equation}
\label{case-1}
L_{loc} \sim \frac{\Delta}{\delta^2}\,, \qquad \delta \ll \Delta \ll 1,
\end{equation}
due to the relation, $2 \cos \mu = \bar \epsilon_2 -\delta$ . In the other limit case with $\bar \epsilon_2=2$ corresponding to the border of stability, we have,
\begin{equation}
\label{case-2}
L_{loc} \sim \frac{1}{\delta}\,, \qquad \delta \ll 1\,,\,\, \, \Delta =0 \,.
\end{equation}
The above two dependencies have been found through different approaches in Refs.\cite{F89,B92,EW93}. Note that similar divergences emerge near another resonance with $E=\bar\epsilon_1$.

The resonances occurring in the dimer model lead to an emergence of a finite region of $E$ near the resonant energies, where the localization length $L_{loc}$ is larger than $L$. In such a way, the borders of energy windows of a perfect transmission may be treated as effective mobility edges. It was argued \cite{DWP90} that the number $M$ of effectively extended states is inversely proportional to the square root of the sample size, $M \sim 1/\sqrt{L}$. Note again that for the dimers of an infinite size the extended states arise for a {\it discrete set} of resonant energies only. Therefore, there is no contradiction with the general statement that ``true" mobility edges do not exist in one-dimensional random potentials.

In Ref.\cite{EW93} a more general dimer model was considered, in which two sites $A$ and $B$ are involved in a single unit cell. In such a chain the $AA, BB, AB$ and $BA$ cells are distributed at random with corresponding probabilities $P_{AA}, P_{AB}, P_{BA}$ and $P_{BB}$. For the case of $P_{AB}=P_{BA}=0$ the conventional dimer model is recovered. As one can see, this model also belongs to the class of models with short-range correlations.

The first experimental verification of an emergence of delocalized state in the dimer model is reported in Ref.\cite{Bo99}. The authors created semiconductor superlattices with 200 periods of wells and barriers and studied the electronic properties of the samples with regular, random and dimer-type configurations of the wells. Their results have confirmed a high transmission for specific values of energy, in accordance with the theoretical and numerical predictions.

\subsection{The localization length}
\label{5.2}

\subsubsection{Basic expressions}
\label{5.2.1}

So far, we discussed the properties of the Anderson model assuming the disorder is of the white noise type, or fully uncorrelated. Now we consider the general case when the random potential $\epsilon_n$ may include any kind of correlations (``colored noise"). In particular, these correlations can be either {\it short-range} or {\it long-range} ones. The Hamiltonian map approach turns out to be very effective
also for the models with correlated disorder.

In order to characterize the correlation properties of the potential, we introduce the normalized pair (binary) correlator $K(m)$ of the sequence $\epsilon_n$,
\begin{equation}
\label{bin-corr}
\langle \epsilon_n \epsilon_{n+m} \rangle=\sigma^2 K(m)\,, \qquad \langle \epsilon_n \rangle =0 \,,\qquad \langle \epsilon_n^2\rangle = \sigma^2 \,.
\end{equation}
In analogy with Eqs.(\ref{1DCP-FTW}) one can write the relations for the correlator $K(m)$ and its Fourier transform ${\cal K}(\mu)$,
\begin{subequations}\label{spect-gen}
\begin{eqnarray}
{\cal K}(\mu)&=& 1+2\sum_{m=1}^{\infty} K(m)\cos(\mu m)\\[6pt]
K(m)&=&\frac{1}{\pi} \int_{0}^{\pi} {\cal K}(\mu)\cos(\mu m) d\mu\,,
\end{eqnarray}
\end{subequations}
with the normalization condition,
\begin{equation}
\frac{1}{\pi} \int_{0}^{\pi} {\cal K}(\mu) d\mu=1\,.
\label{norm-corr}
\end{equation}
By assuming that the disorder is weak, $\sigma^2 \ll 1$, we expand the logarithm, $\ln \left( r_{n+1}/r_n \right)$, in Eq.(\ref{locAnd}) in the second order over $\epsilon_n$, keeping now an additional term that was neglected for uncorrelated potentials (compare with Eq.(\ref{standard})),
\begin{equation}
\label{Layp approx}
L_{loc}^{-1}(E)= \frac{\sigma^2}{8 \sin^2\mu}-\frac{\langle\epsilon_n \sin(2 \theta_n)\rangle}{2 \sin \mu}.
\end{equation}

The second term in Eq.(\ref{Layp approx}) vanishes if in the recursive relation (\ref{smallmap}) we neglect the linear and quadratic terms in the expansion over $A_n=\epsilon_n/\mu$, thus assuming that $\theta_{n}=\theta_{n-1} -\mu$. This can be safely done for a white-noise disorder, however, for the correlated disorder one has to take into account the next order term proportional to $A_{n-1}$,
\begin{equation}
\label{map appr}
\theta_{n} = \theta_{n-1} - \mu + A_{n-1}\, {\sin^2 \theta_{n-1}}.
\end{equation}
Note that there is no need to keep the quadratic term proportional to $A_{n-1}^2$ since it results in the higher order of expansion of the Lyapunov exponent in the parameter $\sigma^2$.

As one can see, the problem of finding the Lyapunov exponent is reduced to the evaluation of the correlator $\langle\epsilon_n \sin(2\theta_n)\rangle$. This can be done with the method described in Ref.~\cite{IK99,TI01a}. First, we introduce the correlator $K_m$,
\begin{equation}
\label{corr-Kl}
K_m= \langle \epsilon_n e^{2i\theta_{n-m}}\rangle\,
\end{equation}
and write the relation between $K_{m-1}$ and $K_m$ with the use of Eq.(\ref{smallmap}),
\begin{equation}
\label{corr-KlKl}
K_{m-1}= \langle \epsilon_n e^{2i\theta_{n-m+1}}\rangle\approx \langle \epsilon_n e^{2i\theta_{n-m}}e^{-2i\mu} \left[1-i A_{n-m}(1-\cos 2\theta_{n-m}) \right] \rangle \,.
\end{equation}
For weak disorder the triple noise-angle correlators can be factorized and the average over the angle variable can be carried out with the uniform distribution. As a results, one gets,
\begin{equation}
\label{corr-3}
K_{m-1}= K_m e^{-2i\mu}-\frac{i}{2\sin \mu} e^{-2i\mu}
\langle \epsilon_n \epsilon_{n-m} \rangle \,.
\end{equation}
Now, multiplying both sides of Eq.(\ref{corr-3}) by $\exp [-2i\mu (m-1)]$ and summing over $m$ from zero to infinity, one can obtain the expression for $K_0$,
\begin{equation}
\label{corr-K0}
K_{0}= -\frac{i}{2\sin \mu} \sum_{m=1}^{\infty}
\langle \epsilon_n \epsilon_{n-m} \rangle e^{2i\mu m}\,.
\end{equation}
Then the correlator $\langle\epsilon_n \sin(2\theta_n)\rangle$ reads,
\begin{equation}
\label{corr-eps-theta}
\langle\epsilon_n \sin(2\theta_n)\rangle = \Im (K_0)= -\frac{1}{2\sin \mu} \sum_{m=1}^{\infty}
\langle \epsilon_n \epsilon_{n-m} \rangle \cos (2\mu m)\,,
\end{equation}
where $\Im$ stand for the imaginary part. By inserting this correlator into Eq.(\ref{Layp approx}), we arrive at the final expression for the Lyapunov exponent \cite{IK99},
\begin{equation}
\label{Lyap corr}
\lambda= \frac{\sigma^2 }{8 \sin^2\mu} \left( 1 + 2 \sum_{m=1}^{\infty} K(m)\cos (2\mu m)\right) =\frac{\sigma^2 }{8 \sin^2\mu}{\cal K}(2\mu).
\end{equation}

The same result was obtained in Refs.~\cite{GF88,L89,TS05} using different approaches. Since it is valid for weakly disordered systems, $\sigma^2 \ll 1$, the localization length covers many periods of the lattice,
\begin{equation}
\label{l>>1}
L_{loc}(E) \gg 1.
\end{equation}
As a function of $E$ the localization length in Eq.(\ref{Lyap corr}) is symmetric with respect to the band center $E=0$. Note that the parameter $\mu$ is, in fact, the wave number that is related to the energy due to the dispersion relation,
\begin{equation}
\label{disp-A}
E=2 \cos \mu \approx 2-\mu^2\,\qquad {\cal E} =2-E \approx \mu^2.
\end{equation}
Here we introduced new energy ${\cal E}=2-E$ and explore the region close to $E=2$, in order to see where the conventional dispersion relation ${\cal E} = \mu^2$ holds, see Eq.~(\ref{1DCP-Lloc}).

One can see that the only difference of Eq.~(\ref{Lyap corr}) for the correlated disorder from the Thouless expression (\ref{standard}), is an additional factor ${\cal K}(\mu)$ that absorbs possible correlations within the first order of perturbation theory. According to Eq.(\ref{spect-gen}), this factor is the $2\mu$-harmonic of the Fourier transform of the binary correlator (\ref{bin-corr}). It is also known as the power spectrum of the potential $\{ \epsilon_n$ \}, see Section~\ref{2.1}. It should be stressed that according to the Bochner's theorem \cite{Y87}, the Fourier transform (or the Fourier series for integer $m$) of the pair correlator of {\it any} sequence $\epsilon_n$ is a non-negative function. This is in consistence with the proportionality between the Lyapunov exponent $\lambda$ (which cannot be negative for the models without absorption) and the function ${\cal K} (\mu)$ in the relation (\ref{Lyap corr}). Note that the pair correlator itself can have both positive and negative values.

The remarkable fact is that in the first order (Born) approximation the localization length is determined only by the binary correlator. The higher moments of the random potential can appear only in the next terms of the expansion of $L_{loc}^{-1}(E)$ over $\sigma^2$. For a white noise disorder all correlators $K(m)$ with $m \neq 0$ vanish and Eq.(\ref{Lyap corr}) recovers the Thouless formula (\ref{standard}). In the other limiting case of a constant potential, $\epsilon_n = \epsilon$, one gets $K(m) = 1$ and
\begin{equation}
\label{periodic}
{\cal K}(\mu) = 2\pi \sum_{m=-\infty}^{\infty} \delta(\mu -2\pi m).
\end{equation}
According to the dispersion relation Eq.(\ref{disp-A}), the value of $\mu$ is restricted by the interval $0< \mu <\pi$, therefore, ${\cal K}(\mu)=0$. Thus, all states within the allowed zone $-2<E<2$ are delocalized ($L_{loc}(E) = \infty$), as expected.

There is a wide class of correlated potentials between the white noise potential, $K(m)=\delta_{m,0}$, and those with $K(m) = 1$. Note that Eq.~(\ref{Lyap corr}) is generically valid for {\it any} random potential, provided it is statistically homogeneous. The latter means that the binary correlator $K(m)$ can be introduced according to Eq.(\ref{bin-corr}).

\subsubsection{Short-range correlations.}
\label{5.2.2}

A simple example of short-range correlations is quite general case of exponentially decaying correlations,
\begin{subequations}\label{short-corr}
\begin{eqnarray}
{\cal K}(\mu)&=&\frac{\sinh \mu_c}{\cosh \mu_c - \cos \mu}
\\[6pt]
K(m)&=&\exp(-\mu_c m)\,.
\end{eqnarray}
\end{subequations}
Such a correlator arises, for example, for a disorder defined by dichotomic Poisson process \cite{K09}. The localization length for this type of correlators reads,
\begin{equation}
\label{exp}
L_{loc}^{-1}(E) = \frac{\sigma^2}{8 \sin^2 \mu}\,\frac{\sinh \mu_c}{\cosh \mu_c - \cos (2 \mu)}.
\end{equation}
As one can see, the localization length increases with an increase of the correlation radius $\mu_c^{-1}$. However, it remains finite for any finite value of the correlation radius. This means that the exponential correlations cannot be used to create extended eigenstates.

Let us now apply the expression (\ref{Lyap corr}) to the random dimer model discussed above. Without loss of generality, we consider the case for which $\bar \epsilon_1= \bar \epsilon$ and $\bar \epsilon_2= -\bar \epsilon$. Note that in order to use the formula (\ref{Lyap corr}) we have to assume a weakness of the potential, $\bar \epsilon \ll 1$. We also assume that the probability to appear of each dimer in the sequence $\epsilon_n$ is the same, $Q=0.5$. As we have shown, fully transparent (resonant) states are known to occur for $E = \pm \bar \epsilon$, see Section~\ref{5.1}. The statistical properties of this model are characterized by the variance $\sigma = \bar \epsilon^2$ and by the pair correlator,
\begin{equation}
\label{corr-dimer}
K(m)= \delta_{m,0}+\frac{1}{2}\delta_{|m|,1}
\,.
\end{equation}
By substituting this correlator into Eq.(\ref{Lyap corr}), we obtain the inverse localization length for the dimer,
\begin{equation}
\label{2mer}
L_{loc}^{-1}(E) = \frac{\sigma^2 E^2}{4(4-E^2)}.
\end{equation}
Note that in this approximation we have $L_{loc}^{-1}(E) \sim \sigma^2$, therefore, the two resonant states $E= \pm\bar\epsilon$ are very close to the band center, $E=0$.

The generalization for the $N-$mer model for which the values $\bar\epsilon$ and $-\bar\epsilon$ appear in randomly distributed blocks of length $N$, can be readily done. For example, for the trimer we have,
\begin{equation}
\label{3mer-1}
L_{loc}^{-1}(E) = \frac{\sigma^2 (E^2-1)^2}{4(4-E^2)}.
\end{equation}
Apart from the resonant values of energy, this expression gives a global dependence of the localization length on energy $E$ that can be obtained from the general formula (\ref{Lyap corr}).

Finally, we consider the model in which the site potential is constructed from the standard Anderson model by substituting each random value $\epsilon_n$ by the sequence of length $N$ with the same value $\epsilon_n$. In this case there are $N-1$ non-vanishing terms for the pair correlators,
\begin{equation}
\label{corr-NA}
K(m) = (N-|m|)/N\,,\,\,\, \mbox {with}\,\,\,\, |m|= 1,2, \dots, N-1.
\end{equation}
Substituting this correlator into Eq.(\ref{Lyap corr}), the result obtained in Ref.\cite{SVE94} by the transfer-matrix method is recovered with ease,
\begin{equation}
\label{nmer}
L_{loc}^{-1}(E) = \left[ L_{loc}^{(0)}(E)\right]^{-1}\frac{\sin^2(\mu N)}{N \sin^2\mu} =\frac{\sigma^2}{8 \sin^2 \mu}\,\frac{\sin^2(\mu N)}{N \sin^2\mu}.
\end{equation}
Here $L_{loc}^{(0)}(E)$ is the localization length for the white noise potential. There are discrete delocalized states at the energies for which $\mu N= s\pi$ with $s=1,2,...,N-1$. For finite disorder the resonant singularities are transformed into peaks that gradually disappear with an increase of disorder \cite{SVE94}.

The general expression (\ref{Lyap corr}) can be also applied to pseudo-random potentials. As an example, let us take almost periodic potential of the following form \cite{T88,GF88},
\begin{equation}
\label{incomm}
\epsilon_n = \sigma \sqrt2 \cos{(2\pi \alpha n^{\gamma})},
\end{equation}
with an irrational $\alpha$. By performing explicitly the average over all sites $n$, one gets the following result for the pair correlator,
\begin{equation}
\label{corr incomm}
K(m) = \lim_{n \rightarrow \infty} \frac{1}{N} \sum_{n=1}^N \cos(2 \pi\gamma \alpha m n^{\gamma-1}).
\end{equation}
One can see that for $0<\gamma<1$ the limit is independent of $m$ and $K(m) =1$. Thus, statistically, this incommensurate potential is equivalent to a periodic potential and all states are delocalized (for the review see, for example, \cite{S82,S85}). If $\gamma =1$, we get the standard Harper model with the incommensurate potential. In this case the pair correlator is a non-vanishing oscillating function, $K(m) = \cos(2\pi \alpha m)$. This is a clear manifestation of the fact that the potential is not a truly random sequence. The calculation of the series in Eq.~(\ref{spect-gen}) gives ${\cal K}(k) =0$. Thus, in the Harper model with weak potential all states are extended \cite{S85,T88,FGP92,FGP93}. Finally, for $\gamma >1$ the correlator $K(m)$ vanishes for $|m|\ne 0$ and the localization length turns out to be the same as for the uncorrelated potential, Eq.~(\ref{standard}).

\subsubsection{Long-range correlations.}
\label{5.2.3}

\begin{figure}[!ht]
\begin{center}
\includegraphics[width=8.6cm,height=7.6cm]{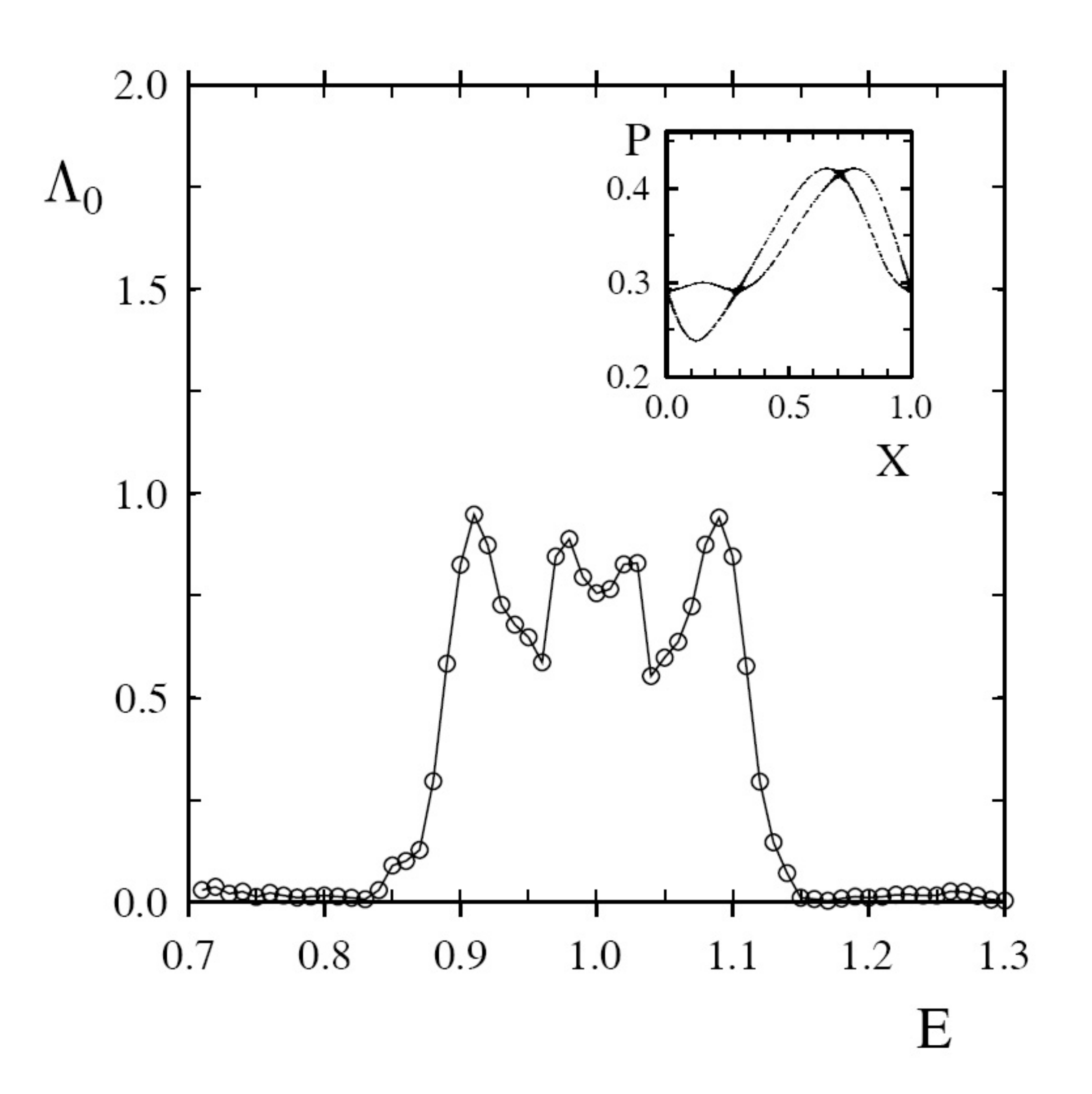}
\end{center}
\caption{Normalized Lyapunov exponent $\Lambda_0(E)$ for the Anderson model (\ref{tb diagonal}) with the site potential $\epsilon_n$ constructed by iteration of the map (\ref{stand map}), see Eq.~(\ref{eps-from}). The length of sequence $\epsilon_n$ is $N = 10^5$ and $\sigma =0.1$. The corresponding trajectory ($X_n,P_n$) for $X_0 = 0, P_0 = 0.292$, and $K = 0.8$ is shown in the inset (after \cite{IK99}).} \label{standmap}
\end{figure}

The slow decay of the binary correlator $K(m)$ means that the terms with $m \gg 1$ cannot be neglected in the Fourier series Eq.~(\ref{spect-gen}). Typically, this happens if $K(m)$ decays as a power law. One possible way to obtain explicitly a sequence of random numbers $\varepsilon_n$ with long-range correlations is to run one of the chaotic nonlinear maps in the region of weak chaos. As a demonstration, we consider the well-known standard map \cite{C79},
\begin{equation}
\begin{array}{ccc}
P_{n+1} & = & P_n +M \sin(2 \pi X_n) \,\,\,\,\, \textrm{\{mod 1\} } \\
X_{n+1} & = & X_n + P_{n+1} \, \qquad \qquad \textrm{\{mod 1\} }.
\end{array}
\label{stand map}
\end{equation}
It is known that for $M < 1$ the trajectory taken from a narrow stochastic layer in the vicinity of separatrix of the nonlinear resonance of period 3, manifests a very slow decay of correlations. Therefore, using the sequence $X_n$ from the standard map as the source of disorder $\epsilon_n$ in the Anderson model, we expect to get a non-standard dependence of the Lyapunov exponent on the energy. The data are shown in Fig.\ref{standmap} where in inset one of such trajectories is presented in the phase space $(P_x, X_n)$ of the model (\ref{stand map}). For this trajectory we constructed the sequence $\epsilon_n$ as follows,
\begin{equation}
\epsilon_n = 2\sigma \sin(2\pi X_n)\,,
\label{eps-from}
\end{equation}
where the small parameter $\sigma$ plays the role of the strength of disorder (another deterministic chaotic sequence was used in Ref.~\cite{PRGM05} where the onset of metal-insulator transition was also observed). The Lyapunov exponent was calculated according to the definition (\ref{loclength}) in which $x_n$ is substituted by $X_n$.

The result is shown in Fig.\ref{standmap} for the normalized Lyapunov exponent,
\begin{equation}
\Lambda_0(E) = \lambda(E)/\lambda_0(E) \,.
\label{Lambda-big}
\end{equation}
Here $\lambda_0(E)$ is the Lyapunov exponent for a white noise disorder with $W =1$. The quantity $\Lambda_0(E)$ is nothing but the factor ${\cal K}(2\mu)$ standing in Eq.(\ref{Lyap corr}), and presented in Fig.~\ref{standmap} as a function of energy $E$. The obtained dependence $\Lambda_0(E)$ clearly indicates the presence of two mobility edges close to $E \approx 0.88$ and $E\approx 1.12$. It is important that the mobility edges are stable with respect to a rather wide variation of the disorder strength $\sigma$ \cite{IK99}.

It is interesting to note that inside the narrow chaotic
region shown in the inset of Fig.\ref{standmap}, the time dependence of the trajectory ($X_n, P_n$) is similar to that known as the so-called intermittency \cite{MO80} which is well studied in 1D maps. The main feature of the ``intermittency" is a kind of combination of a regular motion (rotation around the resonance with period 3), and random jumps occurring when the trajectory reaches the region around the crossing points of the unperturbed separatrix. Therefore, there is a direct link between the correlations in the site potential and the intermittency in the evolution of the corresponding dynamical model which generates this potential (see details in Ref.~\cite{GNS93}). The Lyapunov exponent shown in Fig.~\ref{standmap} is a clear evidence that the long-range correlations may give rise to the mobility edge in the spectrum of a disordered 1D potential. Note that such a mechanism for an emergence of effective mobility edges is different from those previously studied \cite{GF88,DHX88,FGP93} in the models with the incommensurate potentials Eq.~(\ref{corr incomm}), and in the Kronig-Penney model with constant electric field \cite{DSS84,DSS85}.

\subsubsection{Beyond quadratic approximation.}
\label{5.2.4}

We should again to remind that the discussed above approach is perturbative in disorder, in which the small parameter of the theory is the variance $\sigma^2$ of a disorder (in dimensionless units). Since in our calculations we keep the terms up to quadratic ones, the Lyapunov exponent scales as $\sigma^2$. Therefore, an emergence of extended states occurs when the Lyapunov exponent vanishes in the quadratic approximation. This means that the terms ``delocalized states", ``mobility edge", and ``metal-insulator transition" can be used for the samples with the size $L$ less than the localization length calculated within the {\it next approximation}, for $L \ll \sigma^{-4}$. Otherwise, the ``delocalized states" occupying the region $E<1$ should be considered as localized at the distances $\sim \sigma^{-4}$. However, for many practical applications the change of the localization length from $\sim \sigma^{-2}$ to $\sim \sigma^{-4}$, when passing through the ``mobility edge", should be considered as an effective metal-insulator transition.

It is a quite difficult task to obtain analytically the next (quartic) terms for the Lyapunov exponent in the general form. One of few results can be found in Ref.\cite{T02} where the transition from a quadratic to quartic dependence on the strength disorder for the Lyapunov exponent was studied in the continuous 1D model (\ref{1DCP-Ham}) with long-range correlations in the potential. For technical reasons the so-called {\it generalized Lyapunov exponent} \cite{PV87},
\begin{equation}
\tilde \lambda = \lim \limits_{x \rightarrow \infty} \frac{1}{4x}\ln \langle \psi^2(x)k^2+\psi^{\prime 2}(x)\rangle
\label{Lyap-general}
\end{equation}
was analyzed. Although this definition differs from the conventional one \cite{LGP88},
\begin{equation}
\lambda = \lim \limits_{x \rightarrow \infty} \frac{1}{2x}\langle \ln \left( \psi^2(x)k^2+\psi^{\prime 2}(x)\right)\rangle \,,
\label{Lyap-conven}
\end{equation}
both definitions can be used effectively when the main question is how to distinguish between the localized and extended states. It was argued \cite{T02} that to the second order of perturbation theory both definitions give the same value. The general expression for $\tilde \lambda (E)$ was analytically obtained in Ref.\cite{T02} for any kind of weak disorder. The detailed analysis of this expression has shown that the fourth-order terms in the expression for the Lyapunov exponent strongly depends whether the noise is of the Gaussian or non-Gaussian type. Numerical data confirmed the analytical expressions.

The knowledge of the fourth-order term of the Lyapunov exponent is important when the disorder exhibits long-range correlations. Nowadays, this problem is not only of academical interest. Specifically, the long-range correlations emerge in the experiments with cold atoms, see, for example, Refs.~\cite{So07,Bo08,Co08,Ro08,Lo08,M10}. Technically, the disordered potentials in these experiments are introduced with the use of a static speckle intensity generated by a laser beam passing through diffusive plates. As a result, the disorder turns out to have a finite support of the power spectrum, thus, leading to long-range correlations of a specific type. This means that for the observation of the localization effects one has to know how the localization length changes when crossing effective mobility edges. This practical problem has triggered theoretical and numerical studies of the Anderson localization in the presence of long-range correlations \cite{GK09,Lo09} taking into account terms up to the fourth order in the expansion for the Lyapunov exponent.

In Ref.~\cite{GK09} the general expression for the Lyapunov exponent (\ref{Lyap-conven}) was derived for both Gaussian and non-Gaussian disorders, and compared with numerical simulations. The main attention was paid to the transition at the effective mobility edge, at which the leading order dependence of the Lyapunov exponent on the disorder strength crosses from quadratic to quartic dependence. It was found that the drop of the Lyapunov exponent at the mobility edge is steeper for non-Gaussian distribution of speckle intensities which is peculiar to the experiment situation. Another analytical technics was developed in Ref.~\cite{Lo09} where the successive intervals of increasing orders in the expansion for the Lyapunov exponents were analyzed for the potentials close to those used in the experiment. It was, in particular, found that the result is quite sensitive to the asymmetry in the probability distribution of speckle potentials.

\subsection{The inverse problem.}
\label{5.3}

The example of the normalized Lyapunov exponent $\Lambda_0(E)$ with the mobility edges shown in Fig.~\ref{standmap} relies on the uncontrollable generation of the sequence of correlated numbers $\epsilon_n$. However, for the applications, it is desirable to know how to generate a correlated sequence with the use of standard generators of random numbers. Specifically, it is important to have a possibility to construct the potentials with a given dependence $\lambda(E)$. This problem is known as an ``inverse" problem in the theory of localization since it relates to the restoration of the potential $\epsilon_n$ provided the localization length $L_{loc}(E)$ is known.

The normalization condition (\ref{norm-corr}) states that if due to the correlations the localization length $L_{loc}(E)$ increases, as compared to $L_{loc}(0)(E)$, within some interval of energies, it must decrease for the rest of the energy spectrum. In particular, if the correlations lead to the emergence of a wide band of extended states, the rest of the allowed zone is occupied by the states with localization length much shorter than that for a white-noise potential. Thus, the correlations may suppress the localization (even give rise to delocalized states) for some interval of energies, but this is always compensated by the corresponding decrease of $L_{loc}(E)$ for the rest of the states. As one can see, the correlations can be used either to suppress or enhance the localization. These two effects have been confirmed experimentally in the microwave waveguides \cite{KIKS00,KIKSU02,KIK08}. In particular, a strong enhancement of the localization (by a factor of 16) has been recently observed in Ref.~\cite{KIK08}. This effect of localization enhancement by correlations may be important for random lasers \cite{Co99,P03}, where the extremely localized states can provide a higher efficiency.

Since the pair correlations usually give the principal contribution to the observable quantities (e.g. to the scattering cross section), the problem of generation of a random sequence with a prescribed pair correlator, i.e., a colored noise, is of particular interest in physics, engineering, and signal processing. It has been known for a long time that the continuous colored noise with
exponential correlations is generated by the linear Ornstein-Uhlenbeck process, which is based on the integration of a linear Langevin equation driven by a white-noise fluctuating force \cite{K92}. A more general method, valid for the generation of continuous random sequences with an {\it arbitrary} correlator, is based on the convolution of a white noise with the modulation function defined by the pair correlator, see Section~\ref{2.9}. Originally, the convolution method was proposed by Rice \cite{R44}. Applications of the modern versions of this method for the generation of random sequences with specific correlations, including the long-range non-exponential correlations, can be found in Refs. \cite{S88,F88,WO95,CMHS95,MHSS96,RS99,IK99,KIKS00,IM05,CGK06,KIK08}.

The convolution method \cite{IK99,KI99,KIKS00} of the generation of a discrete colored noise $\epsilon_n$ is based on a linear transformation of a white noise sequence $\alpha_n$,
\begin{equation}
\langle \alpha_n\rangle=0\,, \qquad \langle \alpha_n \alpha_{n \prime} \rangle= \delta_{n n \prime}\,,
\label{alpha-def}
\end{equation}
with the use of the modulation function $G(m)$ defined as follows (compare with Eq.~(\ref{1DCP-MF})),
\begin{equation}
\label{colored}
\epsilon_n = \sigma \sum_{m= - \infty}^{\infty} G(m) \alpha_{n+m}\,,
\end{equation}
where $\sigma^2$ is the variance of the sequence $\epsilon_n$. For a homogeneous sequence $\alpha_n$ the generated sequence $\epsilon_n$ is also homogeneous. In order to calculate the modulation function $G(m)$, we substitute Eq. (\ref{colored}) into the definition (\ref{bin-corr}) of the pair correlator. Taking into account that the sequence $\alpha_n$ is uncorrelated, the following relation between the power spectrum $K(\mu)$ and the modulation function $G(m)$ is readily obtained,
\begin{equation}
\label{G-phi}
G(m) = \frac{1}{\pi} \int_0^{\pi} \sqrt{{\cal K}(\mu)} \cos(2\mu m) d\mu.
\end{equation}
This equation defines the function $G(m)$ in terms of the Fourier transform ${\cal K}(\mu)$ of the pair correlator. Since ${\cal K}(\mu)$ is directly related to the localization length, the equations (\ref{colored}) and (\ref{G-phi}) give the solution of the inverse problem due to the relation, ${\cal K}(2\mu) = L_{loc}^0(E)/L_{loc}(E)$.
Here the energy $E$ is substituted through the dispersion relation, $E=2\cos \mu$.

\begin{figure}[!ht]
\begin{center}
\includegraphics[scale=0.85]{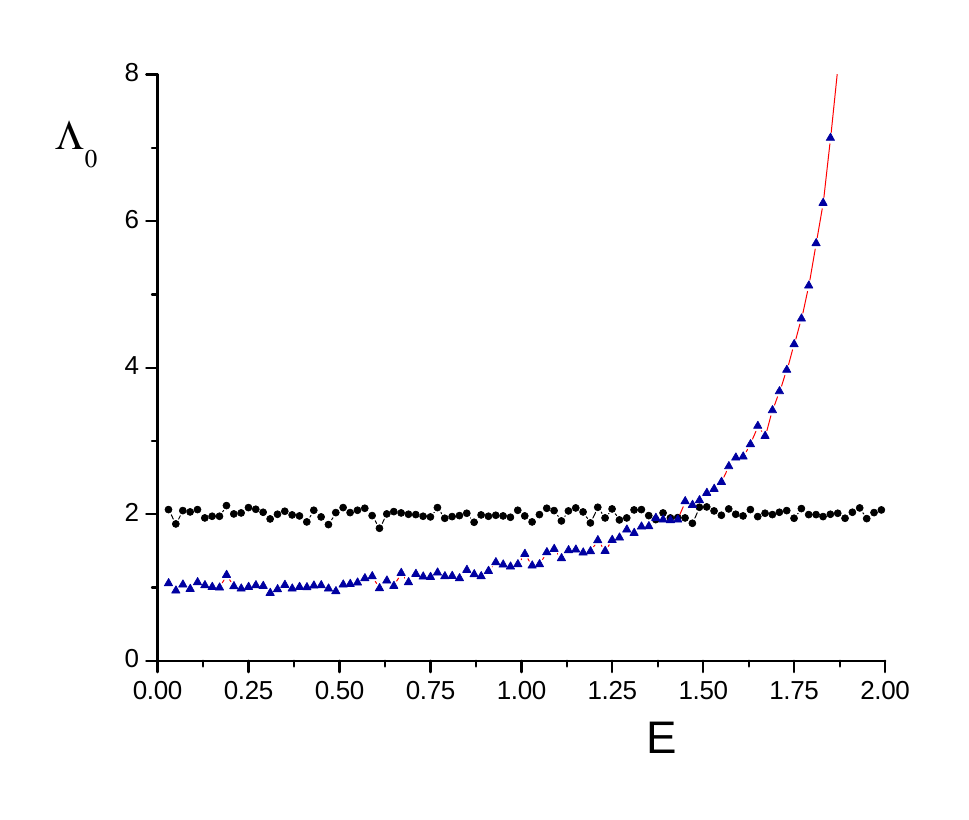}
\end{center}
\caption{Triangles: the Lyapunov exponent $\Lambda_0$ for a white noise disorder with ${\cal K}(2\mu)=1$ in Eq.~(\ref{Lyap corr}), normalized to $1$ for $E=0$. Circles: same for the correlated disorder determined by Eq.~(\ref{const-lyap}). In both cases the length of sequences $\epsilon_n$ is $N=10^5$ and $\sigma^2 \approx 0.1$.}
\label{edge-00}
\end{figure}

In order to demonstrate an effectiveness of the method of construction of the correlated sequence $\epsilon_n$ with a given dependence of the Lyapunov exponent on energy $E$, we consider here the Anderson model in which the Lyapunov exponent is constant in the whole range $-2 < E< 2$ of the energy spectrum. As one can see from Eq.~(\ref{Lyap corr}), in order to create such a situation one needs to have the power spectrum ${\cal K}(2\mu)$ of the form,
\begin{equation}
\label{const-lyap}
{\cal K}(2\mu)=2 \sin^2 \mu \,.
\end{equation}
Then the corresponding pair correlator $K(m)$ is
\begin{equation}
K(m)= \delta_{m,0} -\frac {1}{2}\delta_{|m|,1}\,.
\label{corr-const}
\end{equation}
One can see that in this case the binary correlator $K(m)$ has three components only, $K(0)=1$ and $K(\pm 1)=-1/2$. All other components vanish, $K(m)=0$, for $|m|>1$, thus indicating that the correlations are short-range. The modulation function $G(m)$ defined by Eq.~(\ref{G-phi}) takes the form,
\begin{equation}
\label{beta-flat}
G(m)= \frac{2\sqrt 2}{\pi(1-4m^2)}\,.
\end{equation}
As a result, the localization length $L_{loc}=\lambda ^{-1}$ is independent of the energy, $L_{loc}=4 \sigma^{-2}$. Such a situation may be very effective in applications, see an example in Ref.\cite{DIKR08}. The data in Fig.~\ref{edge-00} clearly demonstrate a very good correspondence with the theory.

\subsection{Effective mobility edge.}
\label{5.4}

The relations (\ref{colored}) and (\ref{G-phi}) give a fast and robust algorithm for the numerical generation of a correlated sequence of any length. It is interesting that the regions of energies where the states are delocalized, $L_{loc}(E) = \infty$, do not contribute to the integral in Eq.(~\ref{G-phi}).
Substituting different white noise sequences $\alpha_n$ in Eq.~(\ref{colored}), an infinite number of correlated sequences $\epsilon_n$ can be generated having the same binary correlator. Therefore, the transport properties of all these sequences are characterized by the same statistical parameter $L_{loc}(E)$, given by Eq.~(\ref{Lyap corr}). The fact that the inverse problem for the localization length does not have a unique solution originates from the Born approximation that depends on the binary correlator only. The higher-order moments of $\epsilon_n$ emerge in the formula for the localization length if the terms higher than quadratic are taken into account in Eq.~(\ref{Layp approx}). Surely, the exact expansion for the localization length contains all the moments of the random potential.

\begin{figure}[!ht]
\begin{center}
\includegraphics[scale=0.75]{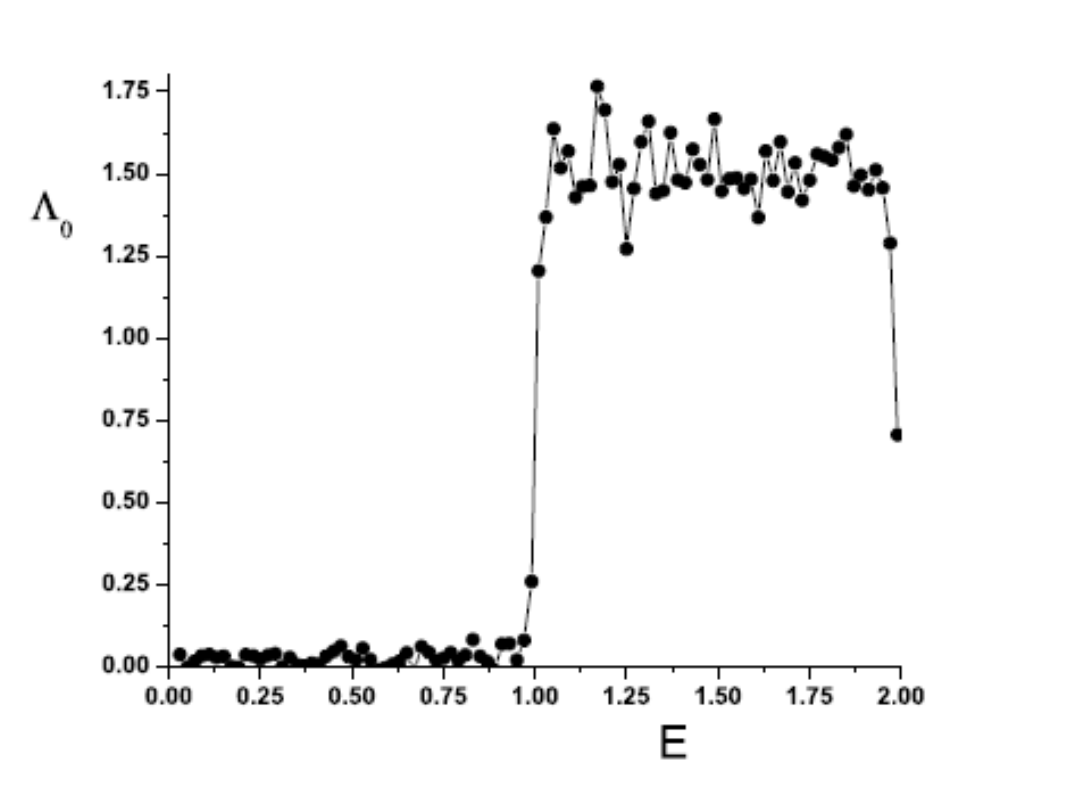}
\end{center}
\caption{The Lyapunov exponent $\Lambda_0$ for sharp mobility edge at $E=1$, with $W=1.5$ in the region $1 \leq E < 2$, and $\sigma\approx 0.33$. The value of $\Lambda_0$ is normalized to be $1$ in the case of white-noise disorder. The length of sequence $\epsilon_n$ is $N=10^5$.}
\label{edge-1}
\end{figure}

The effectiveness of the proposed algorithm can be checked by generating the correlated disorder resulting in a sharp mobility edge in its energy spectrum. Suppose, we would like to have the potential for which the extended states occupy the region $-1 <E<1$, and in the rest of the allowed zone, $1<\mid E \mid <2$ the eigenstates are localized. Then, two sharp mobility edges at $E=\pm 1$ will separate the conducting phase from two insulating phases. Therefore, the pair correlator $K(m)$ and its power spectra ${\cal K}(\mu)$ are,
\begin{subequations}
\begin{eqnarray}
{\cal K}(\mu)&=& \left \{
\begin{array}{rl}
3/2 & \mbox {for}\,\,\, \mu \in [0,\pi/3]\,\,\,\,\mbox{and}\,\,\,\, \mu \in [2\pi/3,\pi] \\
0 & \mbox {for}\,\,\, \mu \in [\pi/3,2\pi/3]
\end{array}\right.
\\[6pt]
K(m)&=& \frac{3}{2\pi m} \sin \frac{2\pi m}{3}\,.
\end{eqnarray}
\label{corr-step}
\end{subequations}
In order to generate such a disorder that provides the corresponding energy dependence $\lambda(E)$, first, one needs to obtain the modulation function $G(m)$ with the use of Eq.~(\ref{G-phi}). It has the form,
\begin{equation}
\label{xi sharp}
 G(m)= \sqrt{\frac{2}{3}}\, \frac{3}{2\pi m}\sin\left(\frac{2 \pi m}{3}\right)\, .
\end{equation}

In order to see how the analytical predictions can be used in practice, the correlated sequence of the length $10^5$ was numerically generated by the algorithm (\ref{colored}) with the modulation function (\ref{xi sharp}) for some value of the variance $\sigma^2=\langle\epsilon_n^2 \rangle$. The normalized Lyapunov exponent $\Lambda_0=\lambda /\lambda_0$ was calculated for the generated sequence according to Eq.~(\ref{colored}), and the result is shown in Fig.~\ref{edge-1} for positive values of energy. The step function dependence is reproduced quite well. The weak fluctuations are due to the finite length of the sequence and a relatively large variance of the disorder. This result is a very convincing evidence that the long-range correlations, indeed, give rise to a band of delocalized states. One can see that the binary correlator $K(m)$ in Eq.~\eqref{corr-step} decays very slowly with $m$, namely, as $1/m$. Such a dependence is quite clear since the binary correlator is a Fourier transform of the function ${\cal K}(\mu)$ and the latter is a step-function. Note that for a sharp ``mobility edge" one has also to have specific correlations described by the oscillating function $\sin (2\pi m/3)$, in addition to a slow decrease of the correlator.

\begin{figure}[!ht]
\begin{center}
\subfigure{\includegraphics[scale=0.7]{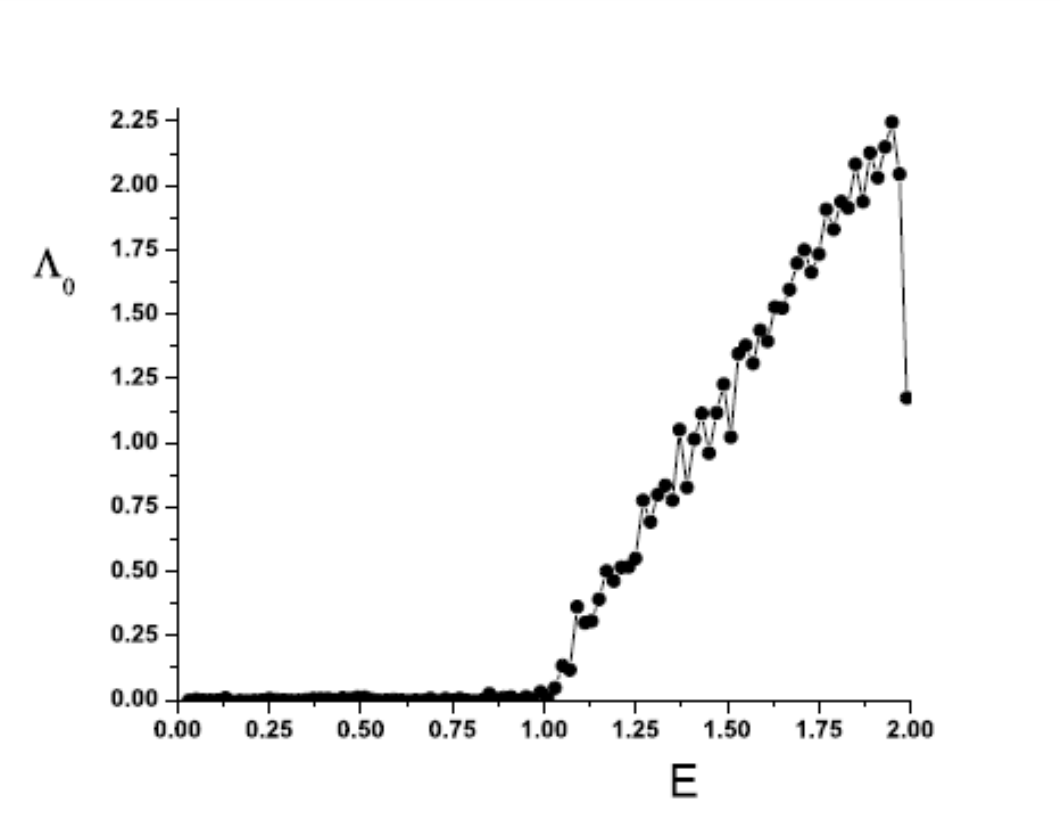}}
\subfigure{\includegraphics[scale=0.7]{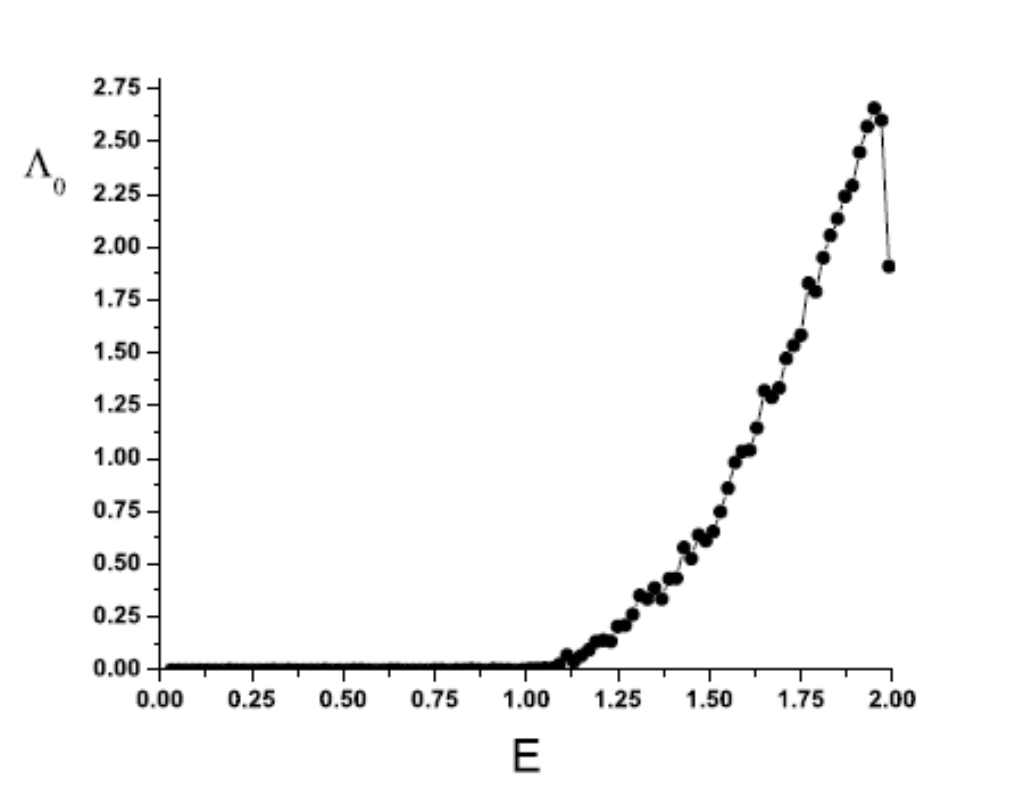}}
\end{center}
\caption{Normalized Lyapunov exponent $\Lambda_0$ for two profiles of ${\cal K}(\mu)$. Left: ``linear mobility edge" (\ref{corr-step1}) for $\sigma \approx 0.29$ and $N=10^5$. Right: ``quadratic mobility edge" (\ref{edge3}) for the same values of $\sigma$ and $N=10^6$.}
\label{edge-23}
\end{figure}

One can also arrange a {\it smooth} mobility edge, for which the Lyapunov exponent is continuous at $|E|=1$. For example, we would like to have the following dependence of ${\cal K}(2\mu)$,
\begin{eqnarray}
{\cal K}(2\mu(E))= \left \{
\begin{array}{rl}
C_0^2\, (|E|-1) & \mbox {for}\,\,\, |E| \in [1,2] \\
0 & \mbox {for}\,\,\, |E| \in [0,1]
\end{array}\right.
\,.
\label{corr-step1}
\end{eqnarray}
In this case the energy dependence of the normalized Lyapunov exponent for $|E|>1$ is linear, see Fig.\ref{edge-23} (left panel). Here the normalization coefficient $C_0^2=\pi/2(\sqrt 3 -\pi/3)$ can be obtained from the normalization condition (\ref{norm-corr}). The modulation function
\begin{equation}
\label{beta-linear}
G(m)=\frac{2}{\pi}\, C_0 \int_0^{\pi/3} \cos {2m\mu}\, \sqrt{2\cos \mu-1}\, d\mu
\end{equation}
defined by Eq.(\ref{G-phi}),
was computed numerically. For comparison, in Fig.\ref{edge-23} (right panel) the data are shown for the mobility edge with a quadratic dependence for $|E|>1$,
\begin{eqnarray}
{\cal K}(2\mu(E))= \left \{
\begin{array}{rl}
C_0^2\, (|E|-1)^2 & \mbox {for}\,\,\, |E| \in [1,2] \\
0 & \mbox {for}\,\,\, |E| \in [0,1]
\end{array}\right.
\,.
\label{edge3}
\end{eqnarray}
In this case we have, $C_0^2=\pi/(2\pi -3 \sqrt 3)$, and the function $G(m)$ is given by,
\begin{equation}
\label{beta-quadr}
G(m) = \frac{2C_0}{\pi} \left\{ \frac{\sin[\pi(2m-1)/3]}{2m-1} + \frac{\sin[\pi(2m+1)/3]}{2m+1} - \frac{\sin(2\pi m/3)}{2m} \right\}\,.
\end{equation}

On can see that the correspondence of the data to the analytical predictions is quite good. Note, however, that in both cases the approach does not work in the vicinity of the band edge, $E\approx 2$. This effect is quite expected since at the band edges the perturbative approach fails. Indeed, close to the band edge, the effective strength of disorder is very large due to the relation, $A_n=\epsilon_n/\sin\mu \rightarrow \infty$ for $\mu \rightarrow 0$ (see Section~\ref{4.2.3}).

In general case when the Lyapunov exponent vanishes near the mobility edge as $(|E|-1)^{n_c}$, the corresponding binary correlator contains the oscillating terms of the order of $1/n$, $1/n^2$, $\dots, 1/n^{n_c+1}$, for $n_c$ integer. As discussed above, power-law decay of the binary correlator (together with specific oscillations) is a typical example of long-range correlations for which the effective mobility edges emerge.

\section{Continuous versus discrete disorder}
\label{6}

\subsection{Stochastic equations approach}
\label{6.1}

Here we discuss the question to what extent the results obtained in Section~\ref{2} for continuous random potentials can be applied to the tight-binding Anderson model (\ref{tb diagonal}). In order to elucidate this problem, we will use one more approach \cite{TI01} which is based on the representation of the Anderson model as a set of stochastic equations related to the classical Hamiltonian (\ref{kickosc}) in which $\xi(t)$ is a continuous and stationary noise. Note that formally the disorder, defined as periodic kicks with random amplitudes, is a non-stationary process. Therefore, the relation between the Anderson model~(\ref{tb diagonal}) and the kicked oscillator~(\ref{kickosc}) does not prove at all the
equivalence of the models~(\ref{tb diagonal}) and~(\ref{kickosc}) but only constitute a hint that such a link may exist.

In our analysis of stochastic oscillators, we focus on the
Hamiltonians represented by Eq.~(\ref{kickosc}), completing the definition of the model by further assuming that the noise
$\xi(t)$ has zero average and that its binary correlator is a
known function,
\begin{equation}
\begin{array}{ccc}
\langle \xi(t) \rangle = 0 & \mbox{and} &
\langle \xi(t) \xi(t+\tau) \rangle = \chi (\tau) .
\end{array}
\label{noisprop}
\end{equation}
In Eq.~(\ref{noisprop}) the symbol $\langle \ldots \rangle$ is used for the time average which is assumed to coincide with the ensemble average for the process $\xi(t)$. Below we do not restrict our consideration to the case of white noise, on the contrary, our interest is in the general case of {\em coloured} noise. As before, we require that the noise $\xi(t)$ is weak, therefore, the fluctuations of the frequency of the oscillator around its average value are assumed to be small.

Our aim is to show that the oscillators of the kind~(\ref{kickosc}), with the above-mentioned noise features, are equivalent to the Anderson model~(\ref{tb diagonal}) if two further conditions are met. First, the correlation function has to be of the form,
\begin{equation}
\chi (\tau) = \frac{\langle A_{n}^{2} \rangle}{T}
\sum_{m=-\infty}^{+\infty} K (m) \; \delta (\tau - m T) ,
\label{corfun}
\end{equation}
where the symbol $K(m)$ stands for the normalized binary correlators,
\begin{equation}
K(m) = \frac{ \langle A_{n+m} A_{n} \rangle}{\langle A_{n}^{2}
\rangle}\,,
\label{bincorr}
\end{equation}
of the random variables $A_{n}$ specified by the second condition. The second requirement is that the unperturbed frequency $\omega$ of the oscillator and the parameters $A_{n}$ are related to the parameters $E$ and $\epsilon_{n}$ of the Anderson model through the identities~(\ref{corres}).

Note that the link established by these two conditions associate the key features of the noise $\xi(t)$ with the corresponding properties of the random potential $\epsilon_{n}$. Indeed, once the random variables $\epsilon_{n}$ and $A_{n}$ are connected by the relation~(\ref{corres}), the correlator~(\ref{bincorr}) become identical to the normalized correlator of the potential $\epsilon_{n}$. Therefore, the spatial correlations of the disorder in the Anderson model
are mirrored by temporal correlations for the noise $\xi(t)$.
In the special case in which the disorder in the Anderson model is {\em uncorrelated} (i.e., $\langle \epsilon_{n+m} \epsilon_{n} \rangle = 0$ for $m \ne 0$), the noise for the random oscillator is {\em white} (i.e., $\langle \xi(t) \xi(t+\tau) \rangle \propto \delta(\tau)$). One can also observe that the case of weak disorder in the Anderson model corresponds to that of weak noise for the random oscillator, since the condition $\sigma^2 = \langle \epsilon_{n}^{2} \rangle \ll 1$ entails the consequence that $\langle A_{n}^{2} \rangle \ll 1$ (except that at the band edge, i.e., for $\omega T \rightarrow 0$, which is a special case where some anomalies are expected to arise, that are not considered here).

In order to obtain the expression for the Lyapunov exponent, we start with Eq.(\ref{divrate}) for which it is convenient to introduce the polar coordinates defined through the standard relations $x = r \sin \theta$, $p = r \cos \theta$. This allows one to cast Eq.~(\ref{lyap}) in the form,
\begin{equation}
\lambda = \lim_{T_{0} \rightarrow \infty} \frac{1}{T_{0}} \int_{0}^{T_{0}}
\frac{\dot{r}}{r} \; dt .
\label{lam-T0}
\end{equation}
To proceed further, we consider the dynamical equations for the
random oscillator in polar coordinates,
\begin{equation}
\dot{\theta} = \omega + \xi(t) \sin^{2} \theta ,
\label{ang}
\end{equation}
\begin{equation}
\dot{r} = - \frac{1}{2} r \xi(t) \sin 2 \theta ;
\label{rad}
\end{equation}
Using the radial Eq.~(\ref{rad}), the expression for the Lyapunov exponent can be finally put into the form,
\begin{equation}
\lambda = - \lim_{T_{0} \rightarrow \infty} \frac{1}{2T_{0}}
\int_{0}^{T_{0}}
\xi(t) \sin \left( 2 \theta(t) \right) \; dt = -\frac{1}{2} \langle
\xi(t) \sin \left( 2 \theta(t) \right) \rangle .
\label{lyap22}
\end{equation}

The problem of computing the Lyapunov exponent~(\ref{lyap}) is thus reduced to that of calculating the noise-angle correlator that appears in Eq.~(\ref{lyap22}). This can be done in the following way which is the extension to the continuum case of the procedure adopted in~\cite{IK99} for the discrete case (see Section 5.2). First, one introduces the noise-angle correlator defined by the relation,
\begin{displaymath}
\gamma(\tau) = \langle \xi(t) \exp \left[ 2 i \theta \left( t +
\tau \right) \right] \rangle .
\end{displaymath}
Starting from this definition, in the limit $\epsilon \rightarrow 0$ one has
\begin{displaymath}
\gamma ( \tau + \epsilon ) = \langle \xi(t) \exp \left[ i 2 \theta \left( t + \tau \right) \right] \left( 1 + 2 i \dot{\theta} \left(t + \tau \right) \epsilon \right) \rangle .
\end{displaymath}
Using the dynamical equation~(\ref{ang}) one can further write,
\begin{equation}
\gamma ( \tau + \epsilon ) = \gamma (\tau) \left( 1 + 2 i \omega \epsilon \right) + 2 i \epsilon \langle \xi(t) \xi(t+\tau)
\exp \left[ 2 i \theta \left( t+\tau \right) \right]
\sin^{2} \theta \left( t+\tau \right) \rangle .
\label{gam-tau}
\end{equation}
For a weak noise one can factorize the correlator that appears in the right hand side of Eq.~(\ref{gam-tau}), and take the average over the angular variable using the flat distribution for $\theta$. Indeed, when $\xi(t) \rightarrow 0$, Eq.~(\ref{ang}) implies that $\dot{\theta} \simeq \omega$ so that after a long time the angular variable is uniformly distributed in the interval $[0,2 \pi]$. As a consequence the noise-angle correlator obeys the relation,
\begin{equation}
\gamma ( \tau + \epsilon ) = \gamma (\tau) \left( 1 + 2 i \omega \epsilon \right) - \frac{i}{2} \chi(\tau) \epsilon , \label{gam1}
\end{equation}
where $\chi(\tau)$ is the correlation function defined by Eq.~(\ref{noisprop}). On the other hand, one can calculate,
\begin{equation}
\gamma ( \tau + \epsilon ) = \gamma (\tau) + \frac{d \gamma(\tau)}{d \tau} \epsilon + O(\epsilon).
\label{gam2}
\end{equation}
Comparing Eqs.~(\ref{gam1}) and~(\ref{gam2}), one obtains the
differential equation,
\begin{displaymath}
\frac{d \gamma(\tau)}{d \tau} = 2 i \omega \gamma(\tau) -
\frac{i}{2} \chi(\tau) ,
\end{displaymath}
whose solution (with the boundary condition $\lim_{\tau \rightarrow
-\infty} \gamma (\tau) = 0$) gives the noise-angle correlator,
\begin{displaymath}
\gamma (\tau) = - \frac{i}{2} \int_{-\infty}^{\tau} \chi(s)
e^{2 i \omega \left( \tau - s \right)} \; ds .
\end{displaymath}
Using this result, the Lyapunov exponent~(\ref{lyap22}) can be
finally written as
\begin{equation}
\lambda = \frac{1}{4} \int_{0}^{+\infty}
\chi(\tau) \cos (2 \omega \tau) \; d \tau
\label{lyap3}
\end{equation}
which implies that the Lyapunov exponent for the stochastic
oscillator~(\ref{kickosc}) is proportional to the Fourier transform $\tilde{\chi}(2\omega)$ of the correlation function at twice the frequency of the unperturbed oscillator. The substitution of the correlation function~(\ref{corfun}) in the
general expression~(\ref{lyap3}) gives,
\begin{equation}
\lambda = \frac{\langle A_{n}^{2} \rangle}{8T}
\left[ 1 + 2 \sum_{m=1}^{+\infty} K (m) \, \cos \left(
2 \omega T m \right) \right] .
\label{lam-luca}
\end{equation}
Taking also into account the relations~(\ref{corres}), one can
finally write the Lyapunov exponent for the random oscillator as follows,
\begin{equation}
\begin{array}{lcl}
\displaystyle
\lambda = \frac{1}{T} \frac{\langle \sigma^{2} \rangle}
{8 \sin^{2} \left( \omega T \right)} \; W \left( 2\omega T \right)\,,
& \mbox{with} &
\displaystyle
W \left( 2\omega T \right) = 1 + 2
\sum_{m=1}^{+\infty} K (m) \, \cos \left( 2 \omega T m \right).
\end{array}
\label{lyap4}
\end{equation}
This expression coincides with Eq.(\ref{Lyap corr}) that is obtained in a different way. Thus, one can conclude that the
Anderson model with {\em correlated} disorder has a classical counterpart represented by a stochastic oscillator with frequency perturbed by a {\em colored} noise.

The above consideration allows one to understand what is the main difference between the model with continuous potentials and the tight-binding Anderson model. One can see that Eq.~(\ref{lyap4}) for the inverse localization length is correct for all energy values inside the unperturbed band {\em except} that at the band center, i.e., for $\omega T = \pi/2$, where an anomaly arises and special methods are required for the analytical investigation (see discussion in Section~\ref{4.2.2}). This anomaly is a resonance effect inherent to a discrete nature of the model~(\ref{tb diagonal}) and cannot therefore be reproduced by the continuous system~(\ref{kickosc}). It should be stressed that other anomalies appear in the Anderson model for the ``rational'' values of the energy (when $\omega T = \pi r/2s$ with $r$ and
$s$ integer numbers), but they do not influence the localization length. The nontrivial effects of the other resonance values emerge when one needs to take into account the orders higher than the second one.

In conclusion, the equivalence between the continuous and discrete models is established for the localization length only, in a whole energy band except the band center and far enough from the band edges. Note that the dispersion relations between the energy $E$ and the wave vector $k$ are different for the two models ($E=k^2$ for the continuous model and $E=2\cos kd$ for the Anderson model with $d$ as the period of an unperturbed potential). Therefore, the continuous model has a single infinite energy band, $0 < E < \infty$, in contrast to the Anderson model that has the finite energy band, $-2 < E < + 2$.

\subsection{Ballistic transport}
\label{6.2}

The method described in the preceding Section turns out to be quite effective, and many of the results discussed previously can be obtained in a more transparent way. Let us start with the ballistic regime for which the localization length $L_{loc}$ is much larger than the size $L$ of a sample. For continuous potentials the transmission coefficient $T_{L}$ can be expressed in terms of two classical trajectories $r_1(t)$ and $r_2(t)$ representing the solutions of Eq.~(\ref{rad}) as follows \cite{LGP88} (compare with Eq.~(\ref{eq:6})),
\begin{equation}
T_{L=t} = \frac{4}{2 + r_{1}^2(t) + r_{2}^2(t)}.
\label{trans1}
\end{equation}
Here $r_{1}(t)$ and $r_{2}(t)$ correspond to two complimentary initial conditions $r(0)=1,\theta(0)=\pi/2$ and $r(0)=1,\theta(0)=0$, respectively. As one can see, the expression (\ref{trans1}) is nothing but the generalization of Eq.~(\ref{eq:6}) discussed in Section~\ref{4.4} on application to the tight-binding Anderson model. It should be emphasized that the formula (\ref{trans1}) is exact and valid {\it both} for continuous and discrete types of a noise in Eq.~(\ref{kickosc}).

In the ballistic regime, i.e., when $L\ll L_{loc}$, one has
$ r_{1,2}(t)\simeq 1$ and Eq.~(\ref{trans1}) can be
written in the form,
\begin{equation}
T_{L} = 1 + \frac{2 - r_{1}^2 - r_{2}^2}{4} + \ldots
\label{trans2}
\end{equation}
Another quantity of physical interest is the resistance which is defined as the inverse of the transmission coefficient,
\begin{equation}
T_{L}^{-1} = \frac{2 + r_{1}^2 + r_{2}^2}{4} .
\label{resis}
\end{equation}

One can see that in order to obtain the mean value of these physical quantities, one has to compute the squared radii $r_{1}^2$ and
$r_{2}^2$, averaged over different disorder realizations.
To do this, one can rely on the Van Kampen's approach developed in Refs.~\cite{K74,K92} and applied to many stochastic models.
This approach is based on the construction of dynamical
equations for the average moments of the position and momentum of a random oscillator. For the second moments one has,
\begin{equation}
\frac{d}{dt} \left( \begin{array}{c} \langle x^{2} \rangle \\
                                     \langle p^{2} \rangle \\
                                     \langle px \rangle
                     \end{array} \right) =
\bf{A} \left( \begin{array}{c} \langle x^{2} \rangle \\
                                     \langle p^{2} \rangle \\
                                     \langle px \rangle
                     \end{array} \right)
\label{momev}
\end{equation}
where the evolution matrix is
\begin{equation}
\bf{A} = \left( \begin{array}{ccc} 0 & 0 & 2 \omega \\
\Upsilon_{1} + \Upsilon_{2} & - \Upsilon_{1} + \Upsilon_{2} & -2 \omega \\
- \omega + \Upsilon_{3} & \omega & - \Upsilon_{1} + \Upsilon_{2}
\end{array} \right)
\label{evmat}
\end{equation}
with
\begin{eqnarray}
\Upsilon_{1} & = &
\int_{0}^{\infty} \chi(\tau) \; d\tau \\
\Upsilon_{2} & = &
\int_{0}^{\infty} \chi(\tau) \cos \left( 2 \omega
\tau \right) \; d\tau \\
\Upsilon_{3} & = &
\int_{0}^{\infty} \chi(\tau) \sin \left( 2 \omega
\tau \right) \; d\tau .
\label{Ups}
\end{eqnarray}
For the general case of colored noise, Eq.~(\ref{momev}) is correct up to order $O(\Upsilon) = O(\xi^{2})$; for the special case of white noise, however, it turns out to be exact.

One can extract an important information from Eq.~(\ref{momev}). In particular, it is possible to obtain the behavior of the average squared radii $r_{1}^2$ and $r_{2}^2$ for $t \rightarrow 0$,
\begin{eqnarray}
\langle r_{1}^{2}(t) \rangle & = & 1 + ( \Upsilon_{1} + \Upsilon_{2} )
t + O(t^{2}) \\
\langle r_{2}^{2}(t) \rangle & = & 1 + (- \Upsilon_{1} + \Upsilon_{2} )
t + O(t^{2}) .
\label{r1-r2}
\end{eqnarray}
As a consequence, one has,
\begin{equation}
\frac{1}{4}\langle r_{1}^{2}(t) + r_{2}^{2}(t) \rangle - \frac{1}{2}=
\frac{1}{2} \Upsilon_{2} t + O(t^{2})
\label{aver}
\end{equation}
Note that these equations are correct up to order $O(t^{2})$, so that it is meaningful to retain the distinction between the parameter $\Upsilon_{1}$ and $\Upsilon_{2}$.
Using the result~(\ref{aver}) and Eq.(\ref{lyap3}) for the Lyapunov exponent $\lambda=L^{-1}_{loc}$, one arrives at the following expressions for the average transmission coefficient and resistance
\begin{equation}
\langle T_{L} \rangle = 1 - 2 \lambda L + O \left(\lambda^2
L^2 \right)
\label{T-1}
\end{equation}
and
\begin{equation}
\langle T_{L}^{-1} \rangle = 1 + 2 \lambda L + O \left(\lambda^2
L^2 \right) .
\label{T-1-1}
\end{equation}
These formulae are in agreement with Eqs.(\ref{1DCP-TResBal}) and (\ref{mean}), (\ref{eq:Re1}) obtained with the use of different approaches.

\subsection{Strong localization}
\label{6.3}

In the localized regime, when $L \gg L_{loc}$, in order to evaluate the average value of the transmission coefficient~(\ref{trans1}) one has to find the probability distribution for the random variable $r$ at large times $L=t \rightarrow \infty$. In this case the radius increases
exponentially and one has,
\begin{equation}
r_{1}(t) \simeq r_{2}(t),
\label{asympt}
\end{equation}
regardless of the initial conditions. As a consequence, we can drop the subscripts $1$ and $2$, and write the
transmission coefficient in the simplified form,
\begin{equation}
\langle T_{L} \rangle \simeq \langle \frac{2}{1 + r^{2}} \rangle .
\label{trans3}
\end{equation}
From the mathematical point of view, the problem of computing the average~(\ref{trans3}) can be better handled by introducing the logarithmic variable $z=\ln r$. The dynamics of the random
oscillator~(\ref{kickosc}) determined by Eqs.(\ref{ang}) and (\ref{rad}), is described by the stochastic differential
equations of the form,
\begin{equation}
\dot{\bf{u}} = {\bf F}^{(0)} ({\bf{u}}) + \alpha {\bf F}^{(1)} ({\bf u}, t).
\label{vankam1}
\end{equation}
Here ${\bf F}^{(0)} ({\bf{u}})$ represents a determinstic function of $\bf{u}$, perturbed by a stochastic function $\alpha {\bf F}^{(1)} ({\bf u}, t)$
with $\alpha \ll 1$.
Indeed, one can reduce Eqs.(\ref{ang}) and (\ref{rad}) to the
form~(\ref{vankam1}) by defining the vectors of Eq.~(\ref{vankam1}) as follows,
\begin{equation}
\begin{array}{ccc}
{\bf{u}} = \left( \begin{array}{c} z\\
                                  \theta
                  \end{array} \right) , &
{\bf{F}}^{(0)} = \left( \begin{array}{c} 0 \\
                                       \omega
                        \end{array} \right) , &
{\bf{F}}^{(1)} = \left( \begin{array}{c}
                 -\frac{1}{2 \alpha} \xi(t) \sin(2 \theta) \\
                  \frac{1}{\alpha} \xi(t) \sin^{2} \theta
                  \end{array} \right)
\end{array}
\label{FFF}
\end{equation}
with $\alpha^2 = \langle \xi^{2}(t) \rangle $.
It is known that a stochastic differential equation of the
form~(\ref{vankam1}) can be associated with a partial differential equation whose solution $P({\bf{u}},t)$ represents the probability distribution for the random variable $\bf{u}$~\cite{K92}. This partial differential equation can be written as
\begin{equation}
\begin{array}{l}
\displaystyle
\frac{\partial P}{\partial t} =
- \sum_{i} \frac{\partial}{\partial u_{i}} \left(
F_{i}^{(0)} P({\bf{u}},t) \right) \\
\displaystyle +
\alpha^{2} \sum_{i,j} \frac{\partial}{\partial u_{i}}
\int_{0}^{\infty}
d \tau \, \langle F^{(1)}_{i} ({\bf{u}},t) \frac{d(u^{-\tau})}{d(u)}
\frac{\partial}{\partial u_{j}^{-\tau}} F^{(1)}_{j} ({\bf u}^{-\tau},
t-\tau) \rangle \frac{d(u)}{d(u^{-\tau})} P({\bf{u}},t) \\
+ O \left( \alpha^{2} \right)
\end{array}
\label{vankam2}
\end{equation}
where ${\bf{u}}^{t}$ stands for the flow defined by the deterministic equation $\dot{{\bf{u}}} = {\bf{F}}^{(0)}({\bf{u}})$, and $d(u^{-\tau})/d(u)$ is the Jacobian of the transformation ${\bf{u}} \rightarrow {\bf{u}}^{-\tau}$. The symbol $O \left( \alpha^{2}
\right)$ represents the omitted terms of high orders. Thus, in the case of weak disorder one can describe the dynamical behavior of the system~(\ref{vankam1}) with an approximate equation of the Fokker-Planck kind.

In our case the approximate Fokker-Planck equation~(\ref{vankam2})
associated with the dynamical system~(\ref{ang})-(\ref{rad}) takes the form \cite{TI01},
\begin{equation}
\begin{array}{ccl}
\displaystyle
\frac{\partial P\left( \theta, z, t \right)}{\partial t} & = &
\displaystyle
- \omega \frac{\partial P}{\partial \theta} + \frac{1}{4} \sin (2 \theta)
\frac{\partial }{\partial \theta} \left\{ \left[ - \Upsilon_{1} +
\Upsilon_{2} \cos (2 \theta) + \Upsilon_{3} \sin (2 \theta) \right]
\frac{\partial P}{\partial z} \right\} \\
\displaystyle
& + & \displaystyle
\frac{1}{2} \frac{\partial }{\partial \theta}
\left\{ \sin^{2} (\theta) \left[ \Upsilon_{3}
\cos (2 \theta) - \Upsilon_{2} \sin (2 \theta) \right]
\frac{\partial P}{\partial z} \right. \\
\displaystyle
& + & \displaystyle
\left. \sin^{2} (\theta)
\frac{\partial }{\partial \theta}
\left[ \left( \Upsilon_{1} - \Upsilon_{2} \cos (2 \theta) - \Upsilon_{3}
\sin (2 \theta) \right) P \right] \right\} \\
& + & \displaystyle
\frac{1}{4} \sin (2 \theta) \left[ \Upsilon_{2} \sin (2 \theta) -
\Upsilon_{3} \cos (2 \theta) \right] \frac{\partial^{2} P}{\partial z^{2}} .
\end{array}
\label{fokpl-n}
\end{equation}
We remark that in the general case of colored noise this equation is correct to the second order in $\xi(t)$. However, in the special case when the noise
$\xi(t)$ is {\em white} it can be shown that
Eq.~(\ref{fokpl-n}) becomes exact.

Although we have the Fokker-Planck equation~(\ref{fokpl-n}) for the general distribution $P(z,\theta,t)$, in order
to evaluate the average of the transmission coefficient~(\ref{trans1}), we actually need only the probability distribution for the radial variable $r$ (or for the equivalent logarithmic variable $z$). Therefore, one can consider the restricted Fokker-Planck
equation,
\begin{equation}
\begin{array}{ccl}
\displaystyle
\frac{\partial}{\partial t} \int_{0}^{2 \pi} P(\theta, z,t) \; d\theta
& = & \displaystyle
\frac{1}{8} \int_{0}^{2 \pi} \left[ \left( 1 - \cos (4 \theta)
\right) \Upsilon_{2} - \sin (4 \theta) \Upsilon_{3} \right]
\frac{\partial^{2} P}{\partial z^{2}} \; d\theta \\
& + & \displaystyle
\frac{1}{4} \int_{0}^{2 \pi} \left[ 2 \Upsilon_{1} \cos (2 \theta)
- \Upsilon_{2} \left( 1 + \cos (4 \theta) \right) - \Upsilon_{3}
\sin (4 \theta) \right] \frac{\partial P}{\partial z} \; d\theta
\end{array}
\label{eq-time}
\end{equation}
obtained by integrating Eq.~(\ref{fokpl-n}) over the redundant
angular variable. To proceed further, we assume that after a transient time the probability distribution takes the form,
\begin{equation}
P(\theta,z,t) \simeq \frac{1}{2 \pi} P(z,t) .
\label{flat}
\end{equation}
This assumption can be justified on the grounds that for a weak noise the dynamics of the angular variable is approximately governed by the equation $\dot{\theta} \simeq \omega$. It is reasonable to suppose that for times $t \gg 2 \pi/ \omega$, the angular distribution is flat (excluding the exceptional case when $\omega \simeq 0$, i.e., when the energy value lies in a neighborhood of the band edge, as well as the case of the center of energy band).

As a consequence of the hypothesis~(\ref{flat}), one eventually gets the reduced Fokker-Planck equation for the $z$ variable,
\begin{equation}
\frac{\partial P(z,t)}{\partial z} = \lambda \left[ -
\frac{\partial P(z,t)}{\partial t} + \frac{\partial^{2} P(z,t)}{\partial z^{2}}
 \right] ,
\label{refokpl}
\end{equation}
where $\lambda$ is the Lyapunov exponent~(\ref{lyap4}).
Eq.~(\ref{refokpl}) has the form of a heat equation with a constant drift. Therefore, the solution is
\begin{equation}
P(z,t) = \frac{1}{\sqrt{2 \pi \lambda t}} \exp \left [
-\frac{\left( z - \lambda t \right)^{2}}{2 \lambda t} \right ].
\label{gauss-n}
\end{equation}
This solution satisfies the initial condition $P(z,t=0) = \delta(z)$, according to which $r=1$ at $t=0$.

The knowledge of the distribution~(\ref{gauss-n}) makes possible to compute the average transmission coefficient in the localized regime. Using Eq.~(\ref{gauss-n}), one can evaluate the expression~(\ref{trans3}) and thus obtain,
\begin{equation}
\langle T_{L} \rangle = \int_{-\infty}^{+\infty} \frac{2}{1 + \exp (2z)}
P(z,L) \; dz \simeq \sqrt{\frac{\pi L_{loc}}{2 L}} \exp \left(
-\frac{L}{2 L_{loc}} \right) .
\label{trans4}
\end{equation}
The formula~(\ref{trans4}) shows that in the localized
regime the transmission coefficient decreases exponentially with $L$,
with the correct rate of exponential decay. It is interesting to note that the obtained expression~(\ref{trans4}) fails to reproduce the correct pre-exponential
factor which actually scales as $(L_{loc}/L)^{3/2}$, see Eq.~(\ref{1DCP-TLoc}). This fact was discussed in Ref.~\cite{LGP88} where it was pointed out that such a discrepancy is originated from an approximate evaluation of Eq.~(\ref{trans1}). As shown in Section~\ref{4.4.3}, in the rigorous treatment one has to take into account the correlations between $r_{1}(t)$ and $r_{2}(t)$. For large times, $t \gg L_{loc}$, one can indeed assume the relation (\ref{asympt}), however, the correlations between $r_{1}(t)$ and $r_{2}(t)$ emerging on the ballistic time scale (for $t \leq L_{loc}$) give a noticeable correction which is manifested by a different prefactor $(L_{loc}/L)^{1/2}$ in Eq.~(\ref{trans4}).

Thus, we have to conclude that in
Eq.~(\ref{trans4}) the exponential factor is determined by
the long-time behavior of the random oscillator (which is correctly
described in our approach), while the pre-exponential factor is
strongly influenced by the short-time dynamics of the
oscillator.

Another quantity of general interest is the mean value of the logarithm of transmission coefficient, $\langle \ln T_{L} \rangle$. In the framework of the present approach, it can be easily computed as follows \cite{TI01},
\begin{equation}
\langle \ln T_{L} \rangle \simeq - \langle \ln \left( r^{2} \right) \rangle + \ln(2) - \langle \ln \left( 1 + \frac{1}{r^{2}} \right) \rangle .
\label{lnT-1}
\end{equation}
Therefore, in the limit $L \rightarrow \infty$ one gets,
\begin{equation}
- \frac{1}{L} \langle \ln T_{L} \rangle = \frac{2}{L}
\langle \ln r(L) \rangle .
\label{lnT-2}
\end{equation}
Substituting in the r.h.s. of this equation the average value of the variable $z=\ln r$, one finally obtains,
\begin{equation}\label{ln-double}
\langle \ln T_{L} \rangle = -2 \lambda L = - 2 \frac{L}{L_{loc}},
\end{equation}
which establishes the relation between the average logarithm of the transmission coefficient and the Lyapunov exponent $\lambda$. This relation was already discussed in Sections \ref{2.8} and \ref{4.4.3}.

As for the resistance~(\ref{resis}), it can be written in the form,
\begin{equation}
\langle T_{L}^{-1} \rangle \simeq \frac{1}{2} \left( 1 + r^{2} \right) .
\label{resist2}
\end{equation}
Starting from this expression and making use of the distribution~(\ref{flat}), it is easy to obtain the expressions (\ref{1DCP-AvVarRes}) and (\ref{eq:Re2}). Here the same result can be proved only for strongly localized regime. However, it turns out to be exact and valid for any ratio $L/L_{loc}$, see discussion in Sections~\ref{2.7} and \ref{4.4.2}.

\section{Diagonal versus off-diagonal disorder}
\label{7}

\subsection{Pure off-diagonal disorder}
\label{7.1}

In this Section we briefly discuss an off-diagonal disorder associated with random hopping amplitudes $\vartheta_{n-1,n}\equiv \vartheta_{n}$ in the Anderson tight-binding model (\ref{tb}).
In many aspects the off-diagonal disorder localizes the quantum states similarly to the diagonal disorder. Namely, the states are exponentially localized independently of the level of disorder, if the hopping amplitudes are uncorrelated \cite{EC71,TC76,SE81}. On the other hand, there is a principal difference between the diagonal and off-diagonal disorder reflected by an anomaly in the energy spectrum near the band center.

Without loosing generality, for pure off-diagonal disorder one can assume that the on-site energy in Eq.(\ref{tb}) vanishes, $\epsilon_n = 0$. Then, the Schr\"odinger equation (\ref{tb}) takes the following form:
\begin{equation}
\label{tb off}
\vartheta_{n+1}\psi_{n+1} + \vartheta_{n}\psi_{n-1} = E \psi_{n} .
\end{equation}
As was first pointed out by Dyson \cite{D53}, the density $\rho(E)$ of states at the band center diverges as
\begin{equation}
\rho(E) \sim \frac{1} {8E \ln^{3} {E} } \,,
\label{dens-center}
\end{equation}
for a special case of random variable $\vartheta_n$ drawn from the generalized Poisson distribution. Later, this result was confirmed in Ref.~\cite{ER78} for any statistical distribution of $\vartheta_n$ .

The transport properties at the band center have been discussed in literature since 1976, when it was suggested \cite{TC76} that at this point the localization length diverges. Near the band center the localization length was calculated using two different approaches in Refs.~\cite{TC76,ER78}. Both results give the logarithmic divergence, $L_{loc} \sim | \ln E^2 |$, in the thermodynamic limit (when $L \rightarrow \infty$ with a constant spacing $d$ between sites). Such a divergence was considered to be a signature of an extended state at the band center. However, it was found \cite{FL77} that this state is nevertheless localized due to the fluctuations of the envelope of eigenstates, that were previously ignored. Later, it was shown that the transmission coefficient $T(E=0)$ vanishes in the thermodynamic limit \cite{SE81}, i.e. the state at $E=0$ cannot be an extended state. This ambiguity clearly demonstrates an anomalous nature of the state. Recently, an analytical non-perturbative treatment was proposed in Ref.~\cite{CFE05}, and it was established that the localization length at the band center diverges, $L_{loc}(E=0) \sim \sqrt L$ with an increase of the length $L$. This divergence, however, is not strong enough for this state to be extended.

For a sequence of a finite length the band center value $E=0$ belongs to the spectrum of eigenenergies only if the number of sites is odd. Otherwise, there is no state at $E=0$ because of the so-called chiral symmetry \cite{Bo02,So03}. Let the sequence with odd number of sites starts at the site $n=0$ and ends at $n=L=2N$. Then, for $E=0$ the ratio ${\psi_{2n}}/{\psi_{0}}$ can be easily obtained by iterating Eq. (\ref{tb off}),
\begin{equation}
\label{iteration}
\frac{\psi_{2n}}{\psi_{0}} = (-1)^n\frac{\vartheta_{2n-1}}{\vartheta_{2n}} \frac{\vartheta_{2n-3}}{\vartheta_{2n-2}} \dots \frac{\vartheta_{1}}{\vartheta_{2}}.
\end{equation}
This simple relation leads to the following result for the inverse localization length \cite{SE81},
\begin{equation}
\label{loclength off}
L_{loc}^{-1}(E) = {\frac{1}{N} \langle\mid F_n \mid \rangle}\,, \quad
\mbox {for} \quad N \gg 1 \,,
\end{equation}
where the function $F$ can be expressed through the ratio of the hopping amplitudes,
\begin{equation}
\label{F-n}
F_n = \ln{\frac{\vartheta_1 \vartheta_2 \dots \vartheta_{2n-1}}{\vartheta_2 \vartheta_4 \dots \vartheta_{2n}}}= \sum_{i=1}^n{\left[ \ln\left(\frac{\vartheta_{2i-1}}{\vartheta_0}\right) - \ln\left(\frac{\vartheta_{2i}}{\vartheta_0}\right) \right]}.
\end{equation}

For uncorrelated values of $\vartheta_i$ the central limit theorem is applicable to the sum of random numbers $\ln\left(\vartheta_{i}/\vartheta_0\right)$ in Eq.(\ref{F-n}). Therefore, for $N>>1$ the distribution of fluctuations of $F$ is the Gaussian \cite{SE81,CFE05},
\begin{equation}
\label{gauss}
P(F) = \frac{1}{\sqrt{2 \pi N \sigma^2}}\exp\left( - \frac{F^2}{2N \sigma^2}\right),
\end{equation}
where $\sigma^2$ is the variance,
\begin{equation}
\label{variance}
\sigma^2 = \langle \ln^2\frac{\vartheta_i}{\vartheta_0} \rangle.
\end{equation}
Using the distribution function (\ref{gauss}) and the definition (\ref{loclength off}), it is straightforward to write the result for the localization length \cite{CFE05},
\begin{equation}
\label{loclength off uncorr}
L_{loc}(E=0) = \frac{\sqrt{\pi N/2}}{\sigma}.
\end{equation}
Thus, the localization length increases with the length of the sequence, however, the state $E=0$ cannot be classified as an extended state. Indeed, the localization length $L_{loc}(E=0)$ is much shorter than the size $L$ of the system, and the corresponding transmission coefficient vanishes in the thermodynamic limit \cite{SE81}.

If the random numbers $\vartheta_i$ are correlated, the central limit theorem is not applicable to the variable $F$. However, it is still possible to calculate the distribution function $P(F)$, assuming that the values $\ln(\vartheta_i/\vartheta_0)$ are normally distributed,
\begin{equation}
\label{ln gauss}
P(\ln(\vartheta/\vartheta_0)) = \frac{1}{\sqrt{2 \pi \sigma^2}} \exp{\left[ - \frac{\ln^2(\vartheta/\vartheta_0)}{2 \sigma^2}\right]}.
\end{equation}
According to the Wick's theorem, the odd moments of $F$ vanish and the even moments, $\langle F^{2n}\rangle$, can be expressed through the second moment $\langle F^{2}\rangle$ \cite{K92}. In this case the distribution of $F$, which is a linear combination of Gaussian random variables, remains the Gaussian, even if $\vartheta_i$ are correlated \cite{F71,KTH91}. Because of the Gaussian nature of the random variable $F$, the localization length is defined by its variance, see Eq.~(\ref{loclength off uncorr}). As a result, the following formula for the localization length is obtained \cite{CFE05},
\begin{equation}
\label{loclength off corr}
L_{loc}(E=0) = \frac{{\sqrt{\pi N /2}}}{ \sigma}\left[ 1 + \sum_{n=1}^{N-1}(-1)^n(1-\frac{n}{N}) g(n)\right]^{-1/2},
\end{equation}
where $g(n)$ is the correlation function of $\ln \vartheta_i$,
\begin{equation}
\label{corr ln}
g(n) = \sigma^2 \langle \ln {\frac{\vartheta_{m+n}}{\vartheta_0}}\ln {\frac{\vartheta_{m}}{\vartheta_0}}\rangle.
\end{equation}

\begin{figure}[!ht]
\begin{center}
\includegraphics[width = 12cm]{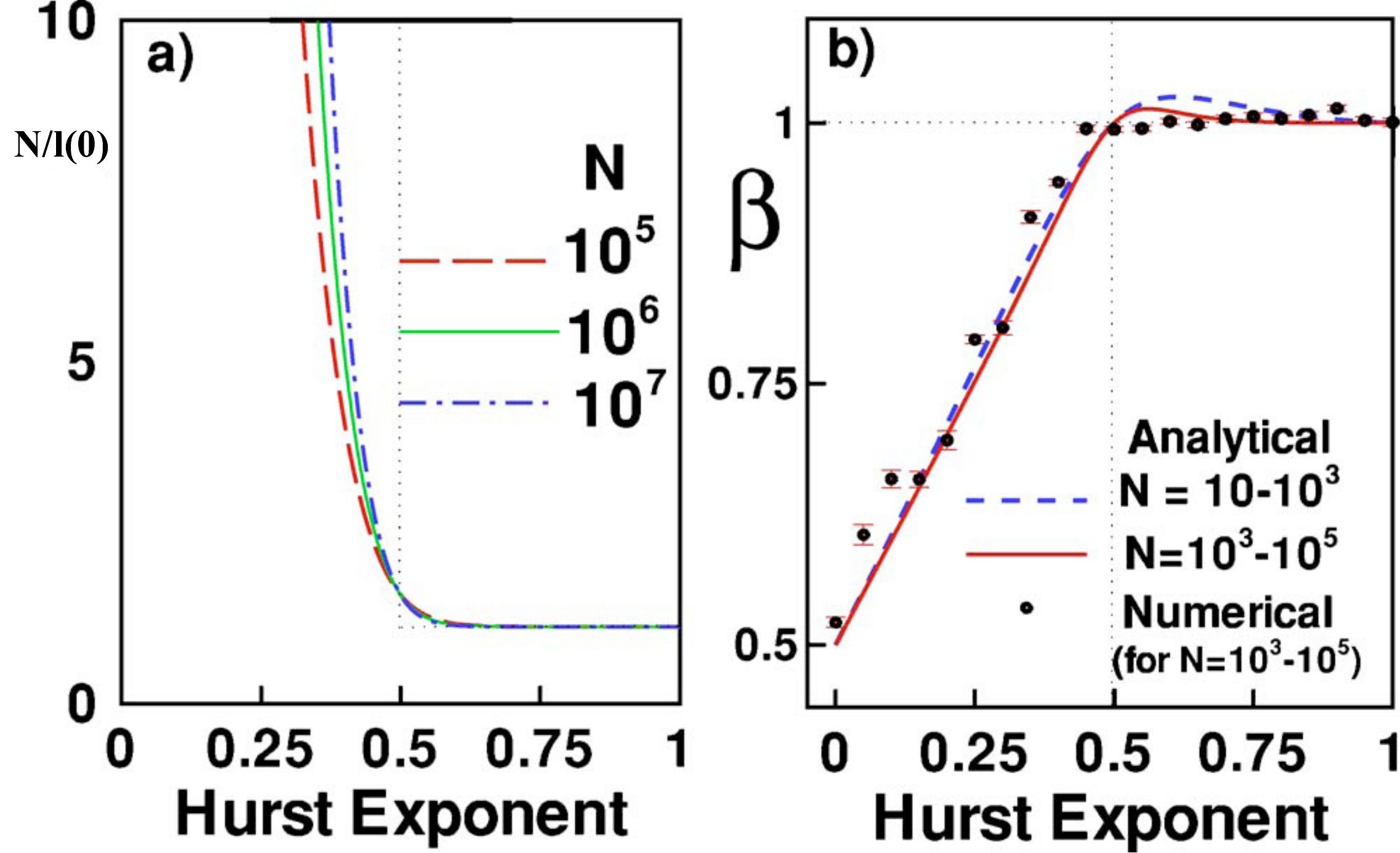}
\end{center}
\caption{Signatures of the localization-delocalization transition for the eigenstate with $E=0$ in a system with off-diagonal correlated disorder (with $N\equiv L$). (a) Normalized Lyapunov exponent $N/L_{loc}(E=0)$ vs Hurst exponent for different lengths of the system. (b) Variation of the exponent $\beta$ in the power-law dependence $L_{loc}(E=0) \propto N^{\beta}$ with the Hurst exponent (after \cite{CFE05}).} \label{hurst}
\end{figure}

 The expression (\ref{loclength off corr}) has been numerically verified for a set of correlated random variables $\ln (\vartheta_n/\vartheta_0)$ with the correlation function of a fractional Brownian motion,
\begin{equation}
\label{frac}
g(n)= 1-\left(\frac{n}{n_c} \right)^{2H},
\end{equation}
where $H$ is the Hurst exponent ($0<H<1$) and $n_c$ is the correlation length. In the numerical simulations the value $n_c$ was always adjusted to the system size $N$, but the variance $\sigma^2$ was $N$-independent \cite{CFE05}. The substitution of the correlation function Eq.~(\ref{frac}) into Eq.~(\ref{loclength off corr}) results in the following expression for the localization length,
\begin{equation}
\label{loclength off corr frac}
L_{loc}(E=0) = \frac{{\sqrt{\pi N /4}}}{ \sigma}\left[ \sum_{n=1}^{N-1}(-1)^{n+1}(1-\frac{n}{N})\left(\frac{n}{N}\right)^{2H}\right]^{-1/2}.
\end{equation}
The relation (\ref{loclength off corr frac}) exhibits a sharp localization-delocalization transition at $H=1/2$. It is well-known \cite{MK00} that at this critical value the fractional Brownian motion becomes a normal diffusion with the variance of the displacement being proportional to time. The data in Fig.~{\ref{hurst}}a show the ratio $L/L_{loc}(E=0)$ vs the Hurst exponent for different lengths of the correlated chain. For $0<H<1/2$ (region of subdiffusion for the fractional Brownian motion) the localization length is shorter than the system length, i.e. the eigenstate at $E=0$ is localized. For $1/2<H<1$ (region of superdiffusion) this eigenstate becomes extended and the ratio $L/L_{loc}(E=0)$ becomes practically independent either of $H$, or of $L$.

At the critical point, $H =1/2$, the scaling of the localization length with the system size changes,
\begin{equation}
L_{loc}(E=0) \propto L^{\beta}\,.
\label{N-beta}
\end{equation}
This tendency is clearly seen in Fig.~\ref{hurst}b where the exponent $\beta$ saturates at the value $\beta=1$ in the region $1/2<H<1$. The scaling for the localized states is characterized by $\beta <1$, and the extended states have $\beta = 1$. The numerical data in Fig.~\ref{hurst} show this transition at $H= 1/2$. This is in agreement with the fact that at $H=1/2$ the correlator (\ref{frac}) vanishes and the localization length in an uncorrelated potential scales as $\sqrt L$ (see Eq.~(\ref{loclength off uncorr})).

\begin{figure}[!ht]
\begin{center}
\includegraphics[width = 6cm]{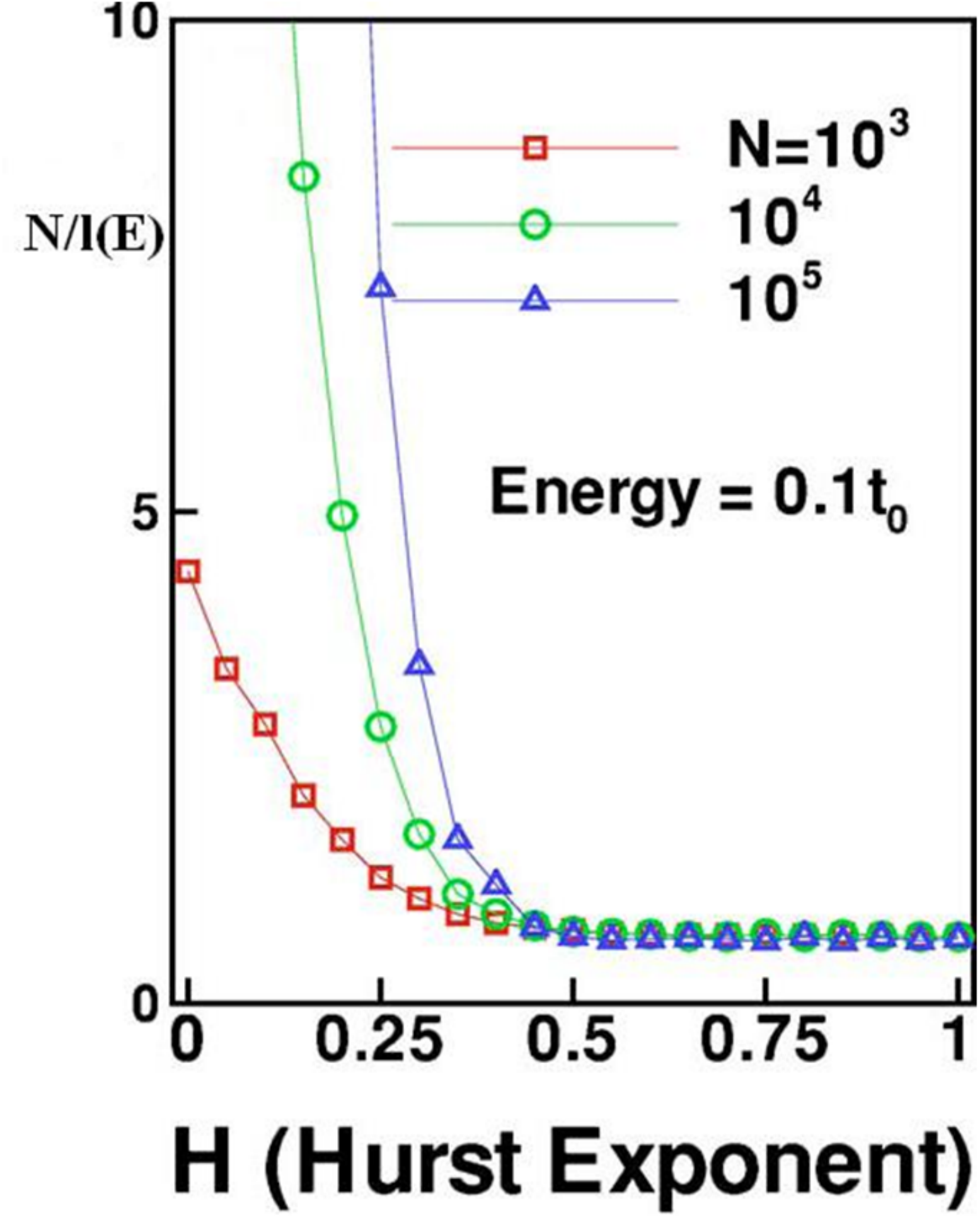}
\end{center}
\caption{ Signatures of the localization-delocalization transition for the eigenstate with $E=0.1 \vartheta_0$ in a system with correlated disorder with $\sigma=0.1$ (after \cite{CFE05}).} \label{hurstE}
\end{figure}

The strength $\sigma$ of disorder does not affect the localization-delocalization transition at $E=0$ where the transition occurs over the value of $H$ only. However, away from the band center ($E>0$) the transition point shifts to the region of subdiffusion $H<1/2$. Here the localization length exhibits $1/\sigma^2$ dependence on the amplitude of disorder \cite{TC76}, unlike $1/\sigma$ dependence at $E=0$. The shift of the transition point for energy $E=0.1 \vartheta_0$ is shown in Fig.~\ref{hurstE}.

The above results indicate that in the case of off-diagonal disorder a special class of long-range correlations may give rise to a band of extended states. Unlike the case of diagonal disorder, here the localization length (\ref{loclength off corr}) is determined by the correlator of logarithm of a random potential, see Eq.~(\ref{corr ln}), and not by the correlator of the potential itself.

\subsection{Mixed disorder}
\label{7.2}

\begin{figure}[!ht]
\begin{center}
\includegraphics[width = 8cm]{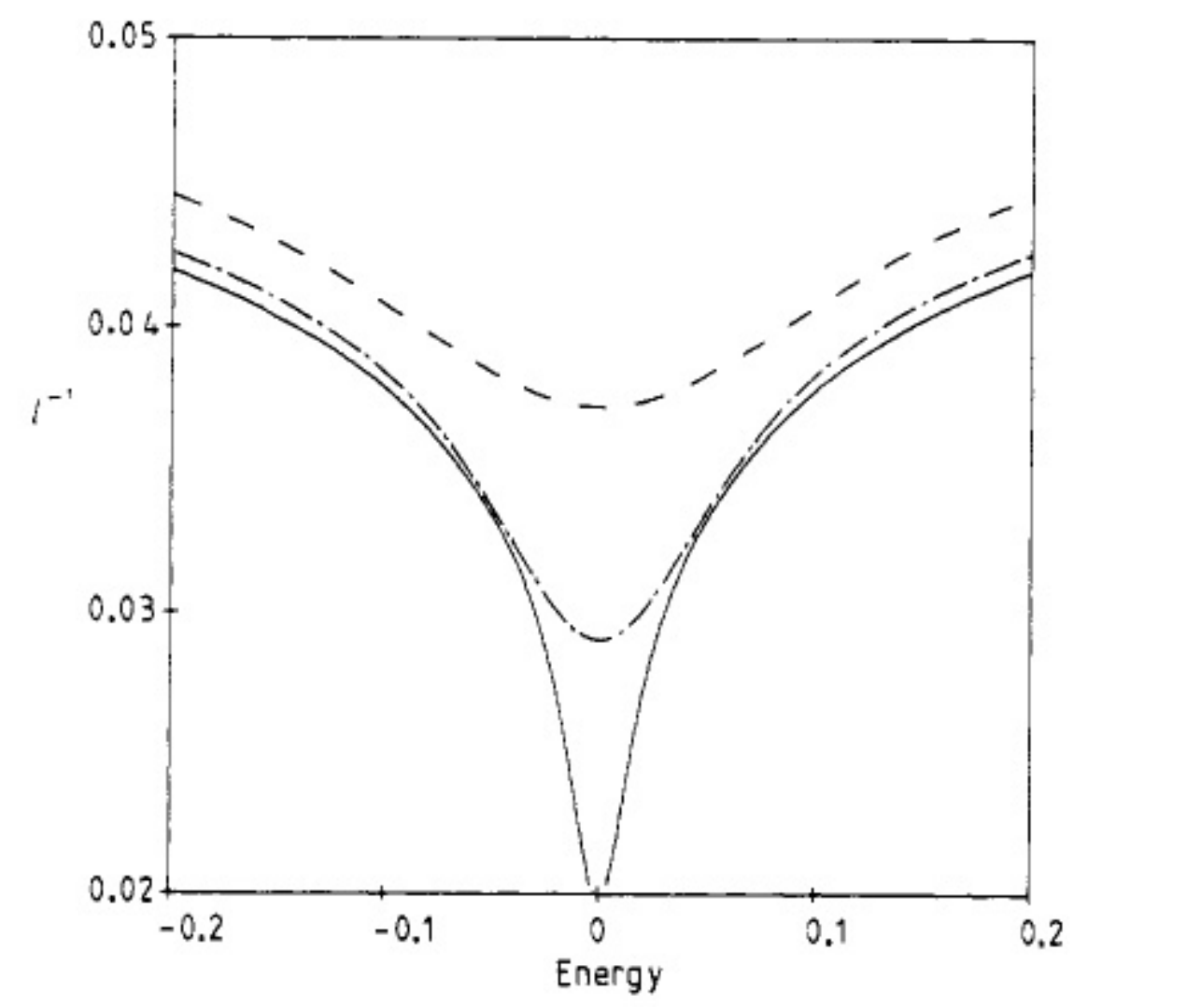}
\end{center}
\caption{ The behavior of the inverse localization length near the band center for uncorrelated disorder: pure off-diagonal (solid curve) and mixed disorder (dashed curve - for stronger and dotted-dashed curve - for weaker diagonal disorder ) (after \cite{SP88}).} \label{mixed}
\end{figure}

Much less is known about the mixed disorder, when the disorder enters both into the diagonal and off-diagonal elements of the Anderson model (\ref{tb}). This case is the most difficult for the analysis. The analytical expression for the localization length is not known even for a weak mixed disorder. It was, however, understood that the presence of the diagonal disorder, in addition to the off-diagonal one, regularizes the divergence of the localization length at $E=0$.
The numerical calculation carried out in Ref.~\cite{SP88} shows that in the case of mixed uncorrelated disorder the inverse localization length $L_{loc}^{-1}(E)$ remains finite at the band center, see Fig.~\ref{mixed}.
Note that in all the cases the band center is the point with the longest localization length.

Formally, the off-diagonal disorder can be eliminated from the Schr\"odinger equation with mixed disorder,
\begin{equation}
\label{mixSch}
\vartheta_{n+1}\psi_{n+1} + \vartheta_{n}\psi_{n-1} = (E-\epsilon_n) \psi_{n},
\end{equation}
by the following substitution proposed in Ref.\cite{F89}:
\begin{equation}
\label{transformation}
\psi_n = \phi_n q_n.
\end{equation}
Here $q_n$ is defined by the recursion relation,
\begin{equation}
\label{recursion}
q_n = 1/(\vartheta_nq_{n-1}).
\end{equation}
It is easy to see that the equation for new function $\phi_n$ does not contain disorder in the off-diagonal terms,
\begin{equation}
\label{newSch} \phi_{n+1} + \phi_{n-1} = q_n^2(E-\epsilon_n) \phi_{n}.
\end{equation}
This exact transformation reduces an eigenvalue problem with mixed disorder to a {\it generalized} eigenvalue problem with pure diagonal disorder \cite{F89}. Note, however, that the factor $q_n$ contains information not only on the $n$th site but on the ($n-1$)th site as well, see Eq. (\ref{recursion}). This means that Eq. (\ref{newSch}) only formally corresponds to the tight-binding model with diagonal disorder. Indeed, the $n$th element of the random sequence $q_n$ can be written as an infinite product containing the information about whole random sequence,
\begin{equation}
\label{inf product}
q_n = \frac{1}{\vartheta_nq_{n-1}}= \frac{\vartheta_{n-1}}{\vartheta_n}q_{n-2} = \frac{\vartheta_{n-1}}{\vartheta_n} \frac{\vartheta_{n-3}}{\vartheta_{n-2}} \frac{\vartheta_{n-5}}{\vartheta_{n-4}} \dots
\end{equation}
Since $q_n$ depends on all the preceding random hopping elements, all the moments of $q_n$ diverge exponentially with the length of the sequence. This singular statistical property limits an application of Eq.(\ref{newSch}).

The generalized eigenvalue problem (\ref{newSch}) exhibits an extended state at the critical energy $E_c$, provided the following relation between the random variables $\vartheta_n$ and $\epsilon_n$ \cite{F89}:
\begin{equation}
\label{relation}
\epsilon_n = E_c (1+ \vartheta_n^2).
\end{equation}
In this case the Lyapunov exponent vanishes linearly, $L_{loc}^{-1}(E) \sim |E-E_c|$, when the energy $E$ approaches the critical energy $E_c$. The relation (\ref{relation}) was obtained in the short-wavelength approximation, when the distance between nearest sites exceeds the wavelength.

Although the relation (\ref{relation}) is deterministic, it does not entail that the sequence $\epsilon_n$ completely determines
the hopping elements $\vartheta_n$. Indeed, the modulus of the value $\vartheta_n$ is defined, however, its sign remains indefinite and may be randomly selected. Thus, Eq.~(\ref{relation}) can be treated as an example of specific correlations between two random sequences, when the absolute values of the fluctuations are synchronized by a deterministic relation, however, the signs of $\vartheta_n$ remain uncorrelated.

\begin{figure}[!ht]
\includegraphics[width = 8cm]{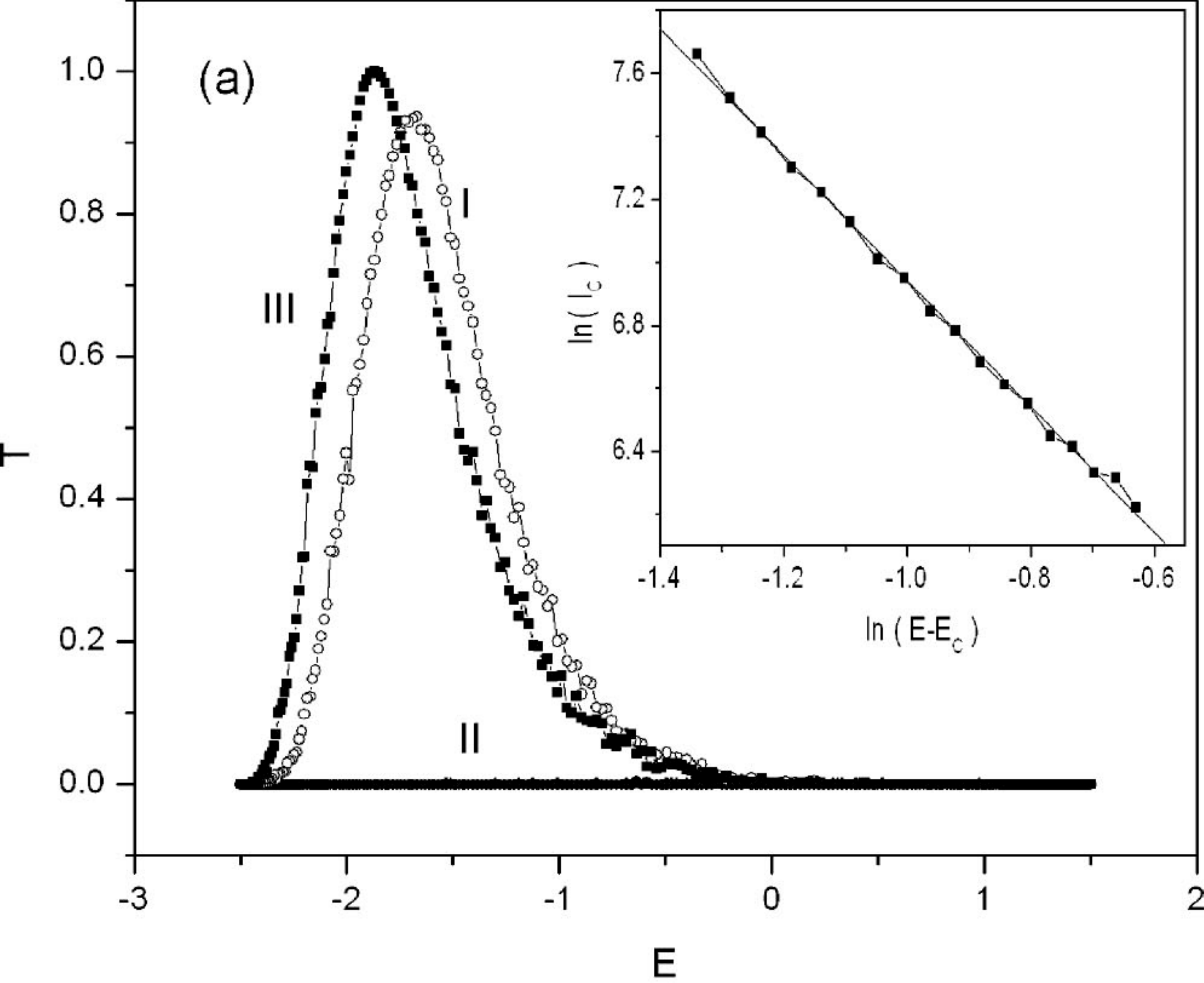}
\includegraphics[width = 8cm]{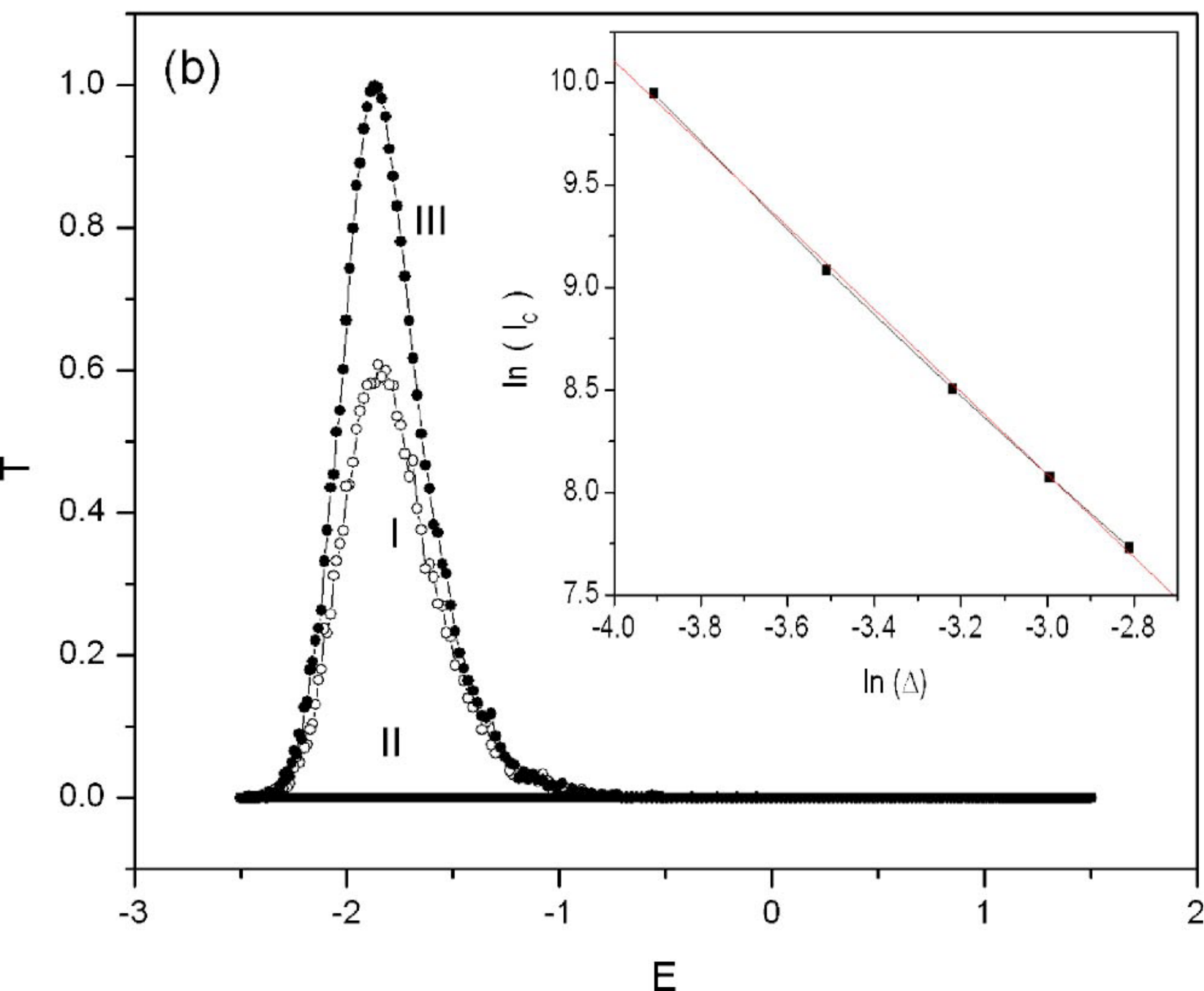}
\caption{ (a) Transmission coefficient vs energy for a system of 1000 sites with $\epsilon_B-\epsilon_A = 1$ averaged over 300 configurations. Curve I (empty symbols) is typical for a diluted system with mixed disorder; hopping constants here $\vartheta_{AA} = 1$, $\vartheta_{BB} = 1.73$, $\vartheta_{AB} =(\vartheta_{AA}+\vartheta_{BB})/2=1.36$. Curve II (solid line close to horizontal axis) is for pure diagonal disorder, with all hopping constants equal, $\vartheta_{AA} = \vartheta_{BB} = \vartheta_{AB}=1$. Curve III (full symbols) stands for system with ''golden correlations', $\vartheta_{AB} = \sqrt{\vartheta_{AA}\vartheta_{BB}}=1.316$, with the resonant transmission at $E_c =-1.9$. Concentration of $B$-type sites is 0.1 for all three curves. (b) Same as in (a), but with concentration of $B$-type sites at 0.5. Inset in (a): Localization length $L_{loc}(E)$ vs $(E-E_c)$ for system with golden correlation. Slope of fitted line is 2. Concentration of $B$-type sites is 0.5. Inset in (b): Localization length $L_{loc}(E_c)$ at
critical energy $E_c$ vs $\Delta =\mid \vartheta_{AB}-\sqrt{\vartheta_{AA}\vartheta_{BB}} \mid$ (after \cite{ZU04}).} \label{Ulloa1}
\end{figure}

An interesting case of mixed disorder in a {\it binary} random sequence has been proposed in Refs.~\cite{ZU04,ZU04a}. Specifically, for a binary sequence the on-site energies in the tight-binding equation (\ref{tb}) take only two values, $\epsilon_n = \epsilon_A$ or $\epsilon_n = \epsilon_B$, thus, there are three values for the hopping constants, $\vartheta_{AA}$, $\vartheta_{BB}$, and $\vartheta_{AB}$. A single impurity $\epsilon_B$ in otherwise perfect sequence of $\epsilon_A$ sites breaks periodicity and reduces the transmission coefficient $T$ from 1 to the following value \cite{ZU04a} :
\begin{equation}
\label{t1}
T = \frac{4 \vartheta_{AB}^4 \sin^2 \mu}{4 \vartheta_{AB}^4 \sin^2 \mu + {\cal D}^2},
\end{equation}
where $E=2 \vartheta_{AA} \cos \mu$ and
\begin{equation}
\label{N1}
{\cal D} = (\epsilon_B-\epsilon_A) \vartheta_{AA} + 2 \left(\vartheta_{AB}^2 - \vartheta_{AA}^2 \right)\cos \mu .
\end{equation}
Therefore, there is a single resonant energy,
\begin{equation}
\label{E1}
E_r = (\epsilon_B-\epsilon_A)\vartheta_{AA}^2/\left(\vartheta_{AA}^2 - \vartheta_{AB}^2 \right)
\end{equation}
for which the transmission reaches unity. It is clear that for a finite-length diluted sequence with a small concentration of randomly distributed $B$-type sites, the transmission coefficient is still close to 1 for the energies near $E_r$, see Fig.~{\ref{Ulloa1}}. However, in the thermodynamic limit, the system, being random, behaves as insulator even at $E=E_r$. The reason is the presence of clusters of $BB$-, $BBB$-, and $BBB\dots$-types, that in the thermodynamic limit localize the states even with the energy $E_r$. For each of these clusters the resonant energies are generically different from $E_r$. The exception is a particular case with the so-called ``golden correlation" with $\vartheta_{AB}= \sqrt{\vartheta_{AA}\vartheta_{BB}}$. In this case the resonant energy $E_c$ for all the clusters turns out to be the same,
\begin{equation}
\label{E_c}
E_c= \epsilon_A+ 2\vartheta_{AA} \cos \mu = \epsilon_B+ 2\vartheta_{BB} \cos \mu.
\end{equation}

The corresponding extended state is a modulated plane wave $\alpha_n \exp(i\mu n)$ with the amplitude $\alpha_n =1$ at the $A$-type sites and $\alpha_n= \sqrt{\vartheta_{AA}/\vartheta_{BB}}$ at the $B$-type sites. The extended state with perfect transmission $T(E_c)=1$ is obtained in the numerical simulations even at high concentration of $B$-type impurities, see Fig.\ref{Ulloa1}. The position of the extended state is independent on the concentration and coincides exactly with that given by Eq.~(\ref{E_c}). The perfect transmission is a result of cancelations of backscattered waves from all the clusters of multi-$B$-type, which occurs only for special golden correlations and only at the resonant energy $E_c$. For all the energies different from $E_c$ the localization length is finite. It scales with energy as $L_{loc}(E) \propto (E-E_c)^{-2}$. The scaling is obtained from the slope of the straight line shown in the insets in Fig.~\ref{Ulloa1}a. Similar mechanism is responsible for two extended states in random dimers \cite{WP90,DWP90}.

Note that the extended state with the energy (\ref{E_c}) emerges due to the {\it deterministic} relation $\vartheta_{AB}=\sqrt{\vartheta_{AA}\vartheta_{BB}}$ between the elements of a random sequence, as in the case of Eq.~(\ref{relation}). This type of correlations is different from the statistical correlations in random sequences to which the theory of correlated disorder is developed and discussed in previous Sections. However, it can be considered as a particular case of the theory since the analytical expressions obtained for statistical correlations can be also applied to deterministic correlations as well, see examples in Sections~\ref{7.2} and \ref{11.5}.

A binary system with mixed disorder exhibits an anomalous dependence of electron transport on the amplitude of disorder. Normally, the localization length becomes shorter with an increase of disorder. Here the extended state exists at $E=E_c$ if Eq.~(\ref{E_c}) has a solution for $\mu_c = \arccos{(\epsilon_B-\epsilon_A)/(\vartheta_{AA} - \vartheta_{BB})}$. For a fixed value of $\epsilon_B-\epsilon_A$ of diagonal disorder, the extended state in the spectrum appears only if the off-diagonal disorder is sufficiently strong,
$\mid \vartheta_{AA} - \vartheta_{BB} \mid > \mid \epsilon_{A} - \epsilon_{B} \mid/2$. With an additional amount of the off-diagonal disorder to a finite-length sequence with pure diagonal disorder, the transmission strongly increases. This tendency is clearly seen when comparing curves II and I in Fig.\ref{Ulloa1}a.

\section{Kronig-Penney models: Delta-like barriers}
\label{8}

Let us now apply the ideas discussed above in connection with the tight-binding Anderson model, to the so-called Kronig-Penney model that is widely used in the solid-state physics. Again, our main interest is in the properties of eigenstates in the presence of specific long-range correlations in random potentials. Originally, the Kronig-Penney model was suggested \cite{KP31} to study the properties of energy spectra in periodic potentials such as those appearing in solid state physics. The model describes an idealized quantum system consisting of an infinite periodic array of square potentials barriers, and can be treated as a model of a periodic crystal lattice. Indeed, the corresponding potential models the realistic one caused by ions in the periodic structure of a crystal, and exhibits the essential features of electronic structure, such as an appearance of allowed and forbidden energy bands. Apart from disorder solids and liquids (see, for example, Ref.\cite{LM66}), the Kronig-Penney model has been also applied to other fields of physics, such as microelectronic devices \cite{J89,TP92}, layered superconductors \cite{TT89}, and quark tunneling in one-dimensional nuclear models \cite{CM90}.

Quite often the width of barriers is much smaller than the distance between them;
this allows to consider the barriers as delta-functions spaced along the $x$-axis in the one-dimensional geometry. In a more general context one can ask the question about the transport properties of one-dimensional structures consisting of these delta-scatterers, however with random amplitudes and spacings between the delta-barriers. Then the corresponding Shr\"odinger equation for stationary states in an infinite array of such delta-barriers can be written as follows,
\begin{equation}
\label{KP-delta}
\psi '' (x) +k^2 \psi (x) = \sum _{n=-\infty}^{+\infty} U_n \psi (x_n) \delta (x-x_n),
\end{equation}
where $x_n$ is the position of the $n-$th barrier and $U_n$ its amplitude. Here Eq.~(\ref{KP-delta}) is written in the unified form, equally valid for the electron and electromagnetic (optic) wave propagation. In the first case we assume the units in which $\hbar^2 /2m=1$, and $k$ stands for the wave number for electrons. In the second case, the wave number $k=\sqrt {E}$ is equal to $\omega/c$ for a classical scalar wave of frequency $\omega$, see Eq.~(\ref{1DCP-Schreq}).

As one can see, the disorder in the model (\ref{KP-delta}) is originated either from random variations of the amplitudes $U_n$ of delta-barriers or from the randomness in their positions $x_n$. In what follows, we term the first case the {\it compositional} (or {\it amplitude}) disorder, in contrast with the second case of the {\it structural} (or {\it positional}) disorder.

\subsection{Kronig-Penney $N$-mers}
\label{8.1}

Let us, first, discuss the situation for which the delta-barriers are spaced periodically, $x_n=nd$, and $U_n$ is the only source of disorder. In the case when $U_n$ are random and delta-correlated numbers (white-noise disorder), all eigenstates are exponentially localized on an infinite $x-$axis. The general solution for the localization length strongly depends on whether the disorder weak or strong, and can be obtained by reducing the Kronig-Penney model to the form similar to the Anderson model (see below). Our main interest, however, is in global properties of localization length, in particular, in the conditions for which the localization length diverges (or becomes anomalously large).

In Refs.~\cite{SMD94,SDBI95} the influence of short-range correlations in the sequence $U_n$ on the localization properties of eigenstates was studied in connection with the discovery of extended states occurring in the Anderson tight-binding models with dimer impurities (see Section~\ref{5.1}). Specifically, it was shown that the correlations induced by the dimers result in an infinite number of resonances. Since these resonance produce the energy bands of extended states for samples of finite size, the effect of short range correlations can be observed experimentally. This was, indeed, done later on, see Ref.~\cite{Bo99}. Below we discuss some of the essential properties of the dimer Kronig-Penney model, that can be easily understood with the use of formal transformation of the Kronig-Penney model into the Anderson model. Note, however, that in the Anderson model the whole energy spectrum is finite and can be treated as the one-band spectrum, see Eq.~(\ref{E-mu}). Contrary, in the periodic Kronig-Penney model (without any disorder) the energy spectrum is infinite and consists of infinite set of allowed and forbidden bands. As a result, in contrast with the one-band A-model, in the Kronig-Penney model a new effect emerges known as the Fabry-Perrot resonances. The influence of these resonances on transport properties will be discussed below (see Section~\ref{10.4}), here we focus on the resonances that are entirely due to short-range correlations in the potential.

Let us show that the equation (\ref{KP-delta}) can be written in the form very similar to that of the Anderson tight-binding model (\ref{tb diagonal}). Indeed, one can easily integrate Eq.~(\ref{KP-delta}) between two successive delta-kicks of the potential. For this, let us define the value of the $\psi$-function and its derivative right {\it before} the $n$-th kick as $x_n\equiv \psi_n$ and $p_n \equiv \psi '_n$ respectively. Then, the integration of the $n$-th kick results in the change of the derivative (or, the ``momentum" in the Hamiltonian map approach), keeping the value of $\psi$-function unchanged,
\begin{equation}
\label{kick}
\tilde p_n= p_n + U_n x_n\,; \qquad \tilde x_n = x_n\,.
\end{equation}
Here $\tilde x_n$ and $\tilde p_n$ are the values of $\psi$-function and its derivative immediately {\it after} the $n$-th kick. Between the $n$-th and $(n+1)$-th kicks the potential is zero, therefore, the integration gives rise to the rotation of $\tilde x_n$ and $\tilde p_n$, according to which one can express the values $x_{n+1}$ and $p_{n+1}$ {\it before} the $(n+1)$-th kick in terms of $\tilde x_n$ and $\tilde p_n$,
\begin{eqnarray}
\label{KP-m}
&& x_{n+1} = \tilde x_n \cos {kd} + \frac{\tilde p_n}{k} \sin {kd}, \\ \nonumber
&& p_{n+1} = -k \tilde x_n \sin {kd} + \tilde p_n \cos {kd}\,.
\end{eqnarray}
By combining Eqs.(\ref{kick}) and (\ref{KP-m}) one gets the two-dimensional map for $x_n$ and $p_n$,
\begin{eqnarray}
\label{KP-map}
&& x_{n+1} = x_n \cos {kd} + \frac{1}{k} (p_n +U_n x_n) \sin {kd}, \\ \nonumber
&& p_{n+1} = -k x_n \sin {kd} + (p_n+U_n x_n) \cos {kd}\,.
\end{eqnarray}
Comparing with the Hamiltonian map (\ref{Ham}) that corresponds to the Anderson model, one can obtain the recursive relation of the form similar to Eq.~(\ref{tb diagonal}),
\begin{equation}
\label{KP-diagonal}
\psi_{n+1} + \psi_{n-1}=\left(2 \cos \mu + \frac{U_n}{k} \sin \mu \right) \psi_n\,,
\end{equation}
where $\mu =k d$ is the phase shift of the $\psi$-function between two successive $\delta$-kicks. This relation can be considered as the reduction of the Shr\"odinger equation to the Poincar\'{e} map, see Refs.~\cite{BFLT82,SJ82}. In a more general context, it can be shown that one can construct the relation between the values of $\psi_n$ at three consecutive points of the set $\{ x_n \}$ for {\it any} one-dimensional potential $V(x)$ (see details in Ref.~\cite{SMD94}). Such a relation involves the two linearly independent solutions of the Shr\"odinger equation in the interval $x_{n}, x_{n+1}$. In our case of the periodic delta-function these solutions are very simple, and we have used them explicitly in Eq.~(\ref{KP-m}).

The formal reduction of the Kronig-Penney model to the recursive relation (\ref{KP-diagonal}) allows us easily to understand the main properties of short-range correlations induced by dimers. Let us consider the sequence $U_n$ in which all values $U_n$ are equal to $\overline{\epsilon}_1$, except two values for which $U_n=U_{n+1}=\overline{\epsilon}_2$. It can be seen that this single dimer does not influence the scattering when the total phase shift $\gamma_n+\gamma_{n+1} = 2\gamma_n$ through the dimer is equal to $\pi$ or to $2\pi$, where $\gamma_n$ is determined by the dispersion relation,
\begin{equation}
\label{gamma-dimer}
2 \cos \gamma_n = 2 \cos \mu + \frac{\overline{\epsilon}_2}{k} \sin \mu \,.
\end{equation}
Note that the value of $\gamma_n$ is confined in the interval $(0, \pi)$, therefore, the sum $\gamma_n+\gamma_{n+1}$ can not be larger than $2\pi$. Also, the value $\gamma_n=\pi$ should be excluded since it corresponds to the band edge of the spectrum defined by Eq.~(\ref{gamma-dimer}). As a result, the resonant energy $E_{cr}=k_{cr}^2$ is defined by $\gamma_n= \pi/2$, thus giving the relation, $-2k_{cr}/\overline{\epsilon}_2= \tan k_{cr}$ (here and below we assume $d=1$). The same arguments can be applied to the second dimer specified by $U_n=\overline{\epsilon}_1$. As a result, for the general case of randomly distributed dimers $\overline{\epsilon}_1$ and $\overline{\epsilon}_2$ there are two resonant values defined via the following relation,
\begin{equation}
\label{k-cr}
k_{cr} = -\frac{\overline{\epsilon}_{1,2}}{2} \tan k_{cr}\,,
\end{equation}
for which either the $\overline{\epsilon}_1$-dimer or $\overline{\epsilon}_2$ dimer do not influence the scattering. Since $\tan k_{cr}$ is a $\pi$-periodic function and takes the values in the whole interval $(-\infty, +\infty )$, there is an infinite set of critical energies $E_{cr}=k_{cr}^2$, with two in each interval $[(2s-1)\pi/2, (2s+1)\pi/2]$ with $s$ integer (see, also, Ref.~\cite{SMD94}). However, for the allowed energy bands defined by the condition $|2 \cos \gamma_n|=2$, the absolute value of disorder has to be smaller than the critical value $|\epsilon_{cr}|=2$ in order to have the resonant states (note that $\epsilon_{cr}<0$).

\begin{figure}[ht]
\begin{center}
\includegraphics[width=8.6cm,height=6.6cm]{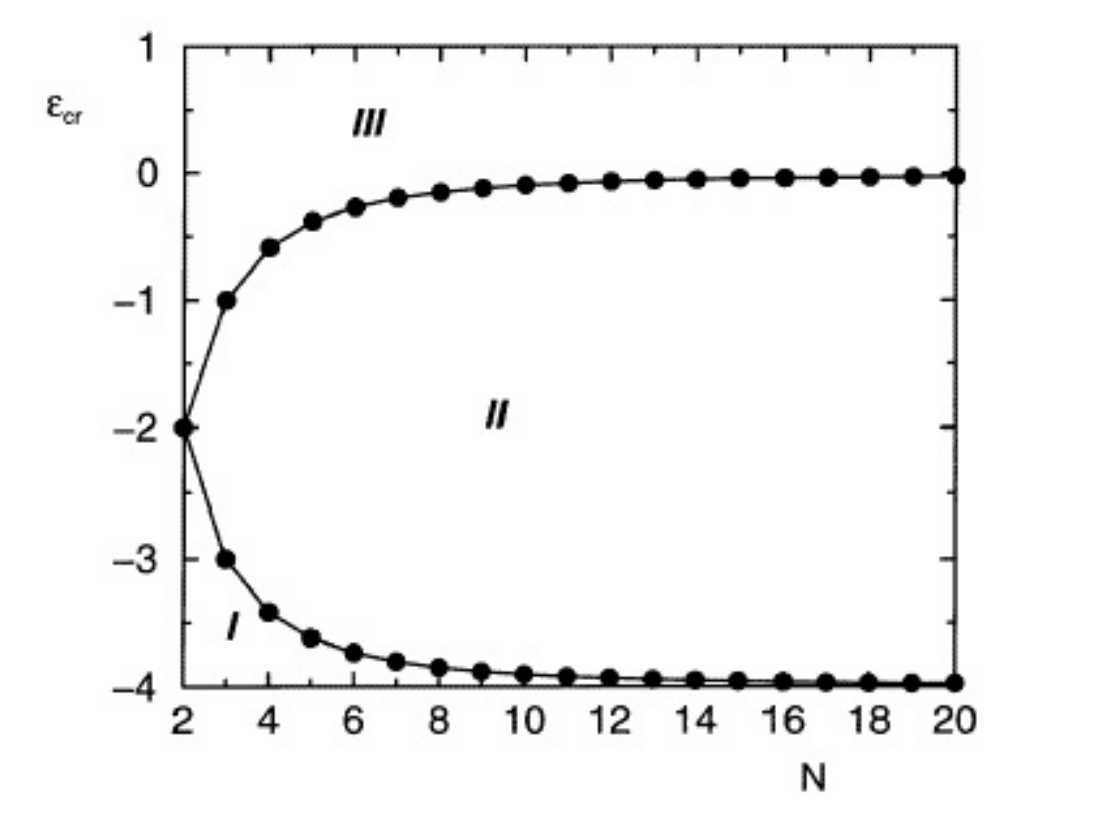}
\end{center}
\caption{Phase diagram demonstrating the critical $\epsilon_{cr}$-values as functions of the block size $N$ (after \cite{KTI97}).}
\label{fig-KPdim}
\end{figure}

The above analysis can be extended to the general case of $N$-mers (when two values $\overline{\epsilon}_1$ and $\overline{\epsilon}_2$ emerge in blocks of length $N$). In this case the resonant energies are defined through the following conditions,
\begin{equation}
\label{N-merKP}
\gamma_N = \frac{\pi}{N}\,, \frac{2\pi}{N}\,, \frac{3\pi}{N}\,, ... \,, \frac{(s+1)\pi}{N}\,,\qquad s=0\,,1\,,2\,, ...\,, N-2\,.
\end{equation}
As one can see, when the block size $N$ increases, the number $2(N-1)$ of resonant states increases as well. In the first zone the strength of disorder has to be smaller than the critical strength $\epsilon_{cr}(N)$ which is defined via the relation, $2\cos k_{cr}+(\overline{\epsilon}_{1,2}/ k_{cr}) \sin k_{cr} = 2\cos \gamma_N$. Equivalently, the values of $\epsilon_{cr}$ are defined as follows,
\begin{equation}
\label{res-KP}
\epsilon_{cr} =- \frac{2k}{\sin k} \left(\cos k -\cos \gamma_N\right)\,.
\end{equation}
The resonances inside the first band appear whenever the derivative of the ratio, $\sin k /(\cos k - \cos \mu_N)$, with respect to $k$ at $k=0$ is less than $-2/\epsilon_{cr}$, and thus,
\begin{equation}
\label{res-KP-1}
\epsilon_{cr}(N)= 2 (\cos \gamma_N -1)\,.
\end{equation}
In the phase diagram, see Fig.\ref{fig-KPdim}, one can distinguish three regions with no resonant states (region I), with some of resonant states (region II), and with all resonant states (region III). The borders of the region II coincide at $\epsilon =-2$ for $N=2$ (random dimers), and are defined by the condition (\ref{res-KP-1}) with $\gamma_N=\pi/N$ (upper bound) and $\gamma_N=\pi (n-1)/N$ (lower bound); they approach the limits $0$ and $-4$ respectively as $N \rightarrow \infty$. We recall that for a perfect (periodic) Kronig-Penney lattice the range of accessible $U_n=U$ values within the first energy band is $[-4, +\infty)$. The lower border corresponds to the appearance of only one resonant state, while as we move towards the upper border, more resonant states emerge. The upper critical curve in Fig.\ref{fig-KPdim} delimits the region III in which all possible resonances exist.

\begin{figure}[!ht]
\begin{center}
\subfigure{\includegraphics[scale=0.65]{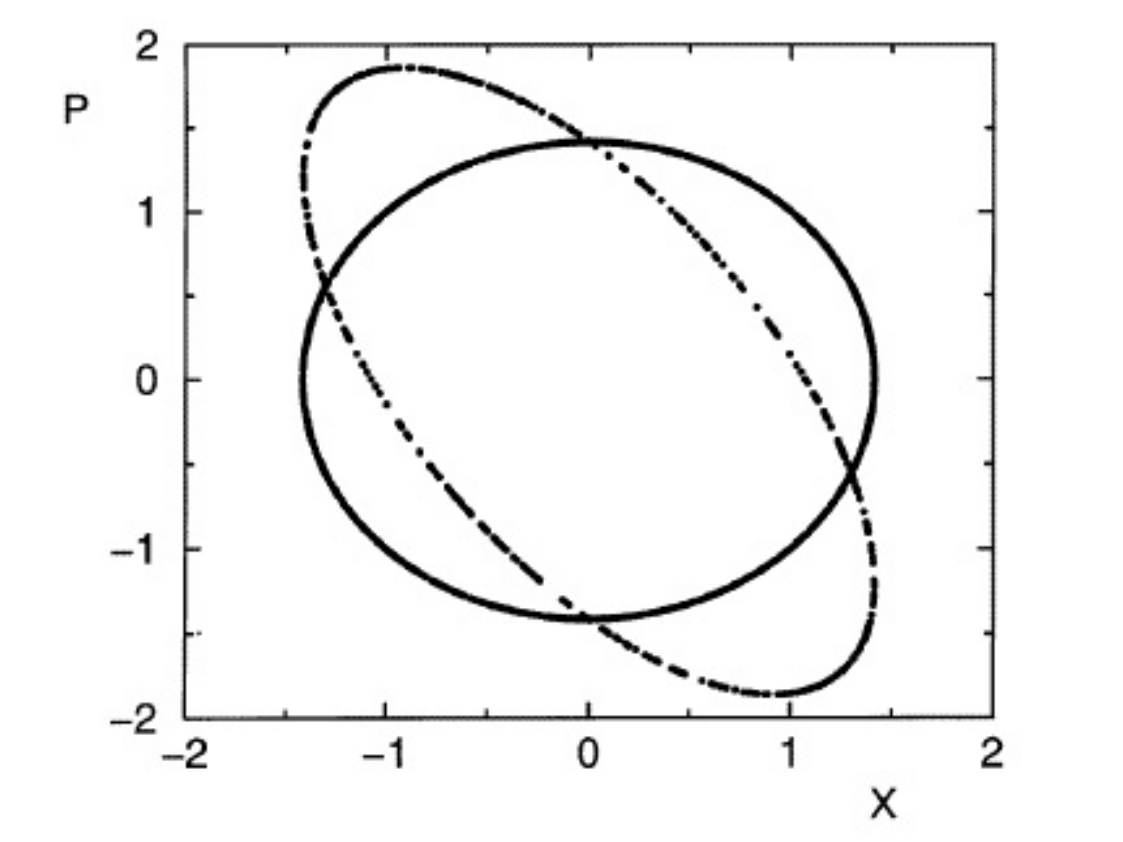}}
\subfigure{\includegraphics[scale=0.65]{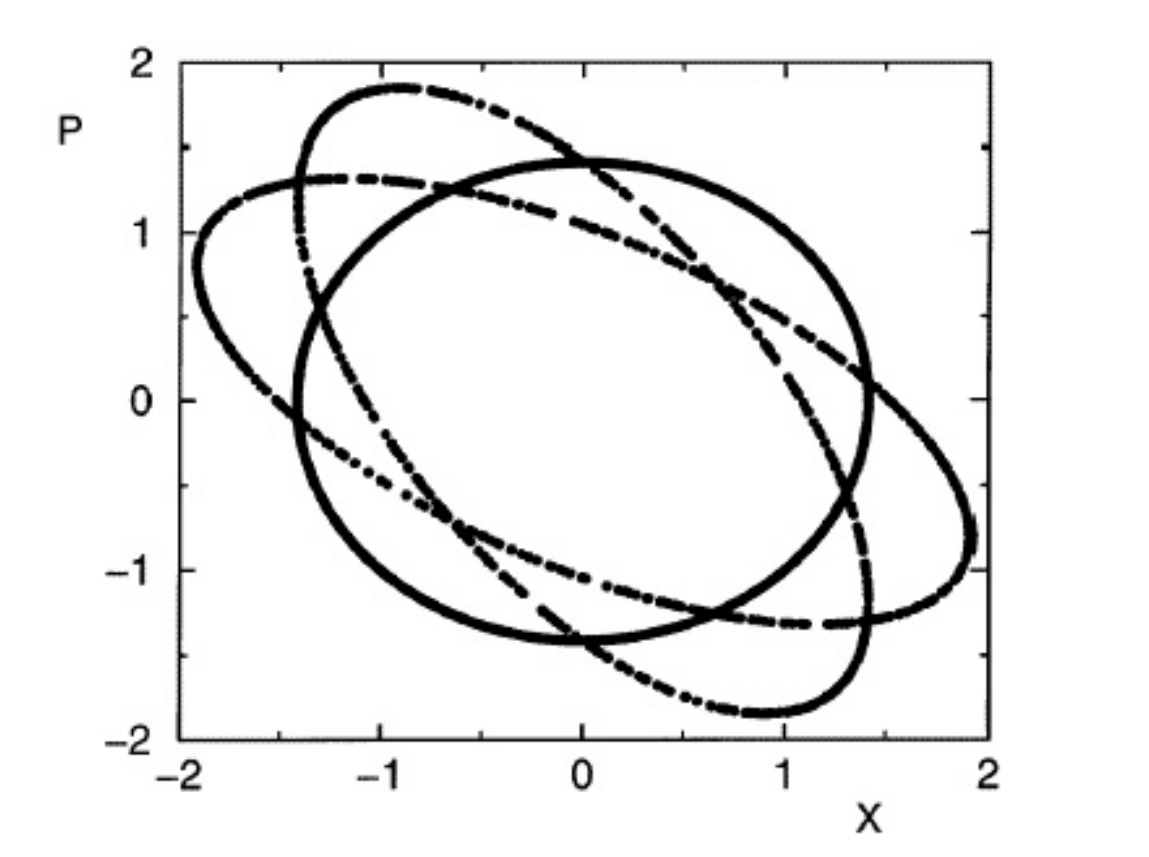}}
\end{center}
\caption{Phase space of the map (\ref{KP-map}) for $x_0=p_0=1$ and $Q=0.5$ with $\overline{\epsilon}_1=0$. Left: the dimer with $N=2, k=k_{cr}=11.4, \overline{\epsilon}_2=9.77885$. Right: the trimer with $N=3, k=k_{cr}=1.489, \overline{\epsilon}_2=1.25$. The length of sequence $\epsilon$ in both cases is $1000$ (after \cite{KTI97}).}
\label{KP-fig3}
\end{figure}

It is interesting to see the form of classical trajectories in the phase space $\{p_n, x_n \}$ corresponding to the $N$-mers. As shown in Ref.~\cite{KTI97}, the iteration of the map (\ref{KP-map}) with initial conditions, $p_0=x_0=1$, create the ellipses in the phase space, each of them correspond to one of the $N$-mers. For example for the case of dimers, $N=2$, there are two ellipses, see Fig.~\ref{KP-fig3}a, that are created by the points $p_n, x_n$ originated from the map (\ref{KP-map}) with either $\overline{\epsilon}_1$ or $\overline{\epsilon}_2$. The probability to appear one of the dimer is chosen to be $Q=0.5$, therefore, with equal probability the trajectory randomly jumps from one ellipse to another. For comparison, the trajectories corresponding to the trimers ($N=3$) are shown in Fig.~\ref{KP-fig3}b where doublets of the values $\overline{\epsilon}_1$ or $\overline{\epsilon}_2$ emerge in the potential $U_n$ with equal probability.

The representation of $N$-mers in the form of classical trajectories can be useful for the analysis of the structure of extended states. It is important to emphasize that the localized states of the original Shr\"{o}dinger equation (\ref{KP-delta}) correspond to unbounded trajectories $p_n, x_n$. The very point is that the exponential increase of the radius $r_n = (x_n^2 + p_n^2)^{1/2}$ is given by the Lyapunov exponent $\lambda$ , and the latter determines the localization length, $L_{loc} = \lambda^{-1}$. This does not mean that the classical trajectories are the eigenstates since the latter are exponentially localized in an infinite space in both directions, $n \rightarrow +\infty$, and $\rightarrow -\infty$, in contrast with the exponential increase of $\psi_n=x_n$ in the Hamiltonian map approach. However, for the resonant values $E_{cr}$ the classical trajectories (specifically, the dependence $x_n$) are statistically equivalent to true extended eigenstates of the Shr\"{o}dinger equation. For this reason, the {\it bounded} trajectories shown in Fig.~\ref{KP-fig3} can be treated as the image of the dimer and trimer eigenstates for resonant energies.

\begin{figure}[!ht]
\begin{center}
\subfigure{\includegraphics[scale=0.65]{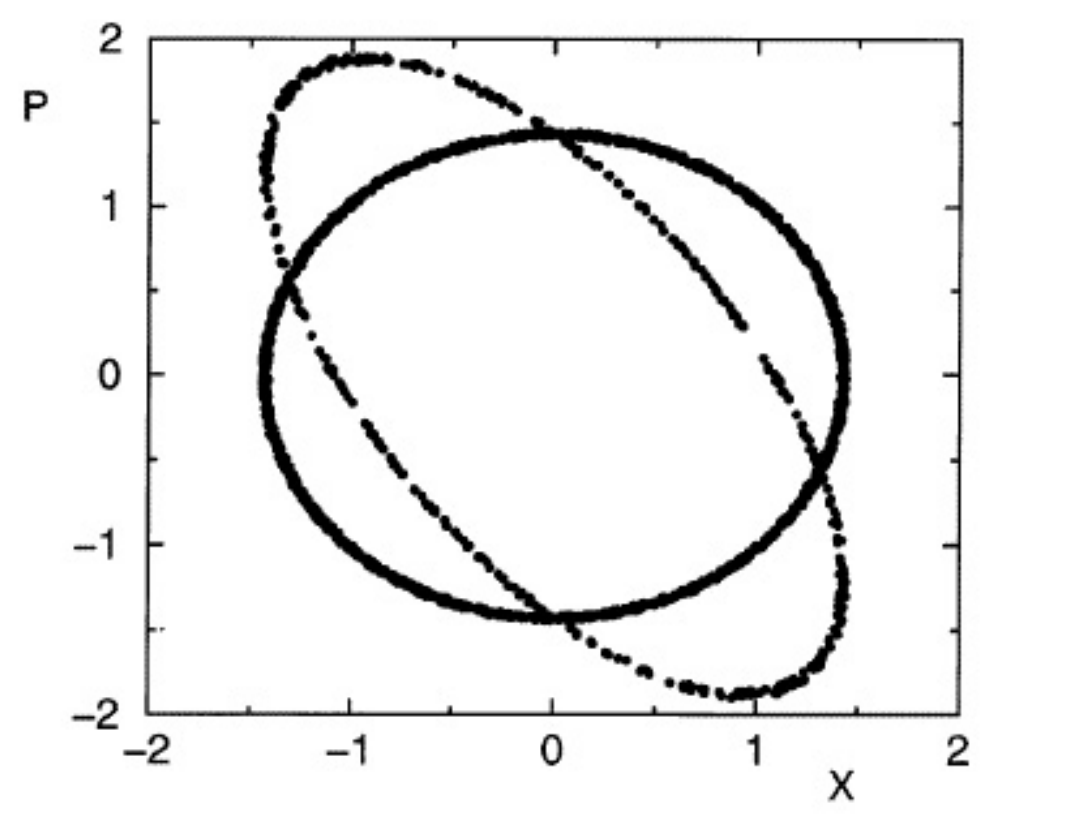}}
\subfigure{\includegraphics[scale=0.65]{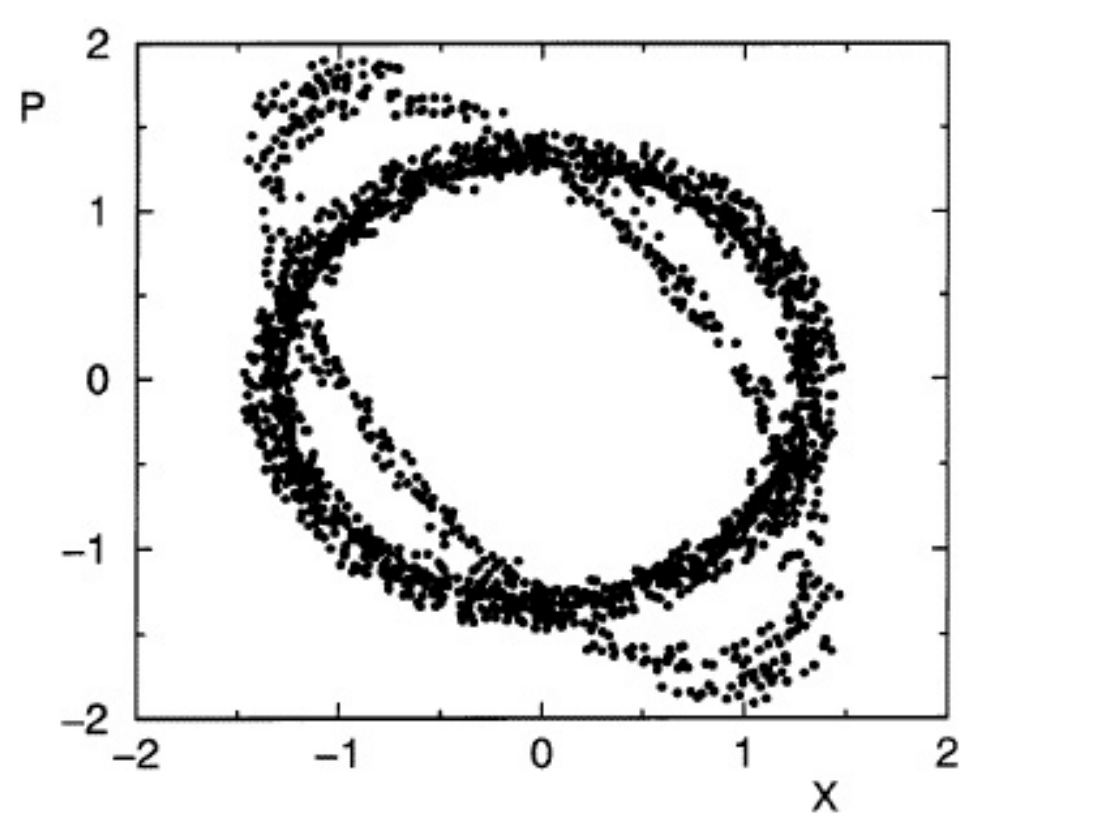}}
\end{center}
\caption{Nearly resonant states for a dimer ($N=2$) for $x_0=p_0=1$ and $Q=0.5$ with $\overline{\epsilon}_1=0$. Left: $k=11.399, \overline{\epsilon}_2=9.75885$. Right: $ k=11.390, \overline{\epsilon}_2=9.75885$. The length of sequence $\overline{\epsilon}$ in both cases is $1000$ (after \cite{KTI97}).}
\label{KP-fig4}
\end{figure}

The representation of the Kronig-Penney model in the form (\ref{KP-diagonal}) that is similar to that appearing for the Anderson model (see Eq.~(\ref{tb diagonal})), allows one to get the expression for the localization length in the vicinity of the resonant energies $E_{cr}$. For this, one needs simply to do the same as for the dimers in the Anderson model (see Section~\ref{5.1}). Specifically, one has to construct an effective map for two successive kicks of the single map (\ref{KP-map}) and neglect the correlations between the phases $\theta_{n+2}$ and $\theta_n$ near the resonance
$k=k_{cr}-\delta \approx k_{cr}$. Applying the resulting two-step map (\ref{r-n2}) obtained in Section~\ref{5.1} to the KP-model, one can estimate the Lyapunov exponent $\lambda$ for $\delta \ll 1$, by expanding $\lambda$ in the parameter $A_n \delta /\sin \mu$, with further averaging over $\theta_n$. This leads to the expression,
\begin{equation}
\label{KP-lyap}
\lambda \approx Q \frac{\delta^2 \overline{\epsilon}_2^2}{\mu^2 \sin^2\mu}\,.
\end{equation}
The direct numerical computation of the Lyapunov exponent, performed in Ref.~\cite{KTI97}, has shown a quite good correspondence between the data and analytical expression (\ref{KP-lyap}). The typical trajectories for nearly resonant values of the energy are shown in Fig.~\ref{KP-fig4}. In comparison with the exact resonance energy, the trajectories are no more bounded in the phase space; initially, for finite times $t_n \equiv n$ they slightly diffuse away from the origin of the phase space, and eventually they diverge exponentially fast.

\subsection{Amplitude disorder}
\label{8.2}

We focus here on the Kronig-Penney model with {\it weak} amplitude disorder. In this case the array of delta-barriers is periodic {\it on average}, namely, with a periodic arrangement of sites, $x_n=nd$, in Eq.~(\ref{KP-delta}), and slightly disordered amplitudes $U_n= U + u_n$. Here $ U = \langle U_n \rangle$ is the mean value of the potential, and $u_n$ is a weak random perturbation with the zero mean, $\langle u_n \rangle =0$, and variance $\sigma^2= \langle u_n^2 \rangle$. The average $\langle ... \rangle $ is performed over the realizations of disorder, and the statistical properties of $u_n$ are assumed to be independent on the site $n$. Our main interest is in the long-range correlations in the sequence $u_n$ that can lead to anomalous transport, specifically, to the emergence of mobility edges, as previously shown to occur in the one-dimensional Anderson model (see Section~\ref{5.4}).

When the disorder is absent, $U_n= U = const $, from Eq.(\ref{KP-diagonal}) one can obtain the dispersion relation,
\begin{equation}
\label{KP-disp}
2 \cos \gamma = 2 \cos \mu + \frac{ U}{k} \sin \mu \,,
\end{equation}
with $E=k^2$ and $\mu=kd$. Here $\gamma=\kappa d$ is nothing but the Bloch phase determined via the Bloch wave number $\kappa$ (in what follows, we put $d=1$). The latter emerges due to the Bloch theorem according to which the solution of Eq.~(\ref{KP-delta}) with the periodic potential, $U(x+d)=U(x)$, has the form,
\begin{equation}
\label{Bloch}
\psi (x) = e^ {i\kappa x} f(x)\,; \qquad f(x+d)=f(x)\,,
\end{equation}
where $f(x)$ is a periodic function determining the structure of eigenstates, therefore, $\psi(x+d)=e^{i\gamma} \psi(x)$. The dispersion relation (\ref{KP-disp}) plays the key role in the theory of transport, since it defines the values of $k=\sqrt E$ for which the Shr\"odinger equation has a solution. Specifically, the allowed energy bands are defined by the condition, $|\cos \gamma|< 1$.

One can see that Eq.~({\ref{KP-diagonal}) is equivalent to the Anderson model
(\ref{tb diagonal}) with the energy
\begin{equation}
E\rightarrow 2\cos
(kd)+\frac {U} {k} \sin (kd)
\label{newE}
\end{equation}
and random potential
\begin{equation}
\epsilon_n\rightarrow \frac{u_n}{k}\sin (kd).
\label{newPot}
\end{equation}
Therefore, the
inverse localization length, measured in the units of the period $d$, can be obtained directly from Eq.~(\ref{Lyap corr}),
\begin{equation}
\lambda= L_{loc}^{-1}(E)=\frac{\sigma^2}{8k^2}\frac{\sin ^2\mu}{\sin ^2\gamma }
{\cal K} (2\gamma )\,,
\label{five}
\end{equation}
where the phase $\,0\leq \gamma \leq \pi $ is given by the dispersion relation (\ref{KP-disp}) and $\mu=kd$. It is instructive that for $U \rightarrow 0$ (therefore, for $\gamma \approx \mu$) the Lyapunov exponent takes the form, $\lambda \approx \sigma^2/(8dk^2){\cal K}(2\mu)$, that is the result (\ref{1DCP-Lloc}) for weak scattering in continuous one-dimensional random potentials. In this case the whole energy range formally is equivalent to one allowed energy band, and the forbidden energy gaps are neglected. Note that in analogy with the Anderson model, the function ${\cal K} (2\gamma )$ in Eq.~(\ref{five}) is the Fourier transform of the pair correlator $K(m)=\langle u_n u_{n+m} \rangle/\sigma^2$ with the variance, $\sigma^2 =\langle u_n^2\rangle$,
\begin{equation} \label{W-KP}
{\cal K}(2\gamma)=1\;+\;2\sum_{m=1}^{\infty}K (m)\cos(2m\gamma)\,.
\end{equation}

The numerical data \cite{KI99,KI00} show that the expression (\ref{five}) works well in a wide range of the parameters of the Kronig-Penney model (\ref{KP-diagonal}). To demonstrate this, let us choose the dependence ${\cal K}(2\gamma)$ to be constant for $\gamma$ in the interval $[\pi/4, 3\pi/4]$ and zero otherwise. Then, with the algorithm discussed in Section~\ref{5.3} (see Eqs.~(\ref{colored}) and (\ref{G-phi})) one obtains the modulation function $G(m)$,
\begin{equation}
\label{pair-corr1}
G(m)=-\frac{2}{\pi m} \sin\left(\frac{\pi m}{2}\right)\,, \qquad G(0)=-1\,.
\end{equation}

The corresponding dependence
$\lambda=L_{loc}^{-1}(k)$ for two allowed zones $k_0 < k <2 \pi$ is shown in Fig.~\ref{KP-fig5}a by a smooth line. The location of mobility edges $k_c$ is defined by the relation,
\begin{equation}
\label{edges}
\cos{k_c} + ( U /2k_c) \sin{k_c}= \pm 1/\sqrt 2 \,,
\end{equation}
and the band edge $k_0$ is the positive root of the following relation,
\begin{equation}
\cos{k_0} + ( U /2k_0) \sin{k_0}= 1 \,.
\label{pos-edges}
\end{equation}
For comparison, in Fig.~\ref{KP-fig5}b
the complementary data are shown, where the regions of localized and
transparent states are interchanged. For this case the coefficients
$G(m)$ have opposite sign for $m\ne 0$,
\begin{equation}
\label{pair-corr2}
G(m)= \frac{2}{\pi m} \sin\left(\frac{\pi m}{2}\right)\,, \qquad G(0)=-1\,.
\end{equation}
The data for the second zone
are not shown because of $1/k^2$ decay of the inverse localization length.
The comparison with numerical data (broken curve) obtained by
the described approach appears to be quite good.
\begin{figure}[!ht]
\begin{center}
\includegraphics[width=11.6cm,height=5.5cm]{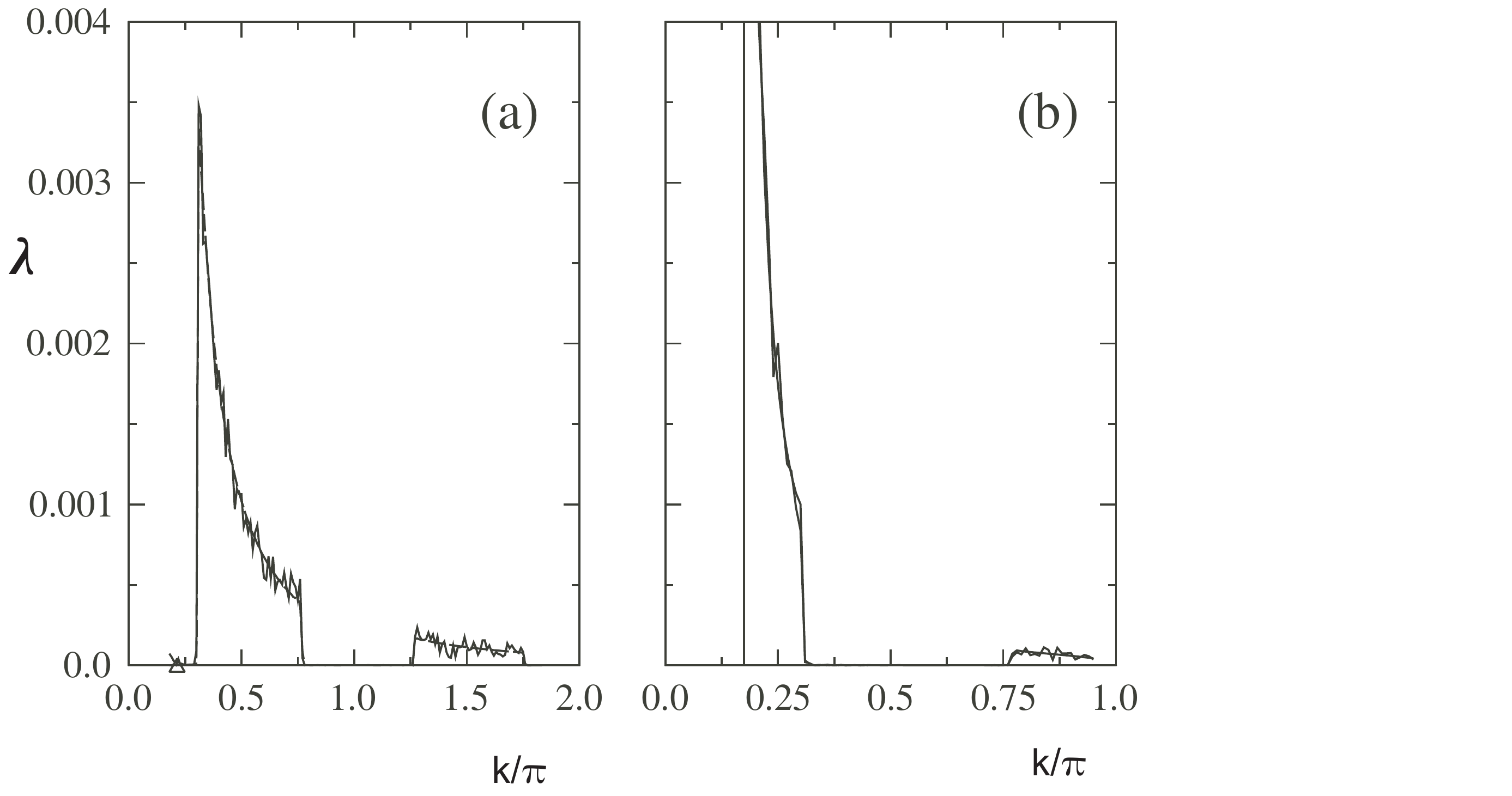}
\end{center}
\caption{Inverse localization length vs. rescaled wave number $k/\pi$
for two complementary potentials with: (a) Eq.~(\ref{pair-corr1}) and (b) Eq.~(\ref{pair-corr2}). Numerical data for $ U=0.3$,
$\sigma=0.1 $ and $n=1, ..., 10^5$ are shown
by broken lines. Analytical dependence (\ref{five}) is
presented by smooth lines. Triangle mark the position of the left edge
of the first allowed zone (after \cite{KI99}).}
\label{KP-fig5}
\end{figure}

In order to see how the correlations influence the transport properties, in Fig.~\ref{KP-fig6} we plot the transmission coefficient $T_N$ for the parameters used in Fig.~\ref{KP-fig5}. Here $T_N$ was computed according to the expression (\ref{eq:6}) with $N=10^3$. The results clearly demonstrate the emergence of the mobility edges.
\begin{figure}[!ht]
\begin{center}
\subfigure{\includegraphics[scale=0.39]{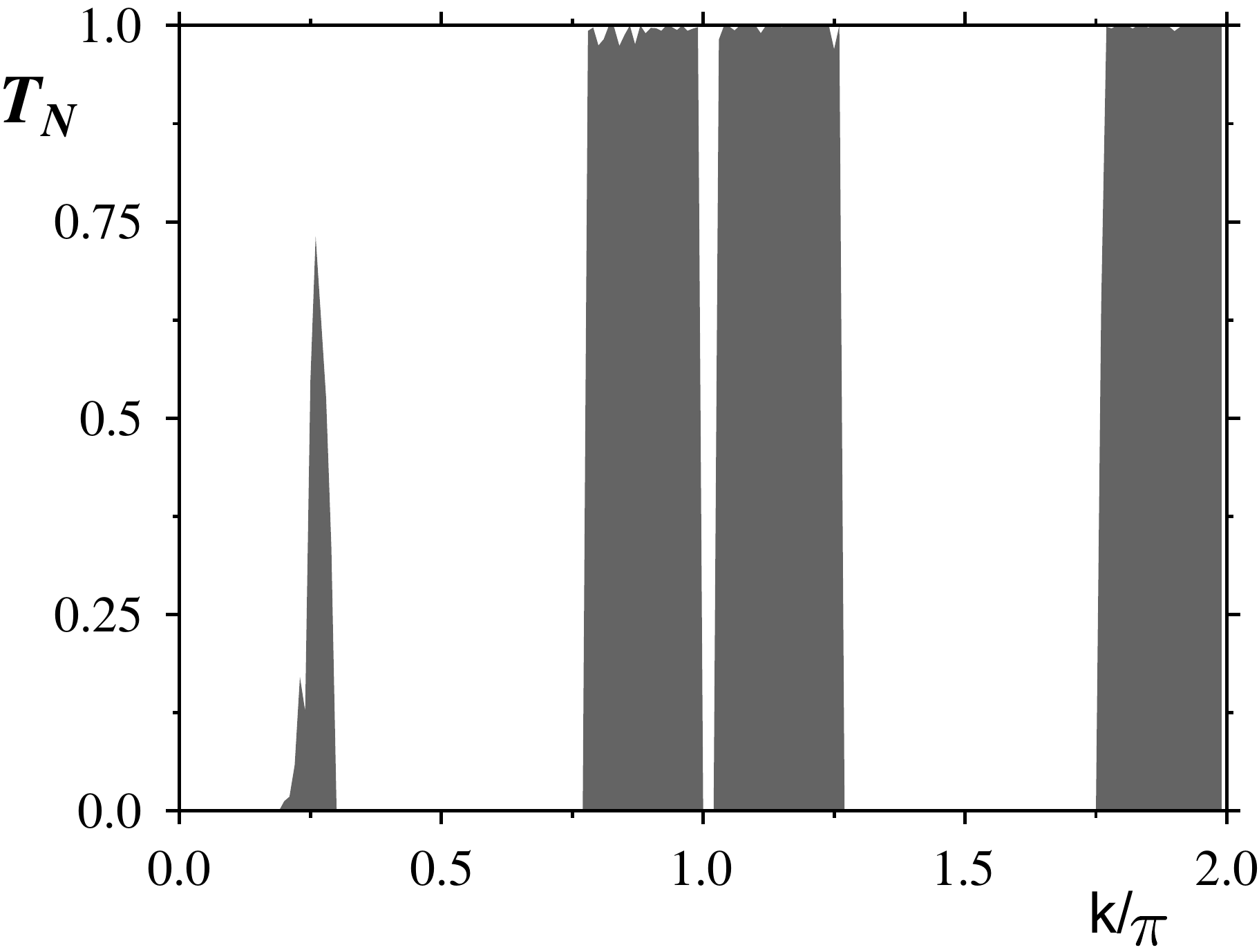}}
\subfigure{\includegraphics[scale=0.39]{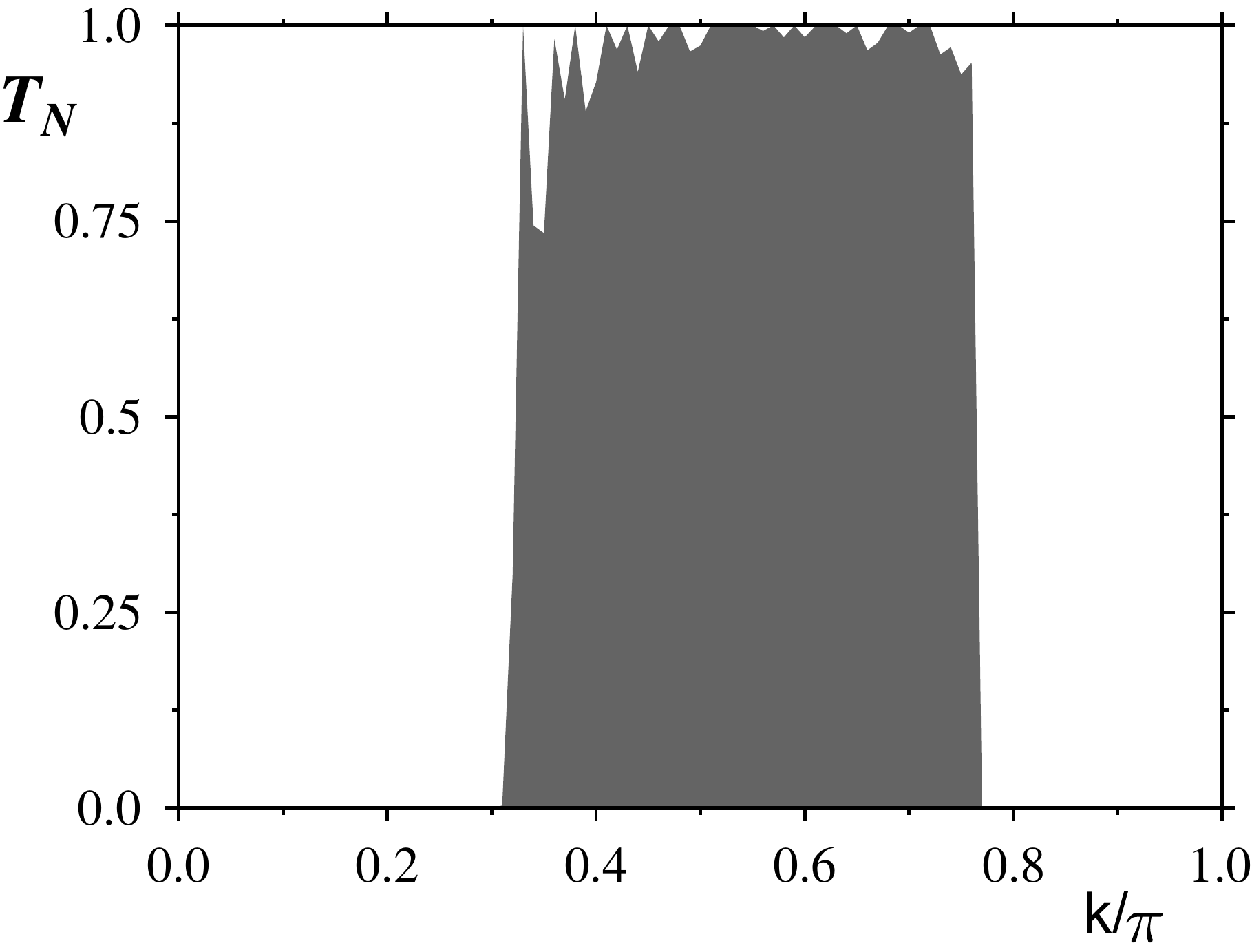}}
\end{center}
\caption{Transmission coefficient versus energy for data of Fig.~\ref{KP-fig5}. (a): correlations in the potential correspond to the dependence
${\cal K}(\mu)=0$ outside the interval $[\pi/4, 3\pi/4]$ in the first zone;
two zones are shown, $0 <k/\pi <2$, see Fig.~\ref{KP-fig5}a. (b): the same for the complementary dependence
${\cal K}(\mu)$ in the first zone only, see Fig.~\ref{KP-fig5}b. The length of sequence $U_n$ in both cases is $N=1000$ (after \cite{KI00}).}
\label{KP-fig6}
\end{figure}

\subsection{Amplitude versus positional disorder}
\label{8.3}

\subsubsection{Generalized expression for localization length}
\label{8.3.1}

Let us consider now the general case of a weak disorder in the Kronig-Penney model with delta-barriers \cite{HIT08}. Specifically, we assume that {\it both} amplitudes, $U_n=U+u_n$, of the barriers and spacings $x_{n+1} - x_{n} = d + \Delta_{n}$ between their positions in Eq.~(\ref{KP-delta}) are slightly perturbed. Moreover, our analysis is not restricted by the white noise disorder; it allows for statistical correlations in the amplitude and position disorders, as well as for cross-correlations between $u_n$ and $\Delta_n$. Here the variables $u_{n}$ represent
fluctuations of the barrier strength around the mean value $U$. As before, we assume the weakness of disorder for which both
variables $u_n$ and $\Delta_{n}$ have zero average, and small variances,
\begin{equation}
\langle u_n \rangle = 0\,, \qquad \langle u_{n}^{2} \rangle \ll U^{2}\,
\label{small-u}
\end{equation}
and
\begin{equation}
\langle \Delta_{n} \rangle = 0 \, \qquad k^{2} \langle \Delta_{n}^{2} \rangle \ll 1\,.
\label{small-delta}
\end{equation}
We remark that the condition
$k^{2} \langle d_{n}^{2} \rangle \ll 1$ implies that the energy $E=k^2$ must be slower than $1/\langle \Delta_{n}^{2} \rangle$.
Apart from the first two moments of the random variables $\Delta_{n}$ and
$u_{n}$, we define the binary correlators,
\begin{equation}
K_{u} (m) = \langle u_{n} u_{n+m} \rangle / \langle u_{n}^{2} \rangle \,, \,\,\,\,
K_{\Delta} (m) = \langle \Delta_{n} \Delta_{n+m} \rangle / \langle \Delta_{n}^{2} \rangle \,, \,\,\,\,
K_{u,\Delta} (m) = \langle u_{n} \Delta_{n+m} \rangle /
\langle u_{n} \Delta_{n} \rangle \, .
\label{bincor}
\end{equation}
We not attribute specific forms to the correlators
$K(m)$, however, we note that they depend only on the index
$m$ because of the spatial homogeneity and isotropy of the potential. For this reason, they are even functions of $m$, and the average $\langle ... \rangle $ over disorder is equivalent to the average along the ``coordinate" $n$. However, the third correlator, $K_{u,\Delta} (m)$, can be non-even, therefore, the model can be considered as only partially isotropic. In this case the final expression for the Lyapunov exponent has to be slightly modified, for details see Ref.~\cite{HIT10a}.

We remind that Eq.~(\ref{KP-delta}) can be treated as
the wave equation for electromagnetic waves in a single-mode
waveguide with the wave number $k=\omega/c$. Therefore, our results
are equally applicable to the classical scattering in optical and
microwave devices of the Kronig-Penney type with correlated
disorder. Our model is also equivalent to the classical oscillator with
a parametric perturbation constituted by a succession of $\delta$-kicks
whose amplitudes and time-dependence are determined by $U_n$ and $\Delta_n$ (see discussion in Section~\ref{6}). The latter correspondence allows one to cast Eq.~(\ref{KP-delta}) in the form of a classical linear oscillator with a noisy frequency,
\begin{equation}
\ddot{x} + \left[ k^{2} - \sum_{n=-\infty}^{\infty} U_{n}
\delta \left( t - t_{n} \right) \right] x = 0\,.
\label{dyneq}
\end{equation}
As one see, in such a description the coordinate $x\equiv \psi$ plays the role of the $\psi$-function, and time $t$ is associated with the positions of delta-barriers, $t \equiv x$.

The analysis performed in Ref.~\cite{HIT08} is based on the Hamiltonian map approach (see Section~\ref{4.2}) according to which the global properties of eigenstates of the
Kronig-Penney model can be analyzed by exploring the time evolution of
the kicked oscillator described by the dynamical
equation~(\ref{dyneq}). Such a dynamical approach considers the
wave equation (\ref{KP-delta}) as an initial-value problem and can be
treated as a modification of the transfer matrix approach.
Integrating the dynamical equation~(\ref{dyneq}) between two successive
kicks, one obtains the map,
\begin{equation}
\begin{array}{ccl}
x_{n+1} & = & \left[ U_{n}k^{-1} \sin \left( \mu + \mu_{n} \right)
+ \cos \left( \mu + \mu_{n} \right) \right] x_{n}
+ k^{-1} \sin \left( \mu + \mu_{n} \right) p_{n} \\
p_{n+1} & = & \left[ U_{n} \cos \left( \mu + \mu_{n} \right) -
k \sin \left(\mu + \mu_{n} \right) \right] x_{n}
+ \cos \left( \mu + \mu_{n} \right) p_{n} \\
\end{array}
\label{hammap1}
\end{equation}
where $\mu=kd$ and $\mu_{n}=k\Delta_{n}$, and the coordinate $x_{n}$ and momentum $p_{n}\equiv \dot{x}_n$
refer to the instant before the $n-$th kick.

The evolution of the dynamical map~(\ref{hammap1}) can be analyzed
as follows. First, we make a weak disorder expansion of
Eq.~(\ref{hammap1}), keeping all terms up to second order.
The expansion is straightforward; the resulting equations, however,
are lengthy and we omit them here.
As a second step, we perform the canonical transformation $(x_{n}, p_{n})
\rightarrow (X_{n}, P_{n})$, such that the unperturbed motion reduces to
a simple rotation in the phase space of the new variables.
Such a trick allows one to eliminate the effect of the periodic kicks
with constant amplitudes $U$. This can be done with the following
canonical transformation,
\begin{equation}
\begin{array}{ccl}
x_{n} & = & \alpha \cos (\mu/2) X_{n} +
(k \alpha)^{-1} \sin (\mu/2) P_{n} \\
p_{n} & = & - k \alpha \sin (\mu/2) X_{n} +
\alpha^{-1} \cos (\mu/2) P_{n}
\end{array}
\label{canonic}
\end{equation}
where the parameter $\alpha$ is defined by the relation
\begin{equation}
\alpha^{4} = k^{-2}
\frac{\sin \mu - \frac{U}{2k} \left( \cos \mu - 1 \right)}
{\sin \mu - \frac{U}{2k} \left( \cos \mu + 1 \right)} .
\label{alpha-4}
\end{equation}
Note that due to the transformation~(\ref{canonic}), the new
variables $X_{n}$ and $P_{n}$ have the same dimension.

In the absence of disorder, i.e., for $u_{n} = 0$ and $\Delta_{n} = 0$,
the rotation angle $\gamma$ between successive kicks in the variables $(X_n, P_n$) is determined by the dispersion relation (\ref{KP-disp}) with $\gamma=\kappa d$.
It should be pointed out that the transformation~(\ref{canonic}) is well-defined for all values of the rotation angle $\gamma$ other than $\gamma = 0$ and $\gamma = \pi$ for which $\alpha$ either
vanishes or diverges. In other words, our approach fails at the
center and at the edges of the first Brillouin zone, i.e., at the edges of the allowed energy bands of the Kronig-Penney model. However, the approach works well {\it near} these critical points, provided the disorder is sufficiently weak.

In order to proceed further, it is useful to pass to the action-angle
variables $(J_{n}, \theta_{n})$, with another canonical transformation,
\begin{equation}
\begin{array}{ccc}
X_{n} = \sqrt{2 J_{n}} \sin \theta_{n}, & & P_{n} = \sqrt{2 J_{n}}
\cos \theta_{n}
\end{array}
\label{XY-transform}
\end{equation}
and represent the Hamiltonian map~(\ref{hammap1}) in terms of the new variables. Leaving aside the mathematical details, one can write,
\begin{equation}
\begin{array}{ccl}
J_{n+1} & = & D_{n}^{2} J_{n} \\
\theta_{n+1} & = & \theta_{n} + \gamma - \frac{1}{2}
\left[ 1 - \cos \left( 2 \theta_{n} + \gamma \right) \right] \tilde{u}_{n}
+ \frac{1}{2}
\left[ \upsilon - \cos \left( 2 \theta_{n} + 2 \gamma \right) \right]
\tilde{\Delta}_{n} \\
\end{array}
\label{hammap5}
\end{equation}
where
\begin{equation}
\begin{array}{ccl}
D_{n}^{2} & = & 1 + \sin \left( 2 \theta_{n} + \gamma \right) \tilde{u}_{n} -
\sin \left( 2 \theta_{n} + 2 \gamma \right) \tilde{\Delta}_{n}
+ \frac{1}{2}
\left[ 1 - \upsilon \cos \left( 2 \theta_{n} + 2 \gamma \right) \right]
\tilde{\Delta}_{n}^{2} \\
& + & \frac{1}{2}
\left[ 1 - \cos \left( 2 \theta_{n} + \gamma \right) \right]
\tilde{u}_{n}^{2}
- \left[ \cos \gamma - \cos \left( 2 \theta_{n} + 2 \gamma \right) \right]
\tilde{u}_n \tilde{\Delta}_{n}.
\end{array}
\label{dnsquared}
\end{equation}
Here
\begin{equation}
\label{upsilon}
\upsilon = \frac{k \sin \gamma}{U} \left( k \alpha^{2} + \frac{1}{k \alpha^{2}}\right)
\end{equation}
and we have introduced the rescaled random variables,
\begin{equation}\label{rescaling}
\tilde{u}_{n} = \frac{\sin\mu}{k\sin\gamma}\,u_{n}\,,\qquad
\tilde{\Delta}_{n}=\frac{U}{\sin\gamma}\Delta_{n}\,.
\end{equation}
In Eqs.~(\ref{hammap5}) and~(\ref{dnsquared}) we have kept only the terms of the weak disorder expansion which are necessary to compute the localization length within the second order approximation (the more general expression can be found in Ref.~\cite{HIT10a}). We remark that the angle variable evolves independently of the action variable.

Note that the action $J_n$ can be expressed in terms of the radius $R_n$ of trajectory in the phase space $(X_n,P_n)$ according to the relation $R_n=\sqrt {2 J_n}$. Therefore, the Lyapunov exponent $\lambda$ can be computed in the same way as for the tight-binding Anderson model, see Eq.~(\ref{locAnd}) in Section~\ref{4.2},
\begin{equation}
\lambda = \frac{1}{2d} \langle \ln D_{n}^{2} \rangle \,,
\label{lyap1}
\end{equation}
where $D_n$ is determined by Eq.~(\ref{dnsquared}).
By expanding the logarithm in Eq.~(\ref{lyap1}), one gets \cite{HIT08,HIT10},
\begin{equation}
\begin{array}{ccl}
\lambda & = & \frac{1}{2d} \Big \langle \left\{
\sin \left( 2 \theta_{n} + \gamma \right) \tilde{u}_{n} -
\sin \left( 2 \theta_{n} + 2 \gamma \right) \tilde{\Delta}_{n}
\right. \\
& + & \frac{1}{4}\left[ 1 -
2 \upsilon \cos\left( 2 \theta_{n} + 2 \gamma \right)+
\cos \left( 4 \theta_{n} + 4\gamma \right) \right] \tilde{\Delta}_{n}^{2} \\
& + & \frac{1}{4} \left[ 1 - 2 \cos \left( 2 \theta_{n} + \gamma \right) +
\cos \left( 4 \theta_{n} + 2 \gamma \right) \right] \tilde{u}_{n}^{2} \\
& - & \frac{1}{2}
\left[ \cos \gamma - 2 \cos \left( 2 \theta_{n} + 2 \gamma \right)
\right.
+ \left. \left.
\cos \left( 4 \theta_{n} + 3\gamma\right) \right]
\tilde{u}_{n} \tilde{\Delta}_{n} \right\} \Big \rangle . \\
\end{array}
\label{logd}
\end{equation}
Now, in order to obtain the Lyapunov exponent $\lambda$, we have to
perform the average over the phase $\theta_n$ and random variables
$u_n$ and $\Delta_n$. To the second order of perturbation theory,
one can neglect the correlations between $\theta_{n}$ and the quadratic
terms $\tilde{u}_n^{2}$, $\tilde{\Delta}_{n}^{2}$, and
$\tilde{u}_{n} \tilde{\Delta}_{n}$. Hence, for the summands in
Eq.~(\ref{logd}) that contain these quadratic terms, one can compute
separately the averages over $\theta_{n}$ and over the random variables
$u_n$ and $\Delta_n$.

In analogy with the Anderson model (see Section~\ref{4.2.1}), it can be shown that for our purposes it is safe to assume that the invariant measure of the phase is a flat
distribution, $\rho(\theta)=1/2\pi$. The assumption holds for all values of $\gamma$, except for $\gamma=\pi/2$ where a small modulation of the invariant measure results in an anomaly for the localization length.
The situation with these values of $\gamma$ is similar to that known for
the standard 1D Anderson model at the center of the energy band, and the
correct expression for $\lambda$ can be obtained following the approach
of Ref.~\cite{IRT98} (see details in Section~\ref{4.2.2}).

It should be stressed that the weak modulations of $\rho(\theta)$ arise also for other ``resonant'' values, $\gamma=s\pi/r$, with $s$ and $r$ integers prime with each other and $r > 2$. However, these modulations do not influence the value of $\lambda$,
because the expression to be averaged in Eq.~(\ref{logd}) has no harmonics higher than $4 \theta$. Thus, our further analysis is valid for all values of $\gamma$ except $\gamma=0$ and $\gamma = \pi$ (i.e., at the edges of the energy bands), and $\gamma=\pi/2$.

After averaging, the expression for the Lyapunov exponent takes the form,
\begin{equation}
\lambda = \displaystyle
\frac{1}{8d}
\left[ \langle \tilde{u}_{n}^{2} \rangle +
\langle \tilde{\Delta}_{n}^{2} \rangle -
2 \langle \tilde{u}_{n} \tilde{\Delta}_{n} \rangle \cos \gamma \right]
+ \displaystyle
\frac{1}{2d} \langle \tilde{u}_{n} \sin(2 \theta_n + \gamma) \rangle -
\frac{1}{2d} \langle \tilde{\Delta}_{n}
\sin \left( 2 \theta_{n} + 2 \gamma \right) \rangle. \\
\label{lyap2}
\end{equation}
One can see that in contrast with the previously studied models with one type of disorder, here we have a much more complicated situation characterized by two types of correlated disorder,
$u_n$ and $\Delta_n$. Moreover, we intend to take into account possible cross-correlations between $u_n$ and $\Delta_n$ as well. In order to compute the binary correlators in Eq.~(\ref{lyap2}),
we generalize the method used in Ref.~\cite{IK99} and discussed in Section~\ref{5.2}. Specifically, we introduce the correlators
$r_{k} = \langle \tilde{u}_{n} \exp \left( i 2 \theta_{n-k} \right) \rangle$ and $s_{k}=\langle \tilde{\Delta}_{n} \exp \left( i 2 \theta_{n-k} \right)\rangle$. Both correlators satisfy the recursive relations that can be obtained by
substituting the angular map of Eq.~(\ref{hammap5}) into the definitions of
$r_{k-1}$ and $s_{k-1}$. The recursive relations allow one to obtain the
correlators $r_{0}$ and $s_{0}$, whose imaginary parts represent the
noise-angle correlators in Eq.~(\ref{lyap2}). As a result, we
arrive at the final expression for the Lyapunov exponent,
\begin{equation}
\begin{array}{ccl}
\lambda & = & \displaystyle
\frac{1}{8d}\left[ \langle \tilde{u}_{n}^{2} \rangle {\cal K}_{u} +
\langle \tilde{\Delta}_{n}^{2} \rangle {\cal K}_{\Delta} -
2 \langle \tilde{u}_{n} \tilde{\Delta}_{n} \rangle \cos \gamma {\cal K}_{u,\Delta} \right] \\
& = & \displaystyle
\frac{\sin^{2}\mu}{8 dk^{2}\sin^{2}\gamma}
\langle u_{n}^{2} \rangle {\cal K}_{u} +
\frac{U^{2}}{8 d \sin^{2}\gamma} \langle \Delta_{n}^{2} \rangle {\cal K}_{\Delta}
- \displaystyle
\frac{1}{4d} \frac{U \sin \mu}{k \sin^{2} \gamma} \cos \gamma \;
\langle u_{n} \Delta_{n} \rangle {\cal K}_{u,\Delta}
\end{array}
\label{invloc}
\end{equation}
where the functions
\begin{equation}
{\cal K}_{i} \equiv {\cal K}_{i} \left( 2 \gamma \right) = 1 + 2 \sum_{m=1}^{\infty}
K_{i}(m) \cos(2\gamma m) \label{ftcorr}
\end{equation}
are the $2\gamma-$harmonics of the Fourier transform of the binary correlators $K_{i} (m)$, see Eq.~(\ref{bincor}). We should stress that Eq.~(\ref{invloc}) has been derived without the assumption of the relation $E \gg U$. Therefore, we take into account the tunneling effects both for $E < U$ and $E > U$. The only constraint is the weakness of both types of disorder, see Eqs.~(\ref{small-delta}) and (\ref{small-u}).

Let us discuss the general expression (\ref{invloc}) for
particular cases. One can see that if there are no correlations between
$u_{n}$ and $\Delta_{n}$ the third term for the Lyapunov exponent is absent, since $\langle u_{n} \Delta_{n} \rangle=0$. In this case there is interplay between the first and second terms in Eq.(\ref{invloc}), and one can establish which one is more important in comparison with the other, depending on the model parameters. However, both terms are positive, this means that there is an enhancement of the localization when two kinds of disorder are present, in comparison with the case of one disorder only. Two limit cases here can be easily obtained, when one kind of disorder, either amplitude or positional one is absent. For example, in the case of the amplitude disorder only, $\Delta_{n}=0$, we have the expression (\ref{five}) discussed in Section~\ref{8.2}. In the other limit case of the positional disorder only, $u_{n}$, we arrive at the expression obtained in Ref.~\cite{IKU01},
\begin{equation}
\label{IKU01}
L_{loc}^{-1}=\frac{U^{2}\langle \Delta_{n}^{2} \rangle}{8 d \sin^{2}\gamma} {\cal K}_{\Delta}(2\gamma)\,.
\end{equation}

The expression (\ref{invloc}) indicates that the correlation between two kinds of disorder, $u_{n}$ and $\Delta_{n}$, can either enhance or suppress the localization in comparison with the case of uncorrelated disorders. This depends on whether the correlation term $\langle u_{n} \Delta_{n} \rangle $ is positive or negative. One can imagine the situation when both the disorders are white-noise type, ${\cal K}_u={\cal K}_\Delta=1$, however, they can be either correlated, ${\cal K}_{u,\Delta}=1$ or anti-correlated, ${\cal K}_{u,\Delta}=-1$. Depending on the sign of the correlations, the localization is either enhanced or suppressed.

The most important conclusion is that specific long-range correlations can make the Fourier transforms~(\ref{ftcorr}) vanish in prescribed energy windows, so that
the Lyapunov exponent~(\ref{invloc}) also vanishes in the same energy
intervals. A method for the construction of random potentials with given binary
correlators $K_{i} (m)$ is described in Section~\ref{5.3}, and tested in a number of papers, see, e.g. \cite{IK99,KI99,KIKS00,IKU01,KIKSU02,KIK08}.
Following the same approach, one can use Eq.~(\ref{invloc}) as
a starting point for the fabrication of devices with prescribed anomalous
transport characteristics.

As is pointed out above, the expression (\ref{invloc}) is not valid at the band edges, where $\gamma= \pi$, as well as at the band center, $\gamma= \pi/2$. In these cases the distribution of phases $\theta_n$ is not flat, and one has to take into account the modulation of $\theta_n$.

\subsubsection{Interplay between two disorders}
\label{8.3.2}

Let us now discuss the problem of how to construct two random sets of variables $u_n$ and $\Delta_n$ with predefined values of the power spectra~(\ref{ftcorr}) in such a way that the Lyapunov exponent (\ref{invloc}) vanishes in some interval of energy. A non-trivial point is that by defining two auto-correlators $K_{u}(m)$ and $K_{\Delta}(m)$, the cross-correlator $K_{u,\Delta}(m)$ can not be chosen independently. Thus, we deal with the ``inverse'' problem which has not a
unique solution. One of the possible algorithms for solving this problem is suggested in Ref.~\cite{HIT10}. Following this suggestion, we consider two sequences $\{X_{n}^{(1)}\}$ and
$\{X_{n}^{(2)}\}$ of independent random variables with the zero mean and delta-like pair correlator,
\begin{equation}
\langle X_{n}^{(i)} \rangle = 0 \,, \qquad
\langle X_{n}^{(i)} X_{n'}^{(j)} \rangle = \delta_{ij} \delta_{nn'}
\label{X-und-Y}
\end{equation}
for $i,j=1,2$ and integer $n,n'$. In terms of these variables one can construct two cross-correlated sequences,
\begin{equation}
\begin{array}{ccl}
Y_{n}^{(1)} & = & X_{n}^{(1)} \cos \eta + X_{n}^{(2)} \sin \eta \\
Y_{n}^{(2)} & = & X_{n}^{(1)} \sin \eta + X_{n}^{(2)} \cos \eta \\
\end{array}
\label{iwn}
\end{equation}
with a real parameter $\eta$ that determines the degree of
inter-correlation between the $Y$ variables.
The mean and variance of these variables are the following,
\begin{equation}
\langle Y_{n}^{(i)} \rangle = 0\,, \qquad
\langle Y_{n}^{(i)} Y_{m}^{(j)} \rangle = \delta_{nm} \left[
\delta_{ij} + \left( 1 - \delta_{ij} \right) \sin \left( 2 \eta \right)
\right]\,.
\label{Y-and-Y-mean}
\end{equation}
One can see that the range of variation of $\eta$
is the interval $[-\pi/4, \pi/4]$, with $\eta = \pi/4$ corresponding to maximal
correlations between the $Y_{n}^{(1)}$ and $Y_{n}^{(2)}$ variables, and
$\eta = -\pi/4$ to maximal anti-correlations. Note that the value $\eta = 0$ corresponds to the absence of cross-correlations.

As we now can control cross-correlations between $\{Y_{n}^{(1)}\}$ and $\{Y_{n}^{(2)}\}$, one needs to find out how to generate a colored noise out
of white-noise sequences. To do this, we introduce two sets of unknown coefficients (or modulation functions, see Section \ref{5.3}),
$G_{u}(m) = G_{u}(-m)$ and $G_{\Delta}(m) = G_{\Delta}(-m)$, and
express the $u_{n}$ and $\Delta_{n}$ random variables as a convolution of the following form (compare with Eq.~(\ref{colored})),
\begin{equation}
u_{n} = \displaystyle
\sum_{m=-\infty}^{\infty} G_{u}(m) Y_{n-m}^{(1)} \,,\,\,\,\,\,\,\,\,
\Delta_{n} = \displaystyle
\sum_{m=-\infty}^{\infty} G_{\Delta}(m) Y_{n-m}^{(2)} .
\label{convol}
\end{equation}
Clearly, the mean values of $u_{n}$ and $\Delta_{n}$ vanish, $
\langle u_{n} \rangle = 0$ and $\langle \Delta_{n} \rangle = 0$. As for the binary correlators, they take the form,
\begin{equation}
\begin{array}{ccl}\label{cross-bin}
K_u(m) & = & \displaystyle
\sum_{n=-\infty}^{\infty} G_{u}(n) G_{u}(n+m)\,, \,\,\,\,\,\,
K_\Delta (m) =
\sum_{n=-\infty}^{\infty} G_{\Delta}(n) G_{\Delta}(n+m)\,, \\
K_{u,\Delta}(m)
& = & \displaystyle \left( \sqrt{\langle u_{n}^{2} \rangle } \sqrt{\langle \Delta_{n}^{2} \rangle } /\sqrt{\langle u_{n} \Delta_{n} \rangle } \right)
\sum_{n=-\infty}^{\infty} \sin \left( 2 \eta \right)G_{u}(n) G_{\Delta}(n+m) .
\end{array}
\end{equation}
From these relations one can easily obtain the normalization conditions,
\begin{equation}\label{norm-for-G}
\sum_{m=-\infty}^{\infty} G_{u}^2(m)= 1 \,, \,\,\,\,\,\,\,\,
\sum_{m=-\infty}^{\infty} G_{\Delta}^2(m)= 1 .
\end{equation}
As a result, the coefficients
$G_{u}(m)$ and $G_{\Delta}(m)$ can be expressed as follows,
\begin{equation}
\begin{array}{ccl}
G_{u}(m) & = & \displaystyle
\frac{1}{\pi} \int_{0}^{\pi} {\cal K}_{u}(k)
\cos \left( 2 m k \right) dk \\
G_{\Delta}(m) & = & \displaystyle
\frac{1}{\pi} \int_{0}^{\pi} {\cal K}_{\Delta}(k)
\cos \left( 2 m k \right) dk . \\
\end{array}
\label{ab}
\end{equation}


Inserting the coefficients~(\ref{ab}) into Eq.~(\ref{convol}), we arrive at the solution of our problem. Specifically, this procedure makes possible to construct two random sequences, $\{ u_{n} \}$ and $\{ \Delta_{n} \}$ with the predefined
power spectra ${\cal K}_{u}$ and ${\cal K}_{\Delta}$.
It should be noted, however, that with this prescription the power
spectrum ${\cal K}_{u,\Delta}$ cannot take a preassigned form of its own, because
it is completely defined in terms of ${\cal K}_{u}$ and ${\cal K}_{\Delta}$. This
can be seen by considering that, once the coefficients $G_{u}(m)$ and
$G_{\Delta}(m)$ are obtained as functions of ${\cal K}_{u}$ and ${\cal K}_{\Delta}$
via Eq.~(\ref{ab}), the power spectrum ${\cal K}_{u,\Delta}$ can be written as
\begin{equation}
{\cal K}_{u,\Delta} =
\frac{\sqrt{ {\cal K}_{u} {\cal K}_{\Delta}}}{\displaystyle
\sum_{m=-\infty}^{\infty} G_{u}(m) G_{\Delta}(m)} .
\label{w3}
\end{equation}

Note that the coefficients $G_{u}(m)$ and $G_{\Delta}(m)$ determine the
cross-correlator $\langle u_{n} \Delta_{n+m} \rangle $ only {\em partially}, see Eq.~(\ref{cross-bin}). As one can see, the free parameter $\eta$ plays an important role in the definition of the cross-correlator $\langle u_{n} \Delta_{n+m} \rangle $. Since the third term in the expression~(\ref{invloc}) is proportional
to ${\cal K}_{u,\Delta}(2\gamma)$, therefore, to the factor $\langle u_{n} \Delta_{n} \rangle$, its value is not completely determined by the choice of ${\cal K}_{u}(2\gamma)$ and ${\cal K}_{\Delta}(2\gamma)$ and also depends on the parameter $\eta$ which sets the degree of cross-correlations between the two disorders.
Taking into account Eqs.~(\ref{w3}) and~(\ref{cross-bin}), one can write the power spectrum ${\cal K}_{u,\Delta}$ in the form,
\begin{equation}
{\cal K}_{u,\Delta} = \frac{\sqrt{\langle u_{n}^{2} \rangle } \sqrt{\langle \Delta_{n}^{2} \rangle}}{ \langle u_{n} \Delta_{n} \rangle}\,
 {\cal K}_{u} {\cal K}_{\Delta}
{\displaystyle } \sin \left( 2\eta \right).
\label{w-u-delta}
\end{equation}
Inserting this result in Eq.~(\ref{invloc}), one obtains,
\begin{equation}
\lambda = \frac{1}{8d} \left[ \langle \tilde{u}_{n}^{2} \rangle
{\cal K}_{u}(2\gamma) + \langle \tilde{\Delta}_{n}^{2} \rangle {\cal K}_{\Delta}(2\gamma) -
2 \sqrt{\langle \tilde{u}_{n}^{2} \rangle
\langle \tilde{\Delta}_{n}^{2} \rangle {\cal K}_{u}(2\gamma) {\cal K}_{\Delta}(2\gamma)} \;
\cos \gamma \sin \left( 2 \eta \right) \right] .
\label{ll}
\end{equation}
With the use of this expression one can practically design the mobility edges in the Kronig-Penney model with the correlated amplitude and positional disorders.

\subsubsection{Examples of mobility edges}
\label{8.3.3}

The validity of the expression (\ref{ll}) has been numerically tested in Ref.~\cite{HIT10}. In order to manifest an emergence of mobility edges with two correlated disorders, the random sequences $\{ u_{n} \}$ and $\{ \Delta_{n} \}$ have been created to produce four mobility edges in the first Brillouin zone of the Bloch number $\kappa=\gamma/d$,
\begin{equation}
\begin{array}{cccc}\label{k1234}
\displaystyle
\kappa_{1} = c_{1} \frac{\pi}{2d}, &
\displaystyle
\kappa_{2} = c_{2} \frac{\pi}{2d}, &
\displaystyle
\kappa_{3} = \frac{\pi}{d} - c_{2} \frac{\pi}{2d}, &
\displaystyle
\kappa_{4} = \frac{\pi}{d} - c_{1} \frac{\pi}{2d} , \\
\end{array}
\end{equation}
with $0< c_{1} < c_{2} < 1$.

As an example, here we consider the case for which the inverse localization length (Lyapunov exponent)
vanishes when the Bloch vector lies in the intervals,
\begin{equation}
\left[0,\kappa_{1} \right], \quad \left[ \kappa_{2},\kappa_{3} \right], \quad
\left[\kappa_{4}, \frac{\pi}{d} \right] .
\end{equation}
In this case the power spectra ${\cal K}_{u}(2\gamma)$ and ${\cal K}_{\Delta}(2\gamma)$ must vanish
in the same intervals (with our construction of $\{u_{n}\}$ and
$\{ \Delta_{n} \}$, this automatically ensures that ${\cal K}_{u,\Delta}(2\gamma) = 0$ in
the same intervals).
In the complementary regions of the Brillouin zone
the power spectrum is assumed to be flat. The constant value of ${\cal K}_{i}(2\gamma)$ in the intervals
$\left[ \kappa_{1},\kappa_{2} \right]$ and $\left[ \kappa_{3}, \kappa_{4} \right]$
can be obtained from the normalization condition
\begin{equation}
\int_{0}^{\pi} {\cal K}_{i}(2\gamma) d\gamma = \pi,
\label{psnorm}
\end{equation}
which follows from the fact that $K_{i}(0) = 1$.
Thus, we have,
\begin{equation}
{\cal K}_{u}(2\gamma) = {\cal K}_{\Delta}(2\gamma) = \left\{ \begin{array}{ccl}
\displaystyle
\frac{1}{c_{2} - c_{1}} &
\mbox{ if } & \kappa \in \left[ \kappa_{1}, \kappa_{2} \right] \mbox{ and }
\kappa \in \left[ \kappa_{3}, \kappa_{4} \right] \\
0 & & \mbox{ otherwise }.
\end{array} \right.
\label{powspe}
\end{equation}
Note that the power spectra~(\ref{powspe}) correspond
to the long-range binary correlators,
\begin{equation}
K_{u}(m) = K_{\Delta}(m) = \frac{1}{c_{2} - c_{1}}\frac{1}{\pi m}
\left[ \sin \left( \pi c_{2} m \right) -
\sin \left( \pi c_{1} m \right) \right] .
\label{lrbincor0}
\end{equation}
Using these expressions for ${\cal K}_{1}(2\gamma)$ and ${\cal K}_{2}(2\gamma)$ in formula~(\ref{ab}),
one obtains the $G_{u}(m)$ and $G_{\Delta}(m)$ coefficients. After,
through Eqs.~(\ref{iwn}) and~(\ref{convol}) the random sequences $\{u_{n}\}$ and $\{\Delta_{n}\}$ can be obtained explicitly.

For the chosen values of the control parameters \cite{HIT10}, $c_{1} = 2/5\,,c_{2} = 4/5\,, U=0.7,\, d=1$, the dispersion relation (\ref{KP-disp}) determines the band edges of the first band,
$k_{min}/\pi = 0.259$ to $k_{max}/\pi = 1.0$, while the mobility edges are located at the points,
\begin{equation}
\begin{array}{cccc}
k_{1}/\pi = 0.327, & k_{2}/\pi = 0.476, & k_{3}/\pi = 0.652, &
k_{4}/\pi = 0.838 .\\
\end{array}
\label{chosen-value}
\end{equation}
As one can see from Fig.~\ref{deloc1}, the numerical results match well the theoretical predictions. The cases of total positive ($\eta = \pi/4$) and negative ($\eta = -\pi/4$) cross-correlations are considered, in comparison with the absence of cross-correlations ($\eta = 0$). The results confirm the prediction that cross-correlations between two disorders can either enhance or suppress the localization, in addition to the effect of specific long-range self-correlations. It is interesting to note that the effect of cross-correlations depends on whether the wave vector is within the left or right part of the band. For example, from Fig.~\ref{deloc1} one can see that the anti-cross correlations enhance the localization for $\kappa/\pi \ge 0.5$ and suppress the localization for $\kappa/\pi \le 0.5$.
\begin{figure}[htp]
\begin{center}
\includegraphics[width=10.0cm,height=7.5cm]{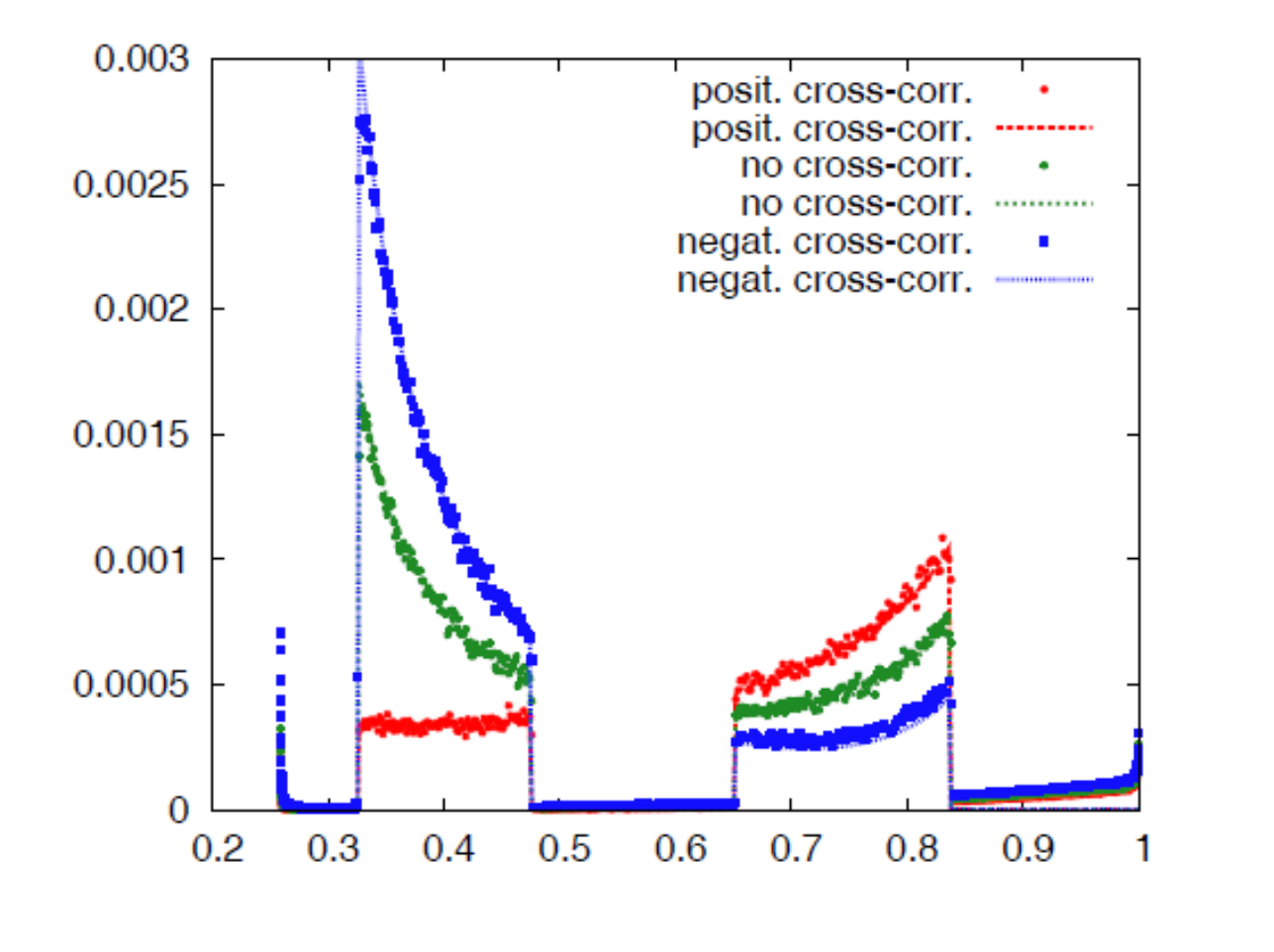}
\end{center}
\caption{(Color online) Inverse localization length versus $k/\pi$. Curves correspond to the theoretical predictions and points to numerical results. The data were
obtained for $U = 0.7, d=1$, and weak disorder, $\sqrt{\langle u_{n}^{2} \rangle} = 0.04$, and $\sqrt{\langle \Delta_{n}^{2} \rangle} = 0.05$ (after \cite{HIT10}). \label{deloc1}}
\end{figure}

In Ref.~\cite{HIT10} it was numerically demonstrated that the enhancement of localization clearly seen in Fig.~\ref{deloc1} for $k_1 < k < k_2$ and $k_3 < k < k_4$, can be increased by squeezing the corresponding energy windows.
This localization enhancement, first discussed in Ref.~\cite{IDKT04}, is a
consequence of the normalization condition (\ref{psnorm}).
To illustrate this effect, we consider the values,
\begin{equation}
\begin{array}{cccc}
k_{1}/\pi_{1} = 0.388, & k_{2}/\pi = 0.402, & k_{3}/\pi = 0.736, &
k_{4}/\pi = 0.755,
\end{array}
\label{mobedges}
\end{equation}
for the same parameters $U, d, \sqrt{\langle u_{n}^{2} \rangle}, \sqrt{\langle \Delta_{n}^{2} \rangle}$ as in Fig.~\ref{deloc1}. The theoretical and numerical results for the inverse localization length,
represented in Fig.~\ref{enhanloc}, display a clear enhancement of
localization in the narrowed energy intervals of localized
states (for experimental realization of this effect see Section~\ref{9.3}). The stronger localization magnifies the effect of the
cross-correlations of the compositional and structural disorder,
enlarging the difference between the extreme cases of total positive
and negative inter-correlations.

\begin{figure}[htp]
\begin{center}
\includegraphics[width=10.0cm,height=7.5cm]{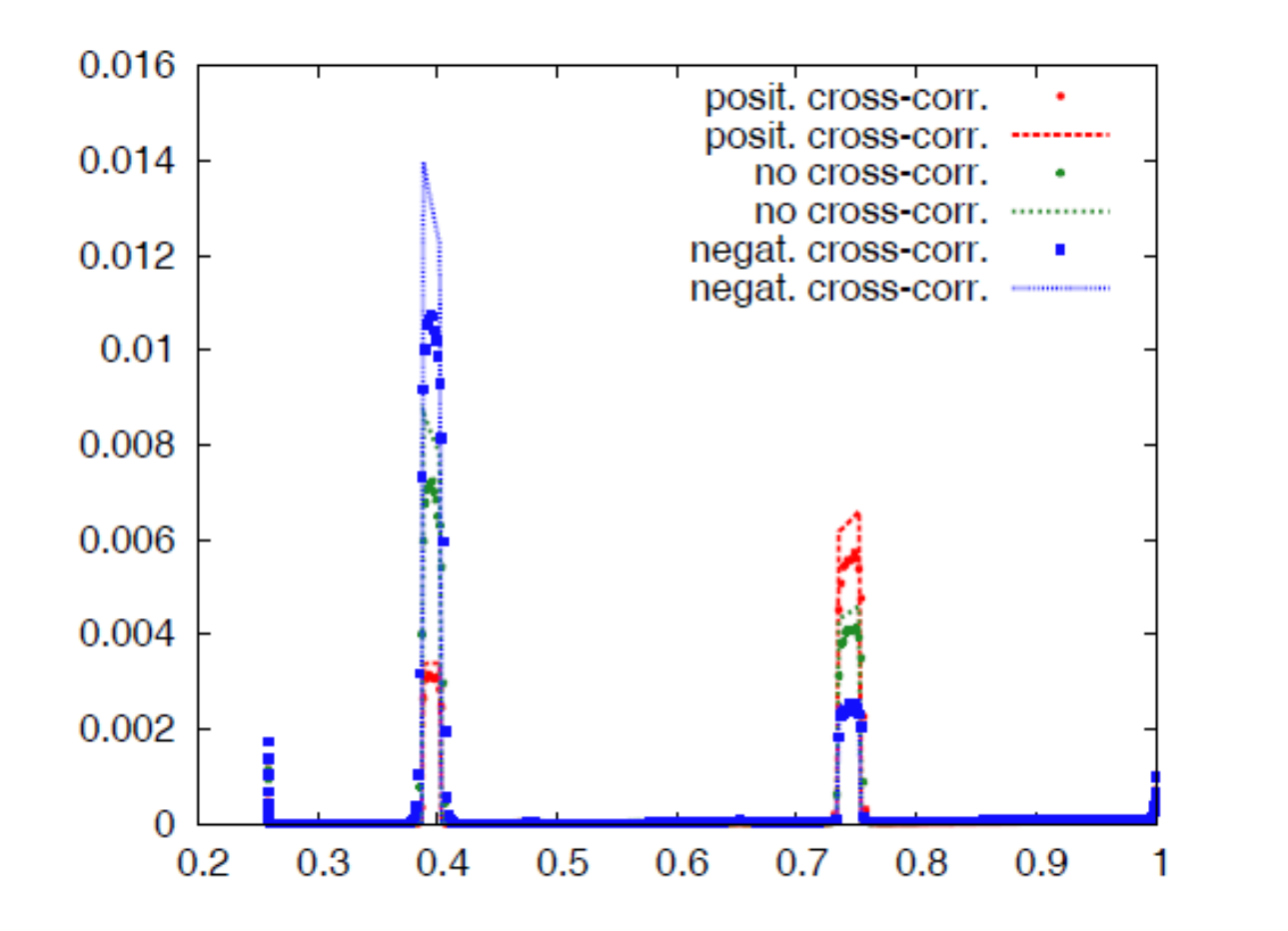}
\caption{(Color online) The same as in Fig.~\ref{deloc1} for $c_1=0.58$ and $c_2=0.62$, determining the values in Eq.~(\ref{mobedges}) (after \cite{HIT10}).\label{enhanloc}}
\end{center}
\end{figure}

Since the analytical results for the localization length are defined for {\it infinite} random sequences, it is important to check how the above considered correlations affect the transmission properties of {\it finite} samples.
For this we consider the random Kronig-Penney model with $N$
lattice sites placed between two semi-infinite perfect leads. From the
mathematical point of view, this means that the variables $u_{n}$ and
$\Delta_{n}$ in the Schr\"{o}dinger equation~(\ref{dyneq}) are defined
as before for $1 \leq n \leq N$, and vanish for $n < 1$ and $n > N$.

Using the numerical approach discussed in Section~\ref{4.4}, one can calculate the transmission coefficient $T_{N}$. In the absence of cross-correlations, only two length scales are relevant for the transport properties of the disordered segment, namely, the length of the segment itself, $L = Nd$ and the localization length $L_{loc}$ of localized states in the infinite random model. In the localized regime, i.e., when the condition $L_{loc} \ll L$
is fulfilled, the numerical data \cite{HIT10} show that the windows of
delocalized states created by long-range self-correlations of the
disorder, survive in the case of disordered segments by manifesting that the value $T_N$ is close to one, see Fig.~\ref{ta_n100}.

\begin{figure}[htp]
\begin{center}
\includegraphics[width=7.5cm,height=8.0cm]{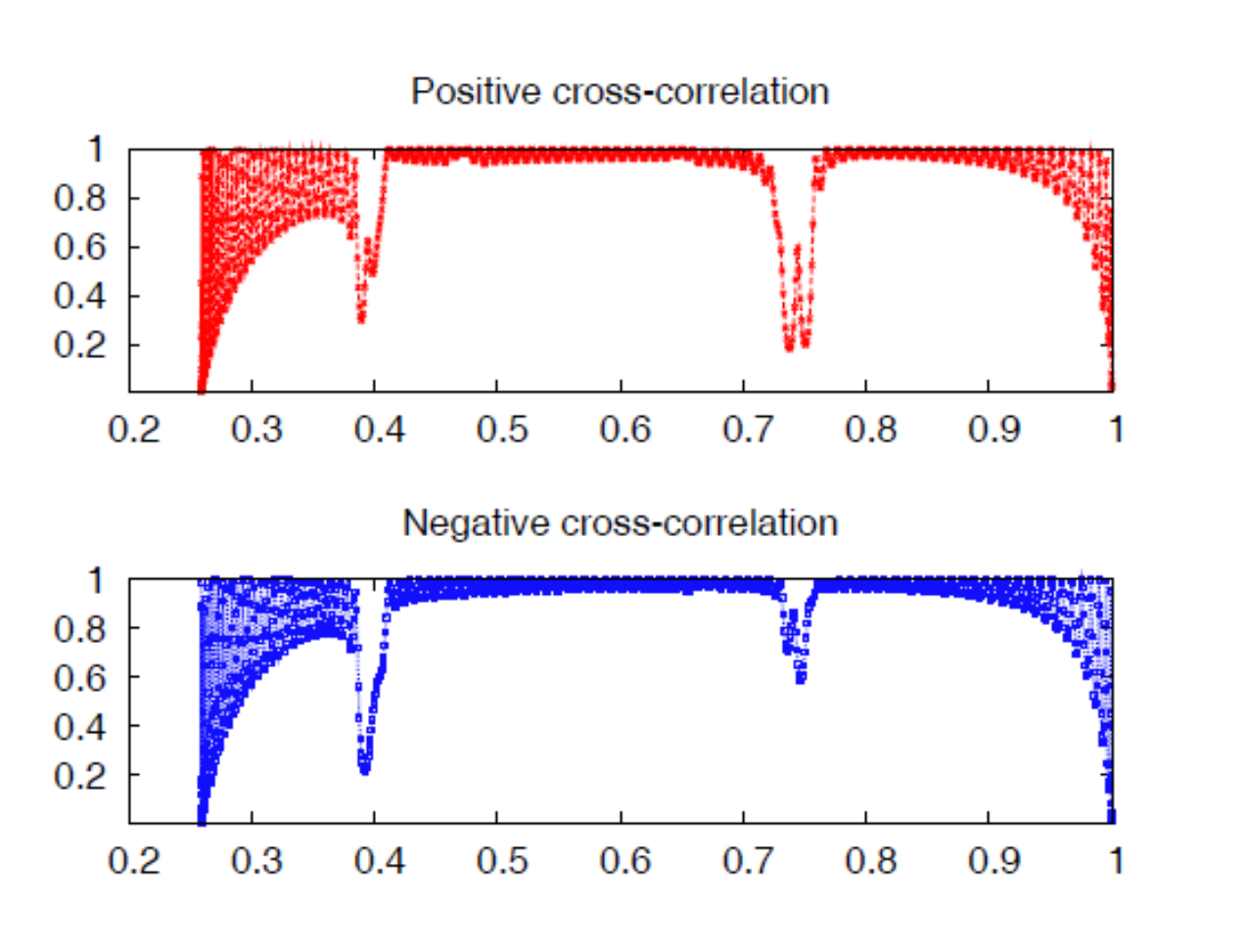}
\caption{(Color online) Transmission coefficient $T_{N}$ versus $k/\pi$ for a random sample of $N = 100$ sites (after \cite{HIT10}).
\label{ta_n100}}
\end{center}
\end{figure}
In this figure the data are given for random Kronig-Penney barriers with the
mean field $U = 0.7$, disorder strength $\sqrt{\langle u_{n}^{2} \rangle} = \sqrt{\langle \Delta_{n}^{2} \rangle} = 0.04$, and windows of localized states $[k_{1},k_{2}]$ and $[k_{3},k_{4}]$, defined by Eq.~(\ref{mobedges}). Each figure represents the transmission coefficients for the cases of total positive and negative cross-correlations. One can clearly see that in the prescribed windows of localized states the transmission coefficient is much smaller than outside of these windows. The non-trivial point is that in spite of a relatively small number of barriers ($N=100$), the effect of long-range correlation is quite strong. This means that the selective transport that arises due to such correlations can be observed experimentally (see also Section 9). As for the dependence of the transmission on whether the cross-correlations are positive or negative, in Ref.~\cite{HIT10} it was found that the transmission in the localized regions is sensitive to the length $N$ of samples.

\subsubsection{Anomaly near the center of energy bands}
\label{8.3.4}

In analogy with the Anderson model one can expect that the Kronig-Penney model also has some anomaly close to the center of energy bands where $\gamma=\pi/2$. In this Section we consider an analytical approach allowing us to describe the properties of this anomaly. We simplify our analysis by discussing the case of a totally uncorrelated disorder \cite{HIT10a}, although the numerical experiments show that the anomaly exists both for correlated and uncorrelated disorder.

We assume a white-noise type of both amplitude and positional disorders, as well as the absence of correlations between $u_{n}$ and $\Delta_{n}$ in Eq.~(\ref{KP-delta}) (or, the same, in the Hamiltonian map (\ref{dyneq})). In this case the expression (\ref{ll}}) for the Lyapunov exponent takes the simple form,
\begin{equation}
\lambda = \frac{1}{8d} \left[ \langle \tilde{u}_{n}^{2} \rangle +
\langle \tilde{\Delta}_{n}^{2} \rangle \right] =
\frac{1}{8d \sin^{2}\gamma} \left[ \frac{\sin^{2}\mu}{k^{2}}
\langle u_{n}^{2} \rangle + U^{2} \langle \Delta_{n}^{2} \rangle \right] .
\label{uncorlyap}
\end{equation}

As known, in the Anderson model with weak disorder the distribution of phases $\theta_n$ at the center of the energy band is not flat, and this is the source of the anomaly in the expression for the localization length. In order to take into account the influence of non-homogeneity of the phase distribution, we should write the expression for the Lyapunov exponent in which the averaging should be performed correctly. This expression for the Kronig-Penney model has the form \cite{HIT10a},
\begin{equation}
\begin{array}{ccl}\label{lam-tilde}
\lambda & = & \displaystyle
\frac{1}{8d} \left[ \langle 1
- 2 \cos \left( 2 \theta_{n} + \gamma \right)
+ \cos \left( 4 \theta_{n} + 2 \gamma \right) \rangle
\langle \tilde{u}_{n}^{2} \rangle \right. \\
& + & \displaystyle \left.
\langle 1 - 2 \upsilon \cos \left( 2 \theta_{n} + 2 \gamma \right)
+ \cos \left( 4 \theta_{n} + 4 \gamma \right) \rangle
\langle \tilde{\Delta}_{n}^{2} \rangle \right] ,
\end{array}
\end{equation}
where $\upsilon$ is determined by Eq.~(\ref{upsilon}) and $\tilde{u}_{n},\, \tilde{\Delta}_{n}$ are the rescaled random variables, see Eq.~(\ref{rescaling}).
Note that for a flat distribution of $\theta_n$, after averaging over $\theta_n$ the expression for the Lyapunov exponent is given by Eq.~(\ref{invloc}). Substituting into Eq.~(\ref{lam-tilde}) the value $\gamma= \pi/2$, one obtains,
\begin{equation}
\lambda =
\frac{1}{8d} \left[ \langle 1 + 2 \sin \left( 2 \theta_{n} \right)
- \cos \left( 4 \theta_{n} \right) \rangle
\langle \tilde{u}_{n}^{2} \rangle +
\langle 1 + 2 \upsilon \cos \left( 2 \theta_{n} \right)
+ \cos \left( 4 \theta_{n} \right) \rangle
\langle \tilde{\Delta}_{n}^{2} \rangle \right].
\label{lyap-four}
\end{equation}

Now we have to find the correct expression for the invariant distribution $\rho(\theta)$ which will allow us to perform the average over the angle variable in the formula for the Lyapunov exponent. To this end, we consider the four-step map for the angle variable, similar to the analysis of the Anderson model at the band center (see Section~\ref{4.2.2}),
\begin{equation}
\begin{array}{ccl}
\theta_{n+4} & = & \displaystyle
\theta_{n} - \frac{1}{2} \left[ 1 + \sin \left( 2 \theta_{n} \right) \right]
\left( \tilde{u}_{n} + \tilde{u}_{n+2} \right)
-\frac{1}{2} \left[ 1 - \sin \left( 2 \theta_{n} \right) \right]
\left( \tilde{u}_{n+1} + \tilde{u}_{n+3} \right) \\
& + & \displaystyle
\frac{1}{2} \left[ \upsilon + \cos \left( 2 \theta_{n} \right) \right]
\left( \tilde{\Delta}_{n} + \tilde{\Delta}_{n+2} \right) +
\frac{1}{2} \left[ \upsilon - \cos \left( 2 \theta_{n} \right) \right]
\left( \tilde{\Delta}_{n+1} + \tilde{\Delta}_{n+3} \right) \\
& + & \displaystyle
\frac{1}{2} \sin \left( 4 \theta_{n} \right)
\left( \langle \tilde{u}_{n}^{2} \rangle -
\langle \tilde{\Delta}_{n}^{2} \rangle \right) .
\end{array}
\label{theta-4}
\end{equation}
In the continuum limit one can replace this map with the It\^{o}
stochastic differential equation,
\begin{equation}
\begin{array}{ccl}
d \theta & = & \displaystyle
\frac{1}{2} \sin \left( 4 \theta \right)
\left( \langle \tilde{u}_{n}^{2} \rangle -
\langle \tilde{\Delta}_{n}^{2} \rangle \right) dt \\
& - & \displaystyle
\sqrt{\frac{\langle \tilde{u}_{n}^{2} \rangle}{2}}
\left[ 1 + \sin \left( 2 \theta \right) \right] d{\mathbb W}_{1}
- \sqrt{\frac{\langle \tilde{u}_{n}^{2} \rangle}{2}}
\left[ 1 - \sin \left( 2 \theta \right) \right] d{\mathbb W}_{2} \\
& + & \displaystyle
\sqrt{\frac{\langle \tilde{\Delta}_{n}^{2} \rangle}{2}}
\left[ \upsilon + \cos \left( 2 \theta \right) \right] d{\mathbb W}_{3}
+ \sqrt{\frac{\langle \tilde{\Delta}_{n}^{2} \rangle}{2}}
\left[ \upsilon - \cos \left( 2 \theta \right) \right] d{\mathbb W}_{4} \\
\end{array}
\label{ito}
\end{equation}
where ${\mathbb W}_{1}(t), \ldots, {\mathbb W}_{4}(t)$ represent four independent Wiener
processes with
\begin{equation}
\begin{array}{ccc}
\langle d{\mathbb W}_{i}(t) \rangle = 0 & \mbox{ and } &
\langle d{\mathbb W}_{i}(t) d{\mathbb W}_{j}(t) \rangle = \delta_{ij} dt .
\end{array}
\label{winer-pro}
\end{equation}
With the It\^{o} equation~(\ref{ito}), one can determine the
conditional probability $p(\theta, t| \theta_{0}, t_{0}) = p$
by solving the Fokker-Planck equation~\cite{G04},
\begin{equation}
\begin{array}{ccl}
\displaystyle
\frac{\partial p}{\partial t} & = & \displaystyle
\frac{1}{2} \left( \langle \tilde{\Delta}_{n}^{2} \rangle -
\langle \tilde{u}_{n}^{2} \rangle \right) \frac{\partial}{\partial \theta}
\left[ \sin \left( 4 \theta \right) p \right] \\
& + & \displaystyle
\frac{1}{4} \frac{\partial^{2}}{\partial \theta^{2}}
\left\{ \left[ \left( 3 \langle \tilde{u}_{n}^{2} \rangle +
\left( 2 \upsilon^{2} + 1 \right) \langle \tilde{\Delta}_{n}^{2} \rangle \right)
+ \left( \langle \tilde{\Delta}_{n}^{2} \rangle -
\langle \tilde{u}_{n}^{2} \rangle \right) \cos \left( 4 \theta \right)
\right] p \right\}. \\
\end{array}
\label{dPdt}
\end{equation}

The stationary solution of this Fokker-Planck equation represents the
desired invariant distribution $\rho(\theta)$. Hence, one has to find the
solution of the ordinary differential equation,
\begin{equation}
\begin{array}{lcl}
\displaystyle
\frac{d}{d\theta} \biggl\{ 2 \left( \langle \tilde{u}_{n}^{2} \rangle
- \langle \tilde{\Delta}_{n}^{2} \rangle \right) \sin \left( 4 \theta \right)
\rho & & \\
\displaystyle
+ \left[ 3 \langle \tilde{u}_{n}^{2} \rangle +
\left( 2 \zeta^{2} + 1 \right) \langle \tilde{\Delta}_{n}^{2} \rangle
+ \left( \langle \tilde{\Delta}_{n}^{2} \rangle -
\langle \tilde{u}_{n}^{2} \rangle \right) \cos \left( 4 \theta \right)
\right] \frac{d \rho}{d \theta} \biggr\} & = & 0
\end{array}
\label{rhoeq}
\end{equation}
that satisfies the normalization condition,
\begin{equation}
\int_{0}^{2 \pi} \rho (\theta) d \theta = 1
\label{normcon}
\end{equation}
and is periodic, $\rho(\theta + 2 \pi) = \rho(\theta)$.

\begin{figure}[thb]
\begin{center}
\includegraphics[width=10.0cm,height=7.5cm]{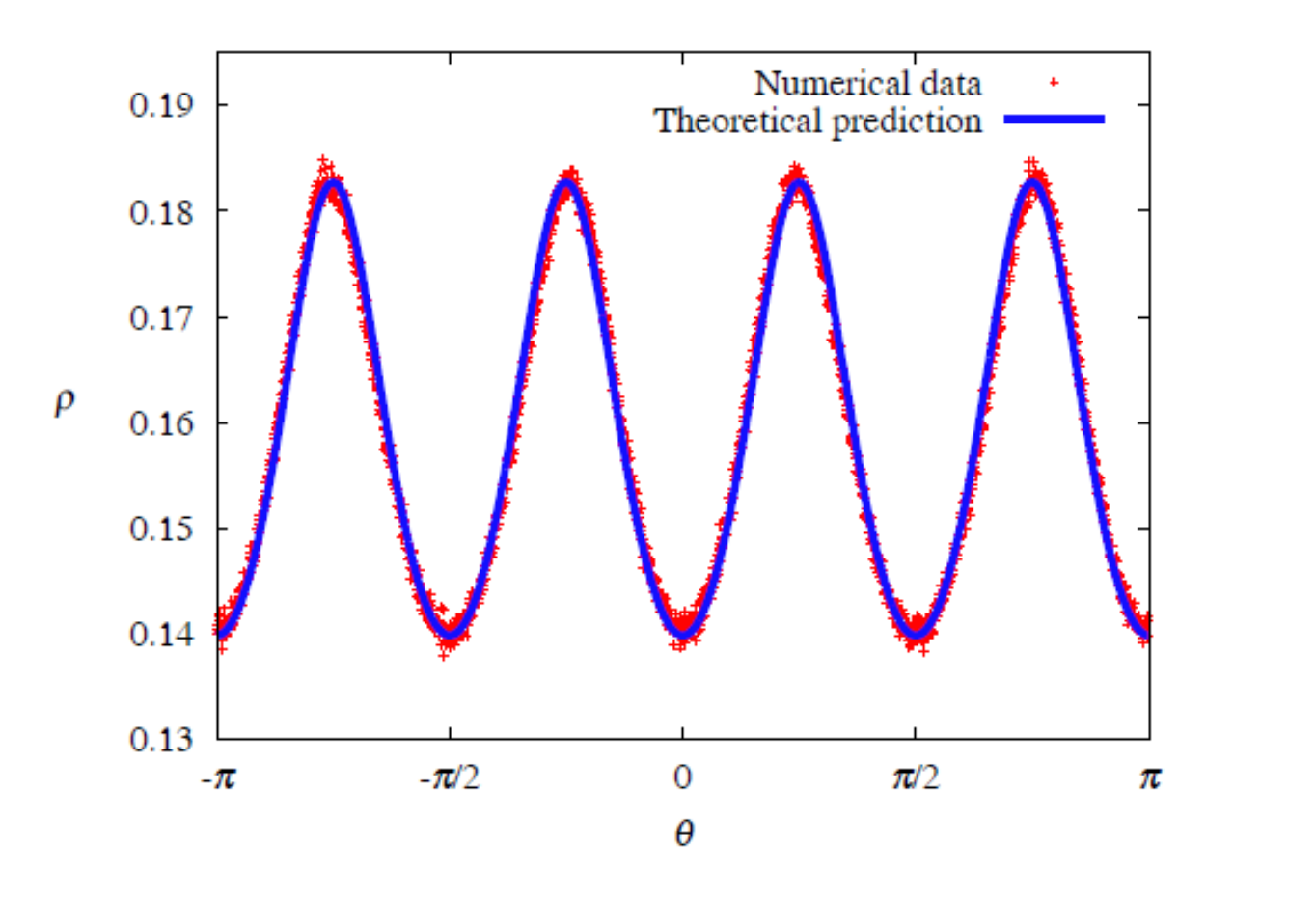}
\caption{(Color online)
Invariant distribution $\rho$ versus $\theta$. The solid line
corresponds to the prediction of Eq.~(\ref{invmea}) while points
represent numerical data (after \cite{HIT10a}). \label{rho}}
\end{center}
\end{figure}

The general solution of Eq.~(\ref{rhoeq}) can be written in the following form,
\begin{equation}
\rho(\theta) = \sqrt{\frac{a - b}{a - b \cos \left( 4 \theta \right)}}
\left[ \rho(0) + \int_{0}^{\theta} \frac{C}{\sqrt{a - b}}
\frac{1}{\sqrt{a - b \cos \left( 4 \varphi \right)}} d \varphi \right]
\label{rho-from-theta}
\end{equation}
where $C$ is an integration constant and
\begin{equation}
\begin{array}{ccl}
a & = & \displaystyle
3 \langle \tilde{u}_{n}^{2} \rangle + \left( 2 \upsilon^{2} + 1 \right)
\langle \tilde{\Delta}_{n}^{2} \rangle \\
b & = & \displaystyle
\langle \tilde{u}_{n}^{2} \rangle - \langle \tilde{\Delta}_{n}^{2} \rangle .
\end{array}
\label{a-und-b}
\end{equation}
In order to fulfil the periodicity condition one has to assign $C=0$. As a result, the invariant distribution takes the form,
\begin{equation}
\rho(\theta) = \frac{\sqrt{a + |b|}}{4 {\bf K} \left( c \right)}
\frac{1}{\sqrt{a - b \cos \left( 4 \theta \right)}}
\label{invmea}
\end{equation}
where
\begin{equation}
{\bf K}(c) = \int_{0}^{\pi/2} \frac{1}{\sqrt{1 - c^{2} \sin^{2} \varphi}}
d \varphi
\label{K-c-ellipt}
\end{equation}
is the complete elliptic integral of the first kind with the argument,
\begin{equation}
c = \sqrt{\frac{2 |b|}{a + |b|}} = \sqrt{ \frac{ 2 \left|
\langle \tilde{u}_{n}^{2} \rangle - \langle \tilde{\Delta}_{n}^{2} \rangle
\right|}{\left|
\langle \tilde{u}_{n}^{2} \rangle - \langle \tilde{\Delta}_{n}^{2} \rangle
\right| + 3 \langle \tilde{u}_{n}^{2} \rangle +
\left( 2 \upsilon^{2} + 1 \right) \langle \tilde{\Delta}_{n}^{2} \rangle}} .
\label{c}
\end{equation}
As expected, the expression ~(\ref{invmea}) shows that when the Bloch number
takes the value $\kappa = \pi/2d$, the invariant measure has a period $\pi/2$.
The numerical data agree well with the formula~(\ref{invmea}) \cite{HIT10a}, see Fig.~\ref{rho}. The data represented in Fig.~\ref{rho} were obtained for the mean field
$U = 8$ and disorder strengths $\sqrt{\langle u_{n}^{2} \rangle} =
\sqrt{\langle \Delta_{n}^{2} \rangle} = 0.02$.

\begin{figure}[thb]
\begin{center}
\includegraphics[width=10.0cm,height=7.5cm]{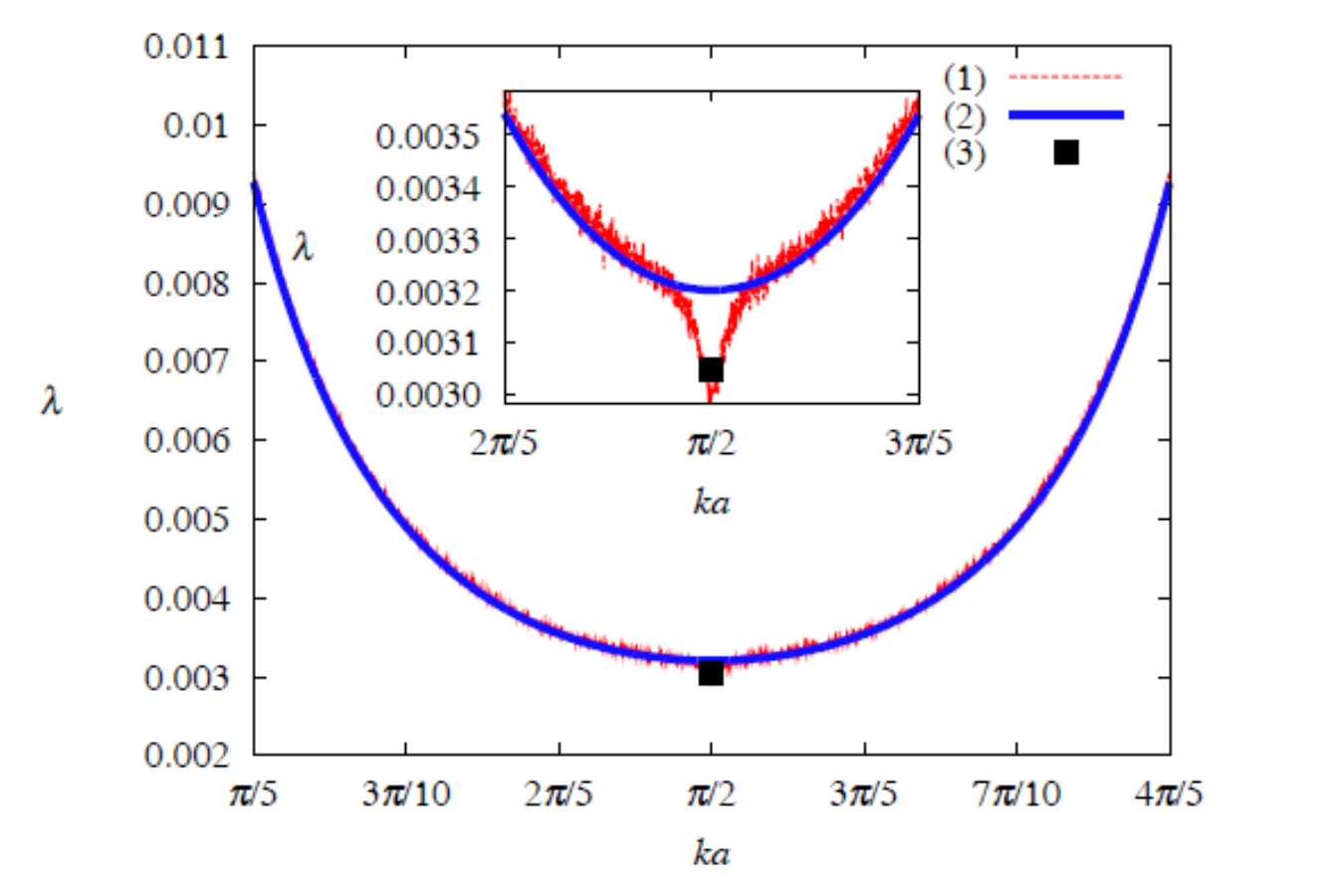}
\caption{(Color online)
Inverse localization length $\lambda$ versus $\gamma$ (uncorrelated disorder) with $ka \equiv \gamma$.
Dashed curve (1) represents numerically obtained values; solid curve
(2) corresponds to the formula~(\ref{uncorlyap}); symbol (3) represents the
anomalous value~(\ref{anolyapp}). The data are given for $U = 4,\, \sqrt{\langle u_{n}^{2} \rangle} = 0.04, \, \sqrt{\langle \Delta_{n}^{2} \rangle} = 0.05$ with $d=1$. The inset magnifies the region close to the band center
(after \cite{HIT10a}).
\label{bc_anomal}}
\end{center}
\end{figure}

Having the invariant distribution, we can now compute the Lyapunov exponent. It follows from Eq.~\eqref{invmea} for $\rho(\theta)$ that the additional contribution to the Lyapunov exponent (in comparison with the case of flat distribution) comes only from the terms with $\cos4\theta$ in Eq.~\eqref{lyap-four}. Note that in the case of flat distribution all trigonometric terms vanish. Thus, after averaging over $\theta$ the expression for the Lyapunov exponent takes its final form,
\begin{equation}
\lambda=
\frac{1}{8d} \bigg\{ \left[ \left| \langle \tilde{u}_{n}^{2} \rangle -
\langle \tilde{\Delta}_{n}^{2} \rangle \right| + 3
\langle \tilde{u}_{n}^{2} \rangle + \left( 2 \upsilon^{2} + 1 \right)
\langle \tilde{\Delta}_{n}^{2} \rangle \right]
\frac{{\bf E}(c)}{{\bf K}(c)} -
2 \left( \langle \tilde{u}_{n}^{2} \rangle + \upsilon^{2}
\langle \tilde{\Delta}_{n}^{2} \rangle \right) \bigg\},
\label{anolyapp}
\end{equation}
where
\begin{equation}
{\bf E}(c) = \int_{0}^{\pi/2} \sqrt{1 - c^{2} \sin^{2} \varphi} d \varphi
\label{E-c}
\end{equation}
is the complete elliptic integral of the second kind with the argument $c$
given by Eq.~(\ref{c}). The numerical computations confirm the existence of an anomaly for $\gamma = \pi/2$
as can be seen in Fig.~\ref{bc_anomal}. The data manifest a small but clear deviation from the value of the Lyapunov exponent predicted by the standard formula~(\ref{uncorlyap}).

\begin{figure}[thb]
\begin{center}
\includegraphics[width=10.0cm,height=7.5cm]{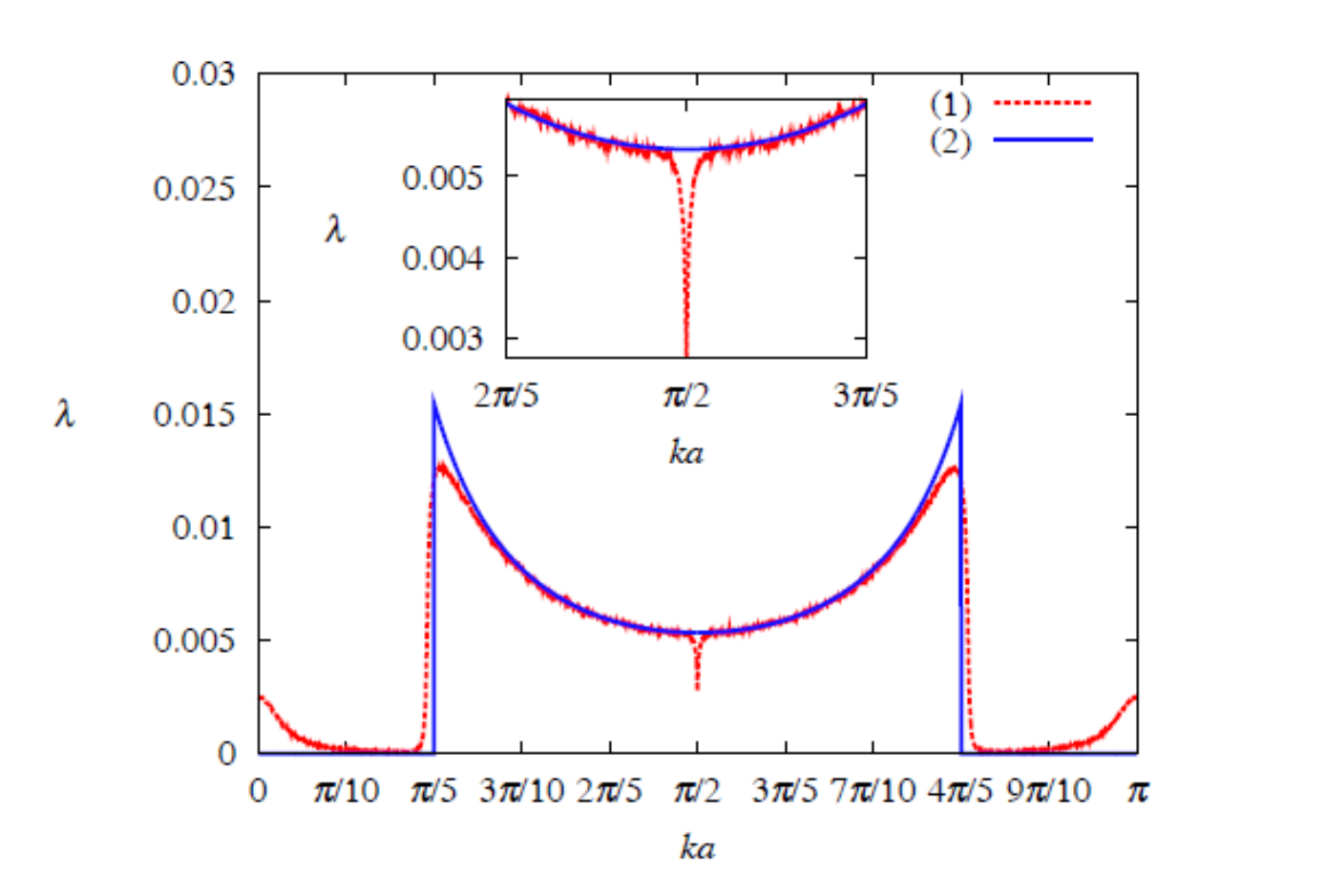}
\caption{(Color online)
Inverse localization length $\lambda$ versus $\gamma$ for the self-correlated
disorder (with $ka \equiv \gamma$). Dashed curve (1) represents numerically computed data, while
solid curve (2) corresponds  to Eq.~(\ref{invloc}).
The inset shows the anomaly in detail (after \cite{HIT10a}).
\label{anomalycorr}}
\end{center}
\end{figure}

One should stress that although we discuss here the case of uncorrelated disorder, there is a numerical evidence that the
correlations can enhance the anomaly of the localization length
near the center of the energy band, in agreement with the theoretical
predictions of Ref.~\cite{TS05}.
As an example, let us consider the case of amplitude and positional
disorder with the self-correlations of the form \cite{HIT10a},
\begin{equation}
K_{u}(m) = K_{\Delta}(m) = \left\{
\begin{array}{lcc}
1 & \mbox{ for} & m = 0 \\
\displaystyle
-\frac{5}{3\pi m} \sin \left( \frac{2}{5} \pi m \right) &
\mbox{ for } & |m| > 0 \,,\\
\end{array} \right.
\label{lrcorr}
\end{equation}
and without the cross-correlations, $K_{u,\Delta}(m) = 0$.
The long-range correlations of the form~(\ref{lrcorr}) create mobility
edges at $\kappa_1 = \pi/5d$ and $\kappa_2 = 4\pi/5d$. In this case in the first energy band the Lyapunov exponent vanishes for $\kappa < \kappa_1$ and $\kappa > \kappa_2$. The numerical data in
Fig.~\ref{anomalycorr} are obtained for this kind of
disorder with $U=8, d=1$ and $\sqrt{\langle u_{n}^{2} \rangle} =
\sqrt{\langle \Delta_{n}^{2} \rangle} = 0.02$.

The data clearly show an enhanced anomaly at $\kappa = \pi/2d$ with respect
to the case of totally uncorrelated disorder. Adding the cross-correlations
does not introduce any significant modification to the picture.

It is interesting to note that although the correlations considered here
enhance the anomaly for $\kappa = \pi/2d$,
other correlations may not produce the same effect.
This can be seen for the case analyzed in Sec.~\ref{8.3.6},
in which the resonance effect for $\kappa = \pi/2d$ is shadowed by a different
kind of anomaly generated by specific cross-correlations for a value of $\kappa$
close, but not identical, to $\pi/2d$.

Finally, we give the expressions for the Lyapunov exponent for
the limit cases of a purely amplitude or positional disorder.
In the first case we have $\langle \tilde{\Delta_{n}^{2}} \rangle = 0$, and one can show that Eq.~(\ref{anolyapp}) reduces to the form,
\begin{equation}
\lambda = \frac{1}{d} \langle \tilde{u}_{n}^{2} \rangle \left[
\frac{\Gamma \left( 3/4 \right)}{\Gamma \left( 1/4 \right)} \right]^{2}
= \frac{1}{8d} \langle \tilde{u}_{n}^{2} \rangle \cdot 0.9139\ldots
\label{lambdad0}
\end{equation}
In the second case, when $\langle \tilde{u}_{n}^{2} \rangle = 0$, one obtains,
\begin{equation}
\lambda = \frac{\langle \tilde{\Delta}_{n}^{2} \rangle}{4d} \left[
\left( \upsilon^{2} + 1 \right)
\frac{{\bf E}\left( 1/\sqrt{\upsilon^{2} + 1} \right)}
{{\bf K} \left( 1/\sqrt{\upsilon^{2} + 1} \right)} - \upsilon^{2} \right] .
\end{equation}
where $\upsilon$ is defined by Eq.~({\ref{upsilon}).

\subsubsection{The band-edge anomaly}
\label{8.3.5}

We now turn our attention to the anomaly for $\gamma \to 0$, i.e., in the neighborhood of the band edge (being inside the energy bands). Because of the similarity with the band-edge anomaly in the Anderson model, we can apply the
method used in Ref.~\cite{IRT98} to the present case. For small values of $\gamma$ the rescaled variables $\tilde{u}_{n}$ and $\tilde{\Delta}_{n}$ (see Eq.~(\ref{rescaling})) can be approximated as
\begin{equation}
\tilde{u}_{n} \approx \frac{\sin \mu_{0}}{k_{0}} \frac{u_{n}}{\gamma} ,\,\,\,\,\,
\tilde{\Delta}_{n} \approx U \frac{\Delta_{n}}{\gamma}.
\label{approxim}
\end{equation}
Here $k_{0}$ stands for the value of the Bloch number corresponding to $\gamma\to 0$. It is determined by the relation (\ref{pos-edges}) which stems from the dispersion relation (\ref{KP-disp}). In the same limit, the parameter $\upsilon$ defined by Eq.~(\ref{upsilon}) can be written as \cite{HIT10a},
\begin{equation}
\upsilon \approx 1 + \frac{2k_{0}^{2}}{U^{2}} \gamma^{2}.
\label{bezeta}
\end{equation}
Taking into account these approximations, one can write the Hamiltonian
map~(\ref{hammap5}) in the form,
\begin{equation}
\begin{array}{ccl}
J_{n+1} & = & D_{n}^{2} J_{n} \\
\theta_{n+1} & = & \displaystyle
\theta_{n} + \gamma - \frac{\xi_{n}}{\gamma}
\sin^{2}\theta_{n} + \frac{\sigma^{2}}{\gamma^{2}} \sin^{3}
\theta_{n} \cos  \theta_{n},
\end{array}
\label{bemap}
\end{equation}
with
\begin{equation}
D_{n} = 1 - \frac{\xi_{n}}{\gamma} \sin  \theta_{n}
\cos \theta_{n}  + \frac{\sigma^{2}}{2\gamma^{2}}
\sin^{4} \theta_{n}.
\label{bedn}
\end{equation}
In Eqs.~(\ref{bemap}) and~(\ref{bedn}), the symbol $\xi_{n}$ represents
the linear combination of structural and compositional disorder
\begin{equation}
\xi_{n} = U \Delta_{n} - \frac{\sin \mu_{0}}{k_{0}} u_{n}
\label{eff-dis}
\end{equation}
with the zero average, $\langle \xi_{n} \rangle = 0$, and variance
\begin{equation}
\langle \xi_{n}^{2} \rangle = \sigma^{2} =
U^{2} \langle \Delta_{n}^{2} \rangle + \frac{\sin^{2}\mu_{0}}{k_{0}^{2}}
\langle u_{n}^{2} \rangle .
\label{var-eff-dis}
\end{equation}
Note that here $\mu_0=k_0 d$.

In the continuum limit one can replace the angular map in
Eq.~(\ref{bemap}) with the stochastic It\^{o} equation,
\begin{equation}
d \theta = \left[ \gamma + \frac{\sigma^{2}}{\gamma^{2}}
\sin^{3} \theta  \cos \theta \right] dt
+ \frac{\sigma}{\gamma} \sin^{2} \theta
d {\mathbb W},
\label{beito}
\end{equation}
whose associated Fokker-Planck equation,
\begin{equation}
\frac{\partial P}{\partial t} = - \frac{\partial}{\partial \theta}
\left\{ \left[ \gamma + \frac{\sigma^{2}}{\gamma^{2}}
\sin^{3} \theta \cos \theta \right] P \right\} +
\frac{1}{2} \frac{\partial^{2}}{\partial \theta^{2}} \left[
\frac{\sigma^{2}}{\gamma^{2}} \sin^{4} \theta P \right],
\label{befp1}
\end{equation}
gives the conditional probability $P(\theta, t | \theta_{0}, t_{0}) = P$
for the stochastic process $\theta(t)$~\cite{G04}.
Introducing the rescaled time,
\begin{equation}
\tau = \frac{\sigma^{2}}{\gamma^{2}} t ,
\label{tau-resc}
\end{equation}
one can cast the Fokker-Planck equation~(\ref{befp1}) in the form
\begin{equation}
\frac{\partial P}{\partial \tau} = - \frac{\partial}{\partial \theta}
\left\{ \left[ \varkappa + \sin^{3} \theta \cos \theta \right] P \right\} +
\frac{1}{2} \frac{\partial^{2}}{\partial \theta^{2}} \left[
\frac{\sigma^{2}}{\gamma^{2}} \sin^{4} \theta P \right]
\label{befp2}
\end{equation}
which contains the noise intensity $\sigma^{2}$ and the distance from the
band edge $\gamma$ combined in the single scaling parameter,
\begin{equation}
\varkappa = \frac{\gamma^{3}}{\sigma^{2}}.
\label{varkappa-resc}
\end{equation}
As one can see, this parameter is similar to that arising at the band edges in the Anderson model. The difference is that here the Bloch number $\gamma$ stands in place of $\mu$ in Eq.~(\ref{varkappa-def}) .

The invariant distribution $\rho(\theta)$ is the stationary solution of
the Fokker-Planck~(\ref{befp2}) that one needs to have in order to derive the Lyapunov exponent. It should satisfy two conditions: it has to be normalizable to one and periodic with period $\pi$. The solution possessing these features has the following form,
\begin{equation}
\rho(\theta) = \frac{1}{N(\varkappa)}
\frac{e^{-f(\theta)}}{\sin^{2}\theta} \int_{\theta}^{\pi}
\frac{e^{f(\phi)}}{\sin^{2}\phi} {\mathrm d}\phi
\label{berho}
\end{equation}
where
\begin{equation}
f(\theta) = 2 \varkappa \left[ \frac{1}{3} \cot^{3} \theta +
\cot(\theta) \right]
\label{function}
\end{equation}
and
\begin{equation}
N(\varkappa) = \sqrt{\frac{2 \pi}{\varkappa}} \int_{0}^{\infty}
\frac{1}{\sqrt{x}} \exp \left[ - 2 \varkappa \left( \frac{x^{3}}{12} + x
\right) \right] {\mathrm d}x .
\label{N-norm}
\end{equation}
The integral representation~(\ref{berho}) defines the invariant measure in
the interval $[0,\pi]$; the dependence of $\rho(\theta)$ can be extended outside of this interval via the relation $\rho(\theta + \pi) = \rho(\theta)$.

To obtain a qualitative understanding of the phase
distribution~(\ref{berho}), it is useful to consider its values at the
edges and at the center of the $[0,\pi]$ interval.
For $\theta \to 0^{+}$ and $\theta \to \pi^{-}$ one has
\begin{equation}
\rho (\theta) \sim
\left\{
\begin{array}{lcl}
\displaystyle
\varkappa^{-5/3} & \mbox{ for } & \varkappa \to 0 \\
(2 \pi)^{-1} & \mbox{ for } & \varkappa \to \infty \\
\end{array} \right. ,
\label{rho-limit-one}
\end{equation}
while for $\theta \to \pi/2$ the invariant distribution behaves as
\begin{equation}
\rho (\theta) \sim
\left\{ \begin{array}{lcl}
\varkappa^{1/3} & \mbox{ for } & \varkappa \to 0 \\
\displaystyle
(2 \pi)^{-1} & \mbox{ for } & \varkappa \to \infty \\
\end{array} \right. .
\label{rho-limit-two}
\end{equation}
These equations show that when the energy moves closer to the band edge on
the scale defined by the disorder strength, i.e., for $\varkappa
\to 0$, the invariant distribution develops two pronounced maxima for
$\theta \sim 0$ and $\theta \sim \pi$.
This conclusion is supported by the direct numerical computation of the
invariant distribution, as shown by Fig.~\ref{beinvmea}.

\begin{figure}[thb]
\begin{center}
\includegraphics[width=10.0cm,height=7.5cm]{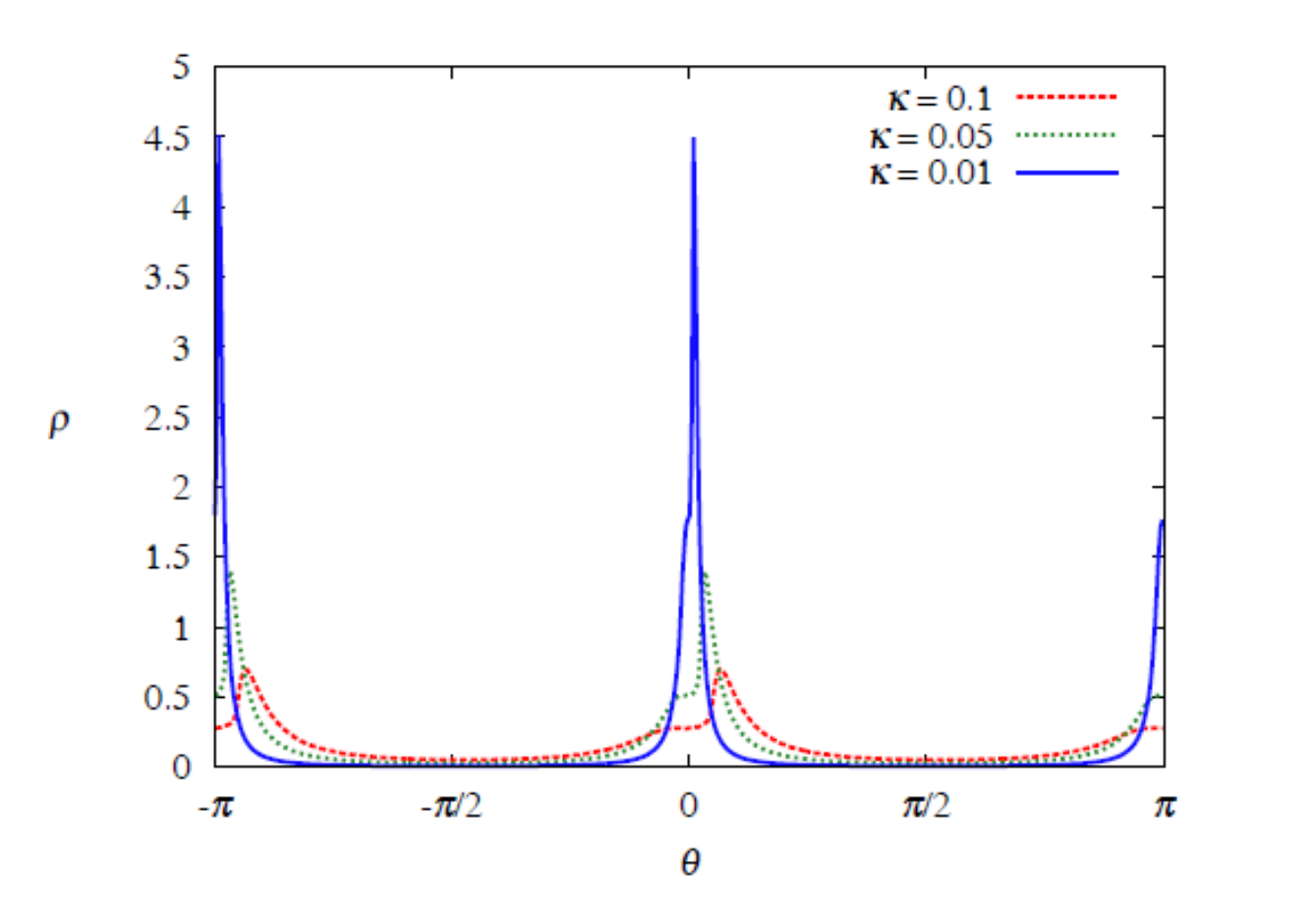}
\caption{(Color online)
Invariant distribution $\rho$ versus $\theta$. The legend shows
the value of $\varkappa$ corresponding to each curve (after \cite{HIT10a}).
\label{beinvmea}}
\end{center}
\end{figure}
The data in Fig.~\ref{beinvmea} were obtained for the mean field $U=8$ and
disorder strengths $\sqrt{\langle u_{n}^{2} \rangle} =
\sqrt{\langle \Delta_{n}^{2} \rangle} = 0.02$.

Therefore, at the band edges the standard assumption of a uniform or slightly modulated distribution is incorrect. In this case it is more convenient to use both terms in the right-hand part of Eq.~(\ref{loclength}), and to evaluate the following expression,
\begin{equation}
\lambda \simeq \frac{1}{d}
\Bigg\langle \log \left[ D_{n} \left| \frac{\sin \theta_{n+1} }
{\sin \theta_{n}} \right| \right] \Bigg\rangle.
\label{belambda}
\end{equation}
Using the angular map in Eq.~(\ref{bemap}), one can write the ratio of two sines in Eq.~(\ref{belambda}) as follows:
\begin{equation}
\begin{array}{ccl}
\displaystyle
\frac{\sin \theta_{n+1} }{\sin \theta_{n} }
& = & \displaystyle
1 + \gamma \cot \theta_{n} +
\frac{\xi_{n}}{\gamma} \sin \theta_{n}
\cos \theta_{n} \\
& + & \displaystyle
\frac{\sigma^{2}}{\gamma^{2}}
\left[ \sin^{2}\theta_{n} \cos^{2}\theta_{n}
- \frac{1}{2} \sin^{4} \theta_{n} \right].
\end{array}
\label{besineratio}
\end{equation}
Substituting the approximate identities~(\ref{bedn}) and~(\ref{besineratio})
in Eq.~(\ref{belambda}), one obtains \cite{HIT10a},
\begin{equation}
\lambda \simeq \frac{\gamma}{d} \Big\langle
\cot \theta_{n} \Big\rangle .
\label{lam-general}
\end{equation}
The average can be now performed with the use of the invariant distribution~(\ref{berho}). The final result takes the form,
\begin{equation}
\lambda = \frac{\gamma}{2d}
\frac{\displaystyle \int_{0}^{\infty} x^{1/2} \exp \left[ -2 \varkappa
\left( \frac{x^{3}}{12} + x \right) \right] {\mathrm d}x}
{\displaystyle \int_{0}^{\infty} x^{-1/2} \exp \left[ -2 \varkappa
\left( \frac{x^{3}}{12} + x \right) \right] {\mathrm d}x} .
\label{beinvloc}
\end{equation}
It can be shown that in the limit case in which $\varkappa \to \infty$, the expression (\ref{beinvloc}) reduces to a quite simple relation,
\begin{equation}
\lambda \simeq \frac{\sigma^{2}}{8 d \gamma^{2}} ,
\label{lam-limit-one}
\end{equation}
that coincides with the standard expression~(\ref{uncorlyap})
for small values of $\gamma $. This result is similar to that known for the standard Anderson model (see Section~\ref{4.2.3}).

In the opposite limit, i.e., for $\varkappa \to 0$, Eq.~(\ref{beinvloc}) gives
\begin{equation}
\lambda \simeq \frac{6^{1/3} \sqrt{\pi}}{2 d \Gamma \left( 1/6 \right)}
\sigma ^{2/3},
\label{lam-limit-two}
\end{equation}
which exhibits the same anomalous scaling found in the Anderson model
at the band-edge~\cite{DG84,IRT98}.
This correspondence is a consequence of the fact that in both models at
the band edge the invariant distribution for the angular variable has
the form~(\ref{berho}) and the ratio $\psi_{n+1}/\psi_{n}$
reduces to the same function of $\theta$.

\subsubsection{Anomalously localized states}
\label{8.3.6}

In this Section we show that specific correlations can give rise to the anomalous states in the Kronig-Penney model~(\ref{KP-delta}) with the amplitude and positional disorders (for details see \cite{HIT10}).
We start by noting that this model can be formally written in the form of the Anderson-type model with both diagonal and off-diagonal disorders. This can be done by eliminating the variables $p_n$ and $p_{n+1}$ from the two-dimensional map (\ref{hammap1}) and reducing this map to the following recurrent relation,
\begin{equation}\label{KP-exact}
\frac{1}{\sin ( \mu + \mu_{n} )} \psi_{n+1} +
\frac{1}{\sin ( \mu + \mu_{n-1} )} \psi_{n-1}=
\left [ \cot ( \mu + \mu_{n} ) +
\cot ( \mu + \mu_{n-1} ) + \frac{1}{k}
\left( U + u_{n} \right) \right ] \psi_{n} .
\end{equation}
We remind that here $\mu=kd$ and $ \mu_n=k\Delta_n$, and
\begin{equation}\label{U-delta}
U_n=U+u_n,\quad x_{n+1}-x_n= d+ \Delta_n.
\end{equation}
For weak disorder the coefficients of this equation can be expanded in powers of $\Delta_{n}$. In the first order approximation we neglect the fluctuations of $\Delta_{n}^{2}$ and make the substitution $\Delta_{n}^{2} \rightarrow \langle \Delta_{n}^{2} \rangle$.
One thus obtains the same form as the Schr\"{o}dinger equation for the
Anderson model with weak diagonal and off-diagonal disorder,
\begin{equation}
\left(1 + \vartheta_{n} \right) \psi_{n+1} +
\left(1 + \vartheta_{n-1} \right) \psi_{n-1} +\epsilon_n \psi_n = E \psi_{n}
\label{andoffdia}
\end{equation}
Here $E$ is a deterministic function of the wave number $k$,
defined by the identity,
\begin{equation}
E(k) = \frac{U/k + 2 \Big\langle \cot \left( \mu + \mu_{n} \right) \Big\rangle}{\Big\langle 1 / \sin \left( \mu + \mu_{n} \right)\Big\rangle} ,
\label{eq}
\end{equation}
while the symbols $\vartheta_{n}$ and $\epsilon_{n}$ stand for the
energy-dependent random variables,
\begin{equation}
\vartheta_{n}(k) = \frac{1 / \sin  \left( \mu + \mu_{n} \right)}
{\Big\langle 1 / \sin \left( \mu + \mu_{n} \right)
\Big\rangle} - 1
\label{gamvar}
\end{equation}
and
\begin{equation}
\epsilon_{n} =
\frac{1}{\Big\langle 1 / \sin \left( \mu +
\mu_{n} \right) \Big\rangle}
\left\{ 2 \Big\langle \cot \left( \mu + \mu_{n} \right)
\Big\rangle - \cot \left( \mu + \mu_{n} \right)
\right.
-  \left.
\cot \left( \mu + \mu_{n-1} \right) - \frac{u_{n}}{k}
\right\} .
\label{epsvar}
\end{equation}
Note that in the absence of structural disorder, the random
variables~(\ref{gamvar}) vanish and the Kronig-Penney's analogue becomes
the ordinary Anderson model with diagonal disorder.

We now focus our attention on the case in which the amplitude disorder
has the form,
\begin{equation}
u_{n} = 2k_{c} \Big\langle \cot \left[ k_{c} \left( d + \Delta_{n} \right)
\right] \Big\rangle -
k_{c} \cot \left[k_{c} \left( d + \Delta_{n} \right) \right] -
k_{c} \cot \left[k_{c} \left( d + \Delta_{n-1} \right) \right],
\label{zerodiagdis}
\end{equation}
where $k_{\rm c}$ represents the wave vector for which $E(k_{c}) = 0 $. Within the allowed energy bands such a value of $k_{c}$ always exists provided the
weak-disorder condition $\langle \Delta_{n}^{2} \rangle U \ll 1$ is satisfied.
Taking into account the dispersion relation~(\ref{KP-disp}),
one can rewrite the relation $E(k_{c})=0$ in the form determining the corresponding value of the Bloch phase $\gamma_c=\kappa_c d$,
\begin{equation}
\cos \gamma_c =
- \frac{\cos(k_{\rm c} d)}{\sin^{2}(k_{\rm c}d)} k_{\rm c}^{2}
\langle \Delta_{n}^{2} \rangle .
\label{con1}
\end{equation}

If the amplitude disorder has the form~(\ref{zerodiagdis}), it
is easy to see that if the electron energy takes the critical value
$k_{c}^{2}$, the equation (\ref{andoffdia}) is reduced to the following one,
\begin{equation}
\left[ 1 + \vartheta_{n}(q_{\rm c}) \right] \psi_{n+1} +
\left[ 1 + \vartheta_{n-1}(q_{\rm c}) \right] \psi_{n-1} = 0 .
\label{offdiaganderson}
\end{equation}
As one can see, it has the form of the Schr\"{o}dinger equation for the Anderson
model with purely off-diagonal disorder and zero energy. As known (see, for example, \cite{FL77,M88,B89,ITA94,CFE05}) and Ref.~\cite{HIT10a}), in this case the Anderson model exhibits anomalous localization at the band center. Specifically, the electronic state at the band center is localized, however, decays away from the
localization center $n_{0}$ in the following way,
\begin{equation}
\psi_{n} \sim \exp \left( - C \sqrt{|n - n_{0}|} \right) .
\label{anodec}
\end{equation}
where $C$ is a constant.

For uncorrelated positional disorder the numerical data for the disorder
strength $\sqrt{\langle \Delta_{n}^{2} \rangle} = 0.05$ and amplitude
$U = 4$ of barriers (with $d=1$) are shown in Fig.~\ref{anoloc}. According to Eq.~(\ref{con1}) the critical value of $\gamma$ is equal to $\gamma_{\rm c} \approx 0.4952$, which is in excellent agreement with the numerically obtained value $\gamma_{\rm c} \approx 0.4953$.
The data in Fig.~\ref{anoloc} clearly manifest the existence of a dip for the Lyapunov exponent when $\gamma=\gamma_c$.

The global dependence of the Lyapunov exponent on the Bloch phase $\gamma$
can be compared with the theoretical
prediction derived from the general result~(\ref{invloc}). In the case for
which the positional disorder is given by Eq.~(\ref{zerodiagdis}),
the formula~(\ref{invloc}) takes the form \cite{HIT10},
\begin{equation}
\lambda = \frac{1}{8d} \langle \tilde{\Delta}_{n}^{2} \rangle
\left[ 1 - 2 \frac{k_{\rm c}^{2} \sin (k d)}{U k \sin^{2} (k_{\rm c} d)}
\cos \gamma \right]^{2} {\cal K}_{\Delta}(2 \gamma).
\label{genthlyap}
\end{equation}
It should be stressed that the theoretical result~(\ref{genthlyap})
fails to describe the behavior of the Lyapunov exponent in a small
neighborhood of the critical value $\gamma_c$.
This failure is due to the fact that both Eq.~(\ref{genthlyap}) and its parent expression~(\ref{invloc}) are not
valid for values of the Bloch number close to $\kappa_{c}=\pi/2d$. At this point, the resonance effect plays an important role. On the other hand, there is a remarkable correspondence between Eq.~(\ref{genthlyap}) and numerical data
for other values of $\kappa=\gamma/d$, as can be seen from Fig.~\ref{anoloc}.

One should emphasize that the Lyapunov exponent also vanishes for a special value $\kappa^{\star}$ of the Bloch number defined by the condition,
\begin{equation}
2 \frac{k_{\rm c}^{2} \sin \left( k^{\star}d \right)}{k^{\star}
\sin^{2} \left( k_{\rm c} d \right)} \cos \gamma^{\star} = U
\label{cond-for-bloch}
\end{equation}
with $\gamma^{\star}=\kappa^{\star}d$. Therefore, when the amplitude disorder has the form~(\ref{zerodiagdis}),
the Kronig-Penney model exhibits both an anomalously localized state for
$\kappa = \kappa_{\rm c}$ and an extended state for $\kappa = \kappa^{\star}$ (at least within the second-order approximation for the Lyapunov exponent).

\begin{figure}[h]
\begin{center}
\includegraphics[width=10.0cm,height=7.5cm]{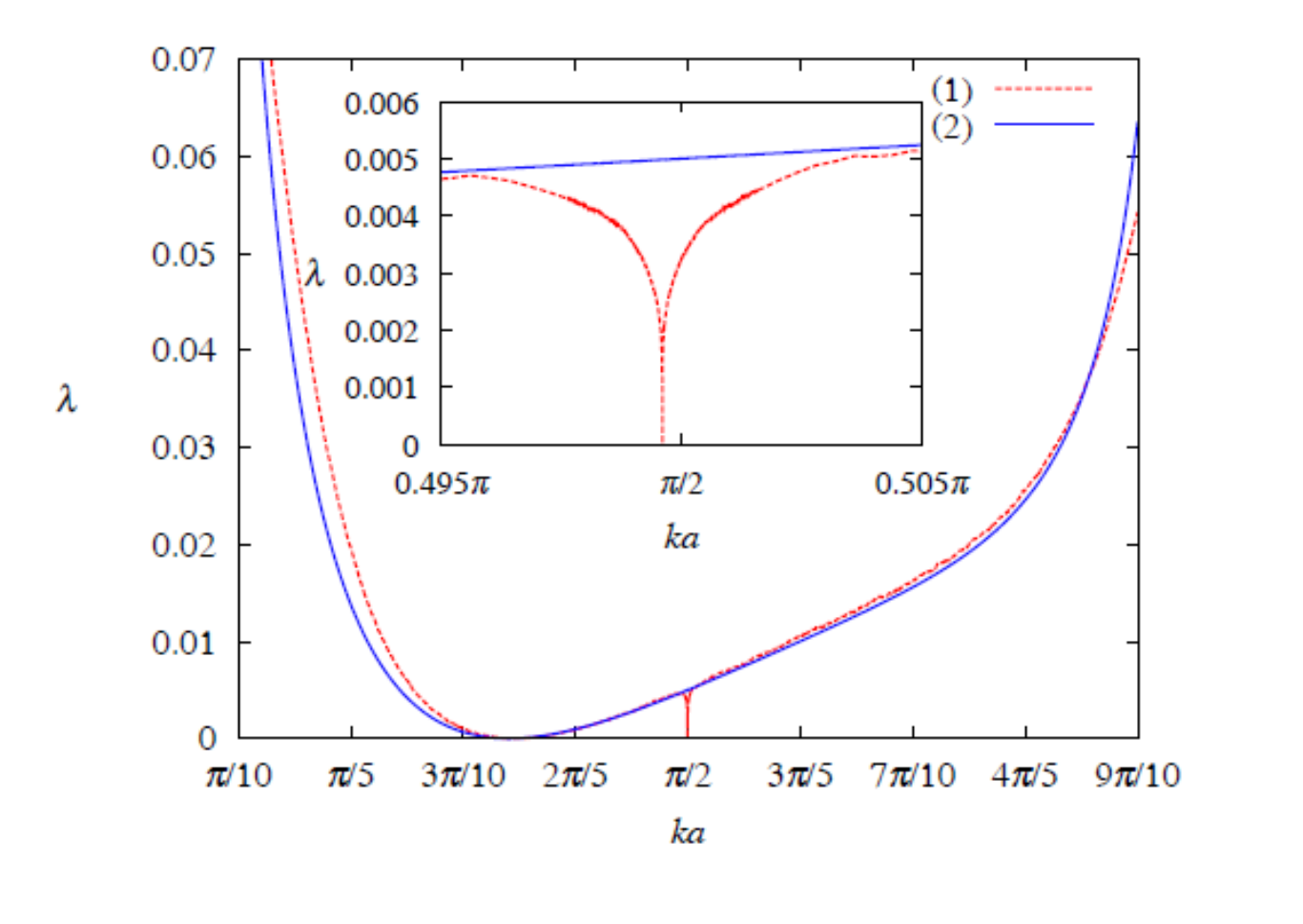}
\caption{(Color online) Inverse localization length $\lambda$ versus
$\gamma$ for uncorrelated positional disorder.
Dashed curve (1) represents numerical data, while solid curve (2)
corresponds to Eq.~(\ref{genthlyap}) with ${\cal K}_{\Delta}(2 \gamma)=1$ . Inset shows the anomaly in
detail (after \cite{HIT10a}).
\label{anoloc}}
\end{center}
\end{figure}

\begin{figure}[thb]
\begin{center}
\includegraphics[width=10.0cm,height=7.5cm]{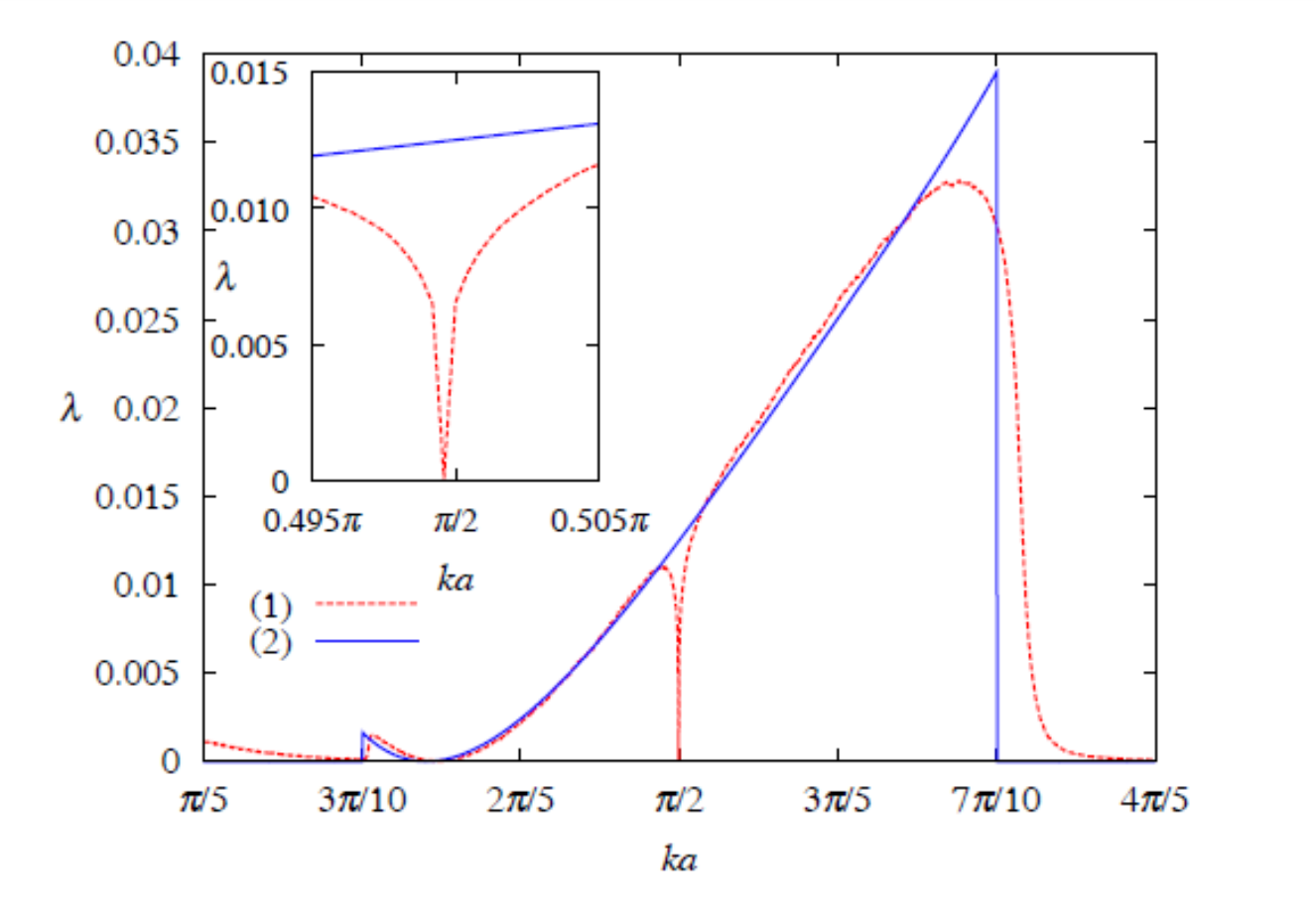}
\caption{(Color online) Inverse localization length $\lambda$ versus $\gamma$ for self-correlated positional disorder. Dashed curve (1) represents numerical data, while solid curve (2) corresponds to Eq.~(\ref{genthlyap}) with ${\cal K}_{\Delta}(2 \gamma)$ defined by Eq.(\ref{powspe}). Inset shows the details of anomaly (after \cite{HIT10a}).
\label{anoloc_cor}}
\end{center}
\end{figure}

In Ref.~\cite{HIT10a} it was argued that the anomalous character of localization at $\gamma=\gamma_{c}$ also emerges for correlated positional disorder. Numerical data that confirm this statement were obtained for specific long-range self-correlations of the form,
\begin{equation}
K_{\Delta}(m) = \frac{1}{c_{2} - c_{1}}\frac{1}{\pi m}
\left[ \sin \left( \pi c_{2} m \right) -
\sin \left( \pi c_{1} m \right) \right]\,.
\label{lrbincor}
\end{equation}
The power spectrum corresponding to such a binary correlator is,
\begin{equation}
{\cal K}(\gamma) = \left\{ \begin{array}{ccc}
\displaystyle
\frac{1}{c_{2} - c_{1}} &
\mbox{ if } & \displaystyle
\gamma \in \left[ c_{1} \frac{\pi}{2}, c_{2} \frac{\pi}{2} \right] \cup
                     \left[ \pi - c_{2} \frac{\pi}{2},
                     \pi - c_{1} \frac{\pi}{2} \right] \\
0 & & \mbox{ otherwise }\,.
\end{array} \right.
\label{powspe1}
\end{equation}
For $c_{1} = 3/5$ and $c_{2} = 1$ the long-range correlations~(\ref{lrbincor})
create two mobility edges at $\gamma = 3 \pi/10$ and $\gamma = 7 \pi/10$. The corresponding numerical data are reported in Fig.~\ref{anoloc_cor}. As one can see, the data follow the analytical predictions relatively well, thus, proving an emergence of mobility edges due to the correlations in the disorder. Another result is that the dip at the critical value of Bloch phase $\gamma_{c}$ does not disappear due to correlations, confirming the expectation of a generic nature of an anomalously localized state.

In order to give an additional evidence for an appearance of the anomaly of type (\ref{anodec}), in Ref.~\cite{HIT10a} a direct numerical study was performed, by establishing the form of tails of the anomalously
localized state. Specifically, Eqs.~(\ref{andoffdia})
and~(\ref{offdiaganderson}) have been solved with the use of the transfer-matrix technique that allows to reconstruct the profile of electronic states.
In the anomalous case one expects that the quantity $\log |\psi_{n}|$ behaves as a random variable with the
zero average and second moment linearly increasing with $n$.
\begin{figure}[!ht]
\begin{center}
\includegraphics[width=10.0cm,height=7.5cm]{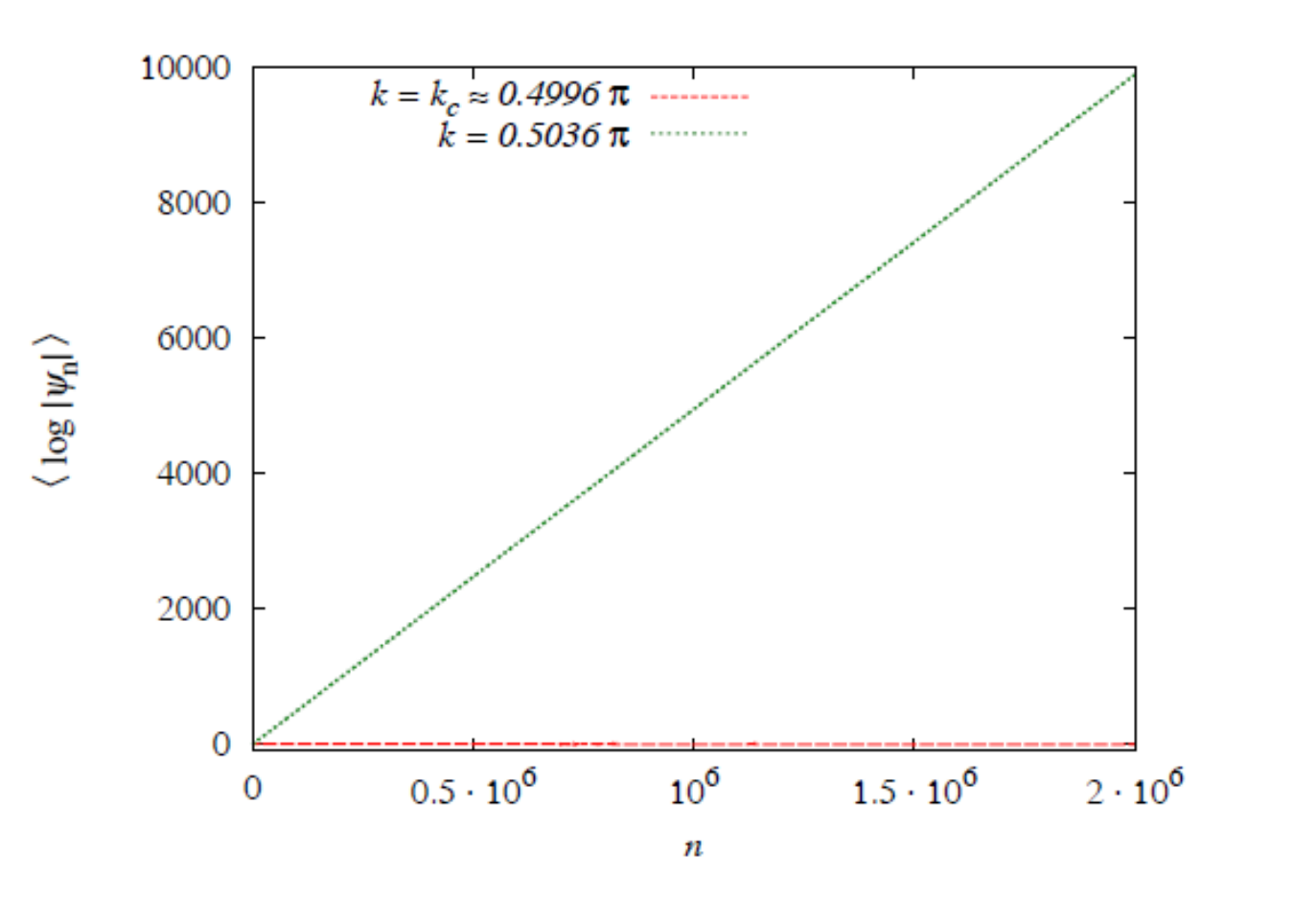}
\caption{(Color online) Dependence
$\langle \log |\psi_{n}| \rangle$ on $n$. Dashed line corresponds
to critical value of the Bloch wave number, $\kappa =\kappa_{\rm c} \simeq 0.4996
\pi$, while dotted line corresponds to $\kappa = 0.5036 \pi$ (after \cite{HIT10a}).
\label{m1}}
\end{center}
\end{figure}
Contrary, for $\kappa \neq \kappa_{\rm c}$ the solution of Eq.~(\ref{andoffdia})
should behave as $|\psi_{n}| \sim \exp (\lambda n)$.
Such a dependence, indeed, was numerically confirmed for first two moments
of the variable $\log |\psi_{n}|$, see Figs.~\ref{m1}
and~\ref{m2}.

\begin{figure}[!ht]
\begin{center}
\includegraphics[width=10.0cm,height=7.5cm]{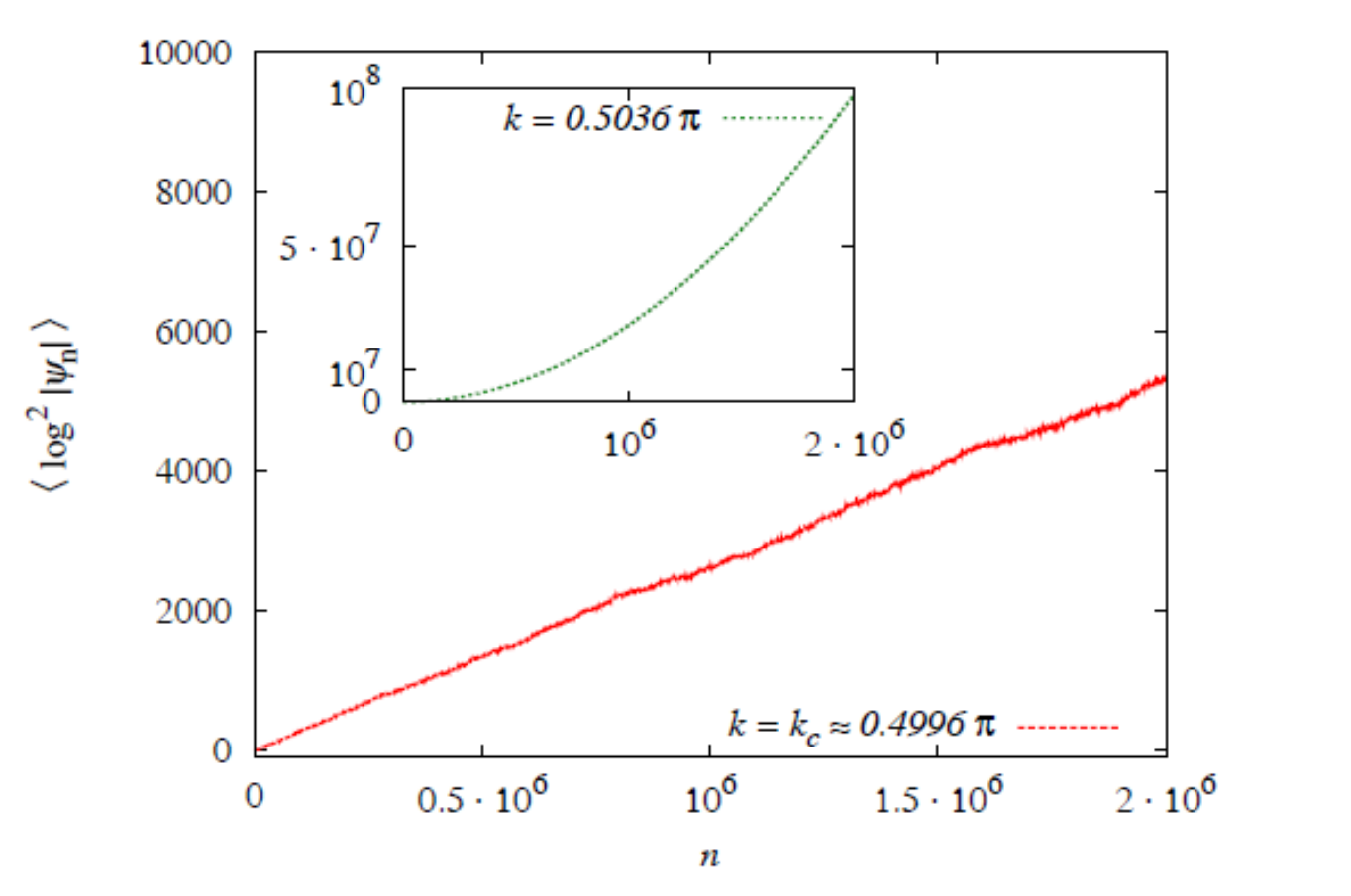}
\caption{(Color online) Dependence
$\langle \log^{2} |\psi_{n}| \rangle$ on $n$. Dashed line corresponds
to critical value of the Bloch wave number, $\kappa = k_{\rm c} \simeq 0.4996 \pi$,
while dotted line in inset corresponds to a Bloch wave number
$\kappa = 0.5036 \pi$ (after \cite{HIT10a}).
\label{m2}}
\end{center}
\end{figure}
In both Figs.~\ref{m1} and~\ref{m2} the behavior of
$\log |\psi_{n}|$ was explored as a function of $n$ for two Bloch wave numbers, for
critical value $\kappa = \kappa_{\rm c} \simeq 0.4996 \pi$, and for
$\kappa = 0.5036 \pi$, which is close to the critical value but not identical
to it. The moments of $\log |\psi_{n}|$ were computed with an average over 1000
disorder realizations. For the sake of simplicity, the
amplitude disorder was used without self-correlations.
Both the first and second moment of $\log |\psi_{n}|$ behave as
expected, corroborating the conclusion that the tails of the electronic
state at the critical point are described by Eq.~(\ref{anodec}).

\section{Single-mode waveguides with correlated disorder}
\label{9}

\subsection{Experimental setup}
\label{9.1}

All theoretical predictions of a possibility to have effective mobility edges in the one-dimensional geometry are based on the analysis of the Lyapunov exponent obtained for weak disordered potentials and infinite samples. On the other hand, in experiments the number of scatterer, either of the delta-type or in the form of finite barriers and wells, can not be very large. Practically, the experimental devices may consists of few hundreds of scatterers or periods, not much more. In addition, the strength of scatterers is not very small as assumed in the theory. Apart from all these restrictions, there are many others, such as an absorbtion of electromagnetic, optic or electron waves, that is unavoidable in practice. One can expect that these experimental imperfections can destroy the coherence effect induced by long-range correlations. Therefore, any experiment with specific long-range correlations in the disorder is if great importance.

As mentioned above, the first experimental realization of dimers in electron superlattices has been reported in Ref.~\cite{Bo99}. Then, in Ref.~\cite{KS98} it was shown that the famous Hofstadter butterfly \cite{H76}, originated from the non-commensurate potentials, can be experimentally observed in one-mode electromagnetic waveguides, in spite of a quite strong absorption in metallic walls. In both these experiments the {\it short-range} correlations arising due to intentionally created scatterers were chosen to manifest the coexistence of extended and localized scattering states in the energy spectrum. The experiments gave a clear evidence that the imperfections do not destroy the coherence effect leading to specific properties of transport induced by short-range correlations. The details of the original experiments on quantum chaos and Anderson localization using microwave waveguides and cavities can be found in Ref.~\cite{S07}.

Here we discuss the first experiments on the emergence of effective mobility edges, that were performed with the use of one-mode waveguide with {\it long-range} correlated disorder \cite{KIKS00,KIKSU02,KIK08}. The setup is a circular waveguide with the transverse dimensions
$a=20$\,mm and $b=10$\,mm, and the circumference of $2.15$ m. At the top of the waveguide there are 100 cylindrical scatterers of radius $r=2.5$\,mm, periodically spaced with the period $d=20.5$\,mm, see Figs.~\ref{view}-\ref{setup}. The transmission and reflection were experimentally measured in the frequency range between $\nu_{\rm min}={c}/{2a} \approx$ 7.5\,GHz and $\nu_{\rm max}={c}/{a}={c}/{2b} \approx$ 15\,GHz, where the only lowest mode was open for propagation. The dispersion relation for an empty
waveguide is given by the standard relation,
\begin{equation}
\label{disp-wave}
k=({2\pi}/{c})\sqrt{\nu^2-\nu^2_{\rm
min}}\,,
\end{equation}
where $c$ is the speed of light. Therefore, the normalized wave
vector $kd/\pi$ ranges from 0 to 1.8. This setup was already used to demonstrate an experimental realization of
the ``Hofstadter butterfly'' \cite{KS98}. By varying the
lengths of the micrometer screws, different scattering arrangements
can be realized. The complete information about the scattering matrix of a given arrangement was extracted with the use of the Wiltron 360B network analyzer, via two antennas located at the ends of the waveguide.
\begin{figure}[!ht]
\begin{center}
\includegraphics[width=13.6cm,height=7.0cm]{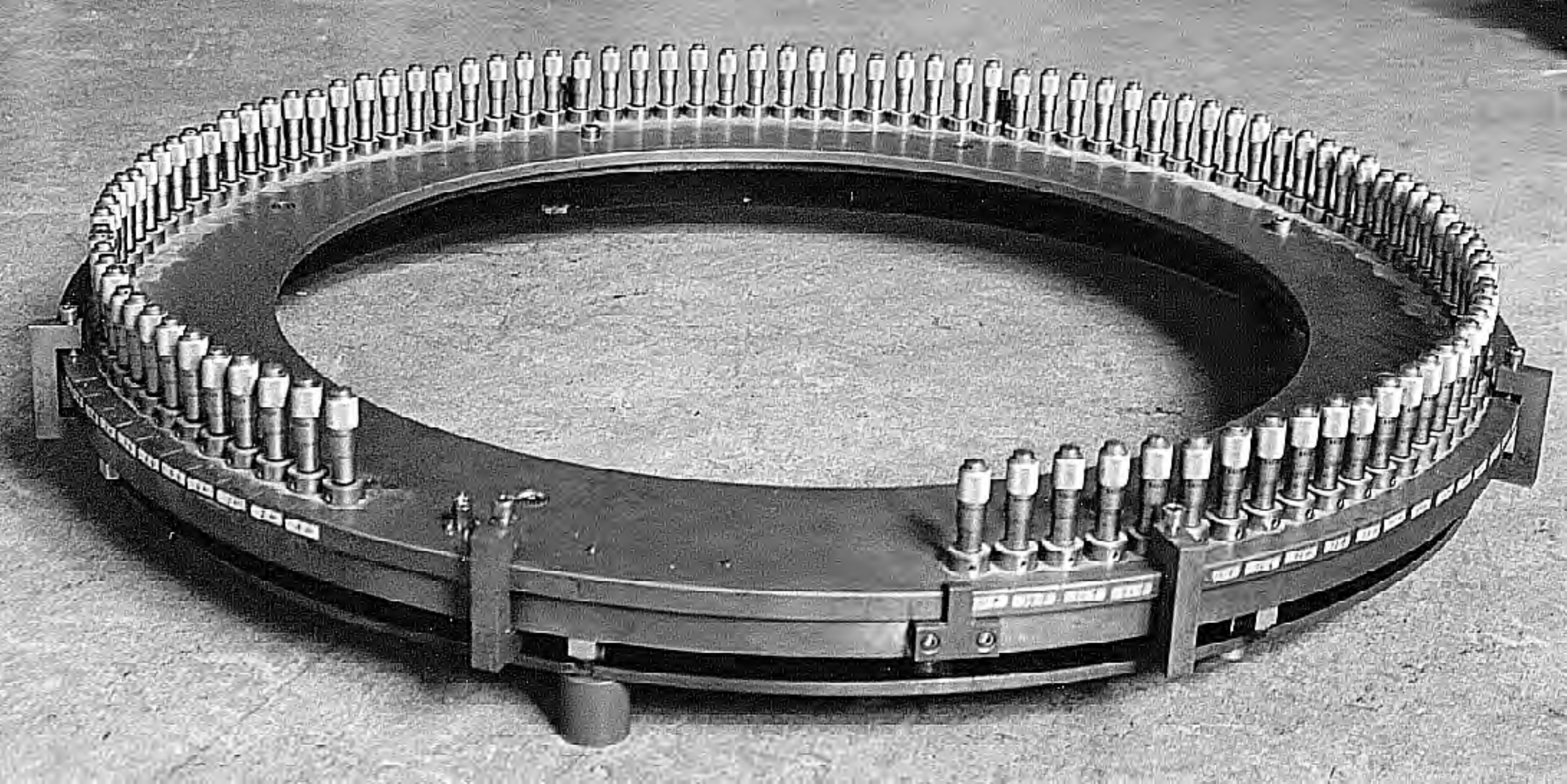}
\end{center}
\caption{
Experimental setup, general view (after \cite{KS98}).}
\label{view}
\end{figure}

\begin{figure}[!ht]
\begin{center}
\includegraphics[width=13.6cm,height=7.0cm]{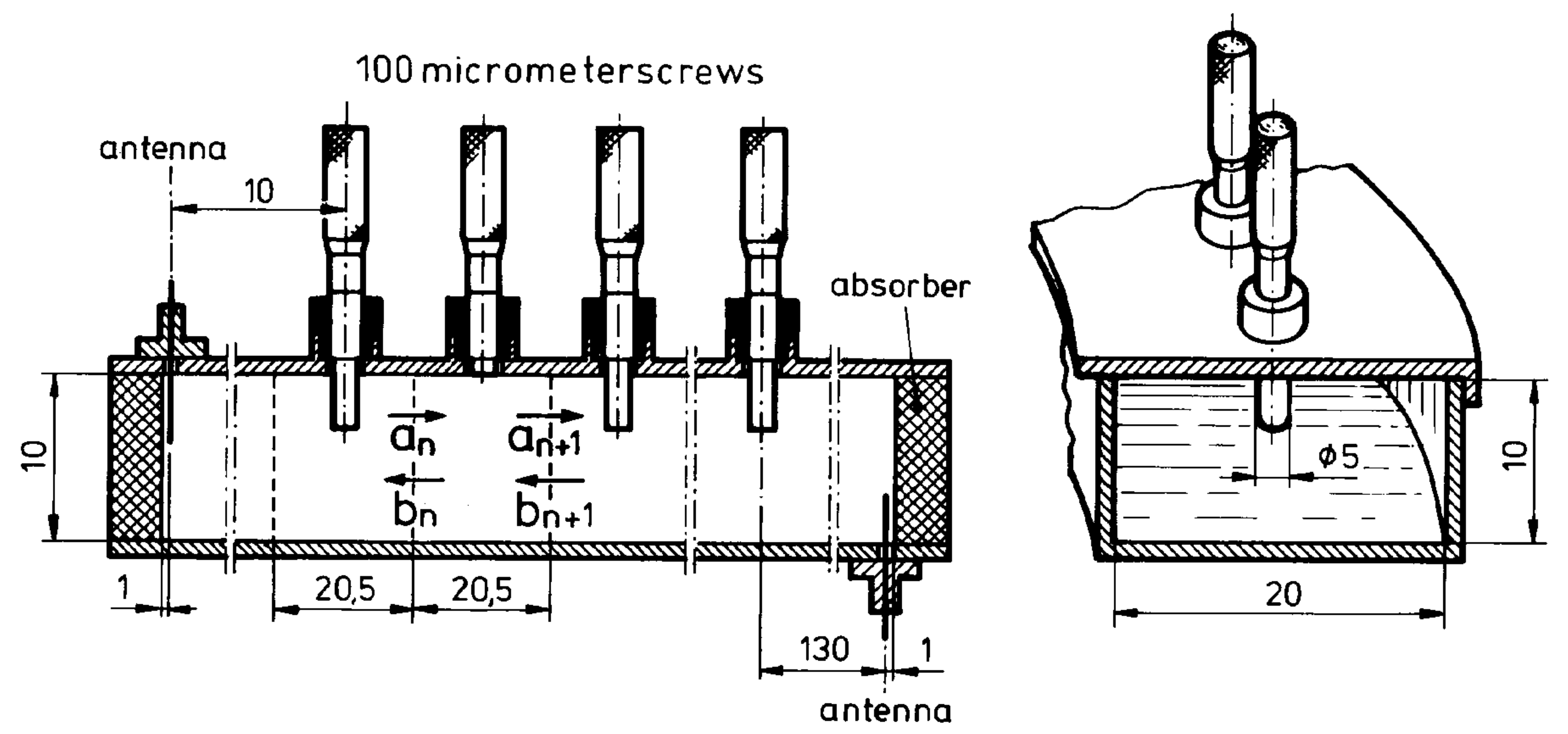}
\end{center}
\caption{
Experimental setup, details. All dimensions are given in mm (after \cite{KIKS00}).}
\label{setup}
\end{figure}

The strength $U_n$ of a single scatterer is associated with its
length embedded inside the waveguide, and varied by means of
the micrometer screws. When $U_n=U=const$, the waveguide is
fully transparent. For a white-noise potential 100 numbers
$u_n =U_n-U$ were drawn as
uncorrelated random numbers with the variance $\sigma^2$. The
corresponding localization length was much larger than the length
$L$ of the waveguide, thus the eigenstates are delocalized within
the waveguide. However, if the white-noise potential $u_n$
is replaced by a correlated sequence, the localization is strongly modified due to correlations.

If the screws are approximated by delta scatterers, the propagation of a single mode through the waveguide can be effectively described by the wave equation (\ref{KP-delta}) for the Kronig-Penney model, in which $U_{n}$ is the amplitude of the $n$th delta-scatterer located at $x=x_{n}=n d$. Therefore, the localization length is defined by the expression (see Eq.~(\ref{five})),
\begin{equation}
\label{eq:locKP}
  L_{loc}^{-1} (E)=\frac{\sigma^2k^2}{8d}\frac{\sin^2(kd)}{\sin ^2\gamma }{\cal K}(2\gamma)\,.
\end{equation}
Since the displacements $u_n$ are proportional to $E=k^2$, here we introduced the variance $\sigma^2 = \left( \langle U_n^2 \rangle - \langle U_n\rangle^2\right)/k^4$ in order to see clearly the $k$-dependence in the expression for the inverse localization length. Note that in comparison with Eq.~(\ref{five}), here the unperturbed amplitude $U$ is replaced by $k^2$, this reflects the electromagnetic nature of perturbation. The Bloch phase $\gamma $ in Eq.~(\ref{eq:locKP}) is given by the Kronig-Penney dispersion relation (\ref{KP-disp}) with $\mu=kd$.

\subsection{Suppression of localization}
\label{9.2}

For a white noise disorder we have, $K(m)=0$ for $|m|\neq 0$, therefore, ${\cal K}(2\gamma) =1$. In order to observe the mobility edges, one needs to suppress the localization, therefore, to have $\lambda=L_{loc}^{-1} = 0$ in a finite energy range. As an example, we consider the situation for which the function ${\cal K}(2\gamma)$ is zero in the regions defined by the parameters $\gamma_1$ and $\gamma_2$, otherwise, it is constant,
\begin{equation}
{\cal K}(2\gamma)=\left\{
\begin{array}{cc}
C_1^2, & \qquad \gamma_1<\gamma<\gamma_2 \\
0, & \qquad \gamma<\gamma_1; \,\,\, \gamma>\gamma_2 \,\,.
\end{array}
\right.
\label{mu12}
\end{equation}
Here, we explore only the region $0<\gamma <\pi/2$ since the function ${\cal K}(2\gamma)$ is symmetric with respect to the point $\gamma=\pi/2$.
The parameter $C_1^2=\pi/2(\gamma_2-\gamma_1)$ is the normalization constant obtained from the condition $K(0)=1$. This dependence exhibits
four sharp mobility edges in the first allowed zone. Their positions are given by two pairs of roots of the dispersion relation (\ref{KP-disp}) with $\gamma=\gamma_1$ and $\gamma=\gamma_2$.

Now the problem is reduced to the construction of potentials with the desired properties. According to the general approach discussed in Section~\ref{5.3}, the disorder $U_n$ resulting in the given dependence (\ref{mu12}) of the function ${\cal K}(2\gamma)$ can be constructed via the convolution of white noise $\alpha_n$ with the modulation function $G(m)$,
\begin{eqnarray}
\label{alg}
u_n &=& \sigma k^2
\sum\limits_{m=-\infty}^{+\infty} G(m) \alpha_{n+m}, \\
\label{Z+gamma}
G(m) &=& \frac 1\pi \int_0^{\pi}\sqrt{W (2\gamma )} \cos(2m\gamma )d\gamma.
\end{eqnarray}
Here $\alpha_n$ are random numbers with the zero mean and variance one. Using Eq.~(\ref{Z+gamma}) one obtains,
\begin{eqnarray}
\label{beta1}
G(m\neq 0)=
\frac {C_1}{\pi m} \left\{ \sin(2m\gamma_2) - \sin(2m\gamma_1) \right\}\,, \qquad G(0)=\frac{2C_1}{\pi} (\gamma_2-\gamma_1)\,.
\end{eqnarray}

Experimentally it is difficult to measure the localization length directly. Instead, the transmission properties were studied, such as the transmission amplitude and its square (transmission coefficient). It is clear that if the localization length is much larger than the sample size, the transmission will be similar to that for a waveguide without disorder. This situation is expected when the Lyapunov exponent vanishes in the first order perturbation theory in $\sigma^2$. Therefore, the emergence of mobility edges can be observed without the knowledge of an exact value of localization length. Specifically, one can expect that in the energy regions where the function ${\cal K}(2\gamma)$ vanishes, the transmission will be much larger that in the regions with finite values of ${\cal K}(2\gamma)$.

For comparison with the theory, the numerical study was done with the use of the Hamiltonian map approach described in Section~\ref{4.4}. In particular, the transmission coefficient $T_N$
can be expressed in terms of the classical map (\ref{KP-map}) and computed by the relation (\ref{eq:6}). The numerical data for the correlated disorder with the function (\ref{mu12}) are shown in Fig.~\ref{band12}a. A sequence of scattering strengths $u_n$ of the length
$N=10^4$ was generated by calculating $G(m)$ from
Eq.~(\ref{beta1}) with $\gamma_1/\pi =0.2$ and $\gamma_2/\pi =0.4$, and
substituted into Eq.~(\ref{alg}). The data present the resulting transmission for
$\sigma=0.1$ and $U=-0.1$. This
value for $U$ was obtained from the dispersion relation (\ref{KP-disp}) by
adjusting the width of the allowed band to the experimental data.

In order to have more convincing results, the data of Fig.~\ref{band12}a were compared with the complimentary situation presented in Fig.~\ref{band12}d. Here the regions of a perfect transmission are interchanged with those for a perfect reflection, see Fig.~\ref{band12}a. For this case the function ${\cal K}(2\gamma)$ is given by the relations,
\begin{equation}
{\cal K}(2\gamma)=\left\{
\begin{array}{cc}
0, & \qquad \gamma_1<\gamma<\gamma_2 \\
C_2^2, & \qquad \gamma<\gamma_1; \,\,\, \gamma>\gamma_2 \,
\end{array}
\right.
\label{mu12-2}
\end{equation}
with $C_2^2=\pi/2(\gamma_1-\gamma_2+\pi/2)$. Correspondingly, one can find,
\begin{eqnarray}
\label{beta2}
G(m\neq 0)= -
\frac {C_2}{\pi m} \left\{ \sin(2m\gamma_2) - \sin(2m\gamma_1) \right\}\,, \qquad G(0)=\frac{2C_2}{\pi} \left(\gamma_1-\gamma_2+\frac{\pi}{2}\right)\,.
\end{eqnarray}

By comparing Fig.~\ref{band12}a with Fig.~\ref{band12}d, one can see a quite good correspondence to the expected dependence of the transmission coefficient $T_N$ on the rescaled wave vector $kd/\pi$. Specifically, very sharp mobility edges emerge in correspondence with the function ${\cal K}(2\gamma)$. The
mobility edges are clearly seen near the points $kd/\pi=0.38,
0.57, 0.76$ which are the roots of the dispersion relation (\ref{KP-disp}) with
$\gamma/\pi=0.4, 0.6, 0.8$, respectively. The transmission spectrum ends
at the band edge, $kd/\pi=0.91$ for $\gamma=\pi$. Data below
$kd/\pi=0.2$ are not shown because of a strong absorption in the waveguide at low frequencies.

The experimental data with $N=100$ scatterers of the waveguide are shown in Fig.~\ref{band12}b and Fig.~\ref{band12}e. In the experiment the mutually complementary random sequences have been used in order to create
the appropriate scattering arrangements in the waveguide. The values of
$u_n$ were mapped into the screw lengths by identifying the
minimum $u_n$ value with a length of 0~mm and the maximum
value with a length of 3~mm. The average screw length of 1.5~mm
determines the width of the Bloch bands according to the dispersion relation. Five measurements were performed with each random
sequence. In one measurement the transmission through a realization
of hundred scatterers have been measured. The dotted lines in
Fig.~\ref{band12}b and Fig.~\ref{band12}e show the results of a
single measurement, the solid lines represent the results of
averaging over all five measurements. As one can see, there is a fingerprint of the windows of high transmission in Fig.~\ref{band12}b, and a quite poor correspondence of experimental data to the numerical ones in Fig.~\ref{band12}e. However, the expected transmission pattern is already visible. Obviously, a quite short length of the sequence $u_n$ in the experiment, $N=100$, is not enough to have good results.

\begin{figure}[!ht]
\begin{center}
\subfigure{\includegraphics[scale=0.78]{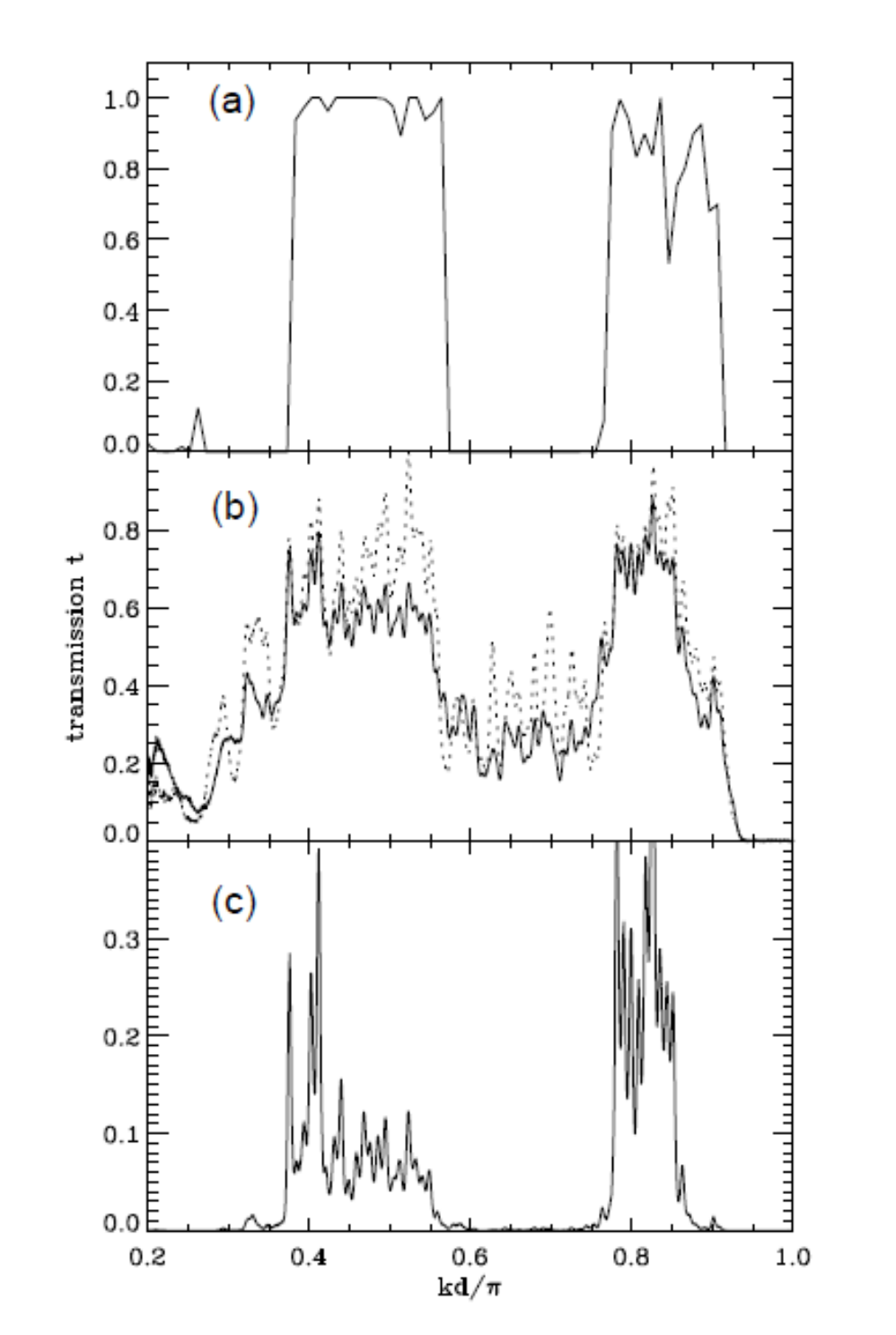}}
\subfigure{\includegraphics[scale=0.78]{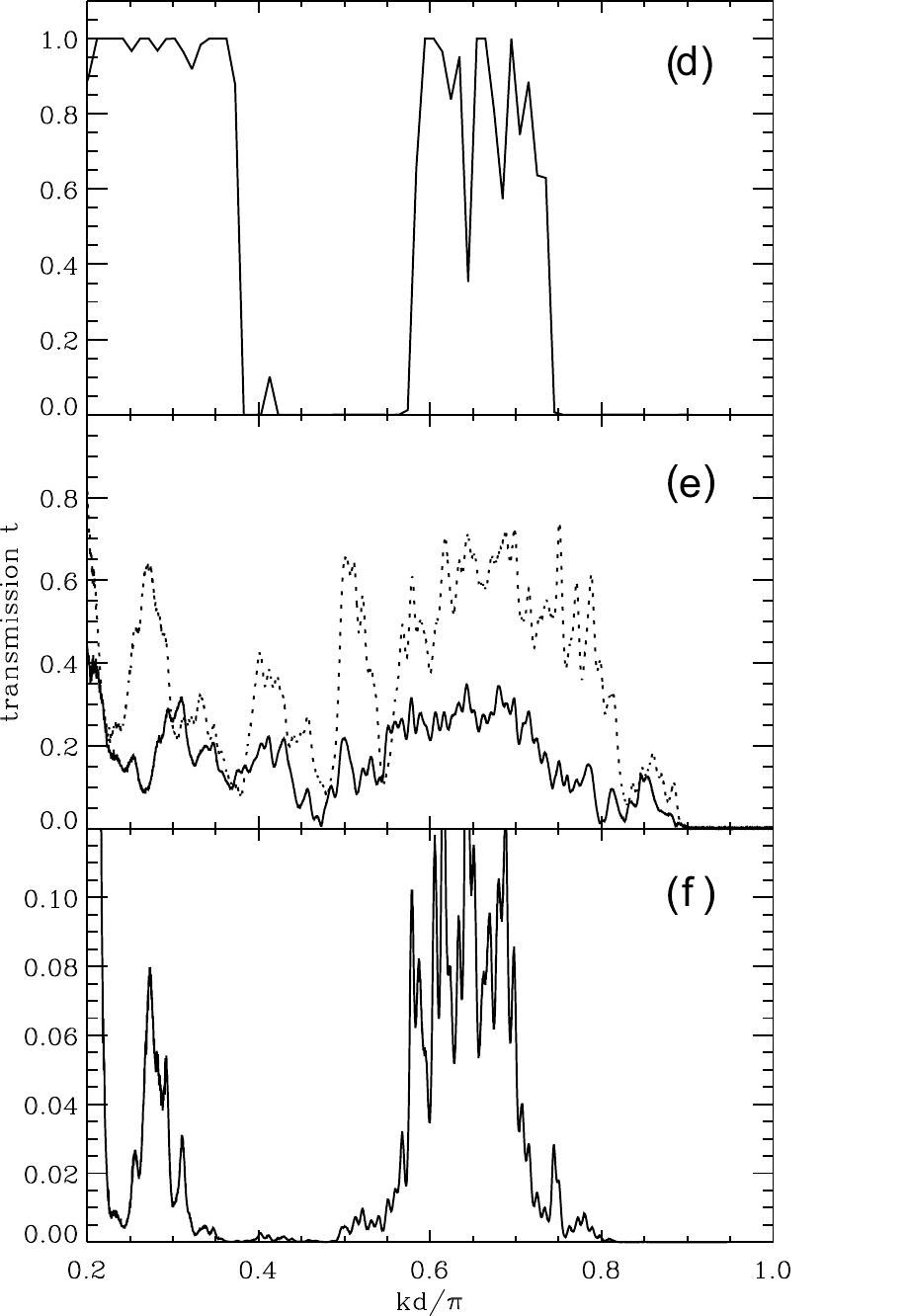}}
\end{center}
\caption{Left panel: Transmission through correlated random
sequences determined by Eq.~(\ref{beta1}); (a)
numerical results for $N=10^4$ scatterers,
(b) microwave transmission through an array of
$N=100$ scatterers (dotted line), and the average over five different
measurements (solid line), (c) microwave transmission through an
array of $N=500$ scatterers obtained by multiplicating the
transfer matrices of five individual measurements. Right panel: (d),(e),(f) are same as (a),(b),(c), respectively, but for the complimentary case defined by Eq.~(\ref{beta2}).}
\label{band12}
\end{figure}

In order to see how the expected pattern of the transmission depends on the length $N$, the sequence $u_n$ of length $N=500$ was created numerically and cutted in five pieces. Then, for each piece of size $N=100$ the measurement of the scattering matrix $S_M(M=1,..,5)$ was performed experimentally, and the total transfer matrix $T$ was found as the matrix product, $T=\prod_{M=1}^5 T_M$. In such a way, it is possible to study the transmission through arbitrary long
sequences of scatterers with a set-up containing only 100 of them.
However, because of absorption this technique is limited in our case to a total number of about 1000 scatterers.
Figures \ref{band12}c and \ref{band12}f show such transmission spectra.
In both cases the expected transparent and non-transparent regions are clearly reproduced in accordance with the theoretical expectations. The data manifest that already for $N=500$ the experiment confirms an emergence of mobility edges due to the energy windows in which the Lyapunov exponent vanishes. As a check, the transmission through the uncorrelated random sequence of 500 scatterers was also studied. The data are shown in Fig.~\ref{random}. Here the transmission is approximately 100 times weaker as for the sequences with the correlated disorder within the transparent region.
\begin{figure}[!ht]
\begin{center}
\includegraphics[scale=0.80]{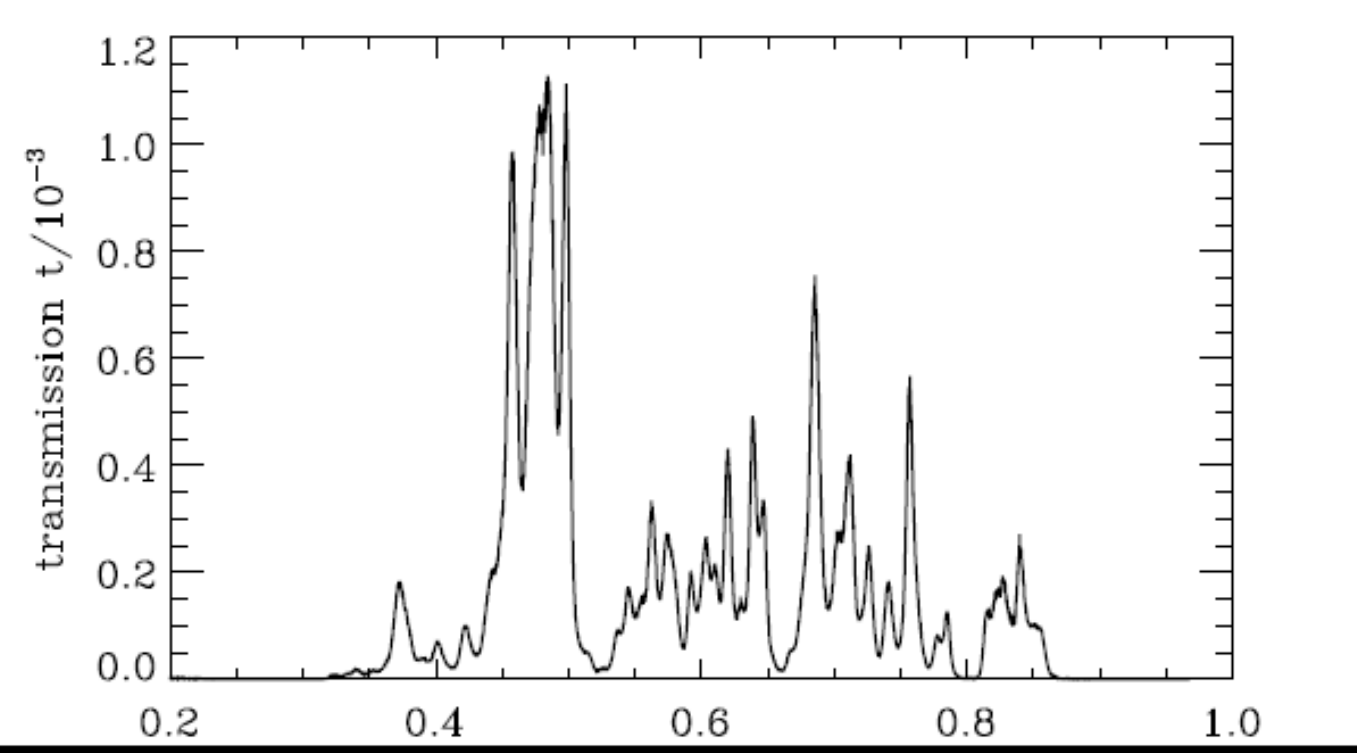}
\end{center}
\caption{
Microwave transmission through a
random arrangement of $N=500$ scatterers obtained by multiplying
the transfer matrices of five individual measurements.}
\label{random}
\end{figure}

\subsection{Enhancement of localization}
\label{9.3}

From the general expression (\ref{five}) for the Lyapunov exponent a quite unexpected conclusion can be drawn about a strong enhancement of the localization. Indeed, the only restriction for the value of the function ${\cal K}(\gamma)$ is due to its normalization,
\begin{equation} \label{eq:norm}
\int_{0}^{\pi} {\cal K}(\gamma) d \gamma =\pi\,.
\end{equation}
This means that the value of ${\cal K}(2\gamma)$ can be very large in some interval of energy, provided this interval is small, and outside the interval ${\cal K}(2\gamma)$ vanishes. Specifically, in order to observe the effect of enhancement
of localization it is necessary to have the function ${\cal K}(2\gamma) ={\cal K}_{max}\gg 1$ within the interval $\Delta \gamma = \gamma_2 - \gamma_1 \ll \pi/2$, see Eq.~(\ref{mu12}). Then, because of the normalization condition (\ref{eq:norm})
the width $\Delta \gamma$ and the enhancement factor ${\cal K}_{max}=C_1^2$ are related by the simple condition, $2{\cal K}_{max}= \pi/\Delta \gamma\ $. In such a case the correlation-induced enhancement of localization within the interval $\Delta \gamma$ is accompanied by a full transparency of the waveguide for all other
frequencies. The Fourier coefficients of the
step-function ${\cal K}(2\gamma) = {\cal K}_{max}$ within the interval
$\Delta \gamma$ and ${\cal K}(2\gamma)=0$ otherwise, is given by the relation,
\begin{equation}\label{eq:corr}
  K(m)=\frac{1}{2m} \frac{\sin(2m\gamma_{2})-\sin(2m\gamma_{1})}{\Delta \gamma}.
\end{equation}
The inverse-power-law decay of $K(m)$ is a signature of
long-range correlations. In order to check experimentally the above predictions, two correlated sequences were generated
using Eqs.~(\ref{alg}) and (\ref{Z+gamma}), see Fig.~\ref{profiles}. In insets, the binary correlators corresponding to these sequences are also show.

\begin{figure}[!ht]
\begin{center}
\includegraphics[scale=0.60]{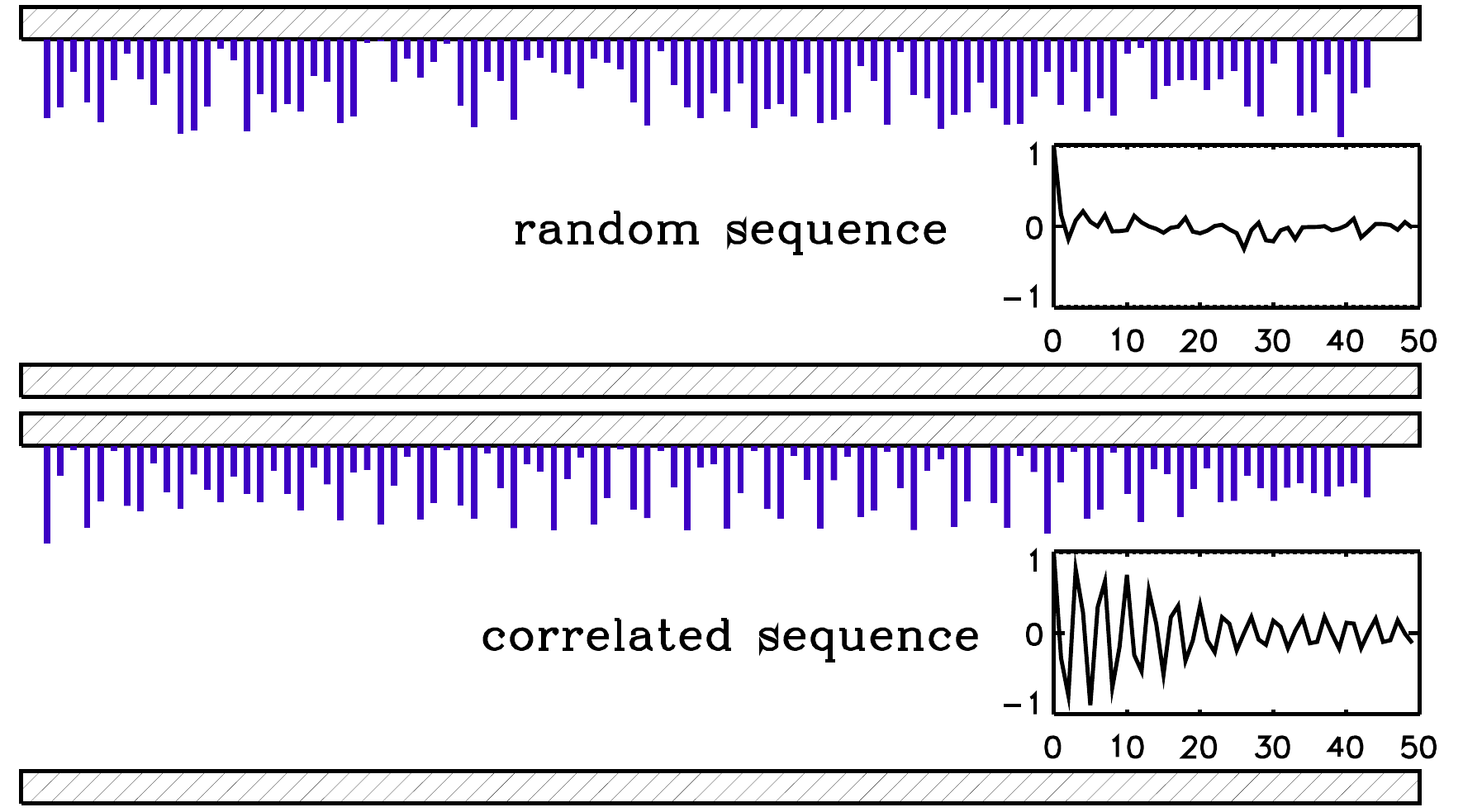}
\end{center}
\caption{(Color online) Profile of intrusion of all 100
micrometer screws into the waveguide for correlated and uncorrelated random sequences $U_n$.
Insets show the corresponding binary correlators calculated from the micrometer screw depths according to Eq.~(\ref{eq:corr}) (after \cite{KIK08}).}
\label{profiles}
\end{figure}

For both sets the elements $S_{12}=S_{21}^*$ and $S_{22}$ of the
scattering matrix were measured as the functions of the antenna position.
Here $S_{12}$ ($S_{22}$) is the transmission (reflection) amplitude of the scattering process when
the fixed antenna is in front of the first scatterer ($n=1$) and
the moving antenna is located between the $n$th and $(n+1)$st
scatterer. The moving antenna emits and receives the signal when
measuring $S_{22}$.

\begin{figure}[!ht]
\begin{center}
\subfigure{\includegraphics[scale=0.5]{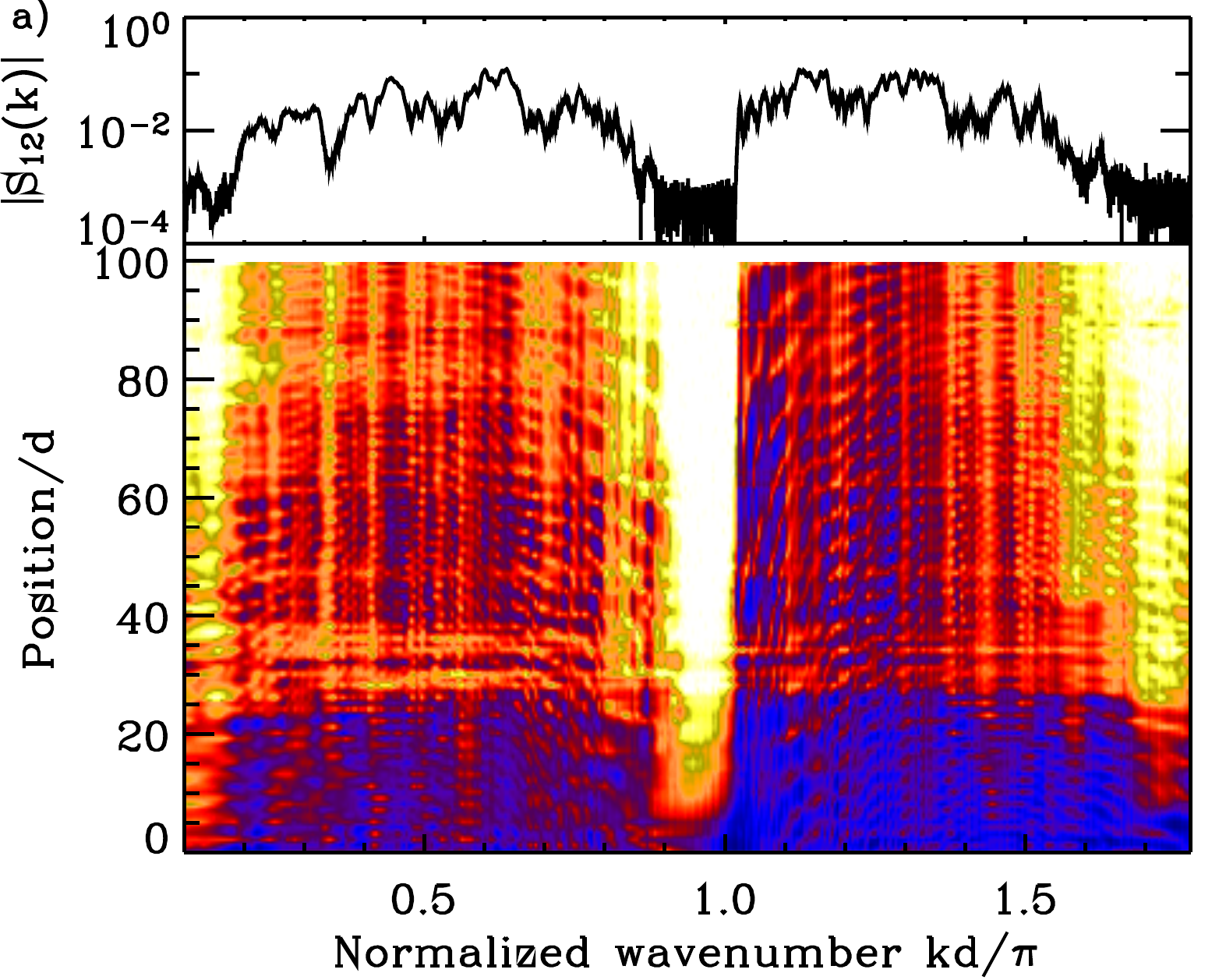}}
\subfigure{\includegraphics[scale=0.5]{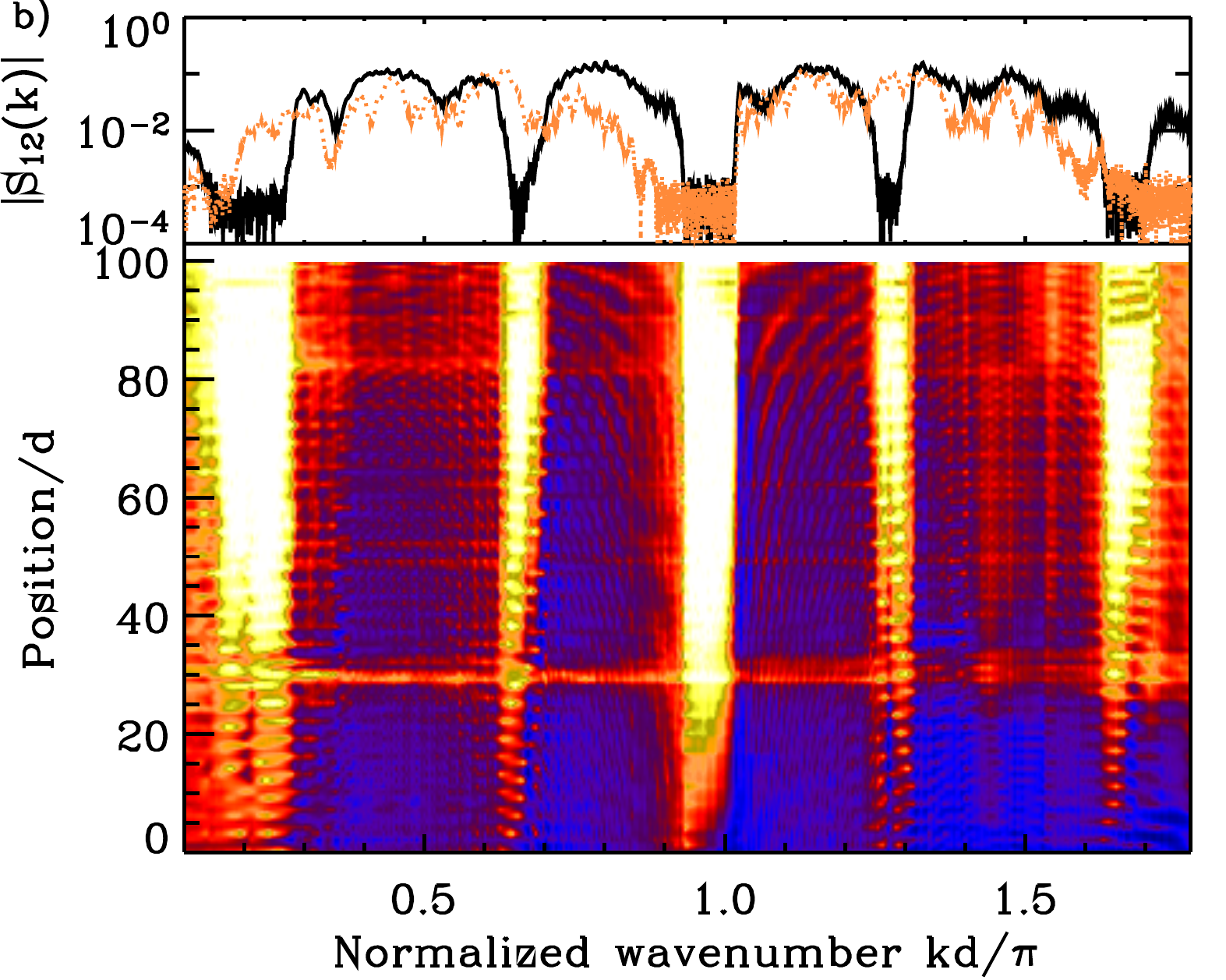}}
\end{center}
\caption{(Color online) The transmission spectrum $|S_{12}(k)|$ versus position $n$ (vertical scale) of the moving antenna is shown for uncorrelated (a) and
correlated (b) disorder. The transmission spectrum $|S_{12}|$
through the whole waveguide is shown by solid curve at the top of
each panel. For comparison, the transmission through uncorrelated disorder
is added on the right panel (b) by dotted curve (after \cite{KIK08}).}
\label{trans12}
\end{figure}

The single-mode transmission patterns $|S_{12}|$ are shown in
Fig.~\ref{trans12} for the purely random (left) and
correlated (right) sequences (shown in Fig.~\ref{profiles}) as
a function of the antenna position (vertical axis) and wave number $k$
(horizonal axis). In addition, on top of each figure,
the dependence of transmission value $|S_{12}|$ through the {\it whole}
waveguide is presented. As one can see, for the uncorrelated disorder there is
a gap close to the band edge, $kd/\pi = 1$. It
originates from the periodic spacing between the scatterers.
For small wave numbers the transmission is small because of
the weak antenna coupling to the waveguide, whereas for the high wave numbers
it is small due to a quite large absorption. In the case of the
correlated disorder there are additional gaps located at $kd/\pi
\approx 0.25, 0.65, 1.25$, and $1.65$. These gaps manifest the
enhancement of localization due to long-range correlations
with $\gamma_1 = 0.2$ and $\gamma_2 = 0.3$ (see Eq.~(\ref{eq:corr})). It is
important that inside these gaps the transmission is practically
absent since the localization length is reduced by a factor of $W_0
\approx 15.7$ and it becomes much less than the length $N=100$ of the
waveguide.

\begin{figure}[!ht]
\begin{center}
\includegraphics[scale=0.60]{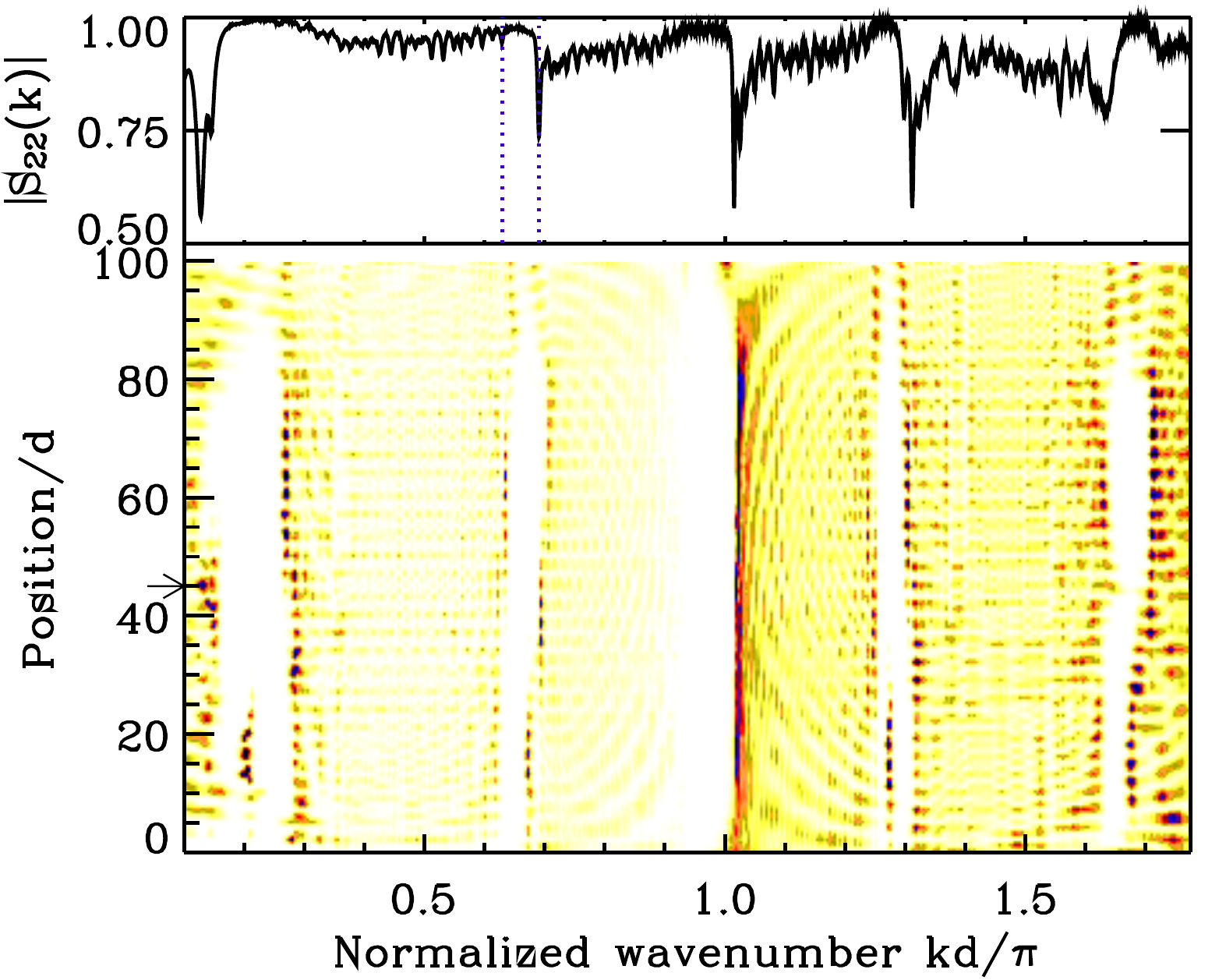}
\end{center}
\caption{ (Color online) The same as in Fig.~\ref{trans12}b but for the reflection pattern $|S_{22}(k)|$. The value $|S_{22}|$ is shown for the antenna located between the 41st and 42nd scatterers (after \cite{KIK08}).}
\label{reflect}
\end{figure}

The emergence of the gaps corresponding to the enhanced localization is also seen in Fig.~\ref{reflect}, where the reflection pattern $|S_{22}|$ is shown for the correlated arrangement of scatterers. On top of the pattern, the reflection at the point between the 41st and 42nd scatterer is shown. Since the localized states usually lead to a strong decrease of $|S_{22}|$, we observe the close to unity reflection within the gaps
caused by the correlations. Note that the fluctuations are
strongly suppressed in comparison to the fluctuations in the band, e.\,g.\ at $kd/\pi=0.55$. This is better seen in Fig.~\ref{reflect} showing a narrow interval of the wave numbers close to the correlation gap at $kd/\pi \approx 0.65$.

Several patterns with relatively high intensity of $|S_{22}|$
correspond to localized states, for which the localization length is much smaller in comparison with the localization length in a white noise
arrangement. In Ref.~\cite{KIK08} these states were termed
the ``enhanced localized states".
They belong to the spectrum of the system with long-range
correlations defined by Eq.~(\ref{eq:corr}). It was checked that these states are neither evanescent modes, nor defect states discussed in Ref.~\cite{LSKS08}. Both types of the states may, in principle, appear within a ``band gap" if some
periodicity is introduced by the maxima of the correlation function. For evanescent modes the reflection $S_{22}$ would be close to 1, independently of the antenna position inside the sample. It should be stressed that the local density of states does not change crucially by the correlations. Since each pair of scatterers brings in one state in the range $0.61 \leq kd/\pi \leq 0.71$ (shown in Fig.~\ref{zoom}), one should expect the emergence of about 10 localized states, assuming that the mean density of states is constant. Accordingly, one can observe about 8 such states, they are seen in Fig.~\ref{zoom}.

\begin{figure}[!ht]
\begin{center}
\includegraphics[scale=0.60]{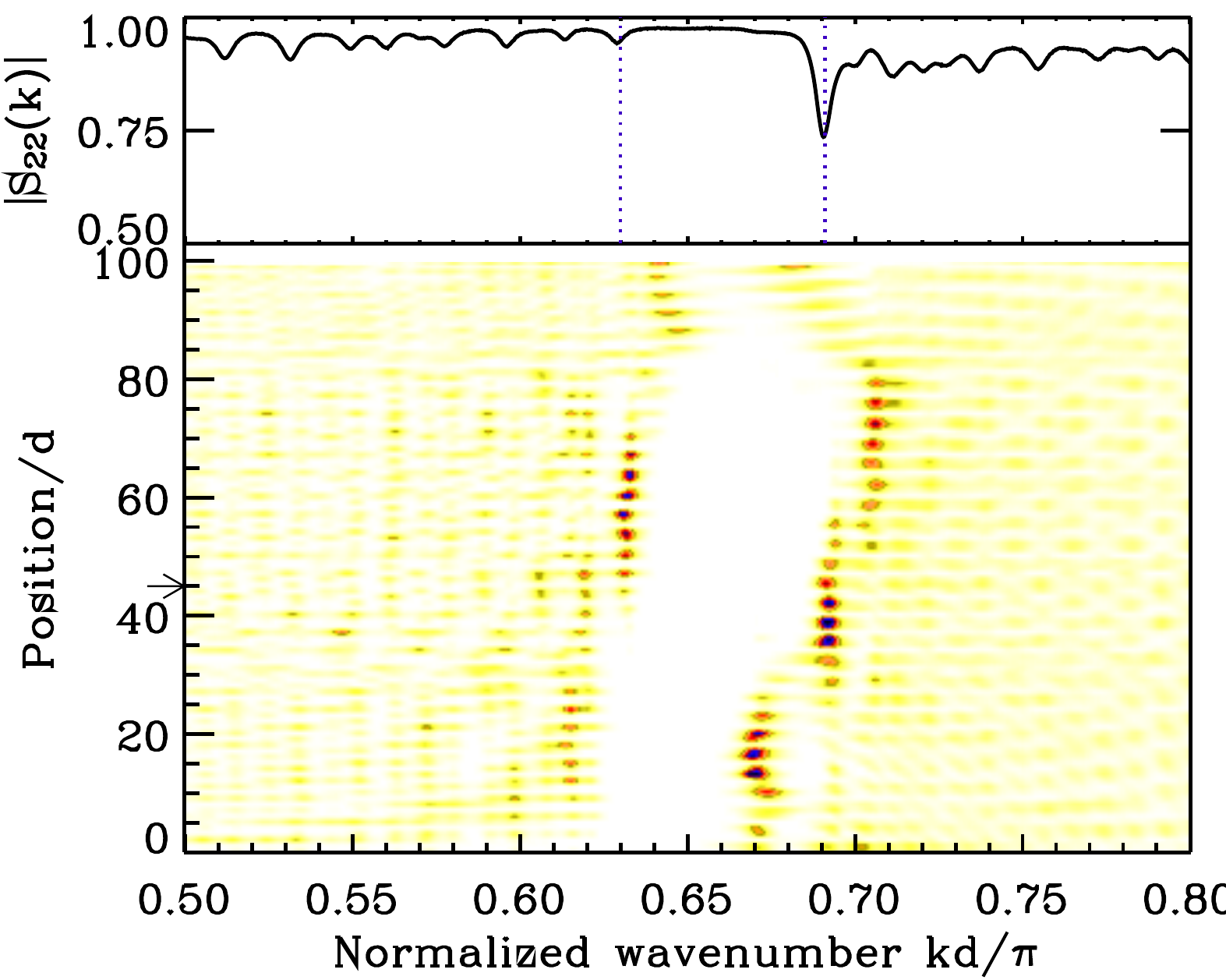}
\end{center}
\caption{(Color online) Details of Fig.~\ref{reflect} for wave numbers between 0.50 and 0.80. In this range several enhanced localized states can be detected. Sharp drop of $|S_{22}|$ at $kd/\pi \approx 0.69$ is due to localized states centered at $n \approx 40$ (after \cite{KIK08}).}
\label{zoom}
\end{figure}

By changing the position of moving antenna it was possible to measure the profiles of the enhanced localized states. Two such states are shown in Fig.~\ref{loc-eigen}. The quantity $1-|S_{22}|$ plotted versus the coordinate of the moving antenna is proportional to the intensity $|\psi|^2$ of the wavefunction \cite{K07}. The exponentially localized states are clearly seen inside the
waveguide, for which the localization length is about 10 spacings of length $d$ between the scatterers. Thus, the emergence of enhanced localized states due
to long-range correlations is found experimentally. It is highly
non-trivial that such a strong enhancement of localization occurs for
relatively weak fluctuations of the potential. Also, the remarkable fact is that the long-range correlations can result in a strong effect on a relatively short scale of $N=100$ scatterers only.

\begin{figure}[!ht]
\begin{center}
\includegraphics[scale=0.60]{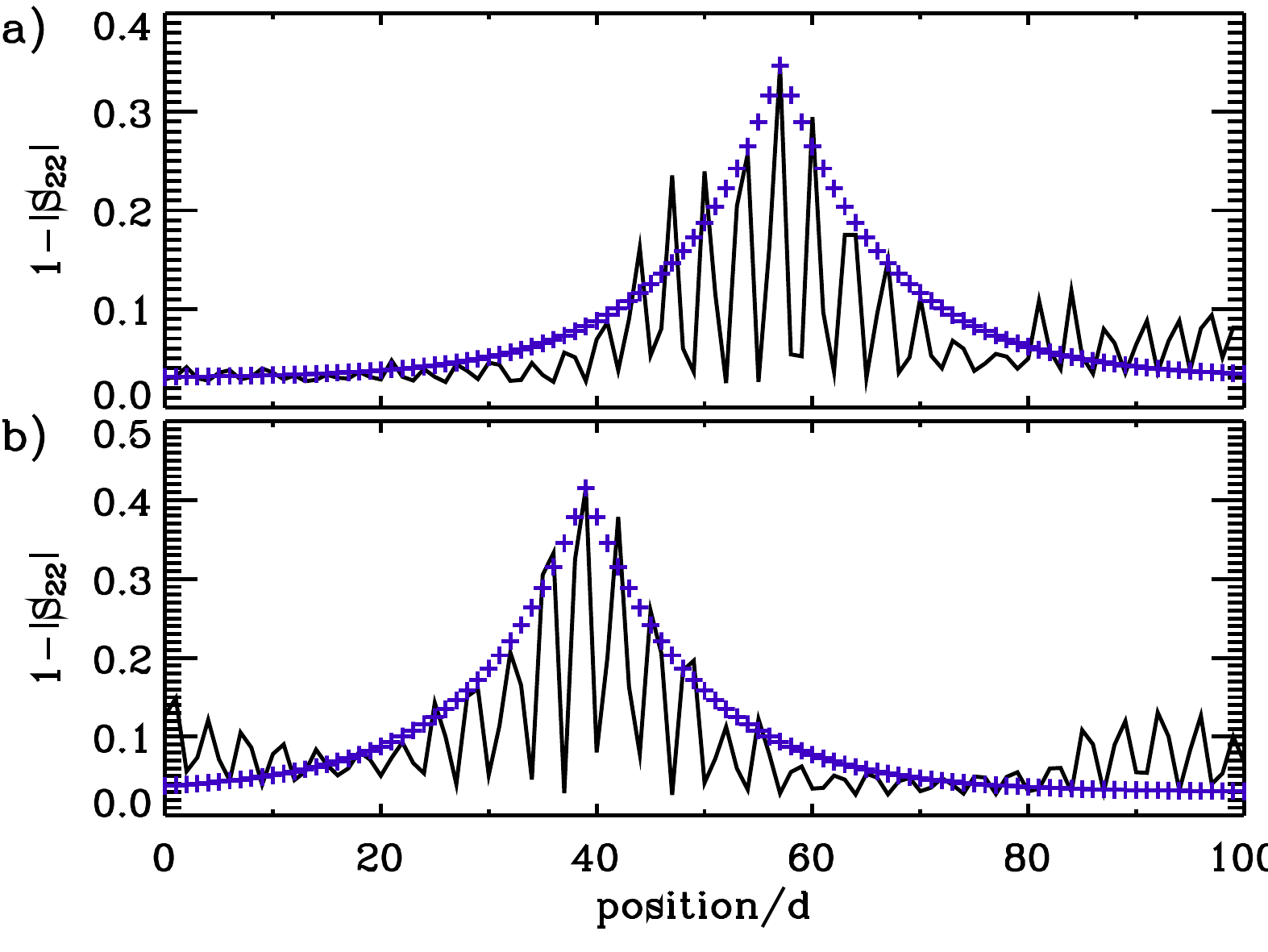}
\end{center}
\caption{(Color online) Profiles of two
enhanced localized states at $kd/\pi \approx 0.63$ (a) and
$0.69$ (b) for the correlated sequence (marked by vertical
dotted lines in Fig.~\ref{reflect} and
\ref{zoom}). Graph (b) is the state responsible
for a sharp decrease of $|S_{22}|$ shown at the top of
Fig.~\ref{zoom}. Crosses show an exponential
decay with a localization length of 10 scatterer spacings (after \cite{KIK08}).}
\label{loc-eigen}
\end{figure}

As demonstrated here, the data obtained for one-mode waveguides with long-range correlated disorder show a quite good agreement between the theory and experiment. It should be stressed that this correspondence was found in spite of the fact that {\it i)} the analytical results are based on the analysis of the Lyapunov exponent for an infinite sample; {\it ii)} the effect of long-range correlations is based on the binary correlator only, that is correct in the first approximation in $\sigma^2$, {\it iii)} the number of scatterers is quite small, {\it iv)} there is the absorption of about 0.04\,dB per unit cell $d$ for the empty waveguide. In addition, the potential was scaled linearly to the micrometer screw depth, which is an approximation.
However, a strong enhancement of localization due to long-range
correlations is clearly seen, indicating that the observed effect is very robust. These experiments can be treated as an indication that the selective transport, emerging due to long-range interactions, can find various applications in the design of 1D structures, especially as such localized states are controlled in the frequency space, a fact that may be important, for example, for random lasing \cite{Co99,P03}.

\section{Kronig-Penney model with finite barriers: Theory versus experiment}
\label{10}

One of the important problems that still remains open is the role of a disorder that cannot be avoided in experimental devices. It is known that the fluctuations of the thickness of layers or variations of the medium parameters, such as dielectric constant, magnetic permeability, or barrier hight for electrons, can strongly influence the transport properties. In spite of remarkable progress in this field, the majority of studies of the wave (electron) propagation through random structures are mainly based on numerical methods \cite{MCMM93,So00,SSS01,Po03,VM03,Eo06,DZ06,No07,Po07,Ao07,NML07,NML08}. It is obvious that giving important results, the numerical approaches suffer from the lack of generality being restricted by specifics of models and parameters. As for existing analytical results, they mainly refer to the simplest models with white-noise disorder \cite{BW85}, or to the patterns with correlated disorder, however, with delta-like potential wells \cite{IK99} or barriers \cite{IKU01,HIT08}.

In this Section we discuss recent experiments with the disordered Kronig-Penney model \cite{LIMKS09} performed in connection with the analytical predictions obtained in previous Sections. Our goal is to explore the properties of transmission in one-dimensional disordered bi-layer arrays of finite size in dependence on the degree of disorder. The results allow one to understand what are restrictions on the theoretical assumptions such as the weakness of disorder, absence of absorption and infinite number of scatterers, that should be taken into account in practical applications. Of special interest is the role of the Fabry-Perot resonances discussed in Refs.~\cite{MIL07,IM09,LM09}. These resonances occur in the periodic systems with finite thicknesses of the layers constituting the unit cell, in contrast with the structures described by (periodic) delta-potentials for which they are absent. We shall demonstrate that the Fabry-Perot resonances can provide a perfect transmission in the so-called \emph{resonance} spectral bands even for a relatively strong disorder, whereas the transmission in other bands is substantially suppressed.

\subsection{Single-mode waveguide with bi-layered filling}
\label{10.1}

As a model of one-dimensional bi-layer structure, we consider an array formed by two alternating dielectric $a$ and $b$ layers with the refractive indices $n_a$ and $n_b$, placed inside an electromagnetic metallic-wall waveguide of constant width $w$ and height $h$, see Fig.~\ref{KPTE-Fig01}. The thicknesses of the $n$th $a$ and $b$ layers are denoted by $d_{a}(n)$ and $d_b(n)$. We consider the {\it positional} disorder emerging due to random thickness variations in the $a$ layer \emph{only}, such that
\begin{equation}\label{KPTE-dandb}
d_a(n)=d_a+\sigma\eta(n),\qquad\langle d_a(n)\rangle=d_a,\qquad d_b(n)=d_b.
\end{equation}
Here $\sigma$ is the root-mean-square deviation of $d_a(n)$ and $\sigma^2$ its variance. Therefore, $\eta(n)$ is a random sequence with the zero average and unit variance, and we consider the case of the white noise disorder,
\begin{equation}\label{KPTE-WNcorr}
\langle\eta(n)\rangle=0,\qquad\langle\eta^2(n)\rangle=1,\qquad\langle\eta(n)\eta(n')\rangle=\delta_{nn'}.
\end{equation}
Here the angular brackets $\langle\ldots\rangle$ stand for a statistical average over different realizations of randomly layered structure. As one can see, such a random structure is \emph{periodic on average} with the period $d=d_a+d_b$.

\begin{figure}[!ht]
\begin{center}
\includegraphics[scale=0.60]{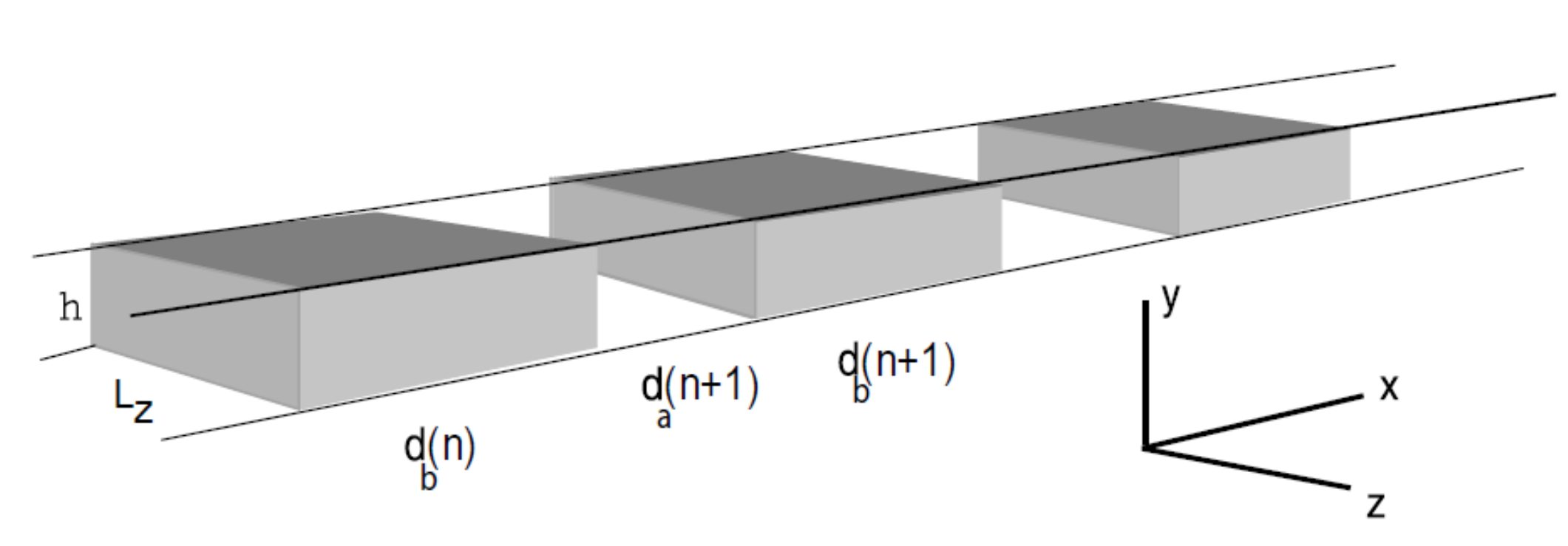}
\caption{(color online). Kronig-Penney model with teflon bars of constant thickness $d_b(n)=d_b(n+1)=d_b$, and air spacings $d_a(n)$ defined by Eq.~\eqref{KPTE-dandb} (after \cite{LIMKS09}).} \label{KPTE-Fig01}
\end{center}
\end{figure}

For the sake of one-dimensionality, we shall treat the lowest TE waveguide mode of frequency $\nu=\omega/2\pi$ whose electric field $\vec{E}(x,z)$ is defined as
\begin{equation}\label{KPTE-EF-def}
E_y=\sin(\pi z/w)\psi(x),\qquad E_x=E_z=0.
\end{equation}
Then, within every $a$ or $b$ layer the function $\psi(x)=\psi_{a,b}(x)$ is governed by the one-dimensional Helmholtz equation with two boundary conditions at the interfaces $x=x_i$ between adjacent layers,
\begin{subequations}\label{KPTE-WaveEqBC}
\begin{eqnarray}
&&\left(\frac{d^2}{dx^2}+k_{a,b}^2\right)\psi_{a,b}(x)=0,\label{KPTE-WaveEq-ab}\\[6pt]
&&\psi_a(x_i)=\psi_b(x_i),\qquad\psi'_a(x_i)=\psi'_b(x_i).\label{KPTE-BC}
\end{eqnarray}
\end{subequations}
The longitudinal wave numbers $k_{a}$ and $k_{b}$ read
\begin{equation}\label{KPTE-kab}
k_{a}=\frac{2\pi}{c}\sqrt{n_{a}^2 \nu^2-(c/2w)^2},\qquad k_{b}=\frac{2\pi}{c}\sqrt{n_{b}^2\nu^2-(c/2w)^2}.
\end{equation}
Note that the second of the boundary conditions in Eq.~\eqref{KPTE-BC} takes into account that the magnetic permeability of any layer equals one, $\mu_{a,b}=1$.
In the following we use the method that is complementary to the Hamiltonian map approach thoroughly discussed in previous Sections. The aim is to show how the common transfer matrix approach can be used to describe transport properties of our model.

To start with, we write the general solution of Eq.~\eqref{KPTE-WaveEq-ab} for the $(a_n,b_n)$ unit cell as a superposition of incoming and outgoing waves,
\begin{subequations}\label{KPTE-PsiAB-exp}
\begin{eqnarray}
\psi_{a}(x)&=&A^{+}_{n}\exp\left[ik_a(x-x_{an})\right]+ A^{-}_{n}\exp\left[-ik_a(x-x_{an})\right]\label{KPTE-PsiAn-exp}\\[6pt]
\mbox{inside}\ &a_n&\mbox{layer, where}\ x_{an}\leq x\leq x_{bn}\,;\nonumber\\[6pt]
\psi_{b}(x)&=&B^{+}_{n}\exp\left[ik_b(x-x_{bn})\right]+ B^{-}_{n}\exp\left[-ik_b(x-x_{bn})\right]\label{KPTE-PsiBn-exp}\\[6pt]
\mbox{inside}\ &b_n&\mbox{layer, where}\ x_{bn}\leq x\leq x_{a(n+1)}\,.\nonumber
\end{eqnarray}
\end{subequations}
Here $A^{\pm}_{n}$ and $B^{\pm}_{n}$ are the complex amplitudes of forward and backward traveling waves, and the coordinates $x_{an}$, $x_{bn}$ denote the left boundaries of the $a_n$ and $b_n$ layers, respectively. Note that $x_{bn}-x_{an}=d_a(n)$ and $x_{a(n+1)}-x_{bn}=d_b$.

With the use of the continuity conditions \eqref{KPTE-BC} for the wave function $\psi_{a,b}(x)$ and its derivative at the interfaces $x_i=x_{bn}$ and $x_i=x_{a(n+1)}$, one can obtain the relation between the amplitudes $A^{\pm}_{n+1}$ and $A^{\pm}_n$ of two adjacent $(a,b)$ cells,
\begin{equation}\label{KPTE-An+1An}
\left(\begin{array}{c}A^{+}_{n+1}\\A^{-}_{n+1}\end{array}\right)=\left(\begin{array}{cc}Q_{11}(n)&Q_{12}(n)\\[6pt]
Q_{21}(n)&Q_{22}(n)\end{array}\right)\left(\begin{array}{c}A^{+}_n\\A^{-}_n\end{array}\right).
\end{equation}
The transfer matrix $\hat{Q}(n)$ has the following elements,
\begin{subequations}\label{KPTE-Q}
\begin{eqnarray}
Q_{11}(n)=Q^*_{22}(n)&=&[\cos(k_bd_b)+i\alpha_+\sin(k_bd_b)] \exp[ik_ad_a(n)],\label{KPTE-Q11}\\[6pt]
Q^*_{12}(n)=Q_{21}(n)&=&i\alpha_-\sin(k_bd_b)\exp[ik_ad_a(n)], \label{Q12}
\end{eqnarray}
\end{subequations}
with the unity of the determinant,
\begin{equation}\label{KPTE-DetQ}
\det\hat{Q}=Q_{11}Q_{22}-Q_{12}Q_{21}=1.
\end{equation}
The asterisk stands for the complex conjugation, and we have introduced the parameters
\begin{equation}\label{alpha-pm}
\alpha_\pm=\frac{1}{2}\left(\frac{k_a}{k_b}\pm\frac{k_b}{k_a}\right),\qquad \alpha_+^2-\alpha_-^2=1.
\end{equation}
Note that the transfer matrix $\hat Q(n)$ differs from cell to cell only by the phase factor $\exp[ik_ad_a(n)]$.

Eq.~\eqref{KPTE-An+1An} provides the transfer matrix equation for the array of $N$ unit $(a,b)$ cells with or without the disorder,
\begin{equation}\label{KPTE-AN+1AN}
\left(\begin{array}{c}A^{+}_{N+1}\\A^{-}_{N+1}\end{array}\right)= \hat{Q}^N \left(\begin{array}{c}A^{+}_1\\A^{-}_1\end{array}\right).
\end{equation}
Here the total transfer matrix $\hat{Q}^N $ is expressed as a product of $N$ transfer matrices for every cell entering the array,
\begin{equation}\label{KPTE-QN}
\hat{Q}^N=\hat{Q}(N)\hat{Q}(N-1)\ldots\hat{Q}(2)\hat{Q}(1).
\end{equation}
Each of the matrices $\hat{Q}(n)$ ($n=1,2,\dots,N$) has the same form \eqref{KPTE-Q} differing in the value of $d_a(n)$ only. Due to the condition \eqref{KPTE-DetQ} of unimodularity for $\hat{Q}(n)$ and the relation \eqref{KPTE-QN}, the transfer matrix $\hat{Q}^N$ is unimodular too,
\begin{equation}\label{KPTE-DetQN}
\det\hat{Q}^N=Q_{11}^NQ_{22}^N-Q_{12}^NQ_{21}^N=1.
\end{equation}

In our numerical simulations as well as in the experimental setup, we have $A^{-}_{N+1}=0$. Thus, the transmittance of $N$ cells reads
\begin{equation}\label{KPTE-TNgen}
T_N\equiv|A^{+}_{N+1}/A^{+}_1|^2=|Q^N_{11}|^{-2}=|S_{12}^N|^2.
\end{equation}
Due to the last equality in Eq.~(\ref{KPTE-TNgen}), the transmittance $T_N$ can be expressed via the element $S_{12}^N$ of the scattering matrix $\hat{S}^N$. The latter is introduced by the well known relation,
\begin{equation}\label{AN+1An-Scat}
\left(\begin{array}{c}A^{-}_1\\A^{+}_{N+1}\end{array}\right)=\left(\begin{array}{cc}S_{11}^N&S_{12}^N\\[6pt]
S_{21}^N&S_{22}^N\end{array}\right)\left(\begin{array}{c}A^{+}_1\\A^{-}_{N+1}\end{array}\right).
\end{equation}
In what follows, we study the frequency dependence of $|S_{12}^N|$ which we call the \emph{transmission spectrum}.

In the case of no disorder, $\eta(n)=0$, the thickness of the $a$ layer does not depend on the cell number $n$, $d_a(n)=d_a$. Therefore, the unperturbed transfer matrix $\hat{Q}^{(0)}$ is described by Eq.~\eqref{KPTE-Q} with $d_a(n)$ replaced by the constant thickness $d_a$. As known (see, e.g., Ref.~\cite{MS08}), the transmittance of the array of $N$ \emph{identical} cells can be written in closed form as follows,
\begin{equation}\label{TN-0}
T^{(0)}_N=|S^{(0)N}_{12}|^2=\Bigg[1+\left|Q^{(0)}_{12} \frac{\sin(N\gamma)}{\sin\gamma}\right|^2\Bigg]^{-1},
\end{equation}
where the Bloch phase $\gamma$ is defined by the dispersion relation,
\begin{equation}\label{KPTE-DR}
\cos\gamma=\cos(k_ad_a)\cos(k_bd_b)-\alpha_+\sin(k_ad_a)\sin(k_bd_b).
\end{equation}

The expression~\eqref{TN-0} describes two kinds of the Fabry-Perot resonances for which the transmission is perfect ($|S^{(0)N}_{12}|=1$) for a whole array of $N$ unit $(a,b)$ cells. The first one is associated with the total length $Nd$ of bi-layer structure. These resonances emerge under the condition $\sin(N\gamma)/\sin\gamma=0$ that gives rise to
\begin{equation}\label{KPTE-FPgamma}
\gamma=m\pi/N,\qquad m=1,2,3,\ldots,N-1 \,.
\end{equation}
At the resonant values \eqref{KPTE-FPgamma} of the Bloch phase $\gamma$ the array of $N$ is perfectly transparent, thus, resulting in $N-1$ oscillations of $T^{(0)}_N$ in dependence on the frequency $\nu$. Note that these oscillations contribute to the profile of the transmission spectrum $|S^{(0)N}_{12}|$ in {\it each} frequency band (see, e.g., Ref.~\cite{MS08}).

The Fabry-Perot resonances of the second type arise in the $b$ layers only, when $|Q^{(0)}_{12}|=\alpha_-\sin(k_bd_b)=0$. It happens when
\begin{equation}\label{KPTE-FPb}
k_bd_b=s\pi,\qquad s=1,2,3,\ldots.
\end{equation}
The nature of these resonances is substantially different. As one can see, they are originated from the condition under which the phase shift of the wave passing through any of $b$ layers is multiple to $\pi$. Therefore, these $b$ layers are ``invisible" for a wave, and the transmission is similar to the case when all the $b$ layers are absent. Clearly, in this case the array is effectively homogeneous and, therefore, its transmission is perfect. The distinctive property of such resonances is that they are distributed all over the spectrum and are not determined by the underlining band structure of the frequency spectrum. Indeed, the condition \eqref{KPTE-FPb} depends on the thickness $d_b$ rather than on the period $d$ determining the band structure of the spectrum. Therefore, any specific frequency band may or may not contain such a Fabry-Perot resonance, depending on specific values of the system parameters. We shall refer to the resonances \eqref{KPTE-FPb} as the \emph{teflon resonances} since in the experiment the $b$ layers are made of teflon, whereas the $a$ layers are just air spacing.

\subsection{Experiments with bi-layer periodic structure}
\label{10.2}

\begin{figure}[!ht]
\subfigure{\includegraphics[height=0.55\columnwidth]{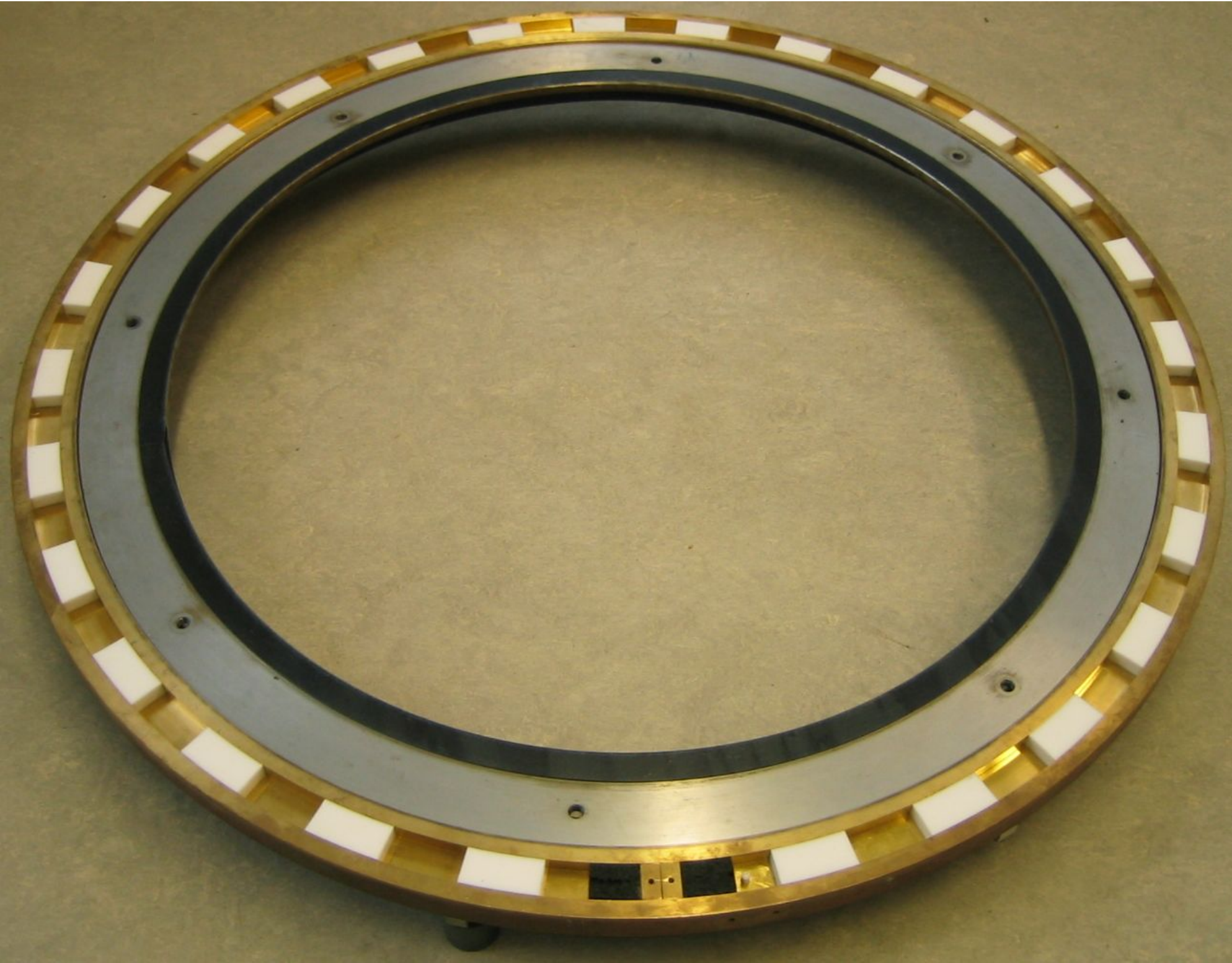}}
\subfigure{\includegraphics[height=0.55\columnwidth]{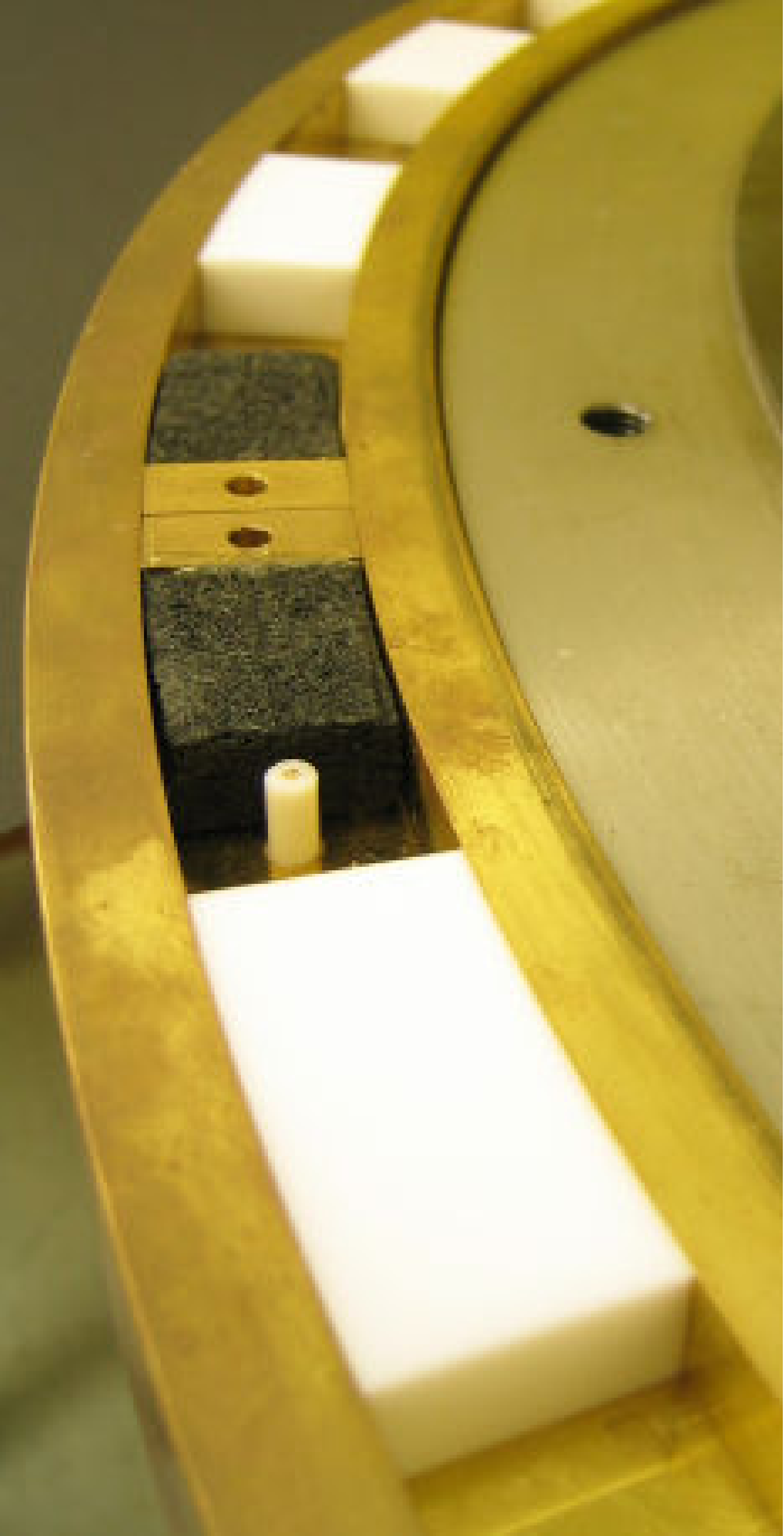}}
\caption{(color online). Left: overview of the waveguide of perimeter $P=234$ cm, with height $h=1$ cm and width $w=2$ cm. Right: an enlargement showing the cylindrical dipole antenna near one of two carbon absorbers (in black). Other antenna is attached to a brass lid (not shown) near the other absorber. White pieces are the teflon bars (after \cite{LIMKS09}).}
\label{KPTE-Fig02ab}
\end{figure}

In Fig.~\ref{KPTE-Fig02ab} we show the experimental setup of a microwave ring guide consisting of $N=26$ unit $(a,b)$ cells, with the teflon bars ($b$ layers) of thickness $d_b=4.078$ cm and refractive index $n_b=\sqrt{2.08}$. The measurement is performed via two electric dipole antennas connected to a network analyzer. In order to mimic the semi-infinite leads connected to each side of the waveguide, two carbon absorbers are placed near both ends of the waveguide (shown as black pieces). The frequency range is $\nu=7.5$ to $\nu=15$\,GHz which
corresponds to the wave lengths from 4 to 2\,cm. This arrangement has been already used in the study of single-impurity effect in the photonic Kronig-Penny model \cite{LSKS08}. Earlier, in an analogous model with the metallic screws instead of teflon pieces, the microwave realization of the Hoftstadter butterfly \cite{KS98,S07} was implemented. The same configuration (metallic screws) was used to investigate the transport properties of a waveguide with on-site correlated disorder \cite{KIKS00,KIKSU02,KIK08} (see Section~\ref{9}).

In comparison with the one-dimensional model introduced in previous Subsection, the waveguide shown in Fig.~\ref{KPTE-Fig02ab} is obviously not rectilinear. Specifically, the teflon bars and air segments are not perfect rectangle, one side is longer than the other by about $5$ percent. However, since the perimeter is much larger than the wave length even in the regime of the first mode, the rectilinear model can be regarded as a good approximation. As shown in Ref.~\cite{LSKS08}, a good quantitative agreement is observed by introducing an {\it effective} thickness of the teflon bars and air segments. In accordance with the results of Ref.~\cite{LSKS08}, for the comparison with the analytical predictions we use the effective thickness $d_b = 4.078\,cm$, although the inner thickness of the teflon bars is 4 cm.

Let us now see how the analytical results derived in preceding Subsection can be used to describe the experimental data obtained with the single-mode waveguide shown in Fig.~\ref{KPTE-Fig02ab}. Our main interest is in the frequency dependence of the transmission spectrum $|S_{12}^N|$ presented in Fig.~\ref{KPTE-Fig03}. In this figure the curve (a) shows the experimentally measured transmission spectrum for the periodic array of 26 cells to be compared with the curve (c) calculated with the use of Eqs.~\eqref{TN-0}, \eqref{KPTE-DR} and \eqref{KPTE-Q}. As one can see, the experimental data for the transmission spectra are about $1/5$ of the theoretical ones. This decrease, which is unavoidable experimentally, can be attributed to a strong absorbtion by the waveguide metallic walls as well as to the influence of antennas. For the latter, it is known that the absorbtion in the antennas decreases with the frequency increase, that is, indeed, reflected by the curve (a). Nevertheless, in spite of such a big difference for the value of the transmission spectrum, the gross-structure of the spectrum clearly manifests the correct positions of the transmission bands in a whole frequency range.

\begin{figure}[!ht]
\vspace{1.0cm}
\includegraphics[width=\columnwidth]{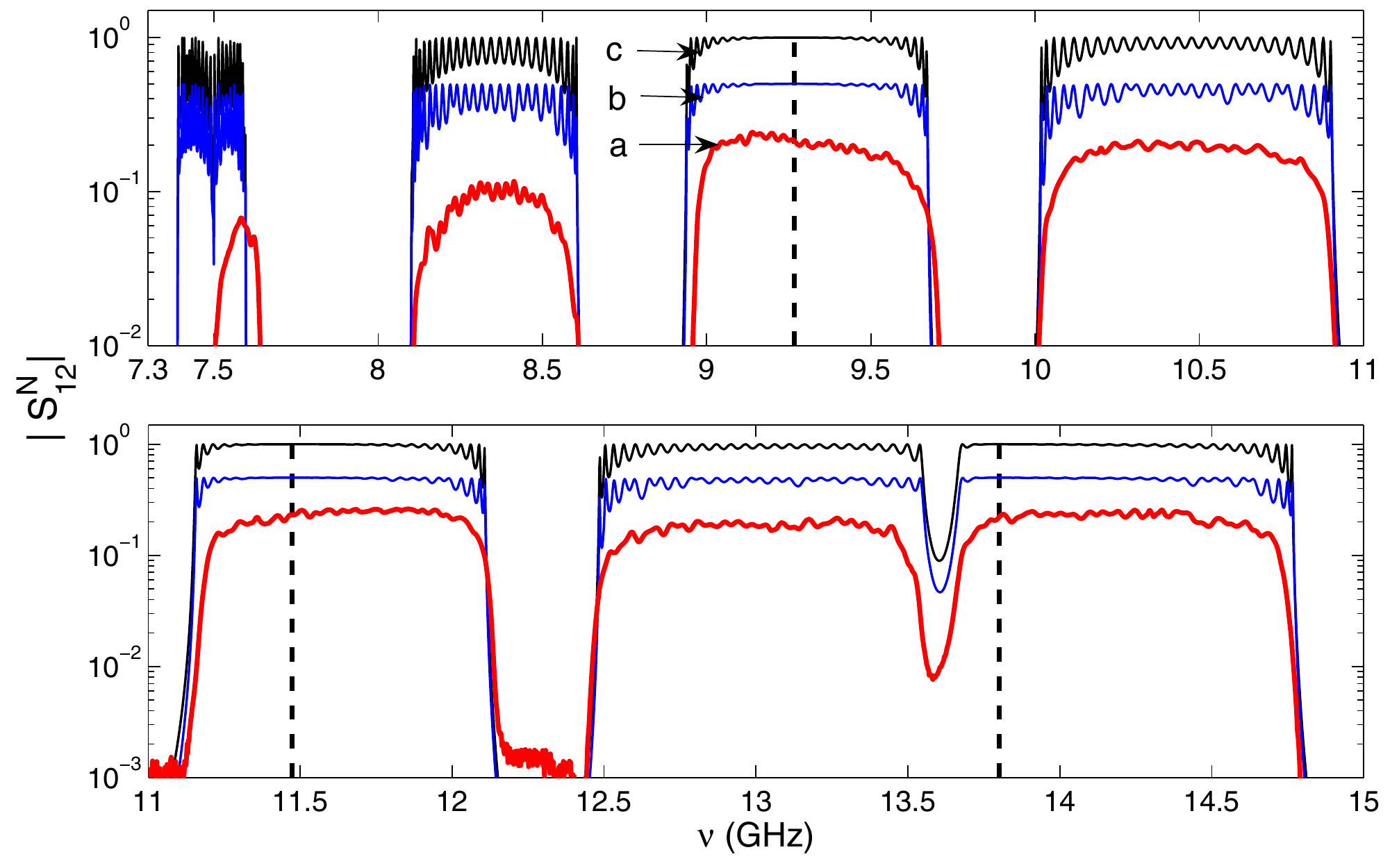}
\caption{(color online). Transmission spectrum for $7.3<\nu<11$ GHz (top) and $11<\nu<15$ GHz (bottom). Curve (a): experimental data for the periodic array. Curve (b): numerical data multiplied by $1/2$ for the randomized array with $\xi=0.0049$. Curve (c): numerical data for perfectly periodic array with $d_a=d_b=4.078$ cm. Dashed vertical lines mark the position of teflon resonances (after \cite{LIMKS09}).}
\label{KPTE-Fig03}
\end{figure}

A close inspection of the theoretical curve (c) in Fig.~\ref{KPTE-Fig03} reveals the effect of the teflon resonances discussed in previous Subsection. Specifically, one can speak about two types of bands. The bands $1,2,4,6$ that are not subjected by the teflon resonances, clearly show the $N-1=25$ Fabry-Perot oscillations \eqref{KPTE-FPgamma}, associated with the total length of bi-layer structure. On the contrary, the transmission spectrum in the bands $3,5,7$, in which the teflon resonances \eqref{KPTE-FPb} arise, is quite flat, and $|S^{(0)N}_{12}|\approx 1$ within these bands. For further discussion, we shall refer to the second type of frequency bands as to the \emph{resonance bands}. Note that these bands disappear when $d_b\to 0$, turning into those with $N-1$ oscillations occurring for delta-function potentials.

According to definition \eqref{KPTE-kab} of the wave numbers $k_a$ and $k_b$, there are two cut-off frequencies. The first one is equal to $\nu_a^{cut}=(c/2wn_a)=7.5$ GHz and corresponds to the $a$ layers with the air between teflon bars. The second one is equal to $\nu_b^{cut}=(c/2w n_b)=5.2$ GHz and corresponds to the $b$ layers (teflon bars). As one can see, the second cut-of frequency is smaller than the first one since $n_a=1<n_b=\sqrt{2.08}$. As for the experimental data, they indicate that the real cut-off frequency (which opens the first transmission band) equals $\nu_1^{bot}=7.387$. It is located within the interval $\nu_b^{cut}<\nu_1^{bot}<\nu_a^{cut}$ where the wave number $k_a$ in air spacings is purely imaginary, $k_{a}=i|k_{a}|$. This indicates that when the first band opens, the transmission in air spacings is due to the tunneling, in contrast with the teflon bars where the wave propagates freely. For this reason the profile of the first band (see numerical curves (b) and (c) in Fig.~\ref{KPTE-Fig03}) is somewhat special: there is a dip in the transmission spectrum right at the frequency $\nu_a^{cut}$ where the first mode opens in the air.

Note that except for the first band, the positions and widths as well as the band profiles of the experimental curve (a) in Fig.~\ref{KPTE-Fig03}, are well reproduced by the transfer matrix calculations described by curve (c). The discrepancy for the first band is understood since for low frequencies the wave length is not sufficiently small in comparison with the perimeter of the circular waveguide, hence, the rectilinear waveguide model fails. For the second and higher bands, the agreement is better and the experimental curve does show the weak $N-1$ oscillations predicted analytically. Clearly, they do not appear as perfectly regular oscillations and this irregularity may be caused by experimental imperfections, such as the uncontrolled variations in the thickness and positions of the teflon bars.

Our estimates show that the maximal deviation $|d_b(n)-d_b|_{max}$ in the thickness of the teflon bars is about 0.01\,cm and the maximum deviation $|d_a(n)-d_a|_{max}$ in the air spacings is about $0.04$ cm. In order to check whether such imprecisions can break the regularity of the oscillations, we performed the numerical simulation of the transmission spectrum \eqref{KPTE-TNgen} assuming that the error in the displacements $d_a(n)-d_a$ is described by Eqs.~\eqref{KPTE-dandb}, \eqref{KPTE-WNcorr} with the random sequence $\eta(n)$ uniformly distributed within the interval $|\sigma\eta(n)|\leqslant0.04$\,cm. The result multiplied by $0.5$ is presented in Fig.~\ref{KPTE-Fig03} by curve (b) . The multiplication was done for a better comparison with the curves (c) and (a) of the perfectly regular case and the experimental data. One can see that with a small amount of disorder the oscillations are less regular, however, remain to be more pronounced in comparison with the experimental data. Probably, another source of destroying the oscillations is the presence of a strong absorbtion.

\subsection{Positional disorder}
\label{10.3}

In order to study experimentally the role of positional disorder in dependence on its strength, we intentionally constructed the disordered bi-layer array in which only the air spacings ($a$ layers) have random thickness $d_a(n)=d_a+\sigma\eta(n)$, while the thickness $d_b$ of all teflon bars ($b$ layers) are constant in accordance with Eqs.~\eqref{KPTE-dandb} and \eqref{KPTE-WNcorr}. Both in experimental measurements and numerical calculations the uncorrelated random entries $\eta(n)$ are uniformly distributed in the interval $[-\sqrt3,\sqrt3]$ in order to have the unit variance. Below, we classify the strength $\xi$ of uncorrelated disorder by the ratio of root-mean-square deviation $\sigma$ to the average cell thickness $d=d_a+d_b$,
\begin{equation}\label{DisMeasure}
\xi\equiv\frac{|d_a(n)-d_a|_{max}}{d}=\sigma\sqrt{3}/d.
\end{equation}

\begin{table}[!ht]
\begin{center}
\begin{tabular}{|c|r|r|c|c|}\hline
case&$\xi/10^{-2}$&$\frac{\sigma}{d}=\frac{\xi}{\sqrt{3}}$&$\left(\frac{\sigma}{d}\right)^2$\\ \hline
very weak&0.49&$0.28\cdot10^{-2}$&$8.0\cdot10^{-6}$\\ \hline
weak&3.00&$1.77\cdot10^{-2}$&$3.0\cdot10^{-4}$\\ \hline
intermediate&12.30&$7.07\cdot10^{-2}$&$5.0\cdot10^{-3}$\\ \hline
strong&49.00&$28.30\cdot10^{-2}$&$8.0\cdot10^{-2}$\\ \hline
\end{tabular}
\caption{Parameter values of random disorder with $d=d_a+d_b=2d_a\approx 8.16$\,cm .}.\label{KPTE-Tab01}
\end{center}
\end{table}

Table~\ref{KPTE-Tab01} presents the values of $\xi$ we consider below, together with the corresponding values of $\sigma/d$ and $(\sigma/d)^2$. The latter we use to simplify the comparison with the analytical results to be obtained below. Note that the case of a very weak disorder with $\xi=0.49\times10^{-2}$ was already discussed above in connection with the imperfections in the experimental setup. In what follows, we refer to the cases $\xi=3.0\times10^{-2}$, $12.3\times10^{-2}$ and $49.0\times10^{-2}$ as to the weak, intermediate, and strong disorder, respectively.

\begin{figure}[!ht]
\includegraphics[width=\textwidth]{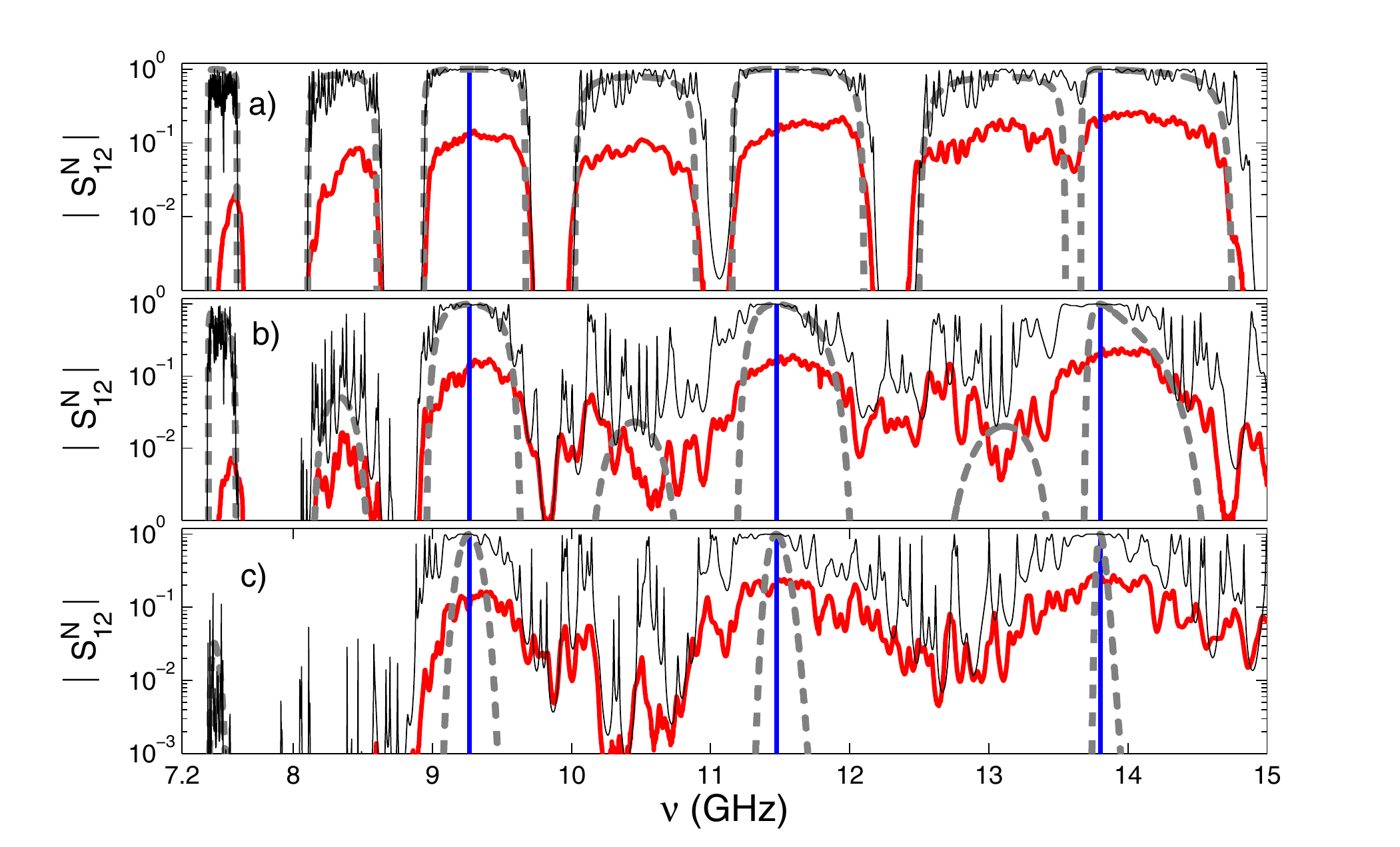}
\caption{(color online). Transmission spectrum $|S_{12}^N|$ for $26$-cells arrays: (a) weak ($\xi=3.0\times10^{-2}$), (b) intermediate ($\xi=12.3\times10^{-2}$), and (c) strong disorder ($\xi=49.0\times10^{-2}$). Transfer matrix calculations and experimental measurements are shown by thin and thick solid curves, respectively. Dashed curves correspond to the analytical expression \eqref{KPTE-S12th}. Vertical lines mark the position of the teflon resonances (after \cite{LIMKS09}).}
\label{KPTE-Fig04}
\end{figure}

The experimental results are summarized in Fig.~\ref{KPTE-Fig04} where the transmission spectrum for the array of $26$ cells with the positional disorder is shown. As to compare the case of intermediate disorder (Fig.~\ref{KPTE-Fig04}b) with that of weak disorder (Fig.~\ref{KPTE-Fig04}a), one can conclude the following. First, for the intermediate disorder only first two gaps are clearly distinguishable; the third only partially. Second, there is no trace of the $N-1$ oscillations in the spectral bands. And finally, the transmission in second, fourth, and sixth bands has decreased substantially in comparison with the weak disorder. However, remnants of the \emph{resonance bands} are still recognized.

As for the strong disorder (Fig.~\ref{KPTE-Fig04}c), the first two transmission bands have disappeared and there is no longer any evidence of the band structure of the unperturbed array. On the other hand, the transmission spectrum is still slightly changed in the vicinity of the teflon resonances, as compared with the case of moderate disorder.

A close inspection of Fig.~\ref{KPTE-Fig04} shows that the transmission spectrum fluctuates more rapidly for the transfer matrix calculations, Eqs.~\eqref{KPTE-TNgen}, \eqref{KPTE-QN}, \eqref{KPTE-Q}, in comparison with the experimental data. As one can see, in the experiment the fast oscillations are strongly suppressed. However, the global dependence of the transmission spectrum on frequency is reproduced by the transfer matrix calculations quite well, at least for weak disorder and partially for stronger disorder. It should be taken into account that the experimental pattern of the transmission spectrum corresponds to a particular realization of the sequence $\eta(n)$, and strongly fluctuates with the change of disorder realization.

The above results refer to the array of $N=26$ cells, thus, the question arises about how the transmission spectrum changes with an increase of $N$. Our experimental setup does not allow for the implementation of a larger number of bi-layer cells. However, given that the transfer matrix calculations are in a good agreement with the experimental data for $N=26$ cells, we now consider only numerically, and after, analytically, the larger arrays for the same three cases: weak, intermediate, and strong disorder.

\begin{figure}[!ht]
\includegraphics[width=\textwidth]{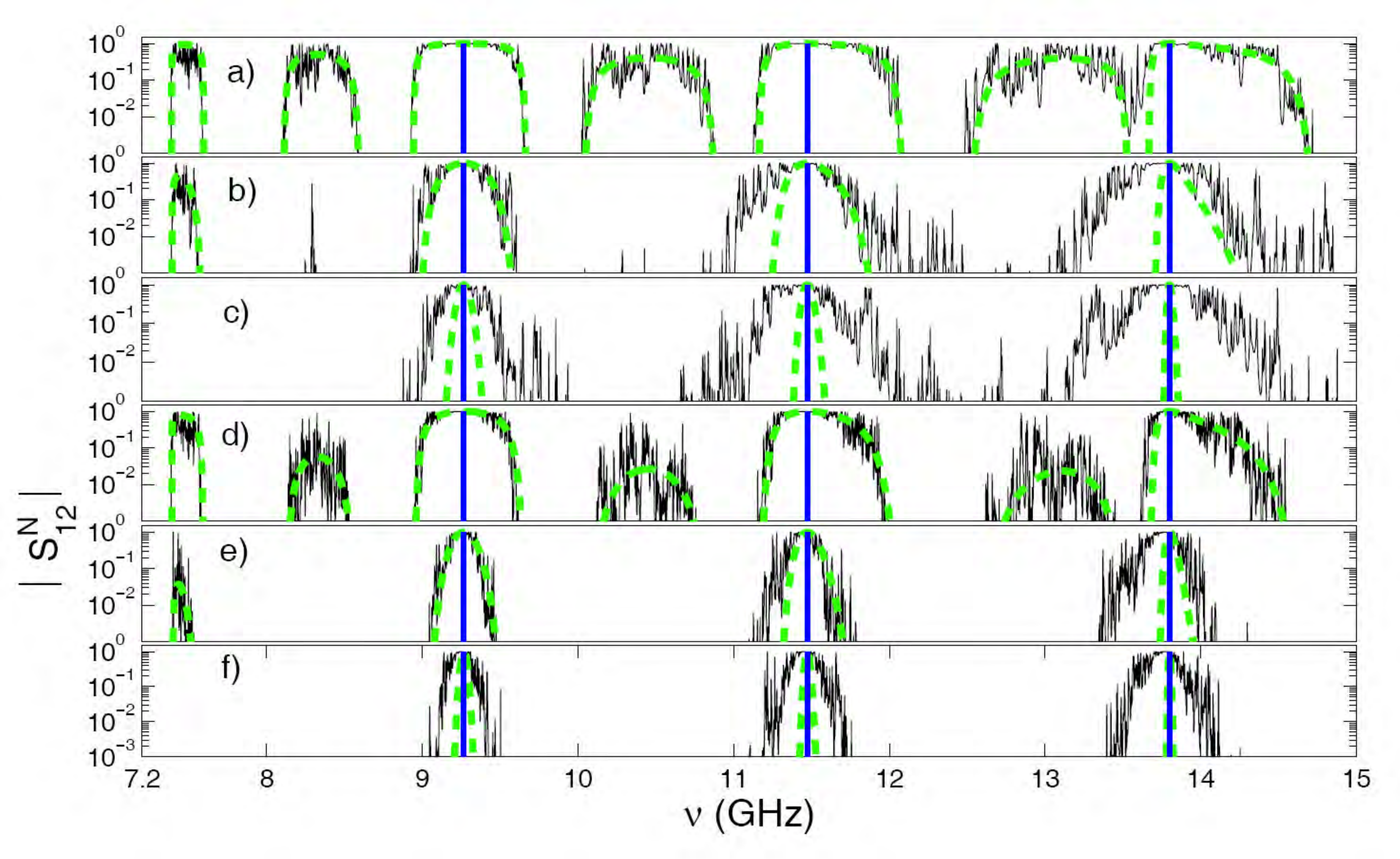}
\caption{(color online). Numerical data for the transmission spectrum \eqref{KPTE-TNgen} (solid curves), and the analytical relation \eqref{KPTE-S12th} (dashed curves) for $N=100$ and $N=400$: (a) to (c) $N=100$ cells; a) weak, b) intermediate and c) strong disorder. d) to f) $N=400$ cells; d) weak, e) intermediate and f) strong disorder (after \cite{LIMKS09}).}
\label{KPTE-Fig05}
\end{figure}

In Fig.~\ref{KPTE-Fig05} we depict the transmission spectrum $|S_{12}^N|$ numerically obtained for an array with $N=100$ and $N=400$ cells for weak, intermediate, and strong disorder. The main results is that with an increase of the structure size $N$ the transmission decreases provided the strength of disorder is fixed, see Figs.~\ref{KPTE-Fig05}a and \ref{KPTE-Fig05}d. However, the decrease occurs only away from the teflon resonance frequencies. Indeed, both for intermediate and strong disorder the transmission spectrum has strongly decreased below $10^{-3}$ for most of the frequencies except around the teflon resonances. As one can see, the teflon Fabry-Perot resonances strongly suppress the localization, and the localization is not homogeneous at all. We postpone the further discussion of this effect, as well as the other ones, and turn again to the analytical approach allowing us to explain the properties of the transmission in more detail.

\subsection{Localization length: Transfer matrix approach}
\label{10.4}

In order to proceed with our analysis, one needs to derive the expression for the localization length and relate it to the numerical and experimental data we have discussed. Below we employ the transfer matrix approach discussed in Ref.~\cite{MIL07}.

In the case of weak positional disorder,
\begin{equation}\label{KPTE-WD}
(k_a\sigma)^2\ll1,
\end{equation}
an analytical expression for $L_{loc}$ can be obtained as follows. First, we expand the transfer matrix $\hat{Q}(n)$ defined by Eq.~\eqref{KPTE-Q},
up to quadratic terms in the perturbation parameter $k_a\sigma\eta(n)$,
\begin{equation}\label{KPTE-Q-expan}
\hat{Q}(n)\approx\left\{1-\frac{(k_a\sigma)^2\eta^2(n)}{2}\right\}\hat{Q}^{(0)}+k_a\sigma\eta(n)\hat{Q}^{(1)}.
\end{equation}
The unperturbed $\hat{Q}^{(0)}$ and first-order $\hat{Q}^{(1)}$ matrices are suitable to be presented in the form
\begin{subequations}\label{KPTE-Q01}
\begin{equation}
\hat{Q}^{(0)}=\left(\begin{array}{cc}u&v^*\\[6pt] v&u^*\end{array}\right),\qquad
\hat{Q}^{(1)}=\left(\begin{array}{cc}iu&-iv^*\\[6pt] iv&-iu^*\end{array}\right);
\end{equation}
\begin{eqnarray}
u=[\cos(k_bd_{b})+i\alpha_+\sin(k_bd_{b})]\exp(ik_ad_{a}),\qquad  v=i\alpha_-\sin(k_bd_{b})\exp(ik_ad_{a}),\\[6pt]
\det\hat{Q}^{(0)}=\det\hat{Q}^{(1)}=|u|^2-|v|^2=1.\label{KPTE-det1}
\end{eqnarray}
\end{subequations}
For further calculations it is also useful to introduce the real and imaginary parts, $u_r\equiv\mathrm{Re}u$ and $u_i\equiv\mathrm{Im}u$,
\begin{equation}\label{UrUi}
u_r=\cos\gamma,\qquad u_i^2=\sin^2\gamma+|v|^2.
\end{equation}
The first equality is identical to the dispersion relation \eqref{KPTE-DR}, while the second one is a direct consequence of the matrix unimodularity \eqref{KPTE-det1}.

In order to extract the effects that are solely due to disorder, it is conventional to perform the following canonical transformation to the Bloch normal-mode representation in the transfer relation~\eqref{KPTE-An+1An},
\begin{equation}\label{KPTE-AA-Blm}
\left(\begin{array}{c}\widetilde{A}^{+}_{n+1}\\ \widetilde{A}^{-}_{n+1}\end{array}\right)=
\hat{P}\hat{Q}\hat{P}^{-1}\left(\begin{array}{c}\widetilde{A}^{+}_n\\ \widetilde{A}^{-}_n\end{array}\right),\qquad\qquad
\left(\begin{array}{c}\widetilde{A}^{+}_n\\ \widetilde{A}^{-}_n\end{array}\right)= \hat{P}\left(\begin{array}{c}A^{+}_n\\A^{-}_n\end{array}\right).
\end{equation}
The matrix $\hat{P}$ is specified by the requirement that the unperturbed matrix $\hat{Q}^{(0)}$ after the transformation becomes diagonal,
\begin{equation}\label{KPTE-Q0-diag}
\hat{P}\hat{Q}^{(0)}\hat{P}^{-1}=\left(\begin{array}{cc}\exp(+i\gamma)&0\\[6pt] 0&\exp(-i\gamma)\end{array}\right).
\end{equation}
Such a representation is in a complete accordance with the Floquet theorem \cite{F883}, or the same, with the Bloch condition \cite{B28}. The solution for the matrix $\hat{P}$ takes the form,
\begin{subequations}\label{KPTE-P}
\begin{eqnarray}
&&\hat{P}=\left(\begin{array}{cc}|v|/\beta_{+}&-iv^*/\beta_{-}\\[6pt] iv/\beta_{-}&|v|/\beta_{+}\end{array}\right),\\[6pt]
&&\beta_{\pm}^2=2\sqrt{1-u_r^2}\,(u_i\mp\sqrt{1-u_r^2})= 2\sin\gamma[u_i\mp\sin\gamma],\\[6pt]
&&\beta_{+}^2\beta_{-}^2=4|v|^2\sin^2\gamma,\qquad \det\hat{P}=\det\hat{P}^{-1}= |v|^2(\beta_{+}^{-2}-\beta_{-}^{-2})=1.
\end{eqnarray}
\end{subequations}

After substituting Eqs.~\eqref{KPTE-Q-expan}, \eqref{KPTE-Q01} and \eqref{KPTE-P} into the relation \eqref{KPTE-AA-Blm}, one can obtain the perturbative recurrent relations for new complex amplitudes,
\begin{eqnarray}
\widetilde{A}^{+}_{n+1}=\left[1-\frac{k_a^2\sigma^2\eta^2(n)}{2}+\frac{ik_a\sigma\eta(n)u_i}{\sin\gamma}\right]\exp(i\gamma)\widetilde{A}^{+}_{n}- \frac{k_a\sigma\eta(n)v^*}{\sin\gamma}\exp(i\gamma)\widetilde{A}^{-}_{n},\label{KPTE-rda-A+}\\[6pt]
\widetilde{A}^{-}_{n+1}=\left[1-\frac{k_a^2\sigma^2\eta^2(n)}{2}-\frac{ik_a\sigma\eta(n)u_i}{\sin\gamma}\right]\exp(-i\gamma)\widetilde{A}^{-}_{n}- \frac{k_a\sigma\eta(n)v}{\sin\gamma}\exp(-i\gamma)\widetilde{A}^{+}_{n}.\label{KPTE-rda-A-}
\end{eqnarray}

From Eqs.~\eqref{KPTE-rda-A+} and \eqref{KPTE-rda-A-} we can observe that one equation can be directly obtained from the other by the complex conjugation, if we suppose that $\widetilde{A}^{+}_{n}=\widetilde{A}^{-*}_{n}$. This means that it is convenient to seek the amplitudes $\widetilde{A}^{\pm}_{n}$ in terms of the radius-angle variables,
\begin{equation}\label{A-RTheta}
\widetilde{A}^{\pm}_{n}=R_n\exp(\pm i\theta_n).
\end{equation}

Now we are in a position to obtain the expression for the inverse localization length $L_{loc}^{-1}$, defined as follows,
\begin{equation}\label{KPTE-Lloc-def}
L_{loc}^{-1}=\frac{1}{2d}\langle\ln\left(\frac{R_{n+1}}{R_n}\right)^2\rangle.
\end{equation}
In order to derive the equation for real amplitude $R_n$, we multiply Eq.~\eqref{KPTE-rda-A+} by Eq.~\eqref{KPTE-rda-A-}. Then, within the second order of approximation in the perturbation parameter $k_a\sigma\eta(n)$, one gets
\begin{eqnarray}\label{KPTE-rda-Rn}
\frac{R^2_{n+1}}{R^2_n}&=&1+\frac{2k_a\sigma\eta(n)|v|}{\sin\gamma}\sin(2\theta_n+k_ad_a)-k^2_a\sigma^2\eta^2(n)\nonumber\\[6pt]
&&+\frac{k^2_a\sigma^2\eta^2(n)}{\sin^2\gamma} [u_i^2+|v|^2+2u_i|v|\cos(2\theta_n+k_ad_a)].
\end{eqnarray}
The logarithm of Eq.~\eqref{KPTE-rda-Rn} that determines the localization length \eqref{KPTE-Lloc-def}, has also to be expanded within the quadratic approximation,
\begin{eqnarray}\label{KPTE-rda-lnRn}
\ln\left(\frac{R_{n+1}}{R_n}\right)^2&=&\frac{2k_a\sigma\eta(n)|v|}{\sin\gamma}\sin(2\theta_n+k_ad_a)\nonumber\\[6pt]
&+&\frac{2k^2_a\sigma^2\eta^2(n)|v|^2}{\sin^2\gamma}\Big[1-\sin^2(2\theta_n+k_ad_a)+\frac{u_i}{|v|}\cos(2\theta_n+k_ad_a)\Big].
\end{eqnarray}
By performing the averaging of Eq.~(\ref{KPTE-rda-lnRn}), the correlations between the phase $\theta_n$ and deviation $\eta(n)$ can be neglected within the considered order of perturbation for a white-noise disorder. This means that one can make the averaging over $\theta_n$ and $\eta(n)$ separately. We also assume that the distribution of phase $\theta_n$ is homogeneous that is correct within the frequency bands, apart from the band edges and center of bands. Therefore, after averaging Eq.~\eqref{KPTE-rda-lnRn} over $\theta_n$, the term linear in $\eta(n)$ and the last term in the brackets vanish, while the term $\sin^2(2\theta_n+k_ad_a)$ is replaced with $1/2$. As a result, we obtain,
\begin{equation}\label{KPTE-Lloc}
L_{loc}^{-1}=(k_a\sigma)^2 \frac{\alpha_{-}^2\sin^2(k_bd_b)}{2d\sin^2\gamma}
\end{equation}
Note that an emergence of the term $\sin^2(k_bd_b)$ in the numerator of Eq.~\eqref{KPTE-Lloc} indicates the presence of teflon resonances \eqref{KPTE-FPb}, at which the localization length $L_{loc}$ diverges. Therefore, the random bi-layer array becomes fully transparent. Moreover, the localization is strongly suppressed in a large neighborhood of these resonances. Indeed, the resonance line-shape of $L_{loc}^{-1}$ is specified by the \emph{smooth} function $\sin^2(k_bd_b)$. If the amount of the disorder, $(\sigma/d)^2$, is not too large, the resonance should be quite broad. This is exactly what is observed in the experimental and numerical transmission spectrum plotted in Figs.~\ref{KPTE-Fig03} -- \ref{KPTE-Fig05}.

As shown in Section~\ref{2.8}, the localization length is directly related to the transmittance $T_N$ for a finite array of the length $L=Nd$, according to the famous relation $\langle\ln T_N\rangle=-2L/L_{loc}$, see Eq.~\eqref{1DCP-AvLn}. Although this relation is rigorously proved for the {\it continuous} potentials with weak disorder, it is reasonable to assume that it is also valid in our case with barriers of finite thickness. In view of this relation and recalling that $T_N=|S_{12}^N|^2$, see Eq.~\eqref{KPTE-TNgen}, we introduce the rescaled dimensionless localization length $N_{loc}$ as
\begin{equation}\label{KPTE-lnS12}
\langle\ln|S_{12}^N|\rangle=-L/L_{loc}\equiv-N/N_{loc}.
\end{equation}
In agreement with Eq.~\eqref{KPTE-Lloc}, the localization length $N_{loc}$ measured in numbers of cells, is defined by
\begin{equation}\label{KPTE-NlocF}
N_{loc}^{-1}=(\sigma/d)^2F(\nu), \qquad F(\nu)=(k_ad)^2\,\frac{\alpha_-^2\sin^2(k_bd_b)}{2\sin^2\gamma}.
\end{equation}
Here we introduced the form-factor $F(\nu)$ that specifies the frequency profile of the inverse localization length and depends only on the parameters of the underlying periodic structure. Note that the rescaled inverse localization length $N_{loc}^{-1}$ defined by Eq.~\eqref{KPTE-NlocF} increases quadratically with $\sigma/d$, as expected in the case of weak disorder.

\begin{figure}[!ht]
\begin{center}
\subfigure{\includegraphics[scale=1.21]{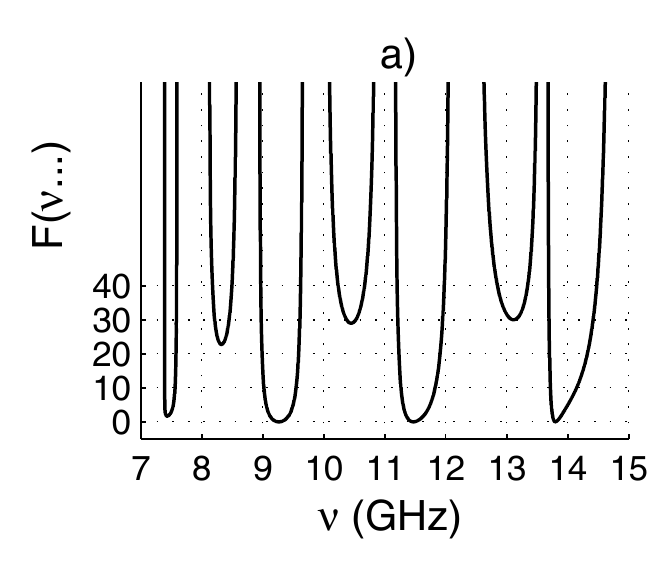}}
\subfigure{\includegraphics[scale=1.21]{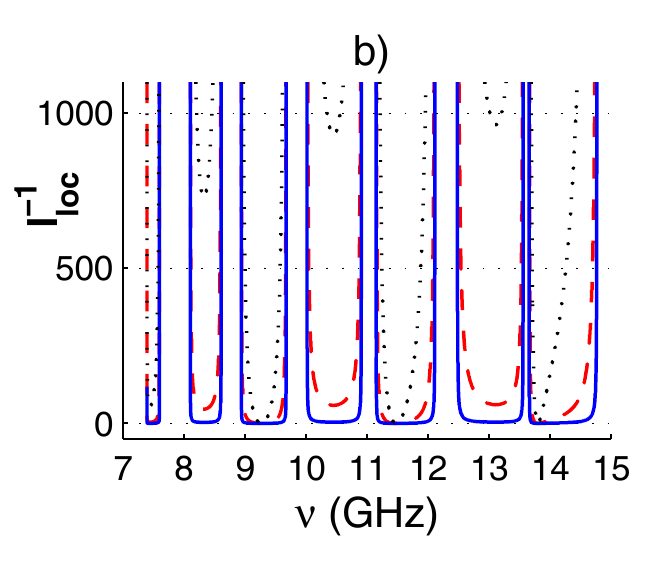}}
\end{center}
\caption{(color online). (a) Form-factor $F(\nu,d_{a,b}, n_{a,b})$ for $d_a=d_b=4.078$ cm, $n_a=1$ and $n_b=\sqrt{2.08}$; (b) ratio $l^{-1}_{loc}=N/N_{loc}$ for $N=400$ for weak (solid curve), intermediate (dashed curve), and strong (dotted curve) disorder (after \cite{LIMKS09}).}\label{KPTE-Fig06ab}
\end{figure}

In Fig.~\ref{KPTE-Fig06ab}a we plot the form-factor $F(\nu)$ for the parameters of our experimental setup, $d_a=d_b=4.078$ cm, $n_a=1$, $n_b=\sqrt{2.08}$, and in Fig.~\ref{KPTE-Fig06ab}b we present the frequency dependence of $N/N_{loc}$ for an array of 400 cells for the cases of weak, intermediate, and strong disorder. The data clearly manifest that the states are completely extended at the teflon resonances that occur in the $3$th $5$th and $7$th bands. This fact explains one of the most interesting features of the transmission reflected by the experimental and numerical data.

In order to compare the details of the analytical results with the experimental and numerical data, we define the theoretical value $|S_{12}^N|_{th}$ of the transmission spectrum as follows,
\begin{equation}\label{KPTE-S12th}
|S_{12}^N|_{th}\equiv\exp(-N/N_{loc})=\exp[-(\sigma/d)^2F(\nu)N].
\end{equation}
As we already discussed in Section~\ref{2.8}, this quantity corresponds to the statistical averaging over the so-called representative (most probable) realizations of the random sequence $\eta(n)$. We expect that such an averaging gives the result closer to our numerical and experimental data obtained for only one random realization.

The expression \eqref{KPTE-S12th} is plotted by dashed curves in Fig.~\ref{KPTE-Fig04} for the cases of weak, intermediate and strong disorder in the $N=26$ cells array, and in Fig.~\ref{KPTE-Fig05} for the array with $N=100$ and $N=400$ unit cells. The inspection of these figures reveals that the theoretical formula \eqref{KPTE-S12th} for the transmission spectrum provides a very good description for the case of weak disorder within the whole frequency range. For intermediate disorder, the agreement is not so good. The higher the frequency the stronger the discrepancy in accordance with the condition \eqref{KPTE-WD} of validity of the analytical calculations.

We would like to stress that the analytical results of this Section are derived with the use of statistical averaging over different realizations of randomly layered structure. On the other hand, in our numerics and in the experiment we had only one random realization, therefore, the average is performed over a relatively small number of layers. Note that the equivalence of these two types of an average arises when the structure is sufficiently long. Therefore, a better correspondence with numerical and experimental data is expected for longer samples. The data in Figs.~\ref{KPTE-Fig04} and \ref{KPTE-Fig05} confirm this expectation. Indeed, by comparing Figs.~\ref{KPTE-Fig04} and \ref{KPTE-Fig05}, we see that the larger the number of cells in the array, the better the agreement between the numerical simulations of Eq.~\eqref{KPTE-TNgen} and the analytical expression \eqref{KPTE-S12th}. Moreover, the quantity \eqref{KPTE-S12th} gives a good description of numerical data even for the case of intermediate disorder up to a quite large frequency, $\nu=11.5$ GHz.

\section{Binary multi-layered periodic-on-average structures}
\label{11}

In this Section we continue with our analysis of how the theoretical approach to specific correlations in random potentials can be applied to realistic models used in various fields of physics. Particulary, we are interested in the wave propagation (or electron transport) through one-dimensional periodic structures whose unit cell consists of two components (see, e.g. \cite{MS08} and references therein). For example, every cell is represented by two different materials in optics and electromagnetism, or by a pair of quantum well and barrier in electronics. The interest to such bi-layer structures is due to many applications in which one needs to create materials, metamaterials, or semiconductor superlattices with given transmission properties.

Below we generalize the Hamiltonian map approach developed in the previous Sections for simpler models. Specifically, we derive the unique analytical expression for the localization length $L_{loc}$ valid for a wide class of periodic-on-average bi-layer structures with weakly disordered thicknesses of both slabs \cite{IM09}. The key point of our approach is that we explicitly take into account possible correlations within two disorders of each layer type as well as between them. The expression for the localization length is analyzed for conventional photonic crystals, metamaterials and semiconductor superlattices.

\subsection{Bi-layer array with positional disorder}
\label{11.1}

We consider the propagation of electromagnetic waves of frequency $\omega$ through an infinite dielectric array (stack) of two alternating $a$ and $b$ layers (slabs). Every kind of slabs is respectively specified by the dielectric constant (permittivity) $\varepsilon_{a,b}$, magnetic permeability $\mu_{a,b}$, corresponding refractive index $n_{a,b}$, impedance $Z_{a,b}$ and wave number $k_{a,b}$,
\begin{subequations}\label{BiL-nZk}
\begin{eqnarray}
&&n_a=\sqrt{\varepsilon_a\mu_a},\qquad Z_a=\mu_a/n_a,\qquad k_a=\omega n_a/c\,;\\[6pt]
&&n_b=\sqrt{\varepsilon_b\mu_b},\qquad Z_b=\mu_b/n_b,\qquad k_b=\omega n_b/c\,.
\end{eqnarray}
\end{subequations}

A disorder is incorporated in the structure via the random
thicknesses of the slabs (\emph{positional disorder}),
\begin{subequations}\label{BiL-dab-n}
\begin{eqnarray}
&&d_a(n)=d_a+\varrho_a(n)\,,\qquad\langle d_a(n)\rangle=d_a\,;\label{BiL-da-n}\\[6pt]
&&d_b(n)=d_b+\varrho_b(n)\,,\qquad\langle d_b(n)\rangle=d_b\,.\label{BiL-db-n}
\end{eqnarray}
\end{subequations}
The integer $n$ enumerates the unit $(a,b)$ cells, $d_a$ and $d_b$ are the average thicknesses of layers, and $\varrho_a(n)$ and $\varrho_b(n)$ stand for small variations of the thicknesses. In the absence of disorder the array of bi-layers is periodic with the period $d=d_a+d_b$. The random entries $\varrho_{a}(n)$ and $\varrho_{b}(n)$ are statistically homogeneous with the zero mean, given variances and binary correlators defined by
\begin{subequations}\label{BiL-CorrDef}
\begin{eqnarray}
&&\langle\varrho_{a}(n)\rangle=0,\qquad\langle\varrho_{a}^2(n)\rangle=\sigma_a^2,\qquad \langle\varrho_a(n)\varrho_a(n')\rangle=\sigma_a^2K_a(n-n')\,;\\[6pt]
&&\langle\varrho_{b}(n)\rangle=0,\qquad\langle\varrho_{b}^2(n)\rangle=\sigma_b^2,\qquad
\langle\varrho_b(n)\varrho_b(n')\rangle=\sigma_b^2K_b(n-n')\,;\\[6pt]
&&\langle\varrho_a(n)\varrho_b(n)\rangle=\sigma_{ab}^2,\qquad
\langle\varrho_a(n)\varrho_b(n')\rangle=\langle\varrho_b(n)\varrho_a(n')\rangle=\sigma_{ab}^2K_{ab}(n-n').
\end{eqnarray}
\end{subequations}
Here the average $\langle ... \rangle$ is performed over the whole array of layers or due to the ensemble averaging, that is assumed to be equivalent. The two-point \emph{auto-correlators} $K_{a}(n-n')$ and $K_{b}(n-n')$ as well as the \emph{cross-correlator} $K_{ab}(n-n')$ are normalized to one, $K_{a}(0)=K_{b}(0)=K_{ab}(0)=1$. Although the variances $\sigma_a^2$ and $\sigma_b^2$ are positive, the cross-average $\sigma_{ab}^2$ can be of arbitrary value (positive, negative or zero). Note that $|\sigma_{ab}^2|\propto\sigma_a\sigma_b$.

We assume the weakness of the positional disorder,
\begin{equation}\label{BiL-WD}
k^2_a\sigma_a^2\ll1,\qquad k^2_b\sigma_b^2\ll1,
\end{equation}
that allows us to develop a proper perturbation theory. In this case all transport properties are entirely determined by the randomness power spectra $\mathcal{K}_a(k)$, $\mathcal{K}_b(k)$, and $\mathcal{K}_{ab}(k)$, defined by the relations (compare with Eq.~(\ref{1DCP-FTW})),
\begin{subequations}\label{BiL-FT-K}
\begin{eqnarray}
\mathcal{K}(k)&=&\sum_{r=-\infty}^{\infty}K(r)\exp(-ikr)=1+2\sum_{r=1}^{\infty}K(r)\cos(kr),\\[6pt]
K(r)&=&\frac{1}{2\pi}\int_{-\pi}^{\pi}dk\mathcal{K}(k)\exp(ikr)=\frac{1}{\pi}\int_{0}^{\pi}dk\mathcal{K}(k)\cos(kr).
\end{eqnarray}
\end{subequations}
By definition \eqref{BiL-CorrDef}, all the correlators $K_{a}(r)$, $K_{b}(r)$ and $K_{ab}(r)$ are real and even functions of the difference $r=n-n'$ between cell indices. Therefore, the corresponding Fourier transforms $\mathcal{K}_a(k)$, $\mathcal{K}_b(k)$ and $\mathcal{K}_{ab}(k)$ are real and even functions of the dimensionless longitudinal wave-number $k$. It should be stressed that the power spectra introduced by Eqs.~\eqref{BiL-FT-K} are non-negative functions of $k$ for any real random sequences $\varrho_a(n)$, $\varrho_b(n)$.

Within every $a$ or $b$ layer the electric field $\psi(x)\exp(-i\omega t)$ of the wave obeys the 1D Helmholtz equation with two boundary conditions at the interfaces between slabs,
\begin{subequations}\label{BiL-WaveEqBC}
\begin{eqnarray}
&&\left(\frac{d^2}{dx^2}+k_{a,b}^2\right)\psi_{a,b}(x)=0,\label{BiL-WaveEq-ab}\\[6pt]
&&\psi_a(x_i)=\psi_b(x_i),\qquad \mu_a^{-1}\psi'_a(x_i)=\mu_b^{-1}\psi'_b(x_i).\label{BiL-BC}
\end{eqnarray}
\end{subequations}
The $x$-axis is directed along the array of bi-layers perpendicular to the stratification, with $x=x_{i}$ standing for the interface coordinate.

\subsection{Generalized Hamiltonian map approach}
\label{11.2}

In order to develop a proper Hamiltonian map approach, the general solution of Eq.~\eqref{BiL-WaveEq-ab} within the same $n$th $(a,b)$ cell should be presented in the real-valued form (compare with Eqs.~\eqref{KPTE-PsiAB-exp}),
\begin{subequations}\label{BiL-Psi-ab}
\begin{eqnarray}
\psi_{a}(x)&=&\psi_{a}(x_{an})\cos\left[k_a(x-x_{an})\right]+k_a^{-1}\psi'_a(x_{an})\sin\left[k_a(x-x_{an})\right]\\[6pt]
\mbox{inside}\,&a_n&\mbox{layer, where}\,x_{an}\leq x\leq x_{bn}\,;\nonumber\\[6pt]
\psi_{b}(x)&=&\psi_{b}(x_{bn})\cos\left[k_b(x-x_{bn})\right]+k_b^{-1}\psi'_b(x_{bn})\sin\left[k_b(x-x_{bn})\right]\\[6pt]
\mbox{inside}\,&b_n&\mbox{layer, where}\,x_{bn}\leq x\leq x_{a(n+1)}\,.\nonumber
\end{eqnarray}
\end{subequations}
The coordinates $x_{an}$ and $x_{bn}$ refer to the left-hand edges of successive $a_n$ and $b_n$ layers. Note that $x_{bn}-x_{an}=d_a(n)$ and $x_{a(n+1)}-x_{bn}=d_b(n)$. The solution \eqref{BiL-Psi-ab} gives a useful relation between the wave function $\psi_{a,b}$ and its derivative $\psi'_{a,b}$ at the opposite boundaries of the same $a$ or $b$ layer,
\begin{subequations}\label{BiL-Map-ab}
\begin{eqnarray}
\psi_a(x_{bn})&=&\psi_a(x_{an})\cos\widetilde{\varphi}_a(n)+k_a^{-1}\psi'_a(x_{an})\sin\widetilde{\varphi}_a(n),\nonumber\\[6pt]
\psi'_a(x_{bn})&=&-k_a\psi_a(x_{an})\sin\widetilde{\varphi}_a(n)+\psi'_a(x_{an})\cos\widetilde{\varphi}_a(n);\\[6pt]
\psi_b(x_{a(n+1)})&=&\psi_b(x_{bn})\cos\widetilde{\varphi}_b(n)+k_b^{-1}\psi'_b(x_{bn})\sin\widetilde{\varphi}_b(n),\nonumber\\[6pt]
\psi'_b(x_{a(n+1)})&=&-k_b\psi_b(x_{bn})\sin\widetilde{\varphi}_b(n)+\psi'_b(x_{bn})\cos\widetilde{\varphi}_b(n).
\end{eqnarray}
\end{subequations}
The corresponding phase shifts $\widetilde{\varphi}_{a}(n)$ and $\widetilde{\varphi}_{b}(n)$ depend on the cell index $n$ via random thicknesses $d_a(n)$ and $d_b(n)$,
\begin{subequations}\label{BiL-phi-ab}
\begin{eqnarray}
&&\widetilde{\varphi}_{a}(n)=\varphi_{a}+\xi_{a}(n),\qquad\varphi_{a}=k_{a}d_a,\qquad\xi_a(n)=k_a\varrho_a(n);\\[6pt]
&&\widetilde{\varphi}_{b}(n)=\varphi_{b}+\xi_{b}(n),\qquad\varphi_{b}=k_{b}d_b,\qquad\xi_b(n)=k_b\varrho_b(n).
\end{eqnarray}
\end{subequations}
By combining Eqs.~\eqref{BiL-Map-ab} with the boundary conditions \eqref{BiL-BC} at $x_i=x_{bn}$ and $x_i=x_{a(n+1)}$, one can write the recurrent relations for the whole $n$th unit $(a,b)$ cell,
\begin{equation}\label{BiL-Map-XY}
X_{n+1}=\widetilde{A}_nX_n+\widetilde{B}_nY_n,\qquad Y_{n+1}=-\widetilde{C}_nX_n+\widetilde{D}_nY_n.
\end{equation}
Here
\begin{equation}\label{BiL-XYdef}
X_n=\psi_a(x_{an}),\qquad\qquad Y_n= k_a^{-1}\psi'_a(x_{an}),
\end{equation}
with the indices $n$ and $n+1$ standing for left and right edges of the $n$th cell. The factors $\widetilde{A}_n$, $\widetilde{B}_n$, $\widetilde{C}_n$, $\widetilde{D}_n$ read
\begin{subequations}\label{BiL-ABCDtilde}
\begin{eqnarray}
\widetilde{A}_n&=&\cos\widetilde{\varphi}_a(n)\cos\widetilde{\varphi}_b(n)- Z_a^{-1}Z_b\sin\widetilde{\varphi}_a(n)\sin\widetilde{\varphi}_b(n),\\[6pt]
\widetilde{B}_n&=&\sin\widetilde{\varphi}_a(n)\cos\widetilde{\varphi}_b(n)+
Z_a^{-1}Z_b\cos\widetilde{\varphi}_a(n)\sin\widetilde{\varphi}_b(n),\\[6pt]
\widetilde{C}_n&=&\sin\widetilde{\varphi}_a(n)\cos\widetilde{\varphi}_b(n)+
Z_aZ_b^{-1}\cos\widetilde{\varphi}_a(n)\sin\widetilde{\varphi}_b(n),\\[6pt]
\widetilde{D}_n&=&\cos\widetilde{\varphi}_a(n)\cos\widetilde{\varphi}_b(n)-
Z_aZ_b^{-1}\sin\widetilde{\varphi}_a(n)\sin\widetilde{\varphi}_b(n).
\end{eqnarray}
\end{subequations}
The ratio between the impedances emerges due to the equality $Z_a/Z_b=k_b\mu_a/k_a\mu_b$.

Note that the recurrent relations \eqref{BiL-Map-XY} can be treated as the Hamiltonian map in the phase space $(X,Y)$ for a linear oscillator subjected to linear parametric force.

\subsubsection{Unperturbed periodic system}
\label{11.2.1}

Without the disorder, $\xi_{a,b}(n)=0$, the constants $A$, $B$, $C$, $D$ do not depend on the cell index (discrete time) $n$ and are defined by Eqs.~\eqref{BiL-ABCDtilde} in which one should replace random phase shifts $\widetilde{\varphi}_{a}(n)$ and $\widetilde{\varphi}_{b}(n)$ with their regular parts $\varphi_{a}$ and $\varphi_{b}$, see Eqs.~\eqref{BiL-phi-ab}. Therefore, the trajectory $X_n,Y_n$ described by the map \eqref{BiL-Map-XY}, creates an {\it ellipse} in the phase space $(X,Y)$ that is an image of the unperturbed periodic motion. The next important step is to perform the canonical transformation,
\begin{subequations}\label{BiL-XY-QP}
\begin{eqnarray}
X_{n}=\upsilon^{-1}Q_n\cos\tau-\upsilon P_n\sin\tau,&\qquad& Q_n=\upsilon X_n\cos\tau+\upsilon Y_n\sin\tau,\\[6pt]
Y_{n}=\upsilon^{-1}Q_n\sin\tau+\upsilon P_n\cos\tau,&\qquad& P_n=-\upsilon^{-1}X_n\sin\tau+\upsilon^{-1}Y_n\cos\tau\,,
\end{eqnarray}
\end{subequations}
to new coordinates $Q_n$ and $P_n$, in which the unperturbed trajectory occupies the \emph{circle},
\begin{equation}\label{BiL-Map-QPunp}
Q_{n+1}=Q_n\cos\gamma+P_n\sin\gamma,\qquad P_{n+1}=-Q_n\sin\gamma+P_n\cos\gamma\,,
\end{equation}
in the phase space $(Q,P)$, see Fig.~\ref{BiL-Fig01}. The variables $\upsilon$, $\tau$ and parameter $\gamma$ can be found from Eqs.~\eqref{BiL-XY-QP}, \eqref{BiL-Map-QPunp} and \eqref{BiL-Map-XY} with the constants $A$, $B$, $C$, $D$ in place of $\widetilde{A}_n$, $\widetilde{B}_n$, $\widetilde{C}_n$, $\widetilde{D}_n$. In particular, one can obtain the relation $2\cos\gamma=A+D$ determining the value of phase $\gamma$,
\begin{equation}
\label{BiL-DR-gamma}
\cos\gamma=\cos\varphi_a\cos\varphi_b-\frac{1}{2}\left(\frac{Z_a}{Z_b}+\frac{Z_b}{Z_a}\right)\sin\varphi_a\sin\varphi_b\,.
\end{equation}
This dispersion relation can be also obtained from the Bloch condition $\psi(x+d)=\exp(i\kappa d)\psi(x)$ defining the Bloch wave number $\kappa=\gamma/d$ for a periodic bi-layer array with the period $d=d_a+d_b$ \cite{B28,F883}. Note that Eq.~\eqref{BiL-DR-gamma} coincides with the dispersion relation \eqref{KPTE-DR} in the case of a non-magnetic media, $\mu_{a,b}=1$.

\begin{figure}[!ht]
\begin{center}
\includegraphics[scale=0.78]{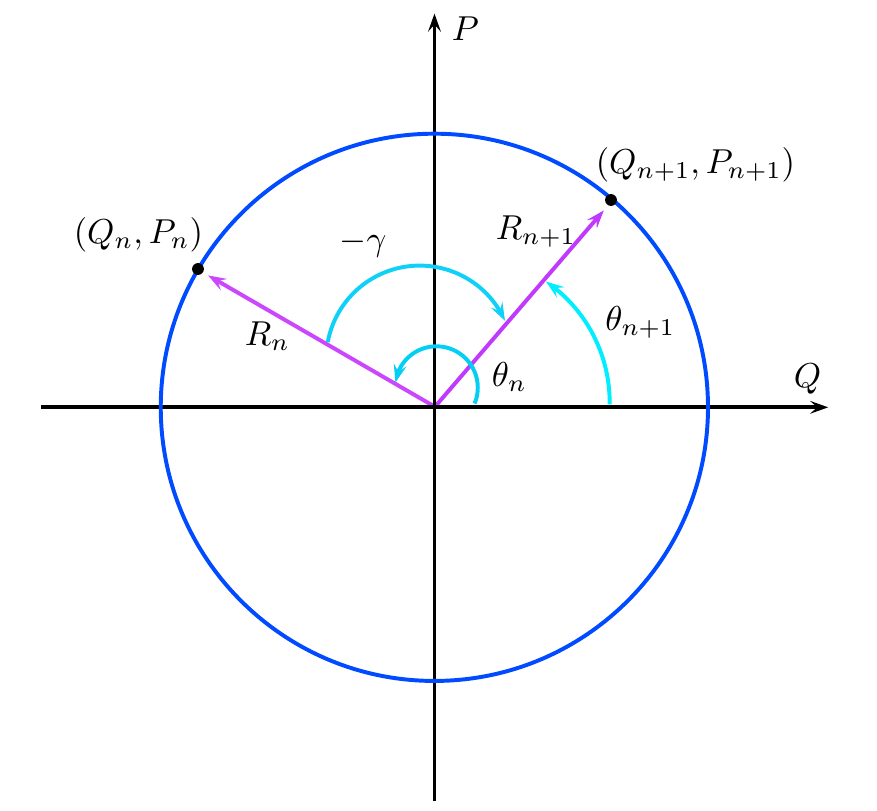}
\end{center}
\vspace{-0.5cm}
\caption{(color online). Unperturbed Hamiltonian map of periodic bi-layer system.}
\label{BiL-Fig01}
\end{figure}

It is suitable now to pass to the radius-angle variables $R_n$ and $\theta_n$ via the standard transformation,
\begin{equation}\label{BiL-QP-RTheta}
Q_n=R_n\cos\theta_n,\qquad P_n=R_n\sin\theta_n.
\end{equation}
By the direct substitution of Eq.~\eqref{BiL-QP-RTheta} into the map \eqref{BiL-Map-QPunp} one can see that for the unperturbed trajectory the radius $R_n$ is conserved, while its phase $\theta_n$ changes by the Bloch phase $\gamma$ in one step of time $n$,
\begin{equation}\label{BiL-Map-RThetaUnp}
R_{n+1}=R_n,\qquad \theta_{n+1}=\theta_n-\gamma.
\end{equation}

\subsubsection{Weak positional disorder}
\label{11.2.2}

Evidently, the weak perturbation of the phase shifts $\widetilde{\varphi}_{a}(n)$ and $\widetilde{\varphi}_{b}(n)$ results in a distortion of the circle \eqref{BiL-Map-RThetaUnp} and can be evaluated in the following way. First, in the initial map \eqref{BiL-Map-XY} we expand the factors $\widetilde{A}_n$, $\widetilde{B}_n$, $\widetilde{C}_n$, $\widetilde{D}_n$ up to the second order in the perturbation parameters $\xi_{a}(n)\ll1$ and $\xi_{b}(n)\ll1$ entering the phase shifts \eqref{BiL-phi-ab}. Then we transform the coordinates $X_n$ and $Y_n$ into $Q_n$ and $P_n$ in accordance with Eqs.~\eqref{BiL-XY-QP} in which the variables $\upsilon$, $\tau$ and Bloch phase $\gamma$ are defined for the unperturbed system. After getting the perturbed map for $Q_n$ and $P_n$, we pass to the radius-angle variables $R_n$ and $\theta_n$ with the use of Eq.~\eqref{BiL-QP-RTheta}. As a result, after a quite cumbersome calculations we get the recurrent relations (perturbed map) for the radius $R_n$ and angle $\theta_n$,
\begin{subequations}\label{BiL-MapRTheta-WD}
\begin{eqnarray}
&&\frac{R_{n+1}^2}{R_n^2}=1+\xi_{a}(n)V_{a}(\theta_n)+ \xi_{b}(n)V_{b}(\theta_n)+\xi_{a}^2(n)W_{a}+ \xi_{b}^2(n)W_{b}+
\xi_{a}(n)\xi_{b}(n)W_{ab},\qquad\quad\label{BiL-MapR-WD}\\[6pt]
&&\theta_{n+1}-\theta_n+\gamma=\xi_{a}(n)U_{a}(\theta_n)+ \xi_{b}(n)U_{b}(\theta_n)\,,\label{BiL-MapTheta-WD}
\end{eqnarray}
\end{subequations}
in which we keep linear and quadratic terms in the perturbation. Here we introduced the functions
\begin{subequations}\label{BiL-VWU}
\begin{equation}\label{BiL-VaVb}
V_{a}(\theta_n)=\frac{\alpha\sin\varphi_b}{\sin\gamma}\sin2\theta_n,\qquad
V_{b}(\theta_n)=-\frac{\alpha\sin\varphi_a}{\sin\gamma}\sin(2\theta_n-\gamma);
\end{equation}
\begin{equation}\label{BiL-WaWbWab}
W_{a}=\frac{\alpha^2\sin^2\varphi_b}{2\sin^2\gamma},\qquad
W_{b}=\frac{\alpha^2\sin^2\varphi_a}{2\sin^2\gamma},\qquad
W_{ab}=-\frac{\alpha^2\sin\varphi_a\sin\varphi_b}{\sin^2\gamma}\cos\gamma;
\end{equation}
\begin{equation}\label{BiL-UaUb}
U_{a}(\theta_n)=\frac{\alpha\sin\varphi_b}{2\sin\gamma}\cos2\theta_n,\qquad
U_{b}(\theta_n)=-\frac{\alpha\sin\varphi_a}{2\sin\gamma}\cos(2\theta_n-\gamma).
\end{equation}
\end{subequations}
Note that Eqs.~\eqref{BiL-WaWbWab} and \eqref{BiL-UaUb} are not exact. We keep here only the terms that contribute to the expression for the localization length $L_{loc}$. The expressions \eqref{BiL-VWU} contain the \emph{mismatching factor},
\begin{equation}\label{BiL-Alpha}
\alpha=\left(\frac{Z_a}{Z_b}-\frac{Z_b}{Z_a}\right)\,,
\end{equation}
characterizing how good two layers, $a$ and $b$, are matched on interfaces. In comparison with the unperturbed relation \eqref{BiL-Map-RThetaUnp}, the variations in the thicknesses of slabs are reflected by perturbations of both the radius $R_n$ and angle $\theta_n$. Below the obtained relations \eqref{BiL-MapRTheta-WD} are considered as the starting point for the derivation of the localization length.

\subsection{Localization length}
\label{11.3}

In the same way as in previous Sections we define the localization length $L_{loc}$ via the Lyapunov exponent $\lambda$,
\begin{equation}\label{BiL-LyapDef}
\frac{d}{L_{loc}}\equiv\lambda=\frac{1}{2}\langle \ln\left(\frac{R_{n+1}}{R_n}\right)^2\rangle\,.
\end{equation}

Now we substitute the recurrent relation \eqref{BiL-MapR-WD} into Eq.~\eqref{BiL-LyapDef} and expand the logarithm within the quadratic approximation in the perturbation parameters,
\begin{eqnarray}\label{BiL-Lyap-VW}
\lambda&=&\frac{1}{4}\langle\xi_{a}^2(n)\rangle\left[2W_a-\langle V_a^2(\theta_n)\rangle \right]
+\frac{1}{4}\langle\xi_{b}^2(n)\rangle\left[2W_b-\langle V_b^2(\theta_n)\rangle \right]\nonumber\\[6pt]
&&+\frac{1}{2}\langle\xi_{a}(n)\xi_{b}(n)\rangle\left[W_{ab}-\langle V_a(\theta_n)V_b(\theta_n)\rangle\right]
+\frac{1}{2}\langle \xi_{a}(n)V_a(\theta_n)\rangle+\frac{1}{2}\langle \xi_{b}(n)V_b(\theta_n)\rangle.\qquad
\end{eqnarray}
Within the discussed approximation the random quantities $\xi_{a}^2(n)$, $\xi_{b}^2(n)$ and $\xi_{a}(n)\xi_{b}(n)$ should be regarded as the terms that are uncorrelated with the functions \eqref{BiL-VaVb} containing the angle variable $\theta_n$. Then, we assume the distribution of phase $\theta_n$ to be homogenous within the first order of approximation in weak disorder. This assumption is correct over the whole spectrum, apart from the vicinity of band edges where $\gamma=0,\pi$, and band center with $\gamma=\pi/2$ \cite{IRT98,HIT08}. The homogeneous distribution of $\theta_n$ implies that
\begin{eqnarray}\label{BiL-theta-aver}
&&2\langle \sin^22\theta_n\rangle=2\langle \cos^22\theta_n\rangle= 2\langle\sin^2(2\theta_n-\gamma)\rangle=1,\nonumber\\[6pt]
&&2\langle\sin2\theta_n\sin(2\theta_n-\gamma)\rangle=\cos\gamma,\qquad\langle\sin4\theta_n\rangle=\langle\cos4\theta_n\rangle=0.
\end{eqnarray}
After substituting the expressions \eqref{BiL-VaVb}, \eqref{BiL-WaWbWab} into Eq.~\eqref{BiL-Lyap-VW}, and performing the averaging over $\theta_n$, we get,
\begin{eqnarray}\label{BiL-Lyap-Exp1}
\lambda&=&\frac{\alpha^2\sin^2\varphi_b}{8\sin^2\gamma}\langle\xi_{a}^2(n)\rangle+ \frac{\alpha^2\sin^2\varphi_a}{8\sin^2\gamma}\langle\xi_{b}^2(n)\rangle
-\frac{\alpha^2\sin\varphi_a\sin\varphi_b}{4\sin^2\gamma}\langle\xi_{a}(n)\xi_{b}(n)\rangle\cos\gamma\nonumber\\[6pt]
&&+\frac{\alpha\sin\varphi_b}{2\sin\gamma}\langle\xi_{a}(n)\sin2\theta_n\rangle- \frac{\alpha\sin\varphi_a}{2\sin\gamma}\langle\xi_{b}(n)\sin(2\theta_n-\gamma)\rangle.
\end{eqnarray}

In order to proceed further, one has to calculate two non-trivial correlators $\langle\xi_{a}(n)\sin2\theta_n\rangle$ and $\langle\xi_{b}(n)\sin(2\theta_n-\gamma)\rangle$ remaining in Eq.~\eqref{BiL-Lyap-Exp1}. To this end, we start with the map \eqref{BiL-MapTheta-WD} which can be rewritten as follows
\begin{equation}
\exp\left[2i\left(\theta_{n'+1}-\theta_{n'}+\gamma\right)\right]=1+\frac{i\alpha\sin\varphi_b}{\sin\gamma}\xi_{a}(n')\cos2\theta_{n'}-
\frac{i\alpha\sin\varphi_a}{\sin\gamma}\xi_{b}(n')\cos(2\theta_{n'}-\gamma).
\end{equation}
Note that here it is sufficient to keep the linear terms in the perturbation, and to neglect the quadratic ones. This equation is presented in a more suitable form,
\begin{eqnarray}\label{BiL-MapTheta-exp}
&&\exp\left[2i\left(\theta_{n'+1}+\gamma\right)\right]=\exp\left(2i\theta_{n'}\right)+
\frac{i\alpha\sin\varphi_b}{\sin\gamma}\xi_{a}(n')\exp\left(2i\theta_{n'}\right)\cos2\theta_{n'}\nonumber\\[6pt]
&&-\frac{i\alpha\sin\varphi_a}{\sin\gamma}\xi_{b}(n') \exp\left(2i\theta_{n'}\right)\cos(2\theta_{n'}-\gamma).
\end{eqnarray}
Now we put $n'=n-r$ in Eq.~\eqref{BiL-MapTheta-exp} and multiply its both sides separately by $\xi_a(n)$ and $\xi_b(n)$, with the subsequent averaging over disorder, taking into account the relations \eqref{BiL-theta-aver}. As a result, we obtain,
\begin{subequations}\label{BiL-xi-theta}
\begin{eqnarray}
&&\langle\xi_{a}(n)\exp\left(2i\theta_{n-r+1}\right)\rangle\exp\left(2i\gamma\right)=\langle \xi_{a}(n)\exp\left(2i\theta_{n-r}\right)\rangle+ \frac{i\alpha\sin\varphi_b}{2\sin\gamma}\langle\xi_{a}(n)\xi_{a}(n-r)\rangle\nonumber\\[6pt]
&&-\frac{i\alpha\sin\varphi_a}{2\sin\gamma}\exp(i\gamma)\langle\xi_{a}(n)\xi_{b}(n-r)\rangle,\\[6pt]
&&\langle\xi_{b}(n)\exp\left(2i\theta_{n-r+1}\right)\rangle\exp\left(2i\gamma\right)=\langle \xi_{b}(n)\exp\left(2i\theta_{n-r}\right)\rangle+
\frac{i\alpha\sin\varphi_b}{2\sin\gamma}\langle\xi_{b}(n)\xi_{a}(n-r)\rangle\nonumber\\[6pt]
&&-\frac{i\alpha\sin\varphi_a}{2\sin\gamma}\exp(i\gamma)\langle\xi_{b}(n)\xi_{b}(n-r)\rangle.
\end{eqnarray}
\end{subequations}
Finally, we multiply the recurrent relations \eqref{BiL-xi-theta} by $\exp(-2i\gamma r)$ and perform the summation over $r$ from one to infinity. Using definitions \eqref{BiL-CorrDef}, \eqref{BiL-phi-ab} we readily get the expressions for two independent correlators,
\begin{subequations}\label{BiL-Corr-xi-theta}
\begin{eqnarray}
&&\langle\xi_{a}(n)\exp\left(2i\theta_{n}\right)\rangle=
\frac{i\alpha\sin\varphi_b}{2\sin\gamma}\langle\xi_{a}^2(n)\rangle\sum_{r=1}^{\infty}K_a(r)\exp(-2i\gamma r)\nonumber\\[6pt]
&&-\frac{i\alpha\sin\varphi_a}{2\sin\gamma}\langle\xi_{a}(n)\xi_{b}(n)\rangle\sum_{r=1}^{\infty}K_{ab}(r)\exp(-2i\gamma r+i\gamma),\\[6pt]
&&\langle\xi_{b}(n)\exp\left(2i\theta_{n}-i\gamma\right)\rangle=
-\frac{i\alpha\sin\varphi_a}{2\sin\gamma}\langle\xi_{b}^2(n)\rangle\sum_{r=1}^{\infty}K_{b}(r)\exp(-2i\gamma r)\nonumber\\[6pt]
&&+\frac{i\alpha\sin\varphi_b}{2\sin\gamma}\langle\xi_{a}(n)\xi_{b}(n)\rangle\sum_{r=1}^{\infty}K_{ab}(r)\exp(-2i\gamma r-i\gamma).
\end{eqnarray}
\end{subequations}
The correlators that we need to find in Eq.~\eqref{BiL-Lyap-Exp1}, are the imaginary parts of the corresponding correlators \eqref{BiL-Corr-xi-theta}.

After the substitution of the obtained results into Eq.~\eqref{BiL-Lyap-Exp1}, together with the use of Eq.~\eqref{BiL-FT-K} for the randomness power spectra, we arrive at the final expression for dimensionless Lyapunov exponent $\lambda=d/L_{loc}$,
\begin{equation}\label{BiL-LyapFin}
\lambda=\frac{\alpha^2}{8\sin^2\gamma}
\Big[k_a^2\sigma^2_a\mathcal{K}_a(2\gamma)\sin^2\varphi_b+k_b^2\sigma^2_b\mathcal{K}_b(2\gamma)\sin^2\varphi_a-
2k_ak_b\sigma^2_{ab}\mathcal{K}_{ab}(2\gamma)\sin\varphi_a\sin\varphi_b\cos\gamma\Big],
\end{equation}
Note that Eq.~\eqref{BiL-LyapFin} is symmetric with respect to the interchange of slab indices $a\leftrightarrow b$. For a white noise disorder in $a$ slabs only ($\sigma_b=\sigma_{ab}=0$ and $\mathcal{K}_a(k)=1$), the obtained expression \eqref{BiL-LyapFin} reduces to Eq.~\eqref{KPTE-Lloc}.

Eq.~\eqref{BiL-LyapFin} should be complemented by the dispersion relation \eqref{BiL-DR-gamma} determining the band-structure in the frequency dependence $\omega(\gamma)$. Clearly, our result for the Lyapunov exponent $\lambda(\omega)$ is reasonable only for real values of the Bloch phase $\gamma$ defining {\it spectral bands}. Inside the \emph{spectral gaps} with $\gamma$ imaginary or complex, the solution corresponds to evanescent Bloch states localized on the scale of the order of $\mathrm{Im}^{-1}\gamma$. This results in the exponentially small transmission even in the case of no disorder, and the presence of weak disorder gives only a small correction for the position of bands. Due to evenness and periodicity of Eqs.~\eqref{BiL-LyapFin} and \eqref{BiL-DR-gamma} with respect to $\gamma$, one can consider only the values of $\gamma$ in the interval $0\leqslant\gamma\leqslant\pi$. It is important to emphasize that the value of the Lypunov exponent \eqref{BiL-LyapFin} is quite sensitive to the parameters of the model, thus resulting in various dependencies of $L_{loc}(\omega)$ inside the energy bands.

From Eq.~\eqref{BiL-LyapFin} one can see that the Lyapunov exponent $\lambda$ consists of three terms. The first two terms are contributed  by the correlations between solely $a$ or solely $b$ slabs, respectively. Therefore, these terms contain the \emph{auto-correlators} $\mathcal{K}_a(2\gamma)$ and $\mathcal{K}_b(2\gamma)$. The third term includes the \emph{cross-correlator}
$\mathcal{K}_{ab}(2\gamma)$ that emerges due to the correlations between two disorders in the $a$ and $b$ layers.

\subsection{Photonic crystals}
\label{11.4}

We discuss here the application of our results to 1D disordered photonic crystals for which all optical parameters $\varepsilon_{a,b}$, $\mu_{a,b}$, $n_{a,b}$, and $Z_{a,b}$ are \emph{positive} constants. The typical frequency dependence of the Lyapunov exponent $\lambda(\omega)$ for the conventional bi-layer stack is shown in Fig.~\ref{BiL-Fig02}. One can see that the dependence $\lambda(\omega)$ is principally different in the first and higher frequency Bloch bands. In particular, the first band starts from $\lambda=0$, while in other bands the Lyapunov exponent diverges at the band edges.

In the first spectral band the localization length $L_{loc}(\omega)=d/\lambda(\omega)$ monotonously decreases from the infinite value at $\omega=0$ where the Bloch phase $\gamma=0$, and vanishes at the band edge where $\gamma=\pi$. For small values of $\omega$ the expressions for the Lyapunov exponent \eqref{BiL-LyapFin} and dispersion relation \eqref{BiL-DR-gamma} are reduced to
\begin{subequations}\label{BiL-Bot}
\begin{eqnarray}
\lambda=\frac{d}{L_{loc}}=\frac{\varphi^2_a\varphi^2_b}{8\gamma^2}\alpha^2\Big[\frac{\sigma^2_{a}}{d_a^2}\,\mathcal{K}_a(2\gamma)+ \frac{\sigma^2_{b}}{d_b^2}\,\mathcal{K}_b(2\gamma)-2\frac{\sigma^2_{ab}}{d_ad_b}\,\mathcal{K}_{ab}(2\gamma)\Big],\label{BiL-Lyap-Bot}\\[6pt]
\gamma^2=(\varphi_a+\varphi_b)^2+\frac{(Z_a-Z_b)^2}{Z_aZ_b}\varphi_a\varphi_b,\qquad\varphi^2_a\ll1,\quad\varphi^2_b\ll1.\label{BiL-DR-Bot}
\end{eqnarray}
\end{subequations}
Typically, the expression in brackets of Eq.~\eqref{BiL-Lyap-Bot} is finite for $\gamma=0$, leading to the conventional frequency dependence of the Lyapunov exponent, see Eq.~\eqref{1DCP-LlocOmega},
\begin{equation}\label{BiL-LyapOmega2}
\lambda=d/L_{loc}\propto\omega^2\qquad\mathrm{when}\quad\omega\to0.
\end{equation}
However, it is not a universal dependence in the presence of correlations in random potentials. As demonstrated in next Subsection, the dependence \eqref{BiL-LyapOmega2} may not hold even when two, $a$ and $b$, disorders are of the white-noise type ($\mathcal{K}_{a}=\mathcal{K}_{b}=1$).

\begin{figure}[!ht]
\begin{center}
\includegraphics[scale=0.4]{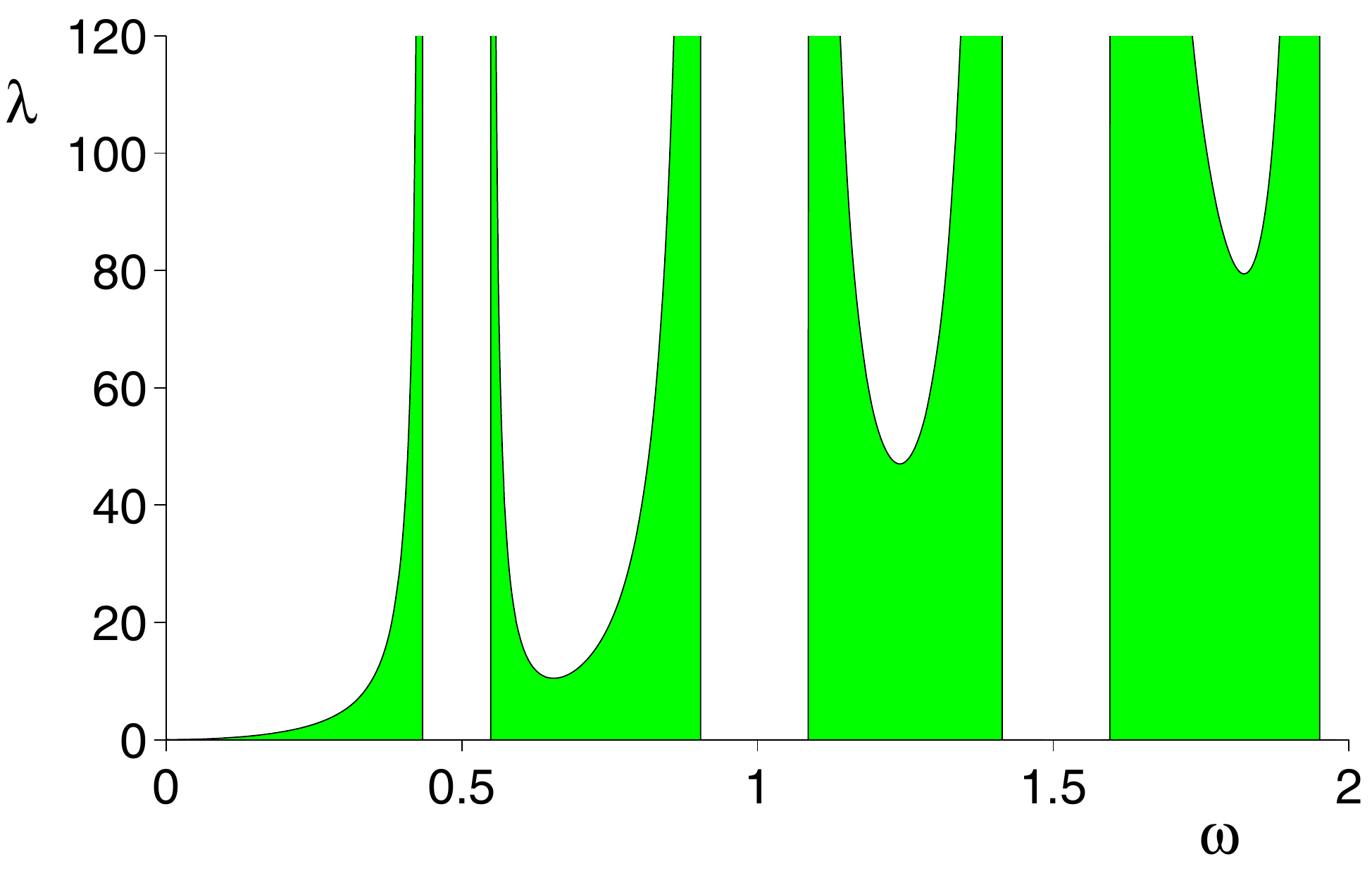}
\end{center}
\vspace{-0.5cm}\caption{(color online). Lyapunov exponent versus frequency in arbitrary units for photonic layered medium with
Eqs.~\eqref{BiL-DR-gamma} and \eqref{BiL-LyapFin}. The parameters are: $\alpha^2n_a^2\sigma^2_{a}/2c^2\approx12.28$,
$\alpha^2n_b^2\sigma^2_{b}/2c^2\approx0.27$ and $n_ad_a/c=1.6$, $n_bd_b/c=0.4$, $\mathcal{K}_{a}=\mathcal{K}_{b}=\mathcal{K}_{ab}=1$ (after \cite{IM09}).}
\label{BiL-Fig02}
\end{figure}

Starting from the second Bloch band the behavior of the localization length $L_{loc}(\omega)$ is somewhat similar. Specifically, $L_{loc}(\omega)$ vanishes at the edges of spectral bands (where $\gamma=0,\pi$) due to the term $\sin^2\gamma$ in the denominator of Eq.~\eqref{BiL-LyapFin}. The localization length achieves its maximal value in the ``middle" of each frequency band, where $\gamma=\pi/2$. It is noteworthy that at $\gamma=\pi/2$, the third term in Eq.~\eqref{BiL-LyapFin} vanishes and the cross-correlations do not influence the localization length. It should be stressed that the spectral properties of $\lambda(\omega)$ mentioned above and displayed in Fig.~\ref{BiL-Fig02}, can be strongly modified by specific correlations occurring in the disorder. For example, the correlations considered in Ref.~\cite{DIKR08}, see Section~\ref{5.3}, can be used to cancel a sharp frequency dependence related to the term $\sin^2\gamma$ in the denominator of Eq.~\eqref{BiL-LyapFin}.

If the impedances of $a$ and $b$ slabs are equal, $Z_a=Z_b$, the mismatching factor $\alpha$ entering Eq.~\eqref{BiL-LyapFin} vanishes and the localization length diverges. Thus, the perfect transparency emerges in spite of the presence of a positional disorder. This conclusion is general \cite{MS08} and does not depend on the strength of disorder. Indeed, since the layers are perfectly matched, there are no waves reflected from the interfaces. According to Eq.~\eqref{BiL-DR-gamma}, in such a case the stack-structure is effectively equivalent to the homogeneous medium with the linear spectrum and average refractive index $\overline{n}$,
\begin{equation}\label{BiL-DR-EqualImp}
\kappa\equiv\gamma/d=\omega\overline{n}/c,\qquad \overline{n}=\frac{n_ad_a+n_bd_b}{d_a+d_b}\,.
\end{equation}
Note that in this case there are no gaps in the spectrum.

Apart from the specific case of equal impedances, the Lyapunov exponent $\lambda(\omega)$ manifests the Fabry-Perot resonances associated with multiple reflections inside $a$ or $b$ slabs from the interfaces. As known, they appear when the thickness $d_{a,b}$ of the corresponding $a$ or $b$ layer equals an integer multiple of half of the wavelength $2\pi c/\omega n_{a,b}$ inside the layer,
\begin{equation}\label{BiL-FPab}
\omega/c=s_a\pi/n_ad_a\qquad\mathrm{or}\qquad\omega/c=s_b\pi/n_bd_b\quad\mathrm{with}\quad s_{a,b}=1,2,3,\dots.
\end{equation}
At the resonances the factor $\sin\varphi_a$ or $\sin\varphi_b$ in Eq.~\eqref{BiL-LyapFin} vanishes giving rise to the resonance increase of the localization length $L_{loc}$ and consequently, to the suppression of localization. It is interesting that in Eq.~\eqref{BiL-LyapFin} the first term, related to auto-correlations of $a$ slabs, displays the Fabry-Perot resonances associated with $b$ layers. Respectively, the second term emerging due to auto-correlations of $b$ slabs, contains the resonances of $a$ layers. The last term related to the cross-correlations provides the resonances of both layers. One has to emphasize that the Fabry-Perot resonances turn out to be quite broad because they are caused by vanishing of smooth trigonometric functions. As a consequence, the localization can be effectively suppressed within a large region of spectral band in which such a resonance occurs.

It is interesting that in the special case for which the ratio of optical thicknesses $n_ad_a$ and $n_bd_b$ of two slabs is a rational number, $n_ad_a/n_bd_b=s_a/s_b$, the corresponding resonances coincide. In such a case one may expect that they give rise to the divergence of $L_{loc}(\omega)$. However, in line with Eq.~\eqref{BiL-DR-gamma}, such a situation can arise only at the edges of spectral bands ($\gamma=0, \pi$) where $\sin^2\gamma$ also vanishes. Therefore, the value of $L_{loc}(\omega)$ can be finite at these special points, instead of diverging (due to resonances) or vanishing (due to band edges).

Of special interest are the long-range correlations leading to the divergence (or to significant decrease) of the localization length $L_{loc}(\omega)$ in the controlled frequency window. This effect is similar to that found in more simple 1D models with correlated disorder (see previous Sections). Here, in the considered model of two disorders this effect is due to a possibility to have the vanishing values of {\it all} Fourier transforms, $\mathcal{K}_{a}=\mathcal{K}_{b}=\mathcal{K}_{ab}=0$, in some intervals of frequency $\omega$. This can be done, e.g., in the way described in Section~\ref{8.3.2}. In such a manner, one can artificially construct an array of random bi-layers with the power spectra abruptly vanishing within a prescribed interval of $\omega$, therefore, resulting in the divergence of the localization length, therefore, {\it suppressing} the localization. On the contrary, with the use of specific correlations one can reduce the localization length, and significantly \emph{enhance} the localization although the disorder remains to be weak.

\subsection{Quarter stack layered medium}
\label{11.5}

The term ``quarter stack" is used for the periodic bi-layer structures for which two basic layers, $a$ and $b$, have the same optical thickness (see, e.g., \cite{MS08}),
\begin{equation}\label{BiL-QSdef}
n_ad_a=n_bd_b.
\end{equation}
Since the unperturbed phase shifts are equal, $\varphi_a=\varphi_b$, the expression \eqref{BiL-LyapFin} for the Lyapunov exponent and the dispersion relation \eqref{BiL-DR-gamma} take a simpler form,
\begin{subequations}\label{BiL-QS-LyapDR}
\begin{eqnarray}
\lambda=\frac{d}{L_{loc}}&=&\frac{(Z_a-Z_b)^2}{Z_aZ_b}\,
\frac{k_a^2\sigma^2_{a}\mathcal{K}_a(2\gamma)+k_b^2\sigma^2_{b}\mathcal{K}_b(2\gamma)- 2k_ak_b\sigma_{ab}^2\mathcal{K}_{ab}(2\gamma)\cos\gamma}{8\cos^2(\gamma/2)}\,,\qquad\label{BiL-QS-Lyap}\\[6pt]
\cos\gamma&=&1-\frac{(Z_a+Z_b)^2}{2Z_aZ_b}\sin^2\varphi_a.\label{BiL-QS-DRgamma}
\end{eqnarray}
\end{subequations}
A specific feature of the band structure described by Eq.~\eqref{BiL-QS-DRgamma} is that starting from the second band, the top of every even band coincides with the bottom of the next odd band at $\gamma=0$, or, the same, at $\omega/c=s\pi/n_ad_a$ ($\varphi_a=s\pi$, $s=1,2,3,\dots$). The gaps arise only at $\gamma=\pi$. Therefore, the Lyapunov exponent is finite at the above values of $\omega/c$, if the numerator in Eq.~\eqref{BiL-QS-Lyap} is also finite. On the other hand, if the numerator does not vanish at $\gamma=\pi$, the Lyapunov exponent diverges at the bottom of each even band and at the top of each odd band.

In order to demonstrate a remarkable sensitivity of the dependence $\lambda(\omega)$ to particular properties of the disorder, it is instructive to analyze two complementary types of correlations. In the first example termed ``\emph{plus correlations}", the random parts of optical thicknesses are equal,
\begin{equation}\label{BiL-PlusCorr}
n_a\varrho_a(n)=n_b\varrho_b(n).
\end{equation}
In this case one can speak about an equality of relative disorders in both layers, $\varrho_a(n)/d_a=\varrho_b(n)/d_b$. As a result, one gets $k_a^2\sigma^2_{a}\mathcal{K}_a(2\gamma)=k_b^2\sigma^2_{b}\mathcal{K}_b(2\gamma) =k_ak_b\sigma_{ab}^2\mathcal{K}_{ab}(2\gamma)$, and the Lyapunov exponent reads
\begin{equation}\label{BiL-Lyap-QSplus}
\lambda_{+}=\frac{(Z_a-Z_b)^2}{2Z_aZ_b}\,k_a^2\sigma^2_{a}\mathcal{K}_a(2\gamma)\tan^2\frac{\gamma}{2}.
\end{equation}
From Eqs.~\eqref{BiL-Lyap-QSplus} and \eqref{BiL-QS-DRgamma} one can see that for finite values of $\mathcal{K}_a(0)$ the Lyapunov exponent is proportional to the fourth power of the frequency $\omega$ at the bottom of the spectrum,
\begin{equation}\label{BiL-Lyap-QSplusBot}
\lambda_{+}\propto\omega^4\qquad\mathrm{when}\quad\omega\to0,
\end{equation}
in contrast with the conventional dependence $\lambda\propto\omega^2$. Moreover, the Lyapunov exponent vanishes at all points at which the top of even bands coincides with the bottom of next odd bands, where $\gamma=0$ and $\varphi_a=s\pi$ ($s=1,2,3,\dots$),
\begin{equation}\label{BiL-Lyap-QSplusCoin}
\lambda_{+}\propto(\omega-s\pi c/n_ad_a)^2\qquad\mathrm{when} \quad\omega\to s\pi c/n_ad_a.
\end{equation}
The dependence of $\lambda_{+}$ on the wave frequency $\omega$ is shown in the left panel of Fig.~\ref{BiL-Fig0405}.

\begin{figure}[!ht]
\begin{center}
\subfigure{\includegraphics[scale=0.5]{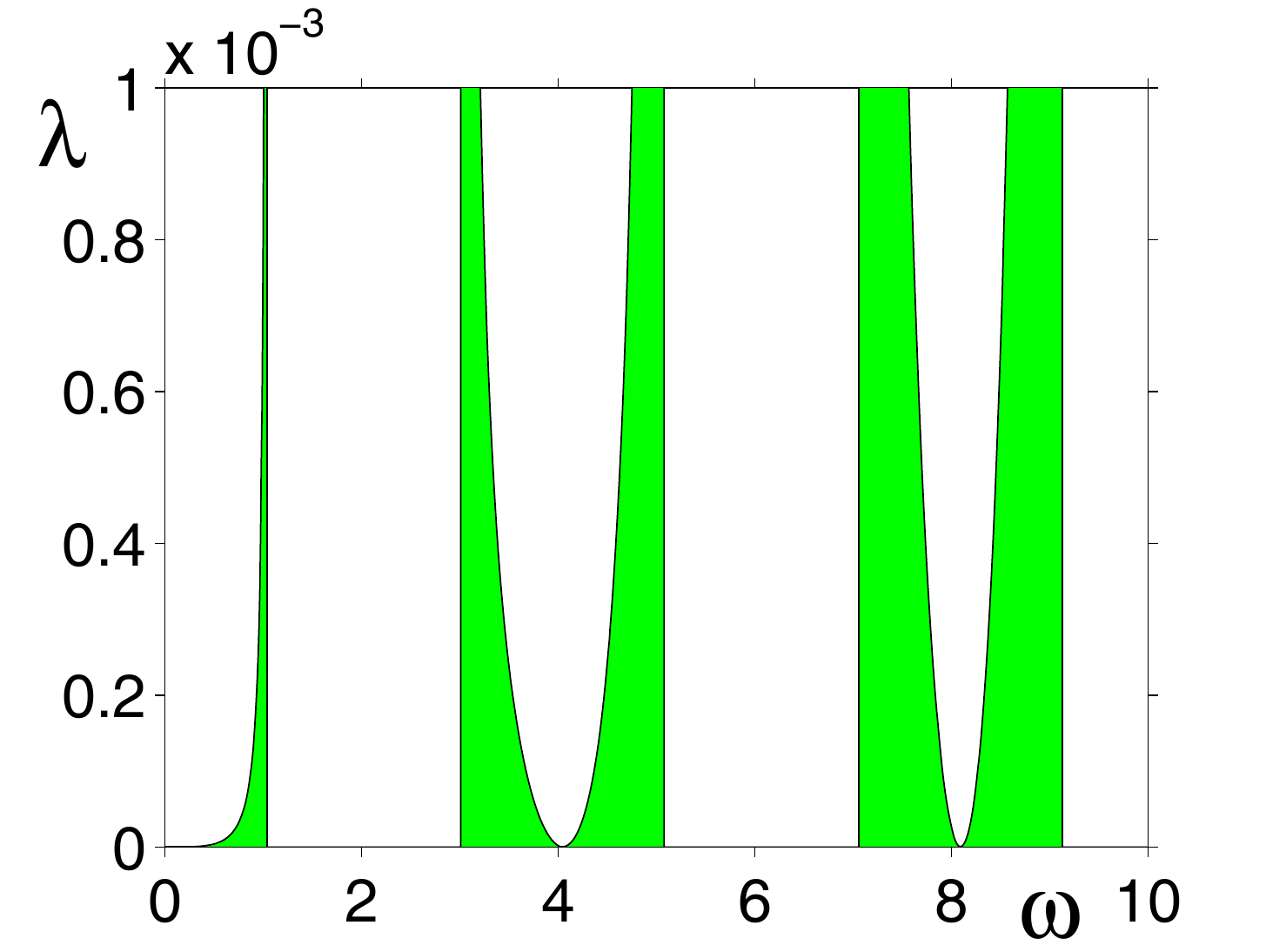}}
\subfigure{\includegraphics[scale=0.3]{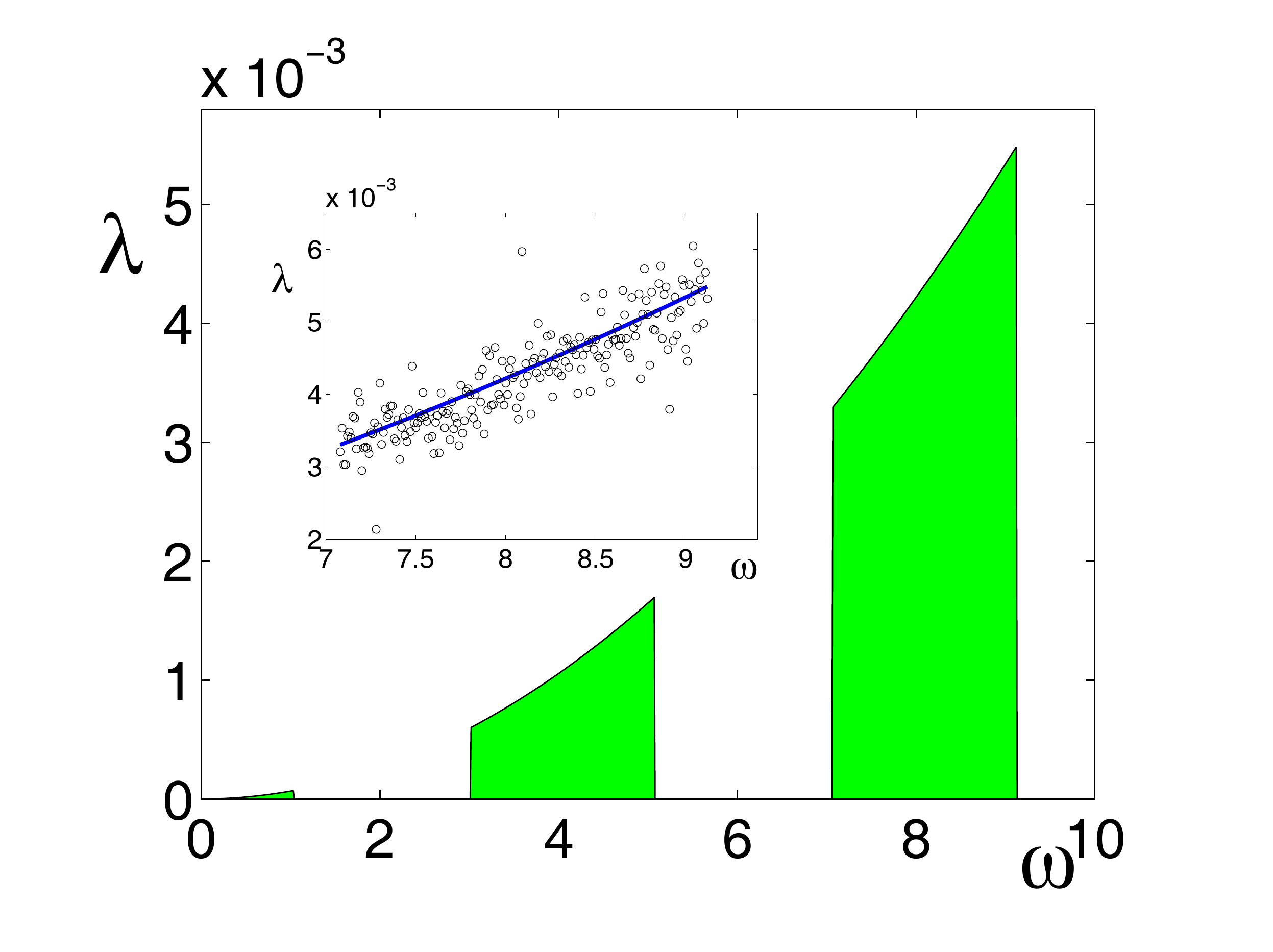}}
\end{center}
\caption{(color online). Lyapunov exponent versus frequency for the quarter stack layered medium, for the plus correlations (left) and minus correlations (right); see Eqs.~\eqref{BiL-QS-DRgamma}, \eqref{BiL-Lyap-QSplus} and \eqref{BiL-Lyap-QSminus}. Here $n_a=3.47$ (silicon), $d=1\mu$m, $d_a=0.224\mu$m, $\sigma^2_{a}=3.75\times 10^{-5}\mu \mathrm{m}^2$, $c=1.0$, and $\mathcal{K}_a=1$ (after \cite{IM09}).}
\label{BiL-Fig0405}
\end{figure}

In the second example termed ``minus correlations", the optical thicknesses of a disorder in $a$ and $b$ slabs have different signs,
\begin{equation}\label{BiL-MinusCorr}
n_a\varrho_a(n)=-n_b\varrho_b(n),
\end{equation}
which is equivalent to $\varrho_a(n)/d_a=-\varrho_b(n)/d_b$. In this case the Lyapunov exponent is described by the expression,
\begin{equation}\label{BiL-Lyap-QSminus}
\lambda_{-}=\frac{(Z_a-Z_b)^2}{2Z_aZ_b}\,k_a^2\sigma^2_{a}\mathcal{K}_a(2\gamma).
\end{equation}
Surprisingly, for the minus correlations and $\mathcal{K}_a(2\gamma)=1$ the Lyapunov exponent is quadratic in $\omega$ ($\lambda_{-}\propto\omega^2$) inside {\it any} spectral band and does not diverge at its edges (see Fig.~\ref{BiL-Fig0405}, right panel). The physical reason of this effect is that in this case the total optical length of every $(a,b)$ cell is constant,
\begin{equation}
n_ad_a(n)+n_bd_b(n)=2n_ad_a,
\end{equation}
although its geometrical thickness $d_a(n)+d_b(n)$ fluctuates randomly. For such correlations the quadratic $\omega$ dependence seems to survive in the nonperturbative regime as well.

In Fig.\ref{BiL-Fig0405} we have used the parameters for a silicon-air stack with plus and minus correlations between two disorders. In order to check our analytical predictions, in the inset we present the numerical solution of Eq.~\eqref{BiL-WaveEqBC} for the Lyapunov exponent in the third frequency band for $N=4\cdot 10^4$ $ab$ cells. The result shows a quite good correspondence, taking into account that such a correspondence is, in fact, of a statistical nature.

\subsection{Metamaterials}
\label{11.6}

Let us now consider \emph{mixed systems} in which the $a$ layer is a conventional right-handed (RH) material, while the $b$ layer is a left-handed (LH) material. This means that the dielectric constant, magnetic permeability and corresponding refractive index in $a$ slabs are positive, whereas they are negative in $b$ slabs. However, in agreement with the definition \eqref{BiL-nZk}, both impedances are positive,
\begin{equation}\label{BiL-MetaDef}
\varepsilon_a,\mu_a,n_a>0,\qquad\varepsilon_b,\mu_b,n_b<0,\qquad
Z_a,Z_b>0.
\end{equation}

Remarkably, in comparison with the conventional stack-structure, the expression \eqref{BiL-LyapFin} for the Lyapunov exponent remains the same. The only difference is that now in the dispersion relation \eqref{BiL-DR-gamma} the sign ``plus" replaces sign ``minus" at the second term,
\begin{equation}\label{BiL-Meta-DRgamma}
\cos\gamma=\cos\varphi_a\cos\varphi_b+\frac{1}{2}\left(\frac{Z_a}{Z_b}+\frac{Z_b}{Z_a} \right)\sin\varphi_a\sin|\varphi_b|,
\end{equation}
since the phase shift $\varphi_{b}$ in the LH slab is negative,
\begin{equation}\label{BiL-Meta-phib}
\varphi_{b}\equiv k_{b}d_b=-\omega|n_b|d_b/c.
\end{equation}
Such a ``minor" correction drastically changes the frequency dependence of the Lyapunov exponent \eqref{BiL-LyapFin}, see Fig.~\ref{BiL-Fig03}. Nevertheless, the Lyapunov exponent typically obeys the conventional dependence $\lambda\propto\omega^2$ when $\omega\to 0$, see Eq.~\eqref{BiL-Bot}.

\begin{figure}[!ht]
\begin{center}
\subfigure{\includegraphics[scale=0.4]{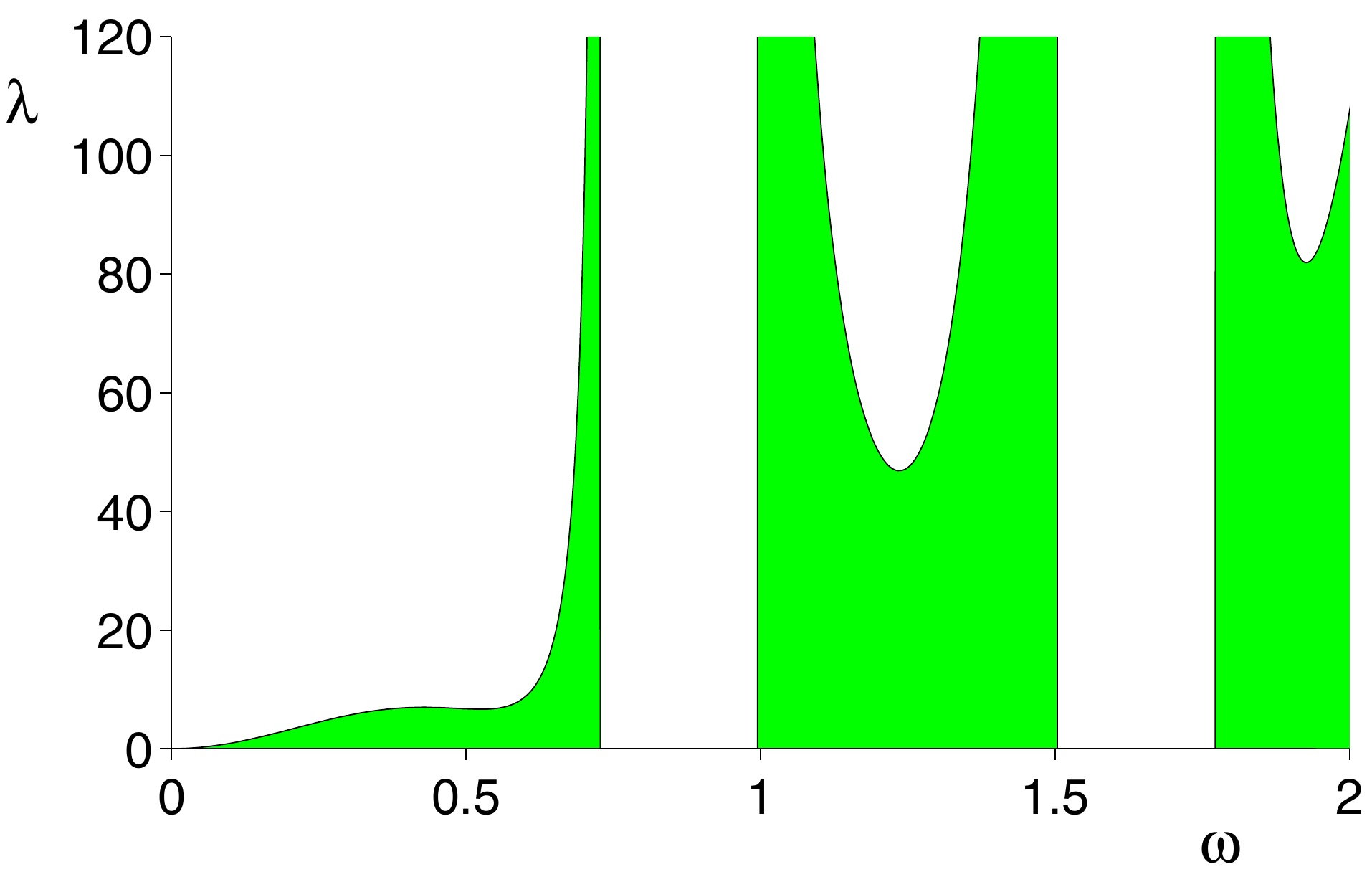}}
\end{center}
\vspace{-0.5cm} \caption{(color online). The same as in Fig.~\ref{BiL-Fig02} for RH-LH bi-layers with ``plus" in Eq.~\eqref{BiL-DR-gamma} (after \cite{IM09}).}
\label{BiL-Fig03}
\end{figure}

Note that the {\it ideal mixed stack} ($\varepsilon_a=\mu_a=n_a=1$ $\varepsilon_b=\mu_b=n_b=-1$, $Z_a=Z_b=1$) has a perfect transmission, $\lambda=0$, even in the presence of a disorder. This effect is due to the perfect matching of the vacuum with the ideal LH material, for which there are no reflecting waves. The ideal mixed stack has a linear spectrum without gaps,
\begin{equation}\label{BiL-DR-IdMeta}
\kappa\equiv\gamma/d=\omega\overline{n}_{id}/c,\qquad \overline{n}_{id}=\frac{|d_a-d_b|}{d_a+d_b}\,.
\end{equation}
This fact directly results from the dispersion relation \eqref{BiL-Meta-DRgamma}, compare with Eq.~\eqref{BiL-DR-EqualImp}. From Eq.~\eqref{BiL-DR-IdMeta} it becomes evident that our conclusions here are physically meaningful only when two basic layers have different unperturbed thicknesses, $d_a\neq d_b$.

One of the interesting features of the mixed bi-layered structures is that for the \emph{RH-LH quarter stack} the average refractive index vanishes,
\begin{equation}\label{BiL-MetaQS}
\overline{n}=(n_ad_a-|n_b|d_b)/(d_a+d_b)=0\qquad\mathrm{when}\quad n_ad_a=|n_b|d_b.
\end{equation}
As a consequence, and in agreement with the dispersion relation \eqref{BiL-Meta-DRgamma}, the spectral bands disappear, therefore, the transmission is absent, probably apart from a discrete set of frequencies for which $\varphi_a=s\pi$ and $\gamma=0$ ($s=0,1,2,\ldots$). However, the expression \eqref{BiL-LyapFin} obtained for the Lyapunov exponent is not valid in such a situation, and an additional analysis is needed.

It is important to note that, practically, in periodic bi-layer metamaterials the permittivity $\varepsilon_b(\omega)$ and permeability $\mu_b(\omega)$ are \emph{frequency-dependent} \cite{MS08}. This fact is crucial for the applications. In particular, the mismatching factor $\alpha$ in Eq.~\eqref{BiL-LyapFin} can vanish only for specific values of frequency $\omega$, thus resulting in a resonance-like dependence for the transmission. Besides, a new specific gap, the so-called $\overline{n}=0$ gap, emerges in the vicinity of $\omega$ where the average refractive index $\overline{n}$ vanishes (see, e.g., Ref.~\cite{MS08} and references therein). Also, for a typical frequency dependence of $\varepsilon_b(\omega)$ and $\mu_b(\omega)$ the refractive index of $b$ slabs can be imaginary in some interval of $\omega$. This gives rise to an emergence of new transmission bands and gaps that are inherent for a dielectric-metal layered array. In particular, a new gap arises at the origin of spectrum, $\omega=0$, in contrast with conventional photonic crystals. Therefore, the question about the behavior of the Lyapunov exponent at $\omega\to0$ has no sense. The unusual localization properties in mixed bi-layer stacks with frequency-dispersive metamaterial were recently discussed in Refs.~\cite{DZ06,Mo10}.

\subsection{Semiconductor superlattices}
\label{11.7}

Our approach can be also applied to the electron transport through the structures with two alternating potential barriers of rectangular profiles having constant amplitudes $U_a$ and $U_b$ and slightly perturbed thicknesses \eqref{BiL-dab-n}. One possible application is the fabrication of nanostructured electronic systems (see, for example, Ref.~\cite{HKM92} and references therein).

The stationary 1D Schr\"{o}dinger equation for the wave functions $\psi_{a,b}(x)$ of an electron with effective masses $m_a$ and $m_b$ inside the barriers and total energy $E$ can be written in the form of Eq.~\eqref{BiL-WaveEq-ab}, where the partial wave numbers $k_a$ and $k_b$ are associated with the barriers,
\begin{equation}\label{BiL-kakb-El}
k_a=\sqrt{2m_a(E-U_a)}/\hbar,\qquad k_b=\sqrt{2m_b(E-U_b)}/\hbar.
\end{equation}
Another change, $\mu_{a,b}\to m_{a,b}$, should be done in the boundary conditions \eqref{BiL-BC} on hetero-interfaces $x=x_i$. With this change, the expression \eqref{BiL-LyapFin} for the Lyapunov exponent conserves its form. In addition, the factor $Z_a/Z_b$ entering the dispersion relation \eqref{BiL-DR-gamma} takes the form, $Z_a/Z_b=k_bm_a/k_am_b$, that also emerges in the expression for the mismatching factor $\alpha$ in Eq.\eqref{BiL-Alpha}. As a result, we have,
\begin{equation}\label{BiL-DR-El}
\cos\gamma=\cos\varphi_a\cos\varphi_b-\frac{1}{2}\left(\frac{k_am_b}{k_bm_a}+\frac{k_bm_a}{k_am_b}\right)\sin\varphi_a\sin\varphi_b,\qquad
\alpha=\left(\frac{k_am_b}{k_bm_a}-\frac{k_bm_a}{k_am_b}\right).
\end{equation}
An example of the energy dependence $\lambda(E)$ in semiconductor superlattices is given in Fig.~\ref{BiL-Fig06}.

\begin{figure}[!ht]
\vspace{-0.2cm}
\begin{center}
\includegraphics[scale=0.6]{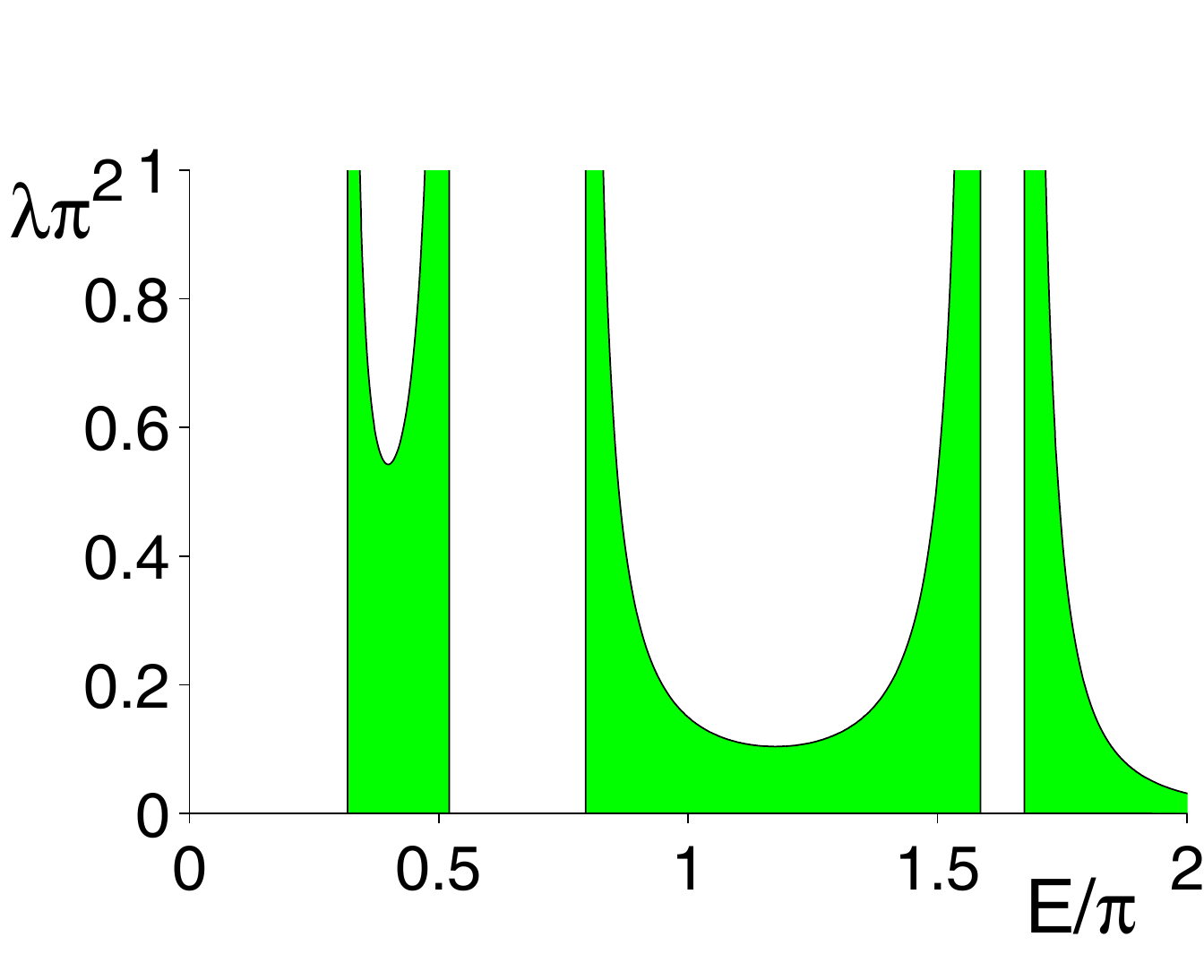}
\end{center}
\vspace{-0.5cm} \caption{(color online). Lyapunov exponent versus energy for electrons in a bi-layered structure. Here $2m_a/\hbar^2 =2m_b/\hbar^2=1$, $U_a=0.15$, $U_b=1.6$, $d_a=0.35\pi$, $d_b=0.65\pi$, $\sigma^2_{a}=(0.1d_a)^2$, and $\sigma^2_{b}=(0.2d_b)^2$. The transition between the tunneling and overbarrier scattering occurs at $E/\pi=0.5$ (after \cite{IM09}).}
\label{BiL-Fig06}
\end{figure}

If the energy $E$ is smaller than the heights of both barriers $E<U_a,\,U_b$, the electron wave numbers are imaginary, $k_a=i|k_a|$ and $k_b=i|k_b|$, and the dispersion relation \eqref{BiL-DR-El} has only the imaginary solution for the Bloch phase $\gamma=i|\gamma|$. As a consequence, the electron states are strongly localized and the structure is not transparent. Correspondingly, there is a gap at the origin of the spectrum $\lambda(E)$.

When the value of $E$ is between the amplitudes of two barriers,
\begin{equation}\label{BiL-ElTun}
U_a<E<U_b,
\end{equation}
the \emph{tunneling propagation} of electrons occurs. In this case the wave number $k_a$ is real while $k_b$ is imaginary. Therefore, the electron moves freely within any $a$ barrier and tunnels through the $b$ barriers. Thus, the expression \eqref{BiL-LyapFin} for the Lyapunov exponent $\lambda$ and the dispersion relation \eqref{BiL-DR-El} for the Bloch phase $\gamma$ have to be modified according to the change $k_b=i|k_b|$,
\begin{subequations}\label{BiL-ElTun-LyapDR}
\begin{eqnarray}
&&\lambda_{tun}=\frac{\alpha^2_{tun}}{8\sin^2\gamma}\Big[k_a^2\sigma^2_a\mathcal{K}_a(2\gamma)\sinh^2|\varphi_b|+ |k_b|^2\sigma^2_b\mathcal{K}_b(2\gamma)\sin^2\varphi_a\nonumber\\[6pt]
&&-2k_a|k_b|\sigma^2_{ab}\mathcal{K}_{ab}(2\gamma)\sin\varphi_a\sinh|\varphi_b|\cos\gamma\Big],\qquad
\alpha_{tun}=\left(\frac{|k_b|m_a}{k_am_b}+\frac{k_am_b}{|k_b|m_a}\right),\qquad\label{BiL-ElTun-Lyap}\\[6pt]
&&\cos\gamma=\cos\varphi_a\cosh|\varphi_b|+\frac{1}{2}\left(\frac{|k_b|m_a}{k_am_b}-\frac{k_am_b}{|k_b|m_a}\right)\sin\varphi_a\sinh|\varphi_b|.
\label{BiL-ElTun-DR}
\end{eqnarray}
\end{subequations}
For a sufficiently large $E$ these equations have real positive solution, so that the bottom of the first transmission band is always inside the interval \eqref{BiL-ElTun} of the tunneling propagation. Note also that the Fabry-Perot resonances arise only in the $a$ barriers playing the role of quantum wells. For this reason the resonance increase of the localization length is due to the second and third terms of Eq.~\eqref{BiL-ElTun-Lyap} only.

From Eqs.~\eqref{BiL-ElTun-LyapDR} one can easily get the Lyapunov exponent for a particular case of an array with delta-like potential barriers and slightly disordered distance between them. This case is analyzed in Ref.~\cite{IKU01} and discussed in Section~\ref{8.3}. The transition to this case can be done in the limit when the thickness of the $b$ barriers vanishes, $d_b\to0$, the height of barriers infinitely increases, $U_b\to\infty$, however, their product remains constant, $U_bd_b=\mbox{const}=V_0d_a$,
\begin{eqnarray}\label{BiL-Lim-Par}
&&|k_b|=\sqrt{2m_bU_b}/\hbar=q\sqrt{m_bd_a/m_ad_b}\to\infty,\qquad|\varphi_b|=q\sqrt{m_bd_ad_b/m_a}\to0,\nonumber\\[6pt]
&&|k_b\varphi_b|m_a/m_b=q^2d_a,\qquad\mathrm{where}\quad q=\sqrt{2m_aV_0}/\hbar.
\end{eqnarray}
Taking into account these relations and neglecting in Eq.~\eqref{BiL-ElTun-Lyap} the second and third terms due to weak-disorder condition \eqref{BiL-WD}, one can readily obtain,
\begin{equation}\label{BiL-LyapDR-Delta}
\lambda=\frac{(qd_a)^4}{8\sin^2\gamma}\frac{\sigma^2_a}{d_a^2}\mathcal{K}_a(2\gamma),\qquad \cos\gamma=\cos\varphi_a+\frac{(qd_a)^2}{2}\frac{\sin\varphi_a}{\varphi_a}.
\end{equation}
These expressions are in a complete correspondence with Eqs.~(13) and (6) from Ref.~\cite{IKU01}, compare also with Eqs.~\eqref{IKU01} and \eqref{KP-disp} from Section~\ref{8}.

For the \emph{overbarrier scattering}, when the electron energy exceeds the amplitudes of the $a$ and $b$ barriers,
\begin{equation}\label{BiL-OvB-El}
U_a<U_b<E,
\end{equation}
both wave numbers $k_a$ and $k_b$ are positive. Then the Lyapunov exponent is described by the general formula \eqref{BiL-LyapFin}, that has to be complemented by the dispersion relation \eqref{BiL-DR-El}. In this case the electron transport is similar to that for the conventional photonic
stack, however, with the dispersive parameters. In Fig.~\ref{BiL-Fig06}, the crossover between the tunneling and overbarrier scattering occurs at $E/\pi=0.5$.

It is interesting that when an electron propagates through the array of barriers with the energy
\begin{equation}\label{BiL-OvB-E}
E=(U_bm_a-U_am_b)/(m_a-m_b),
\end{equation}
its velocity does not change in the barriers, $\hbar k_a/m_a=\hbar k_b/m_b$, although its momentum changes, $\hbar k_a>\hbar k_b$. The Lyapunov exponent vanishes in this case. In agreement with Eq.~\eqref{BiL-DR-El}, in such a situation an electron moves as in free space with the wave number %
\begin{equation}\label{BiL-OvB-kappa}
\kappa\equiv\gamma/d=(k_ad_a+k_bd_b)/(d_a+d_b).
\end{equation}
This effect is similar to that in photonic structures with equal impedances of both basic slabs, see Eq.~\eqref{BiL-DR-EqualImp}.

\section{Bi-layer stack with compositional disorder: Conventional media versus metamaterials}
\label{12}

In the previous section we have considered the effect of {\it positional disorder}, for which the phenomenon of Anderson localization arises due to the fluctuations in the thicknesses of two basic layers. Another kind of the disorder (\emph{compositional disorder}) is originated from the variations of the medium parameters, mainly due to the randomness in the dielectric constants and/or magnetic permeabilities. In this section we derive the localization length $L_{loc}$ for one-dimensional bi-layered structures with random perturbations in the refractive indices for each type of layers. The main attention will be paid to the comparison between the conventional materials and those consisting of mixed right-hand and left-hand materials \cite{Ao07,Ao10,IMT10,MIT10}.

\subsection{Problem formulation. Hamiltonian map}
\label{12.1}

As in previous Section, we consider the propagation of an electromagnetic wave of frequency $\omega$ through an infinite array (stack) of two alternating $a$ and $b$ layers (slabs). Every kind of slabs is respectively specified by the dielectric permittivity $\varepsilon_{a,b}$, magnetic permeability $\mu_{a,b}$, corresponding refractive index $n_{a,b}=\sqrt{\varepsilon_{a,b}\mu_{a,b}}$, impedance $Z_{a,b}=\mu_{a,b}/n_{a,b}$ and wave number $k_{a,b}=\omega n_{a,b}/c$. We address two cases when $a$ slabs contain a conventional right-handed (RH) optic material, while $b$ layers are composed of either RH or left-handed (LH) material. The combination of RH-RH slabs is called the \emph{homogeneous} stack, whereas the array of RH-LH layers is called the \emph{mixed} stack. As known, for the RH medium all optic parameters are positive. On the contrary, for the LH material the permittivity, permeability and corresponding refractive index are negative, however the impedance remains positive. Every alternating slab has the thickness $d_a$ or $d_b$, respectively, so that the size $d$ of the unit $(a,b)$ cell is constant, $d=d_a+d_b$.

As we already discussed, if the impedances of two basic $a$ and $b$ slabs are equal, $Z_a=Z_b$, the localization length diverges and the perfect transparency emerges even in the presence of a positional disorder. In the case of \emph{compositional disorder} the most simple model is a stack-structure whose unperturbed counterpart consists of the perfectly matched basic $a$ and $b$ layers. Specifically, following Ref.~\cite{Ao07,Ao10}, we admit that a disorder is incorporated via the dielectric constants only, so that
\begin{subequations}\label{BiL-CD-kakb}
\begin{eqnarray}
&&\varepsilon_a(n)=[1+\eta_a(n)]^2,\qquad\mu_a=1,\qquad n_a(n)=1+\eta_a(n),\nonumber\\[6pt]
&&Z_a(n)=[1+\eta_a(n)]^{-1},\qquad k_a(n)=\omega\,[1+\eta_a(n)]/c\,; \label{BiL-CD-ka}\\[6pt]
&&\varepsilon_b(n)=\pm[1+\eta_b(n)]^2,\qquad\mu_b=\pm1,\quad n_b(n)=\pm[1+\eta_b(n)],\nonumber\\[6pt]
&&Z_b(n)=[1+\eta_b(n)]^{-1},\qquad k_b(n)=\pm\omega[1+\eta_b(n)]/c\,.\label{BiL-CD-kb}
\end{eqnarray}
\end{subequations}
Here the integer $n$ enumerates the unit $(a,b)$ cells. The upper sign corresponds to the RH material while the lower sign is associated with the LH medium.

In the absence of disorder, $\eta_{a,b}(n)=0$, the homogeneous RH-RH structure is just the air without any stratification, while the mixed RH-LH array represents the so-called \emph{ideal mixed stack} ($\varepsilon_a=\mu_a=1$, $\varepsilon_b=\mu_b=-1$), with perfectly matched slabs ($Z_a=Z_b=1$). Therefore, in accordance with the dispersion relation \eqref{BiL-DR-gamma}, both the unperturbed RH-RH and RH-LH stacks are equivalent to the homogeneous media with the refractive index $\overline{n}$, resulting in the linear spectrum,
\begin{equation}\label{BiL-CD-UnpDR}
\kappa=\omega\overline{n}/c,\qquad\overline{n}=\frac{|d_a\pm d_b|}{d_a+d_b},
\end{equation}
(compare with Eqs.~\eqref{BiL-DR-EqualImp} and \eqref{BiL-DR-IdMeta}).

The random sequences $\eta_{a}(n)$ and $\eta_{b}(n)$ describing the compositional disorder, are statistically homogeneous with the zero mean, given variances and binary correlation functions defined by
\begin{subequations}\label{BiL-CD-CorrDef}
\begin{eqnarray}
&&\langle\eta_{a}(n)\rangle=0,\qquad\langle\eta_{a}^2(n)\rangle=\sigma_a^2,\qquad\langle\eta_a(n)\eta_a(n')\rangle=\sigma_a^2K_a(n-n');\\[6pt]
&&\langle\eta_{b}(n)\rangle=0,\qquad\langle\eta_{b}^2(n)\rangle=\sigma_b^2,\qquad\langle\eta_b(n)\eta_b(n')\rangle=\sigma_b^2K_b(n-n');\\[6pt]
&&\langle\eta_a(n)\eta_b(n)\rangle=\sigma_{ab}^2,\qquad\langle\eta_a(n)\eta_b(n')\rangle=\sigma_{ab}^2K_{ab}(n-n').
\end{eqnarray}
\end{subequations}
The average $\langle ... \rangle$ is performed over the whole array or due to the ensemble averaging, that is assumed to be equivalent. The auto-correlators $K_{a}(r)$ and $K_{b}(r)$ as well as the cross-correlator $K_{ab}(r)$, are normalized to unity, $K_{a}(0)=K_{b}(0)=K_{ab}(0)=1$, and decrease with an increase of the distance $|r|=|n-n'|$. The variances $\sigma^2_{a}$ and $\sigma^2_{b}$ are obviously positive, however, the term $\sigma_{ab}^2$ can be positive, negative or zero. We assume the compositional disorder to be \emph{weak},
\begin{equation}\label{BiL-CD-WD}
\sigma_a^2\ll1,\quad(k_ad_a)^2\sigma_a^2\ll1;\qquad\sigma_b^2\ll1,\quad(k_bd_b)^2\sigma_b^2\ll1;
\end{equation}
that allows one to use the perturbative methods. In this case all transport properties are entirely determined by the randomness power spectra $\mathcal{K}_a(k)$, $\mathcal{K}_b(k)$, and $\mathcal{K}_{ab}(k)$, defined by the standard relations \eqref{BiL-FT-K}. By definition \eqref{BiL-CD-CorrDef}, all the correlators $K_{a}(r)$, $K_{b}(r)$ and $K_{ab}(r)$ are real and even functions of the difference $r=n-n'$ between cell indices. Therefore, the corresponding Fourier transforms $\mathcal{K}_a(k)$, $\mathcal{K}_b(k)$ and $\mathcal{K}_{ab}(k)$ are real and even functions of the dimensionless longitudinal wave number $k$.

As in previous Section, within every $a$ or $b$ layer the electric field $\psi(x)\exp(-i\omega t)$ of the wave is governed by the 1D Helmholtz equation with two boundary conditions at the interfaces between neighboring slabs, see Eqs.~\eqref{BiL-WaveEqBC}. Correspondingly, the general solution of Eq.~\eqref{BiL-WaveEq-ab} within the same $n$th $(a,b)$ cell can be presented in the form \eqref{BiL-Map-ab} that should be combined with the boundary conditions \eqref{BiL-BC}. Since in the case of compositional disorder, the wave number $k_{a,b}$ depends on the cell index $n$, the desired recurrent relations for the whole $n$th unit $(a,b)$ cell is appropriate to formulate for the new ``coordinate" $Q_n$ and ``momentum" $P_n$,
\begin{equation}\label{BiL-CD-QPdef}
Q_n =\psi_{a}(x_{an})\quad\mbox{and}\quad P_n=(c/\omega)\psi'_{a}(x_{an}),
\end{equation}
that differ from the corresponding quantities treated in the previous Section. As a result, for the two opposite edges of the $n$th unit $(a,b)$ cell we have,
\begin{equation}\label{BiL-CD-mapQP}
Q_{n+1}=A_nQ_n+B_nP_n,\qquad P_{n+1}=-C_nQ_n+D_nP_n.
\end{equation}
The factors $A_n$, $B_n$, $C_n$, $D_n$ read
\begin{subequations}\label{BiL-CD-ABCDn}
\begin{eqnarray}
A_n&=&\cos\widetilde{\varphi}_a(n)\cos\widetilde{\varphi}_b(n)-Z_a^{-1}(n)Z_b(n)\sin\widetilde{\varphi}_a(n)\sin\widetilde{\varphi}_b(n),\\[6pt]
B_n&=&Z_a(n)\sin\widetilde{\varphi}_a(n)\cos\widetilde{\varphi}_b(n)+Z_b(n)\cos\widetilde{\varphi}_a(n)\sin\widetilde{\varphi}_b(n),\\[6pt]
C_n&=&Z_a^{-1}(n)\sin\widetilde{\varphi}_a(n)\cos\widetilde{\varphi}_b(n)+Z_b^{-1}(n)\cos\widetilde{\varphi}_a(n)\sin\widetilde{\varphi}_b(n),\\[6pt]
D_n&=&\cos\widetilde{\varphi}_a(n)\cos\widetilde{\varphi}_b(n)-Z_a(n)Z_b^{-1}(n)\sin\widetilde{\varphi}_a(n)\sin\widetilde{\varphi}_b(n).
\end{eqnarray}
\end{subequations}
They depend on the cell index $n$ due to the random refractive indices \eqref{BiL-CD-kakb} that enter into both the impedances $Z_{a,b}(n)$ and phase shifts,
\begin{subequations}\label{BiL-CD-phi-ab}
\begin{eqnarray}
\widetilde{\varphi}_{a}(n)&=&k_a(n)d_a=\varphi_{a}[1+\eta_{a}(n)],\qquad\varphi_{a}=\omega d_a/c\,;\\[6pt]
\widetilde{\varphi}_{b}(n)&=&k_b(n)d_b=\varphi_{b}[1+\eta_{b}(n)],\qquad\varphi_{b}=\pm\,\omega d_b/c\,.
\end{eqnarray}
\end{subequations}
As we discussed before, the recurrent relations \eqref{BiL-CD-mapQP} can be treated as the Hamiltonian map of trajectories in the phase space $(Q,P)$ with discrete time $n$ for a linear oscillator subjected to a time-dependent parametric force.

In the absence of disorder, $\eta_{a,b}(n)=0$, the factors \eqref{BiL-CD-ABCDn} do not depend on the cell index (time) $n$. Therefore, according to the map \eqref{BiL-CD-mapQP}, the trajectory $Q_n,P_n$ creates a circle in the phase space $(Q,P)$ that is an image of the unperturbed motion,
\begin{equation}\label{BiL-CD-mapQPunp}
Q_{n+1}=Q_n\cos\gamma+P_n\sin\gamma,\qquad
P_{n+1}=-Q_n\sin\gamma+P_n\cos\gamma.
\end{equation}
The unperturbed phase shift $\gamma$ over a single unit $(a,b)$ cell is defined as
\begin{equation}\label{BiL-CD-gamma-def}
\gamma=\varphi_a+\varphi_b=\omega(d_a\pm d_b)/c.
\end{equation}
Note, this result can be also obtained from the general dispersion relation \eqref{BiL-DR-gamma} and is in complete accordance with the spectrum \eqref{BiL-CD-UnpDR}, if to take into account that the Bloch wave number is $\kappa=|\gamma|/d$.

Having the circle \eqref{BiL-CD-mapQPunp}, it is suitable to pass to \emph{polar coordinates} $R_n$ and $\theta_n$ via the standard transformation,
\begin{equation}\label{BiL-CD-QP-RTheta}
Q_n=R_n\cos\theta_n,\qquad P_n=R_n\sin\theta_n.
\end{equation}
By direct substitution of Eq.~\eqref{BiL-CD-QP-RTheta} into the map \eqref{BiL-CD-mapQPunp} one gets that for the unperturbed trajectory the radius $R_n$ is conserved, while the phase $\theta_n$ changes by the \emph{Bloch phase} $\gamma$ in one step of time $n$,
\begin{equation}\label{BiL-CD-mapRThetaUnp}
R_{n+1}=R_n,\qquad \theta_{n+1}=\theta_n-\gamma.
\end{equation}

Evidently, the random perturbations, $\eta_{a,b}(n)\neq0$, give rise to a distortion of the circle \eqref{BiL-CD-mapRThetaUnp} that is described by the Hamiltonian map \eqref{BiL-CD-mapQP} with the factors given by Eqs.~\eqref{BiL-CD-ABCDn}. With the use of definition \eqref{BiL-CD-QP-RTheta}, one can readily rewrite this disordered map in the radius-angle presentation. The corresponding \emph{exact} recurrent relations read
\begin{subequations}\label{BiL-CD-mapRTheta}
\begin{eqnarray}
&&\frac{R_{n+1}^2}{R_n^2}=(A_n^2+C_n^2)\cos^2\theta_n+(B_n^2+D_n^2)\sin^2\theta_n+(A_n B_n-C_n D_n)\sin2\theta_n\,,\quad\label{BiL-CD-mapR}\\[6pt]
&&\tan\theta_{n+1}=\frac{-C_n+D_n\tan\theta_n}{A_n+B_n\tan\theta_n}\,.\label{BiL-CD-mapTheta}
\end{eqnarray}
\end{subequations}
Eqs.~\eqref{BiL-CD-mapRTheta} constitute the complete set of equations in order to derive the localization length $L_{loc}$ according to its definition \eqref{BiL-LyapDef} via the Lyapunov exponent $\lambda$. In this respect, we should emphasize the following. In the ideal mixed stack ($\varepsilon_a=\mu_a=1$, $\varepsilon_b=\mu_b=-1$, $Z_a=Z_b=1$) the wave spectrum \eqref{BiL-CD-UnpDR} is singular. Specifically, when the thicknesses $d_a$ and $d_b$ are equal, $d_a=d_b$, the phase velocity $c/\overline{n}$ diverges and the Bloch phase $\gamma$ vanishes. As a result, the circle \eqref{BiL-CD-mapRThetaUnp} presenting the unperturbed map, degenerates into a point, i.e., the unperturbed motion is absent in the phase space $(Q,P)$. As we show below, the perturbation theory has to be developed differently depending on whether the Bloch phase $\gamma$ is finite or vanishing.

\subsection{Bi-layer array with finite Bloch phase}
\label{12.2}

Now we admit the arbitrary relation between the slab thicknesses $d_a$ and $d_b$ for the homogeneous RH-RH bi-layer array, however, for mixed RH-LH stack-structure we assume only different thicknesses of basic slabs, $d_a\neq d_b$. In this case the localization length can be derived in the same manner that was discussed in previous Section \ref{11}. Specifically, we expand the coefficients \eqref{BiL-CD-ABCDn} up to the second order in the perturbation parameters $\eta_{a,b}(n)\ll1$. In doing so, one can expand only the factors containing the impedances $Z_{a,b}(n)$. As to the random phase shifts $\widetilde{\varphi}_{a,b}(n)$, they can be replaced with their unperturbed values $\varphi_{a,b}$, see definitions \eqref{BiL-CD-phi-ab}. This fact becomes clear if we take into account the conclusion obtained in previous Section: The phase disorder contributes to the Laypunov exponent only when the unperturbed impedances are different. These calculations allow us to derive the perturbed map for the radius $R_n$ and angle $\theta_n$,
\begin{subequations}\label{BiL-CD-mapWD}
\begin{eqnarray}
&&\frac{R_{n+1}^2}{R_n^2}=1+\eta_{a}(n)V_{a}(\theta_n)+\eta_{b}(n)V_{b}(\theta_n)+\eta_{a}^2(n)W_{a}+ \eta_{b}^2(n)W_{b}+\eta_{a}(n)\eta_{b}(n)W_{ab},\qquad\qquad\label{BiL-CD-mapR-WD}\\[6pt]
&&\theta_{n+1}-\theta_n+\gamma=\eta_{a}(n)U_{a}(\theta_n)+\eta_{b}(n)U_{b}(\theta_n).\label{BiL-CD-mapTheta-WD}
\end{eqnarray}
\end{subequations}
Here the functions standing at random variables $\eta_{a,b}(n)$ are described by the expressions,
\begin{subequations}\label{BiL-CD-VWU}
\begin{eqnarray}
&&V_{a}(\theta_n)=-2\sin\varphi_{a}\sin(2\theta_n-\varphi_{a}),\qquad V_{b}(\theta_n)=-2\sin\varphi_{b}\sin(2\theta_n-\gamma-\varphi_{a}),\qquad\\[6pt]
&&W_{a}=2\sin^2\varphi_{a},\qquad W_{b}=2\sin^2\varphi_{b},\qquad W_{ab}=4\sin\varphi_{a}\sin\varphi_{b}\cos\gamma;\\[6pt]
&&U_{a}(\theta_n)=-\sin\varphi_{a}\cos(2\theta_n-\varphi_{a}),\qquad U_{b}(\theta_n)=-\sin\varphi_{b}\cos(2\theta_n-\gamma-\varphi_{a}).
\end{eqnarray}
\end{subequations}
Note that in Eqs.~\eqref{BiL-CD-VWU} we keep only the terms that contribute to the localization length $L_{loc}$ in the first (non-vanishing) order of approximation. Since in Eq.~\eqref{BiL-CD-mapR-WD} the factors $V_{a,b}$ containing $\theta_n$ are always multiplied by $\eta_{a,b}(n)$, only the linear terms in these perturbation parameters are needed in the recurrent relation \eqref{BiL-CD-mapTheta-WD} for the angle $\theta_n$.

The perturbed map \eqref{BiL-CD-mapR-WD} is substituted into Eq.~\eqref{BiL-LyapDef} for the Lyapunov exponent. After expanding the logarithm within quadratic approximation in the perturbation parameters, we get the expression similar to Eq.~\eqref{BiL-Lyap-VW} where the averaging should be performed. Within the accepted approximation, we regard the random quantities $\eta_{a}^2(n)$, $\eta_{b}^2(n)$ and $\eta_{a}(n)\eta_{b}(n)$ as the uncorrelated ones, with the factors \eqref{BiL-CD-VWU} containing the angle variable $\theta_n$. It is crucially important that when performing the averaging, we assume the distribution of phase $\theta_n$ to be homogeneous, $\rho(\theta)=1/2\pi$. This assumption is correct for all values of $\gamma$, apart from the case $\gamma=0$ that corresponds to the ideal mixed RH-LH stack-structure with $d_a=d_b$. After some algebra, we arrive at the final expression for the Lyapunov exponent,
\begin{equation}\label{BiL-CD-LyapFin}
\lambda=\frac{d}{L_{loc}}=\frac{1}{2}\sigma_{a}^2\mathcal{K}_a(2\gamma)\sin^2\varphi_{a}+ \frac{1}{2}\sigma_{b}^2\mathcal{K}_b(2\gamma)\sin^2\varphi_{b}+\sigma_{ab}^2\mathcal{K}_{ab}(2\gamma)\sin\varphi_{a}\sin\varphi_{b}\cos\gamma.
\end{equation}
Note that Eq.~\eqref{BiL-CD-LyapFin} is symmetric with respect to the permutation of slab indices $a\leftrightarrow b$.

In accordance with Eq.~\eqref{BiL-CD-LyapFin}, the Lyapunov exponent $\lambda$ consists of three terms. The first two terms contain the \emph{auto-correlators} $\mathcal{K}_a(2\gamma)$ or $\mathcal{K}_b(2\gamma)$. Therefore, these terms are contributed respectively by the correlations between solely $a$ or solely $b$ slabs. The third term includes the \emph{cross-correlator} $\mathcal{K}_{ab}(2\gamma)$ that emerges due to the cross-correlations between two different, $a$ and $b$, disorders.

The expression \eqref{BiL-CD-LyapFin} is universal and applicable for both homogeneous RH-RH and mixed RH-LH stack-structures. The only difference between these cases is the sign in the unperturbed phase shift $\varphi_{b}=\pm\omega d_b/c$. This affects the value \eqref{BiL-CD-gamma-def} of the Bloch phase $\gamma$ and changes the sign at the third cross-correlation term.

For both homogeneous RH-RH and mixed RH-LH stack-structures, the Lyapunov exponent typically obeys the \emph{conventional} frequency dependence,
\begin{equation}\label{BiL-CD-LyapOmega}
\lambda=d/L_{loc}\propto\omega^2\qquad \mathrm{when}\quad\omega\to0.
\end{equation}
However, similar to what is shown in previous Sections, the specific correlations in the disorders of the refractive indices (taken into account in Eq.~\eqref{BiL-CD-LyapFin}) may result in a quite unusual $\omega$-dependence. Of special interest are long-range correlations leading to a significant decrease or increase of the localization length $L_{loc}(\omega)$ in predefined frequency windows. In such a way, one can
enhance or suppress the localization in the systems with compositional disorder.

The Lyapunov exponent $\lambda(\omega)$ exhibits the \emph{Fabry-Perot resonances} occurring when the wave frequency $\omega$ meet the following conditions,
\begin{equation}\label{BiL-CD-FPab}
\omega/c=s_a\pi/d_a\quad\mathrm{or}\quad\omega/c=s_b\pi/d_b,\quad s_{a,b}=1,2,3,\dots.
\end{equation}
At the resonances the factor $\sin\varphi_a$ (or $\sin\varphi_b$) in Eq.~\eqref{BiL-CD-LyapFin} vanishes resulting in the resonance increase of the localization length $L_{loc}$ and, consequently, in the suppression of the localization. When only one type of the basic layers is disordered, i.e., Eq.~\eqref{BiL-CD-LyapFin} contains only one corresponding term, the resonances give rise to the divergence of $L_{loc}(\omega)$. In the special case when the ratio of slab thicknesses $d_a$ and $d_b$ is a rational number, $d_a/d_b=s_a/s_b$, some of the resonances of different types of layers coincide. This also leads to the divergence of the localization length. Remarkably, the Fabry-Perot resonances are quite \emph{broad} because they are caused by the vanishing smooth trigonometric functions. As known from previous Section~\ref{11}, in the case of positional disorder the terms entering Eq.~\eqref{BiL-LyapFin} for the Lyapunov exponent and related to auto-correlations between the same type of slabs, display the Fabry-Perot resonances associated with the other type of layers. On the contrary, in the case of the compositional disorder the corresponding terms in Eq.~\eqref{BiL-CD-LyapFin} manifest the Fabry-Perot resonances belonging to the same type of slabs.

As an example, let us consider a particular case of the white-noise disorders for $a$ and $b$ slabs,
\begin{equation}\label{BiL-CD-CorrFre}
\sigma_{ab}^2=0,\qquad\mathcal{K}_a(k)=\mathcal{K}_b(k)=1.
\end{equation}
Here, the Lyapunov exponent and the inverse localization length turn out to be  exactly the same for {\it both} homogeneous RH-RH and mixed RH-LH stack-structures,
\begin{equation}\label{BiL-CD-LyapWN}
\lambda=\frac{d}{L_{loc}}=\frac{1}{2}(\sigma_a^2\sin^2\varphi_{a}+\sigma_b^2\sin^2\varphi_{b}).
\end{equation}
The numerical results shown in Fig.~\ref{BiL-CD-Fig0102} are obtained with the use of Eqs.~\eqref{BiL-CD-mapRTheta} and without any approximation.
%
\begin{figure}[!ht]
\begin{center}
\subfigure{\includegraphics[scale=0.6]{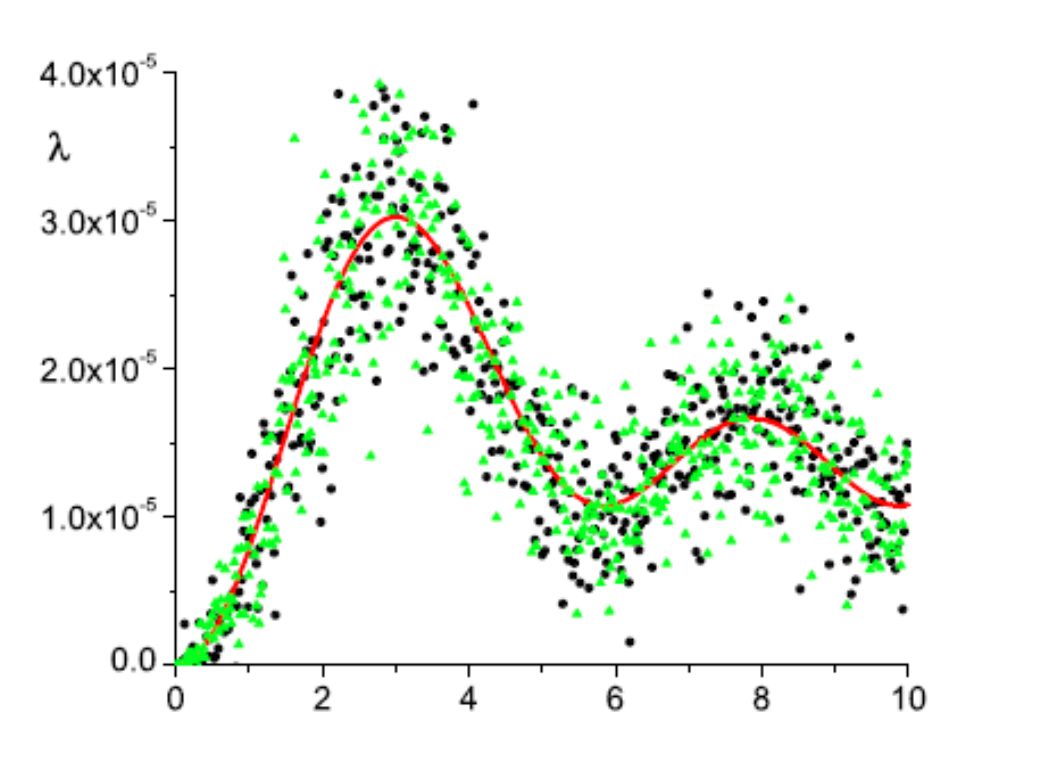}}
\subfigure{\includegraphics[scale=0.6]{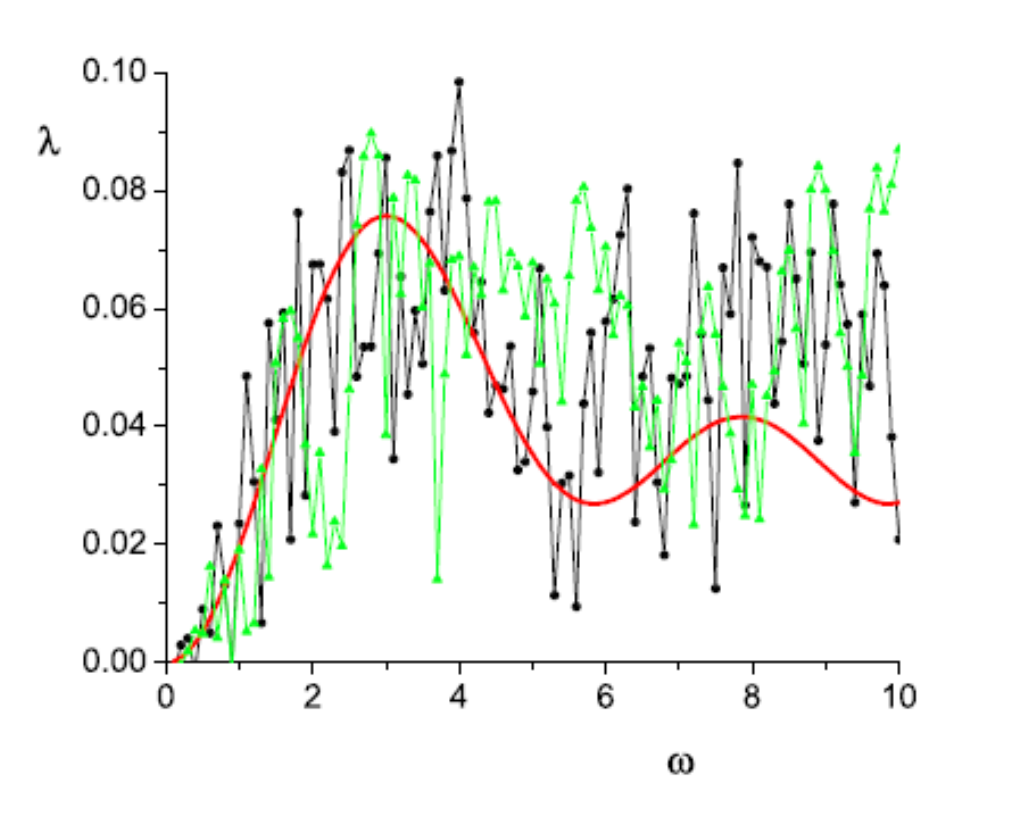}}
\end{center}
\caption{(color online). Lyapunov exponent versus frequency. Left: RH-RH (circles) and RH-LH (triangles) media for $\sigma_a\approx\sigma_b\approx 0.006\,, d_a=0.6\,, d_b=0.4\,, c=1$ and the length of sequence is $N=10^6$. Right: the same for $\sigma_a\approx\sigma_b\approx 0.3$ and the $N=100$. Smooth curve depicts Eq.~\eqref{BiL-CD-LyapWN} (after \cite{IMT10}).}
\label{BiL-CD-Fig0102}
\end{figure}
In the left panel one can see that for a very long sample and weak disorder the analytical expression \eqref{BiL-CD-LyapWN} perfectly corresponds to the data, apart from the fluctuations. For each case only one realization of disorder was used. The fluctuations can be smoothed out by an additional ensemble averaging. In order to see whether our predictions can be confirmed in a real experiment, we also show in the right panel the data for a quite short sample and relatively strong disorder. As one can see, the analytical result is also valid for small frequencies, and gives the qualitatively correct value for the Lyapunov exponent for large values of $\omega$. The more detailed comparison for $\omega d/c\ll1$ and $\omega d/c\gg1$ also shows a nice correspondence.

\subsection{Mixed RH-LH stack with vanishing Bloch phase}
\label{12.3}

Note once more that the result \eqref{BiL-CD-LyapFin} for the Lyapunov exponent is valid for both the homogeneous RH-RH and mixed RH-LH bi-layer array {\it when} the Bloch phase \eqref{BiL-CD-gamma-def} is different from zero. Indeed, the unperturbed map \eqref{BiL-CD-mapRThetaUnp} does not degenerate, and nothing special arises in this case for the evaluation of the Lyapunov exponent. Physically, this means that the unperturbed phase $\theta_n$ rotates when passing any unit $(a,b)$ cell, and through many steps it is randomized provided the Bloch phase $\gamma$ is irrational with respect to $2\pi$. Therefore, the $\theta$-distribution can be regarded as homogeneous even without the disorder. This is why when evaluating the Lyapunov exponent, we assume the homogeneous distribution of phase $\theta_n$, i.e., the corresponding probability density $\rho(\theta)=1/2\pi$.

The situation turns out to be completely different in the case when the wave phase $\gamma=\varphi_a+\varphi_b$ vanishes after passing the unit $(a,b)$ cell. This happens in the ideal mixed RH-LH stack with $d_a=d_b$ because after passing the RH $a$-layer, the phase shift is $\varphi_{a}=\omega d_a/c$, however, it is exactly canceled by another shift, $\varphi_{b}=-\omega d_a/c=-\varphi_a$ in the next LH $b$-slab. As one can see, the unperturbed phase distribution is simply delta-function, therefore, the circle \eqref{BiL-CD-mapRThetaUnp} presenting the unperturbed map, degenerates into a single point in the phase space $(Q,P)$. This means that with a weak disorder, the phase distribution has to be obtained in the next order of perturbation theory, since it is randomized due to the disorder only.

In what follows we consider the ideal mixed stack whose refractive indices of two basic $a$ and $b$ slabs are perturbed by two independent white-noise disorders with the same strength,
\begin{subequations}\label{BiL-CD-FreWN-IMS}
\begin{eqnarray}
&&\sigma_{a}^2=\sigma_{b}^2\equiv\sigma^2,\qquad\sigma_{ab}^2=0,\qquad\mathcal{K}_a(k)=\mathcal{K}_b(k)=1;\label{BiL-CD-FreWN}\\[6pt]
&&d_a=d_b,\qquad\mathrm{i.e.,}\qquad\varphi_{a}=-\varphi_b\equiv\varphi\qquad\mathrm{and}\qquad\gamma=0.\label{BiL-CD-IMS}
\end{eqnarray}
\end{subequations}

The phase distribution $\rho(\theta)$ can be found in the same way as described in Sections~4.2, 4.3 and 8.3. The starting point is the exact recurrent relation \eqref{BiL-CD-mapTheta} between $\theta_{n+1}$ and $\theta_n$. By expanding this expression up to the second order in perturbation parameters $\eta_{a,b}(n)$, one obtains \cite{TIM11},
\begin{equation}\label{BiL-CD-mapTheta-Fre}
\theta_{n+1}-\theta_{n}=-[\eta_a(n)-\eta_b(n)]v(\theta_n)+\sigma^2 v'(\theta_n)v(\theta_n),
\end{equation}
where we introduced the function
\begin{equation}\label{BiL-CD-Vdef}
v(\theta)=\varphi+\sin\varphi\cos(2\theta-\varphi).
\end{equation}
In deriving Eq.~\eqref{BiL-CD-mapTheta-Fre}, we explicitly took into account the conditions $\langle\eta_a(n)\eta_b(n)\rangle=0$, and $\langle\eta_a^2(n)\rangle=\langle\eta_b^2(n)\rangle=\sigma^2$ that directly follows from Eqs.~\eqref{BiL-CD-CorrDef} and \eqref{BiL-CD-FreWN}. This approximation is sufficient in order to obtain the distribution of phases $\theta_n$ in the second order of perturbation. As a result, the expression (\ref{BiL-CD-mapTheta-Fre}) takes the form of the stochastic Ito equation \cite{G04} which can be associated with the Fokker-Plank equation for the distribution function $P(\theta,t)$ (see also Ref.~\cite{IRT98}).

The next step is to obtain the differential equation for the probability density $\rho(\theta)$. In the case when the conditions \eqref{BiL-CD-FreWN-IMS} hold true, the corresponding Fokker-Plank equation takes relatively simple form,
\begin{equation}\label{BiL-CD-FPeq}
\frac{\partial^2P(\theta,t)}{\partial t^2}=\sigma^2\frac{\partial^2}{\partial\theta^2}\left[P(\theta,t)v^2(\theta)\right]- \frac{\sigma^2}{2}\frac{\partial}{\partial\theta}\left[P(\theta,t)\frac{\partial}{d\theta}v^2(\theta)\right]\,.
\end{equation}
Here the ``time" $t$ is the same as the length $N=L/d$ of a sample along which the evolution of phase $\theta$ occurs.

Note that we are interested in the stationary solution of this equation, $\rho(\theta)\equiv P(\theta,t\to\infty)$. The equation for $\rho(\theta)$ reads
\begin{equation}\label{BiL-CD-rho-eq}
\frac{d^2}{d\theta^2}\left[\rho(\theta)v^2(\theta)\right]-\frac{1}{2}\frac{d}{d \theta}\left[\rho(\theta)\frac{d}{d \theta}v^2(\theta)\right]=0\,.
\end{equation}
Here the dependence on the variance $\sigma^2$ has disappeared due to the rescaling of time, $t\to\sigma^2t$. Therefore, in this approximation the phase density $\rho(\theta)$ does not depend on the disorder variance $\sigma^2$. The diffusive equation \eqref{BiL-CD-rho-eq} contains the only function $v(\theta)$, which is periodic with the period $\pi$. Consequently, its solution $\rho(\theta)$ should be also periodic with the same period. In addition, $\rho(\theta)$ should be complemented by the normalization condition,
\begin{equation}\label{BiL-CD-rho-Cond}
\rho(\theta+\pi)=\rho(\theta),\qquad\qquad\int_{0}^{2\pi}d\theta\rho(\theta)=1.
\end{equation}
The solution of Eqs.~(\ref{BiL-CD-rho-eq}), \eqref{BiL-CD-rho-Cond} is
\begin{equation}\label{BiL-CD-rho-sol}
\rho(\theta)=J/v(\theta)\,,\qquad J=\sqrt{\varphi^2 -\sin^2\varphi}\,.
\end{equation}

\begin{figure}[!ht!!]
\begin{center}
\subfigure{\includegraphics[scale=0.7]{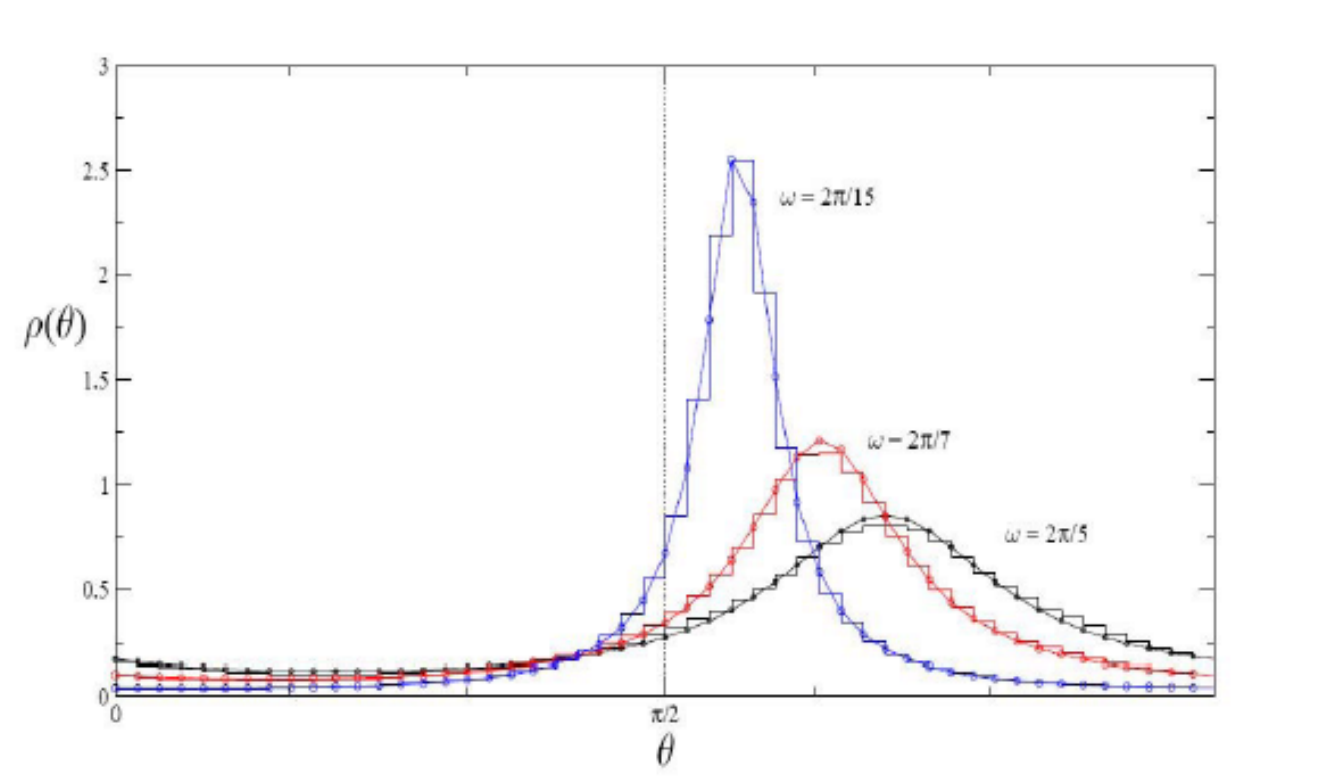}}
\end{center}
\caption{(color online). Stationary distribution $\rho(\theta)$ for various values of $\omega$ noted at the curves in rescaled units $d=1$, $c=1$. Broken curves correspond to numerical data with an ensemble average for $N=10^6$, $10^7$, $10^8$. Smooth curves represent the analytical expression \eqref{BiL-CD-rho-sol} (after \cite{TIM11}).}\label{BiL-CD-Fig03}
\end{figure}

Our results indicate that the phase distribution $\rho(\theta)$ strongly depends on the phase shift $\varphi$, especially, in the limit of small $\omega$, $\varphi\equiv\varphi_{a}=\omega d_a/c\ll1$. Some examples of the distribution function $\rho(\theta)$ for different values of $\varphi$ are shown in Fig.~\ref{BiL-CD-Fig03}. This figure clearly demonstrates that with the decrease of $\omega$ the distribution $\rho (\theta)$ begins to be very sharp in the vicinity of $\theta = \pi/2$. It is worthwhile to note that the situation is somewhat similar to that emerging for the Anderson and Kronig-Penney models. Indeed, in these models after one period of perturbation the unperturbed Bloch phase $\gamma$ also vanishes when approaching the band edges. This leads to a highly non-homogeneous distribution of the perturbed phase, and, as a result, to a non-standard dependence of the Lyapunov exponent on the model parameters. In the considered model of the mixed RH-LH array, the crucial difference is that the situation with $\gamma=0$ emerges {\it independently} of the value of frequency $\omega$, in contrast with the case of Anderson and Kronig-Penney models for which the zero Bloch phase occurs at band edges only, therefore, for specific values of frequency. Thus, one can expect that for mixed RH-LH bi-layer stacks with the specific condition $d_a=d_b$, the dependence of the Lyapunov exponent on the model parameters has to be highly non-trivial. This can be seen from Fig.~\ref{BiL-CD-Fig04} which shows how the dependence of the Lyapunov exponent on frequency $\omega$ is affected by the relation between $d_a$ and $d_b$.

\begin{figure}[!ht]
\vspace{1.5cm}
\begin{center}
\subfigure{\includegraphics[scale=0.7]{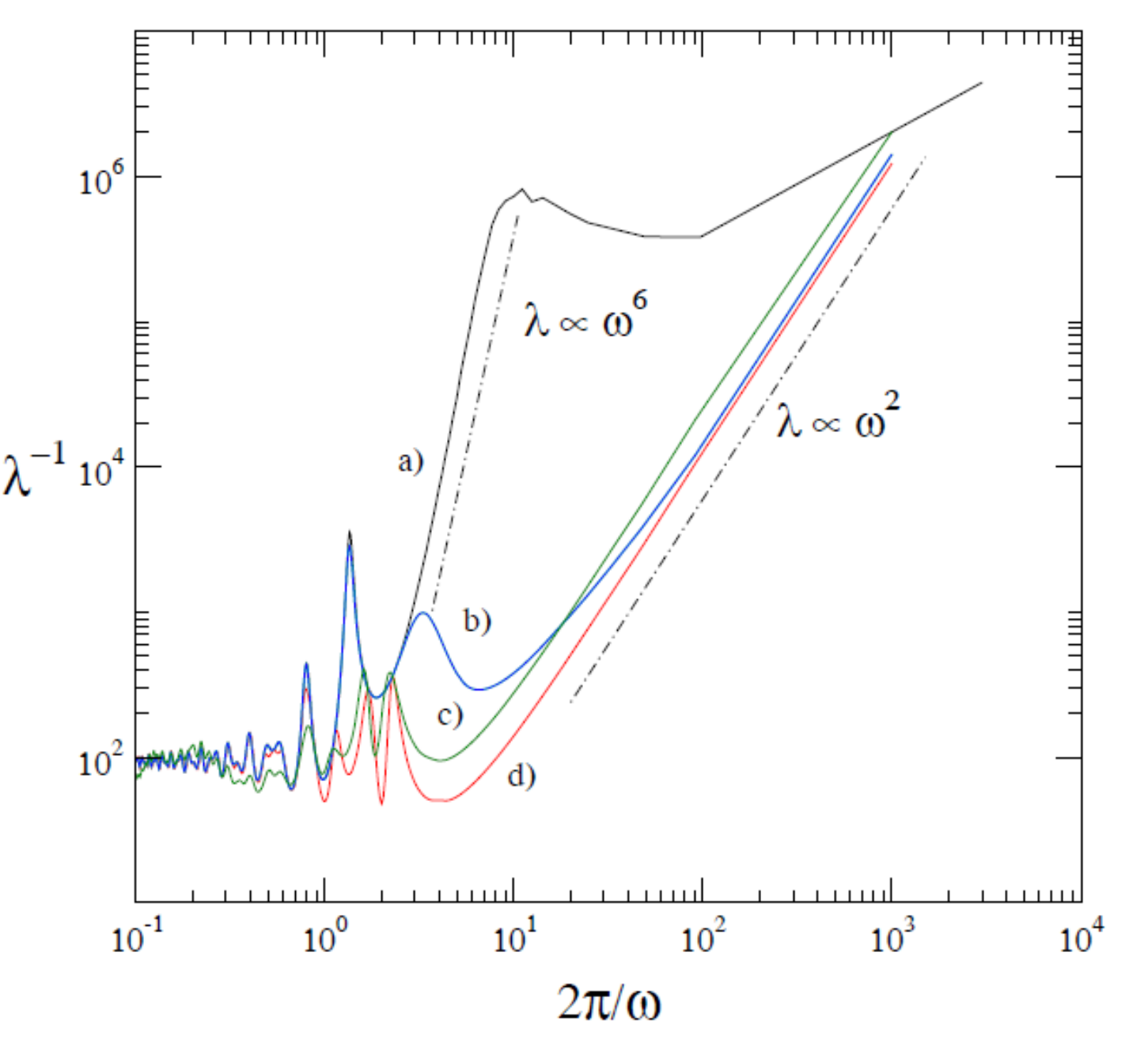}}
\end{center}
\caption{(color online). Numerical data for the inverse Lyapunov exponent versus the rescaled frequency $\omega$ with the variance of disorder $\sigma^2=0.02$ and sequence length $N=10^6$. The data are obtained with an ensemble averaging over 100 realizations of disorder; a) mixed RH-LH array with $d_a=d_b$, b) the same for $d_a=0.99d_b$, c) the same for $d_a=0.1d_b$, d) homogeneous RH-RH array with $d_a=d_b$. For comparison, the dot-dashed lines show two frequencies dependencies discussed in the text (after \cite{TIM11}).}\label{BiL-CD-Fig04}
\end{figure}

The data clearly display that when $d_a$ approaches $d_b$, the standard dependence $\lambda\propto\omega^2$ alternates by a very unusual dependence $\lambda\propto\omega^6$ in the limit $\omega \rightarrow 0$. The latter result was found numerically in Ref.~\cite{Ao07} and later on, was discussed in Refs.~\cite{Ao10,Ao10a,Mo10}. From Fig.~\ref{BiL-CD-Fig04} one can also see that for the homogeneous RH-RH stack-structure the conventional dependence \eqref{BiL-CD-LyapOmega} remains valid even in the specific case of $d_a=d_b$. As shown above, in this case the Lyapunov exponent $\lambda$ is described by the generic expression \eqref{BiL-CD-LyapFin}, that for the homogeneous RH-RH array is valid for {\it any} relation between $d_a$ and $d_b$.

The Lyapunov exponent can be derived according to the definition \eqref{BiL-LyapDef} and exact Hamiltonian map \eqref{BiL-CD-mapR} for the radius $R_n$. Its expansion within the second order of approximation in the disorder $\eta_{a,b}(n)$ gives
\begin{equation}\label{BiL-CD-LyapRHLH}
\lambda=
2\sigma^2\sin\varphi\langle\cos(2\theta_n-\varphi)v(\theta_n)\rangle\,.
\end{equation}
The averaging in this expression has to be performed with the distribution function $\rho(\theta)$ determined by Eqs.~\eqref{BiL-CD-rho-sol} and \eqref{BiL-CD-Vdef}. Since the denominator $v(\theta)$ in Eq.~\eqref{BiL-CD-rho-sol} is the same as the coefficient in Eq.~\eqref{BiL-CD-LyapRHLH}, we come to the remarkable result that the Lyapunov exponent \eqref{BiL-CD-LyapRHLH} vanishes for any value of the phase shift $\varphi$ ! This means that in order to derive the non-vanishing Lyapunov exponent, one has to obtain the expressions for both the phase distribution $\rho(\theta)$ and the ratio $R^2_{n+1}/R^2_n$ in the next order of perturbation, by expanding them up to the fourth order in disorder, that is not a simple task. The Lyapunov exponent for the ideal mixed RH-LH stack with $d_a=d_b$, was analytically found \cite{TIM11} to be proportional to $\sigma^4$, in contrast with the conventional quadratic dependence, $\lambda\propto\sigma^2$. Moreover, the mysterious dependence $\lambda\sim\omega^6$ found numerically in Ref.~\cite{Ao07} was shown to be a non-universal one since it emerges in an {\it intermediate} region of $\omega$ \cite{TIM11a}. As for the correct result for $\omega \rightarrow 0 $, the dependence of the Lyapunov exponent on $\sigma$ and $\omega$ turns out to be $\lambda \propto \sigma^4 \omega^8$.

\section{Tight-binding two-chain model}
\label{13}

\subsection{Transmission matrix}
\label{13.1}

As demonstrated in previous Sections, the correlations in random one-dimensional potential may strongly change the character of localization. In particular, they may give rise to a discrete set of extended states, or even to a continuous band of extended states. Since the onset of Anderson localization in weak random potentials strongly depends on the dimension, it is of great interest to generalize the results obtained in the previous Sections for one-dimensional systems, to multi-channel, quasi-1D, 2D, and 3D systems. Presently, this program is far from being accomplished.

In two-dimensions the tight-binding model with diagonal disorder is represented by the following discrete equation for a stationary eigenstate $\psi_{n,m}$ with energy $E$,
\begin{equation}
\label{tb diagonal 2D}
\psi_{n+1,m} + \psi_{n-1,m}+ \psi_{n,m+1} + \psi_{n,m-1}=(E - \epsilon_{n,m}) \psi_{n,m} \,.
\end{equation}
In a periodic lattice without disorder, $\epsilon_{n,m}= 0$, the eigenfunctions are plane waves $\psi_{n,m}= \exp[i(k_1 n d + k_2 m d]$, where $k_1$ and $k_2$ are the components of the momentum defined by the dispersion relation,
\begin{equation}
E= 2(\cos \mu_1 + \cos \mu_2),\qquad \mu_1=k_1 d, \quad \mu_2=k_2 d\,.
\label{2d-disp}
\end{equation}
Here $\mu_1$ and $\mu_2$ are the phase shifts related to the cell period $d$ of a discrete grid $\{n,m\}$.

If the site potential $\epsilon_{n,m}$ is a set of random numbers, the scaling theory \cite{AALR79} predicts that all eigenstates are exponentially localized. If the site energies are partly regular, e.g. $\epsilon_{n,m}= 0$ for even values of $n$ and the others $\epsilon_{n,m}$ are random for odd values of $n$, a superposition of two plane waves having nodes at the sites with random potential, $\psi_{n,m}= \cos(n\pi/4)\exp[i(km)]$, is an obvious eigenstate with the energy $E=2 \cos k$ \cite{H03}. In a 2D lattice these extended states form a band, therefore, the energy spectrum contains a mobility edge \cite{H03,A92}. In a 1D random sequence ``diluted" by regular sites, the energy spectrum contains a single resonant state with extended wave function \cite{H97,HF97}. It is clear that the nature of extended states in the diluted lattices is related to special symmetry in the location of the impurities over lattice sites \cite{LO01}.

The electron dynamics in a 2D disordered lattice has been studied numerically \cite{MLDM07} for a self-similar correlated site potential biased by a constant electric field and directed along the diagonal of the unit cell.
The authors report clear signatures of the Bloch oscillations of a Gaussian
wave packet in the case of long-range correlations. The oscillations originate from the presence of
two mobility edges predicted in Ref.~\cite{MCLR04}.

So far, the correlated disorder in quasi-one-dimensional models was studied scarcely. In analogy with one-dimensional disordered systems, in the quasi-one-dimensional geometry a continuum of delocalized states is expected for specific long-range correlations only, see discussion in Section \ref{3}.
In particular, it was shown \cite{IM03,TI06} that for a particular case of no dependence of bulk random potential on the transverse coordinate, the problem can be solved analytically by reducing the model to a coset of non-interacting channels for which the localization length has different values. As for the surface scattering in many-mode waveguides, the analytical results obtained in Refs.~\cite{IM03,IMR05,IMR05a,IMR06,RIM07} indicate that with specific long-range correlations one can significantly suppress or enhance the coupling between different channels (see discussion in Ref.~\cite{IM05}).

In order to model few channel wires, in Refs.~\cite{H02,H03} the tight-binding Anderson model with two- and three-coupled chains has been introduced. With the use of the transfer matrix approach the authors where able to derive the expression for the localization length in the case of weak white noise disorder. Below, we extend their approach to the case of correlated disorder for the case of two coupled chains (for details see \cite{BK07}).

The two-chain Anderson model is constructed by two parallel chains of finite length $L=Nd$ of $N$ disordered sites and spacing $d$. The two chains are connected to semi-infinite ideal leads, and coupled to each other by a constant matrix element $h$ allowing an electron to hop from one chain to another. Thus, the Schr\"odinger equation can be written as follows,
\begin{equation}
\begin{array}{c}
\label{Schr1}
     \vartheta(\psi_{1,n+1}+ \psi_{1,n-1}) + h \psi_{2,n}= (E-\epsilon_{1,n}) \psi_{1,n}\,,\\

     \vartheta(\psi_{2,n+1}+ \psi_{2,n-1}) + h \psi_{1,n}= (E-\epsilon_{2,n}) \psi_{2,n}\,.\\
  \end{array}
\end{equation}
In what follows, we omit the hopping element $\vartheta$ by assuming that both the disorder $\epsilon_{i,n}$ with $i=1,2$ and energy $E$ are measured in units of $\vartheta$.

The eigenfunctions for the leads are the plane waves $\exp(\pm
i\mu_in)$ with phase shifts $\mu_1(E)$ and $\mu_2(E)$ defined by two equations,
\begin{equation}
\label{disp}
\begin{array}{c}
    2 \cos \mu_1 = E - h, \\
    2 \cos \mu_2 = E + h \,.\\
\end{array}
\end{equation}
In order to develop a proper perturbation theory, it is convenient to pass to the Bloch representation determining two non-coupled propagating channels in the leads. This can be done with the use of the following transformation,
\begin{equation}
\label{basis} \left( \begin{array}{c}
    \phi_{1,n}\\
    \phi_{2,n}\\
  \end{array}
\right) = \frac{1}{\sqrt{2}}\left(
            \begin{array}{cc}
              1 & 1 \\
              1 & -1 \\
            \end{array}
          \right) \left(
  \begin{array}{c}
    \psi_{1,n} \\
    \psi_{2,n }\\
  \end{array}
\right) \,.
\end{equation}
Here $\phi_{1,n}$ and $\phi_{2,n}$ are introduced instead of $\psi_{1,n}$ and $\psi_{2,n}$.
As a result, in the $\phi$-basis the Schr\"{o}dinger equation takes the following form,
\begin{equation}
\label{Schr2}
\begin{array}{c}
\left(
  \begin{array}{c}
    \phi_{1,n+1} + \phi_{1,n-1} \\
    \phi_{2,n+1} + \phi_{2,n-1}\\
  \end{array}
 \right) =
 \\ \left(
            \begin{array}{cc}
              E- h - \frac{1}{2}(\epsilon_{1,n}+\epsilon_{2,n}) &
               \frac{1}{2}(\epsilon_{2,n}-\epsilon_{1,n}) \\
              \frac{1}{2}(\epsilon_{2,n}-\epsilon_{1,n}) &
               E+ h - \frac{1}{2}(\epsilon_{1,n}+\epsilon_{2,n}) \\
            \end{array}
          \right) \left(
  \begin{array}{c}
    \phi_{1,n} \\
    \phi_{2,n} \\
  \end{array}
\right) .
\end{array}
\end{equation}
As one can see, the channels associated with the functions $\phi_{1,n}$ and $\phi_{2,n}$ are uncoupled for
$\epsilon_{1,n}=\epsilon_{2,n}=0$.

The Lyapunov exponent $\lambda$ can be obtained in the thermodynamic limit as a result of the statistical averaging through the standard relation,
\begin{equation}
\label{conductance}
\lambda=L_{loc}^{-1}(E)=-\lim_{N\rightarrow\infty} \frac{1}{2N} \langle\ln
Tr{(\hat t \hat t^{\dag})}\rangle \,.
\end{equation}
In comparison with Eq.~(\ref{1DCP-LlocLnT}) here we determine the transmission coefficient $T_L=Tr{(\hat t \hat t^{\dag})}$ (or, the dimensional conductance) through the {\it transmission matrix} $\hat t$, and the latter can be expressed via the transfer matrix (see Sections \ref{10.2} and \ref{10.4}).
In order to introduce the transfer matrix, we rewrite Eq.
(\ref{Schr2}) in the form of a four-dimensional map,
\begin{equation}
\label{transfer1}
\begin{array}{l}
\left(
  \begin{array}{c}
    \phi_{1,n+1} \\
    \phi_{1,n}\\
    \phi_{2,n+1} \\
    \phi_{2,n} \\
  \end{array}
\right) \\
=\underbrace{\left(
 \begin{array}{cccc}
    E-h-(\epsilon_{1,n}+\epsilon_{2,n})/2 & -1 & (\epsilon_{2,n}-\epsilon_{1,n})/2 & 0 \\
    1 & 0 & 0 & 0 \\
    (\epsilon_{2,n}-\epsilon_{1,n})/2 & 0 & E+h-(\epsilon_{1,n}+\epsilon_{2,n})/2 & -1\\
    0 & 0 & 1 & 0 \\
  \end{array}\right)}_{\hat{X}_n}
  \left(
  \begin{array}{c}
    \phi_{1,n} \\
    \phi_{1,n-1}\\
    \phi_{2,n} \\
    \phi_{2,n-1} \\
  \end{array}
\right) .
\end{array}
\end{equation}
The matrix $\hat {X}_n$ translates the wavefunction by one
spacing through the site $n$. It is worthwhile to represent this
matrix in the basis of waves
propagating in perfect leads. Since there are no real
scatterers in the leads, the translation through any lead site
changes only the phase of the wavefunction, $\phi_{n+1}=e^{\pm
i\mu} \phi_{n}$, not the modulus. For one site $n$ the matrix $\hat{X}_n$ can be represented in the form,
\begin{equation}
\label{tranmat} \hat{X}_n= \hat X_0 + \hat x_{n}.
\end{equation}
Here the diagonal matrix $\hat X_0$ takes into account the phase shift of the wave function at the translation by one period. The disorder terms responsible for scattering are collected in the
matrix $\hat x_{n}$ which is linear in energy fluctuations
$\epsilon_{i,n}$. The transfer matrix for the whole
sample containing $N$ sites is a product of $N$ single-site
matrices. In this product is sufficient to keep the potential terms up to quadratic ones,
\begin{eqnarray}
\label{appro1} \hat X = \prod_{n=1}^{N} \hat{X}_n \approx (\hat X_{0})^{N} +
\sum_{n=1}^{N} (\hat X_{0})^{N-n} \cdot \hat x_{n} \cdot (\hat X_{0})^{n-1} \\ \nonumber
+\sum_{m>n} (\hat X_{0})^{N-m} \cdot \hat x_{m} \cdot (\hat X_{0})^{m-n-1} \cdot
\hat x_{n} \cdot (\hat X_{0})^{n-1}.
\end{eqnarray}

Since the transmission matrix $\hat t$ and transfer matrix $\hat X$ represent the same physical process, namely, the scattering and the mixing of the modes in the disordered region, they are related to each other. In particular, the transmission matrix is expressed through the elements of the matrix $\hat X$ as follows,
\begin{equation}
\label{transmis}
 \hat{t} = \frac{1}{\delta} \left(
\begin{array}{cc}
X_{44} & -X_{24} \\
-X_{42} & X_{22} \\
\end{array}
\right),
\end{equation}
\begin{equation}
\delta = X_{22}X_{44}-X_{24}X_{42}.
\end{equation}
Using Eq.~(\ref{appro1}), the elements of the matrix $\hat X$ can be calculated explicitly, with a further expansion of $Tr(\hat{t}\hat{t}^{\dagger})$ over weak disorder. These calculations are done separately for the situation when both of the channels are propagating, as well as when one channel is open and the other is closed.

\subsection{Two propagating modes}
\label{13.2}

Now, let us make an expansion of the elements of the matrix $\hat X$ over weak disorder. The linear terms were calculated in Ref. \cite{H02} for uncorrelated potential. In order to take into account possible correlations, one has to keep the quadratic terms in the expansions of $X_{22}$ and $X_{44}$. The elements $X_{24}$ and $X_{42}$ can be calculated in the linear approximation because the quadratic terms lead to higher-order corrections in the expression for $Tr(\hat{t}\hat{t}^{\dagger})$ \cite{BK07},
\begin{equation}
\label{approxX}
\begin{array}{l}
X_{22}= e^{-i \mu_1 L} \left[1 + i \sum_{m} a_{m} - \sum_{m>n}
(1-e^{2 i \mu_1(m-n)})a_{m}a_{n} \right. \\
 \left. - \sum_{m>n}
(e^{i(\mu_1-\mu_2)(m-n)}-e^{i(\mu_1+\mu_2)(m-n)})c_m c_n \right],\\
X_{44}= e^{-i \mu_2 L} \left[1 + i \sum_{m} b_{m} - \sum_{m>n}
(1-e^{2 i \mu_2(m-n)})b_{m}b_{n} \right. \\
 \left. - \sum_{m>n}
(e^{i(\mu_2-\mu_1)(m-n)}-e^{i(\mu_1+\mu_2)(m-n)})c_m c_n \right],\\

X_{24}= - e^{-i\mu_{1}L}[ i \sum_{m}e^{i(\mu_{1}-\mu_{2})m}c_{m} + O(\epsilon^{2})], \\

X_{42}= - e^{-i\mu_{2}L}[ i \sum_{m}e^{i(\mu_{2}-\mu_{1})m}c_{m} +
O(\epsilon^{2})] , \\
\end{array}
\end{equation}
where
\begin{equation}
\label{abc}
a_n = \frac{\epsilon_{1,n}+\epsilon_{2,n}}{4\sin \mu_1}, \,\,\,
b_n = \frac{\epsilon_{1,n}+\epsilon_{2,n}}{4\sin \mu_2}, \,\,\,
c_n = \frac{\epsilon_{2,n}-\epsilon_{1,n}}{4 \sqrt{\sin \mu_1 \sin \mu_2}}.
\end{equation}

Substituting these expansions into Eqs.~(\ref{transmis}) and (\ref{conductance}) and performing the averaging over disorder, after some algebra the following formula for the inverse localization length is obtained \cite{BK07},
\begin{eqnarray}
\label{loclength4}
 \lambda(E) = \frac{\sigma_{1}^2}{64}
\left[\frac{{\cal K}_{11}(2\mu_1)}{\sin^2 \mu_{1}} +
\frac{{\cal K}_{11}(2\mu_2)}{\sin^2 \mu_{2}}+
\frac{2{\cal K}_{11}(\mu_1 + \mu_2)}{\sin\mu_{1}\sin\mu_{2}}
\right] \nonumber \\
+ \frac{\sigma_{2}^2}{64}
\left[\frac{{\cal K}_{22}(2\mu_2)}{\sin^2 \mu_{2}} +
\frac{{\cal K}_{22}(2\mu_1)}{\sin^2 \mu_{1}}+
\frac{2{\cal K}_{22}(\mu_1 + \mu_2)}{\sin\mu_{1}\sin\mu_{2}}
\right] \nonumber \\
\pm \frac{\sigma_{12}^2}{32}
\left[\frac{{\cal K}_{12}(2\mu_1)}{\sin^2 \mu_{1}} +
\frac{{\cal K}_{12}(2\mu_2)}{\sin^2 \mu_{2}}-
\frac{2{\cal K}_{12}(\mu_1 + \mu_2)}{\sin\mu_{1}\sin\mu_{2}}
\right] \,,
\end{eqnarray}
where
\begin{equation}
\sigma_1^2= \langle \epsilon_{1,n}^2\rangle \quad \sigma_2^2= \langle \epsilon_{2,n}^2\rangle \quad \sigma_{12}^2= |\langle \epsilon_{1,n}\epsilon_{2,n}\rangle|\,.
\label{sigma-1-2}
\end{equation}
The expression (\ref{loclength4}) manifests that the localization of an electron occurs due to the elastic backscattering processes in both channels with the change of the momentum by $2\mu_1$ and $2\mu_2$, and due to the inter-channel scattering with the change of the momentum by $\mu_1 + \mu_2$. The scattering of an unperturbed wave, which according to Eq.~(\ref{basis}) is either symmetric or anti-symmetric (with respect to the change of indexes $1$ and $2$), depends on three binary correlators, $K_{11}, K_{22},K_{12}$
defined as follows,
\begin{equation}
 \langle\epsilon_{1,n} \epsilon_{1,n+m}\rangle = \sigma_{1}^2
 K_{11}(m),\,\,
  \langle\epsilon_{2,n}
 \epsilon_{2,n+m}\rangle = \sigma_{2}^2 K_{22}(m), \,\,
 \langle\epsilon_{1,n} \epsilon_{2,n+m}\rangle = \pm \sigma_{12}^2 K_{12}(m) ,
 \end{equation}
Note that the mean value $\langle\epsilon_{1,n}
\epsilon_{2,n}\rangle $ can be either positive or negative,
unlike always positive variances $\epsilon_{1}^2$ and
$\epsilon_{2}^2$. This fact is reflected in Eq.~(\ref{loclength4}) with the non-defined sign of the term corresponding to the inter-correlations between two disorders.

In the expression for the Lyapunov exponent the correlations enter through the functions ${\cal K}_{ij}(\mu)$, that are represented by the Fourier series,
\begin{equation}
\label{phi-n} {\cal K}_{ij}(\mu) = 1 + 2
\sum_{m=1}^{\infty}K_{ij}(m) \cos(\mu m).
\end{equation}

The Lyapunov exponent (\ref{loclength4}) is positively
defined for any energy inside the spectrum. In the case of uncoupled channels, $h=0$, the dispersion relations (\ref{disp}) become equal since $\mu_1 = \mu_2=\mu$, and the cross-correlation term
vanishes in Eq.~(\ref{loclength4}). As a result, Eq.~(\ref{loclength4}) takes the
following form,
\begin{equation}
\label{loclength1}
 \lambda(E) = \frac{1}{\sin^2 \mu}\left[\sigma_{1}^2
{\cal K}_{11}(2\mu)+ \sigma_{2}^2 {\cal K}_{22}(2\mu)\right].
\end{equation}
The standard one-dimensional Anderson model (\ref{Lyap corr}) is recovered if both channels are identical, $\epsilon_{1}=\epsilon_{2}$. Finally, if the two potentials are uncorrelated, then ${\cal K}_{ij}(\mu)
\equiv 1$ and $\sigma_{12}^2=0$. In this case the relation
(\ref{loclength4}) reproduces the result derived in Ref.~\cite{H02},
\begin{equation}
\label{loclength2}
 \lambda(E) = \frac{\epsilon_{1}^2+\epsilon_{2}^2}{64} \left(\frac{1}{\sin \mu_1}+ \frac{1}{\sin \mu_2}
 \right)^2.
\end{equation}

\subsection{One evanescent and one propagating mode}
\label{13.3}

In the energy regions $-2-h<E<h-2$ and $2-h<E<2+h$, one of the
wave numbers $\mu_i$ is necessarily a pure imaginary number. The
corresponding wave function decays exponentially away from the
entering point with the decrement $| \mu_i |$. Since the
transmission matrix is a relation between the propagating waves
modes only, the evanescent modes do not contribute to the
conductance of a long sample if $L>> |\mu_i |^{-1}$.

Let us consider the energy domain $2-h<E<2+h$ where the second mode
is evanescent, $\mu_2=i\kappa$. It was shown in Ref.~\cite{H03} that in the case of
uncorrelated disorder the evanescent mode term does not contribute
to the Lyapunov exponent in Eq.~(\ref{loclength2}), and has to be
omitted. Moreover, the coupling between the propagating and
evanescent modes is strongly suppressed. This results in an extra
factor of 2 (as compared to Eq.~(\ref{loclength2})) in the formula for the Lyapunov exponent,
\begin{equation}
\label{loclength3-n} \lambda(E)=\frac{\epsilon_{1}^{2}+ \epsilon_{2}^{2}}{32 \sin^{2}
\mu_1}.
\end{equation}

It was demonstrated in \cite{BK07} that this scenario of the transition from the propagating to evanescent regime remains unchanged in the case of correlated potentials. It is worthwhile to note that in the weak disorder approximation the
formulas for the Lyapunov exponents are invalid in the vicinity of
the critical energies $E_c = \pm2 \pm h$, where the transition
from the propagating to evanescent regime occurs. At these energies
the perturbation diverges, $\epsilon_i/\sin \mu_i \rightarrow \infty$, and
the Born approximation is invalid. Away from the critical energies $E_c$ and the center of the energy band, the Lyapunov exponent is given by the formula,
\begin{equation}
\label{loclength3} \lambda(E)=\frac{1}{32 \sin^{2}
\mu_1}\left[\epsilon_{1}^{2}{\cal K}_{11}(2\mu_1) +
\epsilon_{2}^{2}{\cal K}_{22}(2\mu_1) \pm 2
\epsilon_{12}^2{\cal K}_{12}(2\mu_1) \right].
\end{equation}
In the counterpart region, $-2-h<E<-2+h$, the first mode becomes
evanescent and the Bloch number $\mu_1$ in Eq.~(\ref{loclength3})
is replaced by $\mu_2$.
\begin{figure}
\begin{center}
\includegraphics[angle = 0,scale = 0.65]{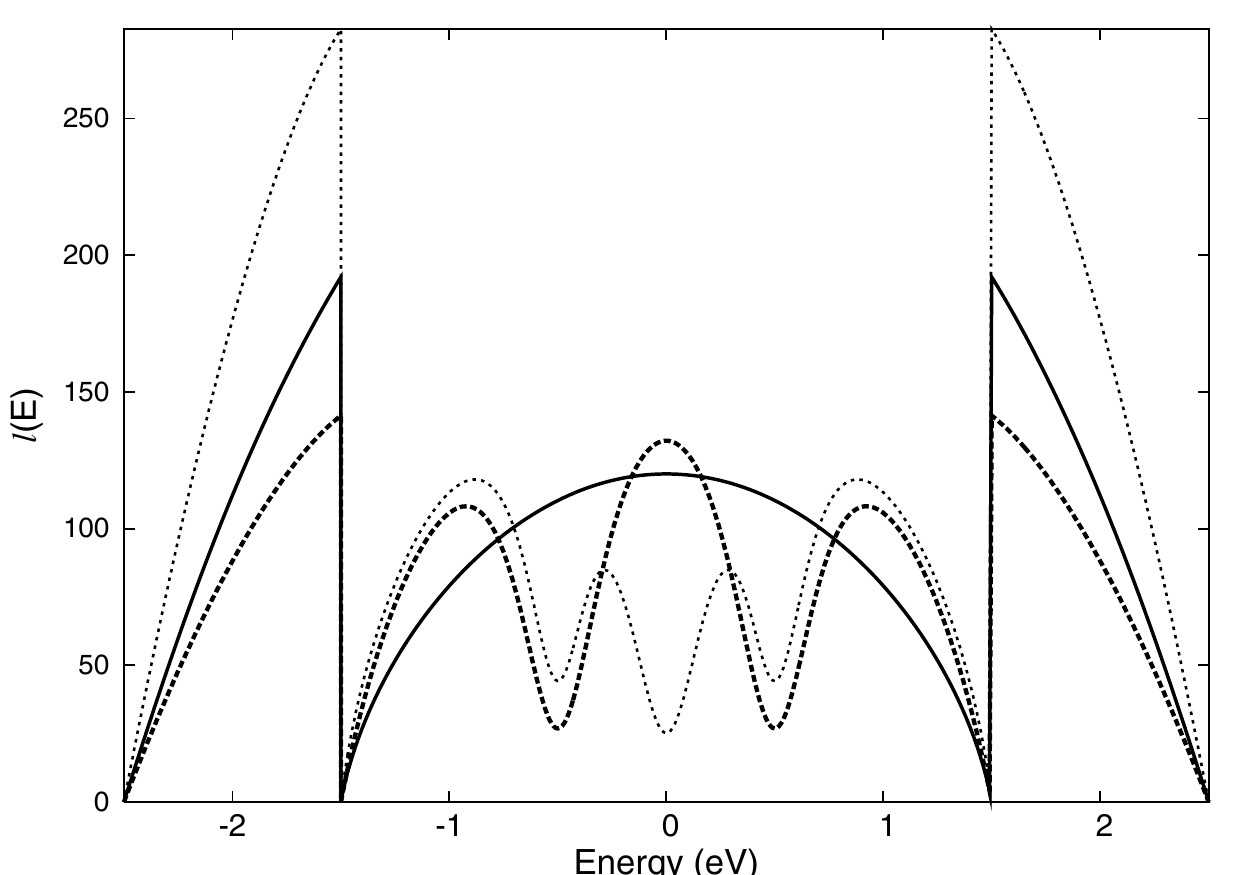}
\caption[cc]{Localization length versus energy. Solid curve corresponds to the
uncorrelated (white noise) potential. Dashed-dotted curve refers to the
exponential intra- and inter-channel correlations, all
three are given by Eq.~(\ref{sampcorr}). Dotted curve corresponds to the
exponential intra-channel correlations and delta-correlated
inter-channel scattering. The parameters of the model are $h=0.5$
eV, $\vartheta=1.0$ eV and $\langle \epsilon^2_1\rangle =\langle
\epsilon^2_2\rangle = \langle
\epsilon^2_{12}\rangle=(0.25)^2$ (after \cite{BK07}).} \label{twochannel}
\end{center}
\end{figure}

The localization length is a complicated linear functional of the binary
correlation functions $K_{ij}(m)$. Unlike single-channel correlated potentials, it is not clear yet what kind of
long-range correlations are sufficient for the mobility edge to
appear in the spectrum of two-channel system. Below we give some examples of two-strand DNA molecules that exhibit the bands of extended states. These bands are associated with long-range correlations in DNA sequences. In Fig.~\ref{twochannel} we show the numerical results \cite{BK07} for the uncorrelated potential and for exponentially decaying correlation function of the following form,
\begin{equation}
\label{sampcorr}
 K_{ij}(m) = (-1)^m e^{-\alpha \mid m \mid}
\end{equation}
where $\alpha$ is the inverse radius of correlations. Here the
correlations alternate with anti-correlations. The localization
length for this specific correlation function shows an oscillatory
behavior in the interval of $E$ where both channels are open. Unlike
this, in the regions where only one of the channels is propagating,
the localization length is a monotonic function of energy. There is
a discontinuous jump for $L_{loc}(E)$ at the critical energies $E =
\pm(2-h)$ where a transition from the propagating to evanescent mode
occurs. These discontinuities is a clear evidence of the fact that the
Born approximation is not valid in the vicinities of the critical
points. The extended states do not appear for this class of short-range
correlations.

Note that in the case of single-channel disordered potentials the binary correlator can be easily reconstructed since it is proportional to the Lyapunov exponent, see Eq.~(\ref{Lyap corr}). Unlike this, in a two-channel system the inverse scattering problem is more complicated since there are three correlators, $K_{11}(m)$,
$K_{22}(m)$, and $K_{12}(m)$, which have to be reconstructed
from the function $\lambda(E)$ given by Eqs.~(\ref{loclength4}) and
(\ref{loclength3}). In any case, it is clear from the structure of
Eqs.~(\ref{loclength4}) and (\ref{loclength3}), and from the results obtained for single-channel case that
the power-decaying correlation functions are necessary in order to
have a mobility edge. If the correlations are of a short-range,
only a discrete set of resonant extended states may appear in the
spectrum.

The existence of an extended state was
predicted for a two-channel random dimer in Ref.~\cite{SO04} where the dichotomous sequences in the both channels are assumed to be identical, $\epsilon_{1,n}= \epsilon_{2,n}$. Specifically, the vanishing Lyapunov exponent at the band center $E=0$ was obtained numerically for non-perturbative values of the random on-site potentials, $\left\langle \epsilon _n\right\rangle=0$, $\left\langle
\epsilon_{1,n}^2\right\rangle = \left\langle
\epsilon_{2,n}^2 \right\rangle = \epsilon _0^2 =1$,
$\epsilon_{12} = \left\langle
\epsilon_{1,n}\epsilon_{2,n}\right\rangle = \epsilon
_0^2 $, and hopping parameter $h=1/2$, see Fig.~\ref{osipov}.
\begin{figure}
\begin{center}
\includegraphics[angle = 0,scale = 0.5]{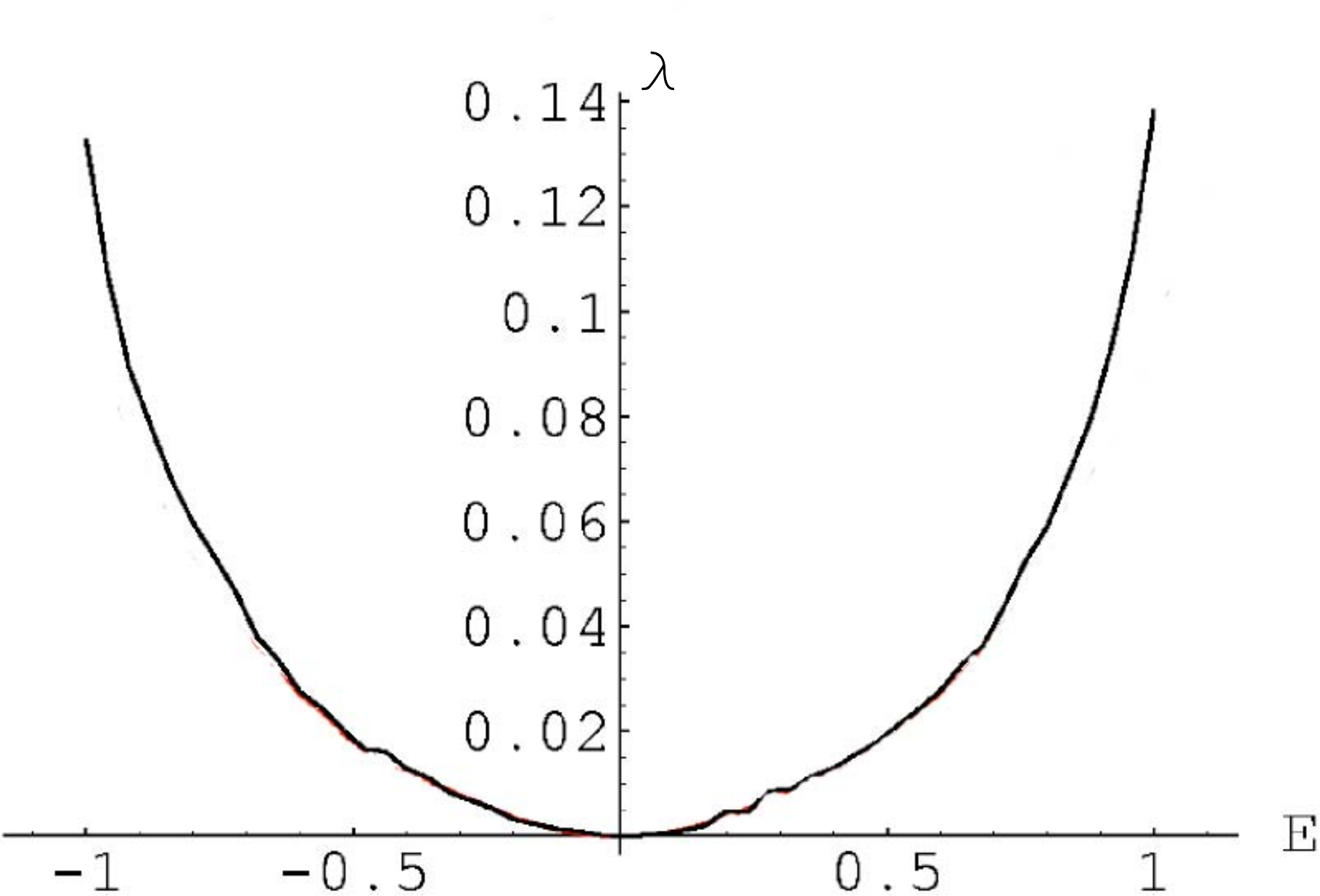}
\caption[cc]{Lyapunov exponent in the vicinity of the band center $E=0$ versus energy.
The length of the sample is $N=4\cdot 10^3$ and 60 values of energies in the interval $-1.1<E< 1.1$ were considered (after \cite{SO04}.)} \label{osipov}
\end{center}
\end{figure}

The extended state do exist in this model for weak scattering potential, $\epsilon_0 \ll 1$. This conclusion follows from Eq.~(\ref {loclength4}), that after the substitution of the correlators
$\left\langle \epsilon _n\epsilon
_{n-1}\right\rangle =1/2$ takes the following form,
\begin{equation}
\label{dimer} \lambda(E)=\frac{\epsilon_0^2}{8}\left(\cot^2\mu_1
+ \cot^2\mu_2 \right).
\end{equation}
This function does not vanish if the inter-channel hopping
parameter $h$ is different from zero, i.e. there is no extended
state in a two-channel dimer. Unlike this, in a single-channel
dimer there are two extended states \cite{DWP90}. Indeed, the Lyapunov exponent defined by Eq.~(\ref{dimer}) vanishes quadratically at $E=0$ ($\mu_1=\mu_2 =\pi/2$).

\section{Electron localization in DNA molecules}
\label{14}

One of the interesting and important applications of the theory developed in the previous sections is related to conductivity of DNA molecules. Each DNA, being a macromolecule, is a long sequence of discrete site potentials -- basic nucleotides. The electron transport along this sequence occurs due to hopping between the neighboring nucleotides. This property justifies the application of the tight-binding model. Since a DNA molecule contains two strands forming the famous helical spiral, an adequate description requires the two-channel model given by Eq.~(\ref{Schr1}). This model accounts for the main biological and informational features of the DNA structure. However, it ignores the presence of the environment and phonons. Thus, our primary interest is to study how the information coded in a sequence of nucleotides, may affect the localization length (therefore, the resistivity) of a given segment of a DNA molecule. The results obtained in the framework of this so-called ladder model should be considered as qualitative, they can be used for the estimation of DNA resistivity by order of magnitude. A question about resistivity of a sequence of nucleotides is of fundamental nature because it may shed light on the problem of DNA sequencing using electrical measurements.

One of the fundamental questions is how the information is transferred along a sequence of nucleotides. For example, if a mutation occurs in the sequence, it is usually healed. This means that some of physical parameters of the DNA molecule are sufficiently sensitive to detect this mutation. The length of a mutation is relatively short ($\sim 10$ base pairs) as compared to the length of a gene ($10^3 - 10^6$ base pairs). Because of small statistical weight of a mutation, the mechanical and thermodynamic characteristics are not sensitive enough for its robust detection. Unlike this, the electrical resistance of a DNA molecule strongly fluctuates even if a single nucleotide in a long sequence is replaced or removed \cite{HXZT05}. This property is a signature of coherent transport that gives rise to universal fluctuations of conductance in mesoscopic samples \cite{A85,LS85}.
\begin{figure}
\includegraphics[width = 8cm]{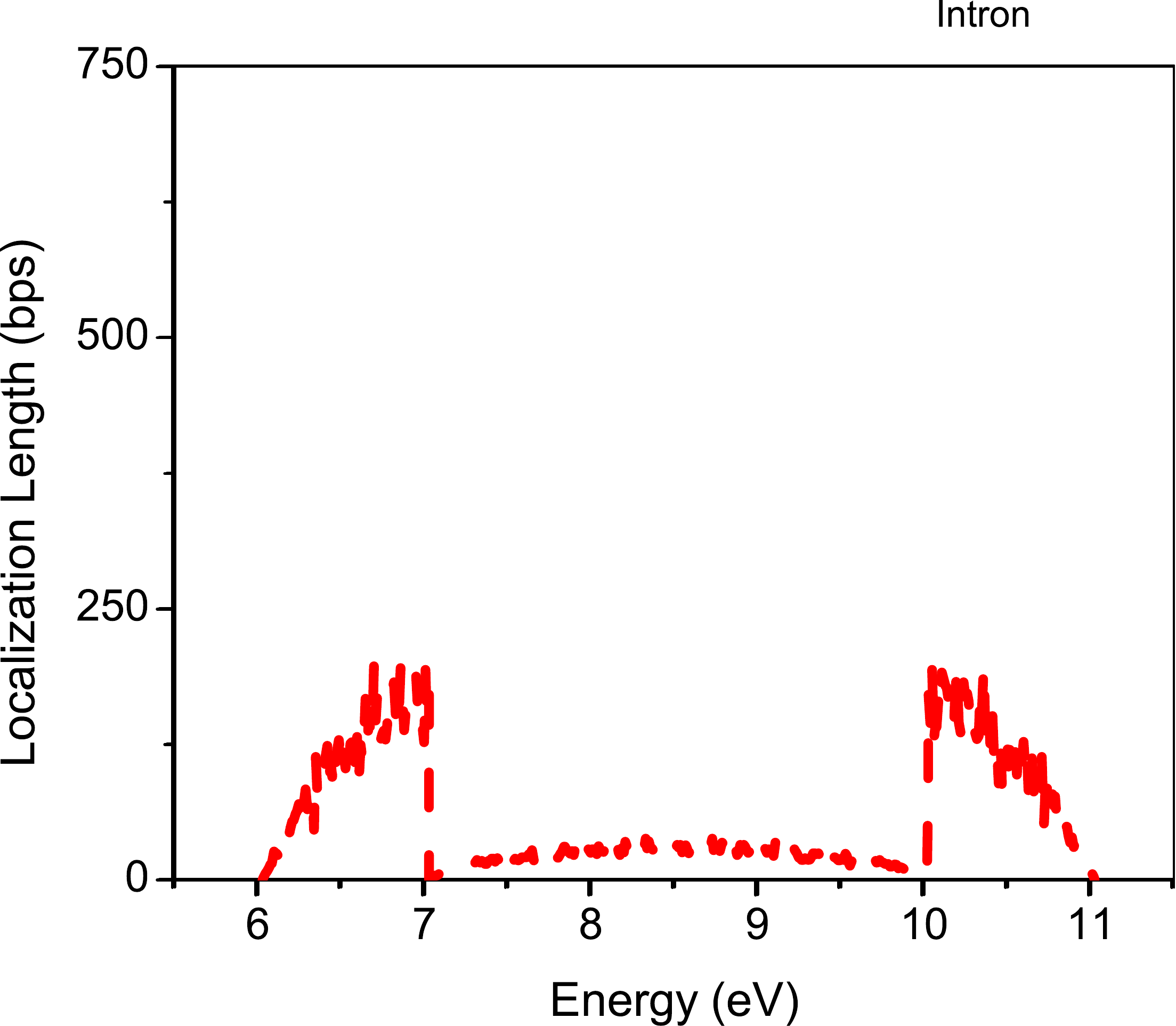}
\includegraphics[width = 8cm]{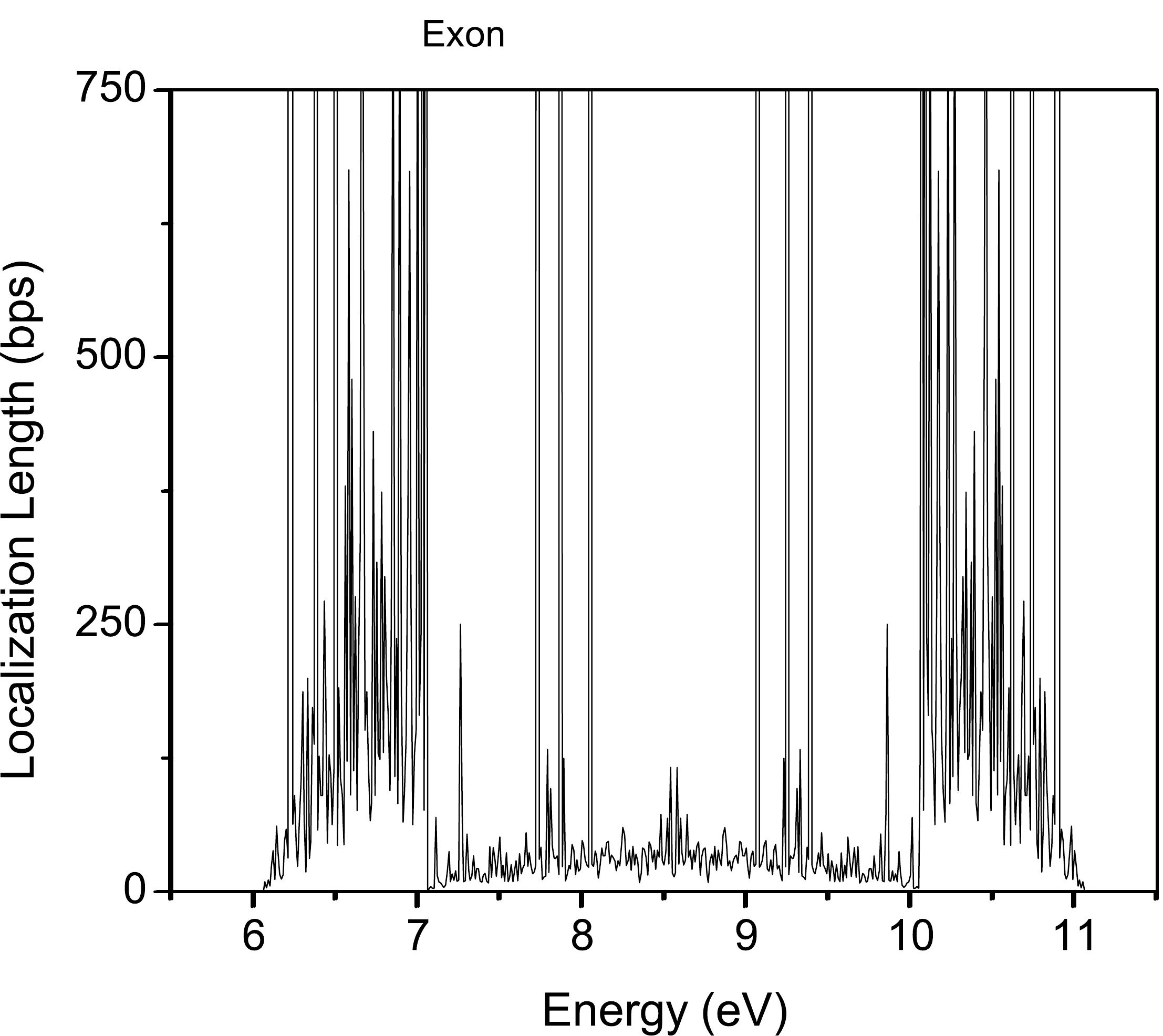}
\caption[cc]{Localization length vs energy for the human
BRCA gene measured in the number of base pairs. The
length of the exon (intron) is 2120 (10 421) bps. Results for
exon and intron are shown by black and gray (red) lines, respectively.
Two channels are propagating if $6.6<E<10.4$ eV. One of the channels becomes evanescent in two symmetric regions, $10.4<E<11.4$ eV and $5.6<E<6.6$ eV, of the width of $2h=1$ eV (after \cite{KBIUY09}).} \label{BRCA}
\end{figure}

\begin{figure}
\includegraphics[width = 8cm]{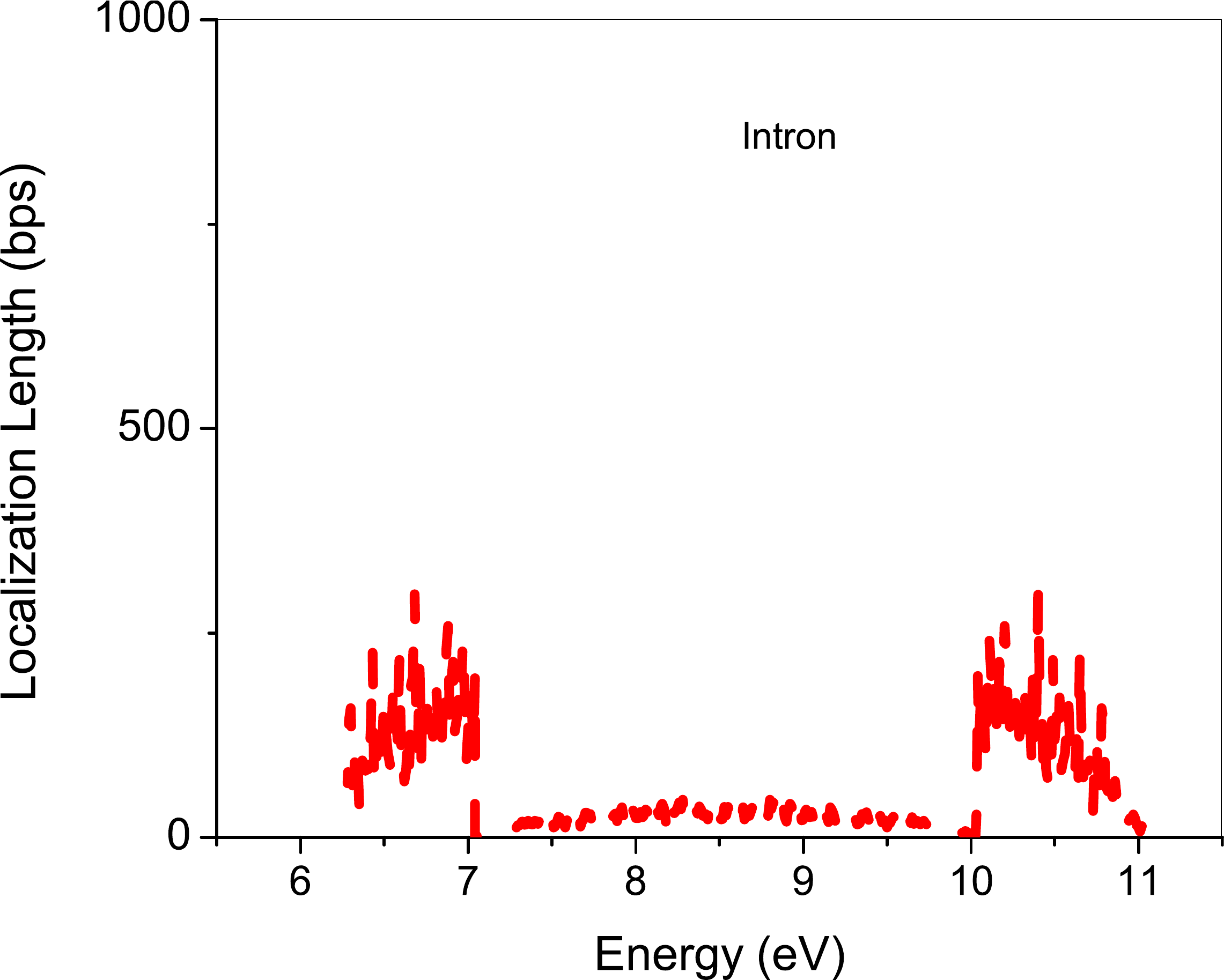}
\includegraphics[width = 8cm]{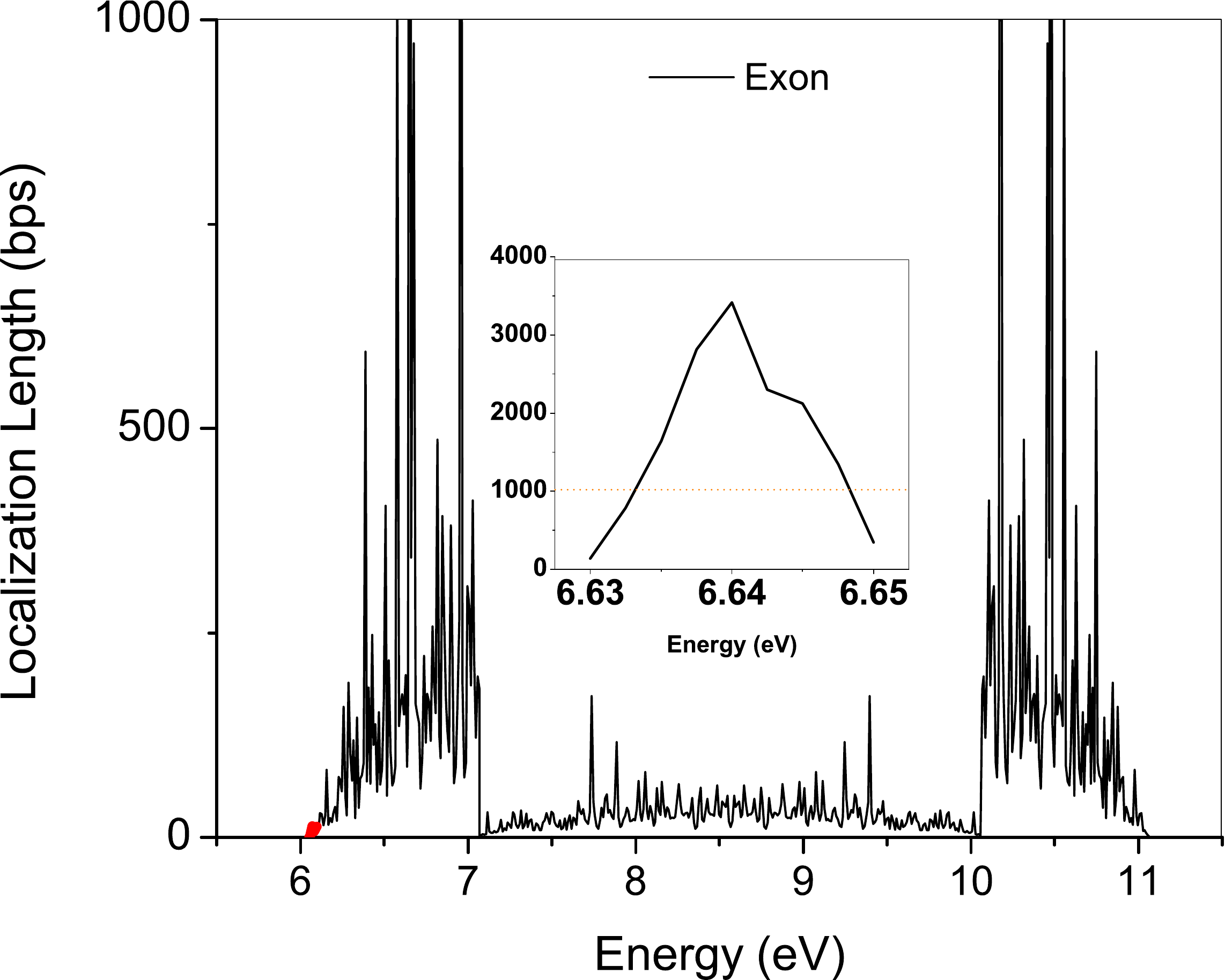}
\caption[cc]{Localization length vs energy for the human
ADAM10 gene. The length of the exon (intron) is 1030
(31 752) bps. Inset shows the fine structure of one of the peaks
(after \cite{KBIUY09}).} \label{ADAM10}
\end{figure}

The genetic information is written by four-letters alphabet. Each letter corresponds to one of four nucleotides: Adenine, Thymine, Cytosine, and Guanine. The ionization energies of the nucleotide are known: $\epsilon_A=8.24$, $\epsilon_T=9.14$, $\epsilon_C=8.87$, and $\epsilon_G=7.75$ eV, and can be used as the values of the site potentials $\epsilon_{1,n}$ and $\epsilon_{2,n}$ in Eq.~(\ref{Schr1}). A conduction band would form if the DNA texts would exhibit some periodicity \cite{B62a}. However, many studies of the DNA texts have revealed rich statistical properties but not the periodicity. One of the suggestions is that a DNA molecule is a stochastic sequence of nucleotides, the main feature of which is long-range correlations \cite{V92,Po92,CL93,LMK94,ECS04}. The nature and role of these correlations are still not clearly understood.

Since the nucleotides in each DNA stand form a long random sequence, one of the methods of detection of correlations is the mapping of a DNA sequence onto a random walk. The long-range correlations are manifested then in an anomalous scaling of the generated classical diffusion \cite{Po92,CL93,SLG02a,SLG02b,RBMK03}.

The quantum transport through a DNA molecule is also strongly affected by the correlations. An uncorrelated sequence of nucleotides localizes all quantum electron states, as occurs in one-dimensional white-noise potential,
making impossible the charge transfer at distances longer than the localization length $L_{loc}(E)$. However, since most of the mutations in DNA are successfully healed, one may assume the existence of charge transport through delocalized states, responsible for the transfer of information at much longer distances \cite{MBB08}. If such delocalized states do exist, they must be located within the coding regions of DNA, the so-called {\it exons}. The coding regions are separated by long segments which apparently do not carry genetic code, the {\it introns}. They may not contain delocalized states. Thus, due to qualitatively different information coded in exons and introns, their transport properties also may be qualitatively different: exons may conduct and introns are insulators.

As shown in previous Sections, a physical reason for delocalization in a random potential can be the statistical correlations. It is natural to assume that the coding and non-coding regions may have principally different binary correlators. Thus, the information itself affects the local conducting properties of DNA. A direct way to check this hypothesis is to calculate the localization length for exons and introns in different DNA molecules. Since a DNA molecule is a double-stranded sequence of nucleotides, it is appropriate to use the two-channel tight-binding model, see Eq.~(\ref{Schr1}). Two random sequences of nucleotides, being random in the longitudinal direction, exhibit regular A-T and C-G matching in the transverse direction. This key-to-lock matching between the strands strongly affects the electron transport in DNA. For a correct evaluation of the localization length in DNA one has to (i) use the two-stranded model; (ii) avoid a simplification of the four-letter DNA alphabet to a binary sequence; and (iii) account for the longitudinal correlations in both strands and in the transversal base pairing.

Since the on-site energies $\epsilon_A$, $\epsilon_T$, $\epsilon_G$, and $\epsilon_C$ do not differ much, it is possible to calculate the Lyapunov exponent of a DNA molecule using the results of the perturbation theory developed in Section~\ref{13} for a two-channel correlated sequence \cite{KBIUY09}. The Lyapunov exponent was calculated according to Eqs.~(\ref{loclength4}) and (\ref{loclength3}) for the following DNA molecules: BRCA, ADAM10, SNAP29, and SUHW1. The results are shown in Figs.~\ref{BRCA} - \ref{SUHW1} where the dependence $L_{loc}(E)$ is given for the exon and intron regions of the aforementioned DNA molecules. The parameters of nucleotide site energies and the hopping amplitudes are the same in all figures, $h=0.5$ eV and $\vartheta=1$ eV.

\begin{figure}
\includegraphics[width = 8cm]{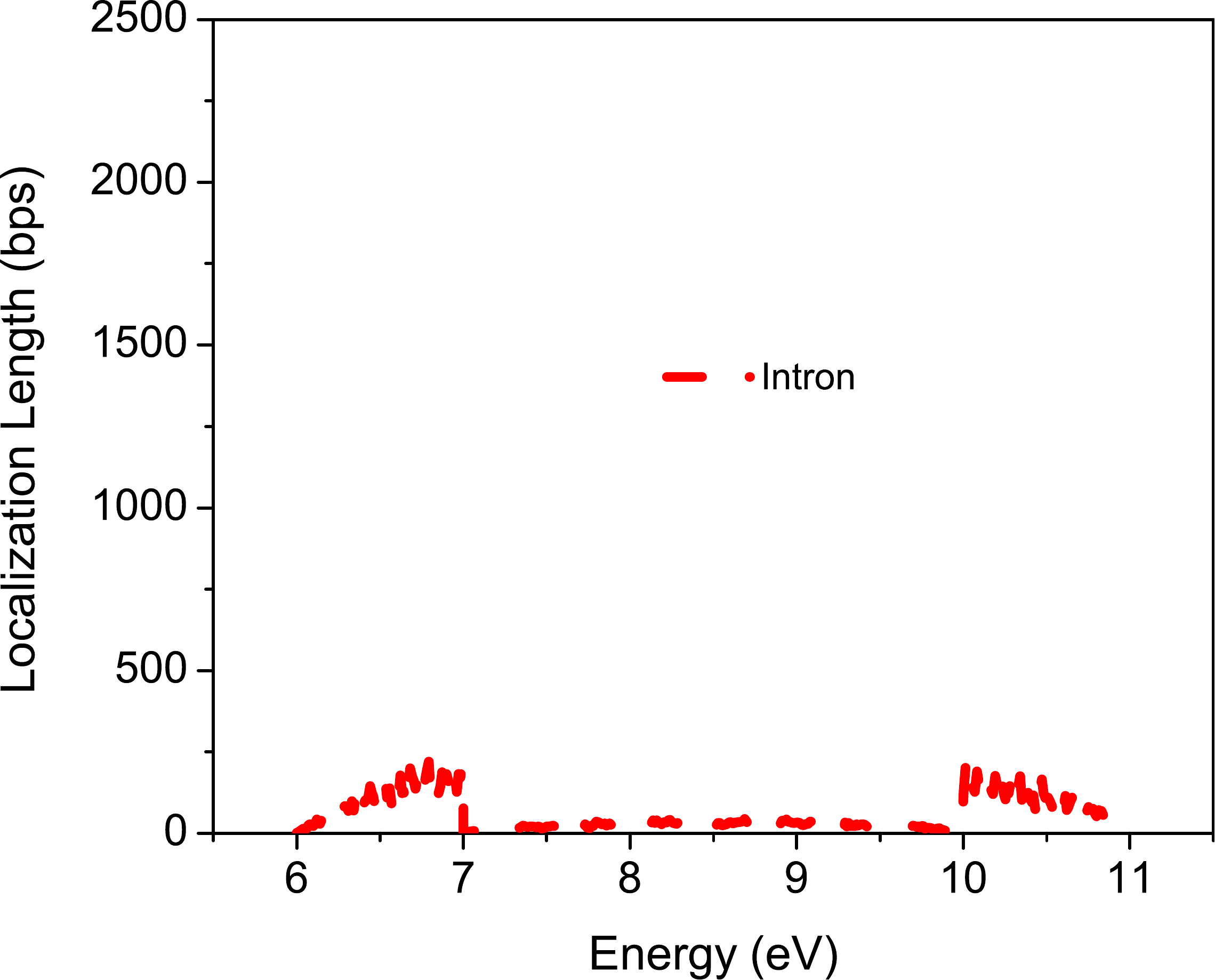}
\includegraphics[width = 8cm]{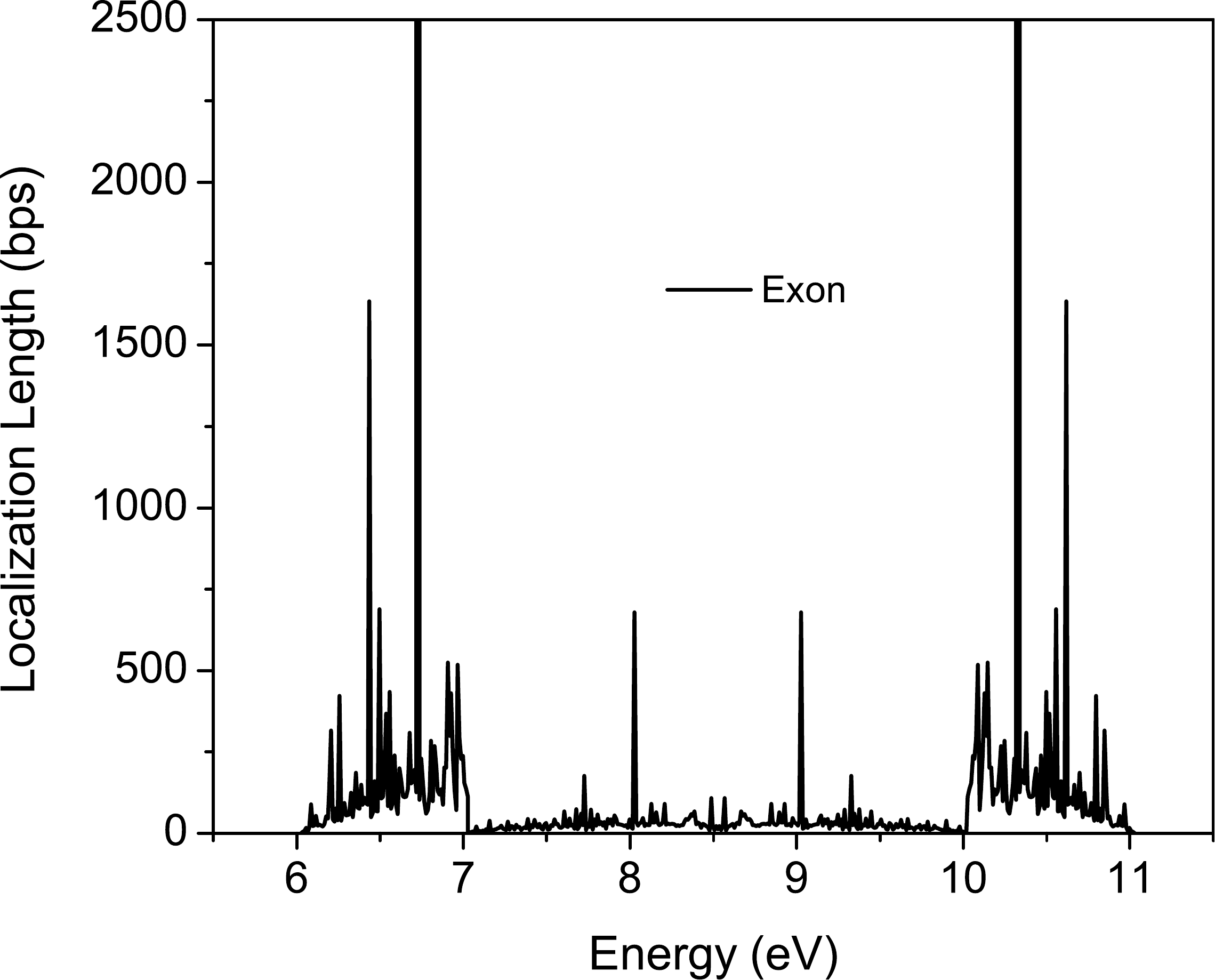}
\caption[cc]{Localization length vs energy for the human
SNAP29 gene. The length of the exon (intron) is 2141
(21701) bps
(after \cite{KBIUY09}).} \label{SNAP29}
\end{figure}

\begin{figure}
\includegraphics[width = 8cm]{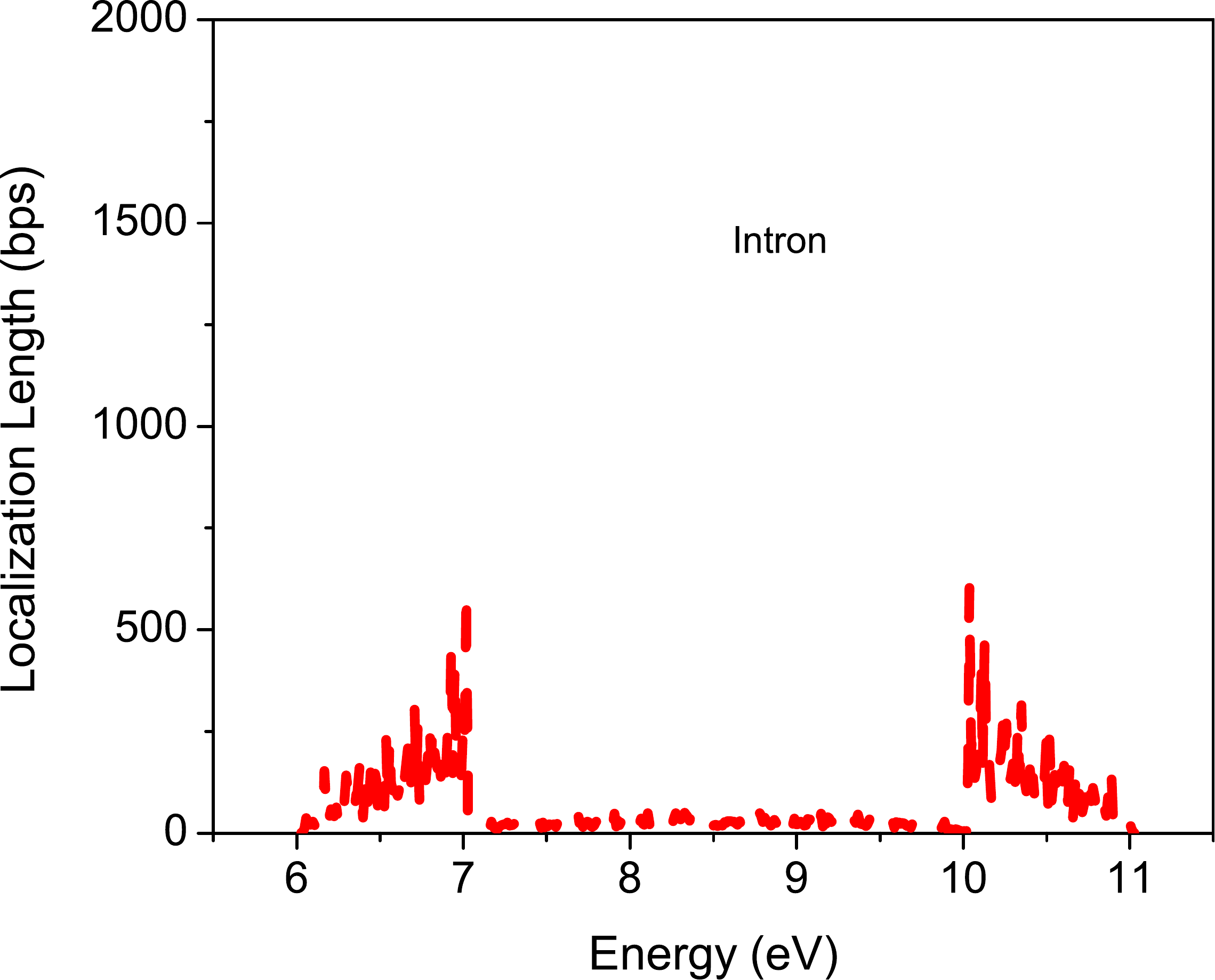}
\includegraphics[width = 8cm]{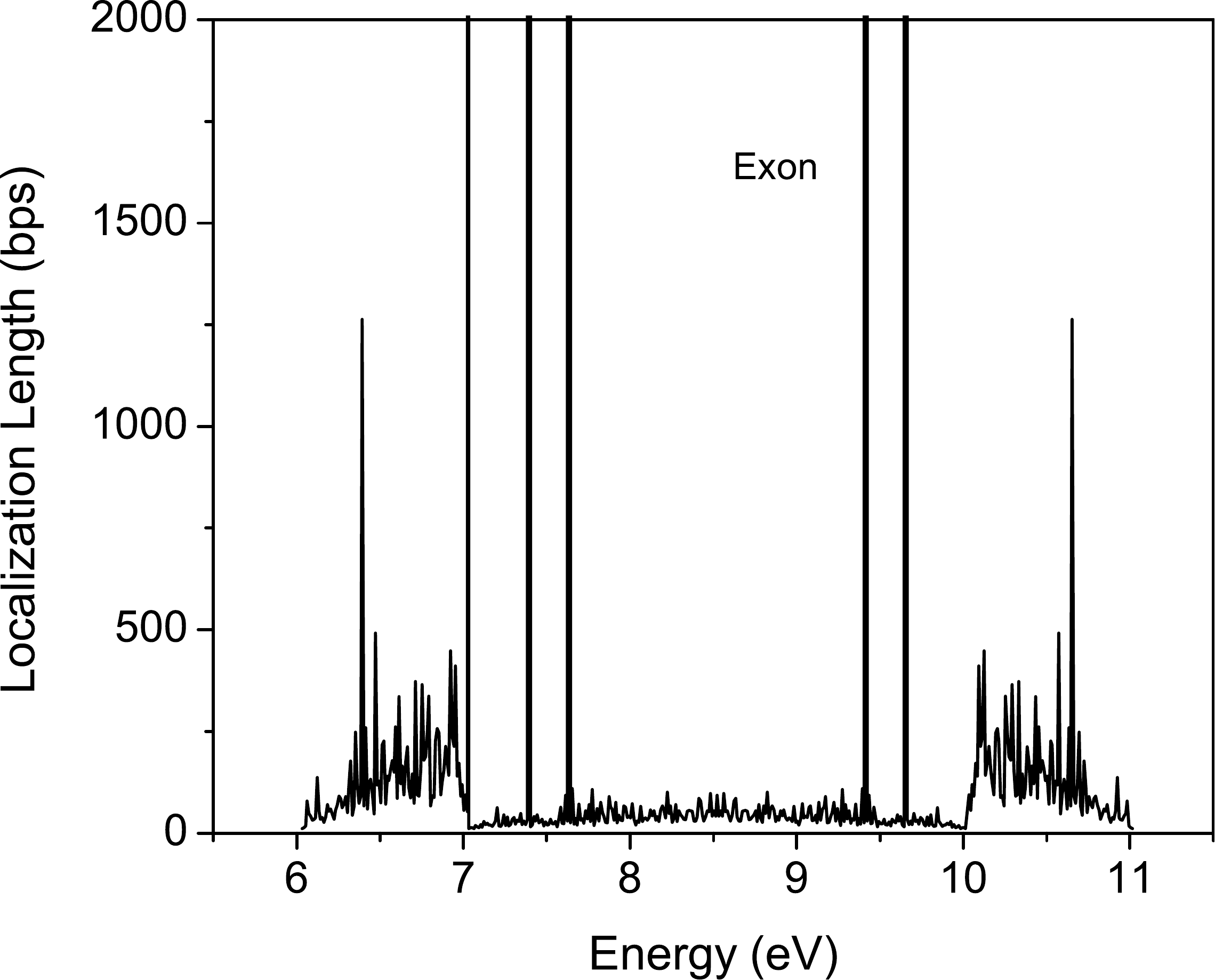}
\caption[cc]{Localization length vs energy for the human
SUHW1 gene. The length of the exon (intron) is 1963
(4405) bps
(after \cite{KBIUY09}).} \label{SUHW1}
\end{figure}

Although the patterns represented in these figures are very different, the common feature is that for most of the energies the localization length inside the exon region exceeds by order of magnitude the localization length inside the intron region. This confirms, by implication, the fact that very different kinds of information are coded in these regions. The vertical axis for each figure is cutted approximately at the length of the corresponding exon
region. There are many peaks in the exon regions with the height that exceeds much the vertical scale, i.e., the states within these peaks are extended. Unlike this, in the intron regions all the states are well localized. The density of the peaks in Figs. \ref{BRCA} and \ref{ADAM10} is much higher than that in Figs.~\ref{SNAP29} and \ref{SUHW1}. Most of the peaks are located in the region of energies where one of the channels is evanescent. Similar sharp peaks in the transmission through the exon regions of Y3 DNA have been numerically detected in Ref.~\cite{S06}. It turns out that this feature is very robust since in that study a single-stranded model of DNA was used. It is naturally to suggest that the peaks are due to long-range correlations in the sequences of nucleotides. Because of A-T and C-G matching the correlation functions in both strands are very similar. This explains why single- or double-stranded DNA sequences exhibit similar patterns of the peaks. Of course, this is true for a general pattern only; the actual positions and amplitudes of the peaks strongly depend on the model (single- or double-stranded DNA).

The fine structure of one of the peaks is shown in the inset of Fig.~\ref{ADAM10}. Since the peaks are of a finite width ($\sim 20$ meV), they are really narrow bands of extended states but not the discrete resonant states predicted and observed in random dimers \cite{DWP90}. The nature of resonant tunneling in random dimers is due to short-range correlations in contrast with specific long-range correlations that are necessary for existence of a continuous band of extended states. In the case of a single channel the width of the band of extended states can be controlled by the parameters of the binary correlator in Eq.~(\ref{spect-gen}). In particular, wide and narrow bands of the extended states have been observed in the experiments with single-mode microwave waveguides \cite{KIKS00}.

For the two-channel system the relation between the positions of the mobility edges and the explicit form of the binary correlator is not known. One may expect that this relation is determined by the relative phase shifts between the Fourier components of the oscillatory correlators $K_{ij}$ in Eq.~(\ref{loclength4}). It is worth mentioning that short- and long-range correlations lead not only to different localization properties but also to a very different classical, as well as quantum, diffusion in DNA \cite{RBMK03,AVLM05}. A pattern $L_{loc}(E)$ is a particular fingerprint of a given DNA sequence and it can be used, in principle, for the classification of DNA molecules. In the previous studies (see, e.g., Refs. \cite{SLG02a,SLG02b,RBMK03,UY03}) the DNA sequences have been characterized by the scaling exponent of the corresponding random walk. We consider that the inverse localization length given by Eq.~(\ref{loclength4}) is more convenient since it characterizes a well-defined physical property, the electrical resistivity. Moreover, Eq.~(\ref{loclength4}) establishes a qualitative relation between the localization length and the informational characteristic (binary correlators) of the DNA sequence.

\begin{figure}
\includegraphics[width = 8cm]{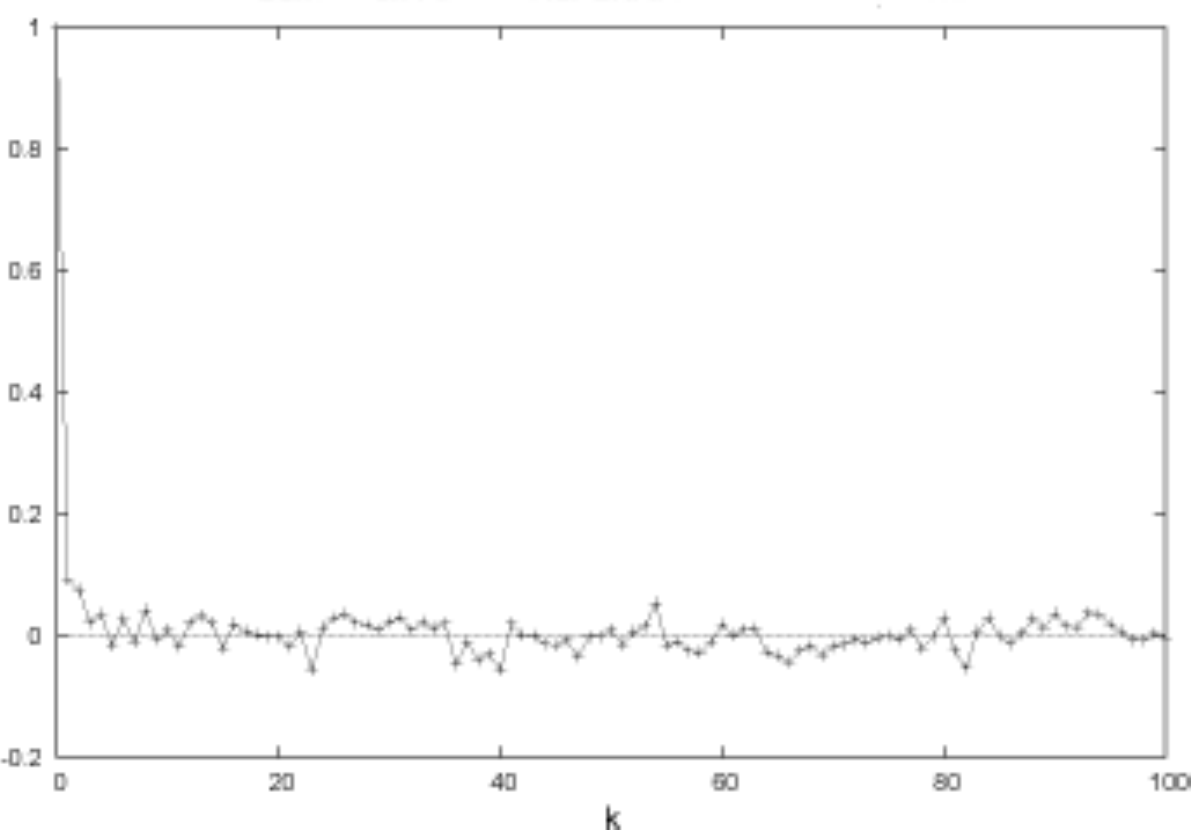}
\includegraphics[width = 8cm]{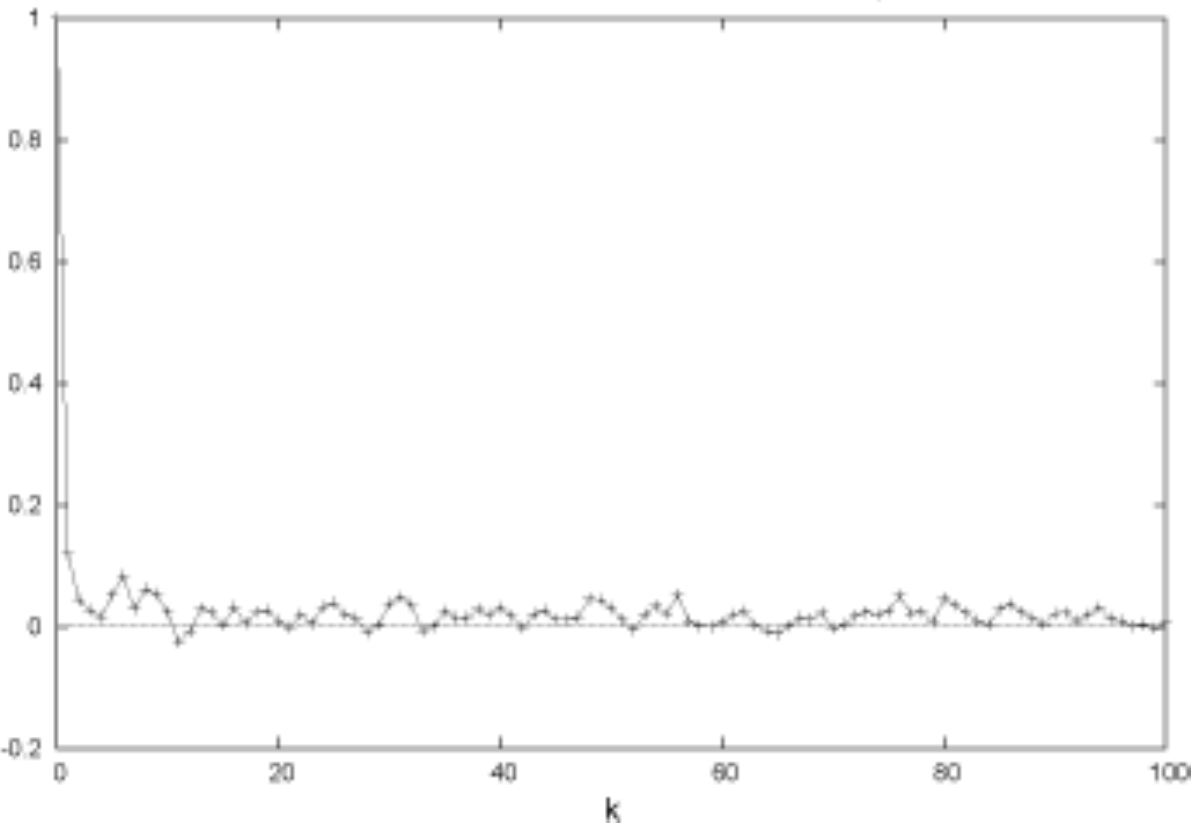}
\caption[cc]{Binary correlator $K_{11}(k)$ for the exon (left panel) and intron (right panel) regions of the human SNAP29 gene.} \label{CorrSNAP29}
\end{figure}

However, by themselves the binary correlators are not very illustrative and informative. To demonstrate this we plot in Figs.~\ref{CorrSNAP29} the correlation function of the exon and intron region of the human SNAP29 gene.
At first glance, these correlators look very similar: they oscillate in a random way with a slowly decaying amplitude. In the case of a single-channel system such a behavior leads to a sharp mobility edge, see Fig.~\ref{edge-1}. However, only the correlator for the exon region (left panel in Fig.~\ref{CorrSNAP29}) gives rise to the mobility edges, see Fig.~\ref{SNAP29}. It is hard to determine what are the main features of the binary correlator that guarantee a mobility edge in the spectrum of two-channel system. We may only mention that the correlator $\propto \sin(a k)/k$ that leads to a sharp mobility edge in Fig.~\ref{edge-1} oscillates around zero, i.e. there is a balance between correlations and anti-correlations. Probably, in the case of introns the balance is shifted too much towards stronger correlations, and this is the reason for qualitatively different patterns of $L_{loc}(E)$ in Fig.~\ref{SNAP29}. Thus, a subtle difference in the correlation functions of exons and introns leads to a well-pronounced difference in their localization length, i.e. in their resistivity. This is one more reason to use the functional Eq.~(\ref{loclength3}) for the classification of DNA and characterization of their local structure.

In 90's an apparent similarity of statistical properties for exons and introns was under a discussion, see Ref.\cite{LMK94}. The authors noted that there is no {\it a priori} reason to believe that the correlation structure should be different between coding and non-coding sequences. On the other hand, they stressed that the divergence between the correlation structure of exons and introns is due to evolutionary mutations, and this may lead to different physical characteristics of these DNA regions. Now we may conclude that a proper quantity which is quite sensitive to different information stored in exons and introns, is the Fourier transform of the binary correlator but not the correlator itself. The former defines the localization length $L_{loc}(E)$ (see Eq.~(\ref{loclength3})). Looking at the plots of $L_{loc}(E)$ in Figs.~\ref{SNAP29} it is easy to discriminate between the exons and the introns regions. Unlike this, the plots of the binary correlator in Fig.~\ref{CorrSNAP29} look very similar.

The presence of extended states in the spectrum of exons regions strongly supports a hypothesis that the information about DNA repair processes transfers in a damaged DNA along a nucleotide sequence due to electrical mechanisms. Since proteins detect the damaged region in a sequence by virtue of transport properties \cite{RJB00,Yo05,HS06}, it is natural that the coding regions (exons) have extended states, while this property is irrelevant for the non-coding regions (introns) \cite{S06}. The role of charge transport in the DNA-repairing deficiency yielding carcinogenesis has been recently reported in Ref.~\cite{SRR08}, where the fluctuations of the transmission coefficient caused by different type of mutations have been analyzed. The $p53$ is a special type of DNA which encodes the $TP53$ protein that suppresses the tumor development in living organisms by activating the DNA repair mechanisms or the cell apoptosis process if the damage of DNA is irreparable. In addition, $p53$ is involved in multiple types of repair processes that follow after a DNA damage occurs. These processes are: the nucleotide excision repair, base excision repair and correction of double strand breaks \cite{BV01}. Mutations of $p53$ gene that reduce the ability to aid in DNA restoration lead to about half of all known cancers \cite{Ao03}. Most of the cancerous mutations of $p53$ are point mutations when a given base pair is substituted by another. The statistical analysis shows that the exons from the 5th to 8th located in the interval from the 13,055th to the 14,588th nucleotide are the most exposed to danger of point mutations \cite {SRR08}.

\begin{figure}[ht]
\begin{center}
\includegraphics[width=3.6in,height=3.4in]{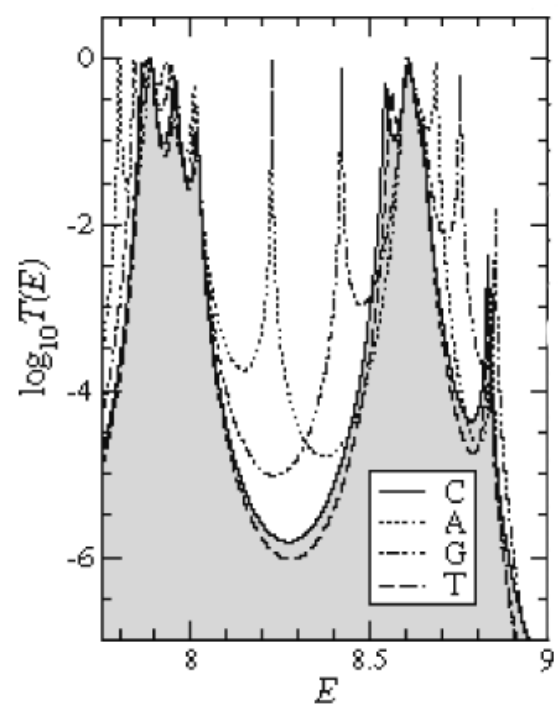}
\caption[cc]{Energy-dependence of logarithmic transmission coefficients
of the original sequence (C shaded solid line)
and mutated (A dotted, G dotted-dashed, T dashed) sequences
with length $L =20$ (from 14 575th to 14 594th nucleotide) of the gene
$p53$ (after \cite{SRR08}).} \label{p53}
\end{center}
\end{figure}

Each mutation gives rise to the corresponding change in charge transmission. Using the tight-binding model, the authors of Ref.~\cite {SRR08} have numerically calculated the variation of the transmission coefficient of $p53$ caused by cancerous and noncancerous mutations. Very commonly the mutations occur at the 14,585th base pair of the $p53$ sequence, normally occupied by Cytosine. The point mutation C$\rightarrow$T is a cancerous one, while the mutations C$\rightarrow$A and C$\rightarrow$G are noncancerous. It was found that cancerous mutations lead to much less changes in the transmission than the noncancerous ones, as it is shown in Fig.~\ref{p53}. Similar results have been obtained for the mutations which occur at different points of the $p53$ sequence. If the damage-repair process is based on the transmission-probe criterion, the cancerous mutations are much harder to detect than the noncancerous ones. Being undetected, such mutations avoid the damage-repair process and thus may develop cancerous tumor. This scenario explains why cancerous mutations are ``invisible" that makes the $p53$ gene ineffective, although it is known to be a ``guardian of the genome."

Here we reviewed some theoretical results for the charge transport though a double-stranded sequence of nucleotides and the effects of statistical correlations. Although a real DNA is much more complex physical system than that represented by the two-channel tight-binding model, we suggest that the obtained results for the localization length reflect the fundamental properties of the exons and introns. A more detailed model of DNA that includes the electron coupling to the backbones of sugar phosphates has been proposed in Refs.~\cite{GCPD02,Z03}. The backbones connections increase the transmission and, in general, reduce the localization length. Some of the interesting results obtained in the framework of this model, as well as the discussion of the role of the environment and finite temperature, can be found in Refs.~\cite{ECS04,KRT05,BK07a,SRR08,ZYU09}.

\section{Localization in self-affine potentials}
\label{15}

In order to study the influence of long-range correlations in the Anderson model (\ref{tb diagonal}), in Ref.~\cite{ML98} it was proposed to generate the site potential $\epsilon_n$ with the use of {\it self-affine} random functions that model the trajectories of fractional Brownian motion. Specifically, the values $\epsilon_n$ of a sequence of length $N$ were obtained as a superposition of incoherent plane waves with all possible wavelengths,
\begin{equation}
\label{self on site}
\epsilon_n = \sum_{m=1}^{N/2} \frac{1}{m^{\alpha/2}} \cos\left(\frac{2 \pi m n}{N} + \phi_m \right),
\end{equation}
where $N$ is even number. Here the source of randomness is the presence of phases $\phi_m $ that are uniformly distributed over the interval $[0,2 \pi]$. The limiting case of white-noise potential is obtained from Eq.~(\ref{self on site}) for $\alpha =0$ (with $N\rightarrow \infty$). Another famous case emerges for $\alpha = 2$ for which the standard Brownian diffusion is recovered. The more correlated potentials emerge for larger values of $\alpha$. As one can see from
Fig.~\ref{self-affine}, the larger $\alpha$ the more smooth are the random profiles.
\begin{figure}[ht]
\begin{center}
\includegraphics[width=3.6in,height=3.4in]{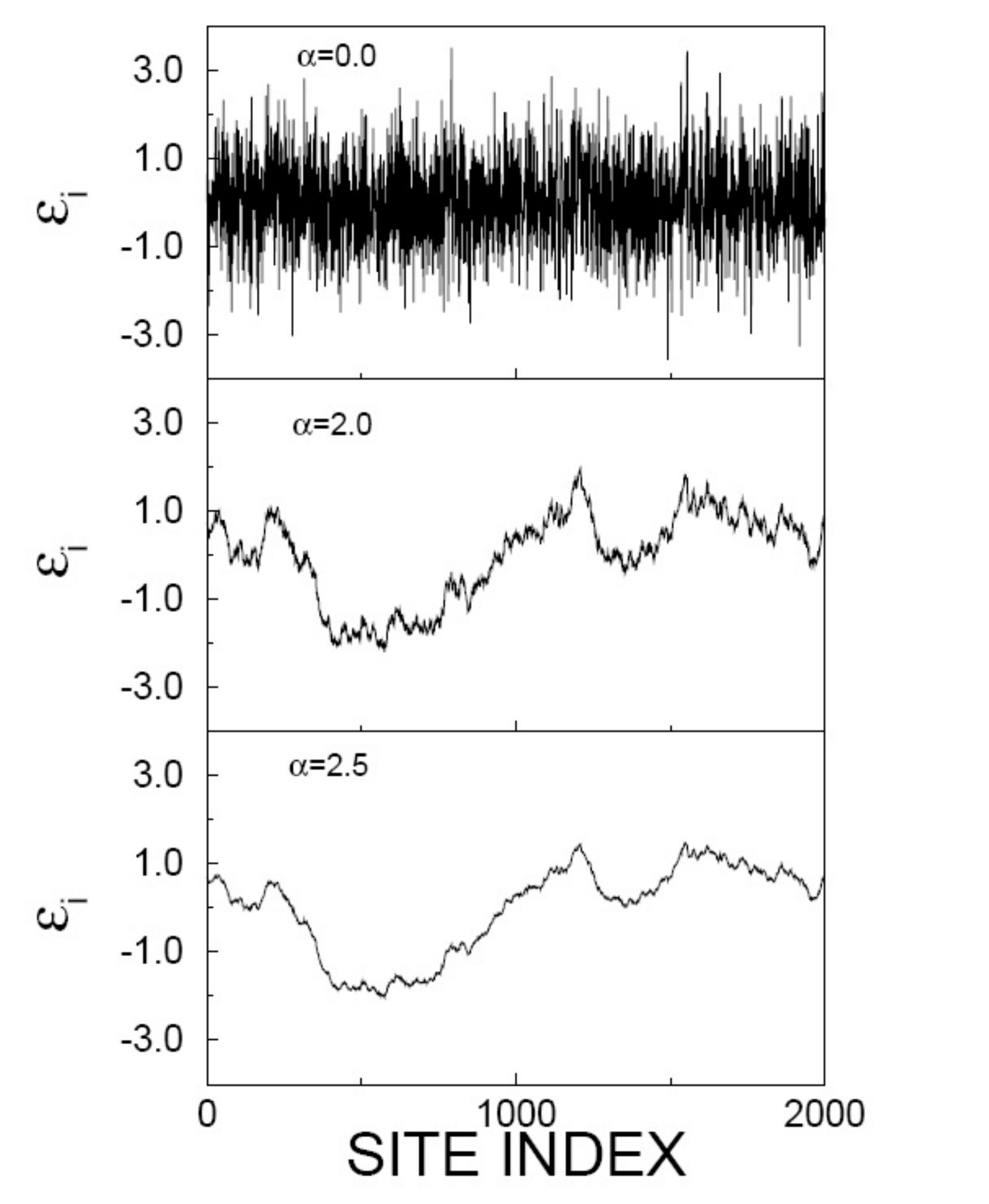}
\caption[cc]{Profile landscapes generated according to Eq. (\ref{self on site}) for different values of $\alpha$ (here $\varepsilon_j \equiv \epsilon_n$). The length of the sequences is $N=4096$ (after \cite{ML98}).} \label{self-affine}
\end{center}
\end{figure}

It is straightforward to obtain that for large $k$ the power spectrum of random potential (\ref{self on site}) is a power-like function, $S(k) \propto 1/k^{\alpha}$. As indicated in Refs.~\cite{Ko00,RKBH01} the fluctuations of self-affine systems increase with the system size. In order to estimate these fluctuations it was noted that the local potentials $\epsilon_n$ are given by the trace of a fractional Brownian particle with the Hurst exponent $H$. Since the fluctuations of the potential (\ref{self on site}) increase with increasing length scale $l$ as
\begin{equation}
\label{fbm}
\langle (\epsilon_{n+l}-\epsilon_n)^2 \rangle \propto l^{2H}\,,
\end{equation}
the fluctuations increase with increasing system size $N$. For $0<H<1/2$ the random process (\ref{self on site}) is known as superdiffusion, and for $1/2<H<1$ as subdiffusion. The Hurst exponent and the exponent $\alpha$ are known to be related, $\alpha = 1 + 2H$, see Refs.~\cite{F88,MK00}.

\begin{figure}[ht]
\begin{center}
\includegraphics[width = 10cm]{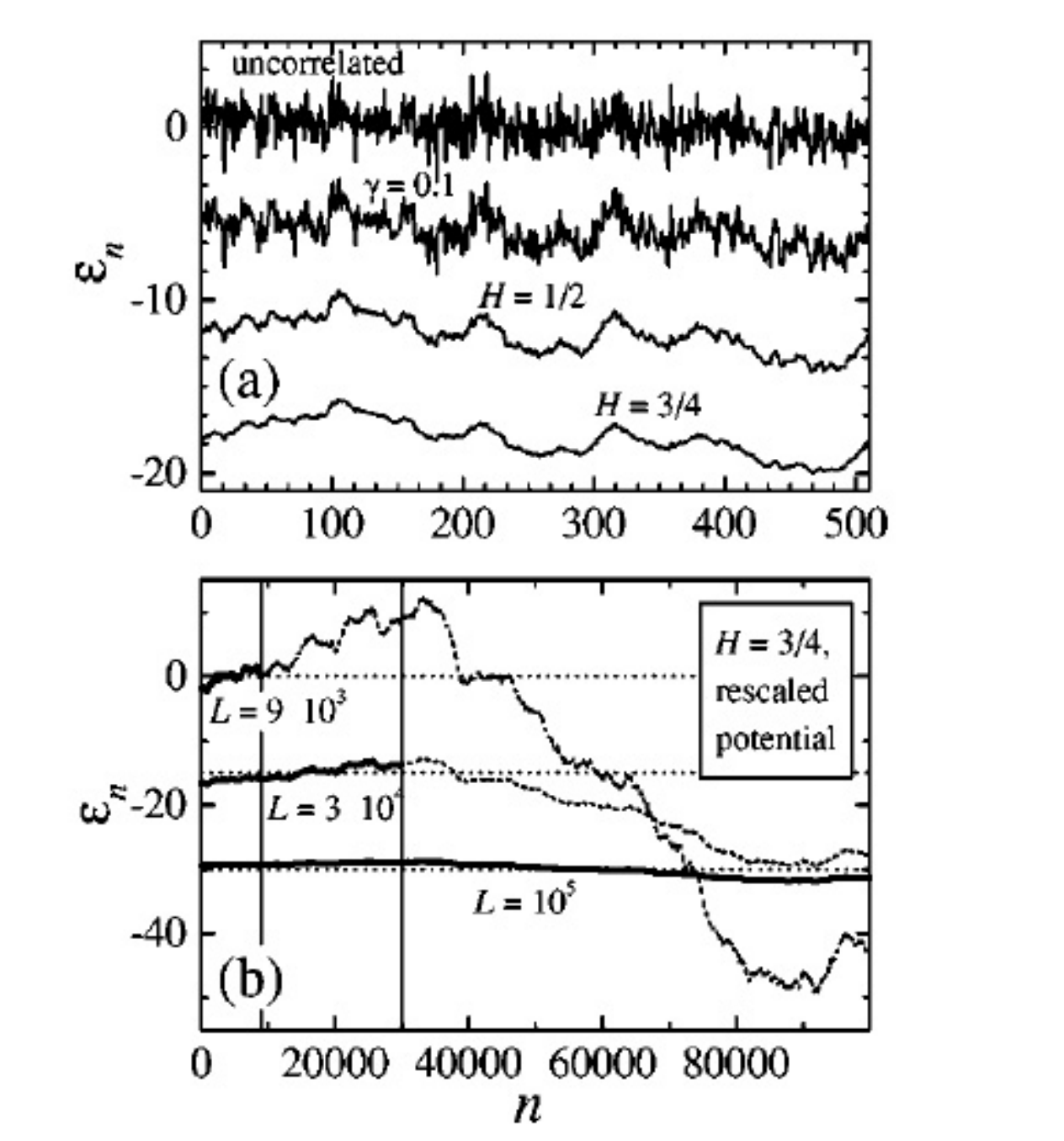}
\caption[cc]{Examples of profiles of sequences $\epsilon_n$. In (a) four types of potentials are shown: Uncorrelated random potential (top curve), correlated potential for which the binary correlator $\langle \epsilon_n \epsilon_{n+l} \rangle$ decays as $l^{-\gamma}$ with $\gamma = 0.1$, and self-affine potentials with $H=1/2$ and $3/4$. The correlated potentials are shifted by integer multiples of 6. In (b) the curves show the same potential landscapes rescaled in such a way that the variance is fixed, $\sigma=1$, for all systems sizes used here ($L \equiv N=9\times 10^3, 3 \times 10^4$, and $10^5$) (after \cite{RKBH01}).} \label{normalization}
\end{center}
\end{figure}

Thus, unlike the convolution algorithm described in Section~\ref{5} that generates stationary correlated sequences, the potential (\ref{self on site}) belongs to a statistical ensemble of {\it non-stationary} sequences. An infinite increase of fluctuations with the length $N$ leads to a confinement of quantum particle between numerous peaks of random potential. Obviously, the nature of such strongly localized states is different from that known for the states localized due to a multiple coherent backscattering process resulting in the Anderson localization, see discussion in Refs.~\cite{RKBH01,K07}.

Since the fluctuations in a self-affine potential increase with the system size, it was proposed \cite{ML98} to introduce a length-dependent normalization factor $C_{\alpha}(N)$ into Eq.~(\ref{self on site}) which normalizes the fluctuations to a finite value,
\begin{equation}
\label{variance1}
\langle \epsilon_{n}^2 \rangle = \sigma^2.
\end{equation}
Taking into account the scaling law of the fractional Brownian motion Eq.~(\ref{fbm}), one obtains that the normalization factor is $C_{\alpha}(N) \propto N^{(1- \alpha)/2} \propto 1/N^H$ \cite{Ko00,RKBH01}. It is clear that due to the rescaling $\epsilon_n \rightarrow \epsilon_n/N^H$ the normalized values of the on-site potential are strongly reduced with the length of the sequence. Various types of potentials are shown in Fig.~\ref{normalization}. It is obvious that the self-affine potentials normalized to the fixed variance,  become weaker and smoother with an increase of length $N$. Nevertheless, the global measure $\sigma$ of disorder which is proportional to the width of the distribution of $\epsilon_n$ remains fixed. This allows to study the effect of correlations on the localization-delocalization transition in the thermodynamic limit ($N \rightarrow \infty$, see Ref.~\cite{ML00}).

The Lyapunov exponent $\lambda(E) = L_{loc}^{-1}(E)$ for distinct values of $\alpha$ was numerically calculated in Ref.~\cite{ML98}, see Fig.~\ref{Lyap self}. The data manifest that all the states remain localized for $\alpha <2$ with the localization length decreasing when approaching the band edges. For the critical value $\alpha = 2$ a delocalized state emerges at the band center $E=0$. Here the Lyapunov exponent vanishes quadratically, $\lambda(E) \propto E^2$. For $\alpha >2$ the Lyapunov exponent vanishes within a finite interval of energies located near the band center, indicating the presence of delocalized states. The energy region with delocalized states becomes wider with an increase of $\alpha$, and the width is practically saturated at $\alpha= 5$ \cite{ML98}. The positions of the mobility edges turn out to be non-symmetric with respect to the band center. This asymmetry is attributed to a particular type of long-range correlations in the sequence Eq.~(\ref{self on site}), see Ref.~\cite{ML98} for detail.
\clearpage
\begin{figure}[ht]
\begin{center}
\includegraphics[width = 10cm]{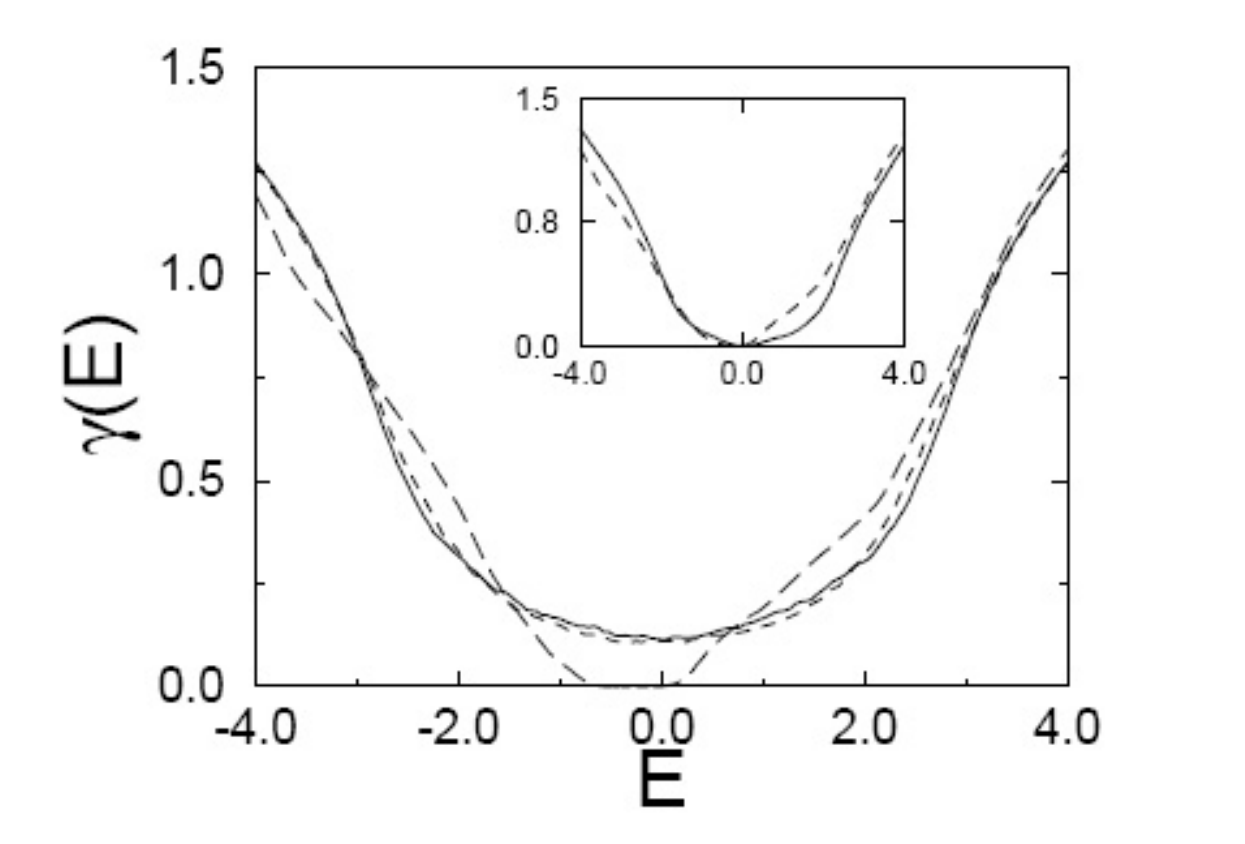}
\caption[cc]{Lyapunov exponent vs energy for the self-affine disorder of length $N = 10^4$. For a white noise potential with $\alpha =0$ the result is shown by solid curve: all the states are localized. Dashed curve corresponds to $\alpha =0$ in Eq.~(\ref{self on site}) with an additional random factor in the amplitudes of plane waves. In contrast, a finite-width band of extended states near $E=0$ emerges for $\alpha = 2.5$ (long-dashed curve). The result for the critical value $\alpha = 2.0$ is shown in inset. In this case the Lyapunov exponent vanishes only at $E=0$ (after \cite{ML98}).} \label{Lyap self}
\end{center}
\end{figure}

The data shown in Fig.~\ref{Lyap self} are calculated for a fixed value $\sigma$ of disorder. For $\alpha < 2$ all states are localized independently of the value of disorder. However, for $\alpha >2$ the localization-delocalization transition may occur also when changing the width $2W$ of the distribution of $\epsilon_n$ defined as $2W =2\sqrt{6} \sigma$. The critical width turns out to be an integer number $W_c = 4$ \cite{SNN04}. This value remarkably coincides with the critical amplitude of disorder obtained for the tight-binding model of random dimers \cite{DWP90} and for the Harper model \cite{H55}. In the latter case the on-site energies are generated from the incommensurate potential
\begin{equation}
\epsilon_n = \frac{W}{2} \cos(2 \pi n \varsigma)\,
\label{harper}
\end{equation}
with an irrational number $\varsigma$. All eigenstates become critical exactly at $W= W_c = 4$ \cite{AA80}. The phase diagram for the eigenstates of the self-affine potential in the $(W,\alpha)$ plane was numerically calculated in Ref.~\cite{SNN04}, see Fig.~\ref{phase}. It is clearly seen that the mixed phase of extended and localized states is separated from the phase of pure localized states by the vertical and horizontal lines. Therefore, the critical parameters characterizing the correlations and the amplitude of disorder are completely independent.

\begin{figure}[ht]
\begin{center}
\includegraphics[width = 8cm]{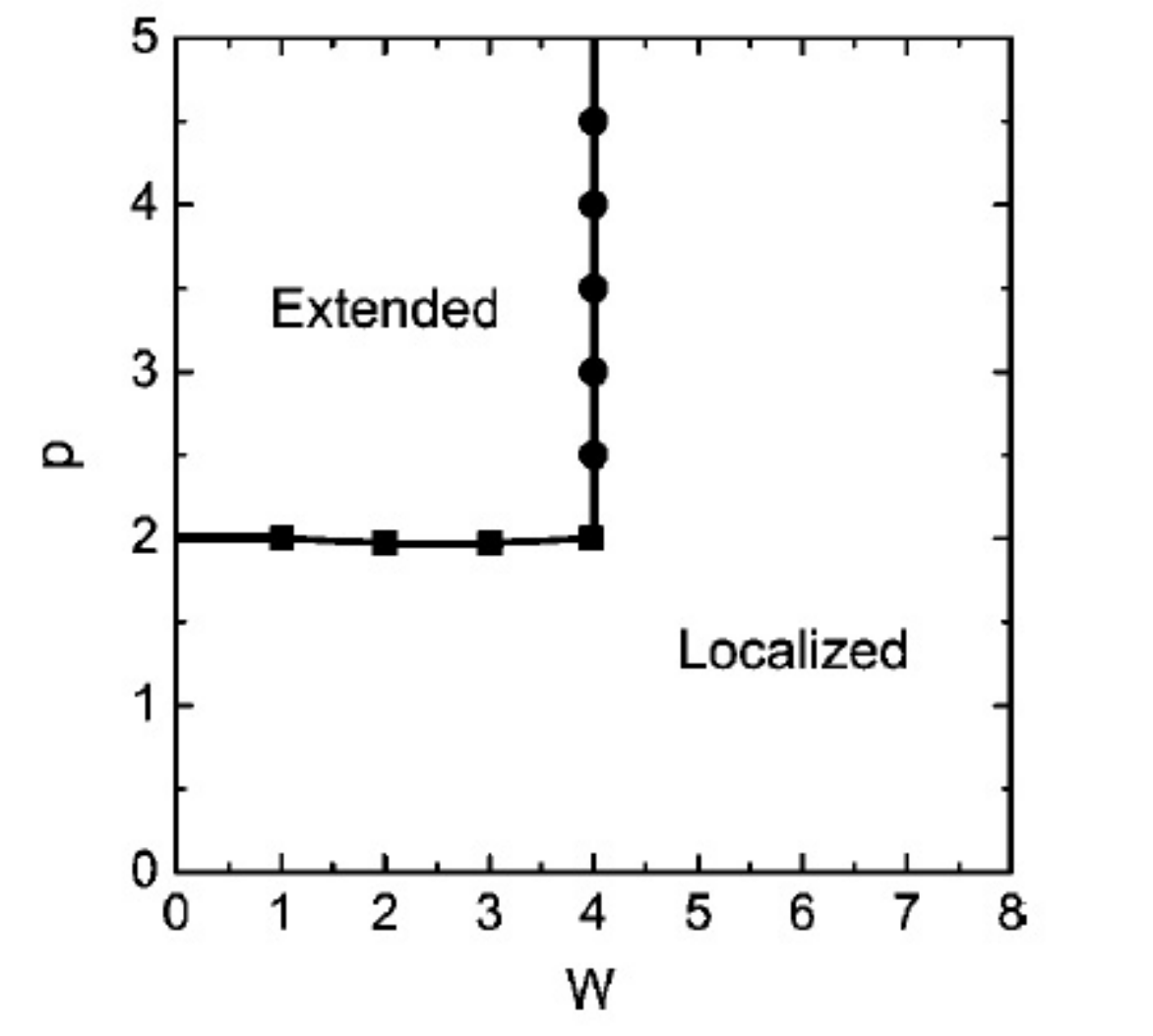}
\caption[cc]{Phase diagram for the self-affine potential with $p\equiv \alpha$ in Eq.~(\ref{self on site}), see details in the text (after \cite{SNN04}).} \label{phase}
\end{center}
\end{figure}

\begin{figure}[ht]
\begin{center}
\includegraphics[width = 8cm]{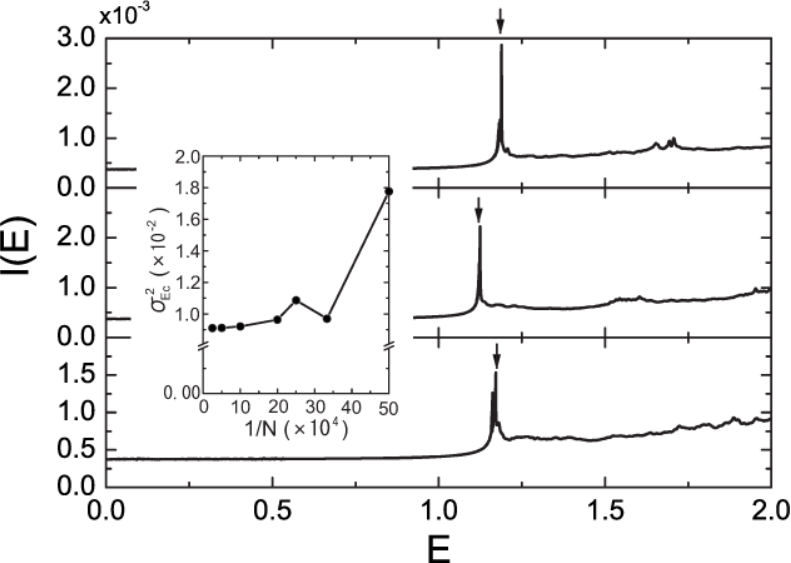}
\caption[cc]{Energy dependence of the inverse participation ratio for three different realizations of self-affine potential (\ref{self on site}) with $\alpha = 4$ and $\sigma= \sqrt2$. Peaks indicated by arrows are the mobility edges. Extended (localized) states located to the left (right) of the peaks. The inset shows the variance of the position of the peaks with the inverse length $1/N$ of the sequence. Extrapolation of the curve to $1/N =0$ manifests non-vanishing realization-to-realization fluctuations (after \cite{NYS09}).} \label{inv}
\end{center}
\end{figure}

Since self-affine potentials exhibit large fluctuations when changing the realization of a potential, apart from the $l$-dependence in Eq.~\ref{fbm}, some physical quantities are not self-averaged. For these quantities realization-to-realization fluctuations may survive in the thermodynamic limit. In particular, in Ref.~\cite{NYS09} it was demonstrated that the position of the mobility edges (similar to that shown in Fig.~\ref{Lyap self} in the region $E>0$ ) exhibits non-vanishing fluctuations in the thermodynamic limit. The position of the mobility edge was obtained not from the calculations of the Lyapunov exponent but from the sharp peak in the plot of the inverse participation ratio $I(N,E)$,
\begin{equation}
\label{inverse}
I(N,E)= \sum_{n=1}^N \mid \psi_n(E) \mid^4,
\end{equation}
where $\psi_n(E)$ is the amplitude of the wave function at the $n$th site. The dependence $I(N,E)$ is shown in Fig.~\ref{inv} for a sequence of length $N=4000$.

Note that the positions of the peak fluctuate between $E=1$ and $E=1.25$ from sample to sample. A mobility edge in the same interval for a self-similar potential with equal parameters was obtained also in Ref.~\cite{RKBH01}. Whereas these fluctuations decrease with the sample size $N$, they tend to the saturation to a non-zero level for $N \rightarrow \infty$. A similar lack of self-averaging and size-dependence was also observed for the energy level-spacing statistics \cite{NYS09,CBI04}.

The absence of the self-averaging leads to qualitatively different graphs for $I(N,E)$ obtained for individual sequences of length $N$, and for the quantity $\langle I(N,E) \rangle$ obtained as a result of averaging over the statistical ensemble. The examples of the energy dependence of $\langle I(N,E)\rangle$ is shown in Fig.~\ref{average}.

\begin{figure}[ht]
\begin{center}
\includegraphics[width = 8cm]{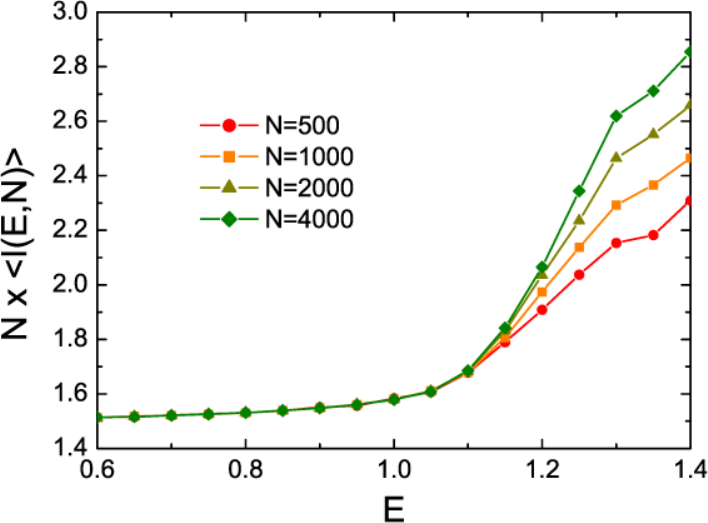}
\caption[cc]{(Color online) Normalized to $N$ inverse participation ratio $N\langle(N,E)\rangle$, averaged over 100 realizations of sequences $\epsilon_n$ of different length (after \cite{NYS09}).} \label{average}
\end{center}
\end{figure}

Due to non-vanishing realization-to-realization fluctuations of the peak (see Fig.~\ref{inv}), the average quantity $\langle I(N,E)\rangle$ does not exhibit a peak. Instead of the peak, there is a smooth crossover behavior of the inverse participation ratio. Since the region of the crossover ($0.9<E<1.1$) as well as the region of the extended states, is independent of the system length, the crossover behavior survives in the thermodynamic limit. The region of localized states ($E>1.1$) exhibits a strong size-effect which is due to large-scale fluctuations. Indeed, the dependence of the sample size $N$ enters explicitly in the definition of self-affine potential Eq.~(\ref{self on site}) through the normalization factor $C_{\alpha}(N)$ added in the considered case. This normalization condition, which is imposed in order to restrict an infinite growth of fluctuations with system size, makes normalized self-affine potentials irrelevant to realistic disordered systems \cite{NYS09}.

\section{Bloch-like oscillations in disordered potentials}
\label{16}

The Bloch oscillations were originally predicted to exist in a periodic lattice in the presence of a dc electric field ${\cal E}$, see \cite{B28,Z34}. Here we briefly discuss the properties of Bloch-like oscillations in a model with correlated potentials giving rise to finite energy bands with extended eigenstates. The model we consider here is the tight-binding 1D Anderson model with a constant electric field ${\cal E}$. Thus, the non-stationary Schrodinger equation takes the following form \cite{DK86,DMML03},
\begin{equation}
\label{dc Schrod}
i\dot{\psi}_n =( \epsilon_n-Fn) \psi_n - \psi_{n+1} - \psi_{n-1}.
\end{equation}
Here the units are used in which the parameter $F$ is $F=e {\cal E} d/\vartheta$ with $e$ as the electron charge and $d$ as the lattice period. The energy is measured in units of the off-diagonal matrix element $\vartheta$, see Eq.~(\ref{disShr}), and time is given in units of the Plank constant $\hbar$.
If the lattice is periodic (with no disorder), an electron which is initially at rest is accelerated and within the allowed energy band its Bloch momentum grows linearly with time, $k(t) = Ft$. At the edge $k = \pi/d $ of the Brillouin zone the electron energy reaches its maximum value at which the group velocity vanishes. Here the electron momentum changes its direction due to Bragg reflection and begins to move  against the electric field to its starting point at the center of the Brillouin zone. After, the process repeats with a period $2 \pi/Fd$. These oscillations are known as the Bloch oscillations (see modern theory in Ref.~\cite{HKKM04}). Since the energy $Fn$ gained from the electric field cannot exceed the width of the allowed energy band $2\vartheta$, the amplitude of the Bloch oscillations in real space is $2\vartheta/F$.

For the lattice model with an external dc electric field there is some similarity between the Bloch states in a periodic potential and extended states in disordered potentials. In both cases an electron gains an energy from the applied electric field ${\cal E}$. The difference is that a Bloch electron is accelerated until its energy reaches the top of the allowed band, however, an electron in a random potential can increase its energy while it is within the energy region of extended states, if such a region exists. Once its energy approaches a mobility edge, the probability of backscattering (which is the source of Anderson localization) strongly increases. Since the scattering is elastic, it excites the lattice eigenstate with the same energy but opposite velocity. After the backscattering the electron is de-accelerated by the electric field to a halt. This process is then repeated with statistical fluctuations caused by a weak disorder.

The above scenario suggests that the oscillations similar to the Bloch oscillations are also allowed in a disordered system if there is a continuous band of extended states \cite{DMML03}. The amplitude of these Bloch-like oscillations is proportional to the bandwidth which in this case is the interval of energies between two mobility edges. Since in a disordered one-dimensional system a continuous band of extended states requires long-range correlations, in Ref.~\cite{DMML03} it was suggested to study the Bloch-like oscillations with the use of self-affine potentials with $\alpha = 3$ in Eq.~(\ref{self on site}), provided it is normalized as discussed above. In this case there are two mobility edges located near the band center $E=0$, similar to those shown in Fig.~\ref{Lyap self} for $\alpha = 2.5$.

\begin{figure}[!hb]
\begin{center}
\includegraphics[width = 8cm]{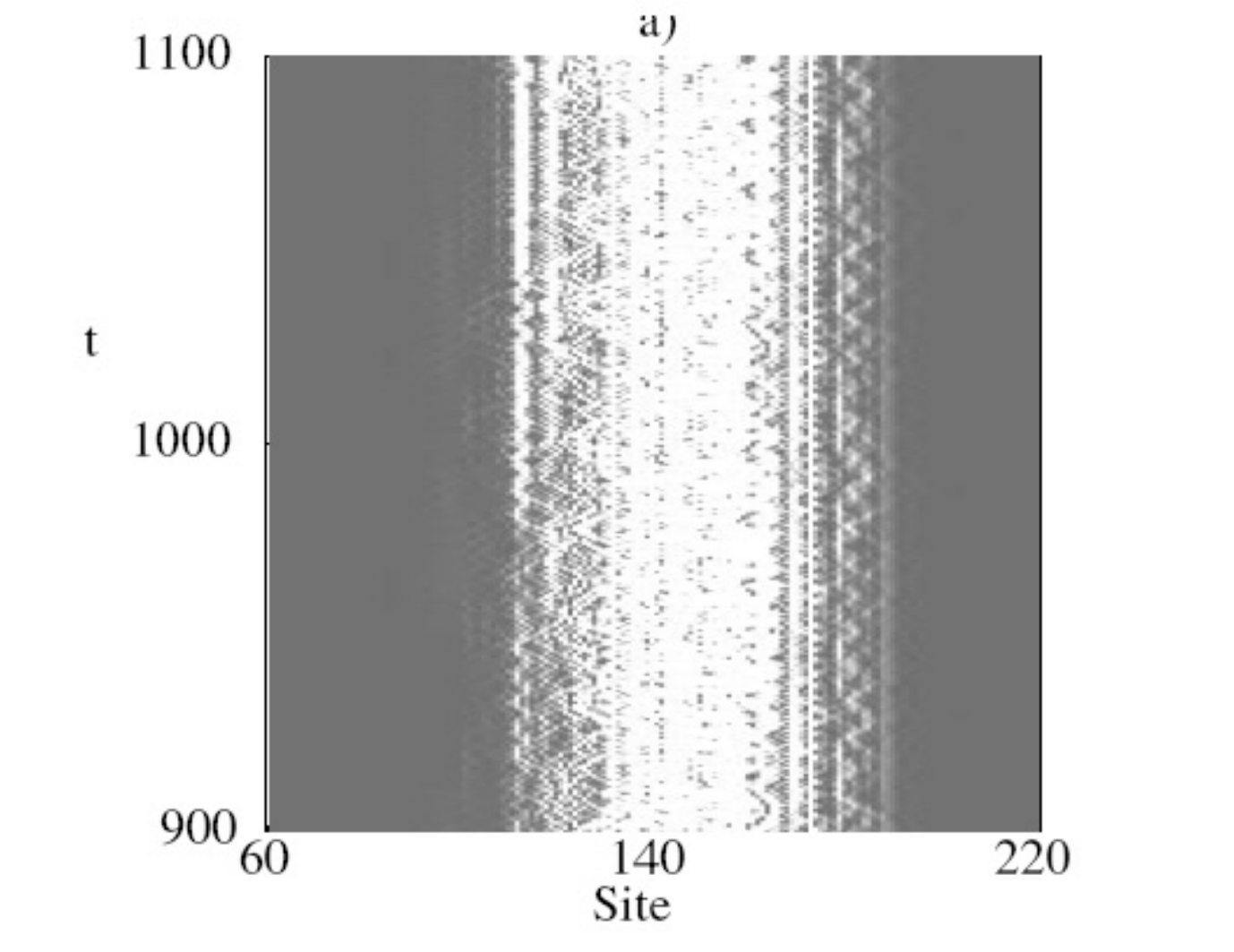}
\includegraphics[width = 7.5cm]{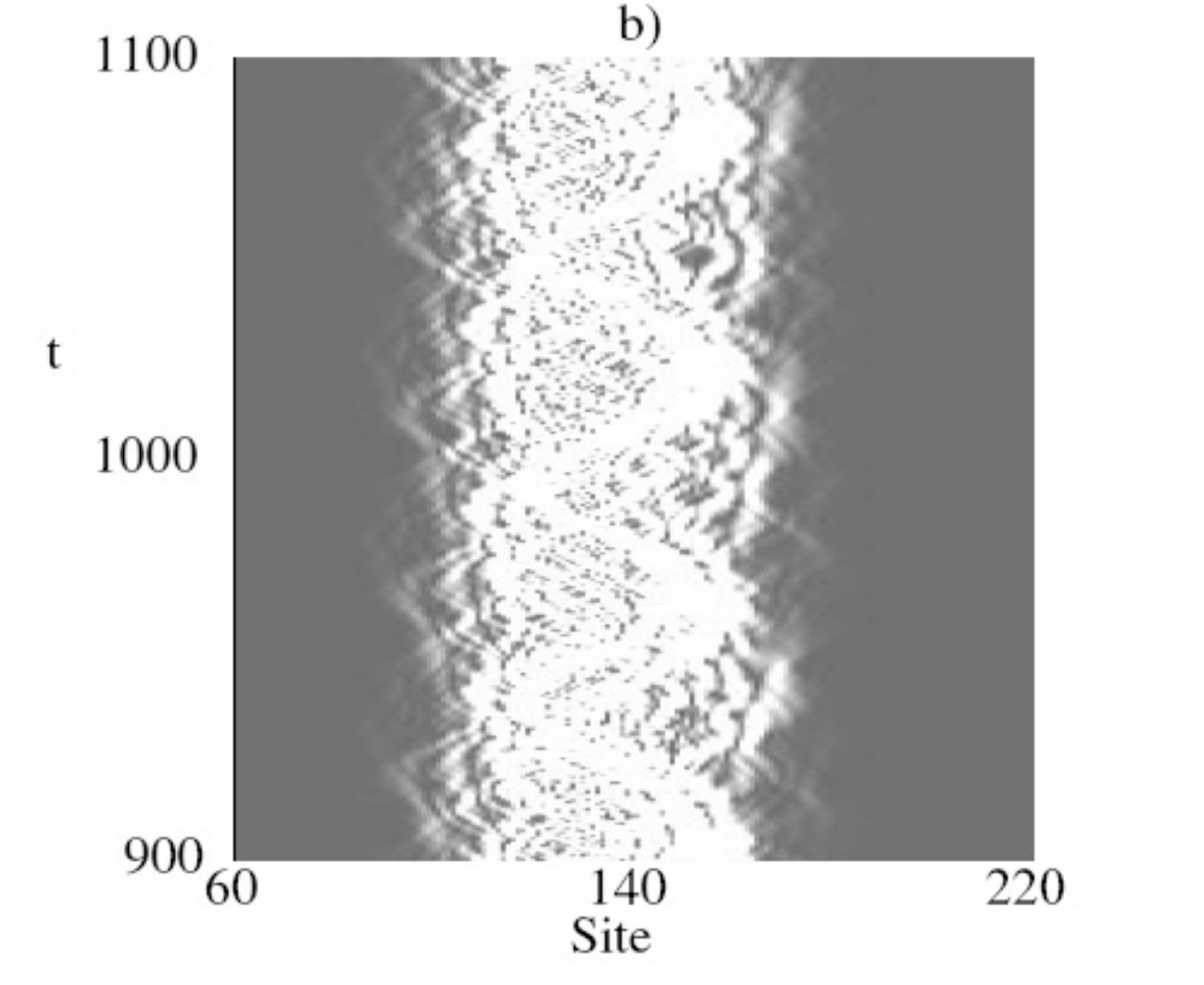}
\caption[cc]{Numerical solution of Eq.~(\ref{dc Schrod}) with the initial condition (\ref{initial}) for $\Delta = 10, F=0.1$ and $N=300$. Higher values of $\psi_n(t)$ are shown by lighter regions. (a) For $\alpha = 0.5$ the phase of extended states is absent and the electric field results in small fluctuations in the initial distribution. (b) For $\alpha = 3$ the probability amplitude oscillates with finite amplitude around the point which is slightly shifted from the initial position at $N=150$. The shift is towards the direction of the force acting on a negative charge (after \cite{DMML03}).} \label{psi}
\end{center}
\end{figure}

Let us assume that initially the electron is represented by a Gaussian wave packet,
\begin{equation}
\label{initial}
\psi_n(t=0) = A \exp{[-(n-n_0)^2/4 \Delta^2]}\,,
\end{equation}
centered at the middle of the sample, for $n_0 = N/2$ . Here $A$ is the normalization factor and $\Delta$ is the width of the packet. With the initial condition (\ref{initial}) the Schr\"odinger equation (\ref{dc Schrod}) was numerically solved for a sample of length $N$ for self-affine sequences (\ref{self on site}) with various values of $\alpha$ \cite{DMML03}.
The evolution of the wave function after the transient period $t \gg 2\pi/\omega_F$ is shown in Fig.~{\ref{psi} for $F= 0.1$ and $\Delta=10$. The pattern in Fig.~\ref{psi}a was obtained for $\alpha = 0.5$ for which all the states are localized. In this case there is no macroscopic electron dynamics: the distribution of the amplitude of the electron density slightly fluctuates in time. On the other hand, for the potential with $\alpha = 3$ the extended states form a continuous band and the time evolution in the form of spatial oscillations is well seen in Fig.~\ref{psi}b.

The more detailed information about the Bloch-like oscillations can be extracted from the temporal evolution of the center of the wave packet,
\begin{equation}
\label{x(t)}
x(t) = \sum_{n=1}^N {(n-n_0) \mid \psi_n(t) \mid^2}.
\end{equation}
The trajectory of the centroid is shown in Fig.~\ref{traj} for two values of the electric field. At the initial stage the amplitude of oscillations is close to $L_F = \delta E_{0}/F$, where $\delta E_{0}=4$ is the bandwidth for a periodic lattice. After a transient time the amplitude is reduced to the value $L_F = \delta E/F$. Here $\delta E \sim 1$ is the interval between the mobility edges, i.e. the bandwidth of the phase of the extended states, see Fig.~\ref{Lyap self}.

\begin{figure}[!hb]
\begin{center}
\includegraphics[width = 8cm]{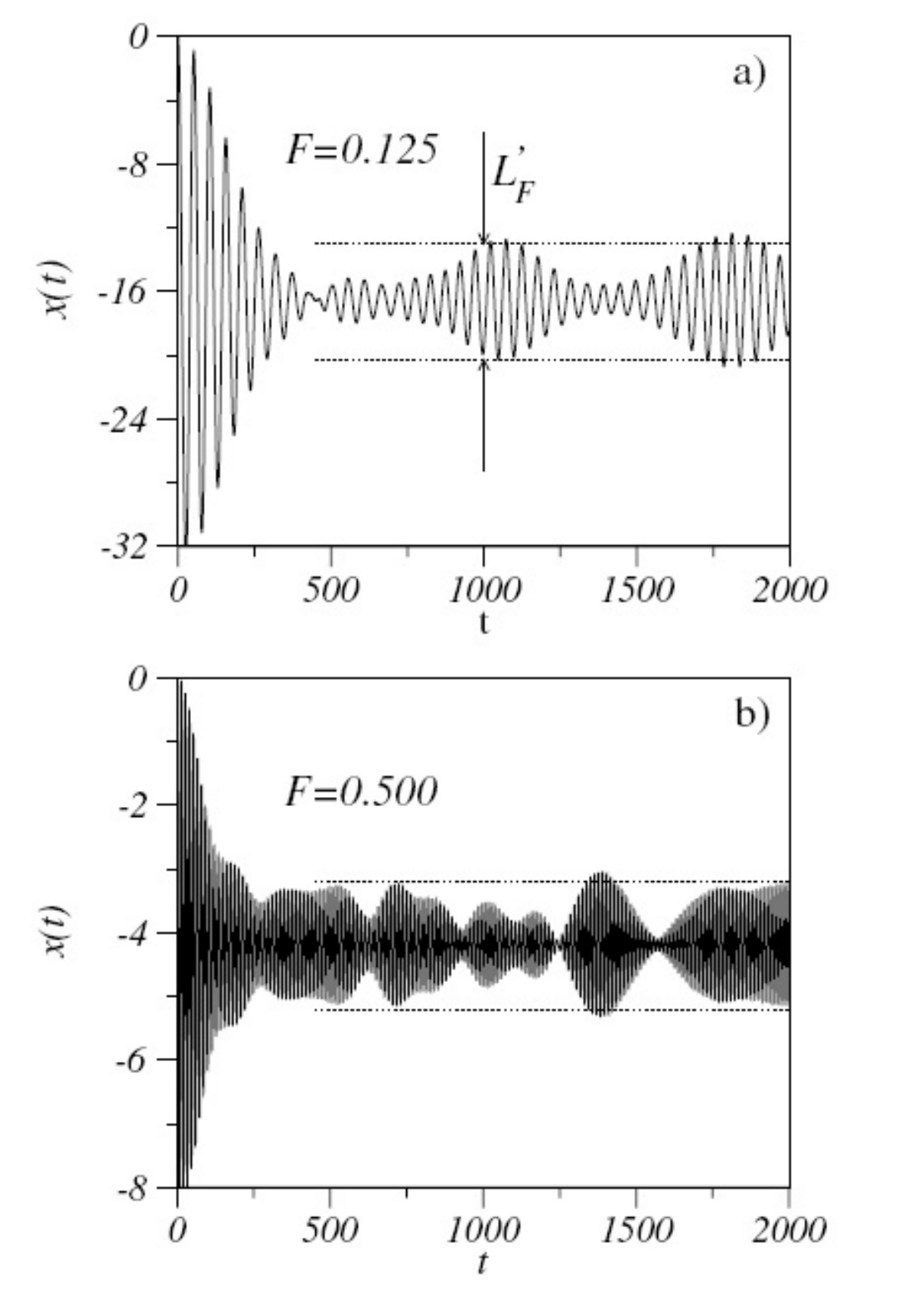}
\caption[cc]{Two trajectories of the centroid of the wave packet with $\Delta =10$ in a lattice of length $N=1000$. For both cases the same self-affine potential with $\alpha = 3$ was generated. Horizontal dotted lines are the boundaries of the region of the Bloch-like oscillations (after \cite{DMML03}).} \label{traj}
\end{center}
\end{figure}

\begin{figure}[ht]
\begin{center}
\includegraphics[width = 8cm]{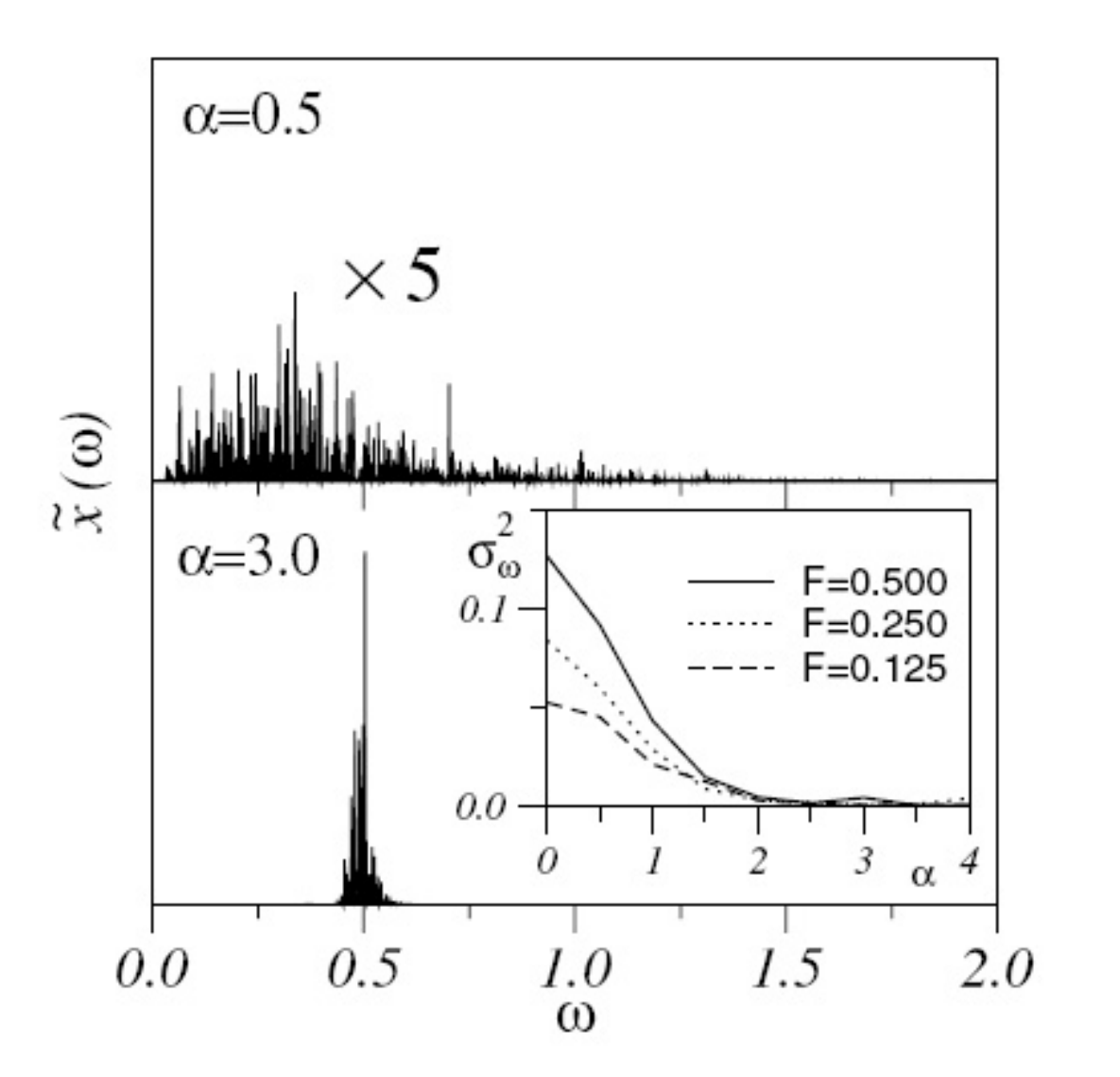}
\caption[cc]{Fourier spectrum $x(\omega)$ of the centroid (\ref{x(t)}) in a lattice with $N=500$. Inset: variance $\sigma_{\omega}^2=\langle x^2(\omega)\rangle$ vs $\alpha$ for different strengths of electric field, with averaging over statistical ensemble of 1000 realizations of the self-similar potentials (after \cite{DMML03}).} \label{self spectrum}
\end{center}
\end{figure}

The data in Fig.~\ref{traj} confirm that the amplitude of the Bloch-like oscillations follows the same scaling, $L_F \sim 1/F$, with electric field as it does in a periodic lattice. The amplitude of oscillations is modulated irregularly which is a consequence of the stochastic nature of the backscattering processes. For each realization of the self-affine potential the modulation pattern looks different but the amplitude remains the same. Due to this invariance the interval between the mobility edges can be evaluated if the amplitude is known, $\delta E = F L_F$.

Because of an irregular modulation of the amplitude, the spectrum of the Bloch-like oscillations contains a quasi-continuous band of frequencies. The center of this band is at $\omega_F = F$, in accordance with the relation known for a periodic lattice. The Fourier transform of the oscillations is shown in Fig.~\ref{self spectrum} for a wave packet with $\Delta =10$ and electric field strength $F=0.5$.
The upper panel shows the spectrum for $\alpha = 0.5$ when there are no extended states in the lattice. A broad spectrum and small amplitude of each component (notice the magnification factor of 5) is an evidence that the wave packet does not move as a whole but fluctuates near its equilibrium position. Unlike this, for $\alpha =3$ the spectrum is much narrower with a sharp peak at $\omega =\omega_F$. Each realization of the self-affine potential (with the same $\alpha$) has its own Fourier spectrum $x(\omega)$. The variance $\sigma_{\omega}^2=\langle x^2(\omega)\rangle$ characterizes the width of the Fourier spectrum. The inset shows that the width sharply drops starting from $\alpha =2$, i.e. from the critical value corresponding to an emergence of extended states. This drop is one more evidence of the Bloch-like oscillations arising at the same value $\alpha =2$ as in the periodic lattice independently of the value of electric field $F$.

It is obvious that the Bloch-like oscillations may occur not only in self-affine potentials but in any disordered system where a phase of delocalized states is separated from the localized states by two mobility edges. In Ref.~\cite{MLDM05} such oscillations were numerically simulated for the tight-binding model with a generalized Harper potential $\epsilon_n = (E_0/2) \cos(2 \pi \omega n^{\nu} )$. This incommensurate potential has a band of extended states between two mobility edges at $E= \pm(2-E_0)$, if $0< \nu < 1$ and $0<E_0<2$ \cite{DHX88,DHX90}. The Bloch-like oscillations in aperiodic potentials, being more regular, reveal a strong similarity with the oscillations in the disordered potentials.

\section{Acknowledgements}

The authors are thankful to their collaborators with whom some of the results discussed in this review have been obtained: V.M.K. Bagci, E. Diez, V. Dossetti-Romero, A. Rodriguez, J.C. Hernandez-Herrejon, I.F. Herrera-Gonzalez, T. Kottos, U. Kuhl, G.A. Luna-Acosta, S. Ruffo, H.-J. St\"{o}ckmann, L. Tessieri, E.J. Torres-Herrera, G.P. Tsironis, S.E. Ulloa, O.V. Usatenko, and V.A. Yampol'skii.

F.M.I is thankful to B.L. Altshuler, F. Borgonovi, H. Cao, A. Crisanti, L.I. Deych, M. Dykman, V.V. Flambaum, V.F. Gantmakher, S.A. Gredeskul, I. Guarneri, I.V. Lerner, A.A. Lisyansky, G. Modugno, B.I. Shklovskii,  G. Shlyapnikov, K.M. Slevin, V.V. Sokolov, and I. Varga for fruitful discussions.

This work is supported by Universidad Aut\'onoma de Puebla, grant EXC08-G and by the US Department of Energy, grant \# DE-FG02-06ER46312.




\end{document}